\documentclass[12pt,a4paper,twoside,sort&compress]{book}

\usepackage{epsfig}
\usepackage[skins]{tcolorbox}
\usepackage{slashed,verbatim,subfigure}
\usepackage[colorlinks=true,linktocpage=true,
linkcolor=blue,citecolor=blue]{hyperref}
\usepackage[a4paper]{geometry}
\usepackage{amsfonts}
\usepackage{amssymb}
\usepackage{amsmath}
\usepackage{setspace}
\usepackage{lscape}
\usepackage[latin1]{inputenc}
\usepackage[TS1,T1]{fontenc}
\usepackage{lmodern}
\usepackage{amsthm}
\usepackage{latexsym}
\usepackage{psfrag}
\usepackage{graphicx}
\usepackage{rotating}
\usepackage{graphics}
\usepackage{subfigure}
\usepackage{mathenv}
\usepackage{amsthm}
\usepackage[titletoc]{appendix}
\usepackage{etoolbox}
\usepackage{tikz}
\newrobustcmd*{\mysquare}[1]{\tikz{\filldraw[draw=#1,fill=#1] (0,0)
rectangle (0.2cm,0.2cm);}}
\newrobustcmd*{\mycircle}[1]{\tikz{\filldraw[draw=#1,fill=#1] (0,0) circle [radius=0.1cm];}}
\newrobustcmd*{\mytriangle}[1]{\tikz{\filldraw[draw=#1,fill=#1] (0,0) --
(0.2cm,0) -- (0.1cm,0.2cm);}}
\usepackage{amssymb}
\usepackage{amsmath}
\usepackage{tcolorbox}
\usepackage{graphics}
\usepackage[final]{pdfpages}
\usepackage{slashed}
\usepackage{amssymb}
\usepackage{epstopdf}

\newcommand{\ignore}[1]{}
\usepackage[latin1]{inputenc}
\usepackage{cite}
\usepackage{subfigure}
\usepackage[small,bf]{caption}
\usepackage{color}
\usepackage[Conny]{fncychap}
\ChNameVar{\LARGE\sffamily\bfseries} \ChNumVar{\Huge} \ChTitleVar{\Huge\sffamily\bfseries}
\ChRuleWidth{0.85pt} \ChNameUpperCase \ChTitleUpperCase %\ChTitleUpperCase
\usepackage{fancyhdr}
\usepackage{amssymb}

\catcode`@=11
\def\seceqaa{\@addtoreset{equation}{section}
           \def\theequation{A\arabic{equation}}}
\def\seceqbb{\@addtoreset{equation}{section}
           \def\theequation{B\arabic{equation}}}
\def\seceqcc{\@addtoreset{equation}{section}
           \def\theequation{C\arabic{equation}}}
\def\seceqdd{\@addtoreset{equation}{section}
           \def\theequation{D\arabic{equation}}}
\def\seceqee{\@addtoreset{equation}{section}
           \def\theequation{E\arabic{equation}}}
\def\seceqff{\@addtoreset{equation}{section}
           \def\theequation{F\arabic{equation}}}
\def\seceqgg{\@addtoreset{equation}{section}
           \def\theequation{G\arabic{equation}}}
\def\seceqhh{\@addtoreset{equation}{section}
           \def\theequation{H\arabic{equation}}}
\def\seceqjj{\@addtoreset{equation}{section}
           \def\theequation{J\arabic{equation}}}
\def\seceqll{\@addtoreset{equation}{section}
           \def\theequation{L\arabic{equation}}}
\catcode`@=11

\begin{document}
\frontmatter
\pagestyle{plain}
%\input{decl}
    %%% Fancy Header %%%%%%%%%%%%%%%%%%%%%%%%%%%%%%%%%%%%%%%%%%%%%%%%%%%%%%%%%%%%%%%%%%
    % Fancy Header Style Options
    \pagestyle{fancy}                       % Sets fancy header and footer
    \fancyfoot{}                            % Delete current footer settings
    \renewcommand{\chaptermark}[1]{         % Lower Case Chapter marker style
      \markboth{\chaptername\ \thechapter.\ #1}{}} %
    \renewcommand{\sectionmark}[1]{         % Lower case Section marker style
      \markright{\thesection.\ #1}}         %
    \fancyhead[LE,RO]{\bfseries\thepage}    % Page number (boldface) in left on even
                                            % pages and right on odd pages
    \fancyhead[RE]{\bfseries\leftmark}      % Chapter in the right on even pages
    \fancyhead[LO]{\bfseries\rightmark}     % Section in the left on odd pages
    \renewcommand{\headrulewidth}{0.3pt}    % Width of head rule
    %%% Clear Header %%%%%%%%%%%%%%%%%%%%%%%%%%%%%%%%%%%%%%%%%%%%%%%%%%%%%%%%%%%%%%%%%%
    % Clear Header Style on the Last Empty Odd pages
    \makeatletter
    \def\cleardoublepage{\clearpage\if@twoside \ifodd\c@page\else%
        \hbox{}%
        \thispagestyle{empty}%              % Empty header styles
        \newpage%
        \if@twocolumn\hbox{}\newpage\fi\fi\fi}
    \makeatother
    %%%%%%%%%%%%%%%%%%%%%%%%%%%%%%%%%%%%%%%%%%%%%%%%%%%%%%%%%%%%%%%%%%%%%%%%%%%%%%%
    %\addcontentsline{toc}{chapter}{CANDIDATES DECLARATION}
  % \clearpage
 
\newpage
\pagestyle{plain}
%\chapter*{COVERPAGE}
\begin{center}
{\Large \bf Aspects of Thermal QCD Phenomenology at Intermediate Gauge/'t Hooft Coupling from String/M-Theory, (HD) Gravity Islands, and Multiverse\footnote{Based on author's Ph.D thesis successfully defended on October 05, 2023}}
\end{center}
\begin{center}
Gopal Yadav\footnote{email: gopalyadav@cmi.ac.in, gyadav@ph.iitr.ac.in}
\end{center}
\begin{center}
{Department of Physics, Indian Institute of Technology, Roorkee- 247 667,\\
Uttarakhand, India\\
Chennai Mathematical Institute,
SIPCOT IT Park, Siruseri 603103, India}
\end{center}

{\bf Abstract:} The holographic dual of thermal QCD-like theories at intermediate coupling was constructed in \cite{HD-MQGP} by including the ${\cal O}(R^4)$ terms in the eleven dimensional supergravity action. First part of the thesis explored the application of \cite{HD-MQGP} and the following issues have been addressed in first part of this thesis.

In \cite{MChPT}, we have computed the low energy coupling constants (LECs) of $SU(3)$ chiral perturbation theory in the chiral limit from the type IIA string dual inclusive of ${\cal O}(R^4)$ corrections. We matched our results with the phenomenological data and found the connection between higher derivative terms and large-$N$ expansion.

 In \cite{McTEQ}, we computed the deconfinement temperature of thermal QCD-like theories at intermediate coupling from ${\cal M}$ theory dual inclusive of ${\cal O}(R^4)$ corrections using Witten's prescription \cite{Witten-Hawking-Page-Tc} in the absence of rotation. In this process, we observed a novel  ``UV-IR'' mixing, ``Flavor Memory'' effect (memory of flavor $D7$ branes considered in parent type IIB string dual was contained in the aforementioned no-braner ${\cal M}$-theory uplift of large-$N$ thermal QCD) and non-renormalization of deconfinement temperature $(T_c)$ beyond one loop in the zero instanton sector from semiclassical computation and entanglement entropy points of view. We also discussed confinement-deconfinement phase transition in thermal QCD-like theories from the entanglement entropy point of view too, where the aforementioned sectors correspond to entanglement entropies of connected and disconnected RT surfaces in gravity dual. Further, we showed compatibility of the results obtained with  M$\chi$PT as obtained in \cite{MChPT}. Then in \cite{Rotation-Tc-M-Theory}, we constructed holographic dual of rotating quark-gluon plasma (QGP) by performing Lorentz transformations in the subspace of eleven-dimensional ${\cal M}$ theory metric on the gravity dual side. We computed the deconfinement temperature ($T_c$) of rotating QGP using semiclassical method \cite{Witten-Hawking-Page-Tc} and found that the deconfinement temperature of rotating QGP is inversely proportional to Lorentz factor, which implies that it decreases when $T_c$  increases and vice-versa. Further, ``UV-IR'' mixing, ``nonrenormanization of $T_c$'' and ``Flavor Memory'' effect were again observed in the holographic study of rotating QGP.

From the applications of the island proposal \cite{AMMZ}, doubly holographic setup \cite{PBD}, and wedge holography \cite{WH-1,WH-2}, we explored the following issues.

 In \cite{RNBH-HD}, we obtained the Page curves of Reissner-Nordstr\"om black hole in the presence of higher derivative terms $({\cal O}(R^2))$ in the gravitational action. We used Island proposal \cite{AMMZ} to do the same and found that Page curves are shifted towards later times or earlier times when Gauss-Bonnet coupling increases or decreases. Scrambling time is affected when we considered general ${\cal O}(R^2)$ terms, whereas it was unaffected in Einstein-Maxwell-Gauss-Bonnet gravity. Based on these results, one can say that the ``dominance of the island'' is affected by the presence of higher derivative terms.
 
 In \cite{HD-Page Curve-2}, we constructed the doubly holographic setup for the non-conformal bath (thermal QCD) in ${\cal M}$-theory background with the inclusion of ${\cal O}(R^4)$ terms. We obtained the Page curve of an eternal neutral black hole by computing the entanglement entropies of Hartman-Maldacena-like and Island surfaces. Our results agree with area computation and Dong's formula for leading order terms in eleven-dimensional supergravity action. Apart from obtaining the Page curve, we showed that entanglement entropy of Hartman-Maldacena-like surface exhibits ``Swiss-Cheese'' structure at ${\cal O}(\beta^0), \beta\sim l_p^6$ in the large-$N$ scenario. We showed that no boundary terms would be required on ETW-``brane'' for higher derivative terms, and the tension of ETW-``brane'' turns out to be non-zero, and therefore it is possible to localize gravity on ETW-``brane'' for massless graviton in our setup. We explicitly showed that obtaining the Page curve of an eternal neutral black hole for massless graviton in our setup is possible. Further, we found a relationship between eleven-dimensional Planckian length and black hole horizon radius to get rid of large-$N$ and IR-enhancement appearing in the ratio of Island surface entanglement entropy and black hole thermal entropy.

 We obtained the Page curve of the Schwarzschild de-Sitter (SdS) black hole using the island proposal from a non-holographic approach in \cite{Gopal+Nitin} for the black hole and de-Sitter patches separately. This was possible because of the insertion of thermal opaque membranes on both sides of the black hole and de-Sitter patches. We obtained the Page curve of the black hole patch by plotting the entanglement entropy of Hawking radiation in the absence and presence of the island surface. In the case of the SdS black hole, we found that island is located inside the black hole in contrast to the universal result that island is located outside the black hole. We also discussed that ``dominance of islands'' and ``information recovery'' takes more time for the low-temperature black hole patch compared to the high-temperature black hole patch.

 We described the Multiverse in \cite{Multiverse} from the application of wedge holography. Multiverse is described as the setup of $2n$ anti de-Sitter (AdS) or de-Sitter branes embedded in the corresponding bulk. Due to different bulks for the AdS and de-Sitter branes, we can't have the Multiverse made up of AdS and de-Sitter branes both. Since all the universes can communicate with each other due to transparent boundary conditions at the defect, one could resolve the ``grandfather paradox'' in this setup. This setup is also used to obtain the Page curve of the Schwarzschild de-Sitter black hole, which has two horizons.

\addcontentsline{toc}{chapter}{Abstract}
\clearpage
\addcontentsline{toc}{chapter}{Acknowledgements}
%Om%Namah%shivaya%Shree%ganeshaya%namah
%Jesus%christ%
%Jai%Mata%Di
\chapter*{Acknowledgments}
\vspace{-1in}
I am very grateful to my supervisor, {\it Prof. Aalok Misra}, for guiding me during my Ph.D. and making me what I am today. I enjoyed all the discussions that we have together. He trained me so that I can now do independent research. Apart from the research part, I learned many things from him, which helped me in the development of my personality in real life and as a researcher. I am very fortunate to have him as my supervisor. I am very thankful to my supervisor and his family, who always treated me like a family member, and for their encouragement, appreciation, and blessings. Without all these, my Ph.D. journey would not have been so smooth and pleasant.

I would like to thank the very humble members of my research committee known as ``Student Research Committee (SRC)'': {\it Prof. Tashi Nautiyal} (chairperson SRC), {\it Prof. Dibakar Roychowdhury} (internal member), {\it Prof. Sugata Gangopadhyay} (external member - Department of Computer Science and Engineering, I.I.T. Roorkee) and {\it Prof. Arnab Kundu} ({Saha Institute of Nuclear Physics (SINP), Kolkata, India}, an external member of me as a research fellow of CSIR), for their
appreciation, encouragement, valuable comments, questions and evaluating me time to time. I am very grateful to {\it Prof. Dibakar Roychowdhury} for the discussions related to the research and post-doctoral position. He always encouraged me not to get tired and never give up.

I would like to thank my collaborators, my supervisor, {\it Dr. Vikas Yadav} and {\it Mr. Nitin Joshi} (Indian Institute of Technology Ropar, Punjab, India) for working on the projects and publishing the papers with me. I would like to thank my colleagues from the string theory and high energy physics group {\it Shivam}, {\it Hemant}, {\it Jitendra}, {\it Salman}, {\it Debarshi}, {\it Pushpa}, {\it Sumit} and {\it Abhishek}  for the valuable discussions, comments and questions. I am very grateful to the two former students from our research group {\it Prof. Pramod Shukla} (Bose Institute, Kolkata, India) and {\it Dr. Karunva Sil} (University of Cyprus) for their very helpful suggestions and guidance to me which helped me in getting a post-doctoral position. I am very grateful to Prof. Keshav Dasgupta for the help related to academia. 

I would like to thank {\it Prof. Juan Maldacena (IAS, Princeton, USA), Prof. Andreas Karch (The University of Texas, Austin, USA), Prof. Tadashi Takayanagi (YITP, Kyoto University, Kyoto, Japan), Prof. Kostas Skenderis (University of Southampton, UK), Prof. Thomas Hartman (Cornell University), Prof. Suvrat Raju (ICTS, Bangalore), Prof. David Berenstein (University of California, Santa Barbara, USA), Prof. Xi Dong (University of California, Santa Barbara, USA), Prof. Arpan Bhattacharyya (IIT Gandhinagar, India), Prof. Christoph Uhlemann (University of Oxford), Prof. Rong-Xin Miao (Sun Yat-Sen University, China), Prof. Xuanhua Wang (UCAS, Wenzhou, China), Prof. Xian-Hui Ge (Shanghai University, China), Prof. Mohsen Alishahiha (IPM, Tehran, Iran), Prof. Sanjay Siwach (BHU, India), Prof. Chethan Krishnan (IISc Bangalore)} for the useful discussions and comments during my PhD. I would also like to thank {\it Prof. Suresh Tiwari, Prof. Rudra Prakash Malik, and Prof. Bhabani Prasad Mandal} from the Banaras Hindu University (BHU), India, who helped me to start my journey into the Theoretical High Energy Physics during my master's. 

I would also like to thank the editors, associate editors, and referees of the Journal of High Energy Physics (JHEP), Physical Review D (PRD), Physics Letters B (PLB), and European Physical Journal C (EPJC) for their comments, criticism, etc. Those comments helped me in learning something new which was missing in my papers. I am very grateful to the organizers (including speakers and participants) of the various conferences/workshops/schools for organizing very interesting events. I enjoyed those events and learned about the many new research topics.

I would like to thank the members of Indian Physics Association (IPA) Roorkee chapter. I enjoyed working with them. The members include, {\it Prof. Akhilesh Kumar Mishra, Prof. Anjani Kumar Tiwari, Ishani, Aalok, Jaikhomba, Nifeeya, Prem Sagar, Priyanshu, Ritika, Shikha, Tsewang, Vanshaj, Vidisha}.

I would like to extend my gratitude to all the faculty members, staff, and to students, and postdocs in the Department of Physics at I.I.T. Roorkee for making this a memorable journey. I would like to thank my colleagues from the {\bf research scholar room 2}: {\it Shakti, Soumya, Monika, Amarjyoti, Sonu, Pooja, Pooja, Manish, Veer and Mukesh} for bearing me and making the workplace vibrant. I also thank my colleagues from the all the research scholar rooms. I thank my friends {\it Shailesh, Rahul, Anurag, Nandeshwar, Sarvesh, Amit, Ramnivas, Shubhajit} for their time to time motivation and encouragement.

I am very thankful to have wonderful seniors who treated me like younger brother: {\it Shubham, Naveen, Aalok, Dibyendu}. I specially thank {\it Prem Sagar} who took care of me always during my stay at IIT Roorkee and {\it Deepak} for the wonderful company. I would like to thank {\it Priyanka, Pooja and Jaikhomba} who helped in various way during pre-synopsis and thesis submission. I am very thankful to {\it Anil, Ramashish, Amarjyoti, Shakti and Soumya} for their real life knowledge, I feel everyone should have friends like them but please don't follow them otherwise you may get bitten at some places. I am also thankful to {\it Shivam and Shubranshu} for their valuable time and discussion on the real life issues. It was a great pleasure to have  {\it Manish, Brij Mohan, Ashutosh, Arun, Harsh, Ishitwa, Shakti} at IIT Roorkee who are the alumni from BHU. I thank my friends {\it Divya, Pavneet, Ankur, Anshu, Hemant, Anju, Rahul, Ashish, Tejaswini, Gagan, Uwais} for their valuable time and discussions (on various topics) during my stay at IITR.

I am very grateful to my family for their continuous support and everything. I am very grateful to my parents ({\it Shri Shyam Narayan Yadav and Shrimati Chinta Devi}) who always left me free to do what I wanted to do. I was very fortunate to learn many things about life from my grandfather ({\it Late Shri Ghurfekan Yadav}). Seeing me as ``Dr. Gopal Yadav'' was his dream, I think he will be very happy for me wherever he will be now. I thank my maternal uncle and aunty, ``Phupha'' and ``Bua'', and brother in laws  who always encouraged me to believe in myself and for their support. I am very fortunate to have {\it bhaiya and bhabhi, Rinki di, Priyanka di, Shaloo, Prity, Sandip, Arvind bhaiya, Santosh bhaiya, Ravi bhaiya, Bunty, Aman, Aditya, Mahesh, Vipul, Pooja, Ranjita di, Anjali, Akriti, Ankita, Shrija, Devansh, Shreya, Shivansh, Tejas} in my life.

I am very grateful to my school teachers, college/university professors, and all of my gurus in various fields for educating me in various aspects of life. Education can bring changes and what I am today is the perfect example of it.

%\input{dedication}
%\newpage
\clearpage
\newpage
~\\
~\\
~\\
~\\
\thispagestyle{empty}
%\begin{flushright}
\begin{center}
\Huge{{\it \textbf{Dedicated to my Family and in the loving memory of my grandfather (Late Shri Ghurfekan Yadav)}}}
\end{center}
~\\
~\\
~\\
\fbox{\begin{minipage}{38em}{\textbf{``HOWEVER DIFFICULT LIFE MAY SEEM, THERE IS ALWAYS SOMETHING YOU CAN DO AND SUCCEED AT. IT MATTERS THAT YOU DON'T JUST GIVE UP.'' - STEPHEN HAWKING} }
\end{minipage}}
%\end{flushright}
\clearpage
%\addtocounter{page}{-1}

%\newpage
~\\
~\\
~\\
~\\
\clearpage
%\addtocounter{page}{-1}

%\input{Hawking}
%\addcontentsline{toc}{chapter}{List of Publications}
%\input{pubs}
\clearpage
%\newpage
%\input{empty1}
\begin{spacing}{1.53}
\addcontentsline{toc}{chapter}{Table of Contents}\tableofcontents
\end{spacing}
\clearpage
%\input{empty1}
%\newpage
\addcontentsline{toc}{chapter}{List of Figures}
\begin{spacing}{1.34}
\listoffigures
\end{spacing}
\addcontentsline{toc}{chapter}{List of Tables}
\begin{spacing}{1.34}
\listoftables
\end{spacing}
\mainmatter
\pagestyle{fancy}
\setcounter{chapter}{0}
\setcounter{section}{0}
\setcounter{subsection}{0}
\setcounter{tocdepth}{3}
\addcontentsline{toc}{chapter}{Part-I}
%\chapter*{PART - I: Top-Down Holographic Study of Thermal QCD-Like Theories at Intermediate Gauge/'t Hooft Coupling}

%\graphicspath{{partA/}{partA/}}
\newpage
%\pagestyle{plain}

%\newpage
%\pagestyle{plain}
%\begin{center}
%{{\huge {Top-Down Holographic Study of Thermal QCD-Like Theories %at Intermediate Gauge/'t Hooft Coupling}}}
%\end{center}
%\newpage
%\thispagestyle{empty}
%\begin{flushright}
~\\
~\\
~\\
~\\
~\\
~\\
\begin{center}
\Huge{{\bf Part-I \\Aspects of Thermal QCD Phenomenology at Intermediate Gauge/'t Hooft Coupling from String/M-Theory} }~\\
\end{center}
~\\
\fbox{\begin{minipage}{38em}{\textbf{``I was an ordinary person who studied hard. There are no miracle people. It happens they get interested in this thing and they learn all this stuff, but they're just people.'' - Richard Feynman }}
\end{minipage}}\\ \\
%\end{flushright}
\newpage
%\clearpage

%\addtocounter{page}{-1}
%\input{empty}
%satyamshivamsundaram
%JaiMataDi
\chapter{Introduction}

\graphicspath{{Chapter1/}{Chapter1/}}
\section{Introduction}
Nature has four kinds of interactions: strong, electromagnetic, weak, and gravitational. These four interactions describe our universe. Strong interaction governs the behavior of quarks inside the nucleons (neutron and proton) or, more broadly, the interaction between nucleons inside the nucleus; electromagnetic interaction mediates the interaction between the charged particles; the nuclear reaction is happening inside the core of the Sun because of weak interaction, gravitational interaction is responsible for the stability of celestial objects. A very nice theoretical framework known as quantum field theory (QFT) has described the three interactions except for gravity. The strong and electromagnetic interactions have been studied from the quantum chromodynamics (QCD) and quantum electrodynamics (QED), respectively; similarly, one is also able to study the electroweak interaction (unified theory of electromagnetic and weak interactions) from QFT. But when we try to study the gravitational interaction from the QFT technique, then we encounter many divergences in theory.

String theory turns out to be a very nice theory because it unifies all the interactions of nature. The same can be seen from the spectrum of the ``string'', which contains all the particles of nature, including graviton. Since graviton is the mediator of gravitational interaction, and hence we can say the sting theory is the quantum theory of gravity. String theory originated in the 1960s to study the QCD, but it could not explain QCD adequately at that time. It was found that this theory unifies all the interactions. Initially, it contained only bosonic degrees of freedom, and the resulting theory was named ``bosonic string theory'' where the dimension of space-time is 26. Fermions were incorporated in string theory using the idea of supersymmetry (SUSY); the theory is known as ``superstring theory''. Five versions of superstring theory were proposed: type IIA, type IIB, type I, Heterotic $E_8 \times E_8$, and Heterotic $SO(32)$. In 1995, it was shown by Witten that these five theories could be unified into a single theory known as ``${\cal M}$-theory'', and these are related to each other via various dualities, e.g., T-duality, S-duality, etc.

In \cite{AdS/CFT}, a very nice duality was proposed between string theory and gauge theory; the AdS/CFT correspondence. This duality relates the strongly coupled gauge theories with weakly coupled gravitational theories; generally, this is called ``gauge-gravity duality''. Using the mapping between the parameters of these theories, one can access the part of the gauge theories which were not earlier possible because of the limitation of the gauge theories. This duality has been very useful in various branches of physics: condensed matter physics, black holes, cosmology, QCD, etc. A very nice return: a theory that was disregarded because of its limitation in the 1960s to explain QCD is now able to explain the many interesting features of QCD very beautifully from the gauge-gravity duality.

\section{AdS-CFT Duality}
\label{intro-ads-cft}
In this section, we will discuss the AdS/CFT duality. This will be accomplished via various subsections. We start with the discussion on the hints of the string dual of QCD in \ref{SQCD}. Maldacena's conjecture will be discussed in \ref{Maldacena-Conjecture}. This conjecture will be realized in detail using \ref{DM-i}, \ref{DM-ii}, \ref{DM-iii}, and \ref{DM-iv}.
\subsection{Hints for String Dual of QCD}
\label{SQCD}
Heuristically, the desire to reformulate QCD as a string theory comes from the presence of flux tubes between quark-antiquark pairs in QCD, which aid in the confinement of these pairs. Regge behaviour is produced when a string is used as a model for the flux tubes. The relationship between the flux tube's mass $M$ and its angular momentum $J$, which is defined as $M^2\propto J$, determines the flux tube's regge behaviour. Regge behaviour was also observed in the spectrum of the mesons. Non-confining gauge theories, however, are not covered by this reasoning. A strong case may be made for the existence of a string counterpart of quantum gravity or any confining gauge theory by taking into account 't Hooft's large-$N_c$ limit. Due to dimensional reduction, the gauge group of QCD, $SU(3)$, does not have an expansion parameter. By changing $SU(3)$ to $SU(N_c)$, taking the limit $N_c\rightarrow\infty$, and expanding in $1/N_c$, 't Hooft generalised QCD. With $i,j=1,...,N_c$ and $a=1,...,N_f$, respectively, the quark and gluon fields $q^i_a$ and $A^i_\mu j$ represent the degrees of freedom of the generalised theory. The number of independent gauge fields is decreased to $N_c^2-1$ since the gauge group is now $SU(N_c)$ rather than $U(N_c)$, but since the working limit is $N_c\rightarrow\infty$, this discrepancy can be disregarded. The number of gluons is therefore assumed to be $N_c^2$. Thus $N_c^2$ is considerably larger than $N_f$, which represents the number of quark degrees of freedom. Gluons therefore control dynamics in the large-$N_c$ limit. One-loop self-energy of the gluon Scale for the Feynman diagram is  $g^{2}_{YM}N_{c}$. If, in addition to limit $N_{c}\rightarrow\infty$, one simultaneously sets a limit on the gauge coupling given as $g_{YM}\rightarrow 0$ while holding the 't Hooft coupling, $\lambda\equiv g^{2}_{YM}N_{c}$, constant, the Feynman diagram has a smooth limit. This is analogous to insisting that the confinement scale, $\Lambda_{QCD}$, in the large-$N_c$ limit stay constant. The one-loop $\beta$ function is specified as,
\begin{eqnarray}
\mu\frac{d}{d\mu}g^{2}_{YM}\propto-N_{c}g^{4}_{YM}.
\end{eqnarray}
It becomes independent of $N_c$ when expressed in terms of $\lambda$.
\begin{eqnarray}
\mu \frac{d}{d\mu}\lambda\propto-\lambda^{2}.
\end{eqnarray}
It is simpler to determine the $N_c$-scaling of Feynman diagrams by using double-line notation. This is done by swapping out the line connected to a gluon for a pair of lines connected to a quark and an antiquark. Feynman diagrams have the expansion in terms of the powers of $\lambda$ and $1/N_c$. To specifically illustrate this, consider the examples of one-, two-, and three-loop vacuum diagrams. While they scale to the same power of $N_c$, they do so at a different power of $\lambda^{l-1}$, where $l$ is the number of loops. Diagram classification is greatly influenced by topology, and non-planar diagrams are suppressed in the large-$N_c$ limit. The Feynman diagrams' relationship to string theory is established by the topology-based classification of the diagrams; the relationship is further clarified by tying a Riemann surface to each diagram. In a Feynman diagram written in double-line notation, each closed loop line can be viewed as the edge of a two-dimensional surface. These surfaces are joined at their boundaries to form a Riemann surface. One adds the point at infinity to the face connected to the diagram's exterior line to produce a compact surface. A non-planar diagram's Riemann surface is a torus as opposed to a sphere for a planar diagram. Scaling for a particular Feynman diagram is as follows: $N^{\chi}_c$, where $\chi$ is the associated Riemann surface's Euler number.
\begin{equation}
\chi=2-2g,
\end{equation}
where the boundaryless, orientable, compact surface is represented by the genus $g$. For the sphere, $\chi=2$ and for the torus, $\chi=0$. Hence, the amplitude of Feynman diagrams can be expressed as
\begin{eqnarray}
{\cal A}=\sum^{\infty}_{g=0}N^{\chi}_{c}\sum^{\infty}_{n=0}c_{g,n}\lambda^{n},
\end{eqnarray}
 where $c,{g,n}$ are the constants. Every confining gauge theory with Yang-Mills fields and matter in adjoint representation is covered by the analysis. A boundary in the Riemann surface that is connected to a Feynmann diagram is introduced when matter or quarks are added to the theory. The Feynmann diagram's power of $N_c$, which stays $N^{\chi}_c$ in the presence of matter, is unaffected by the existence of matter, but the Euler number changes to $\chi=2-2g-b$, where $b$ is the number of boundaries. Open strings are connected to the boundaries. As a result, for a theory that includes both closed and open strings, the sum over the number of boundaries is recognised as an expansion. The genus expansion of a string theory corresponds to the large-$N_c$ expansion of a gauge theory. As a result, {\it the classical limit of the string theory matches the planar limit of the gauge theory. The AdS/CFT correspondence is the result of this duality}.

\subsection{Maldacena's Conjecture}
\label{Maldacena-Conjecture}
According to the Maldacena's conjecture, commonly referred to as the AdS/CFT correspondence, gravity theories in AdS are related to conformal field theories on the boundary of AdS \cite{AdS/CFT}. In the beginning, it was proposed that the type IIB superstring theory on $AdS_{5}\times S^{5}$ is equivalent to ${\cal N}=4$ Super Yang-Mills theory in four dimensions. The $AdS_{5}/CFT_{4}$ model can be simplified down to its most fundamental version, which says that ${\cal N}=4$ SYM theory with gauge group $SU(N_ {c})$ and Yang-Mills coupling constant $g_{YM}$ is dynamically equivalent to type IIB superstring theory with string length $l_{s}=\sqrt{\alpha\prime}$ and coupling constant $g_{s}$ on $AdS_{5}\times S^{5}$ with radius of curvature $R$ and $N$ units of $F_{(5)}$ flux on $S^{5}$. The two free parameters on both sides of the theory are related in the following ways:$$g^{2}_{\rm YM}=2\pi g_{s}$$ and
$$2g^{2}_{\rm YM}N=R^{4}/\alpha^{\prime^{2}}.$$ \par
The conjectured duality yields the equivalence of two theories, which indicates that there is a precise mapping between the gauge invariants and local operators of the gauge side and the states and fields of the string theory. This mapping can be thought of as an exact one-to-one correspondence. Due to the fact that a complete quantum treatment of the superstring cannot be carried out, which limits the use of duality in its most powerful form, one must instead work with the more moderate form of duality, which can be attained by assuming appropriate approximations. Consideration of the large-'t Hooft coupling limit is made for a less robust variant of duality: $N_{c}\rightarrow \infty$ and $g_{\rm YM}^2=g_s\rightarrow 0$ such that $\lambda=g^{2}_{YM}N_{c}$ is very large. In this limit, $1/N$ expansion of the Feynman diagrams in gauge theory are related with the expansion in terms of the string coupling ($g_s$) on the string theory side.\par
The weaker limit of the conjecture that was mentioned earlier continues to present some difficulties to deal with, and we need to go still deeper in order to reach a tractable setting. We are only left with one free parameter for the new limit, and that is $\lambda$; as a result, we are able to investigate the behaviour at both extremes of the parameter range, whether $\lambda$ is extremely tiny or very high. These limits appear naturally in the D-brane picture, which is what inspired the correspondence; as a result, we will provide a quick introduction to this duality from the perspective of this picture in the following section.

\subsection{D-branes and Strings}
\label{DM-i}
In superstring theory, extended objects such as strings are not the only ones that may be specified. The theory also incorporates a wide variety of non-perturbative higher dimensional objects that are referred to as D-branes. For the application of $D3$-brane, see \cite{Ganor-D3}. D-branes can be understood from two distinct vantage points, which are referred to as open string and closed string, respectively. The significance of the string coupling, denoted by $g_s$, which determines the strength of the interaction that takes place between open and closed strings is what establishes which viewpoint is correct. 
\begin{itemize}
\item{\bf Open string perspective}: In this scenario, D-branes can be seen as higher-dimensional objects onto which open strings can terminate. This is one way to think about them. This viewpoint is true for tiny values of the coupling constant, such as $g_s<<1$, for both closed and open string. For low energies $E<<\alpha^{\prime^{-1/2}}$, when massive string excitations are neglected, the dynamics of the open strings can be explained by a supersymmetric gauge theory that is based on the world volume of the D-branes. Scalar field $\phi$ corresponds to open string excitations that are transverse to the D-branes, while gauge field $A_\mu$ corresponds to open string excitations that are parallel to the D-branes. The product $g_s N$ is the effective coupling constant for a stack of it $N$ coincident D-branes with gauge group $U(N)$, and the open string perspective is reliable for $g_{s}N<<1$. 

\item{\bf Closed string perspective}: When viewed from this angle, D-branes can be interpreted as solitary solutions to the problem of supergravity (low energy limit of superstring theory). D-branes are the source of the gravitational field that is responsible for the curvature of the spacetime that is all around them. It is important that the length scale $L$ be somewhat big in order to guarantee the correctness of the supergravity approximation and maintain a low curvature. The expression $L^{4}/\alpha^{\prime^{ 2}}\propto g_{s}N$ appear in the case of a stack of $N$ coincident D-branes. The closed string perspective is only applicable for the case where $g_{s}N>>1$ is being considered.
\end{itemize}
The $AdS_{5}/SCFT_{4}$ correspondence is produced when these two perspectives are applied to a stack of $N$ $D3$-branes that are placed in Minkowski spacetime. The stack of $N$ $D3$-branes extends along the Minkowskian spacetime directions, but it is perpendicular to the other six spatial directions.

%%%%%%%%%%%%%%%%%%%%%%%%%%%%%%%%%%%%%%%%%%%%%%%%%%%%%
%%%%%%%%%%%%%%%%%%%%%%%%%%%%%%%%%%%
\subsection{Open String Perspective}
\label{DM-ii}
The two types of strings in type IIB perturbative string theory for $g_{s}N<<1$ are:
\begin{itemize}
\item Open strings can be thought of as the excitation of a (3+1)dimensional hyperplane, starting and ending on the D3-branes.
\item Closed strings are thought of being the excitation of flat spacetime in the (9+1) dimensions. 
\end{itemize}
For $N$ $D3$-branes in flat spacetime at energies $E<<\alpha^{\prime^{ -1/2}}$, one solely considers massless excitations and disregards all other stringy excitations. The massless closed string states are organised into a ten-dimensional  $\cal{N}=1$ supergravity multiplet, whereas the massless open string excitations are organised into a four-dimensional $\cal {N}=$4 multiplet made up of a gauge field $A_\mu$, six real scalar fields $\phi^i$, and fermionic superpartners. According to the transformation properties of $D3$-branes under their world volume coordinates, the gauge field $A_\mu$ emerges from bosonic massless open string excitations along the $D3$-branes, and six real scalar fields $\phi^i$ emerge from bosonic massless open string excitations transverse to the $D3$-branes. For all massless string modes, the whole effective action is written as
\begin{eqnarray}
& & S=S_{\rm closed}+S_{\rm open}+S_{\rm int}.
\end{eqnarray}
In ten-dimensional supergravity action, $S_{\rm closed}$ is the closed string modes contribution, $S_{\rm open}$ is the open string modes, and $S_{\rm int}$ is the interactions between the closed and open string modes. From the DBI action and the Wess-Zumino term, we can deduce the action for $S_{\rm open}$ and $S_{\rm int}$. For a single $D3$-brane, we have the following Dirac-Born-Infeld action:
\begin{eqnarray}
& & S_{DBI}=-\frac{1}{(2\pi)^{3}\alpha^{\prime 2}g_{s}}\int d^{4}x e^{-\phi}\sqrt{-det({\cal P}[g]+2\pi \alpha^\prime F),}
\end{eqnarray}
where the pullback metric on the $D3$-brane is denoted by the ${\cal P}[g]$. In $\alpha^\prime$, expanding all the way up to the leading order
\begin{eqnarray}
& & S_{open}=-\frac{1}{2\pi g_{s}}\int d^{4}x(\frac{1}{4}F_{\mu\nu}F^{\mu\nu}+\frac{1}{2}\eta^{\mu\nu}\partial_{\mu}\phi^{i}\partial_{\nu}\phi^{i}+{\cal O}(\alpha^\prime)), \nonumber\\
& & S_{int}=-\frac{1}{8\pi g_{s}}\int d^{4}x \phi F_{\mu\nu}F^{\mu\nu}+...
\end{eqnarray}

In the scenario when there are $N$ coincident D3 branes, the  $U(N)$ valued gauge and scalar fields are, $A_{\mu}=A^{a}_{\mu}T_{a}$, and $\phi^{i}=\phi^{ia}T_{a}$ respectively. It is possible to guarantee gauge invariance by performing a trace over the gauge group when generalising the low-energy-effective actions $S_{\rm open}$ and $S_{\rm int}$ for $N$ coincident $D3$-branes. Covariant derivatives take the role of partial derivatives, and one must add a scalar potential V to the action $S_{\rm open}$, with the value of V being provided as
\begin{eqnarray}
& & V=\frac{1}{2\pi g_{s}}\sum_{i,j}Tr[\phi^{i},\phi^{j}]^{2}.
\end{eqnarray}
In the $\alpha^\prime\rightarrow 0$ (or Maldacena) limit, $S_{\rm open}$ is reduced to the ${\cal N}=4$ SYM theory bosonic component provided $2\pi g_{s}=g^{2}_{YM}.$ As a result of the fact that $\kappa\propto\alpha^\prime\rightarrow 0$, one realises that $S_{\rm closed}$ is equivalent to the free supergravity action in Minkowski spacetime of dimension ten. When one performs canonical normalisation by rescaling the dilation $\phi$ by $\kappa$, one discovers that $S_{\rm int}$ disappears and results in the decoupling of open and closed strings.

\subsection{Closed String Perspective}
\label{DM-iii}
In the strong coupling limit $g_{s}N\rightarrow\infty$, the $N$ $D3$-branes must be analysed using a closed string model. In this view, they are massive charged objects that generate different forms of type IIB supergravity, making string theory of this type possible. It can be shown that the ten-dimensional supergravity solution of $N$ $D3$-branes, which preserves the $SO(3,1)\times SO(6)$ isometries of spacetime, and half of the supercharges of type IIB supergravity, is provided by:
\begin{eqnarray}
\label{Closed-metric}
ds^{2}&=&H(r)^{-1/2}\eta_{\mu\nu}dx^{\mu}d^{\nu}+H(r)^{1/2}\delta_{ij}dx^{i}dx^{j},
\end{eqnarray}
\begin{eqnarray}
\label{closed-exp}
e^{2\phi(r)}&=&g^{2}_{s},
\end{eqnarray}
\begin{eqnarray}
\label{closed-four-form}
C_{(4)}&=&(1-H(r)^{-1})dx^{0}\wedge dx^{1}\wedge dx^{2}\wedge dx^{3}+...
\end{eqnarray}
where $i,j=4,...,9$ and $\mu,\nu=0,...,3$, $r^{2}=\sum_{i=4}^{9}x^{2}_{i}$. The type IIB supergravity's eoms and (\ref{closed-exp}) indicate that
\begin{equation}
\label{closed-H}
H(r)=1+\left(\frac{L}{r}\right)^{4},
\end{equation}
where $L^{4}=4\pi g_{s}N\alpha^{\prime^{2}}$. Due to counting the number of coincident $D3$-branes, the flux $F_{(5)}$ through the sphere $S^5$ is quantized. According to the value of r relative to L, the background can be split into two regions:

\begin{itemize}
\item{$r \gg L$}: $H(r)$ can be roughly estimated in this region by 1. The metric (\ref{Closed-metric}) reduces to 10-dimensional flat spacetime for this value of $H(r)$.

\item{$r \ll L$}: A rough approximation of $H(r)$ in this region is $L^4/r^4$. The metric can be roughly described as
    \begin{eqnarray}
    \label{closed-metric-ii}
    ds^{2}&=&\frac{r^{2}}{L^{2}}\eta_{\mu\nu}dx^{\mu}dx^{\nu}+\frac{L^{2}}{r^{2}}\delta_{ij}dx^{i}dx^{j} =\frac{L^{2}}{z^{2}}(\eta_{\mu\nu}dx^{\mu}dx^{\nu}+dz^{2})+L^{2}ds^{2}_{S^{5}},
    \end{eqnarray}
\end{itemize}
where $z=L^{2}/r$, and $\delta_{ij}dx^{i}dx^{j}=dr^{2}+r^{2}ds^{2}_{S^{5}}$. In (\ref{closed-metric-ii}), first term is the $AdS_{5}$ metric in the second line. Therefore, there are two types of closed strings based on the region of interest. 
\begin{itemize}
\item Closed strings moving through a 10-dimensional spacetime that is flat, which is the region corresponding to the limit $r \gg L$.
\item Closed strings that are able to propagate in the near horizon region ($AdS_{5}\times S^{5}$) in the limit $r\ll L$.
\end{itemize}
Both of the types of closed strings that were discussed before are decoupled when the low energy limit is imposed. In conclusion, there are two regions: an asymptotically flat region and a near-horizon region. Both of these regions are distinct from one another. The dynamics of closed strings are described by type IIB supergravity modes in 10-dimensional flat spacetime in asymptotically flat spacetime; however, in the near-horizon region, the dynamics of closed strings are described by string excitations about the solution of type IIB supergravity. Both classes of strings get decoupled in the low energy limit.
\subsection{Combining Both Perspectives}
\label{DM-iv}
We obtain two decoupled effective theories in the low-energy limit from both the open and the closed string viewpoints.
\begin{itemize}
\item{Closed string perspective}: Type IIB supergravity in asymptotically flat $\mathbb{R}^{9,1}$ spacetime and type IIB supergravity on $AdS_{5}\times S^{5}$ background.

\item{Open string perspective}: ${\cal N}=4$ super Yang-Mills (SYM) theory in four dimensional Minkowski spacetime and type IIB supergravity on flat $\mathbb{R}^{9,1}$ spacetime.
\end{itemize}
The conclusion that the two views are dual to one other-exactly what the Maldacena conjectured-is made possible by the fact that physics is unaffected by the perspective chosen and by the existence of type IIB supergravity modes on $\mathbb{R}^{9,1}$ in both perspectives.

\section{Gauge-Gravity Duality}
 The duality proposed by Maldacena in \cite{AdS/CFT} is a special case, and it relates the ${\cal N}=4$ supersymmetric Yang-Mills (SYM) theory and type IIB string theory on the $AdS_5 \times S^5$ background. But the gauge theories can be non-conformal and non-supersymmetric, e.g., QCD. Therefore to construct the gravity dual of such theories, one is required to break the conformal symmetry and supersymmetry. The gauge-gravity duality incorporates all such theories; conformal, non-conformal, supersymmetric, non-supersymmetric, etc.
 
 In \ref{intro-ads-cft}, we argued that one could state the duality in the low energy limit where the open and closed string degrees of freedom decouple. In \cite{AdS/CFT}, this decoupling is responsible for the duality between type IIB superstring theory and the maximally supersymmetric Yang-Mills theory. We can construct the holographic dual from the supergravity solution of the string theories in the absence of stringy corrections. But the problem with the supergravity versions is the absence of decoupling, as mentioned above, of the string degrees of freedom (open and closed). To achieve the same, we have to include the stringy corrections to the supergravity solutions, and it is not an easy task; see related work in \cite{S.Roy-1,S.Roy-2,S.Roy-3,S.Roy-4,S.Roy-5,S.Roy-6,S.Roy-7}.
 
See \cite{KW},\cite{KS},\cite{Nunez-6},\cite{Kruczenski:2003uq},\cite{Bergman:2001rw},\cite{Ballon-Bayona:2018ddm} and \cite{MK-ii} where many techniques have been discussed to obtain the non-conformal and non-supersymmetric holographic dual from the configuration of $D$-branes. We can break the SUSY from the stack of $D$-branes at the conical singularity of the conifold geometry. In these kinds of setups, the gravity dual involves an internal manifold as a conifold in the Calabi-Yau space. For the literature on the large-$N$ QCD defined on $S^3$ and $S^1\times S^3$ where the authors have discussed the various features of QCD, see \cite{H-QCD,H-QCD-1,H-QCD-2,H-QCD-3,H-QCD-4}. For the holographic description of QCD, see \cite{MK-i,A-QCD,Filho-1,Filho-2,Filho-3,Filho-4}. The $D$-branes have been used to study the inflation in \cite{SP-1,SP-2}.
\subsection{Klebanov-Witten Model}\label{KW}
To break the SUSY, authors in \cite{KW} placed $N$ $D3$-branes at the tip of the conifold where the conifold is a cone in six dimensions whose base is five-dimensional ($T^{1,1}$ with topology $S^{2}\times S^{3}$, $T^{1,1}$ has $(SU(2)\times SU(2))/U(1)$ symmetry). The conifold generates a superpotential on the gauge theory side as a result of breaking of SUSY to ${\cal N}=1$. The gauge group in this setup is $SU(N)\times SU(N)$ and this is coupled to four chiral fields $A_{1}, A_{2}, B_{1}, B_{2}$, transformation of $A_{1},A_{2}$ and $B_{1}, B_{2}$ is as $(N,\bar N)$ and $(\bar{N},N)$ respectively, under the aforementioned gauge group. The stack of $N$ $D3$ branes on the gauge theory side give rise to the geometry $AdS_{5}\times T^{1,1}$ space-time with the metric on the gravity dual side
 \begin{equation}
 ds^{2}_{10}=h^{-1/2}\eta_{\mu\nu}dx^{\mu}dx^{\nu}+h^{1/2}ds^{2}_{6}
 \end{equation}
  where the warp factor $h$ is defined as: $h(r)=1+\frac{L^{4}}{r^{4}}$ with $L^{4}=4\pi g_{s}N\alpha^{\prime^{2}}$. The conifold has the following metric
  \begin{equation}
  ds^{2}_{6}=dr^{2}+r^{2}d^{2}_{T^{1,1}},
  \end{equation}
  where
  \begin{equation}
  ds^{2}_{T^{1,1}}=\frac{1}{9}(d\psi+\cos\theta_{1}d\phi_{1}+\cos\theta_{2}d\phi_{2})^{2}+\frac{1}{2}\sum^{2}_{i=1}(d\theta^{2}_{i}+\sin^{2}\theta_{i}d\phi^{2}_{i}).
  \end{equation}
 It has been discussed that there is a relationship of the couplings $g_1$ and $g_2$ related with the product gauge group $S(N)\times SU(N)$ and the integral of $NS-NS$ $B_2$ and R-R two forms $C_2$ over the two-cycle $S^2$ of $T^{1,1}$. If we consider only $D3$ branes, then dilaton and $B_2$ are constant; hence this implies the conformal nature of the gauge theory.

\subsection{Klebanov-Tsytlin Model}
\label{KT-intro}
The attempt to break the conformal symmetry and supersymmetry was done in \cite{KT} by including the additional $M$ fractional $D3$ branes apart from the $N$ $D3$ branes at the tip of the conifold. Fractional $D3$ branes are the $D5$ branes wrapping $S^2$ of the $T^{1,1}$ space mentioned in the previous subsection.
Although these $M$ branes break the conformal symmetry in the gauge theory, we still have ${\cal N}=1$ supersymmetry in theory. Therefore we have a supersymmetric non-conformal gauge theory. Due to the addition of the extra $M$ fractional $D3$ branes into the setup, the gauge group changes to $SU(N+M)\times SU(N)$, and accordingly, the chiral superfields now have the transformations under the product gauge group as, $A_i$ transforms as  $(N+M,\overline{N})$  and $B_i$ transforms as  $(\overline{N+M},N)$. The supergravity solution of this setup has three and five-form fluxes written below
\begin{eqnarray}
\label{KT-BHF}
& &  F_3=M \omega_3 ;\ B_2=3g_s M\omega_2\ln(r/r_0) ; \ H_3=dB_2=3g_s M\frac{1}{r}dr\wedge\omega_2 , \nonumber\\
& & 
    \tilde{F}_{5}={\cal F}_{5}+*_{10}{\cal F}_{5} ; \ {\cal F}_{5}=\left(N+\frac{3}{2\pi}g_{s}M^{2}\ln(r/r_{0})\right){\rm vol}(T^{1,1}),
\end{eqnarray}  
 where $\omega_2$ and $\omega_3$ are the basis of the conifold's base, five form flux includes the backreaction of $F_3$. The effective number of $D3$ branes is now defined as 
\begin{equation}
\label{NEFF_KT}
   N_{eff}(r)=N+\frac{3}{2\pi}g_{s}M^{2}\log\left(\frac{r}{r_{0}}\right)
   \end{equation}  
 (\ref{NEFF_KT}) implies the decrease of $N_{eff}$ when $r\rightarrow 0$
In the Kelbanov-Tsytlin setup, the 10-dimensional warp factor has the following form in the near-horizon limit.
   \begin{equation}
   \label{WF-KT}
   h(r)=\frac{L^{4}}{r^{4}}\left(1+\frac{3g_{s}M^{2}}{2\pi N}\log r\right)
   \end{equation}
From (\ref{KT-BHF}), we see that $B_2$ is not constant anymore, and it depends upon the logarithmic of the radial coordinate. Therefore as discussed in \ref{KW}, due to the non-vanishing nature of $B_2$, the gauge couplings now run with the radial coordinate, and hence we now have a non-conformal theory. Equations (\ref{KT-BHF}) and (\ref{WF-KT}) implying that ${\cal F}_5$ and $h(r)$ will be zero at some radial distance and KT solution is singular in the infra-red (IR).

\subsection{Seiberg Duality Cascade}
The problem with the Klebanov-Strassler (KS) solution is that it is singular in the IR; this singularity should be removed. The KS model also described the confinement on the gauge theory side. As one goes from the UV to IR, the flux through the vanishing $S^2$ $(\int_{S_2}B_2)$ decreases by one unit when r decreases, whereas the five form flux decreases by $M$ units, and hence $N_{\rm eff}\rightarrow N_{\rm eff}-M$. This starts the Seiberg duality in this setup, and gauge group $SU(N+M)\times SU(N)$ changes to $SU(N)\times SU(N-M)$. By performing the repeated Seiberg duality cascade, in the end, we have $SU(M)$ gauge theory in the IR. The series of Seiberg duality performed as above is known as the ``Seiberg Duality Cascade''.

One can interpret this from the gauge theory perspective as follows. When the Seiberg duality cascade ends, $U(1)_R$ symmetry of the chiral fields is broken into $Z_{2M}$ in the presence of $M$ fractional $D3$-branes. In the IR, $Z_{2M}$ symmetry is now spontaneously broken into the group $Z_2$ because of gaugino condensation, which results the transformation of the singular conifold into a deformed conifold. The Klebanov-Strassler solution is a deformed conifold in the IR, which behaves as the Klebanov-Tsytlin solution in the UV. One can remove the singularity of the Klebanov-Tsytlin solution in the IR from the KS solution, but the KS solution is still singular in the UV. That is why one is required to modify the KS solution in the UV, and by doing so, one ends up with a UV complete theory.

\subsection{Resolved Conifold and Flavor $D7$ Brane Embedding}
The singularity appearing in the KT solution has been removed in the KS solution via the deformation of the three cycle ($S^{3}$) in the IR, which results in a deformed conifold. One can remove the singularity from the resolution of the two cycle ($S^2$) as well, and the geometry is known as the resolved conifold, metric for the same is given below.
   \begin{eqnarray}
   ds^{2}_{res}&=&\kappa(r)^{-1}dr^{2}+\frac{\kappa(r)}{9}r^{2}(d\psi+\cos\theta_{1}d\phi_{1}+\cos\theta_{2}d\phi_{2})^{2}+\frac{r^{2}}{6}(d\theta_{1}^{2}+\sin^{2}\theta_{1}d\phi_{1}^{2})\nonumber\\
   &&+\frac{r^{2}+6a^{2}}{6}(d\theta_{2}^{2}+\sin^{2}\theta_{2}d\phi_{2}^{2})
   \end{eqnarray}
where  $\kappa(r)=\frac{r^{2}+9 a^2}{r^{2}+6 a^2}$, ``$r$'' is the radial coordinate, and ``$a$'' being the resolution parameter. In this geometry, metric vanishes on $S^{2}(\theta_1,\phi_1)$ in the limit $r \rightarrow 0$ whereas the metric on resolved $S^2(\theta_2,\phi_2)$ remains finite.$a\rightarrow 0$ limit results in a singular conifold geometry.

 Authors in \cite{Zayas+Tseytlin} constructed the holographic dual from the branes configuration of $N$ $D3$-branes at the tip of the conifold and $M$ $D5$-branes wrapping the blown up $S^2$. The supergravity background includes fluxes and dilaton, but this setup does include the fundamental quarks. One can incorporate the fundamental quarks in holographic theories by including additional flavor branes in the probe approximation in the setup, and it was done in \cite{KT}. In \cite{ouyang}, the author embedded the stack of flavor $D7$-branes via Ouyang embedding. We are interested in the holographic dual constructed in \cite{metrics} where $N_f$ D7-branes were embedded via Ouyang embedding with embedding parameter $\mu$:
   \begin{equation}
   (r^{2}+9 a^{2}r^{4})^{1/4}e^{\frac{\iota}{2}\left(\psi-\phi_{1}-\phi_{2}\right)}\sin\frac{\theta_1}{2}\sin\frac{\theta_2}{2}=\mu
   \end{equation}
 The presence of the  flavor branes affects the dilaton profile as
\begin{equation}
e^{-\phi}=\frac{1}{g_s}-\frac{N_f}{8\pi}\log(r^{6}+9 a^{2}r^{4})-\frac{N_{f}}{2\pi}\log\left(\sin\frac{\theta_1}{2}\sin\frac{\theta_2}{2}\right)
\end{equation}

\section{UV Complete Top-Down Holographic Dual of Thermal QCD}
\label{HG-PC}
This section discusses the construction of holographic dual of thermal QCD-like theories at the intermediate coupling. We will start with the type IIB string dual of large-$N$ thermal QCD in \ref{IIB-intro}. We will discuss the type IIA mirror of \cite{metrics} in \ref{typeIIA-intro} using which we discuss the ${\cal M}$-theory uplift at finite coupling in \ref{M-intro}. Finally we will discuss the ${\cal M}$-theory uplift at intermediate coupling in \ref{Mint-intro} 
\subsection{Type IIB String Dual of Large-$N$ Thermal QCD}
\label{IIB-intro}
The UV complete type IIB string dual of large-$N$ thermal QCD-like theories has been discussed in this section, and this is based on \cite{metrics}. This model has been constructed in such a way that it removes all the drawbacks of \cite{KW}, \cite{KS}, \cite{ouyang}, \cite{Nunez-2} -\cite{Nunez-4} so that one can study the realistic QCD from the gauge-gravity duality. Holographic thermal QCD obtained from this model has the properties: IR confined, fundamental quarks, non-conformality, UV complete, defined for the confined $(T<T_c)$ and deconfined $(T>T_c)$, both phases of QCD and non-supersymmetric. UV completion problem of the Klebanov-Strassler model \cite{KS} has been cured in this model.
\begin{itemize}
\item {\bf Brane/Gauge Picture}: Branes configuration of \cite{metrics} involve $N$ color $D3$ branes, $M\ D5$ and $M\ \overline{D5}$-branes, $N_f$ flavor $D7$ and $\overline{D_7}$-branes. The details about the world volume coordinates of the branes appearing in \cite{metrics} are given in table \ref{table-metrics}. Let us see how one realizes the features of thermal QCD from these branes configuration.
\begin{table}[h]
\begin{center}
\begin{tabular}{|c|c|c|}\hline
&&\\
S. No. & Branes & World Volume \\
&&\\ \hline
&&\\
1. & $N\ D3$ & $\mathbb{R}^{1,3}(t,x^{1,2,3}) \times \{r=0\}$ \\
&&\\  \hline
&&\\
2. & $M\ D5$ & $\mathbb{R}^{1,3}(t,x^{1,2,3}) \times \{r=0\} \times S^2(\theta_1,\phi_1) \times {\rm NP}_{S^2_a(\theta_2,\phi_2)}$ \\
&&\\  \hline
&&\\
3. & $M\ \overline{D5}$ & $\mathbb{R}^{1,3}(t,x^{1,2,3}) \times \{r=0\}  \times S^2(\theta_1,\phi_1) \times {\rm SP}_{S^2_a(\theta_2,\phi_2)}$ \\
&&\\  \hline
&&\\
4. & $N_f\ D7$ & $\mathbb{R}^{1,3}(t,x^{1,2,3}) \times \mathbb{R}_+(r\in[|\mu_{\rm Ouyang}|^{\frac{2}{3}},r_{\rm UV}])  \times S^3(\theta_1,\phi_1,\psi) \times {\rm NP}_{S^2_a(\theta_2,\phi_2)}$ \\
&&\\  \hline
&&\\
5. & $N_f\ \overline{D7}$ & $\mathbb{R}^{1,3}(t,x^{1,2,3}) \times \mathbb{R}_+(r\in[{\cal R}_{D5/\overline{D5}}-\epsilon,r_{\rm UV}]) \times S^3(\theta_1,\phi_1,\psi) \times {\rm SP}_{S^2_a(\theta_2,\phi_2)}$ \\
&&\\  \hline
\end{tabular}
\end{center}
\caption{The Type IIB Brane Construct of \cite{metrics}}
\label{table-metrics}
\end{table}
\begin{itemize}
 \item $N$ $D3$-branes are placed at the tip of conifold geometry, $M\ D5$ and $M\ \overline{D5}$-branes have been placed at the antipodal points of the blown-up $S^2$ with the separation ${\cal R}_{D5/\overline{D5}}$ between them and share the world volume coordinates of vanishing $S^2$. UV, UV-IR, and IR regions are defined as follows in this setup.
  \begin{itemize}
  \item {\bf UV}: $r>|\mu_{\rm Ouyang}|^{\frac{2}{3}}/ {\cal R}_{D5/\overline{D5}}$
  
  \item  {\bf IR/IR-UV interpolating regions}: $r_0/r_h<r<|\mu_{\rm Ouyang}|^{\frac{2}{3}}/{\cal R}_{D5/\overline{D5}}$; $r\sim\Lambda$ corresponds to the deep IR region with confined $SU(M)$ theory.

  \end{itemize}

\item Flavor $D7$-branes are embedded via Ouyang embedding \cite{ouyang} with embedding equation (\ref{Ouyang-definition}). $N_f\ D7$-branes exist in the UV, UV-IR interpolating region, and IR whereas $N_f\ \overline{D7}$-branes exist only in the UV and the UV-IR interpolating regions. Due to this configuration of flavor branes, three-form fluxes vanish in the UV, axion-dilaton modulus remains constant, UV conformality, and absence of Landau poles are ensured in theory. This also implies the realization of chiral symmetry breaking in QCD from $SU(N_f)\times SU(N_f)$ in the UV to $SU(N_f)$ in the IR as no $\overline{D7}$-branes exist in the IR. 

\begin{equation}
\label{Ouyang-definition}
\left(r^6 + 9 a^2 r^4\right)^{\frac{1}{4}} e^{\frac{i}{2}\left(\psi-\phi_1-\phi_2\right)}\sin\left(\frac{\theta_1}{2}\right)\sin\left(\frac{\theta_2}{2}\right) = \mu_{\rm Ouyang},
\end{equation}
for vanishingly small $|\mu_{\rm Ouyang}|$.

\item Similar to the flavor $\overline{D7}$-branes, $\overline{D5}$-branes also exist only in the UV-IR and UV regions. This implies the existence of  $SU(N+M)\times SU(N+M)$ color gauge group in the UV and at $r={\cal R}_{D5/\overline{D5}}$, the aforementioned color gauge group changes to $SU(N+M)\times SU(N)$ as soon as one enters into the IR regime \cite{K. Dasgupta et al [2012]}. The following equation controls the flow of gauge couplings
 
\begin{equation}
\label{RG}
4\pi^2\left(\frac{1}{g_{SU(N+M)}^2} + \frac{1}{g_{SU(N)}^2}\right)e^\phi \sim \pi;\
 4\pi^2\left(\frac{1}{g_{SU(N+M)}^2} - \frac{1}{g_{SU(N)}^2}\right)e^\phi \sim \frac{1}{2\pi\alpha^\prime}\int_{S^2}B_2.
\end{equation}
Equation (\ref{RG}) implies that one can obtain $g_{SU(M+N)}^2=g_{SU(N)}^2=g_{YM}^2\sim g_s\equiv {\rm constant}$ only if $\int_{S^2}B_2 = 0$, and hence UV conformality. This is the reason to include the $M$ $\overline{D5}$-branes at the common boundary of the UV-IR interpolating and the UV regions. The gauge couplings $g_{SU(N+M)}$ and $g_{SU(N)}$ flow in opposite directions (strong and weak, respectively). The RG flow receives a contribution from the $N_f$ flavor $D7$-branes too, and hence $N_f\ \overline{D7}$-branes have been included in the setup.

\item The RG flow from the UV-to-IR triggers the Seiberg-like duality cascade in theory, and under Seiberg duality, $SU(N+M)_{\rm strong}\xrightarrow{\rm Seiberg\ Dual}SU(N-(M - N_f))_{\rm weak}$ in the IR. The application of many Seiberg dualities results in the confined $SU(M)$ gauge theory with $N_f$ flavors in the IR, the finite temperature has been studied in \cite{metrics}, see  \cite{arnabkundu0709.1547},\cite{arnabkundu0709.1554}, \cite{Minwalla-3},\cite{Obers:2008pj} for some related works. The chiral symmetry breaking in the presence of external magnetic field was discussed in \cite{SS-KJ} in top-down Sakai-Sugimoto model.

\end{itemize}

\item {\bf Gravity Picture}: The metric of the ten-dimensional string background \cite{metrics}, also known as the modified Ouyang-Klebanov-Strassler black hole (OKS-BH) background, is written below
\begin{equation}
\label{metric-IIB}
ds^2 = \frac{1}{\sqrt{h}}
\left(-g_1 dt^2+dx_1^2+dx_2^2+dx_3^2\right)+\sqrt{h}\biggl[g_2^{-1}dr^2+r^2 d{\cal M}_5^2\biggr],
\end{equation}
 where the black hole functions $g_i$'s with the horizon $r_h$ are
$ g_{1,2}(r,\theta_1,\theta_2)= 1-\frac{r_h^4}{r^4} + {\cal O}\left(\frac{g_sM^2}{N}\right)$.
where ($\theta_1, \theta_2$) dependence arises due to
${\cal O}\left(\frac{g_sM^2}{N}\right)$ corrections. Compact metric ($d{\cal M}_5^2$) appearing in (\ref{metric-IIB}) has the following form:
\begin{eqnarray}
\label{RWDC}
& & d{\cal M}_5^2 =  h_1 (d\psi + {\rm cos}~\theta_1~d\phi_1 + {\rm cos}~\theta_2~d\phi_2)^2 +
h_2 (d\theta_1^2 + {\rm sin}^2 \theta_1 ~d\phi_1^2) +   \nonumber\\
&&  + h_4 (h_3 d\theta_2^2 + {\rm sin}^2 \theta_2 ~d\phi_2^2) + h_5~{\rm cos}~\psi \left(d\theta_1 d\theta_2 -
{\rm sin}~\theta_1 {\rm sin}~\theta_2 d\phi_1 d\phi_2\right) + \nonumber\\
&&  + h_5 ~{\rm sin}~\psi \left({\rm sin}~\theta_1~d\theta_2 d\phi_1 +
{\rm sin}~\theta_2~d\theta_1 d\phi_2\right),
\end{eqnarray}
$r\gg a, h_5\sim\frac{({\rm deformation\ parameter})^2}{r^3}\ll  1$ for $r \gg({\rm deformation\ parameter})^{\frac{2}{3}}$, i.e. in the UV/IR-UV interpolating region. The $h_i$'s of (\ref{RWDC}) and $M, N_f$ are not constant. They have dependence on $g_s, M, N_f$ up to linear order as given below:
\begin{eqnarray}
\label{h_i}
& & \hskip -0.45in h_1 = \frac{1}{9} + {\cal O}\left(\frac{g_sM^2}{N}\right),\  h_2 = \frac{1}{6} + {\cal O}\left(\frac{g_sM^2}{N}\right),\ h_4 = h_2 + \frac{a^2}{r^2},\nonumber\\
& & h_3 = 1 + {\cal O}\left(\frac{g_sM^2}{N}\right),\ h_5\neq0.
\end{eqnarray}
Equations (\ref{RWDC}) and (\ref{h_i}) indicate the existence of non-extremal resolved warped deformed conifold involving
an $S^2$-blowup (as $h_4 - h_2 = \frac{a^2}{r^2}$), an $S^3$-blowup (as $h_5\neq0$) and squashing of an $S^2$ (because $h_3$ is not strictly unity). The horizon ($r_h$) is a warped squashed $S^2\times S^3$. The ten-dimensional warp factor that includes back-reaction has the following form in the IR
\begin{eqnarray}
\label{eq:h}
&& \hskip -0.45in h =\frac{L^4}{r^4}\Bigg[1+\frac{3g_sM_{\rm eff}^2}{2\pi N}{\rm log}r\left\{1+\frac{3g_sN^{\rm eff}_f}{2\pi}\left({\rm
log}r+\frac{1}{2}\right)+\frac{g_sN^{\rm eff}_f}{4\pi}{\rm log}\left({\rm sin}\frac{\theta_1}{2}
{\rm sin}\frac{\theta_2}{2}\right)\right\}\Biggr],
\end{eqnarray}
where $ L=\left(4\pi g_s N\right)^{\frac{1}{4}}$, and $M_{\rm eff}/N_f^{\rm eff}$ are given as below.
\begin{eqnarray}
\hskip -0.45in M_{\rm eff}/N_{f}^{\rm eff} = M/N_f + \sum_{m\ge n} (a/b)_{mn} (g_sN_f)^m (g_sM)^n.
\end{eqnarray}
Three form fluxes up to ${\cal O}(g_s N_f)$ are given by the following expressions in the IR (with $h_5=0$) \cite{metrics}
\begin{eqnarray}
\label{three-form-fluxes}
& & \hskip -0.4in (a) {\widetilde F}_3  =  2M { A_1} \left(1 + \frac{3g_sN_f}{2\pi}~{\rm log}~r\right) ~e_\psi \wedge
\frac{1}{2}\left({\rm sin}~\theta_1~ d\theta_1 \wedge d\phi_1-{ B_1}~{\rm sin}~\theta_2~ d\theta_2 \wedge
d\phi_2\right)\nonumber\\
&& \hskip -0.3in -\frac{3g_s MN_f}{4\pi} { A_2}~\frac{dr}{r}\wedge e_\psi \wedge \left({\rm cot}~\frac{\theta_2}{2}~{\rm sin}~\theta_2 ~d\phi_2
- { B_2}~ {\rm cot}~\frac{\theta_1}{2}~{\rm sin}~\theta_1 ~d\phi_1\right)\nonumber \\
&& \hskip -0.3in -\frac{3g_s MN_f}{8\pi}{ A_3} ~{\rm sin}~\theta_1 ~{\rm sin}~\theta_2 \left(
{\rm cot}~\frac{\theta_2}{2}~d\theta_1 +
{ B_3}~ {\rm cot}~\frac{\theta_1}{2}~d\theta_2\right)\wedge d\phi_1 \wedge d\phi_2, \nonumber\\
& & \hskip -0.4in (b) H_3 =  {6g_s { A_4} M}\Biggl(1+\frac{9g_s N_f}{4\pi}~{\rm log}~r+\frac{g_s N_f}{2\pi}
~{\rm log}~{\rm sin}\frac{\theta_1}{2}~
{\rm sin}\frac{\theta_2}{2}\Biggr)\frac{dr}{r}\nonumber \\
&& \hskip -0.3in \wedge \frac{1}{2}\Biggl({\rm sin}~\theta_1~ d\theta_1 \wedge d\phi_1
- { B_4}~{\rm sin}~\theta_2~ d\theta_2 \wedge d\phi_2\Biggr)
+ \frac{3g^2_s M N_f}{8\pi} { A_5} \Biggl(\frac{dr}{r}\wedge e_\psi -\frac{1}{2}de_\psi \Biggr)\nonumber  \\
&&  \wedge \Biggl({\rm cot}~\frac{\theta_2}{2}~d\theta_2
-{ B_5}~{\rm cot}~\frac{\theta_1}{2} ~d\theta_1\Biggr), 
\end{eqnarray}
where asymmetry factors appearing in (\ref{three-form-fluxes}) are: $ A_i=1 +{\cal O}\left(\frac{a^2}{r^2}\ {\rm or}\ \frac{a^2\log r}{r}\ {\rm or}\ \frac{a^2\log r}{r^2}\right) + {\cal O}\left(\frac{{\rm deformation\ parameter }^2}{r^3}\right),$ $  B_i = 1 + {\cal O}\left(\frac{a^2\log r}{r}\ {\rm or}\ \frac{a^2\log r}{r^2}\ {\rm or}\ \frac{a^2\log r}{r^3}\right)+{\cal O}\left(\frac{({\rm deformation\ parameter})^2}{r^3}\right)$. \par
As we discussed, the black hole on the gravity dual side is responsible for the finite temperature on the thermal QCD side. This is true when $T>T_c$ in QCD, where $T_c$ is the deconfinement temperature. When $T<T_c$, then the gravitational dual involves a thermal background. We also discussed that we had resolved $S^2$ and the deformation of $S^3$ (responsible for the IR confinement) in this setup. Therefore, we found that {\it gravity dual of large-$N$ thermal QCD as constructed in \cite{metrics} is a resolved warped deformed conifold where fluxes in the IR have replaced branes. The warp factor and fluxes contain the information about back-reactions}.

\end{itemize}

\subsection{Type IIA SYZ Mirror of \cite{metrics}}
\label{typeIIA-intro}
The Strominger-Yau-Zaslow (SYZ) mirror symmetry is triple T-duality along the three isometry directions \cite{syz}. The requirement to apply the SYZ mirror symmetry is that one must have a special Lagrangian (sLag) three-cycle, $T^3$, fibered over a large base to cancel the contributions from open-string disc instantons with boundaries as non-contractible one-cycles in the sLag. Due to absence of the isometry along $\psi$ direction in the metric of \cite{metrics}, we have to work with the coordinates $(x,y,z)$ defined in $T^3(x,y,z)$ which are toroidal analogue of $(\phi_1,\phi_2,\psi)$, and these are related by the following equations \cite{MQGP}:
\begin{equation}
\label{xyz defs}
x = \sqrt{h_2}h^{\frac{1}{4}}sin\langle\theta_1\rangle\langle r\rangle \phi_1,\ y = \sqrt{h_4}h^{\frac{1}{4}}sin\langle\theta_2\rangle\langle r\rangle \phi_2,\ z=\sqrt{h_1}\langle r\rangle h^{\frac{1}{4}}\psi,
\end{equation}
based on \cite{M.Ionel and M.Min-Oo(2008)} authors in \cite{DM-transport-2014,EPJC-2} obtained the following results for the $T^2$-invariant sLag of \cite{M.Ionel and M.Min-Oo(2008)}:
\begin{eqnarray}
\label{sLag-conditions}
& & \hskip -0.2in \left.i^* J\right|_{\rm RC/DC} \approx 0; \  \left.\Im m\left( i^*\Omega\right)\right|_{\rm RC/DC} \approx 0; \ \left.\Re e\left(i^*\Omega\right)\right|_{\rm RC/DC}\sim{\rm volume \ form}\left(T^3(x,y,z)\right),
\end{eqnarray}
for a deformed conifold. Therefore, local $T^3$ of (\ref{xyz defs}) is the sLag to implement the SYZ mirror symmetry when resolved warped deformed conifold is predominantly either resolved or deformed. It is interesting to note that in the ``delocalized limit'' \cite{M. Becker et al [2004]}  $\psi=\langle\psi\rangle$, performing the following coordinate transformation:
\begin{equation}
\label{transformation_psi}
\left(\begin{array}{c} sin\theta_2 d\phi_2 \\ d\theta_2\end{array} \right)\rightarrow \left(\begin{array}{cc} cos\langle\psi\rangle & sin\langle\psi\rangle \\
- sin\langle\psi\rangle & cos\langle\psi\rangle
\end{array}\right)\left(\begin{array}{c}
sin\theta_2 d\phi_2\\
d\theta_2
\end{array}
\right),
\end{equation}
and $\psi\rightarrow\psi - \cos\langle{\bar\theta}_2\rangle\phi_2 + \cos\langle\theta_2\rangle\phi_2 - \tan\langle\psi\rangle ln\sin{\bar\theta}_2$, the term $h_5$ will be changed to as
 $h_5\left[d\theta_1 d\theta_2 - sin\theta_1 sin\theta_2 d\phi_1d\phi_2\right]$, and $e_\psi\rightarrow e_\psi$, i.e., One adds a new isometry along $\psi$ to the existing isometries along $\phi_{1,2}$. It is abundantly evident that this does not hold true on a global scale; the deformed conifold does not have a third global isometry. One must additionally make sure that the $T^3(x,y,z)$ fibration has a large base, as described above (implicating large complex structures of the two two-tori specified above), in order to enable usage of SYZ-mirror duality via three T dualities. This is accomplished through \cite{F. Chen et al [2010]}:
\begin{eqnarray}
\label{SYZ-large base}
& & d\psi\rightarrow d\psi + f_1(\theta_1)\cos\theta_1 d\theta_1 + f_2(\theta_2)\cos\theta_2d\theta_2,\nonumber\\
& & d\phi_{1,2}\rightarrow d\phi_{1,2} - f_{1,2}(\theta_{1,2})d\theta_{1,2},
\end{eqnarray}
for suitably chosen big values of $f_{1,2}(\theta_{1,2})$. The three-form fluxes continue to remain invariant. It was explained in \cite{MQGP} why one might select such large values of $f_{1,2}(\theta_{1,2})$. The guiding premise is that one must have a metric that, at least locally, resembles a non-K\"{a}hler warped resolved conifold after applying the SYZ-mirror transformation to the non-K\"{a}hler resolved warped deformed conifold. Let's look at the effect of three T-dualities along the $(\phi_1,\phi_2, \psi)$ directions.
\begin{itemize} 
\item {\bf T-duality along $\psi$}: First T-duality along $\psi$ directions implies that $N\ D3$-branes are converted into $N\ D4$-branes wrapping the $\psi$ circle, $N_f$ flavour $D7(\overline{D7})$-branes are transformed into $N_f$ flavour $D6(\overline{D6})$-branes, and $M$ fractional $D3(\overline{D3})$-branes are transformed into $M \  D4(\overline{D4})$-branes straddling a pair of orthogonal $NS5$-branes\footnote{See \cite{NAB-1,NAB-2} for where $NS5$ branes have been studied.} ${NS5}_1 (x^{0,1,2,3},\theta_1,\phi_1)$ and ${NS5}_2 (x^{0,1,2,3},\theta_2,\phi_2)$.
\item {\bf T-duality along $\phi_1$}: Second T-duality in the $\phi_1$ direction does not change \\ $NS5_1 (x^{0,1,2,3},\theta_1,\phi_1)$, but it does convert $NS5_2 (x^{0,1,2,3},\theta_2,\phi_2)$ into a Taub-NUT space $(r,\psi,\theta_2,\phi_1)$. In addition, $N\ D4$-branes, $M\ D4(\overline{D4})$-branes, and $N_f$ flavor $D6(\overline{D6})$-branes will be turned into the $N \ D5$-branes, the $M \ D5(\overline{D5})$-branes, and the $N_f$ flavor $D5(\overline{D5})$-branes.
\item {\bf T-duality along $\phi_2$}: The ${NS5}_2 (x^{0,1,2,3},\theta_2,\phi_2)$-brane remains unchanged by the third T-duality along the $\phi_2$ direction, whereas the ${NS5}_1 (x^{0,1,2,3},\theta_1,\phi_1)$ is transformed into a Taub-NUT space$(r,\psi,\theta_1,\phi_2)$. Further, $N\ D5$-branes, $M\ D5(\overline{D5})$-branes and $N_f$ flavor $D5(\overline{D5})$-branes will be transformed into $N\ D6$-branes, $M\ D6(\overline{D6})$-branes and $N_f$ flavor $D6(\overline{D6})$-branes wrapping the non-compact three-cycle $\Sigma^{(3)}(r, \theta_1, \phi_2)$. 
\end{itemize} 
As a result of this, we are able to see that the SYZ type IIA mirror has $N\ D6$-branes, $M\ D6(\overline{D6})$-branes and $N_f$ flavor $D6(\overline{D6})$-branes. Remember that, just like in \cite{metrics}, ranges for the presence of anti-branes are also important in the type IIA mirror. T-dualities were performed along the $T^3(x,y,z)$ with $(x,y,z)$ as the toroidal analogue of $(\phi_1,\phi_2,\psi)$. For simplicity, we have written $(\phi_1,\phi_2,\psi)$.
\subsection{${\cal M}$-theory Uplift of Type IIB Setup}
\label{M-intro}
From the triple T-dual of the type IIB $F_{1,3,5}$ in \cite{MQGP}, we can derive a one-form type IIA potential, from which the following $D=11$ metric ($u\equiv\frac{r_h}{r}$) was obtained:
\begin{eqnarray}
\label{Mtheory met}
& &   ds^2_{11} = e^{-\frac{2\phi^{IIA}}{3}} \left[g_{tt}dt^2 + g_{\mathbb{R}^3}\left(dx^2 + dy^2 + dz^2\right) +  g_{uu}du^2  +   ds^2_{IIA}({\theta_{1,2},\phi_{1,2},\psi})\right] \nonumber\\
& & + e^{\frac{4{\phi}^{IIA}}{3}}\Bigl(dx_{11} + A^{F_1}+A^{F_3}+A^{F_5}\Bigr)^2 \equiv\ {\rm Black}\ M3-{\rm Brane}+{\cal O}\left(\left[\frac{g_s M^2 \log N}{N}\right] \left(g_sM\right)N_f\right),\nonumber\\
& & {\rm where}:\nonumber\\
& & g_{uu}=\frac{3^{2/3}(2\sqrt{\pi g_s N})}{u^2(1-u^4)}\left(1  - \frac{3 g_s^2 M^2 N_f \log(N) \log \left(\frac{r_h}{u}\right)}{32 \pi ^2 N}\right)\nonumber\\
& & g_{tt} = \frac{3^{2/3}(u^4-1)r_h^2}{u^2(2\sqrt{\pi g_{s}N})} \left(\frac{3 g_s^2 M^2 N_f \log(N) \log \left(\frac{r_h}{u}\right)}{32 \pi ^2 N}+1\right)\nonumber\\
& & g_{\mathbb{R}^3} =  \frac{3^{2/3}r_h^2}{u^2(2\sqrt{\pi g_{s}N})} \left(\frac{3 g_s^2 M^2 N_f \log(N) \log \left(\frac{r_h}{u}\right)}{32 \pi ^2 N}+1\right).
\end{eqnarray}
Moreover, in the UV:
{\footnotesize
\begin{eqnarray}
\label{non-conformal-contribution}
& & G_{{x}{x}}^M = G_{{y}{y}}^M = G_{{z}{z}}^M = \frac{3^{2/3}r_h^2}{g_{s}^{2/3}u^2(2\sqrt{\pi g_{s}N})} \left(\frac{3 g_s^2 M^2 N_f \log(N) \log \left(\frac{r_h}{u}\right)}{32 \pi ^2 N}+1\right)\nonumber
\end{eqnarray}
\begin{eqnarray}
& & G_{\phi_1r}^M \sim \nonumber\\
& & \frac{2 {g_s}^{4/3} {N_f}^2 {\sin^2\phi_1} 2\sin\left(\frac{\psi}{2}\right)  \sin ^2({\theta_1}) \left(9 \sin ^2({\theta_1})+6 \cos
   ^2({\theta_1})+4 \cos ({\theta_1})\right) \sqrt[4]{{g_s} N \left(1-\frac{3 {g_s}^2 M^2 {N_f} \log (N)
   \log (r)}{16 \pi ^2 N}\right)} }{3^{5/6} \pi ^{7/4} (\cos
   (2 {\theta_1})-5)^2}\nonumber\\
   & & \times \left(9 {h_5} \sin ({\theta_1})+4 \cos ^2({\theta_1}) \csc ({\theta_2})-2
   \cos ({\theta_1}) \cot ({\theta_2})+6 \sin ^2({\theta_1}) \csc ({\theta_2})\right)\nonumber\\
   & & G_{11\ r}^M\sim \frac{3^{\frac{3}{2}}g_s^{\frac{4}{3}} N_f \sin\phi_1\left( - 8 \cos\theta_1 + 3( - 5 +
   \cos(2\theta_1))\right)\sin\theta_1}{\pi\left( - 5 + \cos(2\theta_1)\right)}.
\end{eqnarray}}
Near $\theta_1\sim\frac{1}{N^{\frac{1}{5}}}, \theta_2\sim\frac{1}{N^{\frac{3}{10}}}$ coordinate patch, one is able to see that:
\begin{eqnarray}
\label{Gphi1r+G11r}
& & G_{\phi_1r}\sim\frac{10 g_s^{\frac{19}{12}}\sin^2\phi_1 \sin\left(\frac{\psi}{2}\right)N^{\frac{3}{20}}}
{2 3^{\frac{5}{6}}\pi^{\frac{7}{4}}} + {\cal O}\left(\frac{1}{N^{\frac{1}{4}}}\right)\ll1\ {\rm for}\ \psi\sim\frac{1}{N^{\alpha\gg\frac{3}{20}}},\nonumber\\
& & G_{11\ r}\sim{\cal O}\left(\frac{1}{N^{\frac{1}{5}}}\right).
\end{eqnarray}

 The five-dimensional subspace $M_5(t,x_{1,2,3},u)$ from the six dimensional internal space $M_6(\theta_{1,2},\phi_{1,2},\psi,x_{10})$ in the MQGP limit around  $\theta_1\sim\frac{1}{N^{\frac{1}{5}}}, \theta_2\sim\frac{1}{N^{\frac{3}{10}}}$. It has been shown that {\it ${\cal M}$-theory uplift is equipped with a $G_2$-structure manifold \cite{MQGP,NPB}.}

\begin{itemize}
\item {\bf MQGP Limit}: The following limit is taken into consideration in \cite{metrics}:
\begin{eqnarray}
\label{limits_Dasguptaetal-i}
&   & \hskip -0.17in {\rm weak string}(g_s){\rm coupling-large\ t'Hooft\ coupling\ limit}:\nonumber\\
& & \hskip -0.17in g_s\ll  1, g_sN_f\ll  1, \frac{g_sM^2}{N}\ll  1, g_sM\gg1, g_sN\gg 1.
 \end{eqnarray}
The ``MQGP'' limit is defined as \cite{MQGP}:
\begin{equation}
\label{MQGP_limit}
g_s\sim\frac{1}{{\cal O}(1)}, M, N_f \equiv {\cal O}(1),\ g_sN_f<1,\ N\gg1,\ \frac{g_s M^2}{N}\ll1,
\end{equation}
The motivation for taking into account the MQGP limit mostly stems from two reasons.

\begin{enumerate}
\item
In contrast to AdS/CFT duality where $g_{\rm YM}\rightarrow0, N\rightarrow\infty$ such that $g_{\rm YM}^2N$ is huge, for strongly coupled thermal QCD like sQGP, we have $g_{\rm YM}\sim{\cal O}(1)$ and $N_c=3$ and hence  $g_{\rm YM}^2N=g_s$ is finite, i.e., $g_s\stackrel{<}{\sim}1$. $N_c=M$ has been derived in \cite{MQGP} from the Seiberg duality cascade and it is allowed to take $N_c=3$. Due to finite string coupling, one is required to address this issue from ${\cal M}$ theory, and the ``M'' in the ``MQGP'' limit implies ${\cal M}$-theory. See \cite{NAB-4} where the authors have studied ${\cal M}$-theory without any uplift. There are also papers available in the related context of ${\cal M}$-theory, e.g., \cite{Ganor-1,Ganor-2,Ganor-3}.
\item
The following are examples of calculational simplification in supergravity, and they make up the second importance of the justifications for considering the MQGP limit (\ref{MQGP_limit}):
\begin{itemize}
\item
In the region of UV-IR interpolation and the UV,
$(M_{\rm eff}, N_{\rm eff}, N_f^{\rm eff})\xrightarrow{\rm MQGP}{\approx}(M, N, N_f)$
\item
Asymmetry Factors $A_i, B_j$(in three-form fluxes)$\xrightarrow{MQGP}1$ in the region where the UV and IR are interpolated, in addition to the UV.

\item
The ten-dimensional warp factor and the non-extremality function in the MQGP limit have the simple forms.
\end{itemize}
\end{enumerate}

\item {\bf IR Color-Flavor Enhancement of Length Scale}: Even with ${\cal O}(1)$ $M$, there is color-flavor enhancement of the length scale in the MQGP limit (\ref{MQGP_limit}) compared to a Planckian length scale in KS-like model in the IR. This is valid if one takes into account terms of higher order in  $g_s N_f$ in the RR and NS-NS three-form fluxes, and NLO in $N$ in the ${\cal M}$-theory metric. It suggests the validity of supergravity calculations and suppresses quantum corrections. This issue has been discussed in detail in \cite{NPB,Misra+Gale}. Now, the effective number of color branes, denoted by the notation$N_{\rm eff}(r)$, will be given by
\begin{eqnarray}
\label{NeffMeffNfeff}
& & N_{\rm eff}(r) = N\left[ 1 + \frac{3 g_s M_{\rm eff}^2}{2\pi N}\left(\log r + \frac{3 g_s N_f^{\rm eff}}{2\pi}\left(\log r\right)^2\right)\right],\nonumber\\
& & M_{\rm eff}(r) = M + \frac{3g_s N_f M}{2\pi}\log r + \sum_{m\geq1}\sum_{n\geq1} N_f^m M^n f_{mn}(r),\nonumber\\
& & N^{\rm eff}_f(r) = N_f + \sum_{m\geq1}\sum_{n\geq0} N_f^m M^n g_{mn}(r).
\end{eqnarray}
The type IIB axion $C_0$ is given as $C_0 =N_f^{\rm eff} \frac{\left(\psi - \phi_1-\phi_2\right)}{4\pi}$,
%the ten-dimensional warp factor h, disregarding the angular part, is given by:
%\begin{eqnarray}
%\label{h}
%h & = & \frac{4\pi g_s}{r^4}\left[N_{\rm eff}(r) + \frac{9g_sM^2_{\rm eff}g_sN_f^{\rm eff}}{8\pi^2}\log r\right].
%\end{eqnarray}
$N_{\rm eff}(r_0\in\rm IR)=0$. The ten-dimensional warp factor is being written as $h \sim \frac{L^4}{r^4}$. The following equation will provide the length scale $L$ in the IR.
\begin{eqnarray}
\label{length-IR}
& & \hskip -0.9in L\sim\sqrt[4]{M}N_f^{\frac{3}{4}}\sqrt{\left(\sum_{m\geq0}\sum_{n\geq0}N_f^mM^nf_{mn}(r_0)\right)}\left(\sum_{l\geq0}\sum_{p\geq0}N_f^lM^p g_{lp}(r_0)\right)^{\frac{1}{4}} L_{\rm KS},
\end{eqnarray}
$L_{KS}\sim \sqrt[4]{g_sM}\sqrt{\alpha^\prime}$. Equation (\ref{length-IR}) suggests that, in comparison to KS, the colour-flavor length scale is enhanced in the IR. This implies the {\it validity of supergravity calculations with $N_c^{\rm IR}=M=3$ and $N_f=2(u/d)+1(s)$ provided one should include the $n,m>1$  terms in $M_{\rm eff}$ and $N_f^{\rm eff}$ in (\ref{NeffMeffNfeff})}.

\item {\bf Obtaining} ${\bf N_c=3}$: 
We will now briefly summarise how to identify the number of colours $N_c$ with $M$ that can be taken as 3 in the ``MQGP limit'' (\ref{MQGP_limit}). This is based on the work done in \cite{Misra+Gale}. Let us write $N_c$ as 
\begin{equation}
\label{N_c}
N_C = N_{\rm eff}(r) + M_{\rm eff}(r), 
\end{equation}
where $N_{\rm eff}(r)$ appears in the following equation,
\begin{equation}
\tilde{F}_5\equiv dC_4 + B_2\wedge F_3 = {\cal F}_5 + *{\cal F}_5,
\end{equation}
via ${\cal F}_5\equiv N_{\rm eff}\times{\rm Vol}({\rm Base\ of\ Resolved\ Warped\ Deformed\ Conifold})$. One can also define $M_{\rm eff}$ as  
\begin{equation}
\label{MEFF}
M_{\rm eff} = \int_{S^3}\tilde{F}_3.
\end{equation}
In (\ref{MEFF}), $S^3$ being dual to $\ e_\psi\wedge\left(\sin\theta_1 d\theta_1\wedge d\phi_1 - B_1\sin\theta_2\wedge d\phi_2\right)$, $B_1$ is an `asymmetry factor' defined in \cite{metrics}; $e_\psi\equiv d\psi + {\rm cos}~\theta_1~d\phi_1 + {\rm cos}~\theta_2~d\phi_2$) and \cite{M(r)N_f(r)-Dasgupta_et_al}: $\tilde{F}_3 (\equiv F_3 - \tau H_3)\propto M(r)\equiv M\frac{1}{1 + e^{\alpha(r-{\cal R}_{D5/\overline{D5}})}}, \alpha\gg1$.
If we define the UV and IR values of a quantity ``A'' as $[A_{UV},A_{IR}]$ then we can define the UV and IR values of $N_{\rm eff}$ and $M_{\rm eff}$ as
\begin{equation}
\label{ranges}
N_{\rm eff} \in[N,0] \hskip 0.2in and \hskip 0.2in N_{\rm eff} \in [0,M].
\end{equation}
Using (\ref{N_c}) and (\ref{ranges}), one can derive the UV and IR values of $N_c$ as given below
\begin{equation}
\label{Nc-UV-IR}
N_c \in [N, M].
\end{equation}
\end{itemize}
Therefore, from (\ref{Nc-UV-IR}), we can easily see that $N_c=M$ in the IR and it was taken as $N_c=M=3$ in the MQGP limit \cite{metrics}. From the various discussions in this section, we found that {\it ${\cal M}$-theory dual constructed from the type IIB string dual of \cite{metrics} has finite number of colors, it is IR confined, it has fundamental quarks, and this dual is valid for all temperatures (both $T>T_c$ and $T<T_c$).}

\subsection{${\cal M}$-Theory Uplift Inclusive of ${\cal O}(R^4)$-Corrections}
\label{Mint-intro}
This section is based on \cite{HD-MQGP}. The first part of the thesis (Part-I) and chapter {\bf 7} are based on this holographic model. We will discuss the inclusion of higher derivative terms in the eleven-dimensional supergravity action and solutions of the metric EOMs. With the inclusion of higher derivative terms on the gravity dual side, we are able to explore the intermediate coupling regime of thermal QCD, which was not possible earlier. The $SU(3)/G_2/SU(4)/Spin(7)$-structure torsion classes of the relevant six-, seven- and eight-folds associated with the ${\cal M}$-theory uplift have been computed in \cite{HD-MQGP}. \par
 The origin of ${\cal O}(R^4)$-corrections to the ${\cal N}=1, D=11$ supergravity action has been discussed in \cite{Green and Gutperle}. This kind of term has been used in the effective
one-loop heterotic action \cite{NAB-3}. The other kind of higher derivative term constructed from Weyl tensor was used in \cite{A-n/s-Weyl} to compute the correction to the viscosity-to-entropy density ration. There are two ways of understanding the origin of the ${\cal O}(R^4)$-corrections to the ${\cal N}=1, D=11$ supergravity action. One is in the context of the effects of $D$-instantons  in IIB supergravity/string theory via the four-graviton scattering amplitude \cite{Green and Gutperle}. The other is $D=10$ supersymmetry \cite{Green and Vanhove}. Let us briefly discuss these before going into the details of the construction of ${\cal M}$-theory dual at intermediate coupling.
\begin{itemize}
\item Let us initially consider interactions generated at leading order in a $D$-instanton (closed-string states of type IIB superstring theory where the entire string is confined at one point in superspace) background within both type IIB supergravity and string descriptions, which includes a one-instanton correction to the tree-level along with the one-loop $R^4$ terms \cite{Green and Gutperle} - the two containing an identical tensorial structure. The coordinates pertaining to the location of the $D$-instanton are used to parameterize the bosonic zero modes. The broken supersymmetries yield the fermionic zero modes.  The Grassmann parameters are fermionic supermoduli that correspond to dilatino zero modes and should be integrated out with the bosonic zero modes. The disk diagram is the most basic open-string world-sheet that occurs in a D-brane procedures.  At the lowest order, an instanton with some zero modes refers to a disk world-sheet along with open-string states connected to the boundary. Consider on-shell amplitudes in the instanton background to infer the one-instanton terms in the supergravity effective action.      The separate fermionic zero modes are absorbed by the integration over the fermionic moduli.  Consider four external graviton amplitudes. In supergravity, the leading term refers to the situation where each graviton is connected with the four fermionic zero modes. A nonlocal four-graviton interaction is produced by integrating over the bosonic zero modes. The world-sheet in the associated string computation is made up of four unconnected disks, each with a single closed-string graviton vertex and four fermionic open-string vertices. By expressing the graviton polarization tensor via $\zeta^{\mu_r\nu_r} =\zeta^{(\mu_r} \tilde \zeta^{\nu_r)}$ and calculating the fermionic integral, the authors in \cite{Green and Gutperle} showed that the four-graviton scattering amplitude is able to be represented via the equation:
\begin{eqnarray}
\label{A}
 &  & C  e^{2i\pi  \tau_0}   \int d^{10}y
e^{i \sum_r
k_r\cdot y}  \nonumber\\
&&\times\left(\hat{t}^{i_1j_1\cdots i_4j_4}\hat{t}_{m_1n_1\cdots
  m_4n_4}-{1\over 4}\epsilon^{i_1j_1\cdots j_4j_4}\epsilon_{m_1n_1\cdots
  m_4n_4}\right)  R_{i_1j_1}^{m_1n_1}
R_{i_2j_2}^{m_2n_2}R_{i_3j_3}^{m_3n_3}
R_{i_4j_4}^{m_4n_4};\nonumber\\
& &
\end{eqnarray}
$t_8$ is specified by:
{\footnotesize
\begin{eqnarray}
\label{t_8}
& &\hat{t}_8^{N_1\dots N_8}   = \frac{1}{16} \biggl( -  2 \left(   G^{ N_1 N_3  }G^{  N_2  N_4  }G^{ N_5   N_7  }G^{ N_6 N_8  }
 + G^{ N_1 N_5  }G^{ N_2 N_6  }G^{ N_3   N_7  }G^{  N_4   N_8   }
 +  G^{ N_1 N_7  }G^{ N_2 N_8  }G^{ N_3   N_5  }G^{  N_4 N_6   }  \right) \nonumber \\
 & &  +
 8 \left(  G^{  N_2     N_3   }G^{ N_4    N_5  }G^{ N_6    N_7  }G^{ N_8   N_1   }
  +G^{  N_2     N_5   }G^{ N_6    N_3  }G^{ N_4    N_7  }G^{ N_8   N_1   }
  +   G^{  N_2     N_5   }G^{ N_6    N_7  }G^{ N_8    N_3  }G^{ N_4  N_1   }
\right) \nonumber \\
& &  - (N_1 \leftrightarrow  N_2) -( N_3 \leftrightarrow  N_4) - (N_5 \leftrightarrow  N_6) - (N_7 \leftrightarrow  N_8) \biggr)
\end{eqnarray}
} ($N_i$ belongs to the ${\bf 8}_v$ of $SO(8)$ or covariantized to $10D$ (or $11D$ that will be used later in {\cal M}-theory)), from \cite{Tseytlin} the light-cone
8D ``zero mode'' tensor $t_8$ is generalized to $10D$: $\hat{t}_8 = t_8 - 1/4 B \epsilon_{10}$ where if one assumes $B_{\rm LC directions}=1$, implies $\hat{t}^{i_1j_1...i_4j_4} = t^{i_1j_1...i_4j_4} - \frac{1}{2}\epsilon^{i_1j_1...i_4j_4}$, and an overall factor of $e^{2\pi i  \tau_0}$, the signature of stringy D-instanton, is computed at $\chi =\Re \tau_0 =0$ in the stringy calculation.

\item Green and Vanhove showed that the eleven-dimensional ${\cal O}(R^4)$ adjustments possess a separate motive based on ten-dimensional supersymmetry \cite{Green and Vanhove}.
The above has been shown by its relationship with the term $C^{(3)}\wedge X_8$ in the ${\cal M}$-theory effective action, that is believed to originate through a number of reasons, including anomaly cancellation \cite{Duff, Horava and Witten}. The formula $X_8$ represents the eight-form in the curvatures obtained via the term in type IIA superstring theory \cite{Vafa and Witten}:
\begin{equation}-   \int d^{10}x  B\wedge  X_8 =  -
{1\over 2}
\int  d^{10}x \sqrt{-g^{A(10)}}\epsilon_{10} B X_8,
\end{equation}
where
\begin{equation}
\label{X_8-def}
X_8 = {1 \over 192} \left( {\rm tr}\ R^4 -
{1\over 4} ({\rm tr}\ R^2)^2\right).
\end{equation}

There exist two distinct ten-dimensional $N=1$ super-invariants with an odd-parity term (\cite{Tseytlin} and previous authors):
$
I_3= t_8 {\rm tr}\ R^4 - {1\over 4} \epsilon_{10}B {\rm tr}\ R^4
$
 and:
$
I_4= t_8 ({\rm tr}\ R^2)^2 - {1\over 4} \epsilon_{10}B ({\rm tr}\ R^2)^2.
$
 Utilizing  that
$
t_8t_8R^4=24t_8{\rm tr}(R^4)-6t_8({\rm tr}\ R^2)^2,
$
as a result, the specific linear combination,
\begin{equation}
I_3 - {1\over 4}I_4 = {1\over 24} t_8 t_8 R^4 - 48 \epsilon_{10}B\ X_8
\end{equation}
includes the ten-form $B \wedge X_8$ as well as $t_8 t_8 R^4$. 
\end{itemize}
Now let us discuss the ${\cal M}$-theory uplift in the presence of ${\cal O}(R^4)$ terms based on \cite{HD-MQGP}. The working action of eleven-dimensional supergravity with ${\cal N}=1$ supersymmetry inclusive of ${\cal O}R^4)$ terms is written below:
\begin{eqnarray}
\label{D=11_O(l_p^6)-intro}
& & \hskip -0.8inS = \frac{1}{2\kappa_{11}^2}\int_M\sqrt{-g}\left[  {\cal R} *_{11}1 - \frac{1}{2}G_4\wedge *_{11}G_4 -
\frac{1}{6}C\wedge G\wedge G\right] + \frac{1}{\kappa_{11}^2}\int_{\partial M} d^{10}x \sqrt{h} K \nonumber\\
& & \hskip -0.8in+ \frac{1}{(2\pi)^43^22^{13}}\left(\frac{2\pi^2}{\kappa_{11}^2}\right)^{\frac{1}{3}}\int d^{11}x\sqrt{-g}\left( J_0 - \frac{1}{2}E_8\right) + \left(\frac{2\pi^2}{\kappa_{11}^2}\right)\int C_3\wedge X_8,
\end{eqnarray}
where $\kappa_{11}^2$ is the eleven-dimensional Planckian length, and four-form field strength is defined as $G=dC$ for the three-form potential $C$. $J_0$ and $E_8$ are define below:
\begin{eqnarray}
\label{J0+E8-definitions}
& & \hskip -0.8inJ_0  =3\cdot 2^8 (R^{HMNK}R_{PMNQ}{R_H}^{RSP}{R^Q}_{RSK}+
{1\over 2} R^{HKMN}R_{PQMN}{R_H}^{RSP}{R^Q}_{RSK})\nonumber\\
& & \hskip -0.8inE_8  ={ 1\over 3!} \epsilon^{ABCM_1 N_1 \dots M_4 N_4}
\epsilon_{ABCM_1' N_1' \dots M_4' N_4' }{R^{M_1'N_1'}}_{M_1 N_1} \dots
{R^{M_4' N_4'}}_{M_4 N_4},\nonumber\\
& & \hskip -0.8in\kappa_{11}^2 = \frac{(2\pi)^8 l_p^{9}}{2};
\end{eqnarray}
Equations of motion of the ${\cal M}$-theory metric and three-form potential are given below.
\begin{eqnarray}
\label{eoms-intro}
& & R_{MN} - \frac{1}{2}g_{MN}{\cal R} - \frac{1}{12}\left(G_{MPQR}G_N^{\ PQR} - \frac{g_{MN}}{8}G_{PQRS}G^{PQRS} \right)\nonumber\\
 & &  = - \beta\left[\frac{g_{MN}}{2}\left( J_0 - \frac{1}{2}E_8\right) + \frac{\delta}{\delta g^{MN}}\left( J_0 - \frac{1}{2}E_8\right)\right],\nonumber\\
& & d*G = \frac{1}{2} G\wedge G +3^22^{13} \left(2\pi\right)^{4}\beta X_8,
\end{eqnarray}
where $\beta$ is defined as \cite{Becker-sisters-O(R^4)}:
\begin{equation}
\label{beta-def}
\beta \equiv \frac{\left(2\pi^2\right)^{\frac{1}{3}}\left(\kappa_{11}^2\right)^{\frac{2}{3}}}{\left(2\pi\right)^43^22^{12}} \sim l_p^6,
\end{equation}
$R_{MNPQ}, R_{MN}, {\cal R}$  in  (\ref{D=11_O(l_p^6)-intro})/(\ref{eoms-intro}) being the Riemann curvature tensor, the Ricci tensor, and the Ricci scalar, respectively, in the elven-dimensions. Ansatz made by authors to solve (\ref{eoms-intro}) in \cite{HD-MQGP} are written as 
\begin{eqnarray}
\label{ansaetze-intro}
& & \hskip -0.8ing_{MN} = g_{MN}^{(0)} +\beta g_{MN}^{(1)},\nonumber\\
& & \hskip -0.8inC_{MNP} = C^{(0)}_{MNP} + \beta C_{MNP}^{(1)}.
\end{eqnarray}
Equation of motion for the three-form potential symbolically written as \cite{HD-MQGP}
\begin{eqnarray}
\label{deltaC=0consistent-intro}
& & \beta \partial\left(\sqrt{-g}\partial C^{(1)}\right) + \beta \partial\left[\left(\sqrt{-g}\right)^{(1)}\partial C^{(0)}\right] + \beta\epsilon_{11}\partial C^{(0)} \partial C^{(1)} = {\cal O}(\beta^2) \sim 0 [{\rm up\ to}\ {\cal O}(\beta)].
\nonumber\\
& & \end{eqnarray}
Authors found that $C^{(1)}_{MNP}=0$ for the consistent set of solutions to (\ref{deltaC=0consistent-intro}) up to ${\cal O}(\beta)$. Therefore, ${\cal M}$-theory receives the ${\cal O}(R^4)$ corrections but not the three-form potential. The correction term of the metric is written as follows.
\begin{eqnarray}
\label{fMN-definitions}
\delta g_{MN} =\beta g^{(1)}_{MN} = G_{MN}^{\rm MQGP} f_{MN}(r),
\end{eqnarray}
without summation in the indices. \par
Let's discuss how to control the higher derivative contributions to the MQGP metric \cite{MQGP}. This was discussed in \cite{HD-MQGP}, and we are going to review it here. It is obvious from (\ref{M-theory-metric-psi=2npi-patch}) in the $\psi=2n\pi,n=0, 1, 2$-coordinate patch  that in the IR: $r = \chi r_h, \chi\equiv {\cal O}(1)$, and up to ${\cal O}(\beta)$:
\begin{equation}
\label{IR-beta-N-suppressed-logrh-rh-neg-exp-enhanced}
f_{MN} \sim \beta\frac{\left(\log {\cal R}_h\right)^{m}}{{\cal R}_h^n N^{\beta_N}},\ m\in\left\{0,1,3\right\},\ n\in\left\{0,2,5,7\right\},\
\beta_N>0,
\end{equation}
where ${\cal R}_h\equiv\frac{r_h}{{\cal R}_{D5/\overline{D5}}}$. Now, $|{\cal R}_h|\ll1$ and as estimated in \cite{Bulk-Viscosity-McGill-IIT-Roorkee} that $|\log {\cal R}_h|\sim N^{\frac{1}{3}}$ implies the competition between Planckian and large-$N$ suppression and infra-red enhancement arising from $m,n\neq0$ in (\ref{IR-beta-N-suppressed-logrh-rh-neg-exp-enhanced}). Let us choose a hierarchy: $\beta\sim e^{-\gamma_\beta N^{\gamma_N}}, \gamma_\beta,\gamma_N>0: \gamma_\beta N^{\gamma_N}>7N^{\frac{1}{3}} + \left(\frac{m}{3} - \beta_N\right)\log N$ such that IR-enhancement does not overpower Planckian suppression that is why authors considered the ${\cal O}(\beta)$ correction to $G^{\cal M}_{yz}$, which had the largest IR enhancement, to set a lower bound on $\gamma_{\beta,N}$/Planckian suppression. Authors concluded that one will go beyond ${\cal O}(\beta)$ if $\gamma_\beta N^{\gamma_N}\sim7N^{\frac{1}{3}}$. Therefore, they restricted themselves at ${\cal O}(\beta)$.

%m

\chapter{SU(3) LECs from Type IIA String Theory}
\graphicspath{{Chapter2/}{Chapter2/}}
This chapter is based on the paper \cite{MChPT}. Some portion of \cite{MChPT} is already present in a thesis of one of the co-authors (VY) \cite{Vikas-Thesis}. Detailed calculations of the relevant quantities appearing in this chapter are given in \cite{Vikas-Thesis}. Hence, we will quote those results in appendix \ref{appendix-MChPT} without going into the details and use those results to calculate the low energy coupling constants (LECs) of $SU(3)$ chiral perturbation theory in the chiral limit.

\section{Introduction}
\label{Intro-MChPT}
The low energy regime of quantum chromodynamics (QCD) is described by chiral perturbation theory ($\chi$PT), which includes the hadrons (mesons and baryons) as degrees of freedom. One constructs the Lagrangian of $\chi$PT based on the symmetry of the theory, and the symmetries are chiral symmetry $SU(N_f)_L\times SU(N_f)_R$, charge conjugation, and parity symmetry. The theory turns out to be renormalized order-by-order in momentum expansion. One obtains the $(N_f^2 -1)$ (pseudo-)Goldstone bosons from the spontaneous symmetry breaking of the chiral symmetry mentioned above, where $N_f$ is the number of QCD flavors. In this chapter, we focused on the $SU(3)$ mesonic chiral perturbation theory, where degrees of freedom considered is $\rho$ vector meson and $\pi$-meson. \\
\ \ \ \ The procedure to include $\rho$ vector meson is the HLS (Hidden Local Symmetry) approach with gauge group is $G_{global}\times H_{local}$, where $G_{global} = SU(N_f)_L\times SU(N_f)_R$ and $H = SU(N_f)_V$ is the HLS \cite{HARADA}. Gauge bosons of HLS are the $\rho$ vector meson and its flavor partners. Expansion of generating functional of QCD is possible in terms of $p/\Lambda_\chi$ or $m/\Lambda_\chi$, where $\Lambda_\chi \sim 4\pi F_\pi \sim 1.1 GeV$ \cite{MG} is the chiral symmetry breaking scale. The $\chi$PT Lagrangian is constructed order by order in the derivative expansion and should be consistent with the symmetries mentioned in the first paragraph. The leading order Lagrangian is $O(p^2)$. At leading order, there are two coupling constants $F_\pi^2$ and $F_\sigma^2$(Where $F_\pi$ and $F_\sigma$ are decay constants of $\pi$ and $\sigma$). $SU(2)$ and $SU(3)$ $\chi$PT Lagrangians were worked out by Gasser and Leutwyler in \cite{GLF} and \cite{GL} respectively. The NLO $SU(3)$ $\chi$PT Lagrangian contains 12 coupling constants ``$(L_{i = 1,2,..10}, H_1, H_2)$'' known as low energy constants (LECs). $L_4, L_5, L_6, L_8$ were calculated from the Lattice simulation in \cite{MILC}, $L_{i = 1,2,..10}$ were obtained at the scale $\mu = M_\rho$ in \cite{Pich},  see \cite{Ecker-2015} for the most update values.\\
\ \ \ \ 
Holographic computation of $SU(3)$ LECs was done in \cite{HARADA} from the top-down Sakai-Sugimoto model \cite{SS1}. The Sakai-Sugimoto model is not UV complete. The Lagrangian obtained in \cite{HARADA} is different from the $SU(3)$ $\chi$PT Lagrangian given in \cite{GL}. The paper \cite{HMM} contains the relationship between the $SU(3)$ LECs and the coupling constants obtained in \cite{HARADA}. We obtained these LECs from a UV complete top-down holographic dual. The gravity dual is type IIA string theory in the presence of higher derivative terms. Type IIA background had been obtained by descending back from the ${\cal M}$-theory to type IIA string theory.

\section{$SU(3)$ Chiral Perturbation Theory Lagrangian and the Phenomenological Values of LECs}
\label{SU(3)-ChPT-LECs}

The term appearing in the $SU(3)\ \chi$PT Lagrangian at ${\cal O}(p^4)$ with one-loop renormalized low energy constants (LECs),$L_i$ and $H_i$, in chiral limit are written as follows \cite{GL}:
\begin{eqnarray}
\label{MChPT-Op4}
& & L_1 \left({\rm Tr}(\nabla_\mu U^\dagger \nabla^\mu U)\right)^2 + L_2\left({\rm Tr}(\nabla_\mu U^\dagger \nabla_\nu U)\right)^2 + L_3{\rm Tr} \left(\nabla_\mu U^\dagger \nabla^\mu U\right)^2\nonumber\\
& & - i L_9 Tr\left({\cal L}_{\mu\nu}\nabla^\mu U \nabla^\nu U^\dagger + {\cal R}_{\mu\nu}\nabla^\mu U \nabla^\nu U^\dagger\right) + L_{10} Tr\left(U^\dagger {\cal L}_{\mu\nu}U{\cal R}^{\mu\nu}\right) + H_1 Tr\left({\cal L}_{\mu\nu}^2 + {\cal R}_{\mu\nu}^2\right),\nonumber\\
& &
\end{eqnarray}
where $\nabla_\mu U\equiv \partial_\mu U - i {\cal L}_\mu U + i U {\cal R}_\mu,\ U=e^{\frac{2i\pi}{F_\pi}}$. LECs, $L_i$ and $H_i$ are defined via the dimensional regularization and renormalizations of the parameters as follow \cite{GL}:
\begin{equation}
\label{Li-GL}
L_i = L_i^r(\mu) + \Gamma_i \lambda(\mu) \ , \qquad
H_i = H_i^r(\mu) + \Delta_i \lambda(\mu) \ ,
\end{equation}
where $\mu$ is the renormalization scale,
and for $SU(3)$ $\chi$PT, $\Gamma_i$ and $\Delta_i$ are given as \cite{GL}:
\begin{equation}
\begin{array}{ccccc}
\Gamma_1 = \frac{3}{32} \ ,
& \Gamma_2 = \frac{3}{16} \ ,
& \Gamma_3 = 0 \ , 
& \Gamma_4 = \frac{1}{8} \ ,
& \Gamma_5 = \frac{3}{8} \ ,
\\
\Gamma_6 = \frac{11}{144} \ ,
& \Gamma_7 = 0 \ ,
& \Gamma_8 = \frac{5}{48} \ ,
& \Gamma_9 = \frac{1}{4} \ , 
& \Gamma_{10} = - \frac{1}{4} \ ;
\\
\Delta_1 = - \frac{1}{8} \ ,
& \Delta_2 = \frac{5}{24} \ .
& & &
\end{array}
\end{equation}
The divergent piece, $\lambda(\mu)$ appearing in (\ref{Li-GL}) is given below.
\begin{equation}
\lambda(\mu) = - \frac{1}{2\left(4\pi\right)^2}
\left[   \frac{1}{\bar{\epsilon}} - \ln \mu^2 + 1 \right]
\ ,
\end{equation}
where
\begin{equation}
\frac{1}{\bar{\epsilon}} = \frac{2}{4-d}
- \gamma_E + \ln 4\pi 
\ ,
\end{equation}
where $d=4$. The phenomenological values of the one-loop renormalized $SU(3)$ LECs appearing in (\ref{MChPT-Op4}) are listed in the table \ref{Table-Lis} \cite{Ecker-2015}. In table \ref{Table-Lis-Source} sources from where the LECs have been extracted is given for the column titled {\bf GL 1985} \cite{GL}.
\begin{table}
\centering
\begin{tabular}{c|c|c|c|c}
\hline
\textbf{LECs} & \textbf{GL 1985} \cite{GL} & \textbf{NLO 2014} & \textbf{NNLO free fit} & \textbf{NNLO BE14} \cite{BE14} \\
\hline
%$10^{3}$ $L_A^{r}=(2L_1^r - L_2^r)$ &  &0.4(2) & 0.68(11) & 0.24(11)\\
%\hline
$10^{3}$ $L_1^{r}$ &0.7(3) & 1.0(1) & 0.64)06 & 0.53(06) \\
\hline
$10^{3}$ $L_2^{r}$ & 1.3(7) & 1.6(2) & 0.59(04) & 0.81(04) \\
\hline
$10^{3}$ $L_3^{r}$ & -4.4(2.5) & -3.8(3) & -2.80(20) & -3.07(20) \\
\hline
$10^{3}$ $L_4^{r}$ & -0.3(5) & 0.0(3) & 0.76(18) & 0.3 \\
\hline
$10^{3}$ $L_5^{r}$ & 1.4(5) & 1.2(1) & 0.50(07) & 1.01(06) \\
\hline
$10^{3}$ $L_6^{r}$ & -0.2(3) & 0.0(4) & 0.49(25) & 0.14(05) \\
\hline
$10^{3}$ $L_7^{r}$ & -0.4(2) & -0.3(2) & -0.19(08) & -0.34(09) \\
\hline
$10^{3}$ $L_8^{r}$ & 0.9(3) & 0.5(2) & 0.17(11) & 0.47(10) \\
\hline
%$F_{0}$(Mev) & 64 & 71
\end{tabular}
\caption{$SU(3)$ NLO LECs $L_i^{r}(i=1,2,...,8)$ from various references.}
\label{Table-Lis}
\end{table}
%%%%%%%%%%%%%%%%%%%%%%%%
\begin{table}
\centering
\begin{tabular}{c|c|c}
 \hline
i & $L_i^{r}(M_{\rho})$ $10^{3}$ & Source \\
\hline
 1 & 0.4 $\pm$ 0.3 & $K_{e4}$, $\pi\pi \rightarrow \pi\pi$\\
  \hline
 2 & 1.4 $\pm$ 0.3 & $K_{e4}$, $\pi\pi \rightarrow \pi\pi$\\
  \hline
 3 & -3.5 $\pm$ 1.1 & $K_{e4}$, $\pi\pi \rightarrow \pi\pi$\\
  \hline
 4 & -0.3 $\pm$ 0.5 & Zweig rule \\
  \hline
 5 & 1.4 $\pm$ 0.5 & $F_K : F_\pi$ \\
  \hline
 6 & -0.2 $\pm$ 0.3 & Zweig rule \\
  \hline
 7 & -0.4 $\pm$ 0.2 & Gell-Mann-Okubo, $L_5$, $L_8$ \\
  \hline
 8 & 0.9 $\pm$ 0.3 & $M_{K0} - M_{K_+}$, $L_5$, $(m_s - \hat{m}):(m_d - m_u)$\\
  \hline
 9 & 6.9 $\pm$0.7 & $<r^2>_V^\pi$\\
 \hline
 10 & -5.5 $\pm$ 0.7 & $\pi \rightarrow e\nu\gamma$ \\
  \hline
\end{tabular}
\caption{Phenomenological Values of the 1-loop renormalised couplings $L_i^{r}(M_{\rho})$ of (\ref{MChPT-Op4}) \cite{Pich}. Last column shows the source to extract this information.}
\label{Table-Lis-Source}
\end{table}
In the chiral perturbation theory that is used in gauge-gravity duality, the holographic renormalization acts as the analogue of the one-loop renormalization. The eleven dimensional on-shell supergravity action including ${\cal O}(\beta)$ correction was worked out in \cite{McTEQ} and written below:
{\footnotesize
\begin{equation}
\label{on-shell-D=11-action-up-to-beta-MChPT}
 S_{D=11}^{\rm on-shell} = -\frac{1}{2}\Biggl[-{2}S_{\rm EH}^{(0)} + 2 S_{\rm GHY}^{(0)}+ \beta\left(\frac{20}{11}S_{\rm EH} - {2}\int_{M_{11}}\sqrt{-g^{(1)}}R^{(0)}
+ 2 S_{\rm GHY} - \frac{2}{11}\int_{M_{11}}\sqrt{-g^{(0)}}g_{(0)}^{MN}\frac{\delta J_0}{\delta g_{(0)}^{MN}}\right)\Biggr].
\end{equation}
}
UV divergent terms that exist in the (\ref{on-shell-D=11-action-up-to-beta-MChPT}) have the following forms:
\begin{eqnarray}
\label{UV_divergences}
& &\left. \int_{M_{11}}\sqrt{-g}R\right|_{\rm UV-divergent},\ \left.\int_{\partial M_{11}}\sqrt{-h}K\right|_{\rm UV-divergent} \sim r_{\rm UV}^4 \log r_{\rm UV},\nonumber\\
& & \left.\int_{M_{11}} \sqrt{-g}g^{MN}\frac{\delta J_0}{\delta g^{MN}}\right|_{\rm UV-divergent} \sim
\frac{r_{\rm UV}^4}{\log r_{\rm UV}}. 
\end{eqnarray}
These UV divergences had been cancelled by using the boundary counter terms in a suitable linear combination and the counter terms are: $\left.\int_{\partial M_{11}}\sqrt{-h}K\right|_{r=r_{\rm UV}}$ and $\left.\int_{\partial M_{11}}\sqrt{-h}h^{mn}\frac{\partial J_0}{\partial h^{mn}}\right|_{r=r_{\rm UV}}$. %\footnote{For consistency, one needs to impose the following relationship between the UV-valued effective number of flavor $D7$-branes of the parent type IIB dual, $N_f^{\rm UV}$ and $\log r_{\rm UV}$: $N_f^{\rm UV} = \frac{\left(\log r_{\rm UV}\right)^{\frac{15}{2}}}{\log N}$.}. 

\section{Holographic Computation of $SU(3)$ LECs}
\label{HC-LECs}
This section will discuss the computation of $SU(3)$ LECs holographically. We considered the gravity dual as type IIA string dual inclusive of ${\cal O}(R^4)$ corrections. Since chiral perturbation theory is the low energy regime of QCD ($T<T_c$), we worked with the thermal metric in our calculations. Mesons are the fluctuations of the flavor $D6$-branes in type IIA string dual of the type IIB string dual constructed from a top-down approach \cite{metrics}. We started with the ${\cal M}$-theory uplift constructed in \cite{HD-MQGP} in the presence of ${\cal O}(R^4)$ terms and then descent back to type IIA string theory. Flavor $D6$-branes is given as: $\iota :\Sigma ^{1,6}\Bigg( R^{1,3},r,\theta_{2}\sim\frac{\alpha_{\theta_{2}}}{N^{\frac{3}{10}}},y\Bigg)\hookrightarrow M^{1,9}$ near the Ouyang embedding (\ref{Ouyang-definition}) for a vanishingly small $|\mu_{\rm Ouyang}|$. The embedding is parametrized by $z=z(r)$. Similar to \cite{Yadav+Misra+Sil-Mesons}, it turns out that the embedding solution is $z$=constant, and if we choose $z=\pm {\cal C}\frac{\pi}{2}$, then $D6/\overline{D6}$-branes will be located at  ``antipodal'' points along the z coordinate. We worked with the redefined radial coordinates $(r,z)$ in terms of the new coordinates $(Y, Z)$ as used in \cite{Yadav+Misra+Sil-Mesons}.
\[r=r_{0}e^{\sqrt{Y^{2}+Z^{2}}} ; \ z={\cal C}\arctan\frac{Z}{Y}. \]
Fluctuations of the $D6$-branes along the world volume correspond to the vector mesons, and fluctuations orthogonal to the $D6$-branes correspond to the scalar mesons\footnote{See \cite{meson-N=2,A-N=2} where the mesons spectrum have been studied for $N=2$ super Yang-Mills theory with fundamental matter using AdS/CFT correspondence. See \cite{Nunez-Mesons}, where the mesonic excitations have been studied from a non Abelian T-dual of $AdS_5 \times S^5$ background.}. The world volume coordinates of the $D6$-branes are ${\Sigma}_7(x^{0,1,2,3},Z,\theta_2,\tilde{y}) = {\Sigma}_2(\theta_2,\tilde{y})\times{\Sigma}_5(x^{0,1,2,3},Z)$. We considered $\tilde{F}$ as the gauge field fluctuation about a small background gauge filed $F_0$ and induced background $i^*(g+B) [i:\Sigma_7\hookrightarrow M_{10}$,  where $M_{10}$ is the type IIA geometry]. The SYZ mirror of the Ouyang embedding appears to be written as $Y=0$ \cite{Yadav+Misra+Sil-Mesons}. We found the following DBI action after retaining the terms up to quadratic order in $\tilde{F}$:
 {\footnotesize
\begin{equation}
\label{DBI action}
{\rm S}^{IIA}_{D6}=\frac{T_{D_6}(2\pi\alpha^\prime)^{2}}{4} \left(\frac{\pi L^2}{r_0}\right){\rm Str}\int \prod_{i=0}^3dx^i dZd\theta_{2}dy \delta\Bigg(\theta_{2}-\frac{\alpha_{\theta_{2}}}{N^{\frac{3}{10}}}\Bigg) e^{-\Phi} \sqrt{-{\rm det}_{\theta_{2}y}(\iota^*(g+B))}\sqrt{{\rm det}_{{\mathbb  R}^{1,3},Z}(\iota^*g)}g^{\tilde{\mu}\tilde{\nu}}\tilde{F}_{\tilde{\nu}\tilde{\rho}}g^{\tilde{\rho}\tilde{\sigma}}\tilde{F}_{\tilde{\sigma}\tilde{\mu}},
\end{equation}
}
where $\tilde{\mu}=i(=0,1,2,3),Z$. Gauge fields in five dimensions are expanded as: \\ $A_\mu(x^\nu,Z) = \sum_{n=1}^\infty \rho^{(n)}_\mu(x^\nu)\psi_n(Z)$ and $A_Z(x^\nu,Z) = \sum_{n=0}^\infty\pi^{(n)}(x^\nu)\phi_n(Z) $. If we consider the lowest lying scalar mode in the aforementioned expansions, which is the pseudo-scalar $\pi$ meson and after the application of gauge transformation:$\rho_{\mu}^{(n)}\rightarrow\rho_{\mu}^{(n)}+{\cal M}_{(n)}^{-1}\partial_{\mu}\pi^{(n)}.$ We obtained the following action written in terms of the $\pi$ meson and $\rho_{\mu}^{(n)}$ vectors mesons.
\begin{eqnarray}
& &{\rm S}^{IIA}_{D6} \sim -\int d^3x\ \ {\rm tr} \left[\frac{1}{2}\partial_{\mu}\pi^{(0)}\partial^{\mu}\pi^{(0)} + \sum_{n\ge 1}\left(\frac{1}{4}\tilde{F}^{(n)}_{\mu\nu}\tilde{F}^{(n)\mu\nu}+\frac{m_{n}^2}{2}\rho^{(n)}_\mu \rho^{(n)\mu }\right)\right].
\end{eqnarray}
Now we integrate out the all higher order vector and axial vector mesons and keep only the lowest vector mesons ($\rho$ meson) in the gauge $A_Z(x^\mu, Z)=0$ \cite{HARADA}:\\ {$V_\mu^{(1)}(x^\mu) = g \rho_\mu(x^\mu) = \left(\begin{array}{ccc}
\frac{1}{\sqrt{2}}\left(\rho_\mu^0 + \omega_\mu\right) & \rho_\mu^+ & K_\mu^{*+} \\
\rho_\mu^- & -\frac{1}{\sqrt{2}}\left(\rho_\mu^0 - \omega_\mu\right) & K_\mu^{*0} \\
K_\mu^{*-} & {\bar K}_\mu^{*0}  & \phi_\mu
\end{array} \right)$} and the lightest pseudo-scalar meson $\pi$ is defined as:  \\{$\pi = \frac{1}{\sqrt{2}}\left(\begin{array}{ccc}
\frac{1}{\sqrt{2}}\pi^0 + \frac{1}{\sqrt{6}}\eta_8 + \frac{1}{\sqrt{3}}\eta_0 & \pi^+ & K^+ \\
\pi^- & -\frac{1}{\sqrt{2}}\pi^0 + \frac{1}{\sqrt{6}}\eta_8 + \frac{1}{\sqrt{3}}\eta_0 & K^0 \\
K^- & {\bar K}^0 & -\frac{2}{\sqrt{6}}\eta_8 + \frac{1}{\sqrt{3}}\eta_0
\end{array}\right)$}. The expression for the gauge field $A_\mu(x^\nu,Z)$ up to ${\cal O}(\pi)$ is as follows:
\begin{eqnarray}
\label{Amu-exp}
& & A_\mu(x^\nu,Z) = \frac{\partial_\mu\pi}{F_\pi}\psi_0(Z) - V_\mu(x^\nu)\psi_1(Z),
\end{eqnarray}
where $\psi_0(z) = \int^Z_0 dZ^\prime \phi_0(Z^\prime), V_\mu^{(1)}(x^\nu) = \rho^{(1)}_\mu - \frac{1}{{\cal M}_{(1)}}\partial_\mu\pi^{(1)}$.
Our motivation was to compute the LECs of $SU(3)$ chiral perturbation theory and we followed the Hidden Local Symmetry (HLS) approach of \cite{HARADA}. To do so, we introduced external vector ${\cal V}_\mu$ and axial vector ${\cal A}_\mu$ fields in addition to the $\pi$ and $\rho$ vector mesons  similar to \cite{HARADA} where we defined  $\frac{1}{F_\pi} \partial_\mu \pi\rightarrow\hat{\alpha}_{\mu \perp}= \frac{1}{F_\pi} \partial_\mu \pi + {\cal A}_\mu - \frac{i}{F_\pi}[{\cal V}_\mu,\pi] + \cdots$ , and $\hat{\alpha}_{\mu ||} \equiv -V_\mu + {\cal V}_\mu - \frac{i}{2F_\pi^2}[\partial_\mu\pi,\pi] + \cdots$. Therefore $\hat{\alpha}_{\mu \perp}$ and $\hat{\alpha}_{\mu ||}$ act as the degrees of freedom in this approach. As we discussed earlier that there are infinite number of fields in the mode expansion gauge and scalar fields. Hence, we truncated the spectrum in such a way that the remaining fields are lowest vector meson $(V_\mu^{(1)} = g \rho_\mu)$ and lightest psuedo-scalar meson ($\pi$ meson)\cite{HARADA}. In terms of the  $\hat{\alpha}_{\mu \perp}$ and $\hat{\alpha}_{\mu ||}$ , the gauge field is written as:
\begin{equation}
A_\mu(x^\mu,z) = \hat{\alpha}_{\mu \perp}(x^\mu) \psi_0(z)
  + (\hat{\alpha}_{\mu ||}(x^\mu) + V_\mu^{(1)}(x^\mu)  )
  + \hat{\alpha}_{\mu ||}(x^\mu)  \psi_1(z),
  \end{equation}
implying
  \begin{eqnarray}
\label{F mu nu}
& &   F_{\mu\nu} = -V_{\mu\nu} \psi_1 +v_{\mu\nu}(1+\psi_1) + a_{\mu\nu} \psi_0 - i[\hat{\alpha}_{\mu ||},\hat{\alpha}_{\nu ||}]\psi_1(1+\psi_1)  + i[\hat{\alpha}_{\mu \perp},\hat{\alpha}_{\nu \perp}](1+\psi_1-\psi_0^{2}) \nonumber\\
  & &
  -i([\hat{\alpha}_{\mu \perp},\hat{\alpha}_{\nu ||}]+[\hat{\alpha}_{\mu ||},\hat{\alpha}_{\nu \perp}])\psi_1 \psi_0,
 \end{eqnarray}
where $\psi_0(\equiv \psi_0(Z)$ and $\psi_1(\equiv \psi_1(Z)$ are the profile functions for the psuedo-scalar $\pi$ meson and $\rho$ vector meson given in appendix \ref{appendix-MChPT}. Based on \cite{HLS-Physics-Reports}, as concerns a chiral power counting, one notices that $M_\rho\equiv{\cal O}(p)$ implying $\hat{\alpha}_{\nu ||}\equiv \frac{{\cal O}(p^3)}{M_\rho^2}\equiv{\cal O}(p),
\hat{\alpha}_{\nu\perp}\equiv {\cal O}(p)$. Further,$V_{\mu\nu}, a_{\mu\nu}$ and $v_{\mu\nu}$ are of ${\cal O}(p^2)$. Therefore, when using (\ref{F mu nu}), $\left(F_{\mu\nu}F^{\mu\nu}\right)^m$  is of ${\cal O}(p^{4m}), m\in\mathbb{Z}^+$. Because of this, one must take into account the kinetic term ($m=1$) at ${\cal O}(p^4)$, which results in the expression shown below:
 \begin{eqnarray}
\label{F_munu F^munu}
& &  F_{\mu\nu}F^{\mu\nu} = \psi_1^2 V_{\mu\nu}V^{\mu\nu} - \psi_1(1+\psi_1) V_{\mu\nu}v^{\mu\nu} - \psi_0\psi_1 V_{\mu\nu}a^{\mu\nu} +i \psi_1^2(1+\psi_1)V_{\mu\nu}[\hat{\alpha}^\mu_{||},\hat{\alpha}^\nu_{ ||}] \nonumber\\
 &&
 - i \psi_1(1+\psi_1 - \psi_0^2)V_{\mu\nu}[\hat{\alpha}^\mu_{\perp},\hat{\alpha}^\nu_{\perp}] + i\psi_0\psi_1^2V_{\mu\nu}([\hat{\alpha}^{\mu}_{\perp},\hat{\alpha}^{\nu}_{||}]+[\hat{\alpha}^{\mu}_{||},\hat{\alpha}^{\nu}_{\perp}]) -  \psi_1(1+\psi_1) v_{\mu\nu}V^{\mu\nu} \nonumber\\
 && + (1+\psi_1)^2 v_{\mu\nu}v^{\mu\nu}+\psi_0(1+\psi_1) v_{\mu\nu}a^{\mu\nu} - i \psi_1(1+\psi_1)^2 v_{\mu\nu}[\hat{\alpha}^\mu_{||},\hat{\alpha}^\nu_{ ||}] \nonumber\\
 && +i(1+\psi_1) (1+\psi_1 - \psi_0^2)v_{\mu\nu}[\hat{\alpha}^\mu_{\perp},\hat{\alpha}^\nu_{\perp}] - i\psi_0\psi_1(1+\psi_1)v_{\mu\nu}([\hat{\alpha}^{\mu}_{\perp},\hat{\alpha}^{\nu}_{||}]+[\hat{\alpha}^{\mu}_{||},\hat{\alpha}^{\nu}_{\perp}]) \nonumber\\
 &&  - \psi_0\psi_1 a_{\mu\nu}V^{\mu\nu}+\psi_0(1+\psi_1) a_{\mu\nu}v^{\mu\nu} +\psi_0^2 a_{\mu\nu}a^{\mu\nu} - i\psi_0\psi_1(1+\psi_1) a_{\mu\nu}[\hat{\alpha}^\mu_{||},\hat{\alpha}^\nu_{ ||}] \nonumber\\
 &&  +i\psi_0(1+\psi_1-\psi_0^2)a_{\mu\nu}[\hat{\alpha}^\mu_{\perp},\hat{\alpha}^\nu_{\perp}] - i\psi_0^2\psi_1 a_{\mu\nu}([\hat{\alpha}^{\mu}_{\perp},\hat{\alpha}^{\nu}_{||}]+[\hat{\alpha}^{\mu}_{||},\hat{\alpha}^{\nu}_{\perp}]) \nonumber\\
 &&
  +i \psi_1^2(1+\psi_1)[\hat{\alpha}_{\mu ||},\hat{\alpha}_{\nu ||}]V^{\mu\nu} - i \psi_1(1+\psi_1)^2[\hat{\alpha}_{\mu ||},\hat{\alpha}_{\nu ||}]v^{\mu\nu} \nonumber\\
 &&  -i\psi_0\psi_1(1+\psi_1)[\hat{\alpha}_{\mu ||},\hat{\alpha}_{\nu ||}]a^{\mu\nu} - \psi_1^2(1+\psi_1)^2[\hat{\alpha}_{\mu ||},\hat{\alpha}_{\nu ||}][\hat{\alpha}^\mu_{||},\hat{\alpha}^\nu_{||}]\nonumber\\
 &&  +\psi_1(1+\psi_1)(1+\psi_1-\psi_0^2)[\hat{\alpha}_{\mu ||},\hat{\alpha}_{\nu ||}][\hat{\alpha}^\mu_{\perp},\hat{\alpha}^\nu_{\perp}] - \psi_0\psi_1^2(1+\psi_1)[\hat{\alpha}_{\mu ||},\hat{\alpha}_{\nu ||}]([\hat{\alpha}^{\mu}_{\perp},\hat{\alpha}^{\nu}_{||}]+[\hat{\alpha}^{\mu}_{||},\hat{\alpha}^{\nu}_{\perp}]) \nonumber\\
 &&  - i\psi_1(1+\psi_1-\psi_0^2)[\hat{\alpha}_{\mu \perp},\hat{\alpha}_{\nu \perp}]V^{\mu\nu} +i(1+\psi_1)(1+\psi_1-\psi_0^2)[\hat{\alpha}_{\mu \perp},\hat{\alpha}_{\nu \perp}]v^{\mu\nu} \nonumber\\
 &&  + \psi_1(1+\psi_1)(1+\psi_1-\psi_0^2)[\hat{\alpha}_{\mu \perp},\hat{\alpha}_{\nu \perp}][\hat{\alpha}^\mu_{||},\hat{\alpha}^\nu_{||}]
  -(1+\psi_1-\psi_0^2)^2[\hat{\alpha}_{\mu \perp},\hat{\alpha}_{\nu \perp}][\hat{\alpha}^\mu_{\perp},\hat{\alpha}^\nu_{\perp}] \nonumber\\
 &&  + \psi_0\psi_1(1+\psi_1-\psi_0^2)[\hat{\alpha}_{\mu \perp},\hat{\alpha}_{\nu \perp}]  ([\hat{\alpha}^{\mu}_{\perp},\hat{\alpha}^{\nu}_{||}]+[\hat{\alpha}^{\mu}_{||},\hat{\alpha}^{\nu}_{\perp}]) \nonumber\\
 &&  +i \psi_0\psi_1^2([\hat{\alpha}_{\mu \perp},\hat{\alpha}_{\nu ||}]+[\hat{\alpha}_{\mu ||},\hat{\alpha}_{\nu \perp}])V^{\mu\nu} - i\psi_0\psi_1(1+\psi_1)([\hat{\alpha}_{\mu \perp},\hat{\alpha}_{\nu ||}]+[\hat{\alpha}_{\mu ||},\hat{\alpha}_{\nu \perp}])v^{\mu\nu} \nonumber\\
 &&  -i\psi_0^2\psi_1([\hat{\alpha}_{\mu \perp},\hat{\alpha}_{\nu ||}]+[\hat{\alpha}_{\mu ||},\hat{\alpha}_{\nu \perp}])a^{\mu\nu} -\psi_0\psi_1^2(1+\psi_1)([\hat{\alpha}_{\mu \perp},\hat{\alpha}_{\nu ||}]+[\hat{\alpha}_{\mu ||},\hat{\alpha}_{\nu \perp}])[\hat{\alpha}^\mu_{||},\hat{\alpha}^\nu_{ ||}] \nonumber\\
 && +\psi_0\psi_1(1+\psi_1-\phi_0^2)([\hat{\alpha}_{\mu \perp},\hat{\alpha}_{\nu ||}]+[\hat{\alpha}_{\mu ||},\hat{\alpha}_{\nu \perp}])[\hat{\alpha}^\mu_{\perp},\hat{\alpha}^\nu_{\perp}]  \nonumber\\
 && - \psi_0^2\psi_1^2([\hat{\alpha}_{\mu \perp},\hat{\alpha}_{\nu ||}]+[\hat{\alpha}_{\mu ||},\hat{\alpha}_{\nu \perp}])([\hat{\alpha}^{\mu}_{\perp},\hat{\alpha}^{\nu}_{||}]+[\hat{\alpha}^{\mu}_{||},\hat{\alpha}^{\nu}_{\perp}]).
  \end{eqnarray}
Parity is defined as: $x^i\rightarrow-x^i$, and $Z\rightarrow-Z$ where $i$ are the conformally Minkowskian spatial directions. When $A_\mu(x, Z)$ is odd then $\alpha_\perp$ is even and $\alpha_{||}$ is odd. Further, when $V_\mu$ is odd then $\psi_0(Z)$ and $\psi_1(Z)$ are odd and even respectively. Since the coupling constants should be scalar and even-$Z$ parity. We found that the terms with $(3\hat{\alpha} _{||}s\ ,\ 1\hat{\alpha}_\perp\ {\rm or}\ 3\hat{\alpha}_\perp s\ ,\ 1\hat{\alpha} _{||} )$ in (\ref{F_munu F^munu}) are odd under the Parity transformation. Similarly, term like $tr(\hat{\alpha}_{\mu \perp}\hat{\alpha}_{||}^{\mu})$ has been dropped because of odd-$Z$ parity at ${\cal O}(p^2)$. At ${\cal O}(p^2)$, we have the following Lagrangian \cite{MChPT,Vikas-Thesis}:
\begin{eqnarray}
\label{Lagrangian-Op2}
 & & \hskip -0.45in {\mathcal{L}}_{(2)}
\ni
F_\pi^2 \,
{\rm tr}[{\hat{\alpha}}_{\mu\perp}{\hat{\alpha}}^{\mu}_{\perp}]
+
a F_\pi^2 \,
{\rm tr}[{\hat{\alpha}}_{\mu ||}{\hat{\alpha}}^{\mu}_{||}]
- \frac{1}{2 g_{YM}^2} \,
{\rm tr}[V_{\mu\nu} V^{\mu\nu}].
\end{eqnarray}
At ${\cal O}(p^4)$, $SU(3)$ chiral perturbation theory Lagrangian is obtained as \cite{MChPT,Vikas-Thesis}:
\begin{eqnarray}
\label{Lagrangian-Op4}
 & & \hskip -0.45in {\mathcal{L}}_{(4)}
\ni
y_1 \,
{\rm tr}[{\hat{\alpha}}_{\mu\perp}{\hat{\alpha}}^{\mu}_{\perp}
{\hat{\alpha}}_{\nu\perp}{\hat{\alpha}}^{\nu}_{\perp}]
+
y_2 \,
{\rm tr}[{\hat{\alpha}}_{\mu\perp}{\hat{\alpha}}_{\nu\perp}
{\hat{\alpha}}^{\mu}_{\perp}{\hat{\alpha}}^{\nu}_{\perp}]
+y_3 \,
{\rm tr}[{\hat{\alpha}}_{\mu||}{\hat{\alpha}}^{\mu}_{||}
{\hat{\alpha}}_{\nu||}{\hat{\alpha}}^{\nu}_{||}]
+y_4 \,
{\rm tr}[{\hat{\alpha}}_{\mu||}{\hat{\alpha}}_{\nu||}
{\hat{\alpha}}^{\mu}_{||}{\hat{\alpha}}^{\nu}_{||}] \nonumber\\
& &
+y_5 \,
{\rm tr}[{\hat{\alpha}}_{\mu\perp}{\hat{\alpha}}^{\mu}_{\perp}
{\hat{\alpha}}_{\nu||}{\hat{\alpha}}^{\nu}_{||}]
+y_6 \,
{\rm tr}[{\hat{\alpha}}_{\mu\perp}{\hat{\alpha}}_{\nu\perp}
{\hat{\alpha}}^{\mu}_{||}{\hat{\alpha}}^{\nu}_{||}]
+y_7 \,
{\rm tr}[{\hat{\alpha}}_{\mu\perp}{\hat{\alpha}}_{\nu\perp}
{\hat{\alpha}}^{\nu}_{||}{\hat{\alpha}}^{\mu}_{||}] \nonumber\\
& &
+y_8 \,
\left\{ {\rm tr}[{\hat{\alpha}}_{\mu\perp}{\hat{\alpha}}^{\nu}_{||}
{\hat{\alpha}}_{\nu\perp}{\hat{\alpha}}^{\mu}_{||}]
+
{\rm tr}[{\hat{\alpha}}_{\mu\perp}{\hat{\alpha}}^{\mu}_{||}
{\hat{\alpha}}_{\nu\perp}{\hat{\alpha}}^{\nu}_{||}\right\}
+y_9 \,
{\rm tr}[{\hat{\alpha}}_{\mu\perp}{\hat{\alpha}}_{\nu ||}
{\hat{\alpha}}^{\mu}_{\perp}{\hat{\alpha}}^{\nu}_{||}]\nonumber\\
& & 
+z_1 \,
{\rm tr}[v_{\mu\nu}v^{\mu\nu}]
+z_2 \,
{\rm tr}[a_{\mu\nu}a^{\mu\nu}]
+z_3 \,
{\rm tr}[v_{\mu\nu}V^{\mu\nu}]
+iz_4 \,
{\rm tr}[V_{\mu\nu}{\hat{\alpha}}^{\mu}_{\perp}
{\hat{\alpha}}^{\nu}_{\perp}]\nonumber\\
& &
+iz_5 \,
{\rm tr}[V_{\mu\nu}{\hat{\alpha}}^{\mu}_{||}
{\hat{\alpha}}^{\nu}_{||}]
+iz_6 \,
{\rm tr}[v_{\mu\nu}
{\hat{\alpha}}^{\mu}_{\perp}{\hat{\alpha}}^{\nu}_{\perp}]
+iz_7 \,
{\rm tr}[v_{\mu\nu}
{\hat{\alpha}}^{\mu}_{||}{\hat{\alpha}}^{\nu}_{||}]
-iz_8 \,
{\rm tr}\left[a_{\mu\nu}
\left({\hat{\alpha}}^{\mu}_{\perp}{\hat{\alpha}}^{\nu}_{||}
+{\hat{\alpha}}^{\mu}_{||}{\hat{\alpha}}^{\nu}_{\perp}
\right)\right].
\end{eqnarray}
where
\begin{equation}
\hat{{\alpha}}_{\mu {||/\perp}}= \frac{i}{2}\left(\xi_R D_\mu \xi^\dag_R  \pm \xi_L D_\mu \xi^\dag_L \right); D_\mu \xi^\dag_{R/L}=\partial_\mu \xi^\dag_{R/L} - i ({\cal R/ \cal L})_\mu \xi^\dag_{R/L} + i \xi^\dag_{R/L} V_\mu.\nonumber
\end{equation}
\begin{equation}
 v_{\mu\nu}
= \frac{1}{2} \left(
\xi_R {\cal R}_{\mu\nu} \xi^\dag_R + \xi_L {\cal L}_{\mu\nu} \xi_L^\dag \right)  \hspace{0.5cm}
{\rm and} \hspace{0.5cm}
a_{\mu\nu}
=  \frac{1}{2} \left(
\xi_R {\cal R}_{\mu\nu} \xi^\dag_R - \xi_L {\cal L}_{\mu\nu} \xi_L^\dag
\right),\nonumber
\end{equation}
${\cal L}_{\mu\nu} = \partial_{[\mu}{\cal L}_{\nu]} - i[{\cal L}_\mu,{\cal L}_\nu]$ and ${\cal R}_{\mu\nu} = \partial_{[\mu}{\cal R}_{\nu]} - i[{\cal R}_\mu,{\cal R}_\nu]$ and ${\cal L}_\mu = {\cal V}_\mu - {\cal A}_\mu$ and ${\cal R}_\mu =  {\cal V}_\mu + {\cal A}_\mu $, and $\xi_L^\dagger(x^\mu) = \xi_R(x^\mu) = e^{\frac{i \pi(x^\mu)}{F_\pi}}$. Coupling constants appearing in (\ref{Lagrangian-Op2}) and (\ref{Lagrangian-Op4}) are given by the following radial integrals.
\begin{eqnarray}
\label{LECs-RI}
& & F_\pi^{2} = -\frac{{\cal V}_{\Sigma_2}}{4}<<\dot\psi_0^{2}>>,
\nonumber\\
& &
 aF_\pi^{2} = -\frac{{\cal V}_{\Sigma_2}}{4}<<\dot\psi_1^{2}>>,
 \nonumber\\
& &
 \frac{1}{g_{\rm YM}^2} = \frac{{\cal V}_{\Sigma_2}}{2}<\psi_1^{2}>,
 \nonumber\\
& &
  y_1 = -y_2 = -\frac{{\cal V}_{\Sigma_2}}{2}<(1+\psi_1-\psi_0)^2>,
  \nonumber\\
& &
  y_3 = -y_4 = -\frac{{\cal V}_{\Sigma_2}}{2}<\psi_1^{2}(1+\psi_1)^2>,
  \nonumber\\
& &
  y_5 = -{\cal V}_{\Sigma_2}<\psi_0^{2}\psi_1^{2}>,
  \nonumber\\
& &
  y_6 = -y_7=-{\cal V}_{\Sigma_2}<\psi_1(1+\psi_1)(1+\psi_1-\psi_0^{2})>,
 \nonumber\\
 & &
  y_8 = -y_9= -{\cal V}_{\Sigma_2}<\psi_0^{2}\psi_1^{2}>,
  \nonumber\\
   & &
   z_1 = -\frac{{\cal V}_{\Sigma_2}}{4}<(1+\psi_1)^2>,
  \nonumber
  \end{eqnarray}
  \begin{eqnarray}
& &
  z_2 = -\frac{{\cal V}_{\Sigma_2}}{4}<\psi_0^2>,
  \nonumber\\
& &
  z_3 = \frac{{\cal V}_{\Sigma_2}}{2}<\psi_1(1+\psi_1)>,
 \nonumber\\
  & &
  z_4 = {\cal V}_{\Sigma_2}<\psi_1(1+\psi_1-\psi_0^{2}),
  \nonumber\\
& &
  z_5 = -{\cal V}_{\Sigma_2}<\psi_1^{2}(1+\psi_1)>,
  \nonumber\\
& &
  z_6 = -{\cal V}_{\Sigma_2}<(1+\psi_1)(1+\psi_1-\psi_0^{2})>,
  \nonumber\\
& &
  z_7 = {\cal V}_{\Sigma_2}<\psi_1(1+\psi_1)^{2}>,
  \nonumber\\
& &
  z_8 = {\cal V}_{\Sigma_2}<\psi_0^{2}\psi_1>,
  \end{eqnarray}
 where we have used the following notations.
  \begin{equation}
  << A >> =  \int_{0}^{\infty}{\mathcal{V}_1(z)} A dz, \nonumber
\end{equation}
\begin{equation}
 < A > =  \int_{0}^{\infty}{\mathcal{V}_2(z)} A dz, \nonumber
\end{equation}
and \begin{equation}
\hskip 1.6in {\cal V}_{\Sigma_2} = -\frac{T_{D_6}(2\pi\alpha^\prime)^{2}}{4}\int dyd\theta_{2} \delta\Bigg(\theta_{2}-\frac{\alpha_{\theta_{2}}}{N^{\frac{3}{10}}}\Bigg).\nonumber
\end{equation}
The simplified expressions of the radial integrals appearing in (\ref{LECs-RI}) are given in the appendix \ref{appendix-MChPT}. Relation between SU(3) LECs of \cite{GL} (equation (\ref{MChPT-Op4})) and couplings appearing from holographic computation (equation (\ref{Lagrangian-Op4})) was obtained in \cite{HMM}:
\begin{eqnarray}
 & & L_1^r=\frac{L_2^r}{2}=-\frac{L_3^r}{6}=\frac{1}{32}\left(\frac{1}{g_{YM}^2}- z_4 + y_2\right),
\\
& & L_9^r=\frac{1}{8}\left(\frac{2}{g_{YM}^2}-2z_3-z_4 -z_6\right),
\\
& & L_{10}^r=\frac{1}{4}\left(-\frac{1}{g_{YM}^2}+2z_3-2z_2+2z_1\right).
\end{eqnarray}

{\bf Matching Holographic Result with the Phenomenological Data}:
We now discuss the matching of the one-loop renormalized ${\cal O}(p^4)$  $SU(3)$ $\chi$PT LECs obtained by us with the phenomenological data given in the literature. We will also discuss the matching of  $F_\pi^2, g_{\rm YM}(\Lambda_{\rm QCD}=0.4 {\rm GeV}, \Lambda=1.1 {\rm Gev}, \mu=M_\rho)$ where ``HLS-QCD'' matching scale is represented by $\Lambda$ \cite{HLS-Physics-Reports}, while the renormalization scale is represented by $\mu$. We showed that $L_{1,9,10}^r$ match exactly whereas $L_{2,3}^r$ match up to the order of magnitude and signs.

\begin{itemize}
\item
{\bf Step 1: Matching $L_{1,2,3}^r$}:
Using (\ref{Vol2}), (\ref{gsq}), (\ref{y_i}), (\ref{z_i}) and $y_2=-y_1$ we obtained:
{
\begin{eqnarray}
\label{L1}
& & \hskip -0.4in  L_1^r = \frac{L_2^r}{2} = - \frac{L_3^r}{6} = \frac{1}{32}\left(\frac{1}{g_{YM}^2}- z_4 + y_2\right) \nonumber\\
& &  \hskip -0.4in = \frac{1}{143360 \sqrt{7} ({f_{r_0}}-1) {f_{r_0}}
   {g_s}^8 \log N  M^4 N_f ^8 \alpha _{\theta _1}^3 {\cal C}_{\psi_1}^{\rm IR}\ ^2 \Omega}\Biggl\{3 \pi  ({f_{r_0}}+1) N^{7/5}\nonumber\\
& & \hskip -0.4in \times \Biggl(-\frac{4822335 \sqrt{2} \sqrt[8]{3} \sqrt[4]{7} \pi ^3 \beta  \left({\cal C}_{zz}^{(1)} - 2 {\cal C}_{\theta_1z}^{(1)} + 2 {\cal C}_{\theta_1x}^{(1)}\right) {\cal C}_{\phi_0}^{\rm IR}\ ^2 \epsilon ^{9/4}
   {f_{r_0}} ({f_{r_0}}+1) {g_s}^4 M^2 N_f ^4 \alpha _{\theta _1}^9 {\cal C}_{\psi_1}^{\rm IR}\  N^{2 {f_{r_0}}+\frac{7}{5}}}{({f_{r_0}}-1)^3
   \left(\log N\right) ^3}\nonumber\\
& & \hskip -0.4in +\frac{430080 \sqrt{21} \beta  \left({\cal C}_{zz}^{(1)} - 2 {\cal C}_{\theta_1z}^{(1)} + 2 {\cal C}_{\theta_1x}^{(1)}\right) \epsilon ^3 ({f_{r_0}}-1) {f_{r_0}} {g_s}^8 \log N  M^4 {m_0}^2 N^{3/5}
   N_f ^8 \alpha _{\theta _1}^5 {\cal C}_{\psi_1}^{\rm IR}\ ^2}{\pi  ({f_{r_0}}+1)}\nonumber\\
& & \hskip -0.4in +\frac{64 \sqrt{\frac{7}{\pi }} ({f_{r_0}}-1)^2 {g_s}^9 \log r_0
   M^4 N_f ^8 {\cal C}_{\psi_1}^{\rm IR}\ ^2 N^{\frac{2 {f_{r_0}}}{3}-\frac{3}{5}}\Omega}{({f_{r_0}}+1)^2} +\frac{26880 \sqrt{7} ({f_{r_0}}-1) {g_s}^9 \log r_0  M^4
   N_f ^8 {\cal C}_{\psi_1}^{\rm IR}\ ^2 N^{\frac{2 {f_{r_0}}}{3}-\frac{2}{5}}\Omega}{({f_{r_0}}+1)^2}\nonumber\\
& & \hskip -0.4in -\frac{8960 \sqrt{7} ({f_{r_0}}-1) {f_{r_0}} {g_s}^9 \log N  M^4 N_f ^8
   \alpha _{\theta _1}^2 {\cal C}_{\psi_1}^{\rm IR}\ ^2 N^{\frac{2 {f_{r_0}}}{3}-\frac{2}{5}} \Omega}{{f_{r_0}}+1}\nonumber\\
& & \hskip -0.4in -\frac{132269760 \sqrt{2} \sqrt[8]{3} \sqrt[4]{7} \pi ^3 {\cal C}_{\phi_0}^{\rm IR}\ ^2
   \epsilon ^{17/4} {f_{r_0}} ({f_{r_0}}+1)^2 {g_s}^4 M^2 N_f ^4 \alpha _{\theta _1}^9 {\cal C}_{\psi_1}^{\rm IR}\  N^{2
   {f_{r_0}}+\frac{7}{5}}}{({f_{r_0}}-1)^3 \left(\log N\right) ^3}\nonumber\\
& & \hskip -0.4in -\frac{5160960 \sqrt{7} \epsilon ^2 ({f_{r_0}}-1) {f_{r_0}} {g_s}^8 \log N  M^4 N^{3/5}
   N_f ^8 \alpha _{\theta _1}^5 {\cal C}_{\psi_1}^{\rm IR}\ ^2}{\pi }\nonumber\\
& & \hskip -0.4in-\frac{195259926456 \sqrt[4]{3} \pi ^7 {\cal C}_{\phi_0}^{\rm IR}\ ^4 \epsilon ^{13/2} {f_{r_0}}
   ({f_{r_0}}+1)^4 \alpha _{\theta _1}^{15} N^{4 {f_{r_0}}+\frac{11}{5}}}{({f_{r_0}}-1)^7 \left(\log N\right) ^7}\Biggr)
\Biggr\} ,
\end{eqnarray}
}
where:
\begin{equation}
\label{Omega}
 \Omega \equiv \left(7 \beta  \left({\cal C}_{zz}^{(1)} - 2 {\cal C}_{\theta_1z}^{(1)} + 2 {\cal C}_{\theta_1x}^{(1)}\right) {f_{r_0}}^2 \gamma ^2 {g_s}^2
   \left(\log N\right) ^2 M^4+3456 \epsilon ^2 ({f_{r_0}}+1) N^2\right).
\end{equation}
Since, we are working up to ${\cal O}(\beta)$ and $\epsilon$ is small and therefore working up to ${\cal O}(\epsilon^2)$, we found  that (\ref{L1}) simplified as:
\begin{eqnarray}
\label{L1_b}
& & \hskip -0.4in  L_1^r = \frac{1}{143360 \sqrt{7} ({f_{r_0}}-1) {f_{r_0}}
   {g_s}^8 \log N  M^4 N_f ^8 \alpha _{\theta _1}^3 }\nonumber\\
& & \hskip -0.4in \times\Biggl\{3 \pi  ({f_{r_0}}+1) N^{7/5} \Biggl(\frac{64 \sqrt{\frac{7}{\pi }} ({f_{r_0}}-1)^2 {g_s}^9 \log r_0
   M^4 N_f ^8 N^{\frac{2 {f_{r_0}}}{3}-\frac{3}{5}}}{({f_{r_0}}+1)^2} \nonumber\\
& & \hskip -0.4in +\frac{26880 \sqrt{7} ({f_{r_0}}-1) {g_s}^9 \log r_0  M^4
   N_f ^8  N^{\frac{2 {f_{r_0}}}{3}-\frac{2}{5}}}{({f_{r_0}}+1)^2} -\frac{8960 \sqrt{7} ({f_{r_0}}-1) {f_{r_0}} {g_s}^9 \log N  M^4 N_f ^8
   \alpha _{\theta _1}^2  N^{\frac{2 {f_{r_0}}}{3}-\frac{2}{5}} }{{f_{r_0}}+1}\nonumber\\
& & \hskip -0.4in -\frac{5160960 \sqrt{7}  ({f_{r_0}}-1) {f_{r_0}} {g_s}^8 \log N  M^4 N^{3/5}
   N_f ^8 \alpha _{\theta _1}^5}{\pi}\left(\frac{\epsilon ^2}{ \Omega}\right)\Biggr)
\Biggr\}.
\end{eqnarray}
Further, we need to consider: $f_{r_0}=1-\kappa, 0<\kappa\ll1$ (from (\ref{f_{r_0}}) and (\ref{fr0-omega}), will be discussed later when we match $L_9^r$ with the phenomenological value). Hence (\ref{L1_b}) further simplified as:
\begin{eqnarray}
\label{L1_c}
& & \hskip -0.4in  L_1^r = \frac{1}{143360 \sqrt{7} {f_{r_0}}
   {g_s}^8 \log N  M^4 N_f ^8 \alpha _{\theta _1}^3 }\Biggl\{3 \pi  ({f_{r_0}}+1) N^{7/5} \Biggl(\frac{26880 \sqrt{7}  {g_s}^9 \log r_0  M^4
   N_f ^8  }{({f_{r_0}}+1)^2} \nonumber\\
& & \hskip -0.4in  -\frac{8960 \sqrt{7}  {f_{r_0}} {g_s}^9 \log N  M^4 N_f ^8
   \alpha _{\theta _1}^2   }{{f_{r_0}}+1} -\frac{5160960 \sqrt{7}  {f_{r_0}} {g_s}^8 \log N  M^4 N^{\frac{1}{3}}
   N_f ^8 \alpha _{\theta _1}^5}{\pi}\left(\frac{\epsilon ^2}{ \Omega}\right)\Biggr)
\Biggr\}.
\end{eqnarray}
In addition, considering that the $\log r_0$ in (\ref{L1_c}) is, in reality, $\log\left(\frac{r_0}{{\cal R}_{D5/\overline{D5}}}\right)$ - ${\cal R}_{D5/\overline{D5}}>r_0$, where ${\cal R}_{D5/\overline{D5}}$ denotes the separation between $D5-\overline{D5}$-branes; we realised from (\ref{L1_c}) that in order to obtain a positive value (which is required by the phenomenological value of $L_1^r$), $\Omega<0$ is necessary. Note that, as demonstrated later in (\ref{alphatheta1}), fitting with the experimentally determined value of the pion decay constant $F_\pi$ demands a $N$-suppression in $\alpha_{\theta_1}$, which implies that the $N$ enhancement in the last term in (\ref{L1_c}) is an artificial enhancement.  Therefore, in order to make certain that one is not picking up a ${\cal O}\left(\frac{1}{\beta}\right)$ contribution in $L_1^r$ from $\frac{\epsilon^2}{\Omega}$ in (\ref{L1_c}), in addition to make sure that the last term in (\ref{L1_c}) required to partially compensate for the initial two terms that are negative within the same (as clarified earlier) results in a positive $L_1^r$, from (\ref{Omega}), we set: 
\begin{equation}
\label{epsilon-sqrtbeta-over-N-lambdaeps}
\epsilon = \lambda_{\epsilon}\frac{\sqrt{\beta}}{N}.
\end{equation}
In conclusion, when all of these findings are taken into account, along with the demand that the value of  $L_1^{\rm exp}=0.64\times10^{-3}$, we get the following:
Finally, combining the above observations with the requirement to match the experimental value $L_1^{\rm exp}=0.64\times10^{-3}$, we implemented the following constraint:
\begin{eqnarray}
\label{CCsO4}
& & \left({\cal C}_{zz}^{(1)} - 2 {\cal C}_{\theta_1z}^{(1)} + 2 {\cal C}_{\theta_1x}^{(1)}\right) = -\frac{493.7 (\delta +1) ({f_{r_0}}+1) \lambda_{\epsilon}^2}{{f_{r_0}}^2 \gamma ^2 {g_s}^2 \left(\log N\right) ^2 M^4},\nonumber\\
& & \delta = \frac{0.053 \alpha _{\theta _1}^3 N^{-\frac{2 {f_{r_0}}}{3}-1}}{{g_s}}.
\end{eqnarray}
{\it Because of this, we may deduce from (\ref{epsilon-sqrtbeta-over-N-lambdaeps}) that the parameter $\epsilon$ serves as an expansion parameter combining the $\frac{1}{N}$ and $\beta$ expansions. In addition, this is the first relationship among large-$N$ and higher derivative corrections in the framework of ${\ cal M}$-theory, which is the dual of large-$N$ thermal QCD-like theories. This connection was made achievable from the use of (\ref{CCsO4}).} \par
When we compare the theoretical values of $L_{2,3}^r$ to their experimental values, as shown in (\ref{L1}), we found that it is possible to obtain a match with the order of magnitude and the sign of the value, but not with the precise numerical value.

\item
{\bf Step 2: Matching $F_\pi^2$}: Now, using (\ref{CCsO4}) and (\ref{Vol2}), we showed that the difference of (\ref{Fpisq}) and the experimental value of $F_\pi^2 = \frac{0.0037 N^{-\frac{2 {f_{r_0}}}{3}-1}}{{g_s}}$\footnote{$F_\pi^2$ can be made to match the experimental value of $92.3 MeV$ wherein from $0^{++}$-glueball mass \cite{Misra+Gale}, one identifies:
$\frac{1700}{4} MeV \equiv \frac{r_0}{\sqrt{4 \pi g_s N}}$.} is zero when:
\begin{eqnarray}
\label{alphatheta1}
& & \alpha_{\theta_1} = \frac{0.03 \sqrt[24]{\beta } {g_s}^{11/3} \sqrt[12]{\lambda_{\epsilon}} \left(\log N\right) ^{2/3} \sqrt[3]{M} N_f ^{2/3}
   \sqrt[3]{{\cal C}_{\psi_1}^{\rm IR}\ } {(1-{f_{r_0}})^{2/3} } N^{-\frac{{f_{r_0}}}{3}-\frac{13}{60}}}{\sqrt[3]{{\cal C}_{\phi_0}^{\rm IR}\ }
   \sqrt[3]{{f_{r_0}}+1}}.
\end{eqnarray}

\item
{\bf Step 3: Matching $L_9^r$}: From \cite{Relationship-Li-yi-zi}, we have:
\begin{equation}
\label{L9}
L_9^r = \frac{1}{8}\left(\frac{2}{g_{\rm YM}^2} - 2z_3 - z_4 - z_6\right).
\end{equation}
By using  (\ref{CCsO4}), (\ref{Vol2}) and (\ref{z_i}), we obtained:
\begin{equation}
\label{L9r-match}
L_9^r = -\frac{0.0031 ({f_{r_0}}-1) {g_s} N^{\frac{2 {f_{r_0}}}{3}+\frac{4}{5}}}{({f_{r_0}}+1) \alpha _{\theta _1}^3}.
\end{equation}
Now, assuming:
\begin{equation}
\label{f_{r_0}}
{f_{r_0}} = 1 - \omega \alpha_{\theta_1}^3,
\end{equation}
where
\begin{equation}
\label{fr0-omega}
\omega = \frac{4.6 N^{-\frac{2 {f_{r_0}}}{3}-\frac{4}{5}}}{{g_s}},
\end{equation}
hence we get a match with the phenomenological/experimental value $L_9^{\rm exp}=6.9\times10^{-3}$. Substituting (\ref{f_{r_0}}) - (\ref{fr0-omega}) into (\ref{CCsO4}), we obtained: 
\begin{equation}
\label{CCsO4-lambda-epsilon}
\left({\cal C}_{zz}^{(1)} - 2 {\cal C}_{\theta_1z}^{(1)} + 2 {\cal C}_{\theta_1x}^{(1)}\right) \approx -\frac{987.4 \lambda_{\epsilon}^2}{ \gamma ^2 {g_s}^2 \left(\log N\right) ^2 M^4}.
\end{equation}
Consistency of (\ref{f_{r_0}}), (\ref{fr0-omega}) and (\ref{alphatheta1}), setting
$\sqrt[24]{\beta}\approx\beta^{0.04}$ to unity (because of the very small exponent of $\beta$), requires:
\begin{eqnarray}
\label{alphatheta1-ii}
\alpha_{\theta_1} = \frac{11.33 \sqrt[3]{{\cal C}^{\rm IR}_{\phi_0}} N^{55/36}}{{g_s} \sqrt[12]\lambda_{\epsilon} \sqrt[3]{M} {N_f}^{2/3} \sqrt[3]{{\cal C}^{\rm IR}_{\psi_1}} \log
   ^{\frac{2}{3}}(N)}.
\end{eqnarray}

\item
{\bf Step 4: Matching $g_{\rm YM}^2(\Lambda_{\rm QCD}=0.4$GeV, $\Lambda=1.1$GeV, $\mu=M_\rho$)}: Similarly, $g_{\rm YM}^2$  can be chosen to match the experimental value $36$ (at $\Lambda_{\rm QCD}=0.3$GeV and the HLS-QCD matching scale ``$\Lambda''=1.1$GeV \cite{HLS-Physics-Reports}) and renormalization scale $\mu=M_\rho$ by imposing:
\begin{equation}
\frac{12 \delta  N}{\alpha _{\theta _1}^2 \left(\frac{855.11 \beta ^{3/2} (\delta +1) \lambda_{\epsilon}^3 {m_0}^2}{{f_{r_0}}^2 \gamma ^2
   {g_s}^2 \left(\log N\right) ^2 M^4}+12 N\right)}=36,
\end{equation}
which was affected by the following:
\begin{equation}
\label{gYMsq-gamma}
\gamma = \frac{ 10^{6} \lambda_{\epsilon}^{3/2} {m_0} \beta ^{3/4} \alpha _{\theta _1}\sqrt{ 8.4(\delta + 1)}}{{f_{r_0}}
   {g_s} \left(\log N\right)  M\sqrt{ N \left(3.27\times 10^9 \delta -1.18\times 10^{11} \alpha _{\theta _1}^2\right)}}
\end{equation}
We obtained an upper bound on ${\cal C}_{\phi_0}^{\rm IR}$ from the expression (\ref{gYMsq-gamma}) by demanding that the argument of the square root in the denominator of (\ref{gYMsq-gamma}) be positive:
\begin{eqnarray}
\label{constant_upper_bound}
& & {\cal C}_{\phi_0}^{\rm IR} < \frac{1.04\times10^{-13} \sqrt[8]{\beta } ({f_{r_0}}-1)^2 {g_s}^2 \sqrt[4]{\lambda_{\epsilon}} M N_f ^2
   {\cal C}_{\psi_1}^{\rm IR}\  N^{-3 {f_{r_0}}-\frac{73}{20}} \log ^2(N)}{{f_{r_0}}+1}.
\end{eqnarray}

\item
{\bf Step 5: Matching $L_{10}^r$}: Now, using (\ref{Vol2}), (\ref{CCsO4}), (\ref{z_i}) and \cite{Relationship-Li-yi-zi}, \cite{Wilson-matching} (This also gives the UV-finite portion of the $rho-pi$ one-loop correction through the use of a ``$a(M_\rho)$'' factor with a value equal to 2), for $N_f=3$, we showed that:
\begin{eqnarray}
\label{L10}
& &  L_{10}^r = \frac{1}{4}\left(-\frac{1}{g_{\rm YM}^2} + 2z_3 - 2z_2 + 2z_1 \right) + \frac{11 N_f a(M_\rho)}{96(4\pi)^2}\nonumber\\
& &  = \frac{0.47 {g_s} N^{\frac{2 {f_{r_0}}}{3}+1}}{\alpha _{\theta _1}}-
\frac{8640. {g_s} N^{5/3} \alpha _{\theta _1}^2}{{g_s} N^{5/3}-1831.63 \alpha _{\theta _1}^3} + 4.4\times10^{-3}.
\end{eqnarray}
Since, $L_{10}^r = -5.5\times10^{-3}$, the following is the value that we were able to achieve for the $\theta_1$ delocalization parameter $\alpha_{\theta_1}$:
\begin{eqnarray}
\label{alphatheta1-L10}
& & \alpha_{\theta_1} = -2.8\times10^{-14} g_s N^{\frac{5}{3}} + \frac{\sqrt{4.2\times10^{-6} - 2.9\times10^{-11}g_s^2N^{\frac{10}{3}}}}{2}.
\end{eqnarray}
This indicates that $N<140$ for the case where $g_s=0.1$. For $g_s=0.1$, this implies $N<140$. For the purpose of the numerical calculation, we set $N = 10^2, g_s = 0.1$, and $M = N_f = 3$. We then obtained the following non-linear relation between ${\cal C}_{\psi_1}^{\rm IR}, \lambda_{\epsilon}$ and ${\cal C}_{\phi_0}^{\rm IR}$ (considered that they also satisfy (\ref{constant_upper_bound}):
\begin{equation}
\label{alphatheta1-L10-ii}
\frac{1.6\sqrt[3]{{\cal C}^{\rm IR}_{\phi_0}} }{\lambda_{\epsilon}^{\frac{1}{12}}  \sqrt[3]{{\cal C}^{\rm IR}_{\psi_1}}} = 10^{-7}.
\end{equation}

\end{itemize}
%%%%%%%%%%%%%%%%%%%%%%%%%%%
%%%%%%%%%%%%%%%%%%%%

\section{Summary}
\label{Summary-LECs}
In this chapter, we computed the LECs of $SU(3)$ chiral perturbation theory from the type IIA string dual inclusive of ${\cal O}(R^4)$ corrections. We constructed the $SU(3)$ Lagrangian and obtained the coupling constants from the radial integrals quoted in appendix \ref{appendix-MChPT}. We already mention in the beginning of this chapter that we have used the results of \cite{MChPT,Vikas-Thesis} to do our calculations in this chapter.  My contribution has been presented in this chapter.

On matching LECs computed holographically with their phenomenological/experimental values, it turns out that a particular combination of integration constants, $({\cal C}_{zz}^{\rm th}-{\cal C}_{\theta_1 z}^{\rm th}+2 {\cal C}_{\theta_1 x}^{\rm th})$,  is appearing in all the LECs of $SU(3)$ chiral perturbation theory Lagrangian up to ${\cal O}(p^4)$. Here, ${\cal C}_{MN}^{\rm th}$ are integration constants appearing in solutions to ${\cal O}(R^4)$ corrections as worked out in \cite{MChPT} to the MQGP metric  for the thermal gravitational dual. This combination, $({\cal C}_{zz}^{\rm th}-{\cal C}_{\theta_1 z}^{\rm th}+2 {\cal C}_{\theta_1 x}^{\rm th})$, encodes information about the compact part of non-compact four cycle around which flavor $D7$-branes are wrapping in type IIB setup. The key results of this chapter are summarized below.
\begin{enumerate}
\item To match our results of the one-loop renormalized $\chi$PT Lagrangian's LECs $L_{1, 2, 3, 9, 10}^r, F_\pi^2, g_{SU(3)}$ with the experimental values, we had to fix the eight non-zero parameters listed below for the  given values of $N, M, N_f$ in the MQGP limit (\ref{MQGP_limit}):
\begin{enumerate}
\item  a linear combination of integration constants ($({\cal C}_{zz}^{\rm th}-{\cal C}_{\theta_1 z}^{\rm th}+2 {\cal C}_{\theta_1 x}^{\rm th})$) appearing in the solutions of the EOMs of the ${\cal O}(R^4)$ ${\cal M}$-theory uplift's metric components $G^{\cal M}_{zz, \theta_1 z, \theta_1x} $\footnote{It turns out that there is a specific combination of only three costants of integration that emerges while matching $\chi$PT LECs up to ${\cal O}(p^4)$. Although in theory there are additional constants of integration that exist in other ${\cal O}(R^4)$ ${\cal M}$-theory metric components, it has been found that there is only one such combination. Even though it is not clear the reason why this particular combination, yet it is naturally obvious that it uses $G^{\cal M}_{zz, \theta_1 z, \theta_1x} $ as these simply belong to the $S^3$, portion of the non-compact four-cycle wrapped around the flavor $D7$-branes in the type IIB dual of \cite{metrics}; these $D7$-flavor branes are (triple) T dualized to the type IIA $D6$ flavor branes. In view of fact, this is a very non-trivial signature of the four-cycle that is enveloped by the type IIB $D7$-branes. It manifests itself as ${\cal O}(R^4)$-corrections to the MQGP background, which leads to an uplift of thermal QCD-like theories.}; 
\item integral constants,
$ {\cal C}^{\rm IR}_{\psi_1,\phi_0}$\footnote{It is possible to self-consistently set the ${\cal C}^{\rm UV}_{\psi_1,\phi_0}$ value to zero occurring in the solutions of the EOMs in the IR of the $\rho$ meson and the $\pi$ meson respectively.}

\item
$\epsilon$ (or equivalently $\lambda_\epsilon$ since  $\epsilon=\lambda_\epsilon\frac{\sqrt{\beta}}{N}), \gamma$ appearing in $a = \left(b + {\bf \gamma} {\cal O}\left(\frac{g_sM^2}{N}\right)\right)r_0$; 

\item delocalization parameter  $\alpha_{\theta_1}$ of $\theta_1$; 

\item tension of $D6$-brane or equivalently $\alpha^\prime$; 

\item
$ f_{r_0}$ as in $r_0 = N^{-\frac{f_{r_0}}{3}}$
\end{enumerate}
All the above points are summarised in the table \ref{Table-FM}.
{\footnotesize
\begin{table}[h]
\begin{center}
\begin{tabular}{|c|c|c|c|}\hline
S. No. & Quantities whose    & Parameters of the holographic dual  & 
Equation numbers \\
& experimental values & used for fitting & \\
& are fitted to & & \\ \hline
1. & $L_{1,2,3}^r$ & Specific linear combination of & (\ref{CCsO4}) \\
& & constants of integration appearing in & [using (\ref{epsilon-sqrtbeta-over-N-lambdaeps})] \\
& & solutions to ${\cal O}(R^4)$ corrections & \\
& & to ${\cal M}$-theory metric components & \\ 
& &   $G^{\cal M}_{zz, \theta_1 z, \theta_1x}$; $f_{r_0}; \gamma; \lambda_\epsilon$; & \\ \cline{3-4}
& & ${\cal O}(R^4)-\frac{1}{N}$ connection: &  (\ref{epsilon-sqrtbeta-over-N-lambdaeps}) \\
& & $a-r_0$ relation must have & \\
& & an $\epsilon\sim\frac{l_p^3}{N}$ contribution & \\
& & at ${\cal O}\left(\frac{g_s M^2}{N}\right)^0$ & \\ \hline
2. & $F_\pi^2;\ L_9^r$ & $f_{r_0}; \alpha_{\theta_1}; {\cal C}^{\rm IR}_{\phi_0}; {\cal C}^{\rm IR}_{\psi_1}; \lambda_\epsilon$ & (\ref{alphatheta1}), (\ref{L9r-match})-(\ref{alphatheta1-ii})[consistency \\
& & & check] \\ \hline
3. & $g_{SU(3)}$ & $\gamma$; upper bound on ${\cal C}^{\rm IR}_{\phi_0}$ &  (\ref{gYMsq-gamma}), (\ref{constant_upper_bound}) \\ \hline
4. & $L_{10}^r$ & $\alpha_{\theta_1}; {\cal C}^{\rm IR}_{\phi_0}; {\cal C}^{\rm IR}_{\psi_1}$ & (\ref{alphatheta1-L10}) - (\ref{alphatheta1-L10-ii}) [even though  \\
& & & specific values of $N, M, N_f, g_s$ \\
& & & chosen, but can find analog of \\
& & & (\ref{alphatheta1-L10-ii}) $\forall N<140$ (\ref{alphatheta1-L10}) \\
& & & respecting (\ref{MQGP_limit})] \\ \hline
\end{tabular}
\end{center}
\caption{An Overview of the agreement between the theoretical and experimental Values of one-Loop Renormalized $\chi$PT LECs, $F_\pi^2$, and $g_{SU(3)}^2$.}
\label{Table-FM}
\end{table}
}

\item  In addition, the normalisation condition of the $\rho$-meson profile function $\psi_1(Z)$ ($n=1$ mode) is used to determine the integration constants ${\cal C}_{\psi_1}^{\rm UV}$ appearing in the solution to $\psi_1(Z)$ EOM in the UV, in terms of ${\cal C}^{\rm IR}_{\psi_1}, f_{r_0}, \gamma, \epsilon, \lambda_\epsilon$ and
$T_{D6}$ or equivalently $\alpha^\prime$ (via ``${\cal V}_{\Sigma_2}$'');

\item Further, the normalization condition of the $\rho$-meson profile function $\psi_1(Z)$ ($n=1$ mode) is used to determine the constant of integration ${\cal C}_{\psi_1}^{\rm UV}$ appearing in the solution to $\psi_1(Z)$ in the UV, in terms of ${\cal C}^{\rm IR}_{\psi_1}, f_{r_0}, \gamma, \epsilon, \lambda_\epsilon$ and
$T_{D6}$ or equivalently $\alpha^\prime$ (via ``${\cal V}_{\Sigma_2}$''); it had been shown that we were able to set ${\cal C}^{\rm UV}_{\psi_1}=0$ in a manner that was self-consistent.

\item If one were to take the expression for $T_{D6}/\alpha^\prime/{\cal V}_{\Sigma_2}$ derived above from the normalisation condition of $\psi_1(Z)$ and substitute it into the normalisation condition for the profile function $\phi_0(Z)$, one would be able to self-consistently set  ${\cal C}^{\rm UV}_{\phi_0}=0$.

\end{enumerate}

%Satyam Shivam Sundaram
%Jai Mata Di
\chapter{Deconfinement Phase Transition in Thermal QCD-Like Theories at Intermediate Coupling in the Absence and Presence of Rotation}

\graphicspath{{Chapter3/}{Chapter3/}}

\section{Introduction} 
 We can compute the corrections to the infinite-'t Hooft-coupling limit as done in \cite{SGS} using gauge/gravity duality, but this time using an AdS background. They used terms quartic in the Weyl tensor to include higher-derivative corrections on the gravity side. From the gauge theory point of view, there is a $SU(N)$ supersymmetric Yang-Mills plasma with ${\cal N}=4$ SUSY at an intermediate 't Hooft coupling. They provided an explanation for the transport peak that appeared in the spectral function of the stress-energy tensor at zero spatial momentum in the small frequency zone. This is a characteristic of perturbative plasma generally.\par
 Different bottom-up models built from gauge/gravity duality have been used to study QCD at strong coupling, whereas perturbation theory has been used to study it at weak coupling. The Sakai-Sugimoto model \cite{SS1} is a well-known top-down holographic type IIA dual, although one that only works in the infrared. \cite{metrics} and its ${\cal M}$-theory uplift \cite{MQGP} are the only UV-complete (type IIB) top-down holographic dual of QCD-like theories at strong coupling that we are aware of. Based on the hard-thermal-loop perturbation theory (HTLpt), authors in \cite{IC} have investigated QCD thermodynamic functions at intermediate coupling. Gauge/gravity duality can also be used to explore the intermediate coupling regime of QCD. The authors obtained $O(l_p^6)$ corrections to the ${\cal M}$-theory metric in the \cite{HD-MQGP}, as explained in chapter {\bf 1}, in order to proceed in this direction.\par
In general theories of gravity, Wald provided a formula to determine black hole entropy \cite{Wald-Entropy}. He thought about the Lagrangian that results from the diffeomorphism invariant classical theory of gravity in $n$ dimensions. Noether charge is a $(n-2)$-form in these theories of gravity, and the $(n-2)$-form Noether charge's integral across its bifurcate killing horizon will yield the black hole's entropy. Therefore, for stationary black holes with bifurcate killing horizons, entropy equals Noether charge. A formula to calculate entropy for dynamical black holes was made in \cite{Wald-Entropy-2}. \par
Every physical phenomenon has an energy/distance scale that defines it.  RG group flow is used to move from UV to IR. However, short-range physics begins to interact with long-range physics in noncommutative field theories and string theory, a process known as UV/IR mixing. UV divergences in real $\phi^4$ theory defined on commutative space are changed into infrared poles in the identical theory defined on noncommutative space as a result of UV/IR mixing, for example. There are further examples that are nicely explained in \cite{UV_IR}. Theoretical models of gravity are nonlocal. As a result, UV/IR mixing might be present in such theories.  In the gauge/gravity duality, the energy scale in the gauge theory side matches with the radial coordinate in gravitational theory. In the context of the AdS/CFT correspondence, the authors looked at the ``I.R.-U.V.'' connection and demonstrated how infrared phenomena in bulk theory are turned into ultravoilet phenomena in boundary theory \cite{Susskind+Witten}. In particular, the job of the ultraviolet regulator in the ${\cal N} = 4$ super Yang Mills theory is played by the infrared regulator in the bulk theory.  It's fascinating that in our work, matching at the deconfinement temperature, ${\cal M}$-theory actions dual to the thermal and black-hole backgrounds at the UV-cut-off, and obtaining a relationship between the ${\cal O}(R^4)$ metric corrections in the IR, a specific type of UV-IR connection manifests itself.\par
Using a semiclassical analysis, we determined the deconfinement temperature of QCD-like theory at the intermediate coupling using gauge/gravity duality, as was first discussed in \cite{Witten-Hawking-Page-Tc}. The confinement-deconfinement phase transition in large $N_c$ gauge theories could also be discussed from the perspective of entanglement entropy \cite{Tc-EE}. In this procedure, one must divide one of the spatial coordinates into a segment of length ${\it l}$ and the other segment in order to calculate the entanglement entropy among both of them. Ryu and Takayanagi provided rules in \cite{RT} on how to calculate entanglement entropy from AdS/CFT correspondence. There are two surfaces-connected and disconnected-as explained in \cite{Tc-EE}. It has a critical value, which is indicated by the symbol ${\it l_{crit}}$. It is the connected surface which takes over entanglement entropy if  we are below the critical point of ${\it l}$, or ${\it l}<{\it l_{crit}}$, and it is the disconnected surface if we are above the critical value of ${\it l}$, or ${\it l}>{\it l_{crit}}$. These two regions correspond to the confining and deconfining phases of large $N_c$ gauge theories, respectively. Therefore, we can interpret this as a phase transition for large $N_c$ gauge theories involving confinement-deconfinement phases.\par
 In noncentral relativistic heavy ion collisions (RHIC), a fluid known as quark-gluon plasma (QGP) is created. According to experimental findings from the RHIC collaboration, the angular momentum and angular velocity of this rotating fluid are on the order of $10^3 \hbar$ and $\omega \sim 6$ MeV, respectively. In \cite{omega-simulation}, the authors applied hydrodynamic simulations of heavy ion collisions and found $\omega \sim 20-40$ MeV. Considering the results mentioned above, a top-down analysis employing gauge-gravity duality of the effect of rotation on the deconfinement temperature of thermal QCD-like theories will be very interesting, with this motivation we obtained deconfinement temperature of thermal QCD from a top-down approach in \cite{Rotation-Tc-M-Theory}. The authors in \cite{R-Tc} examined the consequences of rotating plasma on the deconfinement temperature of QCD in Hard wall and Soft wall models, and they discovered that deconfinement temperature of QCD decreases as angular velocity of the plasma increases. Similar works using a bottom-up method are available in \cite{G-D-Rotation,Kerr-AdS-1,Kerr-AdS-2}.

%This chapter is organized as follows. In section \ref{DPT-WR-i}, we discuss the deconfinement phase transition in thermal QCD-like theories without rotation via \ref{Tc-SS-i} and \ref{Tc-SS-ii}, in \ref{Tc-SS-iii}, we discuss the deconfinement phase transition using the entanglement entropy, and \ref{Tc-SS-iv} is on the compatibility of this paper with \cite{MChPT}. We study the rotating QGP in \ref{DPT-WR-ii} starting with construction of holographic dual of rotating QGP in \ref{Tc-SS-v} and discuss the phase transition in \ref{Tc-vi} similar to \ref{DPT-WR-i}. Results of this chapter is summarised in \ref{Tc-Summary}. There is a supplementary appendix \ref{appendix-McTEQ} on the EOMs and solutions to the ${\cal O}(R^4)$ metric corrections to the MQGP background of \cite{MQGP,NPB}, as well as a discussion on taking the decompactification-limit of a periodic spatial direction  in the thermal background ${\cal M}$-theory uplift used in \cite{MChPT} to obtain the ${\cal M}$-theory thermal background in this paper, and deriving the sign of a linear combination of constants of integration in the solutions to the EOMs of the ${\cal O}(R^4){\cal M}$-theory metric corrections as required from matching of results of \cite{MChPT} with phenomenological values of the LECs at NLO in the $SU(3)\ \chi$PT Lagrangian of \cite{GL}. 

\section{Deconfinement Phase Transition in Thermal QCD-Like Theories in the Absence of Rotation}
\label{DPT-WR-i}
In this section, we will discuss the deconfinement phase transition of thermal QCD-like theories at intermediate coupling without including the effect of rotation.
\subsection{$T_c$ from Semi-classical Method}
\label{Tc-SS-i}
Here, we'll detail how we implemented Witten's proposal \cite{Witten-Hawking-Page-Tc} to obtain the deconfinement temperature thermal QCD by applying the gauge/gravity duality and holographic renormalization of the eleven-dimensional supergravity action via \ref{Tc-G-I}, \ref{Tc-G-II}, and \ref{Tc-G-III}.
\par
The ${\cal M}$-theory dual for high temperatures, that is, temperatures that are higher than the critical temperature $T_c$, will entail a black hole with the metric presented by:
\begin{eqnarray}
\label{TypeIIA-from-M-theory-Witten-prescription-T>Tc}
\hskip -0.1in ds_{11}^2 & = & e^{-\frac{2\phi^{\rm IIA}}{3}}\Biggl[\frac{1}{\sqrt{h(r,\theta_{1,2})}}\left(-g(r) dt^2 + \left(dx^1\right)^2 +  \left(dx^2\right)^2 +\left(dx^3\right)^2 \right)
\nonumber\\
& & \hskip -0.1in+ \sqrt{h(r,\theta_{1,2})}\left(\frac{dr^2}{g(r)} + ds^2_{\rm IIA}(r,\theta_{1,2},\phi_{1,2},\psi)\right)
\Biggr] + e^{\frac{4\phi^{\rm IIA}}{3}}\left(dx^{11} + A_{\rm IIA}^{F_1^{\rm IIB} + F_3^{\rm IIB} + F_5^{\rm IIB}}\right)^2,
\end{eqnarray}
where $A_{\rm IIA}^{F^{\rm IIB}_{i=1,3,5}}$ represent the type IIA RR 1-forms that were derived using the triple T/SYZ-dual of the type IIB $F_{1,3,5}^{\rm IIB}$ fluxes found in the type IIB holographic dual of \cite{metrics}, and the black hole function $g(r) = 1 - \frac{r_h^4}{r^4}$\footnote{See appendix \ref{METRIC_HD_MQGP} for the explicit metric components.}. For temperatures below the critical temperature, denoted by the notation $T<T_c$, the metric is represented by a thermal gravitational dual:
\begin{eqnarray}
\label{TypeIIA-from-M-theory-Witten-prescription-T<Tc}
\hskip -0.1in ds_{11}^2 & = & e^{-\frac{2\phi^{\rm IIA}}{3}}\Biggl[\frac{1}{\sqrt{h(r,\theta_{1,2})}}\left(-dt^2 + \left(dx^1\right)^2 +  \left(dx^2\right)^2 + \tilde{g}(r)\left(dx^3\right)^2 \right)
\nonumber\\
& & \hskip -0.1in+ \sqrt{h(r,\theta_{1,2})}\left(\frac{dr^2}{\tilde{g}(r)} + ds^2_{\rm IIA}(r,\theta_{1,2},\phi_{1,2},\psi)\right)
\Biggr] + e^{\frac{4\phi^{\rm IIA}}{3}}\left(dx^{11} + A_{\rm IIA}^{F_1^{\rm IIB} + F_3^{\rm IIB} + F_5^{\rm IIB}}\right)^2,
\end{eqnarray}
where $\tilde{g}(r) = 1 - \frac{r_0^4}{r^4}$ and $h(r,\theta_{1,2})$ is warp factor in type IIB string theory \cite{metrics, MQGP}. One notes that in (\ref{TypeIIA-from-M-theory-Witten-prescription-T>Tc}), $t\rightarrow x^3,\ x^3\rightarrow t$, and that after performing a double Wick rotation in the newly generated $x^3, t$ coordinates, one finds (\ref{TypeIIA-from-M-theory-Witten-prescription-T<Tc}). In sum, this means: $-g_{tt}^{\rm BH}(r_h\rightarrow r_0) = g_{x^3x^3}\ ^{\rm Thermal}(r_0),$ $ g_{x^3x^3}^{\rm BH}(r_h\rightarrow r_0) = -g_{tt}\ ^{\rm Themal}(r_0)$ in the findings
\cite{VA-Glueball-decay, HD-MQGP}. For further information on type IIA Euclidean/black $D4$-branes, see the work of Kruczenski et al \cite{Kruczenski:2003uq}. We assumed that world volume of the spatial part of the solitonic $M3$ brane (equivalent to solitonic $M5$-brane wrapped around a homologous sum of two-spheres \cite{DM-transport-2014}) is $\mathbb{R}^2(x^{1,2})\times S^1(x^3)$ where $S^1(x^3)$ has the period $\frac{2\pi}{M_{\rm KK}}$ with $M_{\rm KK} = \frac{2r_0}{ L^2}\left[1 + {\cal O}\left(\frac{g_sM^2}{N}\right)\right]$ and $L = \left( 4\pi g_s N\right)^{\frac{1}{4}}$ because $r_0$ is very small IR cut-off for thermal background. Hence, $\lim_{M_{\rm KK}\rightarrow0}\mathbb{R}^2(x^{1,2})\times S^1(x^3) = \mathbb{R}^3(x^{1,2,3})$ implies the four dimensional physics. We took $\tilde{g}(r)=1$ in (\ref{TypeIIA-from-M-theory-Witten-prescription-T<Tc}) for the computation of $T_c$.\par
If we assume that the periods of the thermal circles in the black hole and thermal cal ${\cal M}$-theory backgrounds are $\beta_{\rm BH,Th}$, then at $r={\cal R}_{\rm UV}$: $\beta_{\rm BH}\sqrt{ G^{\rm BH}_{tt}} = \beta_{\rm Th} \sqrt{G^{\rm Th}_{tt}}$, and at $T=T_c$ \cite{Witten-Hawking-Page-Tc},  
\begin{equation}
\beta_{\rm BH}\slashed{\int}_{M_{11}}\left({\cal L}^{\rm BH}_{\rm EH} + {\cal L}^{\rm BH}_{\rm GHY}\delta(r-{\cal R}_{\rm UV}) 
+ {\cal L}^{\rm BH}_{{\cal O}(R^4)}\right) = \beta_{\rm Th}\slashed{\int}_{M_{11}}\left({\cal L}^{\rm Th}_{\rm EH} + {\cal L}^{\rm Th}_{\rm GHY}\delta(r-{\cal R}_{\rm UV}) 
+{\cal L}^{\rm Th}_{{\cal O}(R^4)}\right),
\end{equation}
where $\slashed{\int}$ does not include the coordinate integral of $x^0$ which implies: 
\begin{eqnarray}
\label{actions-equal-Tc-ii}
& & \left(\sqrt{1-\frac{r_h^4}{{\cal R}_{\rm UV}^4}}\right)^{-1}\int_{M_{10}}\left({\cal L}^{\rm BH}_{\rm EH}+{\cal L}^{\rm BH}_{\rm GHY}\delta(r-{\cal R}_{\rm UV}) + {\cal L}^{\rm BH}_{{\cal O}(R^4)}\right) 
\nonumber\\
& & = \int_{\tilde{M}_{10}}\left({\cal L}^{\rm Th}_{\rm EH} + {\cal L}^{\rm Th}_{\rm GHY}\delta(r-{\cal R}_{\rm UV}) 
+ {\cal L}^{\rm Th}_{{\cal O}(R^4)}\right).
\end{eqnarray} 
    
{\bf Derivation of the On-Shell Action}: The following is the structure of the supergravity action in $\kappa_{11}^2=1$-units that corresponds to $cal M$-theory in the presence of ${\cal O}(R^4)$ terms.
\begin{eqnarray}
& & S_{D=11} = \frac{1}{2}\Biggl[\int_{M_{11}} \sqrt{-G^{\cal M}}\left(R - \frac{G_4^2}{2. 4!}
+ \beta\left(J_0 - \frac{E_8}{2}\right) \right) \nonumber\\
& & + 2\int_{\partial M_{11}}\sqrt{-h}K - \frac{1}{6}\int_{M_{11}}C_3\wedge G_4\wedge G_4
+ 4\pi^2\int_{M_{11}}C_3\wedge X_8\Biggr],
\end{eqnarray}
that results in an EOM:
\begin{eqnarray}
\label{GMN_EOM}
& & R_{MN} - \frac{G_{MN}}{2}R - \frac{1}{12}\left(G_{MPQR}G_N^{\ \ PQR} - \frac{G_{MN}}{8}G_4^2\right) \nonumber\\
& & = -\beta\left[\frac{G_{MN}}{2}\left(J_0 - \frac{E_8}{2}\right) + \frac{\delta}{\delta G^{MN}}\left(J_0 - \frac{E_8}{2}\right)\right].
\end{eqnarray}
Using the trace of (\ref{GMN_EOM}), we obtained:
\begin{eqnarray}
\label{trace}
& & -\frac{9}{2}R + \frac{G_4^2}{32} = -\beta\left[\frac{11}{2}\left(J_0 - \frac{E_8}{2}\right) + G^{MN}
\frac{\delta}{\delta G^{MN}}\left(J_0 - \frac{E_8}{2}\right)\right]\nonumber\\
& & \approx -\beta\left[\frac{11}{2}J_0+ G^{MN}
\frac{\delta J_0}{\delta G^{MN}}\right],
\end{eqnarray}
wherever there is a justification for the estimate in \cite{HD-MQGP}. Let us write $R = R^{(0)} + \beta R,
K = K^{(0)} + \beta K, C_3 = C_3^{(0)} + \beta C_3$, etc., in \cite{HD-MQGP}, it is also demonstrated that one is capable of setting $C_3=0$ . Therefore,
\begin{eqnarray}
\label{on-shell-relations}
& & G_4^2 = 144 R^{(0)},\nonumber\\
& &  J_0 \approx \frac{9}{11} R^{(1)} - \frac{2}{11}G^{MN} \frac{\delta J_0}{\delta G^{MN}}.
\end{eqnarray}
Hence, on-shell action up to ${\cal O}(\beta)$ using $X_8(M_{11})=0$ \cite{MQGP} is obtained as:
{\footnotesize
\begin{equation}
\label{on-shell-D=11-action-up-to-beta}
 S_{D=11}^{\rm on-shell} =- \frac{1}{2}\Biggl[-2 S_{\rm EH}^{(0)} + 2 S_{\rm GHY}^{(0)} 
+ \beta \left(\frac{20}{11}S_{\rm EH} - 2 \int_{M_{11}}\left(\sqrt{-G^{\cal M}}\right)^{(1)}R^{(0)}
+ 2 S_{\rm GHY} - \frac{2}{11}\int_{M_{11}}\sqrt{-G^{\cal M}}G^{MN}\frac{\delta J_0}{\delta G^{MN}}\right)\Biggr],
\end{equation}
}
where $\left(\sqrt{-G^{\cal M}}\right)^{(1)}=\frac{G_{\cal M}^{(1)}}{2\sqrt{-G_{\cal M}^{(0)}} }$.

\subsubsection{Black Hole Background Uplift (Relevant to $T>T_c$) and Holographic Renormalization of On-Shell $D=11$ Action}
\label{Tc-G-I}
In this chapter, we worked near the Ouyang embedding implemented by the following coordinate patches \cite{NPB}:
\begin{equation}
\label{small-theta_12}
\theta_1 = \frac{\alpha_{\theta_1}}{N^{\frac{1}{5}}},\ \theta_2 = \frac{\alpha_{\theta_2}}{N^{\frac{3}{10}}}.
\end{equation}
The following expression is what we computed the IR Einstein-Hilbert term:
\begin{eqnarray}
\label{EH-BH-IR}
& & \left.\sqrt{-G^{\cal M}}R\right|_{\rm Ouyang}^{\rm IR}\nonumber\\
& &  \sim  -\frac{b^2 {g_s}^{3/4} {\log N}^3 M {N_f}^3 {r_h}^4 \log
   \left(\frac{{r_h}}{{{\cal R}_{D5/\overline{D5}}^{\rm bh}}}\right) \log \left(1-\frac{{r_h}}{{{\cal R}_{D5/\overline{D5}}^{\rm bh}}}\right) (1-\beta 
   ({\cal C}_{zz}^{\rm bh}-2 {\cal C}_{\theta_1z}^{\rm bh}+3 {\cal C}_{\theta_1x}^{\rm bh}))}{{N^{\frac{5}{4}}} {{\cal R}_{D5/\overline{D5}}^{\rm bh}}^3(r-r_h)
   \sin^3{\theta _1} \sin^2{\theta _2}}.\nonumber\\
\end{eqnarray}
Performing now the angular integration for the previously stated equation,
{\footnotesize
\begin{equation}
\label{EH-global-i}
\int\int d\theta_1 d\theta_2 \frac{1}{\sin^3\theta_1\sin^2\theta_2} = \frac{1}{8} \cot ({\theta_2}) \left(\csc ^2\left(\frac{{\theta_1}}{2}\right)-\sec
   ^2\left(\frac{{\theta_1}}{2}\right)-4 \log \left(\sin \left(\frac{{\theta_1}}{2}\right)\right)+4 \log
   \left(\cos \left(\frac{{\theta_1}}{2}\right)\right)\right)
\end{equation}
}
which simplifies near (\ref{small-theta_12}) as given below:
\begin{equation}
\label{EH-global-ii}
\int_{\theta_1={\alpha_{\theta_1}}{N^{-\frac{1}{5}}}}^{\pi-{\alpha_{\theta_1}}{N^{-\frac{1}{5}}}}\int_{\theta_2= {\alpha_{\theta_2}}{N^{-\frac{3}{10}}}}^{\pi- {\alpha_{\theta_2}}{N^{-\frac{3}{10}}}}
 \frac{d\theta_1 d\theta_2}{\sin^3\theta_1\sin^2\theta_2} \sim \frac{N^{\frac{7}{10}}}{\alpha_{\theta_1}^2\alpha_{\theta_2}} + {\cal O}\left(N^{\frac{3}{10}}\right).
\end{equation}
We derived the following expression for the action density (for Einstein-Hilbert term) in the IR by utilizing equation (\ref{EH-global-ii}) and carrying out radial integration of (\ref{EH-BH-IR}):
\begin{eqnarray}
\label{int-EH-BH-IR}
& & \frac{\left.\int_{r=r_h}^{{\cal R}_{D5-\overline{D5}}^{\rm bh}}\int_{\theta_1={\alpha_{\theta_1}}{N^{-\frac{1}{5}}}}^{\pi-{\alpha_{\theta_1}}{N^{-\frac{1}{5}}}}\int_{\theta_2= {\alpha_{\theta_2}}{N^{-\frac{3}{10}}}}^{\pi- {\alpha_{\theta_2}}{N^{-\frac{3}{10}}}}\sqrt{-G^{\cal M}}R\right|_{\rm Ouyang}}{{\cal V}_4} \sim \nonumber\\
& &  -\frac{b^2 {g_s}^{3/4} {\log N}^3 M {N_f}^3 {r_h}^4 \log
   \left(\frac{{r_h}}{{{\cal R}_{D5/\overline{D5}}^{\rm bh}}}\right) \log \left(1-\frac{{r_h}}{{{\cal R}_{D5/\overline{D5}}^{\rm bh}}}\right) (1-\beta 
   ({\cal C}_{zz}^{\rm bh}-2 {\cal C}_{\theta_1z}^{\rm bh}+3 {\cal C}_{\theta_1x}^{\rm bh}))}{{N}^{\frac{11}{20}} {{\cal R}_{D5/\overline{D5}}^{\rm bh}}^4
   \alpha _{\theta _1}^2 \alpha _{\theta _2}},\nonumber\\
\end{eqnarray}
where ${\cal R}_{D5-\overline{D5}}^{\rm bh}\equiv \sqrt{3}a^{\rm bh}$, with the resolution parameter, $a^{\rm bh} = \left(\frac{1}{\sqrt{3}} + \epsilon^{\rm bh} + {\cal O}\left(\frac{g_sM^2}{N}\right)\right)r_h$ of the blown up $S^2$ \cite{HD-MQGP}. In addition, there is one additional term that appears in the on-shell action as stated in the equation (\ref{on-shell-D=11-action-up-to-beta}), and the simplified version of that term is as follows:
\begin{eqnarray}
\label{sqrtGbetaRbeta0-IR-bh}
& & \frac{\int_{r_h}^{{\cal R}_{D5-\overline{D5}}^{\rm bh}}\int_{\theta_1={\alpha_{\theta_1}}{N^{-\frac{1}{5}}}}^{\pi-{\alpha_{\theta_1}}{N^{-\frac{1}{5}}}}\int_{\theta_2= {\alpha_{\theta_2}}{N^{-\frac{3}{10}}}}^{\pi- {\alpha_{\theta_2}}{N^{-\frac{3}{10}}}}
\left(\sqrt{-G^{\cal M}}\right)^{(1)}R^{(0)}}{{\cal V}_4}\nonumber\\
& & \sim \beta {\cal C}_{\theta_1x}^{\rm bh}\frac{b^2g_s^{\frac{3}{4}}\log^3N M N_f^3r_h^4
\log\left(\frac{r_h}{{\cal R}_{D5/\overline{D5}}^{\rm bh}}\right)\log\left(1 - \frac{r_h}{{\cal R}_{D5/\overline{D5}}^{\rm bh}}\right)}{{{\cal R}_{D5/\overline{D5}}^{\rm bh}}^4N^{\frac{11}{20}}\alpha_{\theta_1}^2\alpha_{\theta_2}}.
\end{eqnarray}
The version of the Einstein-Hilbert term pertaining to the ultraviolet(UV) region is:
{\footnotesize
\begin{eqnarray}
\label{EH-BH-UV}
& & \hskip -0.6in \left.\sqrt{-G^{\cal M}}R\right|_{\rm Ouyang}^{\rm UV} = -\frac{4 \left(34 {\log r}^2+14 {\log r}-1\right) {M_{\rm UV}} r^3}{9 \sqrt{3} \sqrt[4]{\pi } {g_s}^{9/4}
  N^{\frac{5}{4}}
   \sin^3{\theta _1} \sin^2{\theta _2} } \nonumber\\
   & & \times \frac{1}{\left(-{g_s} {N_f^{\rm UV}} (2 {\log N}
   {\log r}+{\log N}+3 (1-6 {\log r}) {\log r})+2 {g_s} (2 {\log r}+1) {N_f^{\rm UV}} \log
   \left(\frac{1}{4} \alpha _{\theta _1} \alpha _{\theta _2}\right)+8 \pi  {\log r}\right)}
   \nonumber\\
& & \hskip 0.5in \approx \frac{4 {M_{\rm UV}} r^3 \left(34 \log ^2(r)+14 \log (r)-1\right)}{9 \sqrt{3} \sqrt[4]{\pi } {g_s}^{13/4}
    {N_f^{\rm UV}}  N^{\frac{5}{4}}
   \sin^3{\theta _1} \sin^2{\theta _2} \log N  (2 \log (r)+1)},
\end{eqnarray}
}
where the correction terms, ${\cal O}\left(\frac{1}{\log N}\right)$, have been removed to make the computation simpler. Since, $\int \frac{r^3 \left(34 \log ^2(r)+14 \log (r)-1\right)}{2 \log (r)+1} = \frac{1}{16} \left(\frac{4 {Ei}(4 \log (r)+2)}{e^2}+r^4 (68 \log (r)-23)\right)$ + constant. 
The following UV finite Einstein-Hilbert term was consequently found:
\begin{eqnarray}
\label{int-EH-BH-UV}
& &\left. \left(1+\frac{r_h^4}{2{\cal R}_{\rm UV}^4}\right)\frac{\int_{{\cal R}_{D5-\overline{D5}}}^{{\cal R}_{\rm UV}}\int_{\theta_1={\alpha_{\theta_1}}{N^{-\frac{1}{5}}}}^{\pi-{\alpha_{\theta_1}}{N^{-\frac{1}{5}}}}\int_{\theta_2= {\alpha_{\theta_2}}{N^{-\frac{3}{10}}}}^{\pi- {\alpha_{\theta_2}}{N^{-\frac{3}{10}}}}
\sqrt{-G^{\cal M}}R}{{\cal V}_4}\right|_{\rm Ouyang}^{\rm UV-Finite}\nonumber\\
& & \sim 
\frac{ {M_{\rm UV}} {r_h}^4 \log \left(\frac{{{\cal R}_{\rm UV}}}{{{\cal R}_{D5/\overline{D5}}^{\rm bh}}}\right) }{ {g_s}^{13/4} {N}^{\frac{11}{20}}  {{\cal R}_{D5/\overline{D5}}^{\rm bh}}^4
   {N_f^{\rm UV}} \alpha _{\theta _1}^2 \alpha _{\theta _2} \log N }.
\end{eqnarray}
The following is the part that make up the EH action's contribution to UV divergent:
\begin{eqnarray}
\label{EH-div}
S_{\rm UV-div}^{\rm EH} \sim \frac{ {M_{\rm UV}} {\cal R}_{\rm UV}^4 \log\left(\frac{{{\cal R}_{\rm UV}}}{{{\cal R}_{D5/\overline{D5}}^{\rm bh}}}\right) }{ {g_s^{\rm UV}}^{13/4}{{\cal R}_{D5/\overline{D5}}^{\rm bh}}^4 {\log N}
    {N_f^{\rm UV}}N^{\frac{11}{20}} \alpha _{\theta _1}^2 \alpha _{\theta _2}}.
\end{eqnarray}
Up to ${\cal O}(\beta)$, the UV-finite boundary Gibbons-Hawking-York contribution is found to be:
\begin{eqnarray}
\label{GHY-BH}
& & \hskip -0.45in \left(1+\frac{r_h^4}{2{\cal R}_{\rm UV}^4}\right)\left.\frac{\left.\int_{\theta_1={\alpha_{\theta_1}}{N^{-\frac{1}{5}}}}^{\pi-{\alpha_{\theta_1}}{N^{-\frac{1}{5}}}}\int_{\theta_2= {\alpha_{\theta_2}}{N^{-\frac{3}{10}}}}^{\pi- {\alpha_{\theta_2}}{N^{-\frac{3}{10}}}} \sqrt{-h^{\cal M}}K\right|_{\rm Ouyang}}{{\cal V}_4}\right|^{r={\cal R}_{\rm UV}} \sim \frac{{M_{\rm UV}} {r_h}^4 \log \left(\frac{{{\cal R}_{\rm UV}}}{{{\cal R}_{D5/\overline{D5}}^{\rm bh}}}\right)}{{g_s}^{9/4} {{\cal R}_{D5/\overline{D5}}^{\rm bh}}^4 {N}^{\frac{11}{20}}
   \alpha _{\theta _1}^2 \alpha _{\theta _2}}.
\end{eqnarray}
The GHY term's UV-divergent contribution is as follows:
\begin{equation}
\label{GHY-UV-div}
S_{\rm UV-div}^{\rm GHY} \sim\frac{ \log\left(\frac{{{\cal R}_{\rm UV}}}{{{\cal R}_{D5/\overline{D5}}^{\rm bh}}}\right) {M_{\rm UV}} {{\cal R}_{\rm UV}}^4}{ \left(g_s^{\rm UV}\right)^{9/4}  {{\cal R}_{D5/\overline{D5}}^{\rm bh}}^4 {N}^{\frac{11}{20}} \alpha _{\theta _1}^2 \alpha _{\theta _2}}.
\end{equation}
We showed that the contribution from the higher derivative term is as follows, up to ${\cal O}(\beta^0)$:
{\footnotesize
\begin{eqnarray}
\label{int-iGdeltaJ0-BH-IR}
& & \hskip -0.4in \frac{\int_{r_0}^{{\cal R}_{D5/\overline{D5}}}\int_{\theta_1={\alpha_{\theta_1}}{N^{-\frac{1}{5}}}}^{\pi-{\alpha_{\theta_1}}{N^{-\frac{1}{5}}}}\int_{\theta_2= {\alpha_{\theta_2}}{N^{-\frac{3}{10}}}}^{\pi- {\alpha_{\theta_2}}{N^{-\frac{3}{10}}}}\sqrt{-G^{\cal M}}\left.G^{MN}\frac{\delta J_0}{\delta G^{MN}}\right|_{\rm Ouyang}}{{\cal V}_4}
\nonumber\\
& & \hskip -0.4in \sim \frac{ M {N_f} {r_h} \left(1-\frac{{r_h}}{{{\cal R}_{D5/\overline{D5}}^{\rm bh}}}\right)^3
    \left(\frac{{r_h}}{{{\cal R}_{D5/\overline{D5}}^{\rm bh}}}\right)  \left({\log N}-9  \left(\frac{{r_h}}{{{\cal R}_{D5/\overline{D5}}^{\rm bh}}}\right) \right) \left[\log N -3  \left(\frac{{r_h}}{{{\cal R}_{D5/\overline{D5}}^{\rm bh}}}\right) \right]^2}{\epsilon ^5
   {g_s} N^{39/20} \log ^2(N)}\nonumber\\
& & \hskip -0.4in \times\int_{\theta_1={\alpha_{\theta_1}}{N^{-\frac{1}{5}}}}^{\pi-{\alpha_{\theta_1}}{N^{-\frac{1}{5}}}}\int_{\theta_2= {\alpha_{\theta_2}}{N^{-\frac{3}{10}}}}^{\pi- {\alpha_{\theta_2}}{N^{-\frac{3}{10}}}}\frac{19683 \sqrt{6} \sin^6{\theta _1}+6642 \sin^2{\theta _2} \sin^3{\theta _1}-40 \sqrt{6}
   \sin^4{\theta _2}}{\sin^7{\theta _1}\sin^4{\theta _2}};
\end{eqnarray}
}
utilizing:
{\footnotesize
\begin{eqnarray}
\label{integrals}
& & -\int dr\frac{r^3 \log (r)}{(2 \log (r)+1)^7}\nonumber\\
& &  = \frac{1}{360} \left(\frac{r^4 \left(256 \log ^5(r)+704 \log ^4(r)+800 \log ^3(r)+488 \log ^2(r)+184 \log
   (r)+19\right)}{(2 \log (r)+1)^6}-\frac{16 {Ei}(4 \log (r)+2)}{e^2}\right);
\nonumber\\
& & -\int dr\frac{r \log (r) \left(4 \log ^2(r)+15 \log (r)+9\right)}{(2 \log (r)+1)^7} \nonumber\\
& & = \frac{2 r^2 \left(7504 \log ^5(r)+22512 \log ^4(r)+30016 \log ^3(r)+23384 \log ^2(r)+8799 \log (r)+833\right)}{2880(2
   \log (r)+1)^6}-\frac{469 {Ei}(2 \log (r)+1)}{5760e};\nonumber\\
& & -\int dr\frac{\log (r) \left(45 a^4 \left(64 \log ^3(r)+208 \log ^2(r)+212 \log (r)+57\right)-8 {r_h}^4 (2 \log
   (r)+5)\right)}{r (2 \log (r)+1)^7}\nonumber\\
& &  = \frac{15 \left(3915 a^4-4 {r_h}^4\right) \log ^2(r)+12 \left(1620 a^4-11 {r_h}^4\right) \log (r)+21600 a^4 \log
   ^4(r)+61200 a^4 \log ^3(r)+1620 a^4-11 {r_h}^4}{30 (2 \log (r)+1)^6}.\nonumber\\
& & 
\end{eqnarray}
}
The higher derivative term that gives rise to the UV finite contribution was found to be:
{\footnotesize
\begin{eqnarray}
\label{int-iGdeltaJ0--UV}
& & \hskip -0.6in  \left.\left(1 + \frac{r_h^4}{2{\cal R}_{\rm UV}^4}\right)\frac{\int_{\theta_1={\alpha_{\theta_1}}{N^{-\frac{1}{5}}}}^{\pi-{\alpha_{\theta_1}}{N^{-\frac{1}{5}}}}\int_{\theta_2= {\alpha_{\theta_2}}{N^{-\frac{3}{10}}}}^{\pi- {\alpha_{\theta_2}}{N^{-\frac{3}{10}}}}\int_{{\cal R}_{D5/\overline{D5}}}^{{\cal R}_{\rm UV}}\sqrt{-G^{\cal M}}\left.G^{MN}\frac{\delta J_0}{\delta G^{MN}}\right|_{\rm Ouyang}}{{\cal V}_4}\right|^{\rm UV-finite}
\nonumber\\
& & \hskip -0.6in = \left.\frac{\int_{\theta_1={\alpha_{\theta_1}}{N^{-\frac{1}{5}}}}^{\pi-{\alpha_{\theta_1}}{N^{-\frac{1}{5}}}}\int_{\theta_2= {\alpha_{\theta_2}}{N^{-\frac{3}{10}}}}^{\pi- {\alpha_{\theta_2}}{N^{-\frac{3}{10}}}}\int_{{\cal R}_{D5/\overline{D5}}}^{{\cal R}_{\rm UV}}\sqrt{-G^{\cal M}}\left.G^{MN}\frac{\delta J_0}{\delta G^{MN}}\right|_{\rm Ouyang}}{{\cal V}_4}\right|^{\rm UV-finite}\nonumber\\
 & & \hskip -0.6in \sim -\frac{ {M_{\rm UV}}}{{g_s^{\rm UV}}^{14/3}
   {\log N}^{11/3} N^{39/20} {N_f^{\rm UV}}^{8/3}}\left(\frac{
   {r_h}^2}{{{\cal R}_{D5/\overline{D5}}^{\rm bh}}^2}+1\right)\nonumber\\
   & & \times \int_{\theta_1={\alpha_{\theta_1}}{N^{-\frac{1}{5}}}}^{\pi-{\alpha_{\theta_1}}{N^{-\frac{1}{5}}}}\int_{\theta_2= {\alpha_{\theta_2}}{N^{-\frac{3}{10}}}}^{\pi- {\alpha_{\theta_2}}{N^{-\frac{3}{10}}}} \frac{19683 \sqrt{6} \sin^6{\theta _1}+6642 \sin^2{\theta _2} \sin^3{\theta _1}-40 \sqrt{6}
   \sin^4{\theta _2}}{\sin^7{\theta _1}\sin^4{\theta _2}}.
\end{eqnarray}
}
The following has been utilized.  The most dominating term in $\sqrt{-G^{\cal M}}G^{MN}\frac{\delta J_0}{\delta G^{MN}}$ is \\ $-\frac{8 {M_{\rm UV}} r^3 \log (r)}{177147 \pi  {g_s^{\rm UV}}^4 (2 {\log r}+1)^7 N^{39/20} {N_f^{\rm UV}}^2 \log ^3(N)}\frac{19683 \sqrt{6} \sin^6{\theta _1}+6642 \sin^2{\theta _2} \sin^3{\theta _1}-40 \sqrt{6}
   \sin^4{\theta _2}}{\sin^7{\theta _1}\sin^4{\theta _2}}$. Since: \\
$\int dr \frac{r^3 \log (r)}{(2 \log (r)+1)^7} = \frac{2 {Ei}(4 \log (r)+2)}{45 e^2}-\frac{r^4 \left(256 \log ^5(r)+704 \log ^4(r)+800 \log ^3(r)+488 \log ^2(r)+184
   \log (r)+19\right)}{360 (2 \log (r)+1)^6}$.\\ Therefore, it is important to highlight that the contribution of UV-finite to: \\ $\left(1 + \frac{r_h^4}{2{\cal R}_{\rm UV}^4}\right)\sqrt{-G^{\cal M}}G^{MN}\frac{\delta J_0}{\delta G^{MN}}$ is:\\
$-\frac{ {M_{\rm UV}}}{{g_s^{\rm UV}}^{14/3}
   {\log N}^{11/3} N^{39/20} {N_f^{\rm UV}}^{8/3}} \frac{r_h^4}{\log\left(\frac{{{\cal R}_{\rm UV}}}{{{\cal R}_{D5/\overline{D5}}^{\rm bh}}}\right)}\left(\frac{
   {r_h}^2}{{{\cal R}_{D5/\overline{D5}}^{\rm bh}}^2}+1\right)\frac{19683 \sqrt{6} \sin^6{\theta _1}+6642 \sin^2{\theta _2} \sin^3{\theta _1}-40 \sqrt{6}
   \sin^4{\theta _2}}{\sin^7{\theta _1}\sin^4{\theta _2}}$,\\ which ends up being the case: ${\cal O}\left(\frac{r_h^4}{\log\left(\frac{{{\cal R}_{\rm UV}}}{{{\cal R}_{D5/\overline{D5}}^{\rm bh}}}\right)}\right)$-suppressed in comparison to \\ $\left.\int_{{\cal R}_{D5/\overline{D5}}}^{{\cal R}_{\rm UV}} \sqrt{-G^{\cal M}}G^{MN}\frac{\delta J_0}{\delta G^{MN}}\right|^{\rm UV-finite}$. Additionally, UV divergence contribution was found from higher derivative term as:
\begin{eqnarray}
\label{int-sqrtGiGdeltaJ0-beta0-UV-div-Tc}
& & \hskip -0.3in \frac{\left.\int \sqrt{-G^{\cal M}}G^{MN}\frac{\delta J_0}{\delta G^{MN}}\right|_{\rm UV-div}}{{\cal V}_4}
\nonumber\\
& & \hskip -0.3in  \sim \int_{\theta_1={\alpha_{\theta_1}}{N^{-\frac{1}{5}}}}^{\pi-{\alpha_{\theta_1}}{N^{-\frac{1}{5}}}}\int_{\theta_2= {\alpha_{\theta_2}}{N^{-\frac{3}{10}}}}^{\pi- {\alpha_{\theta_2}}{N^{-\frac{3}{10}}}}\frac{19683 \sqrt{6} \sin^6{\theta _1}+6642 \sin^2{\theta _2} \sin^3{\theta _1}-40 \sqrt{6}
   \sin^4{\theta _2}}{\sin^7{\theta _1}\sin^4{\theta _2}}\nonumber\\
& & \times \frac{ {M_{\rm UV}} {{\cal R}_{\rm UV}}^4 }{ {{\cal R}_{D5/\overline{D5}}^{\rm bh}}^4 {g_s^{\rm UV}}^{14/3} N^{39/20} {N_f^{\rm UV}}^{8/3} \alpha _{\theta _1}^7
   \alpha _{\theta _2}^4 \log ^{\frac{11}{3}}(N) \log\left(\frac{{{\cal R}_{\rm UV}}}{{{\cal R}_{D5/\overline{D5}}^{\rm bh}}}\right)}.
\end{eqnarray}
In addition, we observe that the surface counter term that neutralizes UV divergence is emerging via higher derivative term is provided by the expression that follows when r is held constant.
{\footnotesize
\begin{eqnarray}
\label{isqrthinvGdeltaJ0-constr}
& & \hskip -0.3in  \beta\left.\frac{\left.\int \sqrt{-h}h^{mn}\frac{\delta J_0}{\delta h^{mn}}\right|_{\rm UV-div}}{{\cal V}_4}\right|_{r=\rm constant} \sim -\beta\frac{\left(\frac{1}{{g_s}^{\rm UV}}\right)^{2/3} {M_{\rm UV}} {{\cal R}_{\rm UV}}^4 }{{g_s^{\rm UV}}^{13/4}{{\cal R}_{D5/\overline{D5}}^{\rm bh}}^4 {\log N}^3 \left\{\log\left(\frac{{{\cal R}_{\rm UV}}}{{{\cal R}_{D5/\overline{D5}}^{\rm bh}}}\right)\right\}^6 N^{11/5} {N_f^{\rm UV}}^2}\nonumber\\
& & \hskip -0.3in \times\int_{\theta_1={\alpha_{\theta_1}}{N^{-\frac{1}{5}}}}^{\pi-{\alpha_{\theta_1}}{N^{-\frac{1}{5}}}}\int_{\theta_2= {\alpha_{\theta_2}}{N^{-\frac{3}{10}}}}^{\pi- {\alpha_{\theta_2}}{N^{-\frac{3}{10}}}} \frac{19683 \sqrt{6} \sin^6{\theta _1}+6642 \sin^2{\theta _2} \sin^3{\theta _1}-40 \sqrt{6}
   \sin^4{\theta _2}}{\sin^7{\theta _1}\sin^4{\theta _2}} + {\cal O}(\beta^2).
\end{eqnarray}
}
Using:
{\footnotesize
\begin{eqnarray}
\label{theta1-theta2-global-intsqrtGiGdeltaJ0-i}
& & \hskip -0.3in \frac{1}{N^{\frac{7}{5}}}\int\int d\theta_1 d\theta_2 \frac{19683 \sqrt{6} \sin^6{\theta _1}+6642 \sin^2{\theta _2} \sin^3{\theta _1}-40 \sqrt{6}
   \sin^4{\theta _2}}{\sin^7{\theta _1}\sin^4{\theta _2}}\nonumber\\
& & \hskip -0.3in  =  \frac{1}{N^{\frac{7}{5}}}\Biggl[-2214 \cot ^3({\theta_1}) \cot ({\theta_2})+2214 \cot ({\theta_1}) \left(2 \csc
   ^2({\theta_1})+1\right) \cot ({\theta_2})+\frac{1}{32 \sqrt{6}}\Biggl\{20 {\theta_2} \csc
   ^6\left(\frac{{\theta_1}}{2}\right)+120 {\theta_2} \csc
   ^4\left(\frac{{\theta_1}}{2}\right)\nonumber\\
& & \hskip -0.3in  +600 {\theta_2} \csc
   ^2\left(\frac{{\theta_1}}{2}\right)+5 {\theta_2} \sec
   ^6\left(\frac{{\theta_1}}{2}\right) \Biggl(-150 \log \left(\sin
   \left(\frac{{\theta_1}}{2}\right)\right)+15 \cos (3 {\theta_1}) \log \left(\cos
   \left(\frac{{\theta_1}}{2}\right)\right)+150 \log \left(\cos
   \left(\frac{{\theta_1}}{2}\right)\right)\nonumber\\
& & \hskip -0.3in +9 \cos ({\theta_1}) \left(-25 \log
   \left(\sin \left(\frac{{\theta_1}}{2}\right)\right)+25 \log \left(\cos
   \left(\frac{{\theta_1}}{2}\right)\right)-8\right)+15 \cos (2 {\theta_1}) \left(-6
   \log \left(\sin \left(\frac{{\theta_1}}{2}\right)\right)+6 \log \left(\cos
   \left(\frac{{\theta_1}}{2}\right)\right)-1\right)\nonumber\\
& & \hskip -0.3in -15 \cos (3 {\theta_1}) \log
   \left(\sin \left(\frac{{\theta_1}}{2}\right)\right)-61\Biggr)-1259712 (\cos (2
   {\theta_2})-2) \cot ({\theta_2}) \csc ^2({\theta_2}) \left(\log \left(\cos
   \left(\frac{{\theta_1}}{2}\right)\right)-\log \left(\sin
   \left(\frac{{\theta_1}}{2}\right)\right)\right)\Biggr\}\Biggr],\nonumber\\
   & &
\end{eqnarray}
}
We found that small values of $\theta_{1,2}$ produced the most significant effect. Therefore, 
\begin{eqnarray}
\label{theta1-theta2-global-intsqrtGiGdeltaJ0-ii}
& &  \int_{\theta_1={\alpha_{\theta_1}}{N^{-\frac{1}{5}}}}^{\pi-{\alpha_{\theta_1}}{N^{-\frac{1}{5}}}}\int_{\theta_2= {\alpha_{\theta_2}}{N^{-\frac{3}{10}}}}^{\pi- {\alpha_{\theta_2}}{N^{-\frac{3}{10}}}} \frac{19683 \sqrt{6} \sin^6{\theta _1}+6642 \sin^2{\theta _2} \sin^3{\theta _1}-40 \sqrt{6}
   \sin^4{\theta _2}}{\sin^7{\theta _1}\sin^4{\theta _2}}\nonumber\\
& &  \sim N^{\frac{9}{10}}\frac{ \left(-6642 {\alpha_{\theta_1}}^3 {\alpha_{\theta_2}}^2+19683 \sqrt{6}
   {\alpha_{\theta_1}}^6 \log \left(\frac{{\alpha_{\theta_1}}}{2 \sqrt[5]{N}}\right)-20
   \sqrt{6} {\alpha_{\theta_2}}^4\right)}{3 {\alpha_{\theta_1}}^6 {\alpha_{\theta_2}}^3}
\sim -\frac{\log N}{\alpha_{\theta_2}^3}N^{\frac{9}{10}}.
\end{eqnarray}
Because of this, the simpler version of the counter term is as follows: 
{\footnotesize
\begin{eqnarray}
\label{ihdeltaJ0-r=rUV}
& & \hskip -0.3in  \beta\left.\frac{\left.\int \sqrt{-h}h^{mn}\frac{\delta J_0}{\delta h^{mn}}\right|_{\rm UV-div}}{{\cal V}_4}\right|_{r=\rm constant} \sim \beta\frac{\left(\frac{1}{{g_s}^{\rm UV}}\right)^{2/3} {M_{\rm UV}} {{\cal R}_{\rm UV}}^4 \log N}{{g_s^{\rm UV}}^{13/4}{{\cal R}_{D5/\overline{D5}}^{\rm bh}}^4 {\log N}^3 \left\{\log\left(\frac{{{\cal R}_{\rm UV}}}{{{\cal R}_{D5/\overline{D5}}^{\rm bh}}}\right)\right\}^6 N^{13/10} {N_f^{\rm UV}}^2\alpha_{\theta_2}^3}.
\end{eqnarray}
}
Further,
\begin{eqnarray}
\label{sqrthJ0-constr}
& & \beta\left.\frac{\int_{r={\cal R}_{\rm UV}} \sqrt{-h} J_0}{{\cal V}_4}\right|^{\rm UV}\nonumber\\
& & \sim \frac{M_{\rm UV}}{g_s^{\rm UV}\log^{\frac{8}{3}}N^{3}N_f^{\rm UV}\ ^{\frac{8}{3}}}\left(\frac{{{\cal R}_{\rm UV}}}{{{\cal R}_{D5/\overline{D5}}^{\rm bh}}}\right)^4\log\left(\frac{{{\cal R}_{\rm UV}}}{{{\cal R}_{D5/\overline{D5}}^{\rm bh}}}\right)\int_{\theta_1={\alpha_{\theta_1}}{N^{-\frac{1}{5}}}}^{\pi-{\alpha_{\theta_1}}{N^{-\frac{1}{5}}}}\int_{\theta_2= {\alpha_{\theta_2}}{N^{-\frac{3}{10}}}}^{\pi- {\alpha_{\theta_2}}{N^{-\frac{3}{10}}}}
 \frac{d\theta_1 d\theta_2}{\sin^3\theta_1\sin^2\theta_2}\nonumber\\
& & \sim \frac{M_{\rm UV}}{g_s^{\rm UV}\log^{\frac{8}{3}}N^{\frac{23}{10}}N_f^{\rm UV}\ ^{\frac{8}{3}}}\left(\frac{{{\cal R}_{\rm UV}}}{{{\cal R}_{D5/\overline{D5}}^{\rm bh}}}\right)^4\log\left(\frac{{{\cal R}_{\rm UV}}}{{{\cal R}_{D5/\overline{D5}}^{\rm bh}}}\right). 
\end{eqnarray}
By matching the results of (\ref{int-sqrtGiGdeltaJ0-beta0-UV-div-Tc}) with those of (\ref{isqrthinvGdeltaJ0-constr}), one is able to impose the following:
\begin{equation}
\label{Nf_UV}
\log ^3N\left\{\log\left(\frac{{{\cal R}_{\rm UV}}}{{{\cal R}_{D5/\overline{D5}}^{\rm bh}}}\right)\right\}^5  {N_f^{\rm UV}}^2 \sim 
 {N_f^{\rm UV}}^{8/3}\log ^{\frac{11}{3}}(N),
\end{equation}
results in:
\begin{equation}
\label{Nf-LogRUV-Tc}
N_f^{\rm UV}\sim\frac{\left(\log\left(\frac{{{\cal R}_{\rm UV}}}{{{\cal R}_{D5/\overline{D5}}^{\rm bh}}}\right)\right)^{\frac{15}{2}}}{\log N}.
\end{equation}
{\bf Holographic Renormalization When $T>T_c$}: It has been found that the form of the UV-divergent part of the on-shell action (\ref{on-shell-D=11-action-up-to-beta}) for the black hole backdrop is as follows: 
{\footnotesize
\begin{eqnarray}
\label{on-shell-UV-divergent-BH}
& & \hskip -0.35in
 S_{\rm D=11}^{\rm on-shell UV-divergent} =- \frac{1}{2}\Biggl[\left(-2\kappa_{\rm EH}^{(0)}(g_s^{\rm UV},N,M_{\rm UV},N_f^{\rm UV};\alpha_{\theta_{1,2}}) + 2\kappa^{(0)}_{\rm GHY}(g_s^{\rm UV},N,M_{\rm UV},N_f^{\rm UV};\alpha_{\theta_{1,2}})\right)\nonumber\\
& & \hskip -0.35in  \times
\left(\frac{{{\cal R}_{\rm UV}}}{{{\cal R}_{D5/\overline{D5}}^{\rm bh}}}\right)^4\log\left(\frac{{{\cal R}_{\rm UV}}}{{{\cal R}_{D5/\overline{D5}}^{\rm bh}}}\right) + \beta\Biggl\{ -\frac{2}{11}\kappa^{(0)}_{\sqrt{-g}g^{MN}\frac{\delta J_0}{\delta g^{MN}}}(g_s^{\rm UV},N,M_{\rm UV},N_f^{\rm UV};\alpha_{\theta_{1,2}})\frac{\left(\frac{{{\cal R}_{\rm UV}}}{{{\cal R}_{D5/\overline{D5}}^{\rm bh}}}\right)^4}{\log\left(\frac{{{\cal R}_{\rm UV}}}{{{\cal R}_{D5/\overline{D5}}^{\rm bh}}}\right)}\Biggr\}\Biggr].
\end{eqnarray}
}
As a result, the following are the necessary counter terms in order to cancel out the UV-divergences:
{\footnotesize
\begin{eqnarray}
\label{CT-BH-UV-divergent}
& &  S_{\rm CT}^{\rm BH}  = - \frac{1}{2}\Biggl[-\frac{\kappa^{(0)}_{\rm EH+GHY}}{\kappa^{(0)}_{{\rm EH}@{\cal R}_{\rm UV}}}\int_{r={\cal R}_{\rm UV}}\sqrt{-h}R\nonumber\\
& &  + \beta N^{\frac{1}{4}} \frac{\left(\frac{2}{11}\kappa^{(0)}_{\sqrt{-g}g^{MN}\frac{\delta J_0}{\delta g^{MN}}}(g_s^{\rm UV},M_{\rm UV},N_f^{\rm UV};\alpha_{\theta_{1,2}})\right)\int_{r={\cal R}_{\rm UV}}\sqrt{-h}g^{mn}\frac{\delta J_0}{\delta g^{mn}}}{\kappa^{(0)}_{\sqrt{-h}g^{mn}\frac{\delta J_0}{\delta g^{mn}}@{\cal R}_{\rm UV}}} \Biggr]_{\log\left(\frac{{{\cal R}_{\rm UV}}}{{{\cal R}_{D5/\overline{D5}}^{\rm bh}}}\right) = \left(N_f^{\rm UV}\log N\right)^{\frac{2}{15}}},\nonumber\\
\end{eqnarray}
}
where the second term of (\ref{CT-BH-UV-divergent}) is worked out in (\ref{ihdeltaJ0-r=rUV}), and
\begin{eqnarray}
\label{sqrtminush}
& & \frac{\left.\int_{\theta_1={\alpha_{\theta_1}}{N^{-\frac{1}{5}}}}^{\pi-{\alpha_{\theta_1}}{N^{-\frac{1}{5}}}}\int_{\theta_2= {\alpha_{\theta_2}}{N^{-\frac{3}{10}}}}^{\pi- {\alpha_{\theta_2}}{N^{-\frac{3}{10}}}} \sqrt{-h}\right|_{r={\cal R}_{\rm UV}}}{{\cal V}_4}\sim \frac{
   \left(\frac{1}{{g_s^{\rm UV}}}\right)^{7/3} {M_{\rm UV}} {\cal R}_{\rm UV}^4\log\left(\frac{{{\cal R}_{\rm UV}}}{{{\cal R}_{D5/\overline{D5}}^{\rm bh}}}\right)}{N^{\frac{3}{10}}  \alpha _{\theta _1}^2
   \alpha _{\theta _2}{{\cal R}_{D5/\overline{D5}}^{\rm bh}}^4}, \nonumber\\
   & & \frac{\left.\int_{\theta_1={\alpha_{\theta_1}}{N^{-\frac{1}{5}}}}^{\pi-{\alpha_{\theta_1}}{N^{-\frac{1}{5}}}}\int_{\theta_2= {\alpha_{\theta_2}}{N^{-\frac{3}{10}}}}^{\pi- {\alpha_{\theta_2}}{N^{-\frac{3}{10}}}}\sqrt{-h}R\right|_{r={\cal R}_{\rm UV}}}{{\cal V}_4} \sim \frac{\left(\frac{1}{{g_s^{\rm UV}}}\right)^{2/3} {M_{\rm UV}} {\cal R}_{\rm UV}^4 \log\left(\frac{{{\cal R}_{\rm UV}}}{{{\cal R}_{D5/\overline{D5}}^{\rm bh}}}\right) }{ {g_s^{\rm UV}}^{3/2} N^{1/5}{{\cal R}_{D5/\overline{D5}}^{\rm bh}}^4 \alpha
   _{\theta _1}^2 \alpha _{\theta _2}}.
\end{eqnarray}

\subsubsection{Thermal Background Uplift (Relevant to $T<T_c$) and Holographic Renormalization of On-Shell $D=11$ Action}
\label{Tc-G-II}
In the limit of large $N$, as well as in the infrared, the $f_{MN}$ EOMs associated with the thermal background become algebraic. Writing below just those components that receive a non-trivial ${\cal O}(\beta)$ contributions, below is the ${\cal O}(\beta)$-corrected MQGP metric for the thermal background in the IR in the $\psi=2n\pi,n=0,1,2$-coordinate patches.
\begin{eqnarray}
\label{MQGP-th-OR4}
& & G^{\cal M}_{rr} = G^{\rm MQGP}_{rr}\left(1-\frac{99 \sqrt{\frac{3}{2}} \beta  {g_s}^{3/2} M \sqrt[5]{\frac{1}{N}} N_f  {r_0} \alpha _{\theta _1}^6
   f_{x^{10}x^{10}}({r_0}) \log ^2({r_0})}{2 \pi ^{3/2} \alpha _{\theta _2}^5}\right),\nonumber\\
& & G^{\cal M}_{x\theta_1} = G^{\rm MQGP}_{x\theta_1}\left(1 - \beta f_{x^{10}x^{10}}(r_0)\right),\nonumber\\
& & G^{\cal M}_{y\theta_1} = G^{\rm MQGP}_{y\theta_1}\left(1 + \beta f_{\theta_1y}(r_0)\right),\nonumber\\
& & G^{\cal M}_{z\theta_1} = G^{\rm MQGP}_{z\theta_1}\left(1+\frac{539 \pi ^3 \beta  N^{2/5} \alpha _{\theta _2}^2 f_{x^{10}x^{10}}({r_0})}{1728 {g_s}^3 M^2 N_f ^2 \log^2({r_0})}\right),\nonumber\\
& & G^{\cal M}_{y\theta_2} = G^{\rm MQGP}_{y\theta_2}\left(1 - 2\beta f_{x^{10}x^{10}}(r_0)\right),\nonumber\\
& & G^{\cal M}_{yy} = G^{\rm MQGP}_{yy}\left(1-\frac{\pi ^{3/2} \beta  N^{2/5} \alpha _{\theta _2}^2 f_{x^{10}x^{10}}({r_0}) \left(1617 \sqrt{3} \pi ^{3/2} {r_0}
   \alpha _{\theta _1}^4-32 \sqrt{2} {g_s}^{3/2} M N_f  \alpha _{\theta _2}\right)}{2592 \sqrt{3} {g_s}^3
   M^2 N_f ^2 {r_0} \alpha _{\theta _1}^4 \log ^2({r_0})}\right),\nonumber\\
& & G^{\cal M}_{yz} = G^{\rm MQGP}_{yz}\left(1 + \frac{\pi ^{3/2} \beta  N^{2/5} \alpha _{\theta _2}^3 f_{x^{10}x^{10}}({r_0})}{81 \sqrt{6} {g_s}^{3/2} M N_f 
   {r_0} \alpha _{\theta _1}^4 \log ^2({r_0})}\right),\nonumber\\
& & G^{\cal M}_{zz} = G^{\rm MQGP}_{zz}\left(1+\frac{539 \pi ^3 \beta  N^{2/5} \alpha _{\theta _2}^2 f_{x^{10}x^{10}}({r_0})}{864 {g_s}^3 M^2 N_f ^2 \log
   ^2({r_0})}\right),\nonumber\\
& & G^{\cal M}_{x^{10}x^{10}} = G^{\rm MQGP}_{x^{10}x^{10}}\left(1 + \beta f_{x^{10}x^{10}}(r_0)\right).
\end{eqnarray}
The Einstein-Hilbert term pertaining to the thermal background, which works in the infrared close to (\ref{small-theta_12}), is:
\begin{eqnarray}
\label{EH-thermal-IR}
& & \left.\sqrt{-G^{\cal M}}R\right|_{\rm Ouyang}^{\rm IR}\sim  -\frac{{g_s}^{3/4} {\log N}^2 \log \left(\frac{r_0}{{\cal R}_{D5/\overline{D5}}^{\rm th}}\right) M {N_f}^3 }{{{\cal R}_{D5/\overline{D5}}^{\rm th}}^3 N^{\frac{5}{4}} \sin^3\theta_1\sin^2\theta_2}\sum_{m=0}^2\kappa_{\rm EH,\ IR}^{\beta^0,m}\ {r_0}^{3-m}\left(r-r_0\right)^{m}\nonumber\\
& &  -\beta  \frac{
   {\log N}  \left(243 \sqrt{6} \sin^3{\theta _1}-8 \sin^2{\theta _2}\right)
   f_{x^{10}x^{10}}({r_0})}{ N^{\frac{1}{4}}\sin^5{\theta _1}{{\cal R}_{D5/\overline{D5}}^{\rm th}}^3 {g_s}^{9/4}  \log \left(\frac{r_0}{{\cal R}_{D5/\overline{D5}}^{\rm th}}\right) M  ({\log N}
   {N_f})^{5/3}}\sum_{m=0}^2\kappa_{\rm EH,\ IR}^{\beta^1,m}\ {r_0}^{3-m}\left(r-r_0\right)^{m},\nonumber\\
\end{eqnarray} 
where ${\cal R}_{D5/\overline{D5}}^{\rm th} = \sqrt{3}a^{\rm th}$, the resolution parameter $a^{\rm th} = \left(\frac{1}{\sqrt{3}} + \epsilon + {\cal O}\left(\frac{g_sM^2}{N}\right)\right)r_0$ \cite{HD-MQGP} of the blown up $S^2$ associated with the thermal background, and we have calculated the numerical prefactors $\kappa_{\rm EH,\ IR}^{\beta^m,n}$, but we will not provide them here because they do not shed any new light on the $\beta^m(r-r_0)^n$ terms in $\sqrt{-G^{\cal M}}R$ that are constrained to the Ouyang embedding of the flavor $D7$-branes in the type IIB dual and in the IR. From (\ref{EH-thermal-IR}), we evaluated $\frac{\left.\int_{r=r_0}^{{\cal R}_{D5-\overline{D5}}}\sqrt{-G^{\cal M}}R\right|_{\rm Ouyang}}{{\cal V}_4}$. At $O(\beta)$, the following terms also appeared in on-shell action:
\begin{eqnarray}
\label{intsqrtGbetaRbeta0}
& & \frac{\int_{r_h}^{{\cal R}_{D5-\overline{D5}}^{\rm bh}}\int_{\theta_1={\alpha_{\theta_1}}{N^{-\frac{1}{5}}}}^{\pi-{\alpha_{\theta_1}}{N^{-\frac{1}{5}}}}\int_{\theta_2= {\alpha_{\theta_2}}{N^{-\frac{3}{10}}}}^{\pi- {\alpha_{\theta_2}}{N^{-\frac{3}{10}}}}
\left(\sqrt{-G^{\cal M}}\right)^{(1)}R^{(0)}}{{\cal V}_4}\nonumber\\
& & \sim \frac{\beta  {g_s}^{3/4} {\log N}^2 M \left(\frac{1}{N}\right)^{11/20} {N_f}^3 {r_0}
  {f_{x^{10}x^{10}}}({r_0}) \left(1-\frac{{r_0}}{{\cal R}_{D5/\overline{D5}}^{\rm th}}\right) \log
   \left(\frac{{r_0}}{{\cal R}_{D5/\overline{D5}}^{\rm th}}\right)}{{{\cal R}_{D5/\overline{D5}}^{\rm th}} \alpha _{\theta _1}^2 \alpha
   _{\theta _2}}.
\end{eqnarray}
In the ultraviolet(UV), the equivalent Einstein-Hilbert term is:
{\footnotesize
\begin{eqnarray}
\label{EH-thermal-UV}
& & \left.\sqrt{-G^{\cal M}}R\right|_{\rm Ouyang}^{\rm UV} =  \kappa_{\rm EH,\ UV}^{\beta^0,0}\ \frac{\Biggl[11-8 {\log \left(\frac{{r}}{{{\cal R}_{D5/\overline{D5}}^{\rm th}}}\right)}\Biggr] {M_{\rm UV}} {N_f^{\rm UV}} {r_0}^4}{{{\cal R}_{D5/\overline{D5}}^{\rm th}}^3 {g_s}^{5/4}  r
  N^{\frac{5}{4}} \sin^3\theta_1\sin^2\theta_2}\nonumber\\
& & +\kappa_{\rm EH,\ UV}^{\beta^0,1}\ \frac{\Biggl[4 {\log \left(\frac{{r}}{{{\cal R}_{D5/\overline{D5}}^{\rm th}}}\right)}-1\Biggr] {M_{\rm UV}} {N_f^{\rm UV}} r {r_0}^2}{{{\cal R}_{D5/\overline{D5}}^{\rm th}}^3 {g_s}^{5/4} N^{\frac{5}{4}} \sin^3\theta_1\sin^2\theta_2}+\kappa_{\rm EH,\ UV}^{\beta^0,2}\ \frac{{M_{\rm UV}}
   {N_f^{\rm UV}} r^3}{18 \sqrt{3} \pi ^{9/4}{{\cal R}_{D5/\overline{D5}}^{\rm th}}^3 {g_s}^{5/4}N^{\frac{5}{4}} \sin^3\theta_1\sin^2\theta_2},\nonumber\\
\end{eqnarray}
}
implying: 
\begin{eqnarray}
\label{int-EH-thermal-UV}
& &  \frac{\left.\int_{\theta_1={\alpha_{\theta_1}}{N^{-\frac{1}{5}}}}^{\pi-{\alpha_{\theta_1}}{N^{-\frac{1}{5}}}}\int_{\theta_2= {\alpha_{\theta_2}}{N^{-\frac{3}{10}}}}^{\pi- {\alpha_{\theta_2}}{N^{-\frac{3}{10}}}} \int_{{\cal R}_{D5-\overline{D5}}}^{{\cal R}_{\rm UV}}dr \sqrt{-G^{\cal M}}R\right|_{\rm Ouyang}^{\rm UV-Finite}}{{\cal V}_4}\nonumber\\
& & \sim 
 -\kappa_{\rm EH,\ UV}^{\beta^0,0}\ \int_{\theta_1={\alpha_{\theta_1}}{N^{-\frac{1}{5}}}}^{\pi-{\alpha_{\theta_1}}{N^{-\frac{1}{5}}}}\int_{\theta_2= {\alpha_{\theta_2}}{N^{-\frac{3}{10}}}}^{\pi- {\alpha_{\theta_2}}{N^{-\frac{3}{10}}}} \frac{{M_{\rm UV}} {N_f^{\rm UV}}
   \left(-\frac{121 {r_0}^4}{16 {{\cal R}_{D5/\overline{D5}}^{\rm th}}^4}-\frac{6 {r_0}^2}{{{\cal R}_{D5/\overline{D5}}^{\rm th}}^2}+2\right)}{ {g_s^{\rm UV}}^{5/4} N^{\frac{5}{4}} \sin^3\theta_1\sin^2\theta_2}.
\end{eqnarray}
Hence, the UV finite Einstein-Hilbert term, after carrying out the radial and angular integrals of the preceding equation is obtained as:
\begin{eqnarray}
\label{EH-thermal-global-iii}
& &  \frac{\left.\int_{\theta_1={\alpha_{\theta_1}}{N^{-\frac{1}{5}}}}^{\pi-{\alpha_{\theta_1}}{N^{-\frac{1}{5}}}}\int_{\theta_2= {\alpha_{\theta_2}}{N^{-\frac{3}{10}}}}^{\pi- {\alpha_{\theta_2}}{N^{-\frac{3}{10}}}} \int_{{\cal R}_{D5-\overline{D5}}}^{{\cal R}_{\rm UV}}dr \sqrt{-G^{\cal M}}R\right|_{\rm Ouyang}^{\rm UV-Finite}}{{\cal V}_4}\nonumber\\
& & \sim 
 -\kappa_{\rm EH,\ UV}^{\beta^0,0}\ \frac{{M_{\rm UV}} {N_f^{\rm UV}}
   \left(-\frac{121 {r_0}^4}{16 {{\cal R}_{D5/\overline{D5}}^{\rm th}}^4}\frac{1}{\alpha_{\theta_1}^2\alpha_{\theta_2}}-\frac{6 {r_0}^2}{{{\cal R}_{D5/\overline{D5}}^{\rm th}}^2}+2\right)}{ {g_s^{\rm UV}}^{5/4} N^{\frac{11}{20}} }\frac{1}{\alpha_{\theta_1}^2\alpha_{\theta_2}}.
\end{eqnarray}
The Einstein-Hilbert divergent term for the thermal background in the UV region is:
\begin{eqnarray}
\label{EH-th-UV-div}
& & \frac{S_{\rm EH - thermal}^{\rm UV-divergent}}{{\cal V}_4} \sim \frac{\kappa_{\rm EH,\ UV}^{\beta^0,2}}{4}\frac{{M_{\rm UV}} {N_f^{\rm UV}} {\cal R}_{\rm UV}^4}{{{\cal R}_{D5/\overline{D5}}^{\rm th}}^4{g_s^{\rm UV}}^{5/4}N^{\frac{11}{20}} }\frac{1}{\alpha_{\theta_1}^2\alpha_{\theta_2}}. 
\end{eqnarray}
For a thermal background up to ${\cal O}(\beta)$, the UV-finite portion of the boundary Gibbons-Hawking-York term is calculated to be:
\begin{eqnarray}
\label{GHY-thermal}
& & \hskip -0.5in \frac{\left.\int \sqrt{-h^{\cal M}}K\right|_{\rm Ouyang}^{r={\cal R}_{\rm UV}\ {\rm UV-finite}}}{{\cal V}_4} \sim\frac{ {r_0}^4 \left( \kappa_{\rm GHY}^{\beta^0, {\rm UV-finite}}\ \frac{{g_s}^3 {M_{\rm UV}}^2 \log
   \left(\frac{{{\cal R}_{\rm UV}}}{{{\cal R}_{D5/\overline{D5}}^{\rm th}}}\right)}{N^{\frac{11}{20}}}\frac{1}{\alpha_{\theta_1}^2\alpha_{\theta_2}}\right)}{{{\cal R}_{D5/\overline{D5}}^{\rm th}}^4 {g_s}^{21/4}
   {M_{\rm UV}} }.
\end{eqnarray}
The Gibbons-Hawking-York term in thermal background has an additional UV divergence component, which is:
\begin{eqnarray}
\label{GHY-UV-divergent}
& & \frac{S_{\rm GHY}^{\rm UV-divergent}}{{\cal V}_4} = -\kappa_{\rm GHY}^{\beta^0,\ {\rm UV-div}}\ \frac{ \log
   \left(\frac{{{\cal R}_{\rm UV}}}{{{\cal R}_{D5/\overline{D5}}^{\rm th}}}\right) {M_{\rm UV}} {{\cal R}_{\rm UV}}^4}{{{\cal R}_{D5/\overline{D5}}^{\rm th}}^4
   {g_s^{\rm UV}}^{9/4} N^{\frac{11}{20}} \alpha_{\theta_1}^2\alpha_{\theta_2}}.
\end{eqnarray}
We also showed that the contribution from the higher derivative term in the IR, up to ${\cal O}(\beta^0)$, is:
\begin{eqnarray}
\label{iGdeltaJ0-thermal-IR}
& & \left.G^{MN}\frac{\delta J_0}{\delta G^{MN}}\right|_{\rm Ouyang}^{\rm IR}\sim\frac{ \left(19683 \sqrt{6} \sin^6{\theta _1}+6642 \sin^2{\theta _2} \sin^3{\theta
   _1}-40 \sqrt{6} \sin^4{\theta _2}\right)}{ \epsilon ^8 {g_s}^{9/4} {\log N}^2 N^{\frac{6}{5}}
   \sin^4\theta_1 \sin^2\theta_2{N_f}^2  \Biggl[{N_f} \left\{{\log N}-3 \log \left(\frac{{r_0}}{{{\cal R}_{D5/\overline{D5}}^{\rm th}}}\right)\right\}\Biggr]^{2/3}},\nonumber\\
\end{eqnarray}
implying:
{\footnotesize
\begin{eqnarray}
\label{int-iGdeltaJ0-thermal-IR}
& &  \frac{\int_{r_0}^{{\cal R}_{D5/\overline{D5}}}\sqrt{-G^{\cal M}}\left.G^{MN}\frac{\delta J_0}{\delta G^{MN}}\right|_{\rm Ouyang}}{{\cal V}_4}
\sim\nonumber\\
& &   \frac{M {N_f} {r_0}^2 \left(1-\frac{{r_0}}{{{\cal R}_{D5/\overline{D5}}^{\rm th}}}\right)^2 \log \left(\frac{{r_0}}{{{\cal R}_{D5/\overline{D5}}^{\rm th}}}\right)
  }{ \epsilon ^8 {g_s}
   {\log N}^2 N^{21/20} {{\cal R}_{D5/\overline{D5}}^{\rm th}}^2 }\frac{ \left(-6642 {\alpha_{\theta_1}}^3 {\alpha_{\theta_2}}^2+19683 \sqrt{6}
   {\alpha_{\theta_1}}^6 \log \left(\frac{{\alpha_{\theta_1}}}{2 \sqrt[5]{N}}\right)-20
   \sqrt{6} {\alpha_{\theta_2}}^4\right)}{3 {\alpha_{\theta_1}}^6 {\alpha_{\theta_2}}^3}\nonumber\\
& & \times  \kappa_{\footnotesize G^{MN}\frac{\delta J_0}{\delta G^{MN}}}^{\beta^0,\ \rm IR}\left\{{\log N}-9 \log \left(\frac{{r_0}}{{{\cal R}_{D5/\overline{D5}}^{\rm th}}}\right)\right\} \Biggl[{\log N}-3 \log\left(\frac{{r_0}}{{{\cal R}_{D5/\overline{D5}}^{\rm th}}}\right)\Biggr]^2.
\end{eqnarray}
}
In addition, the contribution arising from the higher derivative term in the UV near (\ref{small-theta_12}) is:
\begin{eqnarray}
\label{iGdeltaJ0-thermal-UV}
& & \left.G^{MN}\frac{\delta J_0}{\delta G^{MN}}\right|_{\rm Ouyang}^{\rm UV}=  {g_s^{\rm UV}}^{1/4} {\log \tilde{r}} {M_{\rm UV}} \frac{\left(19683 \sin^6{\theta _1}+1107 \sqrt{6}
   \sin^2{\theta _2}\sin^3{\theta _1}-40 \sin^4{\theta _2}\right)}{ {g_s^{\rm UV}}^{17/4} {\log N}^3 (2
   {\log  \tilde{r}}+1)^7 N^{39/20} {N_f^{\rm UV}}^2  \tilde{r} \sin^7{\theta _1}\sin^4{\theta _2}}\nonumber\\
& & \times \Biggl(-\kappa_{\footnotesize G^{MN}\frac{\delta J_0}{\delta G^{MN}}}^{\beta^0,0}\ { a^4 \left(64
   {\log  \tilde{r}}^3+208 {\log  \tilde{r}}^2+212 {\log  \tilde{r}}+57\right)}
\nonumber\\
& & -\kappa_{\footnotesize G^{MN}\frac{\delta J_0}{\delta G^{MN}}}^{\beta^0,1}\ {8 a^2 \left(4
   {\log  \tilde{r}}^2+15 {\log \tilde{r}}+9\right)  \tilde{r}} -\kappa_{\footnotesize G^{MN}\frac{\delta J_0}{\delta G^{MN}}}^{\beta^0,2}\ { \tilde{r}^3}\Biggr),
\end{eqnarray}
where $\tilde{r} \equiv \frac{{r}}{{{\cal R}_{D5/\overline{D5}}^{\rm th}}}$,
Therefore, the UV-finite component of the aforementioned equation, after angular and radial integration, found as:
\begin{eqnarray}
\label{int-iGdeltaJ0-thermal-UV}
& &  \frac{\int_{\theta_1={\alpha_{\theta_1}}{N^{-\frac{1}{5}}}}^{\pi-{\alpha_{\theta_1}}{N^{-\frac{1}{5}}}}\int_{\theta_2= {\alpha_{\theta_2}}{N^{-\frac{3}{10}}}}^{\pi- {\alpha_{\theta_2}}{N^{-\frac{3}{10}}}}\int_{{\cal R}_{D5/\overline{D5}}}^{{\cal R}_{\rm UV}}\sqrt{-G^{\cal M}}\left.G^{MN}\frac{\delta J_0}{\delta G^{MN}}\right|_{\rm Ouyang}^{\rm UV-finite}}{{\cal V}_4}\nonumber\\
& & \sim  - \kappa_{\footnotesize G^{MN}\frac{\delta J_0}{\delta G^{MN}}}^{\beta^0,\ \rm UV-finite}\frac{  {M_{\rm UV}}}{{g_s^{\rm UV}}^{4}
   {\log N}^3 N^{21/20} {N_f^{\rm UV}}^2 }\nonumber\\
& & \times\left(\frac{a^2}{{{\cal R}_{D5/\overline{D5}}^{\rm th}}^2}+1\right) \left(\frac{ \left(-6642 {\alpha_{\theta_1}}^3 {\alpha_{\theta_2}}^2+19683 \sqrt{6}
   {\alpha_{\theta_1}}^6 \log \left(\frac{{\alpha_{\theta_1}}}{2 \sqrt[5]{N}}\right)-20
   \sqrt{6} {\alpha_{\theta_2}}^4\right)}{3 {\alpha_{\theta_1}}^6 
{\alpha_{\theta_2}}^3}\right).
\end{eqnarray}
The UV-differential portion of the equation (\ref{iGdeltaJ0-thermal-UV}) was further deduced as:
\begin{eqnarray}
\label{int-iGdeltaJ0-thermal-UV-div}
& &  \frac{\int_{\theta_1={\alpha_{\theta_1}}{N^{-\frac{1}{5}}}}^{\pi-{\alpha_{\theta_1}}{N^{-\frac{1}{5}}}}\int_{\theta_2= {\alpha_{\theta_2}}{N^{-\frac{3}{10}}}}^{\pi- {\alpha_{\theta_2}}{N^{-\frac{3}{10}}}}\int_{{\cal R}_{D5/\overline{D5}}}^{{\cal R}_{\rm UV}}\sqrt{-G^{\cal M}}\left.G^{MN}\frac{\delta J_0}{\delta G^{MN}}\right|_{\rm Ouyang}^{\rm UV-divergent}}{{\cal V}_4}\nonumber\\
& & =\kappa_{\footnotesize G^{MN}\frac{\delta J_0}{\delta G^{MN}}}^{\beta^0,\ \rm UV-div}\ \frac{{\frac{1}{{g_s^{\rm UV}}}} {M_{\rm UV}} {{\cal R}_{\rm UV}}^4}{{{\cal R}_{D5/\overline{D5}}^{\rm th}}^4{g_s^{\rm UV}}^{9/2} N^{21/20} {N_f^{\rm UV}}^{8/3} \log ^{\frac{11}{3}}(N) \log
   \left(\frac{{{\cal R}_{\rm UV}}}{{{\cal R}_{D5/\overline{D5}}^{\rm th}}}\right)}\nonumber\\
& & \times \frac{ \left(-6642 {\alpha_{\theta_1}}^3 {\alpha_{\theta_2}}^2+19683 \sqrt{6}
   {\alpha_{\theta_1}}^6 \log \left(\frac{{\alpha_{\theta_1}}}{2 \sqrt[5]{N}}\right)-20
   \sqrt{6} {\alpha_{\theta_2}}^4\right)}{3 {\alpha_{\theta_1}}^6 {\alpha_{\theta_2}}^3}.
\end{eqnarray}
We found that:
\begin{eqnarray}
\label{Boundary-CT-th}
& &\left.\frac{\int \sqrt{-h}}{{\cal V}_4}\right|_{r={\cal R}_{\rm UV}} = \kappa_{ \sqrt{-h}}^{\beta^0}\ \frac{ \left(\frac{1}{{g_s^{\rm UV}}}\right)^{7/3} {M_{\rm UV}}  {{\cal R}_{\rm UV}}^4 \log
  \left(\frac{{\cal R}_{\rm UV}}{{\cal R}_{D5/\overline{D5}}}\right)}{N {{\cal R}_{D5/\overline{D5}}^{\rm th}}^4  }\nonumber\\
  & & \hskip 1.2in \times \int_{\theta_1={\alpha_{\theta_1}}{N^{-\frac{1}{5}}}}^{\pi-{\alpha_{\theta_1}}{N^{-\frac{1}{5}}}}\int_{\theta_2= {\alpha_{\theta_2}}{N^{-\frac{3}{10}}}}^{\pi- {\alpha_{\theta_2}}{N^{-\frac{3}{10}}}}
 \frac{d\theta_1 d\theta_2}{\sin^3\theta_1\sin^2\theta_2}\nonumber\\
& & = \kappa_{ \sqrt{-h}}^{\beta^0}\ \frac{ \left(\frac{1}{{g_s^{\rm UV}}}\right)^{7/3} {M_{\rm UV}}  {{\cal R}_{\rm UV}}^4 \log
  \left(\frac{{\cal R}_{\rm UV}}{{\cal R}_{D5/\overline{D5}}}\right)}{N^{\frac{3}{10}}\alpha^2_{\theta_1}\alpha_{\theta_2} {{\cal R}_{D5/\overline{D5}}^{\rm th}}^4  }   \equiv \kappa^{(0)}_{\sqrt{-h}@\partial M_{11}}\ \left(\frac{{\cal R}_{\rm UV}}{{\cal R}_{D5/\overline{D5}}^{\rm th}}\right)^4\log\left(\frac{{\cal R}_{\rm UV}}{{\cal R}_{D5/\overline{D5}}^{\rm th}}\right),\nonumber\\
& & \left.\frac{\int \sqrt{-h}R}{{\cal V}_4}\right|_{r={\cal R}_{\rm UV}}^{\rm UV-div} 
\sim \frac{\left(\frac{1}{{g_s}}\right)^{2/3} {M_{\rm UV}}  {N_f^{\rm UV}} {{\cal R}_{\rm UV}}^4}{N^{\frac{11}{10}}{{\cal R}_{D5/\overline{D5}}^{\rm th}}^4  \sqrt{{g_s}^{\rm UV}} \alpha _{\theta _1}^2 \alpha _{\theta _2}}\equiv\kappa_{\sqrt{-h}R@\partial M_{11}}^{\beta^0}\ {\cal R}_{\rm UV}^4,\nonumber
\end{eqnarray}
\begin{eqnarray}
& & \left.\frac{\int \sqrt{-h}K}{{\cal V}_4}\right|_{r={\cal R}_{\rm UV}}^{\rm UV-div} =-\kappa_{\rm GHY@\partial M_{11}}^{\beta^0,\ {\rm UV-div}}\ \frac{ \log
   \left(\frac{{{\cal R}_{\rm UV}}}{{{\cal R}_{D5/\overline{D5}}^{\rm th}}}\right) {M_{\rm UV}} {{\cal R}_{\rm UV}}^4}{{{\cal R}_{D5/\overline{D5}}^{\rm th}}^4
   {g_s^{\rm UV}}^{9/4} N^{\frac{11}{20}} \alpha_{\theta_1}^2\alpha_{\theta_2}},\nonumber\\
& & 
\nonumber\\
& & \beta\frac{\int\sqrt{-h}\left.G^{mn}\frac{\delta J_0}{\delta G^{mn}}\right|^{\rm Ouyang}_{r={\cal R}_{\rm UV}}}{{\cal V}_4}  \sim {\footnotesize -\frac{ {M_{\rm UV}} {{\cal R}_{\rm UV}}^4 \beta\left(-19683 \sqrt{6} \right)}{{{\cal R}_{D5/\overline{D5}}^{\rm th}}^4 {g_s^{\rm UV}}^{47/12} {\log N}^2 \left\{\log
   \left(\frac{{{\cal R}_{\rm UV}}}{{{\cal R}_{D5/\overline{D5}}^{\rm th}}}\right)\right\}^6 N^{13/10} {N_f^{\rm UV}}^2 
   \alpha _{\theta _2}^3}}\nonumber\\
& & \equiv \beta \kappa^{(0)}_{\sqrt{-h}G^{mn}@\partial M_{11}\frac{\delta J_0}{\delta G^{mn}}}\frac{\left(\frac{{\cal R}_{\rm UV}}{{\cal R}_{D5/\overline{D5}}}\right)^4}{\log^6\left(\frac{{{\cal R}_{\rm UV}}}{{{\cal R}_{D5/\overline{D5}}^{\rm th}}}\right)}.
\end{eqnarray}
Let us assume the following in relation to (\ref{int-iGdeltaJ0-thermal-UV-div}) and the fourth equation in (\ref{Boundary-CT-th}):
\begin{equation}
\label{UV-div-CT-iGdeltaJ0-1}
\left[\log
   \left(\frac{{{\cal R}_{\rm UV}}}{{{\cal R}_{D5/\overline{D5}}^{\rm th}}}\right)\right]^6 N^{-\upsilon + \frac{4}{5}} = \log
   \left(\frac{{{\cal R}_{\rm UV}}}{{{\cal R}_{D5/\overline{D5}}^{\rm th}}}\right) N^{\frac{11}{20}},
\end{equation}
or
\begin{equation}
\label{UV-div-CT-iGdeltaJ0-2}
\log
   \left(\frac{{{\cal R}_{\rm UV}}}{{{\cal R}_{D5/\overline{D5}}^{\rm th}}}\right) = N^{\frac{4\upsilon-1}{20}}.
\end{equation}
Assuming additionally that $\upsilon = \frac{1}{4} + \epsilon$,
\begin{equation}
\label{UV-div-CT-iGdeltaJ0-3}
\log
   \left(\frac{{{\cal R}_{\rm UV}}}{{{\cal R}_{D5/\overline{D5}}^{\rm th}}}\right) \sim 1 +\frac{\epsilon }{5}\log N.
\end{equation}
Given (\ref{Nf_UV}), we can then say
\begin{equation}
\label{NfLogN-more-than-1}
N_f^{\rm UV}=\frac{1}{\log N} + \epsilon_1: 0<\epsilon_1\ll1,
\end{equation}
\begin{equation}
\label{UV-div-CT-iGdeltaJ0-4}
\log
   \left(\frac{{{\cal R}_{\rm UV}}}{{{\cal R}_{D5/\overline{D5}}^{\rm th}}}\right) \sim 1 + \frac{2\epsilon_1}{15}\log N.
\end{equation}
Based on the values found in (\ref{UV-div-CT-iGdeltaJ0-3}) and (\ref{UV-div-CT-iGdeltaJ0-4}), $\epsilon_1  = \frac{3}{2}\epsilon$. \\
{\bf Holographic Renormalization When $T<T_c$}: The UV-divergent contribution of $\frac{1}{2}\Biggl(-\frac{7}{2}S_{\rm EH}^{(0)} + 2 S_{\rm GHY}^{(0)}\Biggr)$ was derived from the preceding discussion as follows:
\begin{eqnarray}
\label{UV-div-on-shell-thermal}
& & \tilde{\kappa}_{\rm UV-div}^{\beta^0,1}\ \left(\frac{{{\cal R}_{\rm UV}}}{{{\cal R}_{D5/\overline{D5}}^{\rm th}}}\right)^4 + \tilde{\kappa}_{\rm UV-div}^{\beta^0,2}\ \left(\frac{{{\cal R}_{\rm UV}}}{{{\cal R}_{D5/\overline{D5}}^{\rm th}}}\right)^4 \log \left(\frac{{{\cal R}_{\rm UV}}}{{{\cal R}_{D5/\overline{D5}}^{\rm th}}}\right).
\end{eqnarray}
The following can be deduced from (\ref{Boundary-CT-th}):
\begin{itemize}
\item The counter term for
$\tilde{\kappa}_{\rm UV-div}^{\beta^0,1}\ \left(\frac{{{\cal R}_{\rm UV}}}{{{\cal R}_{D5/\overline{D5}}^{\rm th}}}\right)^4$ is $\int_{r={\cal R}_{\rm UV}} \sqrt{-h^{(0)}}R^{(0)}$;
\item Counter term to cancel the UV divergence, $ \tilde{\kappa}_{\rm UV-div}^{\beta^0,2}\ \left(\frac{{{\cal R}_{\rm UV}}}{{{\cal R}_{D5/\overline{D5}}^{\rm th}}}\right)^4 \log \left(\frac{{{\cal R}_{\rm UV}}}{{{\cal R}_{D5/\overline{D5}}^{\rm th}}}\right)$ is $\int_{r={\cal R}_{\rm UV}}\sqrt{-h}$.
\end{itemize}

\subsubsection{$T_c$ Inclusive of ${\cal O}(R^4)$ Corrections}
\label{Tc-G-III}
Here, we'll use the on-shell action comparison between black hole and thermal backgrounds to determine the deconfinement temperature using information gathered from \ref{Tc-G-I} and \ref{Tc-G-II}. We will also describe a variant of the UV-IR connection that emerges from the ${\cal O}(R^4)$ terms. \par
We showed that leading order terms in $N, \log\left(\frac{{\cal R}_{\rm UV}}{{\cal R}_{D5/\overline{D5}}}\right)$ and $\frac{r_h}{{\cal R}_{D5/\overline{D5}}}$ corresponding to the black hole and thermal backgrounds have the following forms:
\begin{eqnarray}
\label{bh-action-LO}
& & \left(1+\frac{r_h^4}{2{\cal R}_{\rm UV}^4}\right)S_{D=11,\ {\rm on-shell UV-finite}}^{BH} \sim
\frac{2 \kappa_{\rm GHY}^{\rm bh} {M_{\rm UV}} {r_h}^4 \log \left(\frac{{{\cal R}_{\rm UV}}}{{\cal R}_{D5/\overline{D5}}^{\rm bh}}\right)}{{g_s}^{9/4} N^{11/20}
   \alpha _{\theta _1}^2 \alpha _{\theta _2}}+\Biggl[{2   {\cal C}_{\theta_1x}^{\rm bh} \kappa_{\left(\sqrt{-G^{\cal M}}\right)^{(1)}R^{(0)}}^{\rm IR} }\nonumber\\
& & +\frac{20  \left(-{\cal C}_{zz}^{\rm bh} + 2 {\cal C}_{\theta_1z}^{\rm bh} - 3 {\cal C}_{\theta_1x}^{\rm bh}\right) \kappa_{\rm EH}^{\beta,\ \rm IR} }{11 }\Biggr]\frac{ b^2   {g_s}^{3/4}  M {N_f}^3 {r_h}^4 \log ^3(N) \log
   \left(\frac{{r_h}}{{{\cal R}_{D5/\overline{D5}}}}\right) \log
   \left(1 - \frac{{r_h}}{{{\cal R}_{D5/\overline{D5}}}}\right)}{ N^{11/20} {{\cal R}_{D5/\overline{D5}}}^4 \alpha _{\theta _1}^2 \alpha _{\theta _2}}\beta.\nonumber\\
\end{eqnarray}
%Writing ${\cal R}_{D5/\overline{D5}}^{\rm bh} = \left(1+{\sqrt{3}}\epsilon\right)r_h$, and using (\ref{alphaepsilon}), $ \log\left(1 - \frac{{r_h}}{{{\cal R}_{D5/\overline{D5}}}}\right) = \log\left(\frac{\sqrt{3}\epsilon}{1 + \epsilon}\right)\\ \sim-\left|\log \left(M_{\rm UV}\left(N_f^{\rm UV}\right)^{\frac{2}{15}}\right)\right|\sim-\log\log N$. 
Following is the on-shell action that corresponds to the thermal background uplift:
\begin{eqnarray}
\label{th-action-LO}
& &  S_{D=11,\ {\rm on-shell UV-finite}}^{\rm thermal} \sim \frac{2 \kappa_{\rm GHY}^{{\rm th},\ \beta^0} {M_{\rm UV}} {r_0}^4 \log \left(\frac{{{\cal R}_{\rm UV}}}{{\cal R}_{D5/\overline{D5}}^{\rm th}}\right)}{{g_s^{\rm UV}}^{9/4}
   N^{11/20} {{\cal R}_{D5/\overline{D5}}^{\rm th}}^4 \alpha _{\theta _1}^2 \alpha _{\theta _2}}+\frac{2{g_s}^{3/4} \kappa_{\rm EH, IR}^{{\rm th},\ \beta^0} M {N_f}^3
   {r_0}^2 \log ^2(N) \log \left(\frac{{r_0}}{{\cal R}_{D5/\overline{D5}}^{\rm th}}\right)}{ N^{11/20} {{\cal R}_{D5/\overline{D5}}^{\rm th}}^2 \alpha _{\theta _1}^2 \alpha
   _{\theta _2}}\nonumber\\
& & -\frac{20 \beta  \kappa_{\rm EH, th}^{{\rm IR},\ \beta} {r_0}^3 \left(2 \alpha _{\theta _2}^3-729 \sqrt{6} \alpha _{\theta _1}^3 \alpha _{\theta _2}\right)
   {f_{x^{10}x^{10}}}({r_0})}{11{g_s}^{9/4} M N^{7/20} {N_f}^{5/3} {{\cal R}_{D5/\overline{D5}}^{\rm th}}^3 \alpha _{\theta _1}^4 \log ^{\frac{2}{3}}(N) \log
   \left(\frac{{r_0}}{{\cal R}_{D5/\overline{D5}}^{\rm th}}\right)}\beta.  
\end{eqnarray}  
Now, the following equation must be solved for equality of ${\cal O}(\beta^0)$ terms in (\ref{actions-equal-Tc-ii}):
\begin{eqnarray}
\label{rh-r0-beta0}
& & \frac{2 \kappa_{\rm GHY}^{\rm bh} {M_{\rm UV}} {r_h}^4 \log \left(\frac{{{\cal R}_{\rm UV}}}{{\cal R}_{D5/\overline{D5}}^{\rm bh}}\right)}{{g_s}^{9/4} } = \frac{2 \kappa_{\rm GHY}^{{\rm th},\ \beta^0} {M_{\rm UV}} {r_0}^4 \log \left(\frac{{{\cal R}_{\rm UV}}}{{\cal R}_{D5/\overline{D5}}^{\rm th}}\right)}{{g_s^{\rm UV}}^{9/4}
    {{\cal R}_{D5/\overline{D5}}^{\rm th}}^4}\nonumber\\
& & +\frac{2{g_s}^{3/4} \kappa_{\rm EH, IR}^{{\rm th},\ \beta^0} M {N_f}^3
   {r_0}^2 \log ^2(N) \log \left(\frac{{r_0}}{{\cal R}_{D5/\overline{D5}}^{\rm th}}\right)}{  {{\cal R}_{D5/\overline{D5}}^{\rm th}}^2}.
\end{eqnarray}
Near $(\theta_1,\theta_2)\sim \left(\frac{\alpha_{\theta_1}}{N^{\frac{1}{5}}},\frac{\alpha_{\theta_1}}{N^{\frac{3}{10}}}\right)$, we found that $\frac{\kappa_{\rm GHY}^{{\rm th},\ \beta^0}}{\kappa_{\rm EH, IR}^{{\rm th},\ \beta^0}}\sim10^5$, therefore $\kappa_{\rm EH, IR}^{{\rm th},\ \beta^0}$ term has been dropped.  Hence,
\begin{equation}
\label{r_h-r_0-relation}
r_h = \frac{\sqrt[4]{\frac{\kappa_{\rm GHY}^{{\rm th},\ \beta^0}}{\kappa_{\rm GHY}^{{\rm bh},\ \beta^0}}} {r_0} {{\cal R}_{D5/\overline{D5}}^{\rm bh}} \sqrt[4]{\frac{\log
   \left(\frac{{{\cal R}_{\rm UV}}}{{{\cal R}_{D5/\overline{D5}}^{\rm th}}}\right)}{\log
   \left(\frac{{{\cal R}_{\rm UV}}}{{{\cal R}_{D5/\overline{D5}}^{\rm bh}}}\right)}}}{{{\cal R}_{D5/\overline{D5}}^{\rm th}}}
.
\end{equation}
Since $\frac{r_0}{L^2} =\frac{m^{0^{++}}}{4}$ \cite{Glueball-Roorkee} and utilizing $T_c=r_h/\pi L^2$ \cite{NPB}, the following equation describes the deconfinement temperature:
\begin{equation}
\label{rh-r0-beta0-Tc}
T_c = \frac{\sqrt[4]{\frac{\kappa_{\rm GHY}^{{\rm th},\ \beta^0}}{\kappa_{\rm GHY}^{{\rm bh},\ \beta^0}}} {m^{0^{++}}} {{\cal R}_{D5/\overline{D5}}^{\rm bh}} \sqrt[4]{\frac{\log
   \left(\frac{{{\cal R}_{\rm UV}}}{{{\cal R}_{D5/\overline{D5}}^{\rm th}}}\right)}{\log
   \left(\frac{{{\cal R}_{\rm UV}}}{{{\cal R}_{D5/\overline{D5}}^{\rm bh}}}\right)}}}{4 \pi{{\cal R}_{D5/\overline{D5}}^{\rm th}}}.
\end{equation}
The match at ${\cal O}(\beta)$ utilizing (\ref{bh-action-LO}) and (\ref{th-action-LO}) produces:
\begin{eqnarray}
\label{rh-r0-beta-1}
& & \Biggl[{2    {\cal C}_{\theta_1x}^{\rm bh} \kappa_{\left(\sqrt{-G^{\cal M}}\right)^{(1)}R^{(0)}}^{\rm IR} } +\frac{20  \left(-{\cal C}_{zz}^{\rm bh} + 2 {\cal C}_{\theta_1z}^{\rm bh} - 3 {\cal C}_{\theta_1x}^{\rm bh}\right) \kappa_{\rm EH}^{\beta,\ \rm IR} }{11 }\Biggr]\nonumber\\
& & \times\frac{ b^2   {g_s}^{3/4}  M {N_f}^3 {r_h}^4 \log ^3(N) \log
   \left(\frac{{r_h}}{{{\cal R}_{D5/\overline{D5}}}}\right) \log
   \left(1 - \frac{{r_h}}{{{\cal R}_{D5/\overline{D5}}}}\right)}{ N^{11/20} {{\cal R}_{D5/\overline{D5}}}^4 \alpha _{\theta _1}^2 \alpha _{\theta _2}}\nonumber\\
& & = \frac{20 \kappa_{\rm EH, th}^{{\rm IR},\ \beta} {r_0}^3 \left(2 \alpha _{\theta _2}^3-729 \sqrt{6} \alpha _{\theta _1}^3 \alpha _{\theta _2}\right)
   {f_{x^{10}x^{10}}}({r_0})}{11{g_s}^{9/4} M N^{7/20} {N_f}^{5/3} {{\cal R}_{D5/\overline{D5}}^{\rm th}}^3 \alpha _{\theta _1}^4 \log ^{\frac{2}{3}}(N) \log
   \left(\frac{{r_0}}{{\cal R}_{D5/\overline{D5}}^{\rm th}}\right)},
\end{eqnarray}
this results in,
{\footnotesize
\begin{eqnarray}
\label{rh-r0-beta-3}
& & \hskip -0.5in f_{x^{10}x^{10}}({r_0})
\nonumber\\
& & \hskip -0.5in \sim \left.-\frac{b^2 {g_s}^3 M^2 {N_f}^{14/3}  \left(\frac{r_h}{{\cal R}_{D5/\overline{D5}}^{\rm bh}}\right)^4 \alpha _{\theta _1}^2 \log ^{\frac{2}{3}}(N) \log
   \left(\frac{{r_0}}{{\cal R}_{D5/\overline{D5}}^{\rm th}}\right) \log \left(\frac{{r_h}}{{\cal R}_{D5/\overline{D5}}^{\rm bh}}\right) \log
   \left(1-\frac{{r_h}}{{\cal R}_{D5/\overline{D5}}^{\rm bh}}\right) }{ \beta  {\kappa_{\rm EH}^{\beta,\ \rm IR}} \sqrt[5]{N} \left(\frac{r_0}{{\cal R}_{D5/\overline{D5}}^{\rm th}}\right)^3     \alpha _{\theta _2}^2 \left(729 \sqrt{6} \alpha _{\theta _1}^3-2 \alpha _{\theta _2}^2\right)}\right|_{(\ref{r_h-r_0-relation})}\nonumber\\
& & \hskip -0.5in \times \left(11 \pi ^{17/4} {\cal C}_{\theta_1x}^{\rm bh} {\kappa_{\left(\sqrt{-G^{\cal M}}\right)^{(1)}R^{(0)}}^{\rm IR}}
    \log ^3(N)-10 {\kappa_{\rm EH}^{\beta,\ \rm IR}} \log^3(N)  \left(-{\cal C}_{zz}^{\rm bh} + 2 {\cal C}_{\theta_1z}^{\rm bh} - 3 {\cal C}_{\theta_1x}^{\rm bh}\right)\right).
\end{eqnarray}
}
In the context of this discussion, the term ``UV-IR connection'' refers to the equations (\ref{rh-r0-beta-3}) and (\ref{r_h-r_0-relation}). Because this is providing a connection that connects the integration constant appearing in the ${\cal O}(R^4)$ corrections to the thermal background along the ${\cal M}$-theory circle and a specific combination of integration constants appearing in the ${\cal O}(R^4)$ corrections to the black hole background. The combination figuring in the black hole background is along the compact part $S^3$ of the non-compact four cycle (locally) $\Sigma_{(4)} = \mathbb{R}_+\times S^3$ wrapping the flavor $D7$-branes of the type IIB dual \cite{metrics}, hence this has the information of flavor branes in ${\cal M}$-theory uplift which has no branes and this is known as ``Flavor Memory'' effect in this context.\\
{\bf Non-Renormalization of the Deconfinement Temperature ($T_c$) from Semiclassical Method}: We now turn to Green and Gutperle's argument that a $SL(2,\mathbb{Z})$ completion of an effective $R^4$ interaction results into a fascinating non-renormalization theorem that prohibits perturbative corrections to this term beyond one loop in the zero-instanton sector \cite{Green and Gutperle}. The term that occurs in (\ref{A}) bilinear in the tensor $\hat t$ bears the same structure just like terms that occur in the zero instanton sector via both the one-loop four-graviton amplitude and a $(\alpha')^3$ effect at tree level \cite{wittengross}. Thus, in the Einstein frame, one gets the expression that follows for the total effective $R^4$ action, which can be formally represented by:
\begin{equation}
\label{modular-completion-i}
S_{R^4} = (\alpha')^{-1} \left[ a \zeta(3)\tau_2^{3/2}   + b
\tau_2^{-1/2} + c
e^{2\pi i  \tau} +\cdots \right] R^4 \equiv  (\alpha')^{-1}
f(\tau,\bar{\tau})R^4,
\end{equation}
$a, b$ are understood to be numerical constants, $R^4$ represents the contractions $\hat t \hat t R^4$ in (\ref{A}), and $\cdots$ represents perturbative and nonperturbative corrections to the coefficient of $R^4$ with $\tau = C_{\rm RR} + i e^{-\phi}$ (type IIB) and $\bar \tau$. The first term of (\ref{modular-completion-i}) reflects the $(\alpha^\prime)^3$ tree-level contribution, whereas the second term represents the one-loop compared to the first. The entire equation for $S_{R^4}$, on the other hand, has to be invariant under $SL(2,\mathbb{Z})$ transformations: $\tau\to (a\tau+b)(c\tau+d)^{-1}$ ($a,b,c,d\in\mathbb{Z}: ad-bc=1$). It imposes severe constraints on its form. Because the $R^4$ factor remains invariant, $f(\tau,\bar \tau)$
in (\ref{modular-completion-i}) serves as a scalar under the $SL(2,\mathbb{Z})$ transformations, implying a sum of all instantons and anti-instantons. The authors provided a simple function which fulfills all of these conditions, such as \cite{Green and Gutperle}:
\begin{equation}
\label{modular-completion-ii}
  f(\tau,\bar{\tau}) = \sum_{(p,n )\neq
(0,0)}{\tau_2^{3/2}\over |p+n\tau|^3},
\end{equation}
wherein the summation denotes the sum of all positive and negative $p,n$ values except $p=n=0$. Now:
\begin{equation}\label{modular-completion-iii-f}
  f=2\zeta(3)\tau_2^{3/2} +{\tau_2^{3/2}\over \Gamma(3/2)}\sum_{n\neq 0,p }\int_0^{\infty}dy
y^{1/2}e^{-y|p+n\tau|^2}.
\end{equation}
The sum over $p$, utilizing  the   Poisson
resummation
formula\\ $\left(\sum_{n\rightarrow-\infty}^\infty f(n) = \sum_{m\rightarrow-\infty}^\infty {\rm FT}[f](m); {\rm FT}[e^{-\pi A(p+x)^2}] = \frac{e^{-\frac{M^2}{4 A\pi} - i M x}}{2\pi A}, M=2\pi m\right)$,
provides,
\begin{eqnarray}
\label{modular-completion-iii}
& &     f(\tau,\bar \tau) = 2\zeta(3)\tau_2^{3/2} + {2\pi^{2}\over 3}\tau_2^{-1/2} +
2\tau_2^{3/2}\sum_{m,n\neq 0}\int_0^{\infty}dy
  \exp\left(-{\pi^2 m^2\over y}+2\pi i m n\tau_1 - yn^2 \tau_2^2
\right),\nonumber
\end{eqnarray}
\begin{eqnarray}
& &  = 2\zeta(3)\tau_2^{3/2} +
  {2\pi^2\over
3} \tau_2^{-1/2}  +8 \pi  \tau_2^{ 1/2}  \sum_{m \ne 0 n\ge 1   }
\left|{m\over n}\right|
e^{2\pi i mn\tau_1} K_1 (2\pi |mn|\tau_2)\nonumber\\
&&   = 2\zeta(3)\tau_2^{3/2} + {2\pi^2\over  3} \tau_2^{-1/2}   \nonumber\\
&&  +4\pi^{3/2}  \sum_{m,n \ge 1} \left({m\over n^3}\right)^{1/2}
(e^{2\pi i mn
\tau} + e^{-2\pi i mn \bar  \tau} ) \left(1 + \sum_{k=1}^\infty  (4\pi mn
\tau_2)^{-k} {\Gamma(  k -1/2)\over \Gamma(- k -1/2) } \right).
\end{eqnarray}
In (\ref{modular-completion-iii}), the asymptotic expansion for $K_1(z)$ for large $z$ is used after carrying out a perturbative expansion in $\frac{1}{\tau_2}$ of the non-perturbative instanton contribution of charge $mn$. {\it The perturbative terms in (\ref{modular-completion-iii}) end after the one-loop term, as previously proposed.  The non-perturbative terms are a sum of single multiply-charged instantons and anti-instantons with action proportional to $|mn|$.    The terms in parentheses in (\ref{modular-completion-iii}) reflect a never-ending series of perturbative corrections near the charge instantons $mn$.   As a result, there aren't any perturbative corrections in the action up to ${\cal O}(R^4)$, i.e., $S_{R^4}$, beyond one loop in the zero-instanton sector. Hence, this suggests the ``non-renormalization of the deconfinement temperature ($T_c$)'' beyond one-loop in the zero instanton sector.}

\section{Wald Entropy, $T_c$ from Entanglement Entropy and M$\chi$PT Compatibility}
\subsection{Wald Entropy}
Lets now prove that the black hole entropy that was calculated using (\ref{bh-action-LO}) and Wald's technique are consistent with one another. We takes away from the (\ref{bh-action-LO}) that the formula for calculating the BH entropy, ${\cal S}_{\rm BH}$, up to ${\cal O}(\beta^0)$, is as follows:
\begin{eqnarray}
\label{entropy-semi-classical}
& & {\cal S}_{\rm BH} = \beta_{\rm BH} \frac{\partial S^E_{\rm BH}}{\partial\beta} - S^E_{\rm BH} \sim\frac{\sqrt{N}{M_{\rm UV}} {r_h}^3 \log \left(\frac{{{\cal R}_{\rm UV}}}{{\cal R}_{D5/\overline{D5}}^{\rm bh}}\right)}{{g_s}^{7/4} N^{11/20}
   \alpha _{\theta _1}^2 \alpha _{\theta _2}}\nonumber\\
& & \sim \frac{M_{\rm UV}\left(N_f^{\rm UV}\log N\right)^{\frac{2}{15}}r_h^3}{N^{\frac{1}{20}}}.
\end{eqnarray}
%Note, there is a multiplicative factor of $\beta_{\rm BH}\sim\frac{\sqrt{g_sN}}{r_h}$ in all terms in the action corrsponding to the BH gravity dual, e.g., (\ref{int-EH-BH-IR}), (\ref{int-EH-BH-IR}), (\ref{sqrtGbetaRbeta0-IR-bh}), (\ref{int-EH-BH-UV}), 
%(\ref{GHY-BH}), (\ref{GHY-UV-div}), (\ref{int-iGdeltaJ0-BH-IR}), (\ref{int-sqrtGiGdeltaJ0-beta0-UV-div}), 
%(\ref{isqrthinvGdeltaJ0-constr}), (\ref{ihdeltaJ0-r=rUV}) and (\ref{sqrthJ0-constr}). 
It's interesting to note that the Wald entropy density that results from the ${\cal O}(\beta^0)$ action may be expressed as:
\begin{eqnarray}
\label{Wald-i}
& & \frac{{\cal S}_{\rm BH}^{\rm Wald}}{{\cal V}_3} = \oint_{M_6(\theta_{1,2},x,y,z,x^{10})} \frac{\partial R}{\partial R_{mnpq}}\epsilon^{mn}\epsilon^{pq} \sqrt{g_{x^{1,2,3},\theta_{1,2},x,y,z,x^{10}}}d\theta_1d\theta_2 dx dy dz dx^{10}
\nonumber\\
& & \sim -\frac{\left(3 b^2-1\right) \left(9 b^2+1\right) {g_s}^{5/4} M N^{-1/20} {N_f} {r_h}^3 \log \left(\frac{r_h}{{\cal R}_{D5/\overline{D5}}^{\rm bh}}\right)
   \left(\log N -9 \log \left(\frac{r_h}{{\cal R}_{D5/\overline{D5}}^{\rm bh}}\right)\right) }{ \left(6 b^2+1\right) \alpha _{\theta _1}^2 \alpha _{\theta _2}}\nonumber\\
& & \times \left[{N_f} \left(\log N -3 \log \left(\frac{r_h}{{\cal R}_{D5/\overline{D5}}^{\rm bh}}\right)\right)\right]^{2/3},
\end{eqnarray}
where ${\cal V}_3$ represents $\mathbb{R}^3 (x^{1,2,3}$ coordinates volume). When replacing $b = \frac{1}{\sqrt{3}} + \epsilon, \epsilon\sim \left(|\log r_h\right|)^{\frac{9}{2}} N^{-\frac{9}{10} - \alpha_\epsilon}$, (\ref{Wald-i}) implying:
\begin{eqnarray}
\label{Wald-ii}
& & \frac{{g_s}^{5/4} M {N_f} {r_h}^3 N^{-\frac{19}{20}-{\alpha_\epsilon}}\left(|\log r_h\right|)^{\frac{11}{2}}
   \left(\log N -9 \log \left(\frac{r_h}{{\cal R}_{D5/\overline{D5}}^{\rm bh}}\right)\right)}{\alpha _{\theta _1}^2 \alpha
   _{\theta _2}}\nonumber\\
& &  \times \left[{N_f} \left(\log N -3 \log \left(\frac{r_h}{{\cal R}_{D5/\overline{D5}}^{\rm bh}}\right)\right)\right]^{2/3}.
\end{eqnarray}
Both approaches result in the same conclusion, which is that the black hole entropy is proportional to $r_h^3$. The findings are a perfect match for (for the sake of simplicity, we have disregarded in (\ref{alphaepsilon}) numerical factors and $M, N_f$ which, in the MQGP limit, are ${\cal O}(1)$):
\begin{eqnarray}
\label{alphaepsilon}
& & \hskip -0.3in \alpha_\epsilon = - \frac{9}{10} 
+\frac{\left|\log \left(M_{\rm UV}\left(N_f^{\rm UV}\right)^{\frac{2}{15}}\right)\right|}{\log N}
+ \frac{211\log\log N}{\log N}\sim \frac{\left|\log \left(M_{\rm UV}\left(N_f^{\rm UV}\right)^{\frac{2}{15}}\right)\right|}{\log N}
- \frac{9}{10};
\end{eqnarray}
$\alpha_\epsilon>0$ if $M_{\rm UV}\left(N_f^{\rm UV}\right)^{\frac{2}{15}}<N^{-\frac{9}{10}}$. We demonstrated that in order to compute the entropy of a black hole from ${\cal O}(R^4)$ terms, one must evaluate a total of four classes of terms independently while calculating the $\frac{\partial J_0}{\partial R_{trtr}}$:
\begin{eqnarray}
\label{Wald-J0-i-Tc}
& & \hskip -0.3in (i) \left(G^{rr}\right)^2 \left(G^{tt}\right)^2\left(R_{PrtQ} + \frac{1}{2}R_{PQtr}\right)R_t^{\ \ RSP}R^Q_{\ \ RSr}\nonumber\\
& & \hskip -0.3in \sim \left(G^{rr}\right)^2 \left(G^{rr}\right)^2R_{trrt}\left(R_t^{\ \ z\theta_1t}R^r_{\ \ z\theta_1r} + R_t^{\ \ \theta_1zt}R^r_{\ \ \theta_1zr}\right)\nonumber\\
& & \hskip -0.3in \sim-\frac{\left(\log N -9 \log \left(\frac{r_h}{{\cal R}_{D5/\overline{D5}}^{\rm bh}}\right)\right)^2 \sqrt[3]{{N_f} \left(\log N -3 \log \left(\frac{r_h}{{\cal R}_{D5/\overline{D5}}^{\rm bh}}\right)\right)}}{{g_s}^{3/2} N^{3/2}
   {N_f}^5 \left(\log N -3 \log \left(\frac{r_h}{{\cal R}_{D5/\overline{D5}}^{\rm bh}}\right)\right)^5};\nonumber\\
& & \hskip -0.3in (ii) R^{HrtK}R_H^{\ \ RSt}R^r_{\ \ RSK} + \frac{1}{2}R^{HKtr}R_H^{\ \ RSt}R^Q_{\ \ RSK}\sim R^{trrt} \left(R_t^{\ \ z\theta_1t} R^r_{z\theta_1r} + R_t^{\theta_1zt} R^r_{\ \ \theta_1zr}\right)\nonumber\\
& & \hskip -0.3in \sim -\frac{(\log N -9 \log \left(\frac{r_h}{{\cal R}_{D5/\overline{D5}}^{\rm bh}}\right))^2}{{g_s}^{3/2} N^{3/2} {N_f}^{14/3} \left(\log N -3 \log \left(\frac{r_h}{{\cal R}_{D5/\overline{D5}}^{\rm bh}}\right)\right)^{14/3}};
\nonumber\\
& & \hskip -0.3in (iii)  \left(G^{rr}\right)^2 G^{tt}\left(R_{PrtQ} + \frac{1}{2}R_{PQtr}\right)R_t^{\ \ RSP}R^Q_{\ \ RSr}
\sim \frac{{g_s}^{9/2} M^6 \log ^3({r_h}) (\log N -12 \log \left(\frac{r_h}{{\cal R}_{D5/\overline{D5}}^{\rm bh}}\right))^3}{{N_f}^{5/3} \left(\log N -3 \log \left(\frac{r_h}{{\cal R}_{D5/\overline{D5}}^{\rm bh}}\right)\right)^{14/3}};\nonumber\\
& & \hskip -0.3in (iv)  G^{tt} \left(R^{HMNr}R_{PMNt} + \frac{1}{2}R^{HrMN}R_{PtMN}\right)R_H^{\ \ rtP}
\sim \frac{\sqrt{{g_s}} M^2 \log \left(\frac{r_h}{{\cal R}_{D5/\overline{D5}}^{\rm bh}}\right)}{{N_f}^{11/3} \left(\log N -3  \log \left(\frac{r_h}{{\cal R}_{D5/\overline{D5}}^{\rm bh}}\right)\right)^{14/3}}\nonumber\\
& & \hskip 3.5in \times  \left(\log N -12 \log \left(\frac{r_h}{{\cal R}_{D5/\overline{D5}}^{\rm bh}}\right)\right).
\end{eqnarray}
As a result, we are able to notice that the contribution (iii) is the one that predominates in the MQGP limit, and the black hole entropy from the higher derivative terms is obtained as:
\begin{eqnarray}
\label{S-J0-Wald}
& & \frac{{\cal S}_E^{{\cal O}(R^4)}}{{\cal V}_3}\sim \oint_{\theta_{1,2},x,y,z,x^{10})} \frac{\partial J_0}{\partial R_{rtrt}}\sqrt{-G_9} d\theta_1 d\theta_2 dx dy dz dx^{10}\nonumber\\
& &  \sim \frac{{g_s}^{23/4} M^8 {N_f}^{4/3} \left(\frac{r_h}{{\cal R}_{D5/\overline{D5}}^{\rm bh}}\right)^3 \log ^4\left(\frac{r_h}{{\cal R}_{D5/\overline{D5}}^{\rm bh}}\right) \left(\log N -12 \log \left(\frac{r_h}{{\cal R}_{D5/\overline{D5}}^{\rm bh}}\right)\right)^3}{\sqrt[20]{N} \alpha _{\theta _1}^2 \alpha _{\theta _2} \left(\log N -3 \log \left(\frac{r_h}{{\cal R}_{D5/\overline{D5}}^{\rm bh}}\right)\right)^{8/3}}.
\end{eqnarray}
One notes that (\ref{S-J0-Wald}) has the same dependency on $r_h^3$ and $N^{-\frac{1}{20}}$, and semiclassical result is found at ${\cal O}(\beta)$. This imposes the following restriction, which must be met in order for the latter and the Wald entropy results to coincide exactly at ${\cal O}(\beta)$:
\begin{eqnarray}
\label{Wald-matching}
& & 
\Biggl[{2    {\cal C}_{\theta_1x}^{\rm bh} \kappa_{\left(\sqrt{-G^{\cal M}}\right)^{(1)}R^{(0)}}^{\rm IR} } +\frac{20  \left(-{\cal C}_{zz}^{\rm bh} + 2 {\cal C}_{\theta_1z}^{\rm bh} - 3 {\cal C}_{\theta_1x}^{\rm bh}\right) \kappa_{\rm EH}^{\beta,\ \rm IR} }{11 }\Biggr]
\log ^3(N) \left|\log \left(M_{\rm UV}\left(N_f^{\rm UV}\right)^{\frac{2}{15}}\right)\right|\nonumber\\
& &  \sim \frac{ \log ^3 \left(\frac{r_h}{{\cal R}_{D5/\overline{D5}}^{\rm bh}}\right) \left(\log N -12 \log \left(\frac{r_h}{{\cal R}_{D5/\overline{D5}}^{\rm bh}}\right)\right)^3}{ \left(\log N -3 \log \left(\frac{r_h}{{\cal R}_{D5/\overline{D5}}^{\rm bh}}\right)\right)^{8/3}}.
\end{eqnarray} 
As a result, we found that linear constraints must be placed on the constants of integration that appear in the solutions to the EOMs of the metric corrections at ${\cal O}(R^4)$ in the directions that contain the ``memory'' of the compact part of the non-compact four-cycle ``wrapped'' by the flavor $D7$-branes in the type IIB dual \cite{metrics} of the large-$N$ thermal QCD-like theories. 
\subsection{Deconfinement from Entanglement Entropy}
\label{Tc-SS-iii}
The confinement-deconfinement phase transition in thermal QCD has been explored in this section from the perspective of entanglement entropy, which relies on \cite{Tc-EE}. We have computed the entanglement entropy of regions that have been defined as ``connected'' and ``disconnected'' in an appropriate manner. We shall demonstrate that a phase transition from the confined phase to the deconfined phase will take place at a critical value of the coordinate length ({\it l}).  
%{Authors in \cite{Tc-EE} have investigated entanglement entropy between an interval and its complement in the gravity dual of large $N_c$ gauge theories and they found that there are two RT surfaces - disconnected and connected. Below a critical value of ${\it  l}$ connected surface dominates while above that critical value of ${\it  l}$ disconnected surface dominates. This is analogous to finite temperature deconfinement transition in dual theories. \par}
The quantum entanglement entropy among the regions $A \equiv \mathbb{R}^{d-1}\times l$ and $B \equiv \mathbb{R}^{d-1}\times(\mathbb{R}-l)$, with $l$ representing a period of length ${\it l}$, may be expressed as the following expression \cite{RT}:
\begin{equation}
\label{EE-def}
S_A=\frac{1}{4 G_N^{(d+2)}} \int_{\gamma} d^d\sigma\sqrt{G_{ind}^{(d)}},
\end{equation}
where $G_N^{(d+2)}$ is the Newton constant in the dimension $(d+2)$, and $G_{ind}^{(d)}$ is the value of the determinant corresponding to the induced string frame metric on the co-dimension two minimal surface $\gamma$. The equation (\ref{EE-def}) can be generalized to apply to non-conformal theories as shown in the following \cite{Tc-EE}:
\begin{equation}
\label{EE-def-NCT}
S_A=\frac{1}{4 G_N^{(d+2)}} \int_{\gamma} d^d\sigma e^{-2\phi}\sqrt{G_{ind}^{(d)}},
\end{equation}
where $\phi$ denotes the profile of the dilaton. Due to the absence of a dilaton in a ${\cal M}$-theory dual, the equation (\ref{EE-def-NCT}) will be considered to be replaced with the following expression:
\begin{equation}
\label{EE-def-M Theory}
S_A=\frac{1}{4 G_N^{(11)}} \int_{\gamma} d^9\sigma \sqrt{G_{ind}^{(9)}} .
\end{equation}  
Consider now the gravity dual's metric in string frame, which may be stated as:
\begin{equation}
\label{metric-Tc}
ds^2=\alpha(r)[\sigma(r) dr^2+dx_{\mu}dx^{\mu}]+g_{mn}dx^mdx^n ,
\end{equation}
where $x_\mu(\mu=0,1,2)$ denotes $(2+1)$ Minkowskian coordinates, $r$ represents the radial coordinate, and $x^m(m=3,5,6,7,8,9,10)$ is equivalent to $x^3$ and has six angular coordinates $(\theta_{1,2},\phi_{1,2},\psi,x^{10})$. Observing that $g_{x^3x^3}(r=r_0)=0$, $(x^3,r)$ create a cigar-like shape. The following expressions determine the volume of the seven-fold:
\begin{equation}
\label{V_int}
{\cal V}_{\rm int}=\int \prod_m dx^m \sqrt{g} .
\end{equation}
In this case, the QFT is defined on ${\rm I\!R}^{2+1}$, which has a dimension of $(2+1)$. Let us define two regions in two dimensions, labeled A and B, as shown below:
\begin{eqnarray}
&&
A={\rm I\!R}\times l , \nonumber\\
&& B={\rm I\!R} \times ({\rm I\!R}-l).
\end{eqnarray}
Using the equation $(\ref{EE-def-M Theory})$, we will now determine the entanglement entropy between $A$ and $B$ using the metric $(\ref{metric-Tc})$. The following is an expression for the induced metric on $\gamma$ (note that we replaced $x_1$ by $x$):
\begin{equation}
\label{induced metric}
ds^2|_\gamma=\alpha(r)\left[\left(\sigma(r) + \left(\frac{dx}{dr}\right)^2 \right) dr^2+ dx_2^2\right]+g_{mn}dx^mdx^n .
\end{equation}
Hence,
\begin{equation}
\label{EEC}
\frac{S_A}{{\cal V}_1} =\frac{1}{4 G_N^{(11)}} \int dr\sqrt{H(r)}\sqrt{\sigma(r)+ (\partial_rx(r))^2} ,
\end{equation}
where $H(r)$ is defined as:
\begin{equation}
H(r)= {\cal V}_{int}^2\alpha(r)^2 .
\end{equation}
$H(r=r_0)=0$ is the same as \cite{Tc-EE}. The equation of motion for $x(r)$ could be derived from (\ref{EEC}) as follows:   
\begin{equation}
\label{dx/dr}
\frac{dx}{dr}=\pm \frac{\sqrt{\sigma(r)}\sqrt{H(r_{*})}}{\sqrt{H(r)-H(r_*)}} ,
\end{equation}
where $r_*$ is the radial distance when $dr/dx$ becomes zero. By integrating the previous equation, we found:
\begin{equation}
\label{lrstar}
l(r_*)=2\sqrt{H(r_*)}\int_{r_*}^\infty dr\frac{\sqrt{\sigma(r)}}{\sqrt{H(r)-H(r_*)}} .
\end{equation}
Entanglement entropy will be reduced to the following using equations (\ref{EEC}) and (\ref{dx/dr}).
\begin{equation}
\label{connected-EE}
\frac{S_A}{{\cal V}_1} =\frac{1}{2 G_N^{(11)}} \int_{r_*}^{r_\infty} dr\frac{\sqrt{\sigma(r)} H(r)}{\sqrt{H(r)-H(r_*)}} .
\end{equation}
The UV cut-off is denoted here by $r_\infty$. According to the explanation in \cite{Tc-EE} that $r_*=r_0$, when $l>l_{max}$ the entanglement entropy of the disconnected surface can be expressed as follows:
\begin{equation}
\label{Disconnected Surface}
\frac{S_A}{{\cal V}_1} =\frac{1}{2 G_N^{(11)}} \int_{r_0}^{r_\infty} dr\sqrt{\sigma(r)H(r)} .
\end{equation}
Connected and disconnected surfaces' entanglement entropy differences have the following expression for $l<l_{max}$:
\begin{equation}
\label{DCD}
\frac{2 G_N^{(11)}(S_A^{(conn)}-S_A^{(disconn)})}{{\cal V}_1} = \int_{r_*}^{\infty} dr\left(\frac{\sqrt{\sigma(r)} H(r)}{\sqrt{H(r)-H(r_*)}} - \sqrt{\sigma(r)H(r)}\right)- \int_{r_0}^{r_*} dr\sqrt{\sigma(r)H(r)} .
\end{equation}
Let us discuss the implications of (\ref{DCD}).
\begin{itemize}
\item If ${\it l}$ is small then the entropy is lowest for a connected solution with a big $r_*$.
\item There are two possible outcomes once ${\it l}$ begins to rise. First, until ${\it l}$ reaches the highest possible value, ${\it l}_{max}$, the connected solution can keep being the lowest. Second, the disconnected solution may become the dominant one above a certain value of ${\it l}$, denoted by ${\it l}_{crit}$ and ${\it  l}_{crit}<{\it  l}_{max}$.
\item In the former, the phase transition will occur at ${\it  l}={\it  l}_{max}$, while in the latter, it will occur at ${\it  l}={\it  l}_{crit}<{\it  l}_{max}$.
\end{itemize}
Equations (\ref{TypeIIA-from-M-theory-Witten-prescription-T<Tc}) and (\ref{metric-Tc}) allow us to deduce the following:
\begin{equation}
\alpha(r)=\frac{e^{\frac{-2\phi^{IIA}}{3}}}{\sqrt{h(r,\theta_{1,2})}}; \ \sigma(r)=\frac{e^{\frac{2\phi^{IIA}}{3}}}{\tilde{g(r)}},
\end{equation}
where $\phi^{IIA}$ represents the type IIA dilaton profile, which could be deduced from the ${\cal M}$-theory metric  as follows:
\begin{equation}
\label{type-IIA-dilaton}
G^{\cal M}_{x_{10}x_{10}}=e^{\frac{4\phi^{IIA}}{3}}.
\end{equation}
Based on the equation (\ref{V_int}), the simpler version of the ${\cal V}_{int}$ for the metric given in the equation (\ref{TypeIIA-from-M-theory-Witten-prescription-T<Tc}) is as follows:
\begin{eqnarray}
\label{Vint}
  {\cal V}_{\rm int} \sim \frac{{g_s}^{3/4} M N^{9/20} \sqrt{1-\frac{{r_0}^4}{r^4}} \left(-{N_f} \log \left(9 a^2 r^4+r^6\right)+\frac{16 \pi }{{g_s}}+\frac{\Omega}{g_s}\right){}^{4/3}}{ r \alpha _{\theta _1}^3 \alpha _{\theta _2}^2} \left(\frac{1}{2} \beta  ({\cal C}_{zz}^{\rm th}-2
   {\cal C}_{\theta_1z}^{\rm th})+1\right)\nonumber\\
& & \hskip -6.5in \times  \Biggl[{g_s} \log N  {N_f} \left(3 a^2-r^2\right) (2 \log (r)+1)+\log
   (r) \nonumber\\
   & & \hskip -6.4in \times \left(4 {g_s} {N_f} \left(r^2-3 a^2\right) \log \left(\frac{1}{4} \alpha _{\theta _1} \alpha _{\theta
   _2}\right)-24 \pi  a^2+r^2 (8 \pi -3 {g_s} {N_f})\right)\nonumber\\
& & \hskip -6.4in+2 {g_s} {N_f} \left(r^2-3 a^2\right) \log
   \left(\frac{1}{4} \alpha _{\theta _1} \alpha _{\theta _2}\right)+18 {g_s} {N_f} \left(r^2-3 a^2 (6
   r+1)\right) \log ^2(r)\Biggr],
\end{eqnarray}
%{(strictly speaking the periodicity of $x^3$, $\frac{2\pi}{M_{\rm KK}}$ where $M_{\rm KK} = \frac{2r_0}{L^2}\left[1 + {\cal O}\left(\frac{g_sM^2}{N}\right)\right]$, but this does not influence the computation of $l_{\rm max}$ and $l_{\rm crit}$)}
where  $\Omega\equiv \left({g_s} (2 \log N +3) {N_f}-4 {g_s} {N_f} \log
   \left(\frac{1}{4} \alpha _{\theta _1} \alpha _{\theta _2}\right)-8 \pi \right)$. $\alpha(r)$ and $\sigma(r)$ were derived from the equation (\ref{type-IIA-dilaton}) and $h(r,\theta_{1,2})$ is provided in \cite{metrics, MQGP}; a simpler version of them are presented below:
\begin{eqnarray}
\label{alpha}
\alpha(r) = \frac{3^{2/3} \left(-{N_f} \log \left(9 a^2 r^4+r^6\right)+\frac{16 \pi }{{g_s}}+\frac{\Omega}{g_s}\right){}^{2/3}}{8 \pi ^{7/6} \sqrt{\frac{{g_s} N}{r^4}}}\nonumber \\
   & & \hskip -3.5in  -\frac{27\ 3^{2/3} b^{10} \left(9 b^2+1\right)^4
   \beta  M \left(\frac{1}{N}\right)^{3/4} r \left(19683 \sqrt{6} \alpha _{\theta _1}^6+6642 \alpha _{\theta _2}^2
   \alpha _{\theta _1}^3-40 \sqrt{6} \alpha _{\theta _2}^4\right) \left(6 a^2+{r_0}^2\right) (r-2 {r_0})}{16 \pi ^{13/6} \left(3 b^2-1\right)^5 \left(6 b^2+1\right)^4 \log N ^4 \sqrt{N} {N_f}
   {r_0}^4 \alpha _{\theta _2}^3 \left(9 a^2+{r_0}^2\right) \sqrt{\frac{{g_s} N}{r^4}}}\nonumber \\
   & & \hskip -3.5in \times \left( \log
   ^3({r_0}) \left(-{N_f} \log \left(9 a^2 r^4+r^6\right)+\frac{16 \pi }{{g_s}}+\frac{\Omega}{g_s}\right){}^{2/3}\right).
   \end{eqnarray}
and
\begin{eqnarray}
\label{beta-Tc}
& &\hskip -0.8in \sigma(r) = \frac{4 \left(\frac{\pi }{3}\right)^{2/3} \left(\frac{32 \pi  a^2 {g_s} M^2 {N_f} ({c_1}+{c_2} \log
   ({r_0}))}{N \left(9 a^2+r^2\right) \left(-{N_f} \log \left(9 a^2 r^4+r^6\right)+\frac{16 \pi }{{g_s}}+\frac{\Omega}{g_s}\right)}+1\right)}{\left(1-\frac{{r_0}^4}{r^4}\right) \left(-{N_f} \log
   \left(9 a^2 r^4+r^6\right)+\frac{16 \pi }{{g_s}}+\frac{\Omega}{g_s}\right){}^{2/3}} \nonumber \\
& &  \hskip -0.8in -\frac{18
   \sqrt[3]{\frac{3}{\pi }} b^{10} \left(9 b^2+1\right)^4 \beta  M \left(\frac{1}{N}\right)^{5/4} r^5 \left(19683
   \sqrt{6} \alpha _{\theta _1}^6+6642 \alpha _{\theta _2}^2 \alpha _{\theta _1}^3-40 \sqrt{6} \alpha _{\theta
   _2}^4\right)}{\left(3 b^2-1\right)^5 \left(6
   b^2+1\right)^4 \log N ^4 {N_f} {r_0}^4 \alpha _{\theta _2}^3 \left(9 a^2+{r_0}^2\right)
   \left({r_0}^4-r^4\right)}\nonumber \\
     & & \hskip -0.8in \times \frac{\left(6 a^2+{r_0}^2\right) (r-2 {r_0}) \log ^3({r_0})}{ \left(-{N_f} \log \left(9 a^2 r^4+r^6\right)+\frac{16 \pi }{{g_s}}+\frac{\Omega}{g_s}\right){}^{2/3}},
\end{eqnarray}
$c_{1,2}$ appeared from ${\cal O}\left(\frac{g_sM^2}{N}\right)$-correction $\left(\frac{g_sM^2}{N}\right)(c_1 + c_2 \log r_0)r_0$ from $a-r_0$ relationship (motivated by \cite{EPJC-2,Bulk-Viscosity-McGill-IIT-Roorkee}). We were able to get the following using the equations (\ref{Vint}) and (\ref{alpha}):
\begin{eqnarray}
\label{Hr}
& & H(r) \equiv {\cal V}_{\rm int}^2\alpha(r)^2 = h_0(r) + \beta h_1(r),
\end{eqnarray}
where:
\begin{eqnarray}
& & \hskip-0.7in h_0(r)\sim\frac{\sqrt{\text{gs}} M^2 r^2}{ \sqrt[10]{N} \alpha _{\theta _1}^6
   \alpha _{\theta _2}^4} \left(1-\frac{{r_0}^4}{r^4}\right)
 \lambda_1^2 \lambda_2^4,
   \end{eqnarray}
and,
\begin{eqnarray}
& & \hskip-0.5in h_1(r)\sim\frac{{g_s}^{3/2} M^2 N^{9/10} \left(1-\frac{{r_0}^4}{r^4}\right)}{ r^2 \alpha
   _{\theta _1}^6 \alpha _{\theta _2}^4 \left(9 a^2+{r_0}^2\right)}\lambda_1^2  \lambda_2^{8/3}\left(\frac{\lambda_2^{4/3}}{64 \pi ^{10/3} \left(3 b^2-1\right)^5 \left(6 b^2+1\right)^4 {g_s} \log N^4 N^{3/2} {N_f} {r_0}^4 \alpha
   _{\theta _2}^3 }   \right)\nonumber\\
   & & 
 \hskip-0.5in\times  \left(\frac{3 \sqrt[3]{3} r^4 ({\cal C}_{zz}^{\rm th}-2 {\cal C}_{\theta_1 z}^{\rm th})}{64 \pi ^{7/3} {g_s} N} -81 \sqrt[3]{3} b^{10} \left(9 b^2+1\right)^4 M \left(\frac{1}{N}\right)^{3/4} r^5\Sigma_1 \left(6
   a^2+{r_0}^2\right) (r-2 {r_0}) \log ^3({r_0}) \right),\nonumber\\
\end{eqnarray}
wherein $\lambda_1$ and $\lambda_2$ have been defined as follows:
\begin{eqnarray}
& & \lambda_1 =\Biggl[{g_s} \log N N_f \left(3 a^2-r^2\right) (2 \log (r)+1)\nonumber\\
 & &
 +\log (r) \left(4 {g_s} {N_f} \left(r^2-3
   a^2\right) \log \left(\frac{1}{4} \alpha _{\theta _1} \alpha _{\theta _2}\right)-24 \pi  a^2 + r^2 (8 \pi -3 {g_s}
   {N_f})\right)\nonumber\\
   & & +2 {g_s}
   {N_f} \left(r^2-3 a^2\right) \log \left(\frac{1}{4} \alpha _{\theta _1} \alpha _{\theta
   _2}\right)+18 {g_s}
   {N_f} \left(r^2-3 a^2 (6 r+1)\right) \log ^2(r)\Biggr],\\
   & & 
   \lambda_2=\left(-{N_f} \log \left(9 a^2 r^4+r^6\right)+\frac{8 \pi }{{g_s}}-4 {N_f} \log \left(\frac{\alpha_{\theta_1}\alpha _{\theta _2}}{4\sqrt{N}}\right)\right).
\end{eqnarray}
In a similar manner, we were able to derive $h_0(r^*)$ and $h_1(r^*)$ by exchanging the value of $r$ with $r^*$. Hence,
\begin{eqnarray}
\label{Hrstarbeta}
& & H(r^*) = h_0(r^*) + \beta h_1(r^*).
\end{eqnarray}
After that, rewriting $\sigma(r)$ as follows:
\begin{eqnarray}
\label{sigmauptobeta}
& & \sigma(r) = \sigma_0(r) + \beta \sigma_1(r).
\end{eqnarray}
Utilizing equations (\ref{Hr}), (\ref{Hrstarbeta}) and (\ref{sigmauptobeta}), the integrand that is seen in (\ref{lrstar}) can also be expressed as:
\begin{eqnarray}
& &\sqrt{\frac{\sigma(r)}{H(r) - H(r^*)}} = \sqrt{\frac{\sigma_0(r)}{h_0(r) - h_0(r^*)}} + \beta\frac{\left(\sigma_1(r)(h_0(r) - h_0(r^*)) - \sigma_0(r)(h_1(r) - h_1(r^*))\right)}{2\sqrt{\sigma_0(r)}(h_0(r) - h_0(r^*))^{\frac{3}{2}}};
\nonumber\\
& & 
\end{eqnarray}
where,
{\footnotesize
\begin{eqnarray}
\label{h_0(r)}
 h_0(r^*) \sim \frac{M^2 \left({r^*}^4-{r_0}^4\right) \left({g_s} {N_f} \log \left(9 a^2
   {r^*}^4+{r^*}^6\right)-\Omega-16 \pi
   \right){}^4}{{g_s}^{7/2} \sqrt[10]{N} {r^*}^2 \alpha _{\theta _1}^6
   \alpha _{\theta _2}^4}\nonumber\\
& & \hskip -4.5in \times \Biggl[\log ({r^*}) \left(6 a^2 ({g_s} \log N  {N_f}-4 \pi )+4 {g_s} {N_f}
   \left({r^*}^2-3 a^2\right) \log \left(\frac{1}{4} \alpha _{\theta _1} \alpha _{\theta
   _2}\right)+{r^*}^2 (8 \pi -{g_s} (2 \log N +3) {N_f})\right)\nonumber\\
& & \hskip -4.5in+{g_s} {N_f} \left(3
   a^2-{r^*}^2\right) \left(\log N -2 \log \left(\frac{1}{4} \alpha _{\theta _1} \alpha _{\theta
   _2}\right)\right)+18 {g_s} {N_f} \left({r^*}^2-3 a^2 (6 {r^*}+1)\right) \log
   ^2({r^*})\Biggr].
\end{eqnarray}
}
\begin{itemize}
\item In $H(r^*)$ if we consider
{
\begin{eqnarray}
\label{logr^*-1}
 \log r^*\to  \frac{\sqrt{72 {g_s}^2 {N_f}^2 \left(\log N -2 \log \left(\frac{1}{4} \alpha _{\theta _1}
   \alpha _{\theta _2}\right)\right)+\Omega{}^2}+\Omega }{36 {g_s} {N_f}}\nonumber\\
& & \hskip -3.5in = \frac{{g_s} {N_f} \left(\log N -2 \log \left(\frac{1}{4} \alpha _{\theta _1} \alpha _{\theta
   _2}\right)\right)}{8 \pi }+\frac{{g_s}^2 {N_f}^2 \left(-4 \log \left(\frac{1}{4} \alpha _{\theta _1} \alpha
   _{\theta _2}\right)+2 \log N +3\right)}{64 \pi ^2} \nonumber\\
   & & \hskip -3.5in \times \left(\log N -2 \log \left(\frac{1}{4} \alpha _{\theta _1} \alpha
   _{\theta _2}\right)\right)+\frac{{g_s}^3 {N_f}^3 \left(\log N -2 \log \left(\frac{1}{4}
   \alpha _{\theta _1} \alpha _{\theta _2}\right)\right)}{512 \pi ^3}\nonumber\\
& & \hskip -3.5in \times  \left(16 \log ^2\left(\frac{1}{4} \alpha _{\theta _1} \alpha
   _{\theta _2}\right)+4 \log N ^2-4 (4 \log N -3) \log \left(\frac{1}{4} \alpha _{\theta _1} \alpha _{\theta
   _2}\right)-6 \log N +9\right)
   \nonumber\\
   & &\hskip -3.5in +O\left({N_f}^4\right),
\end{eqnarray}
}
then,
\begin{eqnarray}
\label{sqrth0}
& & \hskip -0.2in \sqrt{h_0}(r^*) = \frac{a M_{\rm UV} {N_f^{\rm UV}}  {r^*} \sqrt{\left| \log N -2 \log \left(\alpha _{\theta _1} \alpha _{\theta_2}\right)+\log (16)\right| }}{24 \sqrt{6} \pi ^{9/4} {g_s}^{3/4} \sqrt[20]{N} \alpha _{\theta _1}^3 \alpha
   _{\theta _2}^2}+O\left({N_f^{\rm UV}} ^2\right). 
\end{eqnarray}
The integral in the variable $l(r^*)$ is now proportional to \\ $\frac{\sqrt[20]{N} \alpha _{\theta _1}^3 \alpha _{\theta _2}^2}{{g_s}^{5/4} M {N_f}^{10/3} r^3 \log (r)
   (\log N -3 \log (r))^{7/3} | \log N -9 \log (r)| }\sim \frac{\alpha _{\theta _1}^3 \alpha _{\theta _2}^2 \left(\frac{1}{\log (N)}\right)^{10/3}}{{g_s}^{5/4} M
   {N_f}^{10/3} r^3 \log (r)}+O\left(\left(\frac{1}{\log N }\right)^{13/3}\right)$. Since $\int \frac{dr}{r^3\log r} = {Ei}(-2 \log (r)) = \frac{-2 \log ^2(r)+\log (r)-1}{4 r^2 \log ^3(r)}+O\left(\left(\frac{1}{r}\right)^3\right) \approx - \frac{1}{2r^2\log r}$. Therefore, we obtained:
\begin{eqnarray}
\label{lr^*-ii}
& & \hskip -0.3in l(r^*) \sim \frac{a M_{\rm UV} {N_f^{\rm UV}}  \left(\frac{1}{\log (N)}\right)^{10/3} \sqrt{-2 \log \left(\alpha _{\theta _1} \alpha
   _{\theta _2}\right)+\log (N)+\log (16)}}{{g_s}^2 M {N_f}^{10/3} {r^*} \log ({r^*})}\stackrel{r^*\rightarrow\infty}{\longrightarrow}0,
\end{eqnarray}
similar to \cite{Tc-EE}. Since it was assumed that $r^*$ was in the UV, and we found that $l(r^*)$ is a function that monotonically decreases in the UV, we found that $l(r^*)$ gains a maximum value at $r^*={\cal R}_{D5/\overline{D5}}^{\rm th}$. The solution of $l(r^*) = \frac{C}{r^*\log r^*}$ where $C\sim \frac{a \left(\frac{1}{\log (N)}\right)^{10/3} \sqrt{-2 \log \left(\alpha _{\theta _1} \alpha _{\theta
   _2}\right)+{\log N}+\log (16)}}{{g_s}^2 {N_f^{\rm UV}}^{7/3}}$. implying
\begin{equation}
\label{r^*=r^*(l)}
r^* = \frac{C}{l(r^*) W\left(\frac{C}{l(r^*)}\right)}.
\end{equation}
\end{itemize}

\begin{figure}
\begin{center}
\includegraphics[width=0.85\textwidth]{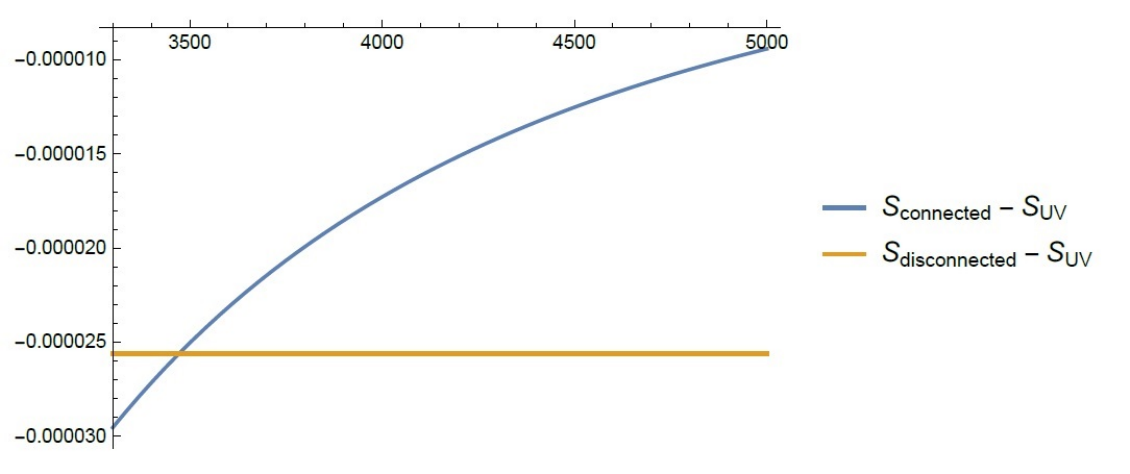}
\end{center}
\caption{Plot $S_{\rm connected}$ (blue) and $S_{\rm disconnected}$(orange) versus $l(r^*)$}
\label{S-cs-l}
\end{figure}

{\bf Entanglement Entropy}: The formula for the connected region entanglement entropy that takes into account both tiny $l$ and large $r$ is as follows:
\begin{eqnarray}
\label{S-connected-large-r}
& & \frac{S}{{\cal V}_1} = \left\{\frac{\int_{{r^*}}^{{\cal R}_{\rm UV}} \frac{\sqrt{\sigma (r)} H(r)}{\sqrt{H(r)-H({r^*})}} \, dr}{2
   {G_N}^{(11)}}\right\}.
\end{eqnarray}
For ${\cal M}$-theory background, we found the following contributions to the entanglement entropies at ${\cal O}(\beta^0)$ and ${\cal O}(\beta)$.
\begin{eqnarray}
\label{S-connected-UV-finite}
& & \frac{S_{\rm connected}^{{\rm UV-finite},\ \beta^0}}{2{\cal V}_1} \sim \frac{\log N ^{4/3} M_{\rm UV} {N_f^{\rm UV}} ^{4/3} \alpha _{\theta _2}^2 \left({r_0}^4 {r^*}^2 \log
   ^3({r^*})+{r^*}^6 \log ^2({r^*})\right)}{\sqrt[12]{{g_s}} \sqrt[20]{N} {r^*}^2 \alpha
   _{\theta _1}^6 \log ({r^*})}\nonumber\\
& & \approx \frac{\log N ^{4/3} M_{\rm UV} {N_f^{\rm UV}} ^{4/3} {r^*}^4 \alpha _{\theta _2}^2 \log
   ({r^*})}{\sqrt[12]{{g_s}} \sqrt[20]{N} \alpha _{\theta _1}^6};\nonumber\\
& & \frac{S_{\rm connected}^{{\rm UV-finite}, \beta}}{2{\cal V}_1} \sim \frac{\beta  \sqrt[3]{\frac{1}{{g_s}}} M_{\rm UV} {r^*}^4 ({\cal C}_{zz}^{\rm th}-2 {\cal C}_{\theta_1z}^{\rm th})
   \left(\frac{1}{\log (N)}\right)^{17/3}}{{g_s}^{7/4} \sqrt[20]{N} {N_f^{\rm UV}} ^{2/3} \alpha _{\theta _1}^3 \alpha
   _{\theta _2}^2}.
\end{eqnarray}
Following is an expression that has been obtained for the disconnected region entanglement entropy:
\begin{eqnarray}
\label{S-disconnected}
& & \frac{S_{\rm disconnected}}{2{\cal V}_1} = \left\{\frac{\int_{{r_0}}^{{\cal R}_{\rm UV}} \sqrt{\sigma (r) H(r)} \, dr}{2 {G_N}{}^{(11)}}\right\}\nonumber\\
& & = \left\{\frac{\left(\int_{{r_0}}^{{\cal R}_{D5/\overline{D5}}} + \int_{D5/\overline{D5}}^{{\cal R}_{\rm UV}}\right) \sqrt{\sigma (r) H(r)} \, dr}{2 {G_N}{}^{11}}\right\}\nonumber\\
& & = \frac{1}{\alpha_{\theta_1}^3\alpha_{\theta_2}^2}g_s^{\frac{5}{4}}N_f^{\frac{8}{3}}
\Biggl[\log^2 {\cal R}_{D5/\overline{D5}}{\cal R}_{D5/\overline{D5}}\left(-6a^2+{\cal R}_{D5/\overline{D5}}^2\right)\left(\log N - 3 \log {\cal R}_{D5/\overline{D5}}\right)^{\frac{5}{3}}\nonumber\\
& &  - \log^2r_0r_0^2\left(-6 a^2 + r_0^2\right)\left(\log N - 3\log r_0\right)^{\frac{5}{3}}\Biggr]\left(1 + \frac{{\cal C}_{zz}^{\rm th} - 2 {\cal C}_{\theta_1z}^{\rm th}}{2}\beta\right) + \frac{M_{\rm UV} \sqrt[20]{\frac{1}{N}} {N_f^{\rm UV}} ^{4/3} }{\sqrt[12]{{g_s}} \alpha _{\theta _1}^3 \alpha _{\theta _2}^2}\nonumber\\
& & \times \Biggl[{{\cal R}_{\rm UV}}^4 \log ({{\cal R}_{\rm UV}}) (\log N -3 \log
   ({{\cal R}_{\rm UV}}))^{4/3} \nonumber\\
   & & \hskip 0.3in - {\cal R}_{D5/\overline{D5}}^4 \log ({\cal R}_{D5/\overline{D5}})(\log N -3 \log
   ({\cal R}_{D5/\overline{D5}}))^{4/3}\Biggr].
\end{eqnarray} 
Using $a = \left(\frac{1}{\sqrt{3}} + \epsilon\right)r_0$, focus in the range, $r\in\left[r_0, {\cal R}_{D5/\overline{D5}}\right]$, entanglement entropy is obtained as:
\begin{eqnarray}
\label{S-disconnected-IR}
-\frac{\epsilon  {g_s}^{5/4} M \sqrt[20]{\frac{1}{N}} {N_f}^{8/3} {r_0}^4 (\beta  ({\cal C}_{zz}^{\rm th}-2
   {\cal C}_{\theta_1z}^{\rm th})+2)}{\alpha _{\theta _1}^3 \alpha _{\theta _2}^2}.
\end{eqnarray}
{\it Due to the highly non-trivial cancellation of ${\cal O}(R^4)$ corrections ${\cal C}_{zz}^{\rm th}-2 {\cal C}_{\theta_1z}^{\rm th} = 0$, see (\ref{f_zz-f_theta1x-f_theta1z}), we are able to deduce that there aren't any ${\cal O}(R^4)$-corrections to the entanglement entropies of the connected and disconnected regions.} It is important to note that radial distances are always calculated in units of ${\cal R}_{D5/\overline{D5}}^{\rm th}$, i.e., $ \log {\cal R}_{D5/\overline{D5}} \equiv 0$, we obtained:
\begin{eqnarray}
\label{final-S-connected+S-disconnected}
& & \frac{S_{\rm disconnected} - S_{\rm UV}^{\rm disconnected}}{2{\cal V}_1} \sim  -\frac{{g_s}^{5/4} M \sqrt[20]{\frac{1}{N}} {N_f}^{8/3} r_0^4 \log ^2(r_0) ({\log N}-3 \log
   (r_0))^{5/3}}{64 \sqrt[3]{2} 3^{5/6} \pi ^{41/12} \alpha _{\theta _1}^3 \alpha _{\theta _2}^2};\nonumber\\
   & & \frac{S_{\rm connected} - S_{\rm UV}^{\rm connected}}{2{\cal V}_1} \sim \frac{{M_{\rm UV}} {N_f^{\rm UV}}^{4/3} \alpha _{\theta _2}^2 \log ^{\frac{4}{3}}(N) \left(r_0^4
   {r^*}^2 \log ^3({r^*})-{r^*}^6 \log ^2({r^*})\right)}{72\ 2^{2/3} 3^{5/6}
   \pi ^{25/12} \sqrt[12]{{g_s}} \sqrt[20]{N} {r^*}^2 \alpha _{\theta _1}^6 \log
   ({r^*})}\nonumber
   \end{eqnarray}
   
\begin{eqnarray}   
& & \frac{S_{\rm connected} - S_{\rm UV}^{\rm connected}}{2{\cal V}_1} \approx -\frac{{M_{\rm UV}} {N_f^{\rm UV}}^{4/3} {r^*}^4 \log ^{\frac{4}{3}}(N) \log ({r^*})}{72\ 2^{2/3} 3^{5/6} \pi
   ^{25/12} \sqrt[12]{{g_s}} \sqrt[20]{N}},
\end{eqnarray}
wherein:
\begin{eqnarray}
\label{SdisconnectedUV+SconnectedUV}
& & S_{\rm UV}^{\rm disconnected} \sim \frac{M_{\rm UV}N_f^{\rm UV}\ ^{\frac{4}{3}} {\cal R}_{\rm UV}^4\left(\log N - 3 \log {\cal R}_{\rm UV}\right)^{\frac{4}{3}}\log{\cal R}_{\rm UV} }{N^{\frac{1}{20}}\alpha_{\theta_1}^3\alpha_{\theta_2}^3},
\nonumber\\
& & S_{\rm UV}^{\rm connected} \sim \frac{M_{\rm UV}N_f^{\rm UV}\ ^{\frac{4}{3}} {\cal R}_{\rm UV}^4\left(\log N \right)^{\frac{4}{3}}\log{\cal R}_{\rm UV} \alpha_{\theta_2}^2}{N^{\frac{1}{20}}\alpha_{\theta_1}^6}.
\end{eqnarray}
One should take attention that in the limit of large $N$, ${\cal R}_{\rm UV}<\left(4\pi g_s N\right)^{\frac{1}{4}}$ and set $\alpha_{\theta_1}^3 =  \alpha_{\theta_2}^4$, we found that $S_{\rm UV}^{\rm disconnected} = S_{\rm UV}^{\rm connected}$. %As $l$ increases, i.e., $r^*$ decreases and reaches ${\cal R}_{D5/\overline{D5}}^{\rm th}$, $S_{\rm connected}$ changes from being negative to vanishing and $S_{\rm disconnected}$ stays negative implying disconnected region has lesser entropy. 
At $r^* = r_{\rm criticial}$, $S_{\rm connected} = S_{\rm disconnected}$ and $r_{\rm critical}$ can be obtain from the equation given below:
\begin{eqnarray}
\label{rstar-gamma-eqn}
& &  \left(r^*\right)^4\log r^* = \gamma,
\end{eqnarray}
wherein $\gamma \sim \frac{{g_s}^{4/3} M {N_f}^{8/3} \alpha _{\theta _1}^3 \log ^2(r_0) (\log (N)-3 \log
   (r_0))^{5/3}}{\pi ^{4/3} {M_{\rm UV}} {N_f^{\rm UV}}^{4/3} \alpha _{\theta _2}^4 \log
   ^{\frac{4}{3}}(N)}$. Equation (\ref{rstar-gamma-eqn}) has the solution, ${r^*}=e^{\frac{1}{4} W(4 \gamma )}\approx \sqrt{2} \sqrt[4]{\gamma}$. Using $M_{\rm UV},N_f^{\rm} \sim \frac{1}{\log N}$, $\log r_0 = -\frac{f_{r_0}}{3}\log N$, and $f_{r_0}\approx 1$ \cite{MChPT}, we obtained:
\begin{equation}
\label{r-critical}
r_{\rm critical} = \frac{\sqrt[3]{{g_s}} M {N_f}^{2/3} \alpha _{\theta
   _1}^{3/4} \log ^{\frac{7}{6}}(N)}{\sqrt[4]{2} \sqrt[3]{\pi }
   \alpha _{\theta _2}} r_0.
\end{equation}
Choosing $M=N_f=3, g_s=0.1-1$; (\ref{r-critical}) implying $r_{\rm critical} \sim {\cal O}(1)\frac{\alpha_{\theta_1}^{\frac{3}{4}}}{\alpha_{\theta_2}}$. By considering the 4D-limit that is achieved through $M_{\rm KK}\rightarrow0$, or, analogously, $r_0\rightarrow0$, one can therefore deduce the following result from the (\ref{r_h-r_0-relation}):
\begin{eqnarray}
\label{consistency}
& & {\cal O}(1)\frac{\alpha_{\theta_1}^{\frac{3}{4}}}{\alpha_{\theta_2}} \stackrel{\scriptsize 4D-{\rm limit}}{\longrightarrow} \frac{\sqrt[4]{\frac{\kappa_{\rm GHY}^{{\rm th},\ \beta^0}}{\kappa_{\rm GHY}^{{\rm bh},\ \beta^0}}}  {{\cal R}_{D5/\overline{D5}}^{\rm bh}} \sqrt[4]{\frac{\log
   \left(\frac{{{\cal R}_{\rm UV}}}{{{\cal R}_{D5/\overline{D5}}^{\rm th}}}\right)}{\log
   \left(\frac{{{\cal R}_{\rm UV}}}{{{\cal R}_{D5/\overline{D5}}^{\rm bh}}}\right)}}}{{{\cal R}_{D5/\overline{D5}}^{\rm th}}},
\end{eqnarray}
proportional to the deconfinement temperature. For the numerical purpose, we choose $N=100, M_{\rm UV} = N_f^{\rm UV} = 0.01, M=N_f=3, \alpha_{\theta_{1,2}}: \alpha_{\theta_1}^3\alpha_{\theta_2}^2=2$, and $r_0=N^{-\frac{f_{r_0}}{3}}, f_{r_0}\approx 1$ \cite{MChPT}, and hence, $l(r_{\rm critical})=3850$. For these values plot of entanglement entropy versus $l(r^*)$ is shown in Fig. \ref{S-cs-l}.
In addition to this, let us consider the possibility that $r^*\in[r_0,{\cal R}_{D5/\overline{D5}}]$, i.e., IR-valued $r^*$, and split the radial integral $\int_{r^*}^{{\cal R}_{\rm UV}}$ into $\left(\int_{r^*}^{{\cal R}_{D5/\overline{D5}}} + \int_{{\cal R}_{D5/\overline{D5}}}^{{\cal R}_{\rm UV}}\right)$, we obtained:
\begin{eqnarray}
\label{rstar-IR-i}
& & \int_{r^*}^{{\cal R}_{D5/\overline{D5}}}dr\frac{\sqrt{\sigma(r)} H(r)}{\sqrt{H(r) - H(r^*)}} \nonumber\\
& & \sim 2 \sqrt{{g_s}} M^2 {N_f}^{13/3} \int_{r^*}^{{\cal R}_{D5/\overline{D5}}} r^6 \sqrt{1-\frac{r_0^4}{r^4}} \log ^2(r) (\log N -3
   \log (r))^{11/3} \, dr\nonumber\\
& & = \int_{r^*}^{{\cal R}_{D5/\overline{D5}}} \left(2 \sqrt{{g_s}} \log N ^{11/3} M^2 {N_f}^{13/3} r^6 \sqrt{1-\frac{r_0^4}{r^4}} \log
   ^2(r)+O\left(\left(\log N\right)^{8/3}\right)\right) dr\nonumber\\
& & = \Biggl[\frac{2}{343} \sqrt{{g_s}} \log N ^{11/3} M^2 {N_f}^{13/3} r^7 \Biggl(7 \log (r) \Biggl(7 \log (r)
   \, _2F_1\left(-\frac{7}{4},-\frac{1}{2};-\frac{3}{4};\frac{r_0^4}{r^4}\right)\nonumber\\
& & -2 \,
   _3F_2\left(-\frac{7}{4},-\frac{7}{4},-\frac{1}{2};-\frac{3}{4},-\frac{3}{4};\frac{r_0^4}{r^4}\right
   )\Biggr)+2 \,
   _4F_3\left(-\frac{7}{4},-\frac{7}{4},-\frac{7}{4},-\frac{1}{2};-\frac{3}{4},-\frac{3}{4},-\frac{3}{4};\frac{r_0^4}{r^4}\right)\Biggr)\Biggr]_{r^*}^{{\cal R}_{D5/\overline{D5}}}\nonumber
   \end{eqnarray}
   \begin{eqnarray}
& & = \frac{4}{343} \sqrt{{g_s}} \log N ^{11/3} M^2 {N_f}^{13/3}-\frac{2 r_0^4
   \left(\sqrt{{g_s}} \log N ^{11/3} M^2 {N_f}^{13/3}\right)}{27
   {{\cal R}_{D5/\overline{D5}}}^4}+O\left(r_0^5\right) \nonumber\\
& & -\frac{2}{7} \left(\sqrt{{g_s}} \log N ^{11/3} M^2 {N_f}^{13/3}  r^*\ ^7 \log
   ^2( r^*)\right)+\frac{1}{3} \sqrt{{g_s}} \log N ^{11/3} M^2 {N_f}^{13/3} r_0^4
    r^*\ ^3 \log ^2( r^*)+O\left(r_0^5\right),\nonumber\\
& & 
\end{eqnarray}
and,
\begin{eqnarray}
\label{rstar-IR-ii}
& & \int_{{\cal R}_{D5/\overline{D5}}}^{{\cal R}_{\rm UV}}dr\frac{\sqrt{\sigma(r)} H(r)}{\sqrt{H(r) - H(r^*)}} \nonumber\\
& & =  \int_{{\cal R}_{D5/\overline{D5}}}^{{\cal R}_{\rm UV}}dr\frac{\sqrt[4]{\frac{1}{{g_s}}} \sqrt[6]{{g_s}} {M_{\rm UV}} {N_f^{\rm UV}}^{4/3} r^3 \alpha _{\theta
   _2}^2 \sqrt{1-\frac{r_0^4}{r^4}} \log (r) (\log N -3 \log (r))^{8/3}}{18\ 2^{2/3} 3^{5/6} \pi
   ^{25/12} \log N ^{4/3} \sqrt[20]{N} \alpha _{\theta _1}^3}+O\left(\left({N_f^{\rm UV}}\right)^2\right)\nonumber\\
& & =\int_{{\cal R}_{D5/\overline{D5}}}^{{\cal R}_{\rm UV}}dr \frac{\log N ^{4/3} {M_{\rm UV}} {N_f^{\rm UV}}^{4/3} r^3 \alpha _{\theta _2}^2
   \sqrt{1-\frac{r_0^4}{r^4}} \log (r)}{18\ 2^{2/3} 3^{5/6} \pi ^{25/12} \sqrt[12]{{g_s}}
   \sqrt[20]{N} \alpha _{\theta _1}^3} + {\cal O}\left(\sqrt[3]{\log N }\right)\nonumber\\
& & = \int_{{\cal R}_{D5/\overline{D5}}}^{{\cal R}_{\rm UV}}dr \frac{\log N ^{4/3} {M_{\rm UV}} {N_f^{\rm UV}}^{4/3} r^3 \alpha _{\theta _2}^2 \log (r)}{18\ 2^{2/3} 3^{5/6}
   \pi ^{25/12} \sqrt[12]{{g_s}} \sqrt[20]{N} \alpha _{\theta
   _1}^3}+O\left(\left(r_0^4\right)^1\right)\nonumber\\
& & = \Biggl[\frac{\log N ^{4/3} {M_{\rm UV}} {N_f^{\rm UV}}^{4/3} r^4 \alpha _{\theta _2}^2 (4 \log (r)-1)}{288\ 2^{2/3}
   3^{5/6} \pi ^{25/12} \sqrt[12]{{g_s}} \sqrt[20]{N} \alpha _{\theta _1}^3}\Biggr]_{{\cal R}_{D5/\overline{D5}}}^{{\cal R}_{\rm UV}}\nonumber\\
& & = \frac{\log N ^{4/3} {M_{\rm UV}} {N_f^{\rm UV}}^{4/3} {{\cal R}_{D5/\overline{D5}}}^4 \alpha _{\theta _2}^2}{288\
   2^{2/3} 3^{5/6} \pi ^{25/12} \sqrt[12]{{g_s}} \sqrt[20]{N} \alpha _{\theta
   _1}^3}+\frac{\log N ^{4/3} {M_{\rm UV}} {N_f^{\rm UV}}^{4/3} {\cal R}_{\rm UV}^4 \alpha _{\theta _2}^2 \log
   ({\cal R}_{\rm UV})}{72\ 2^{2/3} 3^{5/6} \pi ^{25/12} \sqrt[12]{{g_s}} \sqrt[20]{N} \alpha _{\theta _1}^3}.
\end{eqnarray}
\begin{itemize}
\item We found that in $S_{\rm connected}^{r^*\in[{\cal R}_{D5/\overline{D5}},{\cal R}_{\rm UV}]}, S_{\rm disconnected}^{r^*=r_0}, S_{\rm connected}^{r^*\in[r_0,{\cal R}_{D5/\overline{D5}}]}$, ``UV-divergent'' terms are same, represented by $S_{\rm UV}\sim \frac{\log N ^{4/3} {M_{\rm UV}} {N_f^{\rm UV}}^{4/3} {\cal R}_{\rm UV}^4 \log ({\cal R}_{\rm UV})}{\sqrt[12]{{g_s}}
   \sqrt[20]{N}}$. 

\item Additionally,
$S_{\rm connected}^{r^*\in[r_0,{\cal R}_{D5/\overline{D5}}]} - S_{\rm UV}\sim\sqrt{{g_s}} \log N ^{11/3} M^2 {N_f}^{13/3}$.

\item Hence, we obtained (in a vein of some of the examples in \cite{Tc-EE}):
\begin{eqnarray*}
\label{comparison}
& & S_{\rm connected}^{r^*\in[r_0,{\cal R}_{D5/\overline{D5}}]} - S_{\rm UV} > S_{\rm connected}^{r^*\in[{\cal R}_{D5/\overline{D5}},{\cal R}_{\rm UV}]} > S_{\rm disconnected}^{r^*=r_0}  - S_{\rm UV},\ l(r^*)>l(r_{\rm critical}),\nonumber\\
& & S_{\rm connected}^{r^*\in[r_0,{\cal R}_{D5/\overline{D5}}]} - S_{\rm UV} > S_{\rm connected}^{r^*\in[{\cal R}_{D5/\overline{D5}},{\cal R}_{\rm UV}]} - S_{\rm UV} < S_{\rm disconnected}^{r^*=r_0}  - S_{\rm UV},\ l(r^*)<l(r_{\rm critical}).
\end{eqnarray*}

\end{itemize}

\subsection{${\cal M}\chi$PT Compatibility}
\label{Tc-SS-iv}
In chapter {\bf 2}, we obtained one-loop renormalized LECs of the $SU(3)$ chiral perturbation theory Lagrangian up to ${\cal O}(p^4)$ from type IIA string dual of large-$N$ thermal QCD like-theory, including the ${\cal O}(R^4)$ corrections. Due to our chiral limit restriction, several terms in the $SU(3)$ chiral perturbation theory Lagrangian of Gasser and Leutwyler \cite{GL} were missing. As a result, we were able to calculate $g_{YM}$, $L_{1,2,3,9,10}^r$, and $F_\pi$. We compare our findings to the one-loop renormalized LEC phenomenological values listed in \cite{Ecker-2015}. In chapter {\bf 2}, we showed how to match the phenomenological values of the ${\cal O}(p^4)$ $SU(3)$ $\chi$PT LECs ($g_{YM}$, $L_{1,9,10}^r$, and $F_\pi$) in five stages, in addition to the order of the magnitude and signs of $L_{2,3}^r$.
\par
We discovered that in order for $L_9^r$ to match the holographic computation, a particular set of integration constants had to have a definite sign [see equation (\ref{CCsO4-lambda-epsilon})]. In appendix \ref{appendix-McTEQ}, by taking the decompactification limit (i.e. $M_{\rm KK}\rightarrow0$ limit) of the spatial circle $S^1(x^3)$ appearing as part of the ${\cal M}$-theory metric utilized in \cite{MChPT}, i.e., in  $S^1(x^0)\times R^2 \times S^1\left(\frac{1}{M_{KK}}\right)$ to reconstruct a theory similar to 4D QCD limit, we were able to derive values for these integration constants in an explicit manner. In the matter of fact, we used those values to show that the previously mentioned combination of integration constants, which appears in all of the LECs of $SU(3)$ $\chi$PT Lagrangian up to ${\cal O}(p^4)$, actually could be rendered to have a negative sign as needed by comparing with the phenomenological/experimental values of the one-loop renormalized LECs of $SU(3)$ $\chi$PT Lagrangian as discussed previously. This can be seen from (\ref{f_zz-f_theta1x-f_theta1z}), $ {\cal C}_{zz}^{\rm th} - 2 {\cal C}_{\theta_1z}^{\rm th} + 2 {\cal C}_{\theta_1x}^{\rm th} = 2 {\cal C}_{\theta_1x}^{\rm th} \sim \frac{\left(\frac{1}{N}\right)^{7/6} \Sigma _1}{ \epsilon ^{11} {g_s}^{9/4}
   \log N ^4 {N_f}^3 {r_0}^5 \alpha _{\theta _1}^7 \alpha _{\theta _2}^6}$ and this is negative when $\Sigma_1<0$. 

\section{Deconfinement Temperature of Rotating QGP at Intermediate Coupling from ${\cal M}$-Theory}
\label{DPT-WR-ii}
This part is based on \cite{Rotation-Tc-M-Theory}. In this paper, we have obtained the deconfinement temperature of thermal QCD-like theories in the presence of rotation. We will discuss the holographic construction of rotating QGP from a top-down approach in \ref{Rotating-QGP}. Using this holographic dual, we will discuss the calculation of $T_c$ in the presence of rotation in \ref{O-beta0-Tc} and UV-IR mixing, Flavor Memory effect, and non-renormalization of $T_c$ in \ref{UV-IR-mixing}
\subsection{Top-Down Holographic Dual of Rotating QGP}
\label{Rotating-QGP}
When $T>T_c$ on the gravity dual side and $T<T_c$ on the gauge theory side, we could investigate the affect of rotation in thermal QCD-like theories by introducing a rotating cylindrical black hole and thermal backgrounds on the gravity dual side. In order to obtain the background of a rotating cylinder black hole, we need to make $x^3$ periodic by replacing it with $l \phi$, where $l$ represents the length of the cylinder and $0\leq \phi \leq 2 \pi$. This will allow us to obtain the rotating cylinder black hole background. After completing the following Lorentz transformation around the cylinder of length $l$, one can obtain the holographic dual of rotating quark-gluon plasma. This can be done by following \cite{Lorentz-boost-1,Lorentz-boost-2}:
\begin{eqnarray}
\label{Lorentz-boost}
& & t \rightarrow \frac{1}{\sqrt{1-l^2 \omega^2}}\left(t+l^2 \omega \phi \right); \ \phi \rightarrow \frac{1}{\sqrt{1-l^2 \omega^2}} \left(\phi + \omega t\right),
\end{eqnarray}
where $\omega$ represents the angular velocity of the rotating cylindrical black hole in the gravity dual, which is connected to the rotation of quark-gluon plasma by means of the gauge-gravity duality. Only when $\omega l <1$, the Lorentz transformation, also known as the Lorentz-boost, considered to be trustworthy. Following the application of the Lorentz transformation, which is denoted by the equation (\ref{Lorentz-boost}), the equation (\ref{TypeIIA-from-M-theory-Witten-prescription-T>Tc}) is rewritten as follows: 
{\footnotesize 
\begin{eqnarray}
\label{TypeIIA-from-M-theory-Witten-prescription-T>Tc-Rotation-Canonical}
\hskip -6in ds_{11}^2|_{BH} & = & e^{-\frac{2\phi^{\rm IIA}}{3}}\Biggl[\frac{1}{\sqrt{h(r,\theta_{1,2})}}\Biggl(-{\cal Y}_1(r) dt^2 +{\cal Y}_2(r)\left(d \phi+{\cal Y}_3(r) dt\right)^2 
 + \left(dx^1\right)^2 +  \left(dx^2\right)^2 \Biggr)
\nonumber\\
& & \hskip -0.1in+ \sqrt{h(r,\theta_{1,2})}\left(\frac{dr^2}{g(r)} + ds^2_{\rm IIA}(r,\theta_{1,2},\phi_{1,2},\psi)\right)
\Biggr] + e^{\frac{4\phi^{\rm IIA}}{3}}\left(dx^{11} + A_{\rm IIA}^{F_1^{\rm IIB} + F_3^{\rm IIB} + F_5^{\rm IIB}}\right)^2,
\end{eqnarray}
}
where ${\cal Y}_1(r)=\frac{g(r)\left(1-l^2 \omega^2\right)}{\left(1-g(r)l^2\omega^2\right)}; \
{\cal Y}_2(r)=\frac{l^2 \left(1- g(r)l^2 \omega^2\right)}{\left(1-l^2\omega^2\right)}; \
{\cal Y}_3(r)=\frac{\omega \left(1- g(r)\right)}{\left(1-g(r) l^2\omega^2\right)}$.
Note that we have the equation (\ref{TypeIIA-from-M-theory-Witten-prescription-T>Tc-Rotation-Canonical}), we are able to determine the Hawking temperature of the rotating cylindrical black hole by utilizing the formula that appeared in \cite{G-D-Rotation}:
{\footnotesize
\begin{eqnarray}
\label{Hawking-temp-defn}
& &  \hskip -0.3in
T_H(\gamma)=\Biggl|\frac{\kappa}{2 \pi}\Biggr|=\Biggl|\frac{\lim_{r\rightarrow r_h}-\frac{1}{2}\sqrt{\frac{G^{rr}}{-\hat{G_{tt}}}}\hat{G_{tt}},r}{2 \pi}\Biggr|  =\frac{r_h}{\sqrt{3} \pi ^{3/2}   \sqrt{N} \sqrt{g_s}}\left(\frac{1}{\gamma}+\frac{\beta}{2} \left(-{\cal C}_{zz}^{\rm BH} + 2 {\cal C}_{\theta_1z}^{\rm BH} - 3 {\cal C}_{\theta_1x}^{\rm BH}\right)\right),
\end{eqnarray}
}
where $\hat{G_{tt}}=-{\cal Y}_1(r)$, $\hat{G_{tt}},r$ suggests that a derivative of  $\hat{G_{tt}}$ with respect to $r$ is being computed, and Lorentz factor, $\gamma=\frac{1}{\sqrt{1-l^2 \omega^2}}$. Considering that $L^4= 4 \pi g_s N$, the Hawking temperature of the rotating cylindrical black hole is calculated to be:
\begin{eqnarray}
\label{Hawking-temp}
& & T_H(\gamma)= \left(\frac{r_h}{\pi L^2}\right)\left(\frac{1}{\gamma}+\frac{\beta}{2} \left(-{\cal C}_{zz}^{\rm BH} + 2 {\cal C}_{\theta_1z}^{\rm BH} - 3 {\cal C}_{\theta_1x}^{\rm BH}\right)\right) \nonumber\\
& & = T_H(0)\left(\frac{1}{\gamma}+\frac{\beta}{2}\left(-{\cal C}_{zz}^{\rm BH} + 2 {\cal C}_{\theta_1z}^{\rm BH} - 3 {\cal C}_{\theta_1x}^{\rm BH}\right)\right),
\end{eqnarray}
where $T_H(0)$ represents the Hawking temperature for a static black hole that was obtained in \cite{NPB}, the ${\cal O}(\beta)$ term in equations (\ref{Hawking-temp-defn}) and (\ref{Hawking-temp}) is computed in small-$\omega$ limit with $\gamma=1$. The constants of integration ${\cal C}_{zz/ \theta_1z / \theta_1z}^{\rm BH}$ could be found in the ${\cal O}(R^4)$ correction to the black hole background metric \cite{HD-MQGP}. As a consequence of this, the Hawking temperature of the rotating cylindrical black hole will be multiplied by the inverse of the Lorentz factor. Since ${\cal C}_{zz}^{\rm BH}=2 {\cal C}_{\theta_1 z}^{\rm BH}$ and $|{\cal C}_{\theta_1 x}^{\rm BH}| \ll 1$ \cite{McTEQ}, the Hawking temperature does not get any ${\cal O}(R^4)$ correction, which is the same thing as saying that the Hawking temperature is unaffected by the presence of ${\cal O}(R^4)$ terms,
\begin{eqnarray}
\label{Hawking-temp-i}
& &
 T_H(\gamma)= \frac{T_H(0)}{\gamma}=T_H(0)\sqrt{1-l^2\omega^2}.
\end{eqnarray}
In order to produce the rotating cylindrical thermal background, we substitute $x^3$ by $l \phi$ and carry out the Lorentz transformation (\ref{Lorentz-boost}) of (\ref{TypeIIA-from-M-theory-Witten-prescription-T<Tc}) in the same manner as we did to the black hole background. As a result of these steps, the metric of the rotating cylindrical thermal background simplified as:
\begin{eqnarray}
\label{TypeIIA-from-M-theory-Witten-prescription-T<Tc-rotating-cylindrical}
\hskip -0.2in ds_{11}^2|_{Thermal} & = & e^{-\frac{2\phi^{\rm IIA}}{3}}\Biggl[\frac{1}{\sqrt{h(r,\theta_{1,2})}}\left(-dt^2 + \left(dx^1\right)^2 +  \left(dx^2\right)^2 +l^2\left(d\phi\right)^2 \right)
\nonumber\\
& & \hskip -0.5in+ \sqrt{h(r,\theta_{1,2})}\left(dr^2+ ds^2_{\rm IIA}(r,\theta_{1,2},\phi_{1,2},\psi)\right)
\Biggr] + e^{\frac{4\phi^{\rm IIA}}{3}}\left(dx^{11} + A_{\rm IIA}^{F_1^{\rm IIB} + F_3^{\rm IIB} + F_5^{\rm IIB}}\right)^2.
\end{eqnarray}
\subsection{Deconfinement Temperature of Rotating QGP from ${\cal M}$-Theory Dual} \label{O-beta0-Tc}
Using a semi-classical method \cite{Witten-Hawking-Page-Tc}, we will compute the deconfinement temperature of a rotating QGP at intermediate coupling in this section similar to \ref{Tc-SS-i}. On-shell action, in this case, will be the same as (\ref{on-shell-D=11-action-up-to-beta}), and hence we computed the various terms appearing in (\ref{on-shell-D=11-action-up-to-beta}) for the rotating cylindrical black hole (\ref{TypeIIA-from-M-theory-Witten-prescription-T>Tc-Rotation-Canonical}) and thermal (\ref{TypeIIA-from-M-theory-Witten-prescription-T<Tc-rotating-cylindrical}) backgrounds, and integrated over the angular coordinates similar to \ref{Tc-SS-i}. In the upcoming pages, we will not discuss every step, and we will write the final results because the procedure is the same as \ref{Tc-SS-i}. Only metric is different for the rotating cylindrical black hole and thermal backgrounds compared to \ref{Tc-SS-i}. \par
\begin{figure}
\begin{center}
\includegraphics[width=0.70\textwidth]{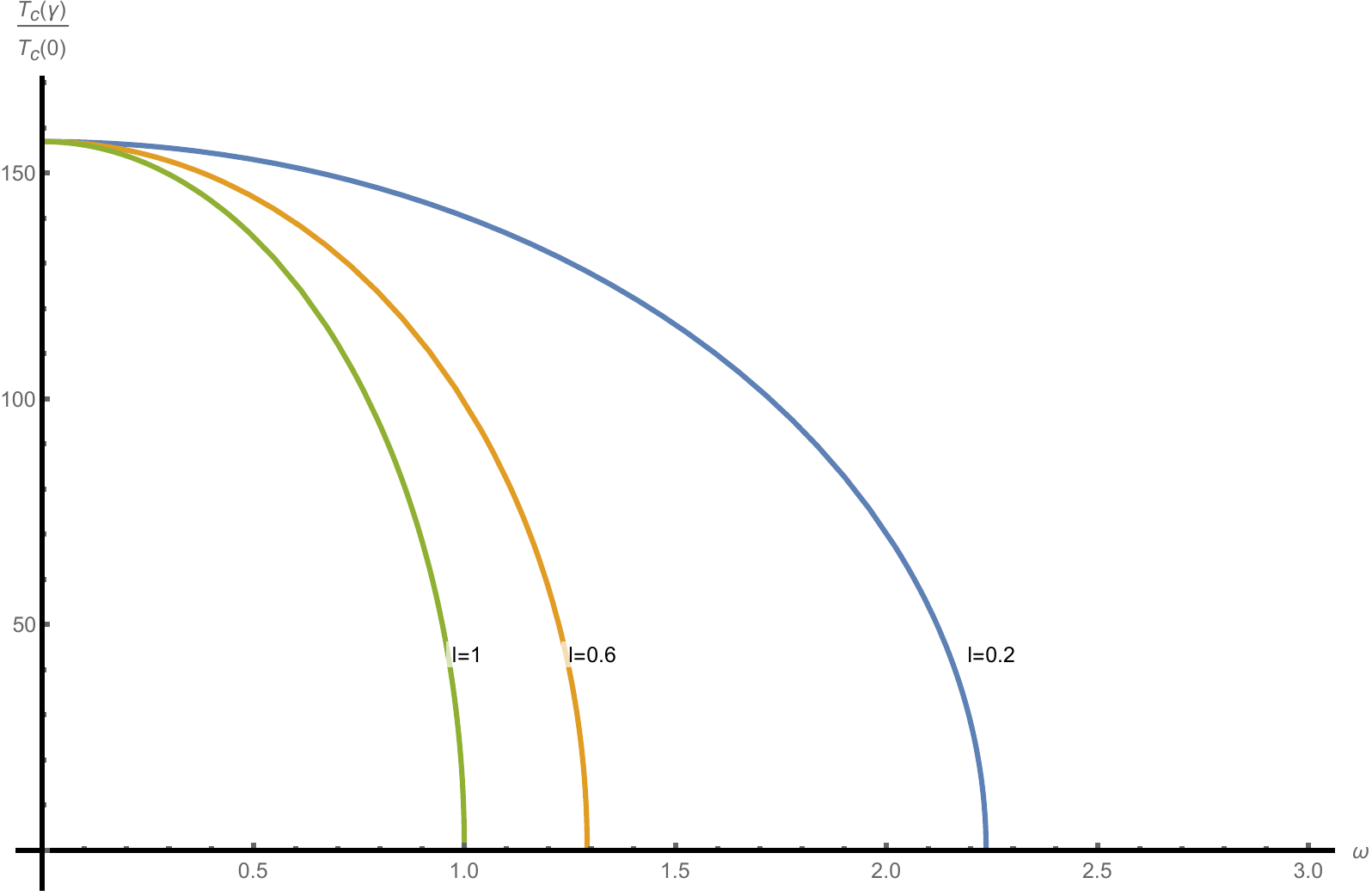}
\end{center}
\caption{Plot of $\frac{T_c(\gamma)}{T_c(0)}$ versus $\omega$ for $l=0.2,0.6,1$.}
\label{Tc-Plot}
\end{figure}

In a procedure analogous to that of \ref{Tc-SS-i}, the UV-finite and holographic IR regularized on-shell action density for the ${\cal M}$-theory rotating cylindrical black hole background (\ref{TypeIIA-from-M-theory-Witten-prescription-T>Tc-Rotation-Canonical}) was calculated as follows: 
{\footnotesize
\begin{eqnarray}
\label{on-shell-BH-rotation}
& &  \left(1+\frac{r_h^4}{2{\cal R}_{\rm UV}^4}\right)\frac{S_{D=11,\ {\rm on-shell\ UV-finite}}^{\rm BH}}{{\cal V}_4} = \lambda_{\rm EH, IR}^{{\rm BH}} \frac{ \epsilon^{\rm BH}  \gamma ^8 \omega^2 M N_f^3 g_s^{3/2} r_h^4 \log ^3(N) \log \left(\frac{r_h}{{\cal R}_{D5/\overline{D5}}^{\rm BH}}\right) \log \left(1-\frac{r_h}{{\cal R}_{D5/\overline{D5}}^{\rm BH}}\right)}{{{\cal R}_{D5/\overline{D5}}^{\rm BH}}^4N^{1/2}} \nonumber\\
& & + \lambda_{\rm EH, UV}^{{\rm BH}} \frac{ \gamma ^8 l M^{\rm UV}   r_h^4  \log ^2\left(\frac{{\cal R}_{\rm UV}}{{\cal R}_{D5/\overline{D5}}^{\rm BH}}\right)}{N^{1/2}{{g_s^{\rm UV}}^{3/2}} {{\cal R}_{D5/\overline{D5}}^{\rm BH}}^4 }  + \lambda_{\rm GHY}^{{\rm BH}}\frac{l M^{\rm UV}  r_h^4  \log\left(\frac{{\cal R}_{\rm UV}}{{\cal R}_{D5/\overline{D5}}^{\rm BH}}\right)}{N^{1/2} {{\cal R}_{D5/\overline{D5}}^{\rm BH}}^4  {{g_s^{\rm UV}}^{3/2}}},
\end{eqnarray}
}
with numerical prefactors $\lambda_{\rm EH, IR}^{{\rm BH}}, \ \lambda_{\rm EH, UV}^{\rm BH}$
and $\lambda_{\rm GHY}^{{\rm BH}}$.  ${\cal V}_4$ being the coordinate volume of $S^1(t)\times_w \mathbb{R}^2(x^{1,2})\times_w S^1(\phi)$ and ${\cal R}_{D5-\overline{D5}}^{\rm {BH/th}}\equiv \sqrt{3}a^{\rm {BH/th}}$, where $a^{\rm {BH/th}}$ represent the resolution parameter of the blown up $S^2$ and have the forms:  $a^{\rm {BH/th}} = \left(\frac{1}{\sqrt{3}} + \epsilon^{\rm {BH/th}} + {\cal O}\left(\frac{g_sM^2}{N}\right)\right)r_{(h/0)}$ \cite{HD-MQGP}. In addition, the UV-finite and holographic IR regularized ${\cal O}(\beta^0)$ contribution of the on-shell action density for the ${\cal M}$-theory rotating cylindrical thermal background (\ref{TypeIIA-from-M-theory-Witten-prescription-T<Tc-rotating-cylindrical}) obtained as follows:
{\footnotesize
\begin{eqnarray}
\label{on-shell-th-rotation}
& & \frac{S_{D=11,\ {\rm on-shell\ UV-finite}}^{\rm thermal}}{{\cal V}_4} =  \frac{ \lambda_{\rm GHY}^{{\rm th}} {M_{\rm UV}}{\it l} {r_0}^4 \log \left(\frac{{{\cal R}_{\rm UV}}}{{\cal R}_{D5/\overline{D5}}^{\rm th}}\right)}{{{g_s^{\rm UV}}^{3/2}}
   N^{1/2} {{\cal R}_{D5/\overline{D5}}^{\rm th}}^4 }+\frac{{g_s}^{3/2} \lambda_{\rm EH, IR}^{{\rm th}} M {N_f}^3{\it l} {r_0}^2 \log^2(N) \log \left(\frac{{r_0}}{{\cal R}_{D5/\overline{D5}}^{\rm th}}\right)}{ N^{1/2} {{\cal R}_{D5/\overline{D5}}^{\rm th}}^2 }\nonumber\\
& &  + \frac{\lambda_{\rm EH,\ UV}^{\rm th}{M_{\rm UV}} {N_f^{\rm UV}}{\it l}  \left(-\frac{121 {r_0}^4}{16 {{\cal R}_{D5/\overline{D5}}^{\rm th}}^4}-\frac{6 {r_0}^2}{{{\cal R}_{D5/\overline{D5}}^{\rm th}}^2}+2\right)}{ {g_s^{\rm UV}}^{1/2} N^{\frac{1}{2}} }, 
\end{eqnarray}  
}
wherein $r_0$ represents the Infrared cut-off of the theory when $T<T_c$ within QCD, and where $\lambda_{\rm GHY}^{{\rm th}}, \ \lambda_{\rm EH, IR}^{{\rm th}}$ and $\lambda_{\rm EH,\ UV}^{\rm th}$ are the numeric factors. The UV-valued parameters are indicated in  (\ref{on-shell-BH-rotation}) and (\ref{on-shell-th-rotation}) by parameters that include the superscript ``${\rm UV}$'' like $M^{\rm UV}, g_s^{\rm UV}$ etc. At the UV cut-off \cite{McTEQ}:
\begin{equation}
\label{SBH=Sth}
\left(1+\frac{r_h^4}{2{\cal R}_{\rm UV}^4}\right)\frac{S_{D=11,\ {\rm on-shell\ UV-finite}}^{\rm BH}}{{\cal V}_4}=\frac{S_{D=11,\ {\rm on-shell\ UV-finite}}^{\rm thermal}}{{\cal V}_4}.
\end{equation}
In (\ref{on-shell-BH-rotation}), $\omega^2<1$, $\lambda_{\rm GHY}^{{\rm BH}} \sim {\cal O}(10^3)\lambda_{\rm EH, IR}^{{\rm BH}} $  and $\lambda_{\rm GHY}^{{\rm BH}} \sim {\cal O}(10)\lambda_{\rm EH, UV}^{{\rm BH}} $, and in (\ref{on-shell-th-rotation}), $\lambda_{\rm GHY}^{{\rm th}} \sim {\cal O}(10^3)\lambda_{\rm EH, IR}^{{\rm th}} $, $\lambda_{\rm GHY}^{{\rm th}} \sim {\cal O}(10^2)\lambda_{\rm EH,\ UV}^{\rm th}$, Consequently, one has to solve the following equation:
\begin{eqnarray}
\label{rh-r0-equation}
& &  \lambda_{\rm GHY}^{{\rm BH}}\frac{l M^{\rm UV}  r_h^4  \log\left(\frac{{\cal R}_{\rm UV}}{{\cal R}_{D5/\overline{D5}}^{\rm BH}}\right)}{N^{1/2} {{\cal R}_{D5/\overline{D5}}^{\rm BH}}^4  {{g_s^{\rm UV}}^{3/2}}}-\frac{ \lambda_{\rm GHY}^{{\rm th}} {M_{\rm UV}}{\it l} {r_0}^4 \log \left(\frac{{{\cal R}_{\rm UV}}}{{\cal R}_{D5/\overline{D5}}^{\rm th}}\right)}{{{g_s^{\rm UV}}^{3/2}} N^{1/2} {{\cal R}_{D5/\overline{D5}}^{\rm th}}^4 }=0.
\end{eqnarray}  
The following is a solution to the equation presented above: 
\begin{eqnarray}
  \label{rh-r0-rotation-beta0}  
 & &r_h= \frac{\sqrt[4]{\frac{ \lambda_{\rm GHY}^{{\rm th}}}{\lambda_{\rm GHY}^{{\rm BH}}}} r_0 {{\cal R}_{D5/\overline{D5}}^{\rm BH}} \sqrt[4]{\frac{\log
   \left(\frac{{{\cal R}_{\rm UV}}}{{{\cal R}_{D5/\overline{D5}}^{\rm th}}}\right)}{\log
   \left(\frac{{{\cal R}_{\rm UV}}}{{{\cal R}_{D5/\overline{D5}}^{\rm BH}}}\right)}}}{{\cal R}_{D5/\overline{D5}}^{\rm th}}. 
\end{eqnarray}
With reference to the numerics: $\lambda_{\rm GHY}^{{\rm th}}=0.08$, $\lambda_{\rm GHY}^{{\rm BH}}=0.23$; Since, ${{\cal R}_{D5/\overline{D5}}^{\rm th/BH}}=\sqrt{3}\left(\frac{1}{\sqrt{3}}+\epsilon^{\rm th/BH} \right)e^{-\kappa_{(h/0)} N^{-1/3}}$, where $\kappa_{(h/0)}\ll 1$ \cite{Bulk-Viscosity-McGill-IIT-Roorkee}, ${\cal R}_{\rm UV} \approx L=(4 \pi g_s N)^{1/4}$ \cite{MChPT} and assuming $g_s=0.1$, $N=100$, $\kappa_{r_{h}}=0.01$, $\kappa_{r_{0}}=0.01$, $\epsilon^{\rm BH}=0.36$ and $\epsilon^{\rm th}=0.1$, the equation  (\ref{rh-r0-rotation-beta0}) can be simplified to:
\begin{eqnarray}
\label{rh-r0-simp}
& & r_h \approx 1.16 r_0.
\end{eqnarray}
In the context of gauge-gravity duality, the Hawking Page phase transition, which occurs at $T=T_H(\gamma)$ on the gravity side  between a thermal and black hole background is dual to the deconfinement phase transition, which occurs at $T=T_c(\gamma)$ in gauge theories , i.e., $T_H(\gamma) \sim T_c(\gamma)$ \cite{Witten-Hawking-Page-Tc},
\begin{eqnarray}
\label{Tc-rotation}
& & \hskip -0.5in
T_c(\gamma)=\frac{1}{\gamma}\Biggl( \frac{r_h}{\pi L^2}\Biggr)=\frac{1}{\gamma}\Biggl( \frac{1.16 r_0}{\pi \sqrt{4 \pi g_s N}}\Biggr)=T_c(0)\sqrt{1-l^2 \omega^2}. 
\end{eqnarray}
Taking $\frac{r_0}{\sqrt{4 \pi g_s N}} = \frac{1700}{4}$MeV \cite{Misra+Gale}, we get $T_c(0) \approx 157$ MeV, which is roughly the value found through holographic research \cite{Tc-HW_SW} and lattice calculations \cite{Tc-L}. The ratio of the deconfinement temperature of QCD in the presence and absence of rotation versus the angular velocity of the rotating quark-gluon plasma for $l=0.2,0.6,1 \ GeV^{-1 }$ is plotted in Fig. \ref{Tc-Plot} which is justified by the equation (\ref{Tc-rotation}).\\
{\bf Comparison with Other Models}: The authors found identical behavior from Nambu-Jona-Lansinio model in \cite{R-Tc3} because of chiral condensate suppression, which was investigated as well from lattice simulations in \cite{Lattice-1,Lattice-2}. Lattice simulations with different kinds of boundary conditions (open, periodic, and Dirichlet) have been carried out in rotating reference frames. In both gluodynamics and the theory of dynamical fermions, the authors investigated the impact of rotation on the critical temperature. The result found that the decrease in deconfinement temperature of thermal QCD from a rise of rotation of quark-gluon plasma was previously studied using gauge-gravity duality from a bottom-up approach in \cite{R-Tc,G-D-Rotation,Kerr-AdS-1,Kerr-AdS-2}. In contrast to our finding and other holographic findings mentioned earlier, the critical temperature for gludynamics exhibits the opposite behavior. However, the authors found that the critical temperature for dynamical fermions decreases as rotation increases. The hard-wall and soft-wall models were used for the bottom-up holographic study in \cite{R-Tc}, and the Einstein-Maxwell-Dilaton system is involved in the gravity dual of \cite{G-D-Rotation}. The authors of \cite{Kerr-AdS-1,Kerr-AdS-2} considered the gravity dual of rotating QGP to be the Kerr-AdS black hole background in five dimensions. The deconfinement temperature of rotating QGP from a top-down model is not covered in any published article. Because top-down models compactify the initial type IIB/IIA string theory or ${\cal M}$-theory on an internal manifold to produce four-dimensional QCD at finite temperature, they are more fundamental than bottom-up models. We successfully studied rotating QGP from a top-down holographic dual for the first time and were able to confirm the non-zero angular velocity of rotating QGP found in hydrodynamic simulation \cite{omega-simulation} and STAR collaboration \cite{RHIC-omega}.
%%%%%%%%%%%%%%%%%%%%%
%%%%%%%%%%%%%%
\subsection{UV-IR Mixing, Flavor Memory Effect and Non-Renormalization of $T_c$ in Rotating QGP}
\label{UV-IR-mixing}
Focusing on the outcomes presented in the appendix \ref{beta}. At ${\cal O}(\beta)$, the UV finite on-shell action densities for the rotating cylindrical black hole and thermal backgrounds are as follows:
{\footnotesize
\begin{eqnarray}
\label{bh-th-action-beta}
& & \left(1+\frac{r_h^4}{2{\cal R}_{\rm UV}^4}\right)\frac{S_{D=11,\ {\rm on-shell \ UV-finite}}^{\beta, \rm BH}}{{\cal V}_4} = \Biggl[{-2 {\cal C}_{\theta_1x} \kappa_{\sqrt{G^{(1)}}R^{(0)}}^{\rm IR} } +\frac{20  \left(-{\cal C}_{zz}^{\rm bh} + 2 {\cal C}_{\theta_1z}^{\rm bh} - 3 {\cal C}_{\theta_1x}^{\rm bh}\right) \kappa_{\rm EH}^{\beta,\ \rm IR} }{11 }\Biggr]\nonumber\\
& & \times\frac{ b^2   {g_s}^{3/2}  M {N_f}^3 {r_h}^4 \log ^3(N) \log
   \left(\frac{{r_h}}{{{\cal R}_{D5/\overline{D5}}}}\right) \log
   \left(1 - \frac{{r_h}}{{{\cal R}_{D5/\overline{D5}}}}\right)}{ N^{1/2} {{\cal R}_{D5/\overline{D5}}}^4}\beta,\nonumber\\
   & & \frac{S_{D=11,\ {\rm on-shell \ UV-finite}}^{\beta,\rm thermal}}{{\cal V}_4} =
 -\frac{20  \kappa_{\rm EH, th}^{{\rm IR},\ \beta} {r_0}^3 N^{1/2}{f_{x^{10}x^{10}}}({r_0})}{11{g_s}^{3/2} M  {N_f}^{5/3} {{\cal R}_{D5/\overline{D5}}^{\rm th}}^3  \log ^{\frac{2}{3}}(N) \log
   \left(\frac{{r_0}}{{\cal R}_{D5/\overline{D5}}^{\rm th}}\right)}\beta.
\end{eqnarray}
}
In a manner analogous to that of the ${\cal O}(\beta^0)$ scenario (\ref{SBH=Sth}), comparing on-shell action density results of ${\cal O}(\beta)$ terms occurring in (\ref{bh-th-action-beta}) at the ultraviolet (UV) cut-off, i.e., 
\begin{equation}
\label{SBH=Sth-i}
\left(1+\frac{r_h^4}{2{\cal R}_{\rm UV}^4}\right)\frac{S_{D=11,\ {\rm on-shell\ UV-finite}}^{\beta, \rm BH}}{{\cal V}_4}=\frac{S_{D=11,\ {\rm on-shell\ UV-finite}}^{\beta, \rm thermal}}{{\cal V}_4}. 
\end{equation}
We got the following results using the equations (\ref{bh-th-action-beta}) and (\ref{SBH=Sth-i}):
{\footnotesize
\begin{eqnarray}
\label{UV-IR mixing-rotation}
& & \hskip -0.5in f_{x^{10}x^{10}}({r_0})= -\frac{b^2 {g_s}^3 M^2 {N_f}^{14/3}  \left(\frac{r_h}{{\cal R}_{D5/\overline{D5}}^{\rm bh}}\right)^4  \log ^{{3}}(N) \log
   \left(\frac{{r_0}}{{\cal R}_{D5/\overline{D5}}^{\rm th}}\right) \log \left(\frac{{r_h}}{{\cal R}_{D5/\overline{D5}}^{\rm bh}}\right) \log
   \left(1-\frac{{r_h}}{{\cal R}_{D5/\overline{D5}}^{\rm bh}}\right)}{  {\kappa_{\rm EH, th}^{\beta,\ \rm IR}} {N} \left(\frac{r_0}{{\cal R}_{D5/\overline{D5}}^{\rm th}}\right)^3}\nonumber\\
& & \hskip 0.3in \times \left(-11  {\cal C}_{\theta_1x} {\kappa_{\sqrt{G^{(1)}}R^{(0)}}^{\rm IR}}
    \log ^3(N)-10 {\kappa_{\rm EH,bh}^{\beta,\ \rm IR}} \left(-{\cal C}_{zz}^{\rm bh} + 2 {\cal C}_{\theta_1z}^{\rm bh} - 3 {\cal C}_{\theta_1x}^{\rm bh}\right)\right).
\end{eqnarray}
}
The ``UV-IR'' mixing that is described in the equation (\ref{UV-IR mixing-rotation}) is almost identical to the UV-IR mixing that is described in \cite{McTEQ} and discussed earlier in this chapter. The combination of  the constants of integration such as $\left(-{\cal C}_{zz}^{\rm bh} + 2 {\cal C}_{\theta_1z}^{\rm bh} - 3 {\cal C}_{\theta_1x}^{\rm bh}\right)$ possesses the memory of flavor $D7$-branes of type IIB string dual referred to as the ``Flavor Memory'' effect, and the identical effect also seen in ${\cal O}(\beta)$ correction of the Hawking temperature (\ref{Hawking-temp}). Because the previously mentioned group of constants of integration is zero, there is no ${\cal O}(\beta)$ correction to the Hawking temperature (\ref{Hawking-temp-i}), and there is also no ${\cal O}(\beta)$ correction to the deconfinement temperature $(T_c)$ due to gauge-gravity duality. This suggests that $T_c$ does not undergo renormalization at ${\cal O}(\beta)$. The holographic renormalization of this part is discussed in appendix \ref{HRN-R-Tc}.

\section{Summary}
\label{Tc-Summary}
We obtained the following results in this chapter.
\begin{itemize}
\item
{\bf UV-IR Mixing and Flavor Memory}: During the computation of $T_c$ from the semiclassical method \cite{Witten-Hawking-Page-Tc}, we obtained ``UV-IR'' mixing described as the connection between the ${\cal O}(\beta)$ corrections to the ${\cal M}$-theory metric along the ${\cal M}$-theory circle in thermal background and the ${\cal O}(\beta)$ correction to a specific combination of the  ${\cal M}$-theory metric components along the compact part of the four-cycle ``wrapped'' by the flavor $D7$-branes of the parent type IIB (warped resolved deformed) conifold geometry - the latter referred to as ``Flavor Memory'' in the ${\cal M}$-theory uplift on equating the ${\cal O}(\beta)$ portion of the on-shell actions of the black hole and thermal backgrounds in the absence and presence of rotation both.

\item
{\bf Non-Renormalization of $T_c$}: 
\begin{itemize}
%%%%%%%%%%%%%%%%
\item
{\bf Semiclassical computation}: We have shown that the leading order result for $T_c$ is still valid after incorporating  the cal ${\cal O}(R^4)$ terms in supergravity action. In the large-$N$ limit, the $t_8t_8R^4$ terms dominate the ${\cal O}(R^4)$ contributions (together with the sub-dominant term, $\epsilon_{11}\epsilon_{11}R^4$), which, from a type IIB viewpoint in the zero-instanton sector, translate to a tree-level contribution at ${\cal O}\left((\alpha^\prime)^3\right)$ and one-loop contribution to the amplitude of four-graviton scattering which is obtained from integration of fermionic zero modes. The $SL(2,\mathbb{Z})$ completeness of such $R^4$ terms \cite{Green and Gutperle}, indicates that they are not perturbatively renormalized beyond one loop of the zero-instanton sector, indicates that they are not renormalized from a type IIB point of view. This shows that $T_c$ is not renormalized at all the loops in ${\cal M}$-theory at ${\cal O}(R^4)$.

\item
{\bf $T_c$ from Entanglement entropy}: We computed the entanglement entropy between two parts by dividing one of the spatial coordinates of the thermal ${\cal M}$-theory background into a segment of length $\it l$ and its complement. This generalization of \cite{Tc-EE} to ${\cal M}$-theory was made possible by an obvious extension of \cite{Tc-EE} to ${\cal M}$-theory. In the same vein as \cite{Tc-EE}, there are two different RT surfaces: connected and unconnected. If one is located below the critical value $it l_crit$, then the connected surface is the one that dominates the entanglement entropy, and if one is located above the critical value ${\it l_{crit}}$, then the disconnected surface is the one that dominates the entanglement entropy; In large $N_c$ gauge theories, the ${\it l}<{\it l_{crit}}$ region belongs to the confining phase, whereas the ${\it l}>{\it l_{crit}}$ region belongs to the deconfining phase of the same theory. In high $N_c$ gauge theories, this phenomenon is regarded as a phase transition between confinement and deconfinement phases.\par
As a result of a precise and delicate cancellation between the ${\cal O}(\beta)$ correction from a subset of the aforementioned metric components, it is remarkable to observe that, $T_c$ receives no ${\cal O}(\beta)$ correction which is consistent with the conjecture discussed above from semiclassical computation.

\item Hawking temperature of rotating cylinder black hole background is proportional to $\left(-{\cal C}_{zz}^{\rm BH} + 2 {\cal C}_{\theta_1z}^{\rm BH} - 3 {\cal C}_{\theta_1x}^{\rm BH}\right)$, as we have obtained in (\ref{Hawking-temp}). To make the correction to the three-form potential zero, it was proved in \cite{McTEQ} that ${\cal C}_{zz}^{\rm BH}=2 {\cal C}_{\theta_1 z}^{\rm BH}$ and $|{\cal C}_{\theta_1 x}^{\rm BH}| \ll 1$. This suggests that even in the presence of rotating Quark Gluon Plasma, there is no ${\cal O}(R^4)$ correction to the deconfinement temperature, and hence the ``Non-Renormalization of $T_c$'' in the rotating QGP as well. 
\end{itemize}
\item
{\bf Deriving ${\cal M}\chi$PT-Phenomenology compatibility}: As demonstrated in chapter {\bf 2}, in order to match the phenomenological value corresponding to the one-loop renormalized low energy constant associated with the ${\cal O}(p^4)$ $SU(3)\ \chi$PT Lagrangian term ``$\left(\nabla_\mu U^\dagger\nabla^\mu U\right)^2$'' with the value found from the type IIA string dual of thermal QCD-like theories including the  ${\cal O}(R^4)$ corrections, we need to impose a constraint, negative sign of the combination of integration constants mentioned above. $\mathbb{R}^2\times S^1(\frac{1}{M_{\rm KK}})$ is the thermal supergravity background counterpart of type IIB (solitonic) $D3$-branes at lower temperatures. In order to restore a boundary four-dimensional QCD-like theory after compactifying on the base of a $G_2$-structure cone, one must take the limit of$M_{\rm KK}\rightarrow0$, and in this limit we derived the values of the aforementioned integration constants explicitly and found that this combination has negative sign.

\item
{\bf Wald Entropy at ${\cal O}(R^4)$}: In the high-temperature ${\cal M}$-theory dual, matching of ${\cal O}(R^4)$ computation of Wald entropy with the black hole entropy puts a linear constraint on the same linear combination of the aforementioned integration constants.

\item {\bf Holographic Dual of Rotating QGP}: By applying Lorentz transformations (\ref{Lorentz-boost}) along $(t, \phi)$ coordinates on the gravity dual side, we are able to derive the rotating QGP on the gauge theory side via gauge-gravity duality. Therefore, for $T>T_c$ and $T<T_c$ on the gauge theory side (where $T_c$ serves as the deconfinement temperature of thermal QCD-like theories), the gravity dual includes a rotating cylinder black hole and thermal backgrounds.

\item {\bf $T_c$ in Rotating QGP}: We found that the deconfinement temperature of rotating QGP (\ref{Tc-rotation}) is inversely proportional to the Lorentz factor, $\gamma$, by equating the ${\cal O}(\beta^0)$ terms for the rotating cylindrical black hole and the cylindrical thermal backgrounds at the UV cut-off (\ref{SBH=Sth}). This means that when the angular velocity of rotating QGP increases, the deconfinement temperature of thermal QCD-like theories decreases and vice-versa. Experimental results from the noncentral Relativistic Heavy Ion Collider (RHIC) have shown that the Quark Gluon Plasma created in these collisions has an angular velocity, and that the deconfinement temperature decreases with increasing rotation \cite{RHIC-omega}. To ensure that the deconfinement temperature of rotating QGP computed using a top-down technique is in agreement with actual results, our study serves as a good check for the top-down duals of \cite{metrics} and \cite{HD-MQGP}.

\end{itemize}

%m

\chapter{Conclusion and Future Outlook }
\graphicspath{{Chapter4/}{Chapter4/}}

We explored the intermediate coupling regime of thermal QCD-like theories from a top-down model constructed in \cite{HD-MQGP}. The model is obtained as the extension of \cite{MQGP} by including higher derivative terms (${\cal O}(R^4)$) in the eleven dimensional supergravity action. We studied the following issues in the context of holographic thermal QCD at intermediate coupling.
\begin{itemize}
\item {\bf $SU(3)$ Chiral Perturbation Theory from IIA String Theory}:
We obtained the low energy coupling constants (LECs) appearing in the $SU(3)$ chiral perturbation theory Lagrangian in the chiral limit from the type IIA string dual inclusive of ${\cal O}(R^4)$ corrections \cite{MChPT}. We matched our results with the phenomenological values of the LECs available in the literature, and in this process we found that resolution parameter of the conifold geometry receives an ${\cal O}\left(\frac{l_p^3}{N}\right)$ correction. We also found the connection between higher derivative terms and $\frac{1}{N}$ correction.

\item {\bf Deconfinement Phase Transition in Thermal QCD-Like Theories at Intermediate Coupling}: We discussed the deconfinement phase transition in thermal QCD from the semi-classical method and entanglement entropy point of view. The semi-classical method is based on the Hawking Page phase transition between thermal and black hole background on the gravity dual side. The procedure is to obtain the on-shell action densities of thermal and black hole backgrounds and equates these two at the UV cut-off. We obtained the deconfinement temperature of thermal QCD by equating the leading order terms. We observed the ``UV-IR'' mixing, ``Flavor Memory'' effect, and ``non-renormalization of $T_c$''. Let us explain these effects in some detail.\\
{\bf UV-IR Mixing and Flavor Memory Effect}: When we equated the on-shell action densities at the UV cut-off, then we obtained a relationship between the integration constants appearing in the black hole background along the compact part of the non-compact four cycle, which is part of the flavor $D7$-branes worldvolume and higher derivative correction to the thermal background along the ${\cal M}$-theory circle. This is called ``UV-IR mixing'' in our setup. Since the aforementioned integration constants are associated with the flavor $D7$-branes of parent type IIB string dual and hence they have the information about the flavor branes in the ${\cal M}$-theory uplift, which has no branes. This is interpreted as the ``Flavor Memory'' effect in our setup.

{\bf Non-Renormalization of $T_c$}: The deconfinement phase transition in thermal QCD corresponds to phase transitions between the entanglement entropies of connected and disconnected surfaces in the language of gauge-gravity duality. We showed that the deconfinement phase transition occurs at the critical value of the length of the interval chosen to compute the entanglement entropy. In this process, we found that entanglement entropies of connected and disconnected surfaces do not receive the ${\cal O}(R^4)$ corrections, and hence $T_c$ does not receive the same. This is known as the non-renormalization of the deconfinement temperature of thermal QCD-like theories.

{\bf ${\cal M}\chi$PT-$T_c$ Compatibility}: While matching one of the one-loop renormalized LECs with the phenomenological data in \cite{MChPT}, we imposed the constraint on the combination of integration constants mentioned earlier but now for the thermal background. The constraint was that one would get the exact matching with phenomenological data provided the combination, as mentioned above, should have a negative sign. In this paper, we explicitly calculated the expressions of integration constants for the thermal background and showed that the constraint is satisfied, and hence we showed the compatibility between our chiral perturbation theory study and deconfinement temperature computation.

\item {\bf Deconfinement Phase Transition in Rotating QGP at Intermediate Coupling}: We constructed the holographic dual of rotating QGP by making one of the spatial coordinates in the ${\cal M}$-theory dual periodic and performing the Lorentz transformation along the temporal and the periodic coordinate mentioned earlier. The gravity dual will be a rotating cylindrical black hole and thermal backgrounds when $T>T_c$ and $T<T_c$ on the thermal QCD side. Since the cylindrical black hole has the angular velocity, this maps to the angular velocity of rotation of QGP in thermal QCD via gauge-gravity duality. We followed the semi-classical method to study the deconfinement phase transition in rotating QGP from a top-down model. We found that the deconfinement temperature of thermal QCD is inversely proportional to the Lorentz factor, implying that as QGP rotation increases, $T_c$ decreases and vice-versa. When we analyze the effect of higher derivative terms similar to \cite{McTEQ}, we found that one again observed ``UV-IR mixing'', ``Flavor Memory effect'', and ``non-renormalization of $T_c$'' even in the rotating QGP. In this paper, we showed the normalization of $T_c$ by calculating the higher derivative correction to the Hawking temperature too. Since, Hawking temperature maps to $T_c$ on the QCD side via Hawking-Page phase transition in the context of gauge-gravity duality. We found that Hawking temperature receives no higher derivative corrections, guaranteeing that $T_c$ is non-renormalized.

\end{itemize}

{\bf Future Physics outlook}: \begin{itemize}
 \item The LEC $H_1$ of (\ref{MChPT-Op4}) and, in general, the LECs of the $\chi$PT Lagrangian at ${\cal O}(p^6)$ \cite{Op6} may be calculated using the given values of the parameters of our ${\cal M}$-theory dual of thermal QCD-like theories.

\item No prior work has applied gauge-gravity duality approaches to the study of the perturbative regime of thermal QCD-like theories in order to explain, for example, the low-frequency peaks expected to appear in spectral functions related to transport coefficients, from ${\cal M}$-theory, in the context of intermediate 't Hooft coupling top-down holography. We would like to explore more application of \cite{HD-MQGP}, e.g., we can obtain the spectral/correlation functions using the techniques available from the gauge-gravity duality with the gravity dual as ${\cal M}$ theory uplift of \cite{MQGP} including the ${\cal O}(R^4)$ corrections. This will allow us to get more insight about the experimental results of the QCD at Relativistic Heavy Ion Collider(RHIC) and test the QCD/${\cal M}$-theory duality used in {\bf part-I} of this thesis. In addition, the dependence on temperature of the speed of sound, attenuation constant, and the bulk viscosity can all be obtained from its solution, along with the ${\cal O}(l_p^6)$ and the non-conformal corrections to them from the same holographic dual. One could attempt to reproduce the well-known weak-coupling result from ${\cal M}$-theory, which states that the ratio of the bulk viscosity to the shear viscosity has a lower bound that is proportional to the square of the deviation of the square of the speed of sound from its conformal value (last reference of \cite{EPJC-2}). In a general sense, the dissipative quasi-normal modes of the spectral functions at low frequencies could be studied to analyze the occurrence of peaks at low frequencies in transport coefficients, providing connections between perturbative QCD results along with QCD plasma RHIC experiments.

\end{itemize}
\addcontentsline{toc}{chapter}{Part-II}
%Satyam Shivam Sundaram
%Jai Mata Di
%\textwidth 16.5 truein
%\textheight 18.5in
%\newpage
%\pagestyle{plain}

\newpage
%\pagestyle{plain}
%\begin{center}
%{{\huge {Top-Down Holographic Study of Thermal QCD-Like Theories %at Intermedihhhate Gauge/'t Hooft Coupling}}}
%\end{center}
%\newpage
%\thispagestyle{empty}
%\begin{flushright}
~\\
~\\
~\\
~\\
~\\
~\\
~\\
~\\
\begin{center}
\Huge{{\bf Part-II \\ (HD) Gravity Islands, and Multiverse}}~\\
\end{center}
~\\
%\end{flushright}
%This part of the thesis is based on the resolution of information paradox in higher derivative theories of gravity, black holes with the multiple horizons and description of the Multiverse. 
%\\
~\\

\fbox{\begin{minipage}{38em}{\textbf{``God does not play dice.'' - Albert Einstein \\
 ``God not only plays dice,
he sometimes throws the dice where they cannot be seen.'' - Stephen Hawking \\
 ``If God throws dice where they cannot be seen, they cannot affect us.'' - Don Page}}
\end{minipage}}\\ \\

\clearpage
%\addtocounter{page}{-1}

%\input{empty}
%satyamshivamsundaram
%JaiMataDi
\chapter{Introduction}
\graphicspath{{Chapter5/}{Chapter5/}}
In this chapter, we present the introduction of the materials needed to understand the information paradox and its resolution from holography. We start with the discussion on entanglement entropy in section \ref{HEE-intro}, we discuss the information paradox and the Page curve in section \ref{IP-Page-Curve-Intro} and finally we discuss the resolution of the information paradox in \ref{IP-Resolution-Holography} from the island proposal, doubly holographic setup and wedge holography in \ref{Island-intro}, \ref{DHS-introduction} and \ref{WH-intro} respectively.
\section{Holographic Entanglement Entropy: Ryu-Takayanagi and Dong's Proposals}
\label{HEE-intro}
{\bf Entanglement Entropy in Quantum Mechanics (QM)}: Let us first discuss the entanglement entropy in quantum mechanical system. Let us consider a system whose state is denoted by $|\psi \rangle$. The density matrix of the system is defined as:
\begin{equation}
\label{dm-qm}
\rho_{AB} = {|\psi \rangle}_{AB}  {\langle \psi|}_{AB}.
\end{equation}
Entanglement entropy is measured by the von-Neumann entropy. For this, first we have to partition the system into two subsystems $A$ and $B$. States in the subsystems $A$ and $B$ are denoted by ${|\psi \rangle}_A$ and ${|\psi \rangle}_B$ such that $|\psi \rangle={|\psi \rangle}_{AB}={|\psi \rangle}_A \otimes {|\psi \rangle}_B$. Reduced density matrix of the subsystem $A$ is obtained by tracing over the degrees of freedom of the subsystem $B$ and vice-versa.
\begin{equation}
\label{red-dm-A}
\rho_{A}={\rm Tr}_B\left(\rho_{AB}\right)={\rm Tr}_B\left({|\psi \rangle}_{AB}  {\langle \psi|}_{AB}\right) = {|\psi \rangle}_A  {\langle \psi|}_A.
\end{equation}
Now, the von-Neumann entropy is defined as:
\begin{eqnarray}
\label{vN-entropy-defn}
S_{\rm EE}=- {\rm Tr} \left(\rho_A {\rm ln}\rho_A \right).
\end{eqnarray}

{\bf Entanglement Entropy in Quantum Field Theory (QFT)}: It is not easy to compute the entanglement entropy in quantum field theories (QFTs) by factoring the system into subsystems because factorization is not always possible in QFTs. Entanglement entropy in QFT is calculated using the replica trick. First, let us define the Renyi entropy:
\begin{eqnarray}
\label{Renyi-entropy}
S_{A}^{(n)}=\frac{1}{1-n} \log\left({\rm Tr} \rho_A^n\right),
\end{eqnarray}
from the Renyi entropy, von-Neumann entropy is defined as: $S_{\rm EE}= {\rm Limit}_{n \rightarrow 1}S_{A}^{(n)}$. In QFTs, usually we calculate the Renyi entropy and then take the limit $n \rightarrow 1$. $\rho_A^n$ is obtained from taking the $n$ copies of the same subsystem $A$, tracing in (\ref{Renyi-entropy}) corresponds to the gluing of the $n$-sheeted surfaces where the subsystem $A$ is defined. It is found that: 
\begin{eqnarray}
{\rm Tr}\rho_A^n=\frac{Z_n(A)}{Z_1^n},
\end{eqnarray}
where $Z_n(A)$ is the partition function of the $n$-sheeted surface. Therefore Renyi entropy is defined as:
\begin{eqnarray}
\label{Renyi-entropy-QFT}
S_{A}^{(n)}=\frac{1}{1-n} \log\left(\frac{Z_n(A)}{Z_1^n}\right).
\end{eqnarray}
Hence, if we can obtain the von-Neumann entropy when $n\rightarrow 1$ in (\ref{Renyi-entropy-QFT}).

{\bf Holographic Entanglement Entropy (HEE)}: As we saw in the earlier discussion that once we know the partition function of the $n$-sheeted surface then we are able to compute the entanglement entropy in QFT but calculation of the partition function itself is a cumbersome process. The AdS/CFT correspondence make this job easier because of the Ryu-Takayanagi proposal to compute the entanglement entropy in dual CFT from the gravity side \cite{RT}. Suppose, we are dealing with $AdS_{d+1}/CFT_d$ duality and the bulk is denoted by ${\cal M}_{d+1}$, subsystem on the boundary CFT is denoted by $A$ with boundary $\partial A$ then the proposal states that
\begin{itemize}
\item We need to find out a co-dimension two surface $(\epsilon_A)$ in the bulk ${\cal M}_{d+1}$ which is anchored on $\partial A$.

\item There is possibility of many surfaces but we have to consider the one which satisfy the homology constraint, i.e., $\epsilon_A$ is smoothly retractable to the boundary region $A$. 

\item Out of those surfaces which satisfy homology constraint, we need to pick the one with the minimal area then the entanglement entropy is defined as:
\begin{eqnarray}
\label{HEE-RT-defn}
S_A=\frac{{\rm min} \left({\rm Area}\epsilon_A\right)}{4 G_N^{(d+1)}}.
\end{eqnarray}
\end{itemize}

The Ryu-Takayangi formula has certain limitation, it is applicable to the the time-independent backgrounds. For the time dependent background, one is required to use the HRT formula \cite{HRT} where HRT stands for Hubney, Rangamani and Takayanagi. Quantum corrections to all order in $\hbar$ to the Ryu-Takayanagi formula was incorporated in \cite{EW} where one is required to extremize the generalised entropy. Surfaces which extremize the generalised entropy are known as quantum extremal surfaces(QES). If there are more than one quantum extremal surfaces then we need to consider the one with minimal area. In \cite{AMMZ}, authors generalised the QES prescription to island surfaces where we are required to extremize the generalised entropy like functional which includes contribution from the island surfaces. In this case, extremal surfaces are known as quantum extremal islands. Since, in this thesis, we are confining ourselves to the time independent backgrounds and therefore we will not discuss the HRT formula.

{\bf Dong's Proposal to Compute HEE in Higher Derivative Theories of Gravity}: One can obtain the HEE in in general higher derivative theories of gravity by using the formula given by X. Dong in \cite{Dong}. In the context of $AdS_{d+1}/CFT_{d}$ correspondence the formula is written below:
\begin{equation}
\label{HD-Entropy-defn}
S_{EE}=\int d^{d-1}y \sqrt{-g}\Biggl[\frac{\partial {\cal L}}{\partial R_{z\bar{z}z\bar{z}}}+\sum_{\alpha}\left(\frac{\partial^2 {\cal L}}{\partial R_{zizj}\partial R_{\bar{z}m\bar{z}l}}\right)_{\alpha} \frac{8 K_{zij}K_{\bar{z}ml}}{(q_{\alpha}+1)}\Biggr],
\end{equation}
where $(a,b)$ and $(i,j,k,l)$ denote the normal and tangential directions, $g$ is the induced metric's determinant on the co-dimension two surface in the bulk, $K_{zij}=\frac{1}{2}\partial_z G_{ij}$ with trace $K_z=K_{zij}G^{ij}$. (\ref{HD-Entropy-defn}) consists of two terms: Wald entropy like term and the second term can be calculated by following these steps:
\begin{itemize}
\item Let us label the each term as $\alpha$th term after getting the final expression of the second term which is obtained from the differentiation of Lagrangian with respect to Riemann tensor twice.

\item We need to perform the following transformations on certain components of the Riemann tensors:
\begin{eqnarray}
\label{riemann-transformations-intro}
R_{abij}=r_{abij}+ g^{ml}\left(K_{ajm}K_{bil}-K_{aim}K_{bjl}\right),\nonumber\\
& & \hskip -3.2in R_{aibj}=r_{aibj}+g^{ml} K_{ajm} K_{bil}- Q_{abij},\nonumber\\
& & \hskip -3.2in R_{ijml}=r_{ijml}+g^{ab}\left(K_{ail}K_{bjm}-K_{aij}K_{bml} \right),
\end{eqnarray}
where $Q_{abij}=\partial_a K_{bij}$.
\item In the $\alpha$th term, the number of $Q_{aaij}$ and $Q_{bbij}$ be denoted by $\gamma$ and the number of $K_{aij}$, $R_{abci}$ and $R_{aijk}$ will be denoted $\delta$. Then $q_\alpha$ corresponding to the $\alpha$th term is defined as: 
\begin{equation}
\label{q-alpha}
q_{\alpha}=\gamma + \frac{\delta}{2}.
\end{equation}
\item From (\ref{riemann-transformations-intro}), we can obtain $r_{abij},r_{aibj},r_{ijml}$ in terms of $R_{abij},R_{aibj},R_{ijml}$, substitutions of the same in (\ref{HD-Entropy-defn}) results the entanglement entropy in terms of original Riemann tensors i.e. in terms of $R_{abij},R_{aibj}$ etc. 
\end{itemize}
The reason for discussing these proposals is that when we compute the Page curve of black holes in doubly holographic setup and wedge holography then these proposals will be useful. The holographic entanglement entropy has also been computed from the holographic stress tensor and surface terms in \cite{A-HEE} and \cite{A-HEE-1} respectively.

\section{Hawking's Information Paradox and Page Curve}
\label{IP-Page-Curve-Intro}
Hawking's black hole information paradox is a long-time puzzle that started with his papers \cite{Hawking,Hawking1}. When matter collapses to form a black hole, the whole matter is stored in the singularity. The horizon of the black hole covers the black hole singularity. Initially, the system is in a pure state. 
Hawking studied the creation of particles in pairs with negative and positive energy in the presence of quantum effects, and he found that a particle with negative energy gets trapped inside the black hole, whereas the particle with positive energy scattered off to infinity is what we receive in Hawking radiation. We can get the radiation from the black hole due to quantum mechanics, which allows the possibility of quantum tunneling through a potential barrier. In the case of a black hole, the horizon acts as the potential barrier. Hawking calculated the spectrum of the particles coming out of the black hole and found that the spectrum behaves as the spectrum of thermal radiation with a temperature known as Hawking temperature, which implies a mixed state. This means that the black hole evolves from the pure state to the mixed state, and hence unitary evolution of quantum mechanics is not preserved. This leads to the famous ``information paradox''.

Page suggested that when we include the quantum effects, then the black hole must follow the unitary evolution \cite{Page}. If we consider the black hole and radiation region as a single system, then one should get the Page curve to resolve the paradox. For the evaporating black hole, entanglement entropy of the Hawking radiation first increases linearly with time up to the Page time and then falls back to zero \cite{Page}. We are interested in eternal black holes, and for these black hoes, instead of falling to zero of the entanglement entropy, one obtains the constant entanglement entropy after the Page time, and this constant value is equal to the twice the thermal entropy of the black holes.

In this part of the thesis, we focus on getting the Page curve of eternal black holes using the recent proposals given in the literature, e.g., island proposal, doubly holographic setup, and wedge holography. Apart from getting the Page curve, we have also obtained other exciting results, which are discussed in the upcoming chapters.

\section{Resolution of Information Paradox from Holography}
\label{IP-Resolution-Holography}
Following three proposals are available in the literature which started with the idea of holography to resolve the black hole information paradox.
\subsection{Island Proposal and its Extension to HD Gravity}
\label{Island-intro}
Authors in \cite{AMMZ} proposed a method to resolve the information paradox which is equivalent to getting the Page curve. Idea is that at early times we get only contribution from the radiation region which gives the divergent part of entanglement entropy at late times because the entanglement entropy of Hawking radiation turns out to be proportional to the time. According to \cite{AMMZ} at early times situation remains the same whereas at late times interior of the black holes becomes part of the entanglement wedge and hence at late times entanglement entropy receives the contributions from the radiation as well as interior of the black holes. The part of the interior the black holes which contributes to the entanglement entropy is known as ``island''. \par
Island rule was proposed from a setup where we couple the evaporating JT(Jackiw Teitelboim) black hole plus conformal matter on the Planck brane with the two-dimensional CFT bath \cite{AMMZ}. The black hole is contained on the Planck brane, and the Hawking radiation is collected in the 2D conformal bath. This setup has the following three descriptions.
\begin{itemize}
\item {\bf 2D-Gravity:} The Planck brane is coupled to the external CFT bath, which acts as the sink for the Hawking radiation.

\item {\bf 3D-Gravity:} Two-dimensional conformal field theory has the three-dimensional gravity dual with metric $AdS_3$ via AdS/CFT correspondence.

\item {\bf QM:} The boundary of the external CFT bath is one dimensional where quantum mechanics (QM) is present.
\end{itemize} 
The island formula was derived from the gravitational path integral using the replica trick for special JT black holes in \cite{rw-1,rw-2}. The authors obtained the Page curve from the disconnected and connected saddles. One obtains the linear time dependence in the Page curve from the disconnected saddles, whereas connected saddles produce the finite part of the Page curve. The discussion of \cite{rw-1} holds for the replica wormholes with $n$ boundary as well. The generalised entropy in the presence of island surface is written as follows:
\begin{eqnarray}
\label{Island-proposal}
S_{\rm gen}(r)={\rm Min}_{\cal I} \Biggl[ {\rm Ext}_{\cal I}\Biggl(\frac{Area(\partial {\cal I})}{4 G_N} + S_{\rm matter}({\cal R} \cup {\cal I})\Biggr)\Biggr],
\end{eqnarray}
where ${\cal R}$, $G_N$ and ${\cal I}$ are representing the radiation region, Newton constant and the island surface. Equation (\ref{Island-proposal}) contains two terms: island surface's area and the matter contribution from the radiation and island regions. From (\ref{Island-proposal}), we can easily see that when island surface is absent then $S_{\rm gen}(r)= S_{\rm matter}({\cal R})$. It has been shown in the literature that island surface emerges at late times and hence initially one obtains the linear time dependence in the Page curve and at late times, when island surface's contribution dominates then one obtains the fall of the entanglement entropy for the evaporating black holes where as constant part (twice of their thermal entropies) for the eternal black holes. Hence, when we include these contributions, we obtain the Page curve. If there are more than one island surfaces then we have to consider the one with the minimal area. We have followed this proposal to obtain the Page curve of Schwarzschild de-Sitter black hole in \cite{Gopal+Nitin} and discussed in detail in the chapter {\bf 8} of this thesis. See \cite{Swansea,Swansea1,Swansea5} for the application of island proposal in the context of JT gravity and other issues \cite{Swansea2,Swansea3,Swansea4}.\par
Island proposal was extended for higher derivative gravity in \cite{NBH-HD}. The proposal is exactly similar to \cite{AMMZ} but we have to replace the first term of (\ref{Island-proposal}) by the term which can give the information about the entanglement entropy of higher derivative gravity and the formula for the same was proposed by X. Dong in \cite{Dong} and hence the island proposal in the presence of higher derivative terms in the gravitational action is written as \cite{NBH-HD}
\begin{eqnarray}
\label{S-total-general}
S_{\rm total} = S_{\rm gravity}+ S_{\rm matter},
\end{eqnarray}
where $S_{\rm matter}$ is the same as the $S_{\rm matter}({\cal R} \cup {\cal I})$ of (\ref{Island-proposal}) and $S_{\rm gravity}$ will be calculated using the Dong's formula \cite{Dong}.
For the $AdS_{d+1}/CFT_d$ correspondence the Dong's formulas is given below\footnote{We have already written the formula in (\ref{DHS-intro}), here we are writing the covariant form of (\ref{DHS-intro}). In this formula, $a$ and $i,j$ represents the tangential and normal directions.}.
\begin{eqnarray}
\label{EE-HD-Formula-intro}
& &
\hskip -0.3in S_{\rm gravity}= \frac{1}{4 G_N} \int d^{d-1} y \sqrt{h} \Biggl[ -\frac{\partial {\cal L} }{\partial R_{{\mu_1} {\nu_1} {\rho_1} {\sigma_1}}} \epsilon_{{\mu_1} {\nu_1}} \epsilon_{{\rho_1} {\sigma_1}}+\sum_\alpha \left(\frac{\partial^2{\cal L} }{\partial R_{{\mu_3} {\nu_3} {\rho_3} {\sigma_3}} \partial R_{{\mu_1} {\nu_1} {\rho_1} {\sigma_1}}}\right)_\alpha \nonumber\\
& &
 \frac{2 K_{ {\lambda_3} {\rho_3} {\sigma_3}}K_{{\lambda_1} {\rho_1} {\sigma_1}}}{q_{\alpha}+1}  [\left(n_{{\mu_3}{\mu_1}}n_{{\nu_3}{\nu_1}} -\epsilon_{{\mu_3}{\mu_1}}\epsilon_{{\nu_3}{\nu_1}} \right)n^{{\lambda_3}{\lambda_1}} +\left(n_{{\mu_3}{\mu_1}}\epsilon_{{\nu_3}{\nu_1}} +\epsilon_{{\mu_3}{\mu_1}}n_{{\nu_3}{\nu_1}} \right)\epsilon^{{\lambda_3}{\lambda_1}}]\Biggr],
\end{eqnarray}
where
\begin{eqnarray}
& &
n_{\mu \nu}=n_\mu^{(i)}n_\nu^{(j)}g_{ij}, \nonumber\\
& & \epsilon_{\mu \nu}=n_\mu^{(i)}n_\nu^{(i)}\epsilon_{ij},\nonumber\\
& & \epsilon_{\mu \nu}\epsilon_{\rho \sigma}=n_{\mu \rho}n_{\nu \sigma}-n_{\mu \sigma}n_{\nu \rho}, \nonumber\\
& & K_{\lambda \mu \nu}=n_\lambda^{(i)}m_\mu^{(a)}n_\nu^{(b)} K_{iab},
\end{eqnarray}
where $m_\mu^{(a)}$ and $n_\mu^{(i)}$ are the unit normal vectors along the tangential ($y^a$) and normal directions, , $h$ is the determinant of the induced metric on co-dimension two surface in the bulk. $q_\alpha$ can be obtained similar to (\ref{q-alpha}).

\subsection{Doubly Holographic Setup}
\label{DHS-introduction}
\begin{figure}
\begin{center}
\includegraphics[width=0.80\textwidth]{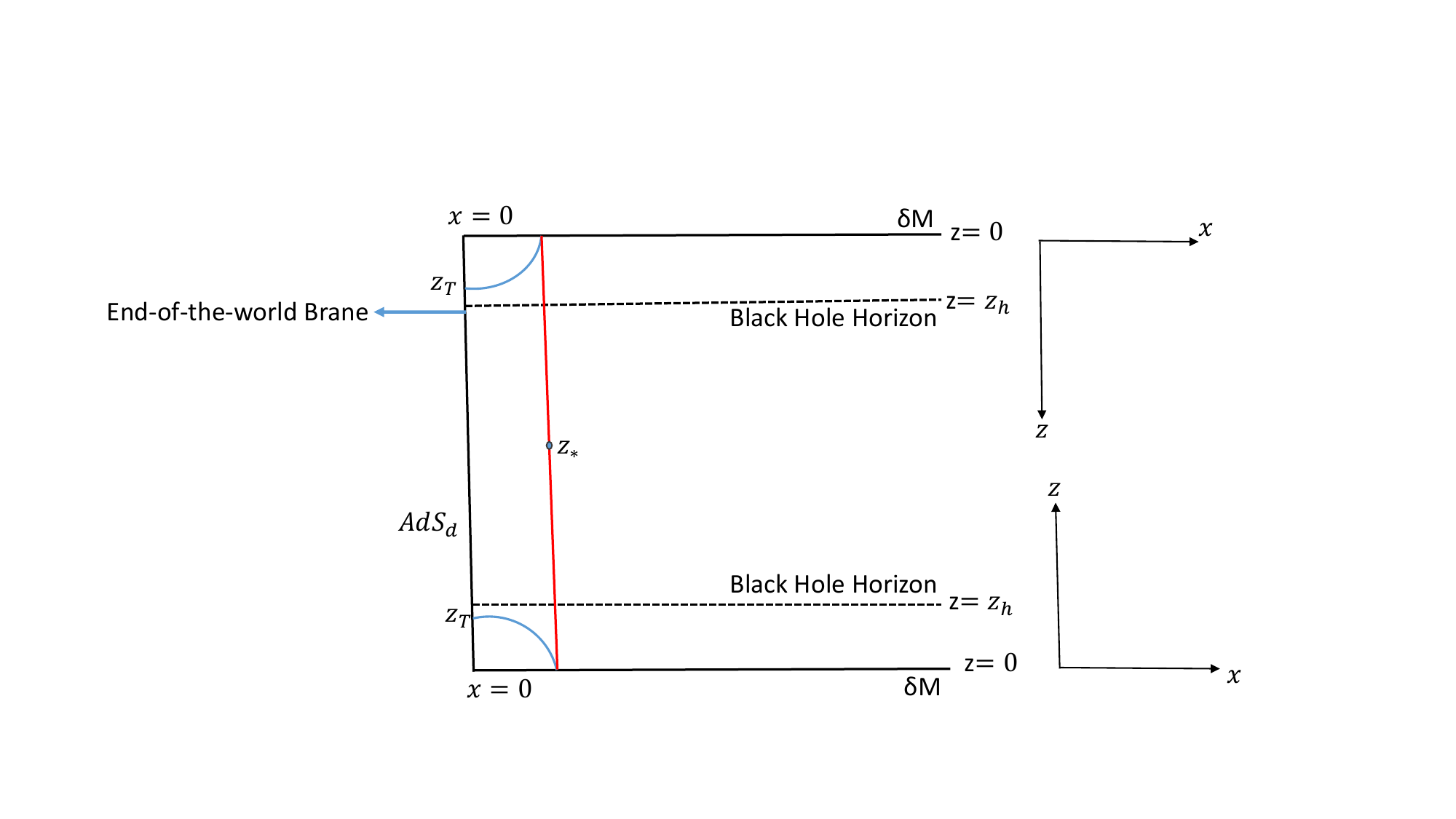}
\end{center}
\caption{Description of doubly holographic setup. Blue curves are the island surfaces and red curve is the Hartman-Maldacena surface. $\delta M$ is the conformal boundary, $z_*$ and $z_T$ are the turning points of Hartman-Maldacena and island surfaces.}
\label{DHS-intro}
\end{figure}

The doubly holographic setup is a nice setup to calculate the Page curve of black holes. As its name suggests, it is the double copy of the usual holography proposed by J. Maldacena. First, we need to take the bulk and truncate the geometry along one of the spatial coordinates \cite{KR1,KR2}. By doing so, one generates $d$-dimensional geometry embedded in the ($d+1$)-dimensional bulk. The $d$-dimensional geometry is known as end-of-the-world brane or Karch-Randall brane in the literature, and this holography is called ``braneworld holography''. The doubly holographic setup is obtained by joining the two copies of the Karch-Randall model. The setup consists of an eternal black hole living on the brane and two baths where we can collect the Hawking radiation. These two baths behave as thermofield double states because these are like two copies of the boundary conformal field theory (BCFT). Let us discuss the double holography in the context of $AdS_{d+1}/BCFT_d$ correspondence using a bottom-up approach, and the setup is shown in figure \ref{DHS-intro}.

The doubly holographic setup has the three descriptions summarised below.
\begin{itemize}
\item {\bf Boundary description}: $d$-dimensional BCFT at the conformal boundary of the bulk $AdS_{d+1}$. The boundary of ${BCFT}_d$ is the ($d-1$) dimensional defect.

\item {\bf Intermediate description}: Gravity on the end-of-the-world brane is coupled to BCFT via transparent boundary condition at the defect.

\item {\bf Bulk description}: The holographic dual of ${BCFT}_d$ is $AdS_{d+1}$ spacetime.
\end{itemize}

The intermediate description is very crucial to resolving the information paradox. Because in this description black hole living on the end-of-the-world brane directly couples with the external CFT bath. Define $S({\cal R})$ as the von Neumann entropy of the subregion ${\cal R}$ on a constant time slice in description {\bf 1}. One can obtain the $S({\cal R})$ in second description from the island rule \cite{AMMZ}:
\begin{equation}
S({\cal R}) = {\rm min}_{\cal I}~{\rm ext}_{\cal I}~S_{gen}({\cal R} \cup {\cal I}),
\end{equation}
where generalised entropy functional $(S_{gen}({\cal R} \cup {\cal I}))$ is \cite{EW}:
\begin{equation}
S_{gen}({\cal R} \cup {\cal I})= \frac{A(\partial {\cal I})}{4 G_N}+ S_{matter}({\cal R} \cup {\cal I}),
\end{equation}
 A doubly holographic setup is advantageous in the sense that we can obtain $S({\cal R})$ very easily using the classical Ryu-Takayanagi formula \cite{RT}. When bulk is $(d+1)$ dimensional then \cite{RT}:
\begin{equation}
\label{DHS-RT}
S_{gen}({\cal R} \cup {\cal I})= \frac{A(\gamma)}{4 G_N^{(d+1)}},
\end{equation}
where $\gamma$ is the minimal co-dimension of two surface in bulk. \par

In figure \ref{DHS-intro}, there are two BCFTs on the conformal boundary of the bulk. The vertical line is the end-of-the-world brane which contains the black hole. The CFT bath collects the Hawking radiation emitted by the black hole. This setup has two possible extremal surfaces: Hartman-Maldacena \cite{Hartman-Maldacena} and island surfaces. The Hartman-Maldacena surface connects the two BCFTs; it starts at the CFT bath, crosses the horizons, reaches up to the turning point, and then meets the thermofield double partner of BCFT. The entanglement entropy is divergent at late times for the Hartman-Maldacena surface, which implies Hawking's information paradox. The island surface starts at the external CFT bath and lands on the end-of-the-world brane. The island surface's entanglement entropy turns out to be a constant value (twice of thermal entropy of the black hole). Therefore, one recovers the Page curve by combining the contributions of the entanglement entropies of both these extremal surfaces. See \cite{PBD,B-N-P-1,B-N-P-2,NGB,Ling+Liu+Xian,Omiya+Wei,Phase-BCFT,critical-islands,IITK1,IITK2,Liu et al,Li+Yang,Deng+An+Zhou,Island-IIB-1,Island-IIB-2,Island-IIB-3,Geng+Nomura+Sun,JIITK} for the extensive literature on the doubly holographic setup. \par

Some authors found that gravity is massive on the end-of-the-world brane \cite{massive-gravity,GB-2,GB-3,GB-4} when we couple the brane to the external CFT bath. In some papers, it was shown by the authors that we could construct the doubly holographic setup with massless gravity on the brane \cite{Massless-Gravity,critical-islands,HD-Page Curve-2,C1}. We constructed the doubly holographic setup from a top-down approach in \cite{HD-Page Curve-2} and details are given in chapter {\bf 7}. We have a non-conformal bath ($QCD_{2+1}$) and the holographic dual is ${\cal M}$ theory inclusive of ${\cal O}(R^4)$ corrections \cite{HD-MQGP}. The reason for the existence of massless graviton in our setup is that we required the wave function of the graviton to be normalized, the second reason is due to the Dirichlet boundary condition on the wave function of the graviton, and the third reason is that end-of-the-world brane had non-zero tension and hence localization of graviton is possible on the brane in a ``volcano''-like potential. We obtained the Page curve with massless gravity in our setup, which was impossible in other doubly holographic setups without the DGP term. One alternate method to deduce the massless gravity on the brane is to include the Dvali-Gabadadze-Porrati (DGP) term \cite{DGP-2} on the brane \cite{Massless-Gravity}.

\subsection{Wedge Holography}
\label{WH-intro}
\begin{figure}
\begin{center}
\includegraphics[width=0.9\textwidth]{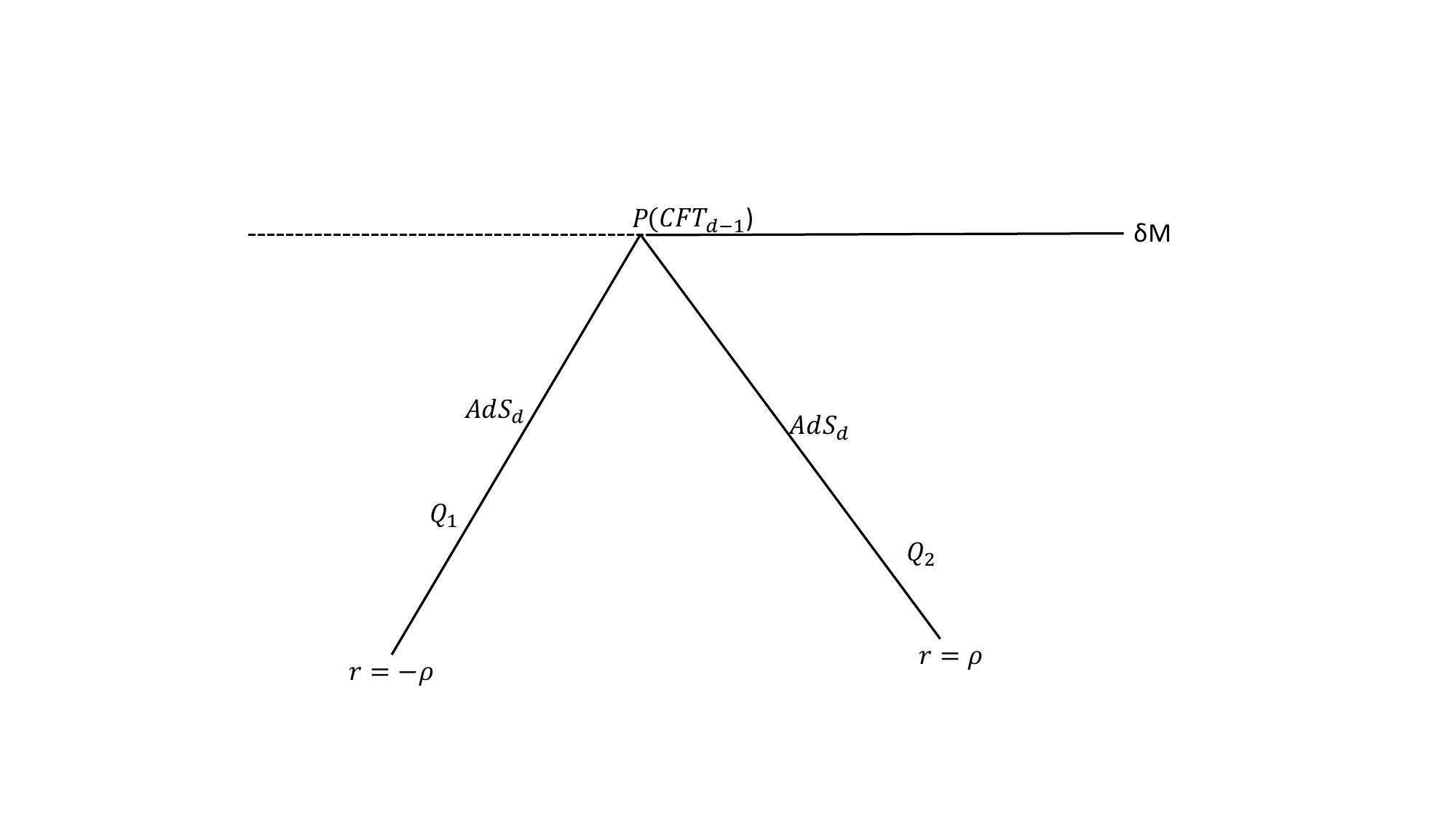}
\end{center}
\caption{Description of wedge holography. Two $d$-dimensional Karch-Randall branes joined at the $(d-1)$ dimensional defect, Karch-Randall branes are embedded in $(d+1)$-dimensional bulk. }
\label{WHS}
\end{figure}

In the doubly holographic setup, the external bath is a fixed CFT bath. In some of the papers, it was found that gravity is massive on the end-of-the-world brane and island prescription is not valid in the masssless gravity. Some of the authors considered the bath as gravitating too \cite{WH-1,WH-2,Wedge-JT,GB-3}. This setup is known as wedge holography in the literature. It was also argued that in wedge holography, Hartman-Maldacena surface does not exist and hence no Page curve in wedge holography. In \cite{Multiverse}, we showed that entanglement entropy of Hartman-Maldacena surface is non-zero for the AdS and Schwarzschild black hole and it is zero for the de-Sitter black hole. This implies that one could get the Page curve for the AdS and Schwarzschild black hole but not for the de-Sitter space using wedge holography. See Fig. \ref{WHS} for the pictorial description of the wedge holography.
One can get the Page curve or not in wedge holography is a debatable topic. Some progress in this direction has been made in \cite{Massless-Gravity}. It was shown by the author that we can get the Page curve with massless gravity localize on the Karch-Randall brane provided we have to include the DGP term on the Karch-Randall brane, see \cite{Massless-Gravity-1,Massless-Gravity-2} for the detailed analysis with examples.\par
Take into consideration the following action, \cite{WH-1,WH-2,Wedge-JT}, to describe the mathematical description of wedge holography:
\begin{eqnarray}
\label{bulk-action}
& & 
S=-\frac{1}{16 \pi G_N^{(d+1)}} \int d^{d+1}x \sqrt{-g_{\rm bulk}} \left(R[g_{\rm bulk}] - 2 \Lambda\right) \nonumber\\
& &-\frac{1}{8 \pi G_N^{(d+1)}} \int_{\alpha=1,2} d^dx \sqrt{-h_\alpha}\left({\cal K}_\alpha-T_\alpha\right),
\end{eqnarray}
where the first term serves as the Einstein-Hilbert term with the cosmological constant $\left(\Lambda=-\frac{d(d-1)}{2}\right)$ and the second term being the boundary terms on the Karch-Randall branes with tensions $T_{\alpha=1,2}$. The bulk action's Einstein equation (\ref{bulk-action}) is as follows:
\begin{equation}
\label{Einstein-equation}
R_{\mu \nu}-\frac{1}{2}g_{\mu \nu} R =\frac{d(d-1)}{2} g_{\mu \nu}.
\end{equation}
The above equation has the following solution \cite{WH-2}:
\begin{equation}
\label{metric-bulk}
ds_{(d+1)}^2=g_{\mu \nu} dx^\mu dx^\nu=dr^2+\cosh^2(r) h_{ij}^{\alpha} dy^i dy^j,
\end{equation}
where $h_{ij}^{\alpha}$ denote the induced metric on Karch-Randall branes. The Neumann boundary condition (NBC) obtained from the variation of the bulk action (\ref{bulk-action}) with respect to the metric on Karch-Randall branes $h_{ij}^{\alpha}$ is as follows:
\begin{equation}
\label{NBC}
{\cal K}_{ij}^{\alpha}-({\cal K}^{\alpha}-T^{\alpha})h_{ij}^{\alpha}=0.
\end{equation}
The wedge holography exist if the metric (\ref{metric-bulk}) is the solution of the Einstein equation (\ref{Einstein-equation}) and the bulk metric (\ref{metric-bulk}) satisfies the NBC (\ref{NBC}) on the branes, i.e., at $r=\pm \rho$. For $-\rho_1 \leq r \leq \rho_2$ with $\rho_1 \neq \rho_2$ \cite{WH-2}, the branes will have different tensions \cite{WH-2}. It is also required that the induced metric $h_{ij}^{\alpha}$ on the branes should be the solution of Einstein equation with a negative cosmological constant to have the anti de-Sitter branes in wedge holography.
\begin{eqnarray}
\label{Brane-Einstein-equation}
R_{ij}^{\alpha}-\frac{1}{2}h_{ij}^{\alpha} R[h_{ij}]^{\alpha} =\frac{(d-1)(d-2)}{2} h_{ij}^{\alpha},
\end{eqnarray}
Similar to the double holography, wedge holography has also the three descriptions:
\begin{itemize}
\item {\bf Boundary description}: ${BCFT}_d$ on the conformal boundary of the bulk $AdS_{d+1}$ with the $(d-1)$ dimensional defect.
\item {\bf Intermediate description}: two gravitating systems are connected with each other via the transparent boundary condition at the defect.
\item {\bf Bulk description}: the holographic dual of ${BCFT}_d$ is classical gravity $AdS_{d+1}$ spacetime.
\end{itemize}
The wedge holographic dictionary for the $(d+1)$- dimensional bulk is stated as: {\it holographic dual of the $(d-1)$-dimensional defect conformal field theory is the classical gravity in $(d+1)$- dimensions}. Therefore it is a co-dimension two holography.
Now let us understand the how this duality exists. 
\\ \\
\fbox{\begin{minipage}{38em}{\it 
Classical gravity in $(d+1)$-dimensional bulk\\ $\equiv$ (Quantum) gravity on two $d$-dimensional Karch-Randall branes with metric $AdS_d$\\ $\equiv$ CFT living on $(d-1)$-dimensional defect.}
\end{minipage}}\\ \\
 Braneworld holography \cite{KR1,KR2} relates first and second line whereas AdS/CFT correspondence \cite{AdS/CFT} between the dynamical gravity on the Karch-Randall brane and defect CFT connects second and third line. Therefore, {\it classical gravity in $(d+1)$ bulk is dual to $CFT_{d-1}$ at the defect.}
Wedge holography helps us in getting the Page curve of the black holes similar to doubly holographic setup discussed in \ref{DHS-introduction}. One is required to compute the entanglement entropies of Hartman-Maldacena and island surfaces and plot of these entropies with time will give the Page curve.

%%%%%%%%%%%%%%%%%%%%%%%%%%%%
%%%%%%%%%%%

%m

\chapter{Page Curves of Reissner-Nordstr{\"o}m Black Hole in HD Gravity}
\graphicspath{{Chapter6/}{Chapter6/}}

\section{Introduction and Motivation}
\label{I-M}
Applying the island proposal described in \ref{Island-intro}, we are able to compute the Page curves of eternal black holes \cite{AMMZ}. In the papers \cite{Island-RNBH,Yu-Ge,Island-SB,Omidi}, the Page curves of the Reissner-Nordstr{\"o}m black hole, charged dilaton black hole, Schwarzschild black hole, and hyperscaling violating black branes were found. In the research referred to as \cite{CLDBH}, the page curve in the charged linear dilaton model for both the non-extremal black hole and the extremal black hole was analyzed. Islands in Kerr-de Sitter spacetime and generalized dilaton theories were the subject of research in the article \cite{Islands-KdS,Tian}. In the research referred to as \cite{NBH-HD}, the effect that mutual knowledge between subsystems on the Page curve was analyzed. Calculations of the page curve used in higher derivative gravity theories could be found in the papers \cite{NBH-HD,HD-Page Curve-2}. Page curves of Schwarzschild black holes were obtained by the authors of the paper mentioned above \cite{NBH-HD} when higher derivative terms, known as ${\cal O}(R^2)$ and Gauss-Bonnet terms, were present in the gravitational action.
 \par
In the papers \cite{Charged-GB-BH-App,Zhang-Li-Guo,BBB,ZZZY,CGLY,BB,LNZ} researchers looked at charged black holes in Einstein-Gauss-Bonnet gravity in four dimensions. See \cite{EGB-Review} for a review of the Einstein-Gauss-Bonnet theory of gravity when applied to four dimensions. Since we were dealing with higher derivative theories of gravity, we employed \cite{Dong} to compute the entanglement entropy for those cases where it was necessary. The non-extremal black holes are what we'll be discussing in this chapter. For extremal black holes, see \cite{Kim+Nam,CLDBH,Yu et al}. {\it It has come to our attention that research on the effect of the Gauss-Bonnet coupling on the Page curve of a charged Einstein-Gauss-Bonnet black hole had not been done. As a result, it would be quite fascinating to investigate how the Page curves of the Reissner-Nordstr\"{o}m black hole will change when the Gauss-Bonnet term is present. As a result of this inspiration, we have computed the Page curves of charged black holes in higher derivative gravity with ${\cal O}(R^2)$ terms and for charged Einstein-Gauss-Bonnet black holes in this work}.
\section{Review}
\label{review}
This section has been broken up into two different subsections.
We will provide a concise review of the charged black hole in Einstein-Guass-Bonnet gravity in the vanishing cosmological constant limit \cite{Charged-GB-BH} and the Page curve calculation of the Reissner-Nordstr\"{o}m black hole  \cite{Island-RNBH} in four dimensions in {\ref{review-CEGB-BH} and {\ref{review-RNBH-Page-curve}.
\subsection{Brief Review of Charged Black Hole in Einstein-Gauss-Bonnet Gravity}
\label{review-CEGB-BH}
It has been demonstrated by the authors of the work \cite{Glavan+Lin} that the Gauss-Bonnet term in four dimensions can be made dynamical by rescaling the Gauss-Bonnet coupling to the form $\alpha \rightarrow \frac{\alpha}{D-4}$. The authors in \cite{AGM} produced a consistent theory of Gauss-Bonnet gravity in $(d+1)$-dimensions with all of the necessary degrees of freedom. The action of Aoki, Gorji, and Mukhohyama (AGM) theory can be written as follows:
\begin{eqnarray}
\label{action-AGM}
S=\frac{1}{2 \kappa^2}\int d^{(d+1)}x \sqrt{-g}\left(R[g]+\alpha {\cal L}_{\rm GB}\right),
\end{eqnarray} 
where ${\cal L}_{\rm GB}=R_{\mu\nu\rho\sigma}[g]R^{\mu\nu\rho\sigma}[g]-4 R_{\mu\nu}[g]R^{\mu\nu}[g]+R[g]^2$ represents Gauss-Bonnet term. The limit $d \rightarrow 3$ of (\ref{action-AGM}) implies consistent theory of four dimensional neutral Gauss-Bonnet black hole. By scaling the Gauss-Bonnet coupling, the author of \cite{Charged-GB-BH} produced a charged Einstein-Gauss-Bonnet black hole solution in four dimensions. In the vanishing cosmological constant limit, we discuss about this solution. 
The Maxwell term, along with Einstein-Gauss-Bonnet gravity, has the following action:
\begin{eqnarray}
\label{action-GB}
S=\frac{1}{2 \kappa^2}\int d^4x \sqrt{-g}[R[g]+\alpha {\cal L}_{\rm GB}-F_{\mu \nu}F^{\mu \nu}],
\end{eqnarray}
where $R[g]$ represents the Ricci scalar, $\alpha$ denotes the Gauss-Bonnet coupling, ${\cal L}_{\rm GB}$ represents the Gauss-Bonnet term, and $F_{\mu\nu}$ denotes the field strength tensor associated with $A_\mu$ gauge field, which are described as follows:
\begin{eqnarray}
\label{LGB-F}
& &
{\cal L}_{\rm GB}=R_{\mu\nu\rho\sigma}[g]R^{\mu\nu\rho\sigma}[g]-4 R_{\mu\nu}[g]R^{\mu\nu}[g]+R[g]^2, \nonumber\\
& & F_{\mu\nu}=\partial_\mu A_\nu -\partial_\nu A_\mu.
\end{eqnarray}
The gravitational action (\ref{action-GB})has the following black hole solution:
\begin{eqnarray}
\label{metric-GB}
ds^2=-F(r)dt^2+\frac{dr^2}{F(r)}+r^2 d\Omega^2,
\end{eqnarray}
where the black hole function $F(r)$ is defined as:
\begin{equation}
\label{F(r)}
F(r)= 1+\frac{r^2}{2 \alpha}\left(1 \pm\sqrt{1+4\alpha\left(\frac{2 M}{r^3}-\frac{Q^2}{r^4}\right)}\right).
\end{equation}
If we do a small expansion in $\alpha$ with a negative sign chosen in (\ref{F(r)}), then we get the following result:
\begin{equation}
\label{Fr-small-alpha}
F(r)=1-\frac{2 M}{r}+\frac{Q^2}{r^2}+\frac{\left(Q^2-2Mr\right)^2}{r^6}\alpha.
\end{equation}
Solving $F(r)=0$ with negative sign of (\ref{F(r)}) yields the black hole's horizons, which are presented in the following form:
\begin{eqnarray}
\label{horizons}
r_{\pm}=M\pm \sqrt{M^2-Q^2-\alpha},
\end{eqnarray}
where the physical horizon of the charged black hole is denoted by the symbol $r_+$ and the Cauchy horizon is denoted by the symbol $r_-$. \par
Let us now examine the relationship between the Gauss-Bonnet coupling $\alpha$ and the black hole horizon $r_+$. We have depicted it in figure {\ref{rplus-alpha-variation}} for $M=1$ and for a variety of different values for the black hole charges. When the Gauss-Bonnet coupling ($\alpha$) is zero, as shown by the figure {\ref{rplus-alpha-variation}} and the equation (\ref{horizons}), we get the typical Reissner-Nordstr\"{o}m black hole. The decrease of Gauss-Bonnet coupling results in the increase of the black hole horizon and vice-versa.
\begin{figure}
\begin{center}
\includegraphics[width=0.60\textwidth]{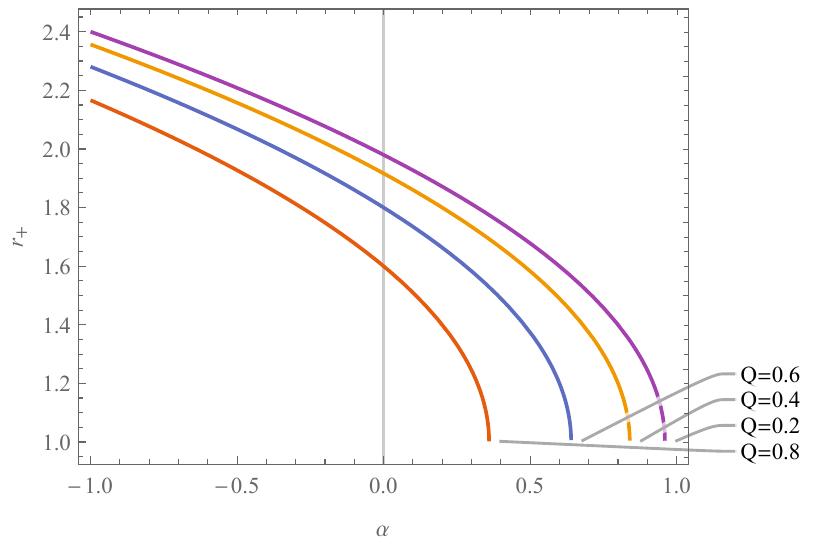}
\end{center}
\caption{Plot of $r_+$ versus $\alpha$}
\label{rplus-alpha-variation}
\end{figure}
When $M=1$, the following values for the Gauss-Bonnet coupling $\alpha$ have been specified by the authors of \cite{Zhang-Li-Guo,CGLY}:
\begin{itemize}
\item {\bf A:}$Q^2-4-2\sqrt{4-2Q^2}< \alpha < 1-Q^2, \ \ {\rm when} \ \ 0<Q<\frac{\sqrt{3}}{2}$.

\item {\bf B:}$Q^2-4-2\sqrt{4-2Q^2}< \alpha <Q^2-4+2\sqrt{4-2Q^2}, \ \ {\rm when} \ \ 0<Q<\sqrt{2}$.
\end{itemize}
In contrast to the region {\bf B}, which only contains the physical horizon ($r_+$), the region {\bf A} contains both horizons ($r_{\pm}$). Since we're looking for the non-extremal black hole, the region {\bf A} is of interest to us. Using $Q=0.8$, we found that for the region {\bf A}, $\alpha \in (-6.66,0.36)$.
\subsection{Review of Page Curve of Reissner-Nordstr{\"o}m Black Hole}
\label{review-RNBH-Page-curve}
Here, we take a look back at \cite{Island-RNBH}'s computation of the Reisnner-Nordstr\"{o}m black hole's Page curve. The action that describes the Einstein-Maxwell gravity is given as:
\begin{eqnarray}
I=\frac{1}{16 \pi G_N} \int_{\cal M} d^4x\sqrt{-g}\left(R-\frac{1}{4}F_{\mu\nu}F^{\mu\nu}\right)+I_{\rm matter},
\end{eqnarray}
where $R$ represents the Ricci scalar, $F_{\mu\nu}$ denotes the field strength tensor associated with the $A_\mu$ gauge field, and $I_{\rm matter}$ represents the matter part of the action. The Reissner-Nordst{\"o}rm black hole has the following metric:
\begin{eqnarray}
\label{metric-RNBH-original}
ds^2=-F(r)dt^2+\frac{dr^2}{F(r)}+r^2 d\Omega^2,
\end{eqnarray}
where, $d\Omega^2=d\theta^2+ \sin^2\theta d\phi^2$ and black hole function:
\begin{equation}
\label{F(r)-RNBH-original}
F(r)=1-\frac{2 M}{r}+\frac{Q^2}{r^2}.
\end{equation}
The following are the black hole horizons that result from solving $F(r)=0$:
\begin{eqnarray}
\label{horizons-RNBH-original}
r_{\pm}=M\pm \sqrt{M^2-Q^2}.
\end{eqnarray}
Below is a version of metric (\ref{metric-RNBH-original}) written in Kruskal coordinates:
\begin{eqnarray}
\label{metric-kruskal-coordinates}
ds^2=-\frac{r_+ r_-}{r^2 \kappa_+^2}\left(\frac{r_-}{r-r_-}\right)^{\frac{\kappa_+}{\kappa_-}-1} e^{-2 \kappa_+ r} dUdV+r^2 d\Omega^2,
\end{eqnarray}
where, 
\begin{equation}
\label{U-V}
U = - e^{-\kappa_+(t-r_*)}, V=e^{\kappa_+(t+r_*)},
\end{equation}
 and the tortoise coordinate $r_*$ is given by:
\begin{eqnarray}
\label{rstar-U}
& & r_*=r+\frac{r_+^2}{r_+ - r_-} \log|r-r_+|-\frac{r_-^2}{r_+ - r_-} \log|r-r_-|.
\end{eqnarray}
For both $r_+$ and $r_-$, we have the following expressions for the surface gravities $\kappa_\pm$:
\begin{eqnarray}
\label{surface-gravity}
& & \kappa_{\pm} =\frac{r_{\pm}-r_{\mp}}{2 r_{\pm}^2}.
\end{eqnarray}
The following are definitions of the charged black hole's Hawking temperature and Bekenstein-Hawking entropy:
\begin{eqnarray}
& & T_{\rm RN}=\frac{\kappa_+}{2 \pi}, \nonumber\\
& & S_{\rm BH}^{(0)} = \frac{\pi r_+^2}{G_N}.
\end{eqnarray}
The conformal factor in the metric, expressed in Kruskal coordinates, is defined as follows:
\begin{eqnarray}
\label{conformal-factor}
g^2(r)=\frac{r_+ r_-}{r^2 \kappa_+^2}\left(\frac{r_-}{r-r_-}\right)^{\frac{\kappa_+}{\kappa_-}-1} e^{-2 \kappa_+ r}.
\end{eqnarray}
As a result, the metric can be expressed as follows:
\begin{eqnarray}
ds^2=-g^{2}(r) dUdV+r^2 d\Omega^2.
\end{eqnarray}
The Penrose diagrams of an eternal Reissner-Nordstr{\"o}m black hole are shown in figures \ref{p1} and \ref{p2}. In figures \ref{p1} and \ref{p2}, the $R_-$ and $R_+$ denote the left and right wedges of radiation regions, the $b_-$ and $b_+$ denote the boundaries of $R_-$ and $R_+$, and the $a_-$ and $a_+$ denote the boundaries of island surface in the left and right wedges, respectively.
\begin{figure}
\centering
\begin{minipage}{.5\textwidth}
  \centering
  \includegraphics[width=.7\linewidth]{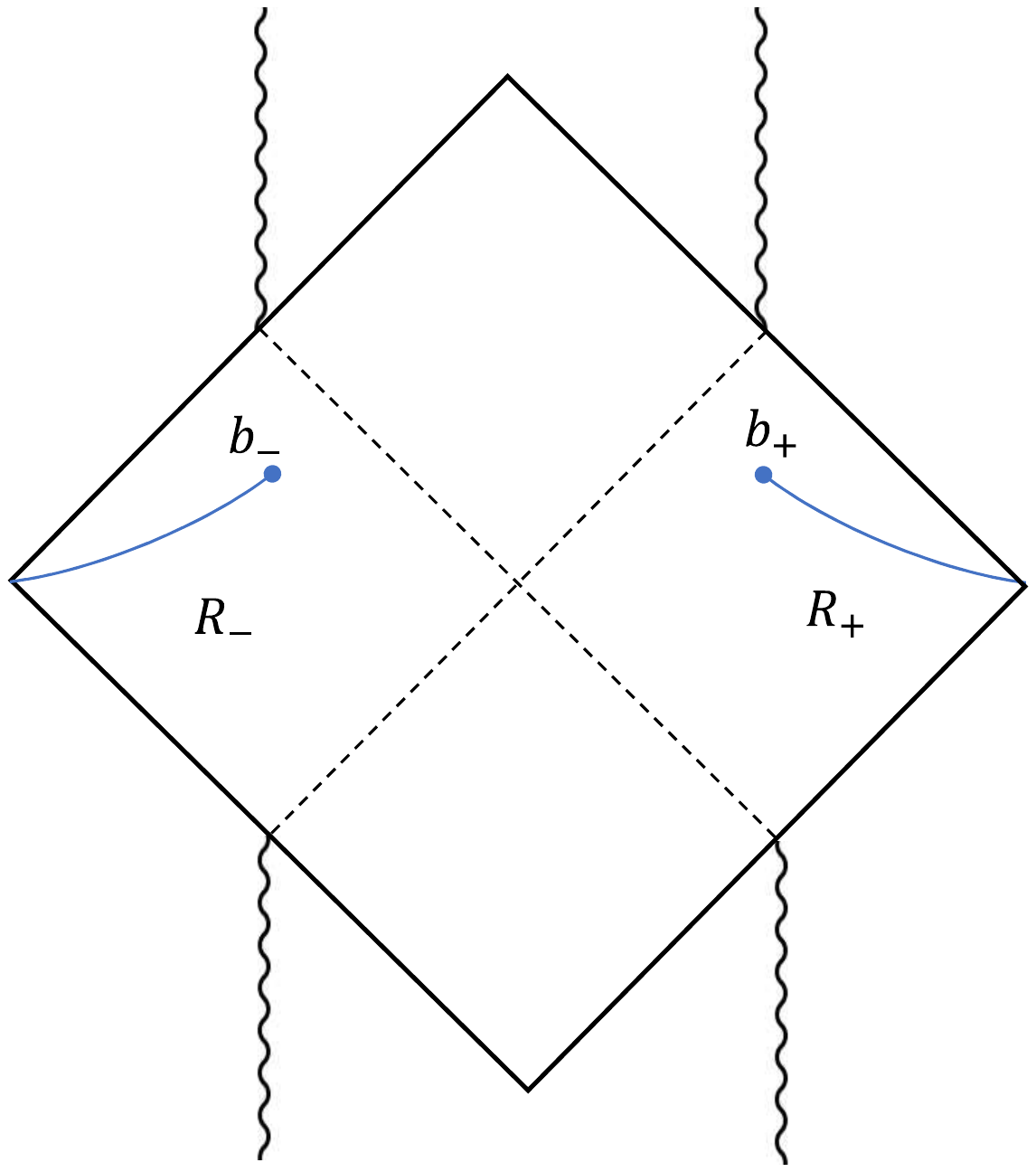}
  \caption{Penrose diagram of an eternal \\ Reissner-Nordstr{\"o}m black hole \cite{Island-RNBH} in the \\absence of island surface.}
  \label{p1}
\end{minipage}%
\begin{minipage}{.5\textwidth}
  \centering
  \includegraphics[width=0.7\linewidth]{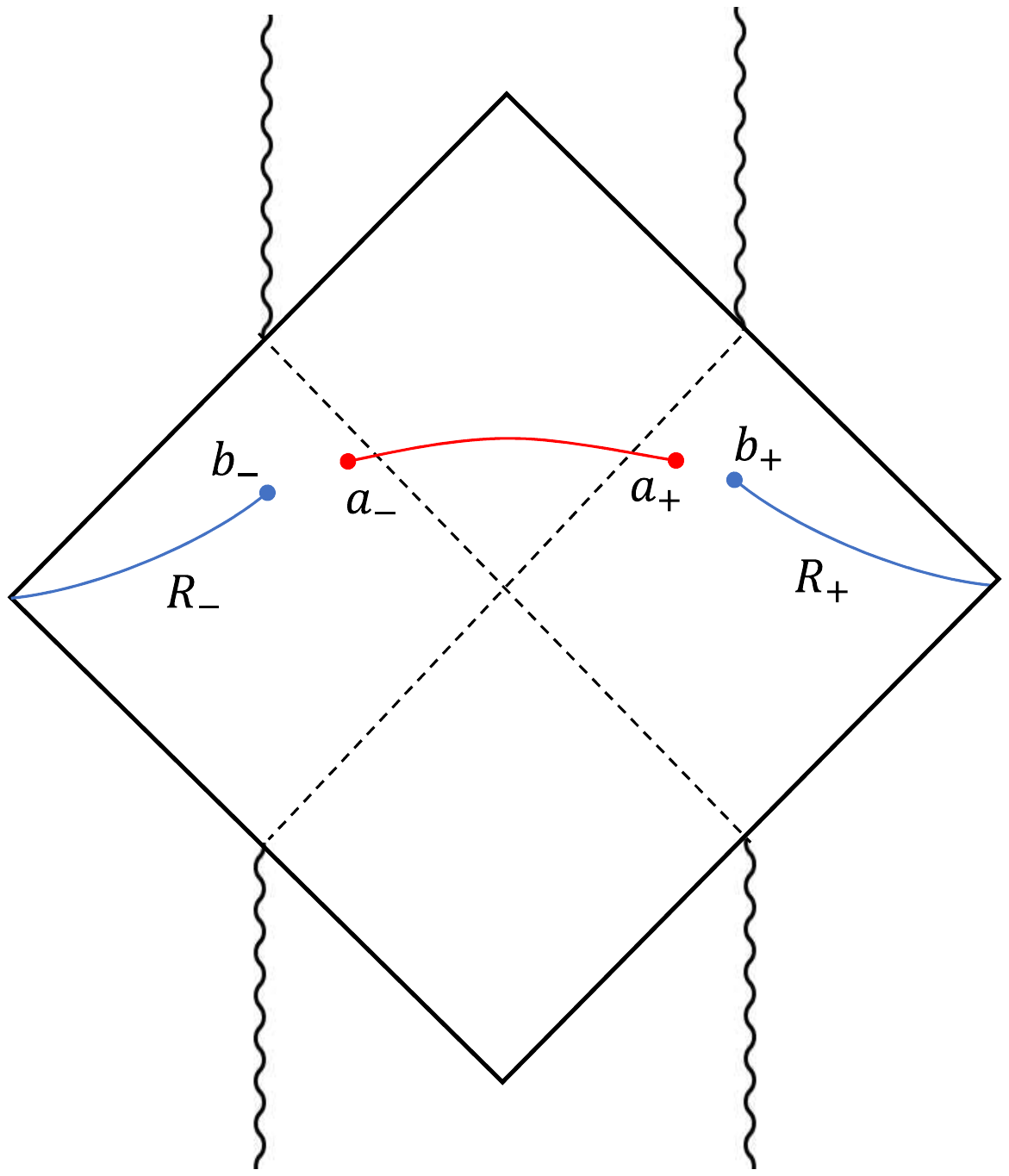}
  \caption{Penrose diagram of an eternal \\ Reissner-Nordstr{\"o}m black hole \cite{Island-RNBH} in the\\ presence of island surface.}
  \label{p2}
\end{minipage}
\end{figure}
\subsubsection{Entanglement Entropy without Island}
Since there is no island surface to begin with, the entanglement entropy of the Hawking radiation will be obtained by applying the formula shown below \cite{CC,CC-1,Yu-Ge}:
\begin{eqnarray}
\label{EE-formula-no-island}
S = \frac{c}{6} \log[d^2(b_+,b_-)],
\end{eqnarray}
where $b_+(t_b,b)$ and $b_-(-t_b+\iota \frac{\beta}{2},b)$ represent the boundaries of the radiation regions in the right and left wedges of the Reissner-Nordstrom black hole, respectively.  The value of $\beta$, which is defined as $\beta=2 \pi/\kappa_+$, represents the inverse of the Hawking temperature. Equation (\ref{EE-formula-no-island}) with no island is applicable to two-dimensional vacuum CFT. Higher-dimensional spacetime's entanglement entropy formula is unknown. However, by assuming that the observer/cut-off surface is located at a large distance from the black hole horizon ($r_+$) ($b_{\pm}\gg r_+$), we can use the (\ref{EE-formula-no-island}) for higher-dimensional spacetime in the s-wave assumption. Therefore, the angular component of the metric may be ignored in the s-wave approximation, leaving us with an asymptotically flat black hole for which we can utilize the (\ref{EE-formula-no-island}). The formula for calculating the geodesic distance among two points $(l_1,l_2)$ is as follows:
\begin{equation}
\label{d-RNBH-HD}
d(l_1,l_2)=\sqrt{g(l_1)g(l_2)(U(l_2)-U(l_1))(V(l_1)-V(l_2))}.
\end{equation}
Now, utilizing equations (\ref{U-V}), (\ref{rstar-U}), (\ref{conformal-factor}), (\ref{EE-formula-no-island}) and (\ref{d-RNBH-HD}), the entanglement entropy of Hawking radiation in the absence of an island surface has been simplified to \cite{Island-RNBH}:
\begin{eqnarray}
\label{EE-without-island}
S = \frac{c}{6}\log[4 g(b)^2 e^{2 \kappa_+ r_*(b)} \cosh^2 \kappa_+t_b],
\end{eqnarray}
If $t_b \rightarrow \infty$, i.e., in late time approximation, we can write $ \cosh \kappa_+t_b \sim e^{\kappa_+t_b}$. The time-dependent component of the equation (\ref{EE-without-island}) can be simplified as:
\begin{eqnarray}
\label{EE-ET-ET-1}
S_{\rm EE}^{\rm WI} \sim \frac{c}{3} \kappa_+ t_b .
\end{eqnarray}
Therefore, we can observe that the entanglement entropy of the Hawking radiation in  without an island surface is increasing linearly with time and approaches infinite at later times via the equation (\ref{EE-ET-ET-1}). This is what causes the information paradox for the Reissner-Nordstr\"{o}m black hole.
% In the next subsection we will see that at late times island appears and entanglement entropy of the Hawking radiation in the presence of island surface will be constant and dominates after the Page time. On combining the both contributions, we obtain the Page curve.
\subsubsection{Entanglement Entropy with Island}
Here, we will go through the computation of the entanglement entropy of the Hawking radiation when an island surface is present. We are able to compute it by utilizing the formula \cite{CC,CC-1,Yu-Ge}, which is as follows:
\begin{eqnarray}
\label{EE-formula-island}
S({\cal R}\cup {\cal I}) = \frac{c}{3} \log\left(\frac{d(a_+,a_-)d(b_+,b_-)d(a_+,b_+)d(a_-,b_-)}{d(a_+,b_-)d(a_-,b_+)}\right),
\end{eqnarray}
where $a_+(t_a,a)$ and $a_-(-t_a+\iota \frac{\beta}{2},a)$ represent the boundaries of the island surface in the right and left wedges of Reissner-Nordstr\"{o}m geometry, respectively. In the s-wave approximation ($b_{\pm}\gg r_+$), the entanglement entropy formula (\ref{EE-formula-island}) can be applied to higher-dimensional spacetime similar to (\ref{EE-formula-no-island}). In late time assumption ($t_a,t_b \gg b >r_+$) utilizing (\ref{U-V}), (\ref{rstar-U}), (\ref{conformal-factor}) and (\ref{d-RNBH-HD}); (\ref{EE-formula-island}) simplified to the following form \cite{Island-RNBH}:
\begin{eqnarray}
\label{Smatter-LT-RNBH}
& & 
S({\cal R}\cup {\cal I})^{\rm late \ times} =\frac{c}{6}\log[g^2(a)g^2(b)]+\frac{2 c}{3} \kappa_+ r_*(b)\nonumber\\
& & +\frac{c}{3}\Biggl[-2 e^{-\kappa_+(b-a)}\Biggl|\frac{a-r_+}{b-r_+}\Biggr|^{1/2}\Biggl|\frac{a-r_-}{b-r_-}\Biggr|^{-(r_-^2/r_+^2)} -e^{\kappa_+(r_*(b)-r_*(a)-2 t_b)} \Biggr].
\end{eqnarray}
Therefore, generalized entropy with inclusion of an island surface at late times is as follows \cite{Island-RNBH}:
\begin{eqnarray}
\label{EE-LT-RNBH}
& & 
S_{gen}^{\rm late \ times}(a)=\frac{2 \pi a^2}{G_N}+\frac{c}{6}\log[g^2(a)g^2(b)]+\frac{2 c}{3} \kappa_+ r_*(b)\nonumber\\
& & +\frac{c}{3}\Biggl[-2 e^{-\kappa_+(b-a)}\Biggl|\frac{a-r_+}{b-r_+}\Biggr|^{1/2}\Biggl|\frac{a-r_-}{b-r_-}\Biggr|^{-(r_-^2/r_+^2)}- e^{\kappa_+(r_*(b)-r_*(a)-2 t_b)} \Biggr].
\end{eqnarray}
By taking the variation of (\ref{EE-LT-RNBH}) with respect to $a$, i.e., $\frac{\partial S_{gen}^{\rm late \ times}(a)}{\partial a}=0$, we have to find the solution of the equation given below:
\begin{eqnarray}
\frac{4 \pi  r_+}{G_N}+\frac{c \left(\frac{r_+-r_-}{b-r_-}\right){}^{-\frac{r_-^2}{r_+^2}} e^{\kappa _+ \left(r_+-b\right)}}{3 \left(r_+-b\right) \sqrt{\frac{a-r_+}{b-r_+}}}=0,
\end{eqnarray}
which has solution:
\begin{eqnarray}
\label{a-RNBH}
a \approx r_+ + \Biggl[\frac{c^2 \left(\frac{r_+-r_-}{b-r_-}\right){}^{-\frac{2 r_-^2}{r_+^2}} G_N^2 e^{2 \kappa _+ \left(r_+-b\right)}}{144 \pi ^2 r_+^2 \left(b-r_+\right)} \Biggr],
\end{eqnarray}
Substitution of $a$ using (\ref{a-RNBH}) in (\ref{EE-LT-RNBH}), results in Hawking radiation's total entanglement entropy at late times:
\begin{eqnarray}
\label{EE-RUI}
& &
S_{\rm total}^{(0)}=\frac{2\pi r_+^2}{G_N}+\frac{c}{3}  \log \left(\frac{r_- e^{2 \kappa _+ \left(b-r_+\right)} \left(\frac{r_-^2}{\left(r_+-r_-\right) \left(b-r_-\right)}\right){}^{\frac{1}{2} \left(\frac{\kappa _+}{\kappa _-}-1\right)}}{b
   \kappa _+^2}\right)+small,\nonumber
   \end{eqnarray}
   \begin{eqnarray}
   & & 
   S_{\rm total}^{(0)}=2 S_{\rm BH}^{\rm (0)}+\frac{c}{3}  \log \left(\frac{r_- e^{2 \kappa _+ \left(b-r_+\right)} \left(\frac{r_-^2}{\left(r_+-r_-\right) \left(b-r_-\right)}\right){}^{\frac{1}{2} \left(\frac{\kappa _+}{\kappa _-}-1\right)}}{b
\kappa _+^2}\right)+small.
\end{eqnarray}
We may deduce from the preceding equation that the total entanglement entropy remains the same, which means that it takes precedence after the Page time, and as a result, we have the Page curve of an eternal Reissner-Nordstr\"{o}m black hole.
\subsubsection{Page Curve}
We showed that the entanglement entropy without an island surface increases linearly with time by using (\ref{EE-ET-ET-1}), and we found that the entanglement entropy remains constant at late times by using (\ref{EE-RUI}). Island surface appears after the Page time, which saturates the linear growth of entanglement entropy and attains a fix value, which equals twice of the Bekenstein-Hawking entropy of the Reissner-Nordstr\"{o}m black hole, we are able to get the Page curve of the Reissner-Nordstr\"{o}m black hole. \par
We used only the leading order term in the $G_N$ in equation (\ref{EE-RUI}) when we drew the Page curve of the Reissner-Nordstr\"{o}m black hole (shown in figure {\ref{Page-curve-RNBH}}) with the value $M=1,Q=0.8,G_N=1$. The blue line in figure {\ref{Page-curve-RNBH}} refers to the linear time growth of the entanglement entropy equation (\ref{EE-ET-ET-1})), and the orange line refers to the entanglement entropy with island surface (equation ({\ref{EE-RUI})), which takes over after the Page time and results in the Page curve. 
\begin{figure}
\begin{center}
\includegraphics[width=0.50\textwidth]{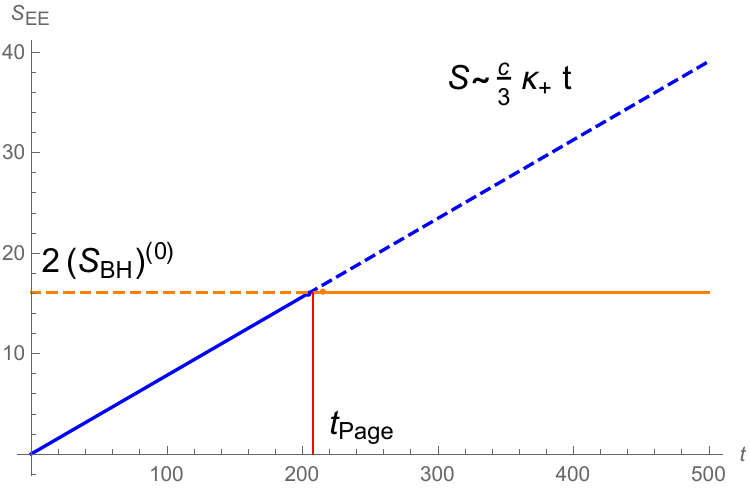}
\end{center}
\caption{Page curve of an eternal Reissner-Nordstr{\"o}m black hole by considering only leading order term in $G_N$.}
\label{Page-curve-RNBH}
\end{figure}
\par
{\bf Page time:} Page time is described as the point in time at which the entanglement entropy of the Hawking radiation begins to approach zero for a black hole that is evaporating and reaches a value that is constant for a black hole that is eternal. From matching the equations (\ref{EE-ET-ET-1}) and (\ref{EE-RUI}), we are able to obtain the Page time, which is represented by the following:
\begin{eqnarray}
\label{Page-time-RNBH}
t_{Page}^{(0)} \sim \frac{6 \pi  r_+^2}{c \kappa _+ G_N} =\frac{12 \pi r_+^4}{c\ G_N (r_+ - r_-)} = \frac{3 S_{\rm BH}^{(0)}}{2 \pi c T_{\rm RN}}.
\end{eqnarray}

{\bf Scrambling time:} The term ``scrambling time'' refers to the time when we are able to retrieve the information that was lost when it was sucked into the black hole in the form of Hawking radiation \cite{scrambling-time-1,scrambling-time-2}. Information from a black hole that has evaporated halfway can be quickly retrieved as Hawking radiation, as detailed in \cite{scrambling-time-1}. If the black hole hasn't evaporated halfway, we'll have to wait until it does so that we can swiftly retrieve the information we've lost. Scrambling time, as used in the terminology of entanglement wedge reconstruction \cite{scrambling-time-EW}, is the amount of time it takes the information from the cut-off surface ($r=b_+$) to arrive at the island surface's boundary ($r=a_+$). Once the information thrown into the black hole has traveled to the boundary of the island surface ($r=a_+$), the entanglement entropy of Hawking radiation begins to include contribution from the island surface, as the degrees of freedom of the island become part of the entanglement wedge of the Hawking radiation. Time it takes the information to travel from the cut-off surface ($r=b$) to the boundary of the island surface ($r=a$) when sent into a black hole is provided by:
\begin{eqnarray}
\label{delta-t}
t_{scr}^{(0)} =r_*(b)-r_*(a)= b-a+\frac{r_-^2}{r_+ -r_-} \log\left(\frac{a-r_-}{b-r_-}\right)-\frac{r_+^2}{r_+ -r_-} \log\left(\frac{a-r_+}{b-r_+}\right).
\end{eqnarray} 
Scrambling time of the Reissner-Nordstr\"{o}m black hole is found to be when the value of $a$ from equation (\ref{a-RNBH}) is substituted into the previous equation and the obtained expression is written as:
\begin{eqnarray} 
\label{tscr}
t_{scr}^{(0)}\sim \frac{2 r_+^2}{(r_+ - r_-)} \log\left(\frac{\pi r_+^2}{G_N}\right) + small=\frac{1}{2 \pi T_{\rm RN}} \log\left(S_{\rm BH}^{(0)}\right) + small.
\end{eqnarray}
This result is comparable to \cite{scrambling-time-2}, which suggests that black holes are the most efficient scramblers. We can observe this by looking at the equation (\ref{tscr}), which shows that the scrambling time is logarithmic of the thermal entropy of charged black hole analogous to \cite{scrambling-time-2}.

\section{Page Curves of Charged Black Hole in HD Gravity up to ${\cal O}(R^2)$}
\label{Page-curve-R^2}
We studied generic higher derivative terms of ${\cal O}(R^2)$ for the gravitational action as that of considered in \cite{NBH-HD}, and we obtained the Page curves of an eternal Reissner-Nordstr\"{o}m black hole. The gravitational action of the Einstein-Maxwell theory with inclusion of ${\cal O}(R^2)$ is given as below:
\begin{eqnarray}
\label{action-R^2}
S=\frac{1}{16 \pi G_N}\int d^4x \sqrt{-g}\left(R[g]+\lambda_1 R^2[g]+\lambda_2 R_{\mu \nu}[g] R^{\mu \nu}[g]+\alpha {\cal L}_{GB}[g]-F_{\mu \nu}F^{\mu \nu}\right),
\end{eqnarray}
where $R[g]$ denotes the Ricci scalar, $R_{\mu\nu}[g]$ represents the Ricci tensor, and ${\cal L}_{GB}$ denotes the Gauss-Bonnet term defined in (\ref{LGB-F}). The formula to calculate the Wald entropy in higher derivative theories of gravity is given as \cite{Wald-Entropy,Wald-Entropy-2}:
\begin{eqnarray}
\label{Wald-Entropy-defn}
S_{\rm Wald}=\frac{1}{4 G_N}\oint \sqrt{h} \left(\frac{\partial {\cal L}}{\partial R_{\mu\nu\rho\sigma}}\right)\epsilon_{\mu\nu}\epsilon_{\rho\sigma} d^2x,
\end{eqnarray}
where $h$ denotes the determinant of the induced metric of the co-dimension two surface. The metric ({\ref{metric-RNBH-original}) is rewritten as follows:
\begin{equation}
\label{metric-redefined}
ds^2=-\left(\frac{(r-r_+)(r-r_-)}{r^2}\right)dt^2+\frac{dr^2}{\left(\frac{(r-r_+)(r-r_-)}{r^2}\right)}+r^2\left(d\theta^2+\sin^2\theta d\phi^2\right),
\end{equation}
We found the following for the action (\ref{action-R^2}):
\begin{eqnarray}
\label{der-Lag}
& & 
\frac{\partial {\cal L}}{\partial R_{\mu\nu\rho\sigma}}=\Biggr[g^{\nu\sigma}g^{\mu\rho}(1+2\alpha R[g])+2\alpha R^{\mu\nu\rho\sigma}[g]-8\alpha g^{\mu\rho}R^{\nu\sigma}[g]\Biggl]\nonumber\\  & &+2 \lambda_1 R[g] g^{\nu\sigma}g^{\mu\rho}+2\lambda_2 g^{\mu\rho}R^{\nu\sigma}[g],
\end{eqnarray}
For the metric (referred to as (\ref{metric-redefined})), the Wald entropy that corresponds to the action (referred to as (\ref{action-R^2})) can be calculated utilizing the equation (referred to as (\ref{der-Lag})) as follows:
\begin{eqnarray}
\label{Wald-Entropy-R^2-gravity}
& &
S_{\rm Wald}^{(R^2)}=S_{\rm BH}^{(\alpha)}=\frac{1}{4 G_N}\oint \sqrt{h} \left(\frac{\partial {\cal L}}{\partial R_{trtr}}\right) d\theta d\phi \nonumber\\
& &= \frac{\pi}{G_N}\left(r_+^2+\frac{4 \alpha(r_+ +6 r_-)}{r_+}-2\lambda_2\left(\frac{r_- r_+}{r_+^2}\right)\right).
\end{eqnarray}
Given that $r_+ \gg r_-$, the preceding equation can be simplified to read as follows:  
\begin{eqnarray}
\label{Wald-Entropy-R^2-gravity-simp}
& &
S_{\rm Wald}^{(R^2)}=S_{\rm BH}^{(\alpha)}\sim \frac{\pi}{G_N}\left(r_+^2+4 \alpha\right).
\end{eqnarray}
The formula which was given in \cite{Dong} and is described in chapter {\bf 5}, can be utilized to determine the holographic entanglement entropy that is associated with higher derivative gravity theories. According to what was discussed in chapter {\bf 5}, we are able to write the total entanglement entropy in the presence of higher derivative terms as follows \cite{NBH-HD}:
\begin{eqnarray}
\label{S-total-defn}
S_{\rm total}=S_{\rm gravity}+S_{\rm matter}(R \cup {\cal I}).
\end{eqnarray}
The result of calculating gravity's contribution to the entanglement entropy corresponding to the action (\ref{action-R^2}) is as follows \cite{NBH-HD}:
\begin{eqnarray}
\label{EE-gravity-R^2}
S_{\rm gravity} = \frac{A[\partial {\cal I}]}{4 G_{N, \rm ren}}+\frac{1}{4 G_{N, \rm ren}} \int_{\partial  {\cal I}} \left(2 \lambda_{1, \rm ren} R[g]+ \lambda_{2, \rm ren} \sum_{i=1}^2[R_{\mu\nu}[g]n^{\mu}_in^\nu_i-\frac{1}{2}K_i K_i]+2 \alpha_{\rm ren} R[\partial  {\cal I}]\right),\nonumber\\
\end{eqnarray}
where $i$ refers to normal directions, $K_i$ representing the trace of extrinsic curvature, $K_{i,\mu\nu}=-h^\alpha_\mu h_{\nu\beta}$, and $h_{\mu\nu}$ being the induced metric on island surface's boundary. In addition, $G_{N, \rm ren}, \lambda_{1, \rm ren},\lambda_{2, \rm ren}$ and $\alpha_{\rm ren}$ are representing the renormalized Newton constant and the coupling constants that appear in higher derivative gravity action (\ref{action-R^2}). These are what are utilized in the process of absorbing the UV divergences of the von Neumann entropy of matter field that was covered in \cite{NBH-HD}. We can calculate the matter's contribution to the entanglement entropy of the Hawking radiation in the presence of the island surface by utilizing the equations (\ref{U-V}), (\ref{rstar-U}), (\ref{conformal-factor}), (\ref{d-RNBH-HD}) and (\ref{EE-formula-island}), which is written as \cite{Island-RNBH}\footnote{The contribution of matter to the entanglement entropy, denoted by the notation $S_{\rm matter}(R\cup {\cal I})$, is comparable, at least to some extent, to the result given in \cite{Island-RNBH}. Because of this, we currently wrote the result until that point.}:
\begin{eqnarray}
& & 
S_{\rm matter}(R \cup {\cal I})=\frac{c}{6}\log[2^4 g^2(a)g^2(b)\cosh^2(\kappa_+ t_a)\cosh^2(\kappa_+ t_b)]+\frac{c}{3} \kappa_+(r_*(a)+r_*(b))\nonumber\\
& & +\frac{c}{3}\Biggl[\frac{\cosh\left(\kappa_+(r_*(a)-r_*(b)) \right)-\cosh\left(\kappa_+(t_a-t_b)\right)}{\cosh\left(\kappa_+(r_*(a)-r_*(b)) \right)+\cosh\left(\kappa_+(t_a-t_b) \right)} \Biggr].
\end{eqnarray}  
In late time approximation, when $t_a,t_b \gg b > r_+$, the preceding equation can be simplified into the following simpler version:
\begin{eqnarray}
\label{EE-matter-late-times-R^2}
& & 
S_{\rm matter}^{\rm late \ time}(R \cup {\cal I})=\frac{c}{6}\log[g^2(a)g^2(b)]+\frac{2 c}{3} \kappa_+ r_*(b)\nonumber\\
& & +\frac{c}{3} \Biggl[-2 e^{-\kappa_+(b-a)} \Biggl|\frac{a-r_+}{b-r_+}\Biggr|^{1/2}\Biggl|\frac{a-r_-}{b-r_-}\Biggr|^{-(r_-^2/r_+^2)}-e^{\kappa_+\left(r_*(b)-r_*(a)-2t_b\right)} \Biggr].
\end{eqnarray}
Because of this, the generalized entropy of the action (\ref{action-R^2}) is as follows \cite{NBH-HD}:
{
\begin{eqnarray}
\label{gen-entropy-HD-gravity}
& & 
S_{gen}(r)={\rm Min}_{\cal I} \Biggl[{\rm Ext}_{\cal I}\Biggl(\frac{A[\partial {\cal I}]}{4 G_{N, \rm ren}}+\frac{1}{4 G_{N, \rm ren}} \int_{\partial  {\cal I}} \Biggl(2 \lambda_{1, \rm ren} R[g]+ \lambda_{2, \rm ren} \sum_{i=1}^2[R_{\mu\nu}[g]n^{\mu}_in^\nu_i-\frac{1}{2}K_i K_i]\nonumber\\
 & & \hskip 1in +2 \alpha_{\rm ren} R[\partial  {\cal I}]\Biggr)  
 + S_{\rm matter}(R\cup {\cal I})\Biggr)
\Biggr].
\end{eqnarray}
}
As we have the metric (\ref{metric-RNBH-original}), we obtained:
$R[g]=0, R[\partial  {\cal I}]=\frac{2}{a^2}, n_1^{t}=\frac{1}{\sqrt{F(r)}},n_2^{r}=\sqrt{F(r)}$,$K_1=0$ and $K_2=-\frac{2}{r}\sqrt{F(r)}$,
implying
\begin{eqnarray}
& & 
\sum_{i=1}^2 R_{\mu\nu}[g]n^{\mu}_in^\nu_i =0,\nonumber\\
& & 
K_1 K_1=0, \nonumber\\
& & K_2 K_2=\frac{4}{a^2}\left(1-\frac{2 M}{a}+\frac{Q^2}{a^2}\right).
\end{eqnarray}
The gravitational component of the entanglement entropy of the Hawking radiation (\ref{EE-gravity-R^2}) can therefore be simplified as follows:
\begin{eqnarray}
\label{EE-gravity-simp-R^2}
& &
S_{\rm gravity}(a)=\frac{2 \pi}{G_{N,\rm ren}}\Biggl[a^2-2 \lambda_{2,\rm ren} \left(1-\frac{2 M}{a}+\frac{Q^2}{a^2}\right)+4 \alpha_{\rm ren}\Biggr],
\end{eqnarray}
Thus, the total generalized entropy is going to be calculated as the sum of the contributions to the entanglement entropy from the gravity component (\ref{EE-gravity-simp-R^2}) and the matter part (\ref{EE-matter-late-times-R^2}), which is given as follows:
\begin{eqnarray}
\label{generalised-EE-R^2}
& & 
S_{\rm gen}(a)=S_{\rm gravity}(a)+S_{matter}^{\rm late \ times}(R \cup {\cal I})(a) \nonumber\\
& & S_{\rm gen}(a) \sim \frac{2 \pi}{G_{N,\rm ren}}\Biggl[a^2-2 \lambda_{2,\rm ren} \left(1-\frac{2 M}{a}+\frac{Q^2}{a^2}\right)+4 \alpha_{\rm ren}\Biggr]+\frac{c}{6}\log[g^2(a)g^2(b)]+\frac{2 c}{3} \kappa_+ r_*(b)\nonumber\\
& & \hskip 0.75in  +\frac{c}{3} \Biggl[-2 e^{-\kappa_+(b-a)} \Biggl|\frac{a-r_+}{b-r_+}\Biggr|^{1/2}\Biggl|\frac{a-r_-}{b-r_-}\Biggr|^{-(r_-^2/r_+^2)}\Biggr],
\end{eqnarray}
where $r_*(b)$, $g(a)$, and $g(b)$ can be replaced from the equations (\ref{rstar-U}) and (\ref{conformal-factor}) in the equation (\ref{generalised-EE-R^2}). The small $t_b$ dependent contribution is ignorable in (\ref{EE-matter-late-times-R^2}). The position of the island surface can be determined by extremizing the equation shown above with respect to the variable $a$, which can be written as:
\begin{eqnarray}
\frac{\partial S_{\rm gen}(a) }{\partial a} \sim \frac{4 \pi  \left(-2 M r_+ \lambda _{2,\rm ren}+2 Q^2 \lambda _{2,\rm ren}+r_+^4\right)}{r_+^3 G_N}+\frac{c \left(\frac{r_+-r_-}{b-r_-}\right){}^{-\frac{r_-^2}{r_+^2}} e^{\kappa _+
   \left(r_+-b\right)}}{3 \left(r_+-b\right) \sqrt{\frac{a-r_+}{b-r_+}}}=0,
\end{eqnarray}
solving above equation, we obtained:
\begin{eqnarray}
\label{a-R^2-gravity}
& & 
a \approx r_+ +\frac{c^2 r_+^6 \left(\frac{r_+-r_-}{b-r_-}\right){}^{-\frac{2 r_-^2}{r_+^2}} G_N^2 e^{2 \kappa _+ \left(r_+-b\right)}}{144 \pi ^2 \left(b-r_+\right) \left(2 \lambda _{2,\rm ren} \left(Q^2-M r_+\right)+r_+^4\right){}^2}.
\end{eqnarray}
Because the island is located beyond the event horizon of the black hole, this creates a causality paradox. It was demonstrated in \cite{island-o-h} that this conclusion occurs in all two-sided eternal black holes or black holes in the Hartle-Hawking state, and it is possible to restore causality using quantum focusing conjecture (QFC) \cite{QFC}. When we separate the black hole from the bath, a finite quantity of energy flux is generated. This energy flux pulls the black hole horizon outwards, and as a result, the island always stays behind the horizon. It was pointed out in \cite{island-coupled} that a limited quantity of energy is also generated even when we couple the black hole to bath, so such energy flux drives the horizon outwards, that suggests that island lies behind the horizon similar to the way that the decoupling process works, and that we are able to get rid of the causality paradox.\par
When the value of ``a'' from the equation (\ref{a-R^2-gravity}) is substituted into the equation (\ref{generalised-EE-R^2}), the generalized entropy equation is simplified to the following form:
\begin{eqnarray}
\label{S-total-HD}
& &
S_{\rm total}^{{\cal O}(R^2)}=S_{\rm gravity}+S_{\rm matter}({\cal R}\cup {\cal I}), \nonumber\\
& & S_{\rm total}^{{\cal O}(R^2)} \sim \frac{2 \pi  \left( r_+^2+4 \alpha_{\rm ren}\right)}{G_N}+\frac{c}{3}  \log \left(\frac{r_- e^{- \kappa _+ \left(b+r_+\right)} \left(\frac{r_-^2}{\left(r_+-r_-\right) \left(b-r_-\right)}\right){}^{\frac{1}{2} \left(\frac{\kappa _+}{\kappa _-}-1\right)}}{b
   \kappa _+^2}\right)+ {\cal O}(G_N), \nonumber\\
   & &S_{\rm total}^{{\cal O}(R^2)} =2 S_{\rm BH}^{(\alpha)}+\frac{c}{3}  \log \left(\frac{r_- e^{- \kappa _+ \left(b+r_+\right)} \left(\frac{r_-^2}{\left(r_+-r_-\right) \left(b-r_-\right)}\right){}^{\frac{1}{2} \left(\frac{\kappa _+}{\kappa _-}-1\right)}}{b \kappa _+^2}\right)+ {\cal O}(G_N).
\end{eqnarray}
If we only look at the leading order term in $G_N$, we are able to see that the total entanglement entropy of an eternal Reissner-Nordstr\"{o}m black hole with inclusion of ${\cal O}(R^2)$ terms in the gravitational action approaches a constant value that is compatible with the literature. In the case where $\alpha \rightarrow 0$, the equation (\ref{S-total-HD}) can be simplified to:
\begin{equation}
S_{\rm total}^{(0)}=\frac{2 \pi r_+^2}{G_N}+\frac{c}{3}  \log \left(\frac{r_- e^{- \kappa _+ \left(b+r_+\right)} \left(\frac{r_-^2}{\left(r_+-r_-\right) \left(b-r_-\right)}\right){}^{\frac{1}{2} \left(\frac{\kappa _+}{\kappa _-}-1\right)}}{b
   \kappa _+^2}\right). 
\end{equation}
It is nearly identical to what is seen in \cite{Island-RNBH}. The Page curves are obtained by combining \ref{EE-ET-ET-1}) and (\ref{S-total-HD}), respectively. Taking into account only the leading order term in (\ref{S-total-HD}), and then replacing $r_+=M+\sqrt{M^2-Q^2}$, the following is what we get:
\begin{eqnarray}
\label{S-total-HD-simp}
& &
S_{\rm total}= \frac{2 \pi  \left( \left(M+\sqrt{M^2-Q^2}\right)^2+4 \alpha\right)}{G_N},
\end{eqnarray}
where $\alpha$ is $\alpha_{\rm ren}$ in (\ref{S-total-HD-simp}) and in this chapter, where we've just used $\alpha$ for the purpose of simplicity. In figure {\ref{SEE-versus-alpha}}, we displayed the Page curves of the charged black hole in with the presence of ${\cal O}(R^2)$ terms for the parameters $M=1,Q=0.8,G_N=1$ for a range of different values of the Gauss-Bonnet coupling ($\alpha$). The diagonal green line indicates (\ref{EE-ET-ET-1}) and the constant red, green and blue horizontal lines are representing the twice of thermal entropy of the black hole when $\alpha=0.2,0,-0.2$ with Page times $t_{1_{\rm Page}},t_{2_{\rm Page}}$ and $t_{3_{\rm Page}}$. Figure \ref{SEE-versus-alpha} represents the shifting of Page curves with Gauss-Bonnet coupling. When $\alpha$ rises then Page curves shift towards later times and vice-versa.
\begin{figure}
\begin{center}
\includegraphics[width=0.70\textwidth]{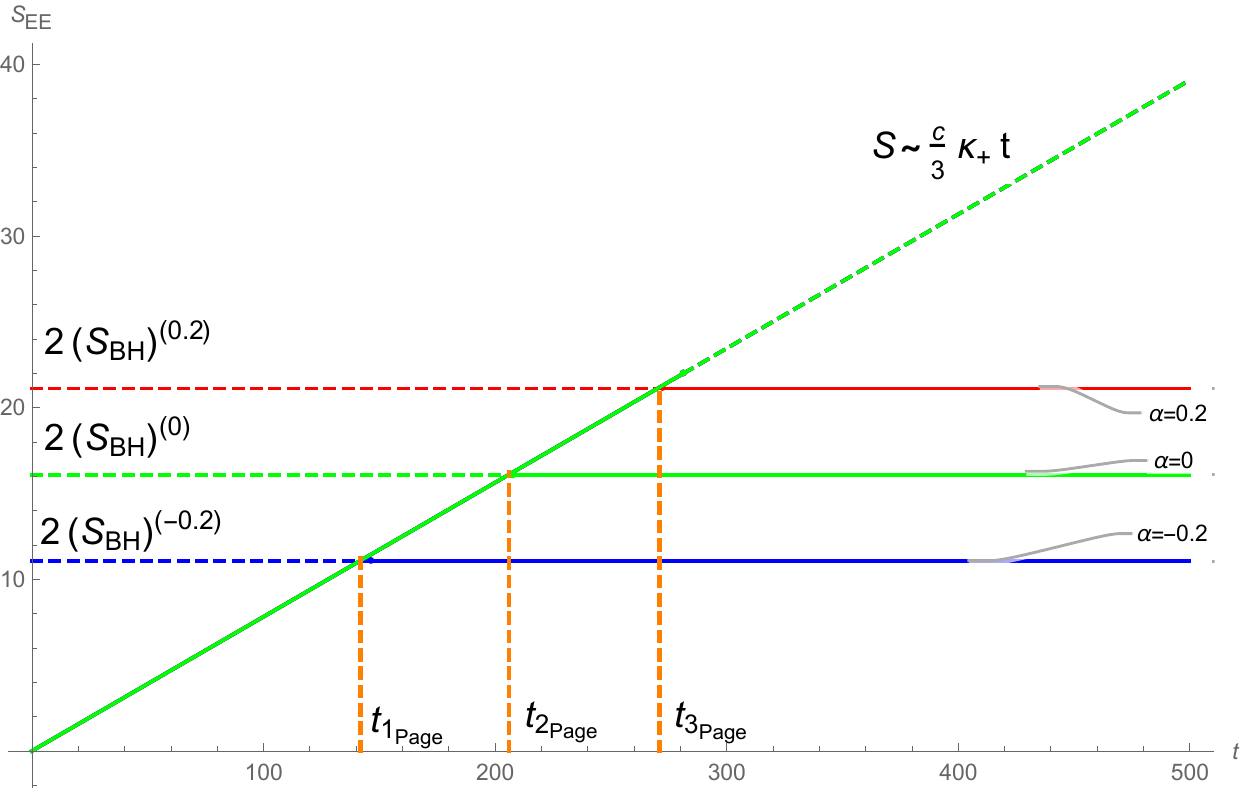}
\end{center}
\caption{Page curves of an eternal Reissner-Nordstr\"{o}m black hole using a number of different Gauss-Bonnet couplings.}
\label{SEE-versus-alpha}
\end{figure}
\par
{\bf Page Time:} As at $t=t_{\rm Page}^{\rm HD}$, $S_{\rm EE}^{\rm WI}=S_{\rm total}$, thus (\ref{EE-ET-ET-1}) and (\ref{S-total-HD}) implying the following  Page time:
\begin{eqnarray}
\label{Page-time-HD}
& & 
t_{\rm Page}^{\rm HD} \sim  \frac{6 \pi  \left(r_+^2+4 \alpha\right)}{c \kappa _+ G_N}=\frac{3 S_{\rm BH}^{(0)}}{2 \pi c T_{\rm RN}}+\frac{12 \alpha}{c \ G_N T_{\rm RN}}=t_{\rm Page}^{(0)}+\frac{12 \alpha}{c \ G_N T_{\rm RN}}.
\end{eqnarray}
The preceding equation makes it abundantly evident that in the case when $\alpha \rightarrow 0$ is being considered, the Page time of the Reissner-Nordstr\"{o}m black hole with the presence of ${\cal O}(R^2)$ terms simplifies to the Page time of the Reissner-Nordstr\"{o}m black hole (\ref{Page-time-RNBH}) when higher derivative terms are removed.
\begin{eqnarray}
t_{\rm Page}^{(0)} \sim \frac{6 \pi r_+^2}{c \kappa _+ G_N}.
\end{eqnarray}
As a result of the fact that the Gauss-Bonnet coupling ($\alpha$) appears with a positive sign in (\ref{S-total-HD}) and (\ref{Page-time-HD}), we receive the Page curves at a later or earlier times when $\alpha$ increases or decreases.\\
{\bf Scrambling time:} The scrambling time of the charged black hole with the presence of ${\cal O}(R^2)$ terms can be found by putting $a$ from (\ref{a-R^2-gravity}) into (\ref{delta-t}). The resulting equation is as follows:
\begin{eqnarray} 
\label{tscr-HD}
t_{scr}^{\rm HD}\sim \frac{2 r_+^2}{r_+-r_-}{ \log \left(\frac{\pi \left(2 \lambda _{2,\rm ren} \left(Q^2-M r_+\right)+r_+^4\right)}{r_+^2 G_N}\right)}+ small .
\end{eqnarray}
In the previous equation, if we consider the limit of $\lambda _{2,\rm ren} \rightarrow 0$, we will get the scrambling time for the Reissner-Nordstr\"{o}m black hole (\ref{tscr}) when higher derivative terms are not taken into account. The small $\lambda _{2,\rm ren}$ expansion of (\ref{tscr-HD}) results in: 
\begin{eqnarray}
\label{tscr-small-lambda}
& &
t_{\rm scr}^{\rm HD} \sim \frac{2 r_+^2}{r_+-r_-} {\log \left(\frac{\pi r_+^2}{G_N}\right)}+\frac{2 \lambda _{2,\rm ren} \left(Q^2-M r_+\right)}{\left(r_+-r_-\right) r_+^2},\nonumber\\
& & =\frac{1}{2 \pi T_{\rm RN}}  {\log \left(S_{\rm BH}^{(0)}\right)}+\frac{2 \lambda _{2,\rm ren} \left(Q^2-M r_+\right)}{\left(r_+-r_-\right) r_+^2}=t_{\rm scr}^{(0)}+\frac{2 \lambda _{2,\rm ren} \left(Q^2-M r_+\right)}{\left(r_+-r_-\right) r_+^2}.
\end{eqnarray}
Hence, scrambling time of the charged black hole with inclusion of ${\cal O}(R^2)$ terms may increase or decrease based on the sign of second term of (\ref{tscr-small-lambda}).

\section{Page Curves of Charged Black Hole in Einstein-Gauss-Bonnet Gravity}
\label{Page-curves-CEGB-BH}
We now obtain the Page curves of charged black hole in Einstein-Gauss-Bonnet gravity. The gravitational action of Einstein-Maxwell-Gauss-Bonnet gravity  without cosmological constant has the following form \cite{Charged-GB-BH}:
\begin{eqnarray}
\label{action-EGB}
S=\frac{1}{2 \kappa^2}\int d^4x \sqrt{-g}[R[g]+\alpha {\cal L}_{\rm GB}-F_{\mu \nu}F^{\mu \nu}].
\end{eqnarray}
The Wald entropy of the charged black hole in Einstein-Gauss-Bonnet gravity has been obtained utilizing (\ref{Wald-Entropy-defn}),(\ref{metric-redefined}),(\ref{der-Lag}) and (\ref{action-EGB}) and the obtained results is written below:
\begin{eqnarray}
\label{Wald-Entropy-EGB}
S_{\rm Wald}^{(\rm EGB)}=S_{\rm BH}^{(\alpha)}= \frac{\pi}{G_N}\left(r_+^2+\frac{4 \alpha(r_+ +6 r_-)}{r_+}\right)
\sim  \frac{\pi}{G_N}\left(r_+^2+4 \alpha\right).
\end{eqnarray}  
The following is a result of the holographic entanglement entropy of the gravitational action (\ref{action-EGB}) that was done in \cite{BSS,Dong,BS}:
{\footnotesize
\begin{eqnarray}
\label{EE-GB}
& &
 S_{\rm gravity}= \frac{1}{4 G_N} \int d^2y \sqrt{h} \Biggl[1+ \alpha \Biggl( 2 R[g] - 4 \left( R_{\mu\nu}[g]n^{\nu i} n^{\mu}_i -\frac{1}{2}K^{i}K_{i}\right) +2 \left(R_{\mu\nu\rho\sigma}[g]n^{\mu i} n^{\nu j} n^{\rho}_i n^{\sigma}_j - K^{i}_{ab}K_{i}^{ab} \right)\Biggr) \Biggr], \nonumber\\
\end{eqnarray}
}
where $h$ denotes the determinant of a given induced metric on a surface with constant coordinates ($t,r=a$):
\begin{eqnarray}
\label{induced-metric}
ds^2=h_{ab}dx^adx^b=a^2\left(d\theta^2+ \sin^2\theta d\phi^2\right).
\end{eqnarray}
Now equation (\ref{EE-GB}) was simplified by the use of Gauss-Codazzi equation \cite{NBH-HD,BS}:
\begin{eqnarray}
\label{GC-equation}
R[g]=R[\partial {\cal I}]-R_{\mu\nu\rho\sigma}[g]n^{\mu i} n^{\nu j}n^{\rho}_i n^{\sigma}_j+ 2 R_{\mu\nu}[g]n^{\nu i} n^{\mu}_i  - K^i K_i+K^i_{ab}K_i^{ab}.
\end{eqnarray}
Holographic entanglement entropy in Einstein-Gauss-Bonnet gravity can be simplified to the following form through the use of the equations (\ref{EE-GB}) and (\ref{GC-equation}), which represent the gravitational contribution to the generalized entropy:
\begin{eqnarray}
\label{EE-EGB-gravity}
S_{\rm gravity}= \frac{1}{4 G_N}  \int d^2y \sqrt{h} \left(1+ 2 \alpha R[\partial {\cal I}] \right),
\end{eqnarray}
and contribution of matter to the entanglement entropy is same as given in \ref{Page-curve-R^2}.  For the induced metric (\ref{induced-metric}), we found that:
$ R[\partial  {\cal I}]=\frac{2}{a^2}$. Hence, (\ref{EE-EGB-gravity}) has the following form for the metric (\ref{induced-metric}):
\begin{eqnarray}
\label{EE-gravity-simp}
S_{\rm gravity}=\frac{2 \pi  a^2}{G_N}\left(1+\frac{4 \alpha}{a^2}
\right),
\end{eqnarray}
Hence, similar to \ref{Page-curve-R^2}, the generalised entropy for this case is obtained as:
\begin{eqnarray}
\label{total-EE-EGB}
& &  S_{\rm gen}(a) \sim \frac{2 \pi  a^2}{G_N}\left(1+\frac{4 \alpha}{a^2}
\right)+  \frac{c}{6}\log[g^2(a)g^2(b)]+\frac{2 c}{3} \kappa_+ r_*(b) \nonumber\\
& &+\frac{c}{3} \Biggl[-2 e^{-\kappa_+(b-a)} \Biggl|\frac{a-r_+}{b-r_+}\Biggr|^{1/2}\Biggl|\frac{a-r_-}{b-r_-}\Biggr|^{-(r_-^2/r_+^2)} \Biggr]. 
\end{eqnarray}
The variation of (\ref{total-EE-EGB}) with respect to $a$ leads to the following equation:
\begin{eqnarray}
\label{der-sgen-a-EGB}
\frac{\partial S_{\rm gen}(a) }{\partial a}\sim  \frac{4 \pi  r_+}{G_N}+\frac{c \left(\frac{r_+-r_-}{b-r_-}\right){}^{-\frac{r_-^2}{r_+^2}} e^{\kappa _+ \left(r_+-b\right)}}{3 \left(r_+-b\right) \sqrt{\frac{a-r_+}{b-r_+}}} =0,
\end{eqnarray}
from the previous equation, we found the location of the island surface as written below:
\begin{eqnarray}
\label{a-GB}
& & 
a \approx r_+ +\frac{c^2 \left(\frac{r_+-r_-}{b-r_-}\right){}^{-\frac{2 r_-^2}{r_+^2}} G_N^2 e^{2 \kappa _+ \left(r_+-b\right)}}{144\pi ^2 r_+^2 \left(b-r_+\right)}.
\end{eqnarray}
Using (\ref{a-GB}), the total entanglement entropy of the charged Einstein-Gauss-Bonnet black hole (\ref{total-EE-EGB}) in late time approximation is obtained as:
\begin{eqnarray}
\label{total-EE-EGB-simp}
& &
S_{\rm total}^{\rm EGB}\sim \frac{2 \pi  \left(4 \alpha +r_+^2\right)}{G_N}+\frac{c}{3}  \log \left(\frac{r_- e^{- \kappa _+ \left(b+r_+\right)} \left(\frac{r_-^2}{\left(r_+-r_-\right) \left(b-r_-\right)}\right){}^{\frac{1}{2} \left(\frac{\kappa _+}{\kappa _-}-1\right)}}{b
   \kappa _+^2}\right),\nonumber\\
   & & S_{\rm total}^{\rm EGB}  = 2 S_{BH}^{(\alpha)} +\frac{c}{3}  \log \left(\frac{r_- e^{-\kappa _+ \left(b+r_+\right)} \left(\frac{r_-^2}{\left(r_+-r_-\right) \left(b-r_-\right)}\right){}^{\frac{1}{2} \left(\frac{\kappa _+}{\kappa _-}-1\right)}}{b
   \kappa _+^2}\right).
\end{eqnarray}
Since the total entanglement entropy of the charged black hole in Einstein-Gauss-Bonnet gravity at late times (\ref{total-EE-EGB-simp}) is the same as the total entanglement entropy of the charged black hole in higher derivative gravity with the ${\cal O}(R^2)$ terms specified in the equation (\ref{S-total-HD}), it may be concluded that the two quantities are identical. Because of this, the Page curves and Page times of the charged black hole in Einstein-Gauss-Bonnet gravity will turn out to be the same as the Page curves and Page times of the charged black hole in higher derivative gravity with ${\cal O}(R^2)$ terms , figure \ref{SEE-versus-alpha}.\\
{\bf Scrambling time:} In this particular instance, the scrambling time is not influenced by the higher derivative terms and is identical to the scrambling time of the Reissner-Nordstr\"{o}m black hole \cite{Island-RNBH}:
\begin{eqnarray} 
\label{tscr-GB-term}
\hskip -0.2in t_{scr}^{\rm EGB}\sim \frac{2 r_+^2}{(r_+ - r_-)} \log\left(\frac{\pi r_+^2}{G_N}\right) + small = \frac{1}{2 \pi T_{\rm RN}}  \log\left(S_{\rm BH}^{(0)}\right) + small=t_{\rm scr}^{(0)}+small.
\end{eqnarray}

\section{Conclusion and Discussion}
\label{Summary}
We obtained the Page curves of an eternal Reissner-Nordstrom black hole in this chapter. We concentrated on the non-holographic model and made the assumption that the observer is located a large distance from the black hole. As a result, we were able to employ the s-wave approximation to compute the entanglement entropy of Hawking radiation utilizing the formula associated with two-dimensional conformal field theory. The affect of higher derivative terms on the Page curve was studied in \cite{NBH-HD} for the neutral black hole. The researchers in that paper took into account the eternal black hole and the evaporating black hole both. They were solely concerned with the more generic ${\cal O}(R^2)$ terms. We considered the general ${\cal O}(R^2)$ terms as well as Gauss-Bonnet term both as the higher derivative terms for the non-extremal eternal Reissner-Nordstr\"om black hole and found that Page time, scrambling time, and the Page curves of eternal black holes are all affected when higher derivative terms are present in the gravitational action. The key results that have been found are summarized below.
\begin{enumerate}
\item In the first scenario, we took into account the general terms (${\cal O}(R^2)$) in the gravitational action described in \cite{NBH-HD} and calculated the Page curves. We came to the conclusion that, because there is initially no island surface, the entanglement entropy of the Hawking radiation will continue to increase in a linear fashion with the passage of time indefinitely. As a result, we ended up at an information paradox for the charged black hole. At later times, an island will appear, and the entanglement entropy of the Hawking radiation will attain a constant value. This value will be equal to double the Bekenstein-Hawking entropy of the black hole, and it will allow us to derive the Page curves for fixed values of the Gauss-Bonnet coupling ($\alpha$). The higher derivative terms shifts the Page curve and hence Page time. {\it Page curves move towards later times as the Gauss-Bonnet coupling ($\alpha$) grows, and Page curves shift towards earlier times as the Gauss-Bonnet coupling ($\alpha$) drops.}

\item In the next scenario, we solely included the Gauss-Bonnet term as the higher derivative term because it is crucial for the investigation of the charged black hole in Einstein-Gauss-Bonnet gravity \cite{Charged-GB-BH}. In a manner analogous to the first scenario, we observe linear time growth of the entanglement entropy of the Hawking radiation at the beginning. This is followed by the emergence of an island at later times, that dominates the linear time growth of the Hawking radiation. As a result, the entanglement entropy of the Hawking radiation eventually reaches exactly two times the Bekenstein-Hawking entropy of the black hole, and we obtained the Page curve exactly similar to the first case. It is interesting that we found that the Page curves with Gauss-Bonnet coupling behave in a manner that is comparable to that which was explained earlier in the first case.

\item In the first scenario, the higher derivative terms have an effect on how long it takes the charged black hole to scramble its information. If the correction term is positive, it will make it so that it is larger, but if it is negative, it will make it so that it is smaller. In addition, if we take the coupling that appears in scrambling time and set it to zero, then we are able to get back the scrambling time of the Reissner-Nordstr\"{o}m black hole. Scrambling time is unaffected by the higher derivative term in the second scenario, which is when we look at solely the Gauss-Bonnet term as the higher derivative term. In this scenario, scrambling time remains identical, just like the scrambling time associated with the Reissner-Nordstr\"{o}m black hole.

\item  When the Gauss-Bonnet coupling goes to zero, we are able to recover the Page curve of the Reissner-Nordsrt\"{o}m black hole, which was obtained in \cite{Island-RNBH}. Further, all of our results reduces to that of \cite{Island-RNBH} in the same limit.

\end{enumerate}

%satyamshivamsundaram
%JaiMataDi  (ERROR IN boundary term and Page Curve with HD terms required)
\chapter{Entanglement Entropy and Page Curve from the ${\cal M}$-Theory Dual of Thermal QCD Above $T_c$ at Intermediate Coupling}
\graphicspath{{Chapter7/}{Chapter7/}}

\section{Introduction and Motivation}
\label{introduction}
In doubly holographic approaches, as described in \ref{DHS-introduction} of chapter {\bf 5}, a gravitational dual black hole is connected with an external CFT bath \cite{Island-HD}. As an illustration, gravity in $d$-dimensions is connected with the external bath in $d$-dimensions, wherein $d$-dimensional exterior bath possesses a corresponding holographic dual in $(d+1)$-dimensions. In such models, we take into consideration two variants of the previously mentioned configuration. We look at two distinct types of extremal surfaces in these models. The very first one corresponds to the Hartman-Maldacena-like surface \cite{Hartman-Maldacena}, that begins from the position at which gravity interacts with with an external bath, i.e., at the defect, and traverses the black hole horizon to get to the defect of the thermofield double associated with doubly holographic setups, and the entanglement entropy involvement via the aforementioned surface possesses linear time dependency, leading to information paradox at later times. The other type of surface includes the island surface, that begins at the exterior bath and ends on the Karch-Randall brane \cite{KR1,KR2}.  The contribution of entanglement entropy coming from the island surface becomes time independent and takes over beyond the Page time. The Page curve is obtained through the combination of the entanglement entropy contributions obtained from both the extremal surfaces.\par
{\it As we discussed in \ref{DHS-introduction}, in some cases, the gravity becomes massive on the Karch-Randall branes, and one was unable to get the Page curve with massless gravity localized on the Karch-Randall branes. In this chapter, we have looked at this major issue of doubly holographic setup from ${\cal M}$-theory perspective in the presence of ${\cal O}(R^4)$ terms in the supergravity action. We have a different situation from the literature that we have a non-conformal bath (QCD bath), and the doubly holographic setup is being constructed from a top-down approach\footnote{See \cite{Island-IIB-1,Island-IIB-2,Island-IIB-3}, for some papers from the top-down approach where the authors have worked with type IIB string theory.} with the inclusion of higher derivative terms in the supergravity action. Interestingly, the issue of the massive graviton is absent in our setup, and one can get the Page curve with massless gravity on the end-of-the-world(ETW) brane.}\par
We cannot utilize the island approach for non-conformal backgrounds because currently there is no recognized Cardy-like formula for the non-conformal theories. As a result, we used the previously mentioned doubly holographic frameworks technique to arrive at the Page curve. In the context of our scenario, on the gravity dual side, there is a ${\cal M}$-theory dual consisting of higher derivative terms, whereas on the gauge theory side, we find non-conformal thermal QCD-like theories at the intermediate coupling, the model is discussed in \ref{HG-PC} of chapter {\bf 1}. In the present chapter, we created a doubly holographic setup analogous to \cite{PBD,Bath-WCFT}, where the researchers utilize a conformal theory on their gauge theory side. There are two types of potential surfaces to think about: Hartman-Maldacena-like and island surfaces. The Hartman-Maldacena-like surface has been accountable for the gradual linear temporal growth of Hawking radiation's entanglement entropy. Above the Page time, the entanglement entropy component via the island surface takes over and this is not dependent upon time, so we extract the Page curve using the ${\cal M}$-theory dual, which includes ${\cal O}(R^4)$ corrections. As a remark, we used the formula of \cite{Dong} to compute the entanglement entropy with the inclusion of higher derivative terms and when higher derivative terms are absent then we used the Ryu-Takayanagi formula to compute the entanglement entropy \cite{RT}.

\section{Doubly Holographic Setup in ${\cal M}$ Theory}
\label{DHS-M-Theory}
We generalized these types of setups to ${\cal M}$-theory background after being inspired by the double holographic setup presented in \ref{DHS-introduction}. After a double Wick-rotation, we have ${\rm QCD}_{2+1}$ at r=0, along $x^{1,2}$ and Wick-rotated $x^3$. One can think of a fluxed ETW-hypersurface or ``brane" $M_{10}=\mathbb{R}^2(x^{2,3}) \times_w M_8$ where $M_8$ is the (non-compact) $SU(4)/Spin(7)$-structure (\cite{HD-MQGP}, Table 2) eight-fold $M_8^{SU(4)/Spin(7)}(r,t,\theta_{1,2},\phi_{1,2},\psi,x^{10})$ at $x^1=0$ which has a black hole. The ETW-hypersurface can be interpreted as $\mathbb{R}^2(x^{2,3})\times_w M_8^{SU(4)/Spin(7)}(t,r,\theta_{1,2},\phi_{1,2},\psi,x^{10})$ containing black $M5$-brane  at $x^1=0$ with world volume $\Sigma^{(6)}=S^1(t)\times_w\mathbb{R}_{\geq0}(r)\times\Sigma^{(4)}$ where $\Sigma^{(4)}= n_1S^3(\theta_1,\phi_1,\psi) \times[0,1]_{\theta_2}+n_2S^2(\theta_1,\phi_1)\times S^2(\theta_2,x^{10})$ with $n_1$ determined by $\int_{S^3\times[0,1]}G_4$ and $n_2$ by $\int_{S^2\times S^2}G_4$; the ${\rm QCD}_{2+1}$ can be thought of as living on $M2$-branes with world volume $\Sigma^{(3)}(x^{1,2,3})$. ${\rm QCD}_{2+1}$ at $r=0$ would interact gravitationally via the pull-back of the ambient $M_{11} =\mathbb{R}^{1,2,3}\times_w M_8 = \mathbb{R}(x^1)\times_w M_{10}$ metric used to contract the non-abelian field strength in the gauge kinetic term obtained as part of the $M2$-brane($x^{1,2,3}$) world-volume action. This has some similarity with points {\bf 2} and {\bf 3} of the doubly holographic setup as discussed in \ref{DHS-introduction}. The doubly holographic setup constructed by us is shown in Fig. \ref{Doubly Holographic Setup}.  Let us be more specific and briefly describe the three equivalent descriptions alluded to towards the beginning of this sub-section. The doubly holographic setup constructed from the bottom-up approach (usually followed in the literature) as described at the beginning of this sub-section, has the following ${\cal M}$-theory description (the one we follow) of top-down double holography with $QCD_{2+1}$ bath  with the numbering matches the one used in the aforementioned three equivalent description:

\begin{enumerate}
\item {\bf Boundary-like Description:} $QCD_{2+1}$ (could be thought of as supported on an $M2$-brane with world-volume $\Sigma^{(3)}(x^{1, 2, 3})$, and) is living at the tip ($r=0$) of a non-compact seven-fold of $G_2$ structure which is a cone over a warped non-K\"{a}hler resolved conifold. The two-dimensional ``defect'' $\Sigma^{(2)}\cong\Sigma^{(3)}(x^{2, 3}; x^1=0)\cong \mathbb{R}^{2}(x^{2,3})$.  

\item {\bf Non-Conformal Bath-ETW Interaction Description:} Fluxed End-of-The-World (ETW)  hypersurface  $M_{10}\cong \mathbb{R}^2(x^{2,3})\times_w M_8^{SU(4)/Spin(7)}(t,r,\theta_{1,2},\phi_{1,2},\psi,x^{10})$ containing black $M5$-brane  at $x^1=0$ coupled to  $QCD_{2+1}$ bath living on $M2$ brane with world volume $\Sigma^{(3)}(x^{1, 2, 3})$ along the defect $\left.\Sigma^{(2)}\cong
M_{10}\cap \Sigma^{(3)}(x^{1, 2, 3})\right|_{x^1=0}$ via exchange of massless graviton.

\item {\bf Bulk Description:}  $QCD_{2+1}(x^{1, 2, 3})$ has holographic dual which is eleven dimensional ($M_{11}\cong \mathbb{R}(x^1)\times_w M_{10}\cong S^1(t)\times_w\mathbb{R}^3(x^{1, 2, 3})\times_w M_7^{G_2}(r,\theta_{1,2},\phi_{1,2},\psi,x^{10})$) ${\cal M}$-theory background (compactified on a seven-fold with $G_2$ structure) containing the fluxed ETW-hypersurface $M_{10} (x^1=0)$.
\end{enumerate}

\begin{figure}
\begin{center}
\includegraphics[width=1.05\textwidth]{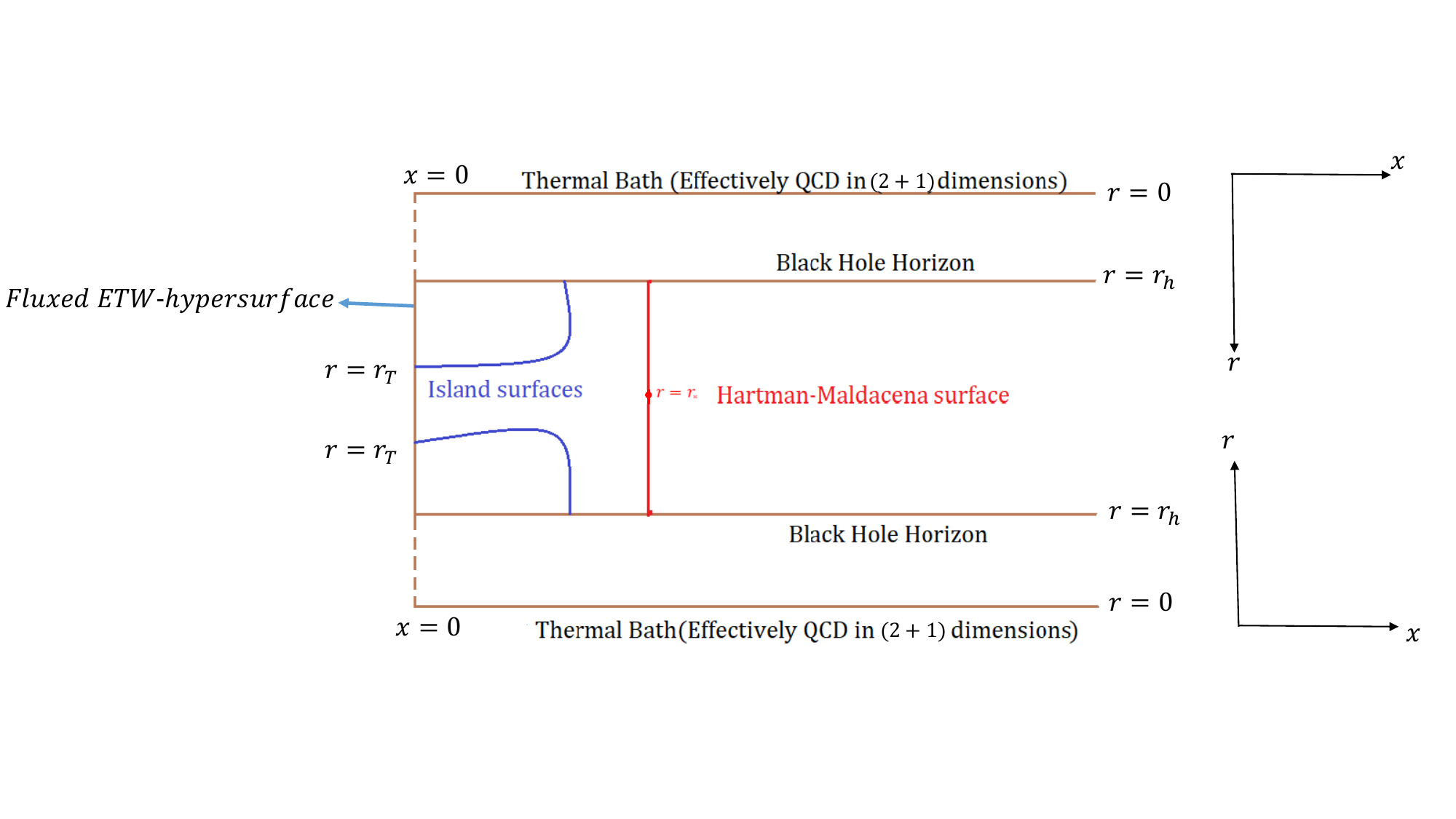}
\end{center}
\caption{Doubly holographic setup in ${\cal M}$-theory dual. ETW-black ``brane''  is coupled the thermal bath (at the tip of seven-fold of $G_2$-structure which is a cone over a warped non-K\"{a}hler resolved conifold), where  the Hawking radiation is collected, which effectively is thermal $QCD_{2+1}$ along $x^{1, 2, 3}$ after Wick-rotation along $x^3$ at $r=0$.   Blue curves correspond to the island surfaces and red curve corresponds to the Hartman-Maldacena-like surface.}
\label{Doubly Holographic Setup}
\end{figure} 
The pictorial representation of the aforementioned ETW-brane(containing a black $M5$-brane)/$M2$-brane(supporting the non-conformal ``bath'' - $QCD_{2+1}$)-setup focusing only on the $x^1-r$-plane, along with the Hartmann-Maldacena-like surface and island surface, is given in Fig. \ref{Doubly Holographic Setup}. Unlike \cite{AMMZ}, \cite{Island-HD} and \cite{GB-3}, the (non-conformal) bath is at $r=0$ instead of the UV cutoff $r=r_{\rm UV}$\footnote{The cut-off, unlike most references in the literature is not at infinity but is such that $r_{\rm UV}\stackrel{<}{\sim}L\equiv\left(4\pi g_s N\right)^{1/4}$ thereby justifying dropping the ``1'' in the 10-dimensional warp factor $h$  in \cite{metrics}, \cite{MQGP}, \cite{HD-MQGP} [that appears in (\ref{TypeIIA-from-M-theory-Witten-prescription-T>Tc})], which otherwise would have been $1 + \frac{L^4}{r^4}\left[1 + {\cal O}\left(\frac{g_sM^2}{N}\right)\right]$.}. In the high temperature ($T>T_c$) ${\cal M}$-theory dual (\ref{TypeIIA-from-M-theory-Witten-prescription-T>Tc}), evidently $r\geq r_h$ (the metric component $g_{tt}^{\cal M}$ being proportional to a warp factor $e^B$ where $B\sim\log \left(1 - \frac{r_h^4}{r^4} + {\cal O}\left(\frac{g_sM^2}{N}\right)\right)$ from the solutions to the supergravity EOM implies that for $B\in\mathbb{R}, r\geq r_h$ in the MQGP limit (\ref{MQGP_limit})).\footnote{Further, as will be shown later, the area/entanglement entropy (inclusive of ${\cal O}(R^4)$ contributions) of the HM surface (see (\ref{AHM_i}), (\ref{Page-curve-beta0}), (\ref{SEE-HM-beta0-simp}), (\ref{SEE-HM-simp-beta0}) and (\ref{SEEHMbeta-iii})) as well as the IS (see 
(\ref{AIS-i}), (\ref{EE-IS-simp}), (\ref{on-shell-L-beta0}) and (\ref{SEEISbeta-iii})) are proportional to a positive power of $M$ - the number of fractional $D3$-branes in the parent type IIB dual \cite{metrics}. Now, the contribution to $r$ integral from $r\in$ UV, i.e., $r>\sqrt{3}a\sim \biggl[1 + {\cal O}\left(\frac{\sqrt{\beta}}{N}\right) + {\cal O}\left(\frac{g_sM^2}{N}\right)\biggr]r_h\sim\left(1 + \epsilon\right)r_h, 
|\epsilon|\ll1$ \cite{MChPT} will involve replacing $M$ by $M^{\rm UV}\equiv M_{\rm eff}(r\in{\rm UV}\Leftrightarrow r>\sqrt{3}a)$ - the UV-valued effective number of fractional $D3$-branes in the aforementioned parent type IIB dual - which is vanishing small ensuring UV conformality. Therefore all integrals $\int^{r_{\rm UV}}_{r_*\ {\rm or}\ r_T}$ of integrands relevant to the HM/IS areas or entanglement entropies, will vanish as $r_{*, T}>r_h$ though being nearer to $r_h$ than $\sqrt{3}a$.} \par
From the work of \cite{Dong}, we could compute the entanglement entropy associated with holographic dual theories with the inclusion of higher derivative terms. We could express the analogue of the generalised entropy functional described in equation (\ref{DHS-RT}) using \cite{Dong} for ${\cal M}$-theory dual in the presence of ${\cal O}(R^4)$ terms as follows:
 \begin{equation}
\label{S-gen-HD}
S_{gen}({\cal R} \cup {\cal I})= S_{\rm gravity},
\end{equation}
where 
\begin{equation}
S_{\rm gravity} = \frac{S_{\rm EE}}{4 G_N^{(11)}},
\end{equation}
where $S_{\rm EE}$ is defined in (\ref{HD-Entropy-defn}). The Page curve is obtained from (\ref{S-gen-HD}) and (\ref{HD-Entropy-defn}) for the Hartman-Maldacena-like and island surfaces in \ref{Page-curve-HD}.
%%%%%%%%%%%%%%%%%%%%%%%%%%%%%%%%%%
%%%%%%%%%%%%%%%%%%%%%%%
\section{ETW-``brane'' Embedding}
\label{ETW-sub}
End of the World (ETW)-``brane'' embedding is a crucial issue associated with the presence of islands in the doubly-holographic approach. The situation is widely recognized at the stage of Einstein gravity at LO (i.e. Einstein-Hilbert(EH)  and the boundary Gibbons-Hawking-York surface terms), but little is understood for higher derivative gravity, e.g. $D=11$ supergravity at ${\cal O}(R^4)$; in top-down frameworks like the one taken into account by us, the ETW-``brane'' is a  fluxed hypersurface ${\cal W}$. We tackled the situation in two subsections within this section: first we obtained the ETW embedding exactly the stage of EH+GHY action, and the subsequent one demonstrated how the aforementioned LO embedding obtains no modifications in the MQGP limit (\ref{MQGP_limit}).
%%%%%%%%%%%%%REPLACED :\hookrightarrow by \rightarrow%%%%%%%%%
The ETW-``brane'' ${\cal W}: AdS_4^\infty\times_w M_6$ with $G_4$ fluxes threading a homologous sum of four-cycles $S^3\times[0,1]$ and $S^2\times S^2$ in $M_6=M_5(\theta_{1,2}, \phi_{1,2},\psi)\times S^1(x^{10})\rightarrow M^{SU(4)/Spin(7)}_8(t,r,\theta_{1,2}, \phi_{1,2},\psi,x^{10})$ (in the large-$N$ MQGP limit) implies: $\partial M_{11}=\left\{r=r_{\rm UV}\right\}\cup {\cal W}$. Considering the end-of-the-world (ETW)-``Brane'' embedding as,
\begin{equation}
\label{ETW_i}
x^1 = x^1(r),
\end{equation}
and inserting it into (\ref{symbolic-metric}) yields:
\begin{equation}
\label{ETW_ii}
ds^2=\alpha(r)\left[-g(r) dt^2+\left(\sigma(r)+ \left(\frac{dx^1}{dr}\right)^2\right) dr^2+\sum_{\mu=2}^3dx_{\mu}dx^{\mu}\right]+g_{mn}dx^mdx^n. 
\end{equation}
\subsection{At ${\cal O}(\beta^0)$}
\label{ETW-sub-sub-i}
If $T$ is the tension associated with the ETW-``brane'', and if ${\cal K}$ and ${\cal K}_{\tilde{m}\tilde{n}}, \tilde{m}, \tilde{n}=r, t, \mu, m$ are the extrinsic curvature scalar and tensor associated with the end-of-the-world (ETW)-``brane'', then \cite{Takayanagi-ETW},
\begin{equation}
\label{ETW_iii}
{\cal K}_{mn} - {\cal K} h_{mn} = - 9 T h_{mn},
\end{equation} 
where $h_{mn}$ represents the induced metric on the end-of-the-world brane. Focus on the $rr$-component of (\ref{ETW_iii}) in the IR. We could argue that in the IR, the embedding function $x^1(r)=x(r)$ always appears as $x'(r),  x''(r)$ in (\ref{ETW_iii}),  and writing $a = \left(b + {\cal O}\left(\frac{g_sM^2}{N}\right)\right)r_h$ \cite{EPJC-2}, 
the terms LO and NLO in $N$ and
{\footnotesize
\begin{eqnarray}
\label{ETW_iv}
  {\cal K}\left(N, g_s; r_h\right) & \sim & -\kappa_K\frac{\left(3 b^2-1\right) \sqrt{ {r_h}} ( \log N -9 \log ( {r_h}))}{\sqrt[4]{ {g_s}} \sqrt[4]{N}
     \sqrt{r- {r_h}} \left(\left(3 b^2-1\right)  \log N +9 \log ( {r_h})\right) \sqrt[3]{  {N_f}
   ( \log N -3 \log ( {r_h}))}}\nonumber\\
%   & &  + {\cal O}\left(\sqrt{r-r_h}\right),\nonumber\\
\hskip -0.1in  {\cal K}_{rr}\left(N, M, N_f, g_s; r_h; x'(r);x''(r)\right) & \sim & -\kappa_{K_{rr}}^{(1)}\frac{\sqrt{{g_s} N} \Sigma(r_h; N, N_f)}{r_h (r- {r_h})^2} \nonumber\\
   & & 
+\kappa_{K_{rr}}^{(2)} \frac{{g_s}^3 M^2 {N_f} \log ({r_h}) (\log N -12 \log ({r_h})) \Sigma(r_h; N, N_f)}{ {r_h}^2 \sqrt{{g_s} N}
   (r-{r_h})^2}\nonumber\\
   & & +\kappa_{K_{rr}}^{(3)}\frac{{r_h} x'(r) \Sigma(r_h; N, N_f)}{ \sqrt{{g_s} N}} +\kappa_{K_{rr}}^{(4)}\frac{{r_h}^2 x''(r)\Sigma(r_h; N, N_f)}{
   \sqrt{{g_s} N}},\nonumber\\
 h_{rr}\left(N, M, N_f, g_s; r_h; x'(r)\right) & \sim & -2\kappa_{K_{rr}}^{(1)}\frac{  \sqrt{g_s N} \Sigma(r_h; N, N_f)}{r_h (r- {r_h})}
  -\kappa_{K_{rr}}^{(3)} \frac{{r_h}^2 x'(r) \Sigma(r_h; N, N_f)}{ \sqrt{{g_s} N}} 
  \nonumber\\
    & & -2 \kappa_{K_{rr}}^{(2)}\frac{3 {g_s}^3 M^2 {N_f} \log
   ({r_h}) (\log N -12 \log ({r_h})) \Sigma(r_h; N, N_f)}{ {r_h} \sqrt{{g_s} N} (r-{r_h})},
\end{eqnarray}
}
wherein $\Sigma(r_h; N, N_f)\equiv \left(6  \log N    {N_f}-3   {N_f} \log \left(9 a^2
    {r_h}^4+ {r_h}^6\right)\right)^{2/3}$, numerical pre-factors appear in $\kappa_{K}, \kappa_{K_{rr}^{(i=1,2,3,4)}}$. Since $b=\frac{1}{\sqrt{3}}+\epsilon$, (e.g., in the $\psi=2n\pi, n=0, 1, 2$-coordinate patches, $|\epsilon|\sim r_h^2\left(|\log r_h|\right)^{9/2}N^{-9/10-\alpha_b}, \alpha_b>0$ \cite{HD-MQGP}),  (\ref{ETW_iii}) may be approximated with $K_{mn} \sim -9 T h_{mn}$. Consider $m=n=r$ component of the same, we found that unlike  ${\cal K}_{rr}\left(N, M, N_f, g_s; r_h; x'(r);x''(r)\right)$, there is no  $x''(r)$ term in $ h_{rr}\left(N, M, N_f, g_s; r_h; x'(r)\right)$. Hence,
\begin{equation}
\label{x''=0}
x''(r)=0.
\end{equation}
urthermore, the LO-in-$N$ terms in the IR ($r={\cal O}(1)r_h$ of) ${\cal K}_{rr}\left(N, M, N_f, g_s; r_h; x'(r); x''(r)\right)$ are proportional to $h_{rr}\left(N, M, N_f, g_s; r_h;x'(r)\right)$ with a proportionality constant ``-2''. This implies that ETW-``brane'' has the following tension:
\begin{equation}
\label{T_ETW}
T\sim\frac{1}{r_h}.
\end{equation}
However, at NLO in $N$, the coefficients are proportional to one another, although having a constant for proportionality of ``-1''. This is balanced by $x'(r)=0$, i.e.,
\begin{equation}
\label{ETW_vii}
x^1(r)={\rm constant},
\end{equation}
and we assume the constant is zero.
\subsection{No Boundary Terms Generated at ${\cal O}(R^4)$}
\label{ETW-sub-sub-ii}
The eleven-dimensional supergravity action, which includes ${\cal O}(R^4)$ terms, is presented by the following:
\begin{eqnarray}
\label{D=11_O(l_p^6)}
& &  S = \frac{1}{2\kappa_{11}^2}\int_{M_{11}} \sqrt{-G^{\cal M}}\left[  {\cal R} *_{11}1 - \frac{1}{2}G_4\wedge *_{11}G_4 -
\frac{1}{6}C\wedge G\wedge G\right] + \frac{1}{\kappa_{11}^2}\int_{r=r_{\rm UV}} d^{10}x \sqrt{h} K \nonumber\\
& & \hskip 0.2in+ \frac{1}{(2\pi)^43^22^{13}}\left(\frac{2\pi^2}{\kappa_{11}^2}\right)^{\frac{1}{3}}\int_{M_{11}} d^{11}x\sqrt{-G^{\cal M}}\left( J_0 - \frac{1}{2}E_8\right) + \left(\frac{2\pi^2}{\kappa_{11}^2}\right)\int C_3\wedge X_8 \nonumber\\
& & \hskip 0.2in
+\frac{1}{\kappa_{11}^2}\int_{\cal W} d^{10}x \sqrt{\gamma} ({\cal K}-9T),\nonumber\\
& & 
\end{eqnarray}
where the terms $J_0$ and $E_8$ have been given in (\ref{J0+E8-definitions}). Because of the hierarchical structure, $|t_8^2G^2R^3|<|E_8|<|J_0|$ \cite{HD-MQGP} in the MQGP limit (\ref{MQGP_limit}), we worked with $J_0$; $\gamma$ , ${\cal K}$ and $T$ in in (\ref{D=11_O(l_p^6)}) have already been defined. From \cite{HD-MQGP}, 
\begin{eqnarray}
\label{ETW_OR4_i}
& & \delta J_0 \sim -\delta g_{\tilde{M}\tilde{N}}\Biggl[g^{M\tilde{M}}R^{H\tilde{N}NK} + g^{N\tilde{N}}R^{HM\tilde{M}K} + g^{K\tilde{M}}R^{HMN\tilde{N}}\Biggr]\tilde{\chi}_{HMNK}\nonumber\\
& & -\frac{1}{2}\Biggl[g^{H\tilde{H}}\left(D_{[N_1}\Biggl(D_{K_1]}\delta g_{M_1\tilde{H}} + D_{M_1}\delta g_{|K_1]\tilde{H}} - D_{\tilde{H}}\delta g_{|K_1]M_1}\Biggr)\right)\chi_H^{M_1N_1K_1}\Biggr],
\end{eqnarray}
where,
\begin{eqnarray}
\label{ETW_OR4_ii}
& & \tilde{\chi}_{HMNK} \equiv R_{PMNQ} R_H^{\ RSP}R^Q_{\ RSK} - \frac{1}{2}R_{PQKN}R_H^{\ RSP}R^Q_{\ RSM};\nonumber\\
& & \chi_H^{M_1N_1K_1} \equiv R_{P\ \ \ \ Q}^{\ M_1N_1}R_H^{\ RSP} R^{Q\ \ \ K_1}_{\ RS} - \frac{1}{2}R_{PQ}^{\ \ M_1N_1}R_H^{\ RSP}R^{Q\ \ \ K_1}_{\ RS}.
\end{eqnarray}
As a result, we are able to identify that a usual boundary term including covariant derivatives of metric fluctuations that would have to be cancelled out by a suitable boundary term (using Stokes theorem) is:
\begin{eqnarray}
\label{ETW_OR4_iii}
& & \int_{{\cal W}}D_{[K_1|}\delta g_{M_1\tilde{H}} \chi^{\tilde{H}M_1[N_1K_1]}d\Sigma_{|N_1]} = \int_{{\cal W}}D_{[K_1|}\delta g_{M_1\tilde{H}} \chi^{\tilde{H}M_1[N_1K_1]}\eta_{|N_1]}\sqrt{-h}d^{10}y.
\end{eqnarray}
The (dual to the) unit normal vector to ${\cal W}$ provided by:
\begin{equation}
\label{ETW_OR4_iv}
\eta_M = \left(\eta_r, \eta_x, \eta_{M\neq r, x}\right) = \frac{\left(-\frac{dx^1(r)}{dr}\sqrt{G^{rr}_{\cal M}}, \sqrt{G^{x^1x^1}_{\cal M}},{\bf 0}\right)}{\sqrt{\left(G^{xx}_{\cal M}\right)^2 + \left(G^{rr}_{\cal M}\right)^2\left(\frac{dx^1(r)}{dr}\right)^2}}. 
\end{equation}
Hence,
\begin{equation}
\label{ETW_OR4_v}
\left.\eta_M\right|_{x^1(r)={\rm constant}} = \left(0, \sqrt{G^{xx}_{\cal M}},{\bf 0}\right).
\end{equation} 
According to (\ref{ETW_OR4_v}), one will be needed to considera $\chi^{HM[x^1 N]}$. Using the fact that the only non-vanishing linearly independent Riemann curvature tensor for the metric (\ref{symbolic-metric}) with one index along $x^1$ is $R_{x^1tx^1t}$, and (\ref{ETW_OR4_ii}), we obtained:
\begin{eqnarray}
\label{ETW_OR4_vi}
& & \chi^{tx^1tx^1} = \frac{1}{2}\left(g_{x^1x^1}^{\cal M}\right)^2\left(R_t^{\ x^1x^1t}\right)^2R^{tx^1x^1t}.
\end{eqnarray}  
Utilizing,
\begin {eqnarray*}
\label{ETW_OR4_vii}
& & R_t^{x^1x^1t}\sim -\frac{ \left(9 a^2+{r_h}^2\right)}{ {N_f} {r_h}^2 \left(6 a^2+{r_h}^2\right)
   (\log N -3 \log ({r_h})) \sqrt[3]{\frac{12 \pi }{{g_s}}+3 \log N  {N_f}-9 {N_f} \log
   ({r_h})}};\nonumber\\
   & & R^{tx^1tx^1} \sim \frac{ \left(9 a^2 ({g_s} \log N  {N_f}+4 \pi )-3 {g_s} {N_f} \left(9
   a^2+{r_h}^2\right) \log ({r_h})+{g_s} \log N  {N_f} {r_h}^2\right)}{ \sqrt{{g_s}}
   {N_f}^3 {r_h}^3 \left(6 a^2+{r_h}^2\right) (r-{r_h}) (\log N -3 \log ({r_h}))^3}.
\end {eqnarray*}
We found that up to leading order in $N$ and $\log N, |\log r_h|$, and in the IR (near $r=r_h$),
\begin {eqnarray*}
\label{ETW_OR4_viii}
& & \left.\beta \chi^{tx^1tx^1}\right|_{x^1={\rm constant}}\sim \frac{\beta}{N}\frac{1}{\sqrt{g_s}N_f^{25/3}r_h^3(r-r_h)\left(\log N - 3 |\log r_h|\right)^{10/3}}.
\end {eqnarray*}
As a result, we can observe from (\ref{ETW_OR4_iii})($\chi^{ttx^1t}=\chi^{x^1x^1x^1t}=0$):
\begin{eqnarray}
\label{ETE_OR4_ix}
& &  \left.\int_{{\cal W}}D _{[K_1|}\delta g^{\cal M}_{M_1\tilde{H}} \chi^{\tilde{H}M_1[N_1K_1]}d\Sigma_{|N_1]}\right|_{x^1={\rm constant}} \sim \int_{{\cal W}}D_{t}\delta g^{\cal M}_{x^1t}
\chi^{tx^1[x^1t]}\eta_{x^1}\sqrt{-h}d^{10}y.
\end{eqnarray}
Since, $g^{\cal M}_{x^1t}=0$. As a result, confined to $x^1=$constant, there isn't any surface term produced if $\delta g^{\cal M}_{x^1t}=0$ is chosen.
\section{Page Curve without Higher Derivative Terms }
\label{Page-curve-WHD}
Here, we obtained the Page curve using the doubly holographic setup constructed in \ref{DHS-M-Theory}. In this part, we do not consider the ${\cal O}(R^4)$ terms in the supergravity action. Therefore, we used the Ryu-Takayanagi formula to compute the entanglement entropies of Hartman-Maldacena-like and Island surfaces in \ref{EE-HM-WHD} and \ref{EE-IS-WHD}. The Page curve has been obtained in \ref{Page-curve-WHD-terms}. In the absence of ${\cal O}(R^4)$ terms, the ${\cal N}=1, D=11$ supergravity action has been provided as follows:
\begin{eqnarray}
\label{D=11}
& & S = \frac{1}{2\kappa_{11}^2}\int_{M_{11}} \sqrt{-G^{\cal M}}\left[  {\cal R} *_{11}\wedge 1 - \frac{1}{2}G_4\wedge *_{11}G_4 -
\frac{1}{6}C\wedge G\wedge G\right] \nonumber\\
& & 
+\frac{1}{\kappa_{11}^2}\int_{r=r_{\rm UV}} d^{10}x \sqrt{h} K +\frac{1}{\kappa_{11}^2}\int_{\cal W} d^{10}x \sqrt{\gamma} ({\cal K}-9T).
\end{eqnarray}
$\gamma$ , ${\cal K}$ and $T$ in (\ref{D=11}) correspond to the induced metric, trace of extrinsic curvature (${\cal K}_{mn}$), and tension of the ETW brane(${\cal W}$: $\partial M_{11} = \left\{r=r_{\rm UV}\right\}\cup {\cal W}$). For $T>T_c$ on the gauge theory side, the black hole metric associated with the ${\cal M}$-theory dual is provided via the following:
\begin{eqnarray}
\label{TypeIIA-from-M-theory-Witten-prescription-T>Tc-BHIP-ch7}
\hskip -0.1in ds_{11}^2 & = & e^{-\frac{2\phi^{\rm IIA}}{3}}\Biggl[\frac{1}{\sqrt{h(r,\theta_{1,2})}}\left(-g(r) dt^2 + \left(dx^1\right)^2 +  \left(dx^2\right)^2 +\left(dx^3\right)^2 \right)
\nonumber\\
& & \hskip -0.1in+ \sqrt{h(r,\theta_{1,2})}\left(\frac{dr^2}{g(r)} + ds^2_{\rm IIA}(r,\theta_{1,2},\phi_{1,2},\psi)\right)
\Biggr] + e^{\frac{4\phi^{\rm IIA}}{3}}\left(dx^{11} + A_{\rm IIA}^{F_1^{\rm IIB} + F_3^{\rm IIB} + F_5^{\rm IIB}}\right)^2, \nonumber\\
\end{eqnarray}
where $A_{\rm IIA}^{F^{\rm IIB}_{i=1,3,5}}$ represent the type IIA RR one-forms produced from the type IIB triple T/SYZ-dual. $F_{1,3,5}^{\rm IIB}$ fluxes in \cite{metrics}, and $g(r) = 1 - \frac{r_h^4}{r^4}$. We could write down the metric ((\ref{TypeIIA-from-M-theory-Witten-prescription-T>Tc-BHIP-ch7}) near the $\psi=2n\pi, n=0, 1, 2$-coordinate patch in the following form: %(exact form of this metric in terms of the model parameters is possible to read off via appendix {\bf A} of \cite{McTEQ}):
\begin{equation}
\label{symbolic-metric}
ds^2=\alpha(r)[-g(r) dt^2+\sigma(r) dr^2+dx_{\mu}dx^{\mu}]+g_{mn}dx^mdx^n ,
\end{equation}
where $x_\mu(\mu=1,2,3)$ denotes spatial coordinates, $r$ denotes radial coordinates, and $x^m(m=5,6,7,8,9,10)$ denotes six angular coordinates $(\theta_{1,2},\phi_{1,2},\psi,x^{10})$ within the conifold geometry. We found the following results through the comparison of equations (\ref{TypeIIA-from-M-theory-Witten-prescription-T>Tc-BHIP-ch7}) and (\ref{symbolic-metric}):
\begin{eqnarray}
\label{alpha+sigma}
& &
\alpha(r, \theta_{1,2})=\frac{e^{-\frac{2\phi^{\rm IIA}}{3}}}{\sqrt{h(r,\theta_{1,2})}},\nonumber \\
& &  \sigma(r, \theta_{1,2})=\frac{h(r,\theta_{1,2})}{g(r)},
\end{eqnarray}
where $\phi^{\rm IIA}$ represents the type IIA dilaton profile, which could be described as $G_{x_{10}x_{10}}^{\cal M}=e^{\frac{4\phi^{\rm IIA}}{3}}$. The following equation has been used to calculate the volume of the compact six-fold:
\begin{equation}
\label{V_int-BHIP}
{\cal V}_{\rm int}=\int \prod_m dx^m \sqrt{-g} .
\end{equation}

\subsection{Entanglement Entropy Contribution from Hartman-Maldacena-like Surface}
\label{EE-HM-WHD}
\vspace{-0.4cm}
To calculate the time-dependent entanglement entropy associated with a Hartman-Maldacena-like surface, we first look at the induced metric on a constant $x^1$ slice, which was calculated utilizing (\ref{symbolic-metric}) as follows:
\begin{equation}
\label{metric-HM-t(r)}
ds^2|_{x^1=x_R}=\alpha(r)\Biggl[\left(-g(r)  \dot{t}(r)^2 +\sigma(r)\right)dr^2+(dx^2)^2+(dx^3)^2\Biggr]+g_{mn}dx^mdx^n.
\end{equation}
The area density functional for a Hartman-Maldacena-like surface could be calculated using (\ref{metric-HM-t(r)}) as follows:
\begin{equation}
\label{Area-HM}
{\cal A}_{HM}(t_b)=\frac{A}{{\cal V}_2}=\int dt \sqrt{\left(-g(r)H(r) +\sigma(r)H(r) \dot{r}(t)^2\right)} \equiv \int dt {\cal L},
\end{equation}
where $H(r)={\cal V}_{int}^2\alpha(r)^3$ and ${\cal V}_2=\int \int dx^2 dx^3$ . The following is the constant of motion E (which in turn is the energy of the minimum surface) since there is no explicit $t$ dependency in the Lagrangian:
\begin{equation}
E = \frac{\partial {\cal L}}{\partial \dot{r}(t)} \dot{r}(t) -{\cal L},
\end{equation}
where $\dot{r}(t)=\frac{dr(t)}{dt}$. Using equation (\ref{Area-HM}), we can reduce the preceding equation as follows:
\begin{equation}
E = \frac{g(r)H(r)}{\sqrt{\left(-g(r)H(r) +\sigma(r)H(r) \dot{r}(t)^2\right)}}.
\end{equation}
When we solved the preceding equation for $\dot{r(t)}$, we got:
\begin{equation}
\label{r-dot}
\dot{r}(t)=\pm \sqrt{\left(\frac{g(r)}{\sigma(r)}\left(1+\frac{g(r)H(r)}{E^2}\right)  \right)}.
\end{equation}
At $r=r^*$, $\dot{r}(t)|_{r=r^*}=0$. As a result, equation (\ref{r-dot}) is going to be rewritten as follows:
\begin{equation}
\label{r_*}
\frac{g(r^*)}{\sigma(r^*)}\left(1+\frac{g(r^*)H(r^*)}{E^2}\right)=0.
\end{equation}
implying
\begin{equation}
E^2=-g(r^*)H(r^*),
\end{equation}
The highest value of $r$ for the a surface having energy $E$ is $r^*$. In the entire geometry, a Hartman-Maldacena-like surface begins at  $x^1=x_{\cal R}$, rises to $r^*$, and then descends to its thermofield double partner. The equation (\ref{Area-HM}) has been simplified as follows:
\begin{equation}
\label{AHM-sim}
{\cal A}_{HM}(t_b)=\int dr \sqrt{-\frac{g(r)H(r)}{\dot{r}(t)^2}+\sigma(r)H(r)}.
\end{equation}
Using (\ref{r-dot}) and (\ref{AHM-sim}), we obtained:
\begin{equation}
\label{AHM-simplified}
{\cal A}_{HM}(t_b)=2 \int_{r_h}^{r^*} dr {\frac{\sqrt{\sigma(r)g(r)}H(r)}{E\sqrt{\left(1+\frac{g(r)H(r)}{E^2}\right)}}}.
\end{equation}
Now we identify time via the integral:
\begin{equation}
\label{tdiff}
\int_{r_h}^{r^*} \frac{dr}{\dot{r}(t)}=t_*-t_b,
\end{equation}
where $t_*=t(r^*)$ and $t_b$ represents the boundary time at $r=r_h$. The boundary time as expressed in the form of energy could be calculated using equations (\ref{r-dot}) and  (\ref{tdiff}) as follows:
\begin{equation}
\label{t_b}
t_b=- P \int_{r_h}^{r^*} \frac{dr}{\sqrt{\frac{g(r)}{\sigma(r)}\left(1+\frac{g(r)H(r)}{E^2}\right)}}.
\end{equation}
As a result, the Hartman-Maldacena-like surface's entanglement entropy is:
\begin{equation}
\label{SHM}
{\cal S}_{HM}(t_b) = \frac{{\cal A}_{HM}(t_b)}{4 G_N^{(11)}}.
\end{equation}
\subsubsection{Hartman-Maldacena-like Surface Analytics/Numerics}
After integrating all of the angular coordinates and consequently including a $(2\pi)^4$ emerging from integration w.r.t. $\phi_{1,2}, \psi, x^{10}$, the area corresponding to the Hartman-Maldacena-like surface is obtained as:
\begin{eqnarray}
\label{AHM_i}
& &  \hskip -0.3in A_{\rm HM} \sim (2\pi)^4 \int_{r_h}^{r^*} dr \Biggl({\pi ^9 E M^2 \sqrt[10]{N} \log ^2(2) (\log (64)-1)^2 N_f^4 g_s^{9/4} \log ^2(r) (\log (N)-3 \log (r))^4 }\nonumber\\
& & \times \left(N_f g_s \left(r^2 (2 \log (N)-18 \log (r)+3)-2 r_h^2 (\log (N)-54 r \log (r))\right)+8 \pi 
   r_h^2\right){}^2 \Biggr).
\end{eqnarray}
Consider $\left|\log r_h\right|\gg\log N$ \cite{Bulk-Viscosity-McGill-IIT-Roorkee} in (\ref{AHM_ii}), we obtained:
\begin{equation}
\label{AHM-i}
A_{\rm HM}\sim {\cal O}(1)\times10^5M^2 \sqrt[10]{N} N_f^6 g_s^{17/4} r_h^4 \left(r_*-r_h\right) \log ^4\left(r_h\right) \left(\log (N)-3 \log \left(r_h\right)\right)^4.
\end{equation}
Now, the principal value of the following integral gives $t_b$:
{\footnotesize
\begin{eqnarray}
\label{tb_i}
& &  t_b \sim{\cal P}\int_{r_h}^{r_*} dr \Biggl(\frac{ E^2\sqrt{N} r^2 \sqrt{g_s} \left(N_f g_s \left(r^2 (2 \log (N)-18 \log (r)+3)-2 r_h^2 (\log (N)-54 r \log (r))\right)+8 \pi 
   r_h^2\right){}^2}{\left(r^4-r_h^4\right) \left(N_f g_s \left(\left(2 r^2-6 a^2\right) \log (N)+3 r \left(108 a^2 \log (r)+r-6 r \log (r)\right)\right)+24 \pi  a^2\right)^2}\Biggr)\nonumber\\
   & & \hskip 0.2in \sim \lim_{\epsilon_1\rightarrow0}\left[4 \pi ^{33/2} E^2 \sqrt{{g_s}} \sqrt{N} \left(\frac{\log ({r_h}-r)}{4 {r_h}}-\frac{\log (r+{r_h})}{4
   {r_h}}+\frac{\tan ^{-1}\left(\frac{r}{{r_h}}\right)}{2 {r_h}}\right)\right]_{r=r_h+\epsilon_1}^{r=r_*}\nonumber\\
   & & \hskip 0.2in -\frac{ E^2 \sqrt{{g_s}} \sqrt{N} (\log (-\epsilon )-\log ({r_h}-{r_*}))}{{r_h}}+\frac{\pi
   ^{33/2} E^2 \sqrt{{g_s}} \sqrt{N} ({r_*}-{r_h})}{2
   {r_h}^2} + {\cal O}\left(({r_*}-{r_h})^2\right).
\end{eqnarray}
}
The Principal value necessitates: $r_* = r_h + \epsilon_1$. Let $a=\left(\frac{1}{\sqrt{3}}+\epsilon\right)r_h$, upto the leading order in $\epsilon$, we obtained:
\begin{eqnarray}
\label{tb}
& & t_b= \frac{ E^2\pi^{33/2} \sqrt{N} \sqrt{g_s} \left(r_*-r_h\right)}{2 r_h^2}.
\end{eqnarray}
Considering $\epsilon_1=\tilde{\epsilon}_1r_h$, we found that for (i) $\tilde{\epsilon}_1\sim0.5$, we have to include the terms up to ${\cal O}(\tilde{\epsilon}_1^2)$ in $t_b$, (ii) $\tilde{\epsilon}_1\sim\frac{1}{\sqrt{2}}$, we have to include the terms up to ${\cal O}(\tilde{\epsilon}_1^3)$ in $t_b$, (iii) $\tilde{\epsilon}_1\sim1$, we have to include the terms up to ${\cal O}(\tilde{\epsilon}_1^4)$ in $t_b$, and (iv) $\tilde{\epsilon}_1\sim\sqrt{5}$, we have to include the terms up to ${\cal O}(\tilde{\epsilon}_1^5)$ in $t_b$. We found that all the way to the Page time,$\tilde{\epsilon}_1\ll1$, and consequently it is justified in preserving just linear terms in $\tilde{\epsilon}_1$ as in (\ref{tb}). We calculated the entanglement entropy associated with a Hartman-Maldacena-like surface using the equations (\ref{AHM-i}) and (\ref{tb}) as follows:
\begin{eqnarray}
\label{Page-curve-beta0}
& & S_{\rm HM}^{\beta^0} = \frac{A_{\rm HM}}{4G_N^{(11)}}\sim \frac{{\cal O}(1)\times10^{-4}M^2  N_f^6 g_s^{15/4} r_h^6 \log ^4\left(r_h\right) \left(\log (N)-3 \log \left(r_h\right)\right){}^4}{E^2 G_N^{(11)} N^{2/5}}t_b.
\end{eqnarray}
As a result, the entanglement entropy related with a Hartman-Maldacena-like surface grows linearly with time.

\subsection{Entanglement Entropy Contribution from Island Surface}
\label{EE-IS-WHD}
We take a constant $t$ slice to calculate the entanglement entropy for an island surface. As a result, utilizing equation (\ref{symbolic-metric}), we expressed the induced metric associated with the island surface as follows:
\begin{equation}
\label{induced-metric-IS}
ds^2|_{constt- time}=\alpha(r)\Biggl[\left(\sigma(r) +\dot{x}(r)^2\right)dr^2+(dx^2)^2+(dx^3)^2\Biggr]+g_{mn}dx^mdx^n,
\end{equation}
where $x^1(r)\equiv x(r)$, and $\dot{x}(r) \equiv \frac{dx(r)}{dr}$. We computed the area density functional associated with the island surface as shown below:
\begin{equation}
\label{AIS-M-Theory}
{\cal A}_{IS}=\frac{A}{{\cal V}_2} = \int dr \sqrt{\left( H(r) \sigma(r)+ H(r)\dot{x}(r)^2\right)} \equiv \int dr {\cal L}  ,
\end{equation}
wherein
\begin{equation}
\label{H(r)}
H(r)= {\cal V}_{int}^2\alpha(r)^3 .
\end{equation} 
Because $x(r)$ represents a cyclic coordinate, the conjugate momentum related to $x(r)$ is $p_{x(r)}=\frac{\partial {\cal L}}{\partial \dot{x}(r)}$.The constant of motion is written as follows:
\begin{equation}
\label{px}
p_{x(r)}={\cal C},
\end{equation}
where ${\cal C}$ is the constant. We simplified the expression (\ref{px}) utilizing equation (\ref{AIS-M-Theory}) as given below: 
\begin{equation}
\label{dx/dr-BHIP}
\frac{H(r)\dot{x}(r)}{{\cal L}} = {\cal C}.
\end{equation}
When we solved the preceding equation for $\dot{x}(r)$, we were provided with:
\begin{equation}
\label{dotx}
\dot{x}(r)=\pm {\cal C}\sqrt{\frac{\sigma(r)}{H(r)-{\cal C}^2 \sigma(r)}}
\end{equation}
At the turning point, $r=r_T$:
\begin{equation}
\label{dr/dx}
\left(\frac{dr}{dx}\right)_{r=r_T} =0.
\end{equation}
Constant ${\cal C}$ could be calculated using the equations (\ref{dotx}) and (\ref{dr/dx}), as shown below:
\begin{equation}
{\cal C}=\pm \sqrt{\frac{H(r_T)}{\sigma(r_T)}}.
\end{equation}
Taking the previously stated expression of ${\cal C}$, the equation (\ref{dx/dr-BHIP}) reduces to the following structure:
\begin{equation}
\label{dx/dr-1}
\frac{dx(r)}{dr}=\pm \sqrt{\frac{H(r_T) \sigma(r_T)}{H(r)\sigma(r_T)-H(r_T)\sigma(r)}} .
\end{equation}
Now, the area density functional associated with the island surface has been  simplified using the equations (\ref{AIS-M-Theory}) and (\ref{dx/dr-1}).
\begin{equation}
\label{AIS-integral}
{\cal A}_{IS}=2 \int_{r_h}^{r_T} \sqrt{H(r) \sigma(r) + \frac{H(r) H(r_T)\sigma(r)}{H(r)\sigma(r_T)-H(r_T)\sigma(r)}}.
\end{equation}
Hence, the entanglement entropy associated with the island surface is obtained as follows:
\begin{equation}
\label{EE-IS-WHD-TERMS}
{\cal S}_{IS} = \frac{{\cal A}_{IS}}{4 G_N^{(11)}}
\end{equation}
Since, there is no time dependence in (\ref{EE-IS-WHD-TERMS}), and hence entanglement entropy of the Hawking radiation for the island surface is constant.

\subsubsection{Island Surface Analytics}
When (\ref{AIS-integral}) is evaluated, the entanglement entropy is as follows:
\begin{eqnarray}
\label{AIS-i}
& &
{\cal A}_{\cal IS}=\int_{r_h}^{r_T} \Biggl(\frac{2 \sqrt{2} \pi  M N^{3/10} r N_f^2 g_s^{11/8} \log (r) (\log (N)-3 \log (r))^2}{\sqrt{r^4-r_h^4}} \nonumber\\
& & \times \left(N_f g_s \left(r^2 (2 \log (N)-18 \log (r)+3)-2 r_h^2 (\log (N)-54 r \log (r))\right)+8 \pi 
   r_h^2\right)\Biggr),
\end{eqnarray}
implying
%{\scriptsize
%\begin{eqnarray}
%& &
%{\cal A}_{\cal IS}= \Biggl[\sqrt{2} \pi  M N^{3/10} N_f^3 g_s^{19/8} \log \left(r_h\right) \left(\log (N)-3 \log \left(r_h\right)\right)^2 \Biggl(216 r_h \tilde{r}_T  \log \left(r_h\right) \,
 %  _2F_1\left(-\frac{1}{4},\frac{1}{2};\frac{3}{4};\frac{1}{\tilde{r}_T^4}\right)\nonumber\\
%  & & -r_h  \left( \log (N) \left(2 \log \left(\sqrt{1-\frac{1}{\tilde{r}_T^4}}+1\right)-\log
 %  \left(\frac{1}{\tilde{r}_T^4}\right)\right)  +\sqrt{\tilde{r}_T^4-1} \left(18 \log \left(r_h\right)-2 \log (N)-3\right)+73 \sqrt{\pi} r_h \log \left(r_h\right)\right)\Biggr) \Biggr],\nonumber\\
 %  & &
%\end{eqnarray}
%}
%where $\tilde{r}_T\equiv\frac{r_T}{r_h} > 1$. Therefore entanglement entropy from the island surface (\ref{EE-IS-WHD-TERMS}) turns out to be:
{\scriptsize
\begin{eqnarray}
\label{EE-IS-simp}
& & {\cal S}_{IS} \sim \frac{1}{4 G_N^{(11)}} \Biggl[  M N^{3/10} N_f^3 g_s^{19/8} \log \left(r_h\right) \left(\log (N)-3 \log \left(r_h\right)\right){}^2  \Biggl(216 r_h  \tilde{r}_T  \log \left(r_h\right) \,
   _2F_1\left(-\frac{1}{4},\frac{1}{2};\frac{3}{4};\frac{1}{\tilde{r}_T^4}\right)\nonumber\\
  & & -r_h  \left(\log (N) \left(2 \log \left(\sqrt{1-\frac{1}{\tilde{r}_T^4}}+1\right)-\log
   \left(\frac{1}{\tilde{r}_T^4}\right)\right)  +\sqrt{\tilde{r}_T^4-1} \left(18 \log \left(r_h\right)-2 \log (N)-3\right)+73 \sqrt{\pi} r_h \log \left(r_h\right)\right)\Biggr) \Biggr],\nonumber\\
   & &
\end{eqnarray}
}
\noindent where $\tilde{r}_T\equiv\frac{r_T}{r_h} > 1$; $r_T$ is calculated in (\ref{r_T-i}) - (\ref{r_T-iii}).
\subsection{Page curve at LO from Areas of Hartman-Maldacena-like and Island Surfaces}
\label{Page-curve-WHD-terms}
According to (\ref{Page-curve-beta0}), for the numerical values of $g_s, M, N_f, N, r_h$ chosen in appendix \ref{a-2}, $S_{\rm EE}^{\rm HM}=\frac{5\times10^{-4}t_b}{E^2}$, and from equation (\ref{EE-IS-simp}), the entanglement entropy portion of the island's surface remains independent of time, and can be calculated as 421. This contribution takes precedence after the Page time. Hence, we obtained the Page curve as shown in Fig. \ref{Page-Curve-Areas}. The Page time is calculated as $t_{\rm Page}=8.5\times10^5 E^2$. To produce the same $t_{\rm Page}$ with the insertion of ${\cal O}(\beta)$``anomaly terms'' as in \ref{Page-curve-plot-HD}, $E=1.1$. Using this $t_{\rm Page}$, one can observe that $\tilde{\epsilon}_1\sim 10^{-5}\ll1$, as indicated below (\ref{tb}).
\begin{figure}
\begin{center}
\includegraphics[width=0.60\textwidth]{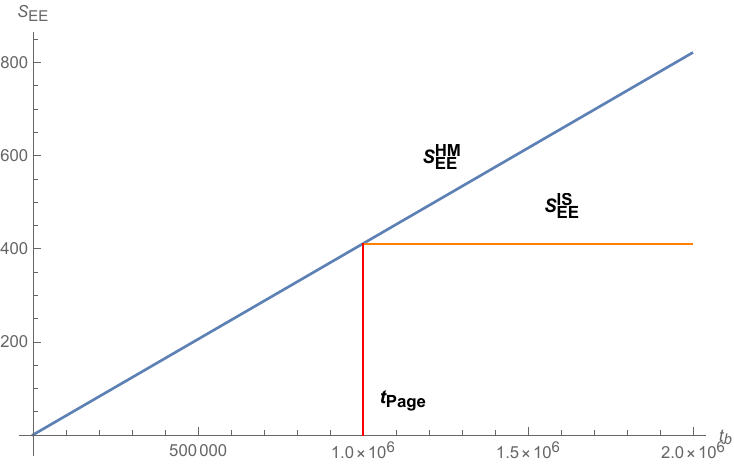}
\end{center}
\caption{Page curve up to ${\cal O}(\beta^0)$ of an eternal neutral black hole in ${\cal M}$-theory dual via doubly holographic setup. The blue line in the figure indicates the area of the Hartman-Maldacena-like surface, and the orange line refers for the area of the island surface; $E$ in (\ref{Page-curve-beta0}) is set to $1.1$ to obtain the same order of the magnitude of $t_{\rm Page}$ as in Fig. \ref{Page-time-vs-SBH-plot}.}
\label{Page-Curve-Areas}
\end{figure}
 
\section{Page Curve with Higher Derivative Terms}
\label{Page-curve-HD}
We obtained the Page curve corresponding to a neutral black hole in with the inclusion of higher derivative terms that are quartic in the Riemann curvature tensor in this section. This section has been broken into four subsections. We calculated the entanglement entropy associated with the HM-like surface in subsection \ref{EE-HM-HD}, discussed the ``Swiss-Cheese'' structure of the identical surface in subsection \ref{Swiss-Cheese-i}, computed the entanglement entropy of island surface in \ref{EE-IS-HD}, and at last obtained the Page curve corresponding to an eternal black hole in subsection \ref{Page-curve-plot-HD} utilizing the outcomes that were obtained from previous subsections.\par
We conducted the same analysis as in \ref{Page-curve-WHD}. \cite{Dong} was used to compute the entanglement entropy associated with Hartman-Maldacena-like surface and island surface in with the inclusion of higher derivative terms. In generic higher derivative gravity theories, we are able to calculate the holographic entanglement entropy (\ref{HD-Entropy-defn}) as follows:
\begin{itemize}
\item When the holographic dual consists of a $(d+1)$ dimensional gravitational background, then with the help of the embedding function, compute the induced metric associated with the co-dimension two surface.

\item Using the above-mentioned induced metric, compute (\ref{HD-Entropy-defn}). This yields the holographic entanglement entropy as an expression of the embedding function and its derivatives.

\item Find the solution to the embedding function's equation of motion.

\item In higher derivative gravity theories, putting the solution found in the previous step inside the action yields the holographic entanglement entropy.
\end{itemize}
{\bf Contribution from $J_0$ term:} In addition to the leading order term in the gravitational action, we need to consider the $J_0$ term to compute the holographic entanglement entropy. Let us see, how $J_0$ term contribute to the holographic entanglement entropy.\par
We derived the four terms via the Wald entanglement entropy formula in (\ref{HD-Entropy-defn}), which is $\frac{\partial J_0}{\partial R_{z\bar{z}z\bar{z}}}$ for the two extremal surfaces. They are listed in \ref{PT-HM} pertaining to the Hartman-Maldacena-like surface and \ref{PT-IS} for the island surface.
\begin{eqnarray}
\label{Wald-J0-c-r}
\frac{\partial J_0}{\partial R_{z\bar{z}z\bar{z}}}=-4 x^2 \left(\frac{\partial J_0}{\partial R_{txtx}}\right).
\end{eqnarray}
To determine the second part in the formula (\ref{HD-Entropy-defn}), we must first compute the four types of derivatives for the Hartman-Maldacena-like and island surfaces, which are listed below:
\begin{eqnarray}
\label{second-der-J0}
& &
\left(\frac{\partial^2 {J_{0}}}{\partial R_{zizj}\partial R_{\bar{z}m\bar{z}l}}\right) K_{zij}K_{\bar{z}ml}=\Biggl(\frac{\partial^2 {J_0}}{\partial R_{xixj}\partial R_{xmxl}}+x^4\frac{\partial^2 {J_0}}{\partial R_{titj}\partial R_{tmtl}} -2 x^2 \frac{\partial^2 {J_0}}{\partial R_{titj}\partial R_{xmxl}}\nonumber\\
& & +4 x^2 \frac{\partial^2 {J_0}}{\partial R_{tixj}\partial R_{xmtl}} \Biggr)\left(\frac{1}{x^2}K_{tij}K_{tml}+K_{xij}K_{xml} -\frac{i}{x} K_{tij} K_{xml} +\frac{i}{x} K_{xij} K_{tml}\right),
\end{eqnarray}
The numerator and denominator coefficients in the preceding equation are $x=x_R$ for the Hartman-Maldacena-like surface and $x=x(r)$ for the island surface. All four types of variations occurring in equation (\ref{second-der-J0}) for the Hartman-Maldacena-like surface in appendix \ref{PT-HM} and island surface in appendix \ref{PT-IS} have been computed and listed.

\subsection{Entanglement Entropy Contribution from Hartman-Maldacena-like Surface}
\label{EE-HM-HD}
Hartman-Maldacena-like surface in ${\cal M}$-theory dual corresponds to a co-dimension two surface situated at $x^1=x_R$. Utilizing equation (\ref{HD-Entropy-defn}), we are able to write the mathematical expression that describes the entanglement entropy associated with a Hartman-Maldacena-like surface as follows:
\begin{equation}
\label{HD-Entropy-HM}
S_{EE}=\int dr dx^2 dx^3  d\theta_1 d\theta_2dxdydz dx^{10} \sqrt{-g}\Biggl[\frac{\partial {\cal L}}{\partial R_{z\bar{z}z\bar{z}}}+\sum_{\alpha}\left(\frac{\partial^2 {\cal L}}{\partial R_{zizj}\partial R_{\bar{z}m\bar{z}l}}\right)_{\alpha} \frac{8 K_{zij}K_{\bar{z}ml}}{(q_{\alpha}+1)}\Biggr],
\end{equation}
where $g$ denotes determinant of the induced metric defined in (\ref{metric-HM-t(r)}) for the co-dimension two surface. For the metric (\ref{metric-HM-t(r)}), the entanglement entropy associated with the Hartman-Maldacena-like surface using (\ref{HD-Entropy-HM}) and appendix \ref{PT-HM} after the angular integrations is obtained as\footnote{The angular integrations have been performed using (\ref{angular-integrations-2}) and (\ref{angular-integrations-1}). We have taken the most dominant terms in the leading order in $N$ from \ref{PT-HM} for the ${\cal O}(\beta^0)$ and ${\cal O}(\beta)$ contributions.}:
\begin{equation}
\label{Lag-total-HM}
S_{\rm EE}^{\rm total,\ {\rm HM}}=\int dr {\cal L}_{\rm {Total}}^{\rm {HM}}=\int dr \left({\cal L}_{0}^{\rm HM}+ {\cal L}_{\cal W}^{\rm HM}+ {\cal L}_{\cal A}^{\rm HM}\right),
\end{equation}
where
\begin{eqnarray}
\label{Lag0waldA}
& & {\cal L}_{0}^{\rm HM}=  \left(  x_R^2 \lambda (r) \sqrt{\alpha (r) \left(\sigma (r)-\left(1-\frac{r_h^4}{r^4}\right) t'(r)^2\right)}\right),\nonumber\\
& & {\cal L}_{\cal W}^{\rm HM}= \left(-4 x_R^2 (\lambda_1(r)+\lambda_2(r)) \sqrt{\alpha (r) \left(\sigma (r)-\left(1-\frac{r_h^4}{r^4}\right) t'(r)^2\right)}\right), \nonumber\\
& & {\cal L}_{\cal A}^{\rm HM} = \frac{1}{x_R^2} \Biggl(Z(r) {\cal L}_1+ x_R^4 W(r){\cal L}_2 -2 x_R^2 U(r){\cal L}_3 +4 x_R^2 \left(U(r)+ 2 V(r)\right){\cal L}_4 \Biggr),
\end{eqnarray}
where $x_R$ being a constant, various $r$ dependent functions appearing in (\ref{Lag0waldA}) are listed in (\ref{alpha-sigma-lambda}), and appendix \ref{complex-to-real}, equations (\ref{dJ0overdR+d2J0overdR2}), (\ref{dJ0overdR+d2J0overdR2-i}) and (\ref{dJ0overdR+d2J0overdR2-ii}) have been used to calculate (\ref{Lag-total-HM}). Utilizing (\ref{DLag-tprime}) - (\ref{EOM-HM}), and writing $t(r) = t_0(r) + \beta t_1(r)$, the equation of motion for the embedding $t(r)$ up to ${\cal O}(\beta^0)$ is as follows:
\begin{eqnarray}
\label{EOM-beta0}
& & N^{3/10} (r-r_h)^{5/2} p_1\left(r_h\right) {t_0}''(r)+\frac{5}{2} N^{3/10} (r-r_h)^{3/2} p_1\left(r_h\right) {t_0}'(r)=0. 
\end{eqnarray}
The equation (\ref{EOM-beta0}) has the following solution:
\begin{eqnarray}
\label{solution-t0}
& & {t_0}(r) = c_2-\frac{2 c_1}{3 \left(r-r_h\right){}^{3/2}}. 
\end{eqnarray}
substitution of (\ref{solution-t0}) in (\ref{EOM-HM}) yields:
{\footnotesize
\begin{eqnarray}
\label{solution-t1}
& & \hskip -0.3in {t_1}(r)=c_3+\frac{2 r_h^{3/2} \sqrt{\kappa _{\sigma }} \left(11 {r_h}^7 \left(p_8^{\beta }\left(r_h\right)+p_9^{\beta }\left(r_h\right)\right)-3 \left(r-r_h\right){}^7 \left(p_8^{\beta
   }\left(r_h\right)+p_9^{\beta }\left(r_h\right)\right)-11 c_1{}^3 c_2 p_1\left(r_h\right)\right)}{33 c_1{}^3 M N_f^{5/3} g_s^{7/3} \left(r-r_h\right){}^{3/2} \sqrt{\kappa _{\alpha }} \kappa _{\lambda }
   \log \left(r_h\right) \left(\log (N)-9 \log \left(r_h\right)\right) \left(\log (N)-3 \log \left(r_h\right)\right){}^{5/6}}.
\end{eqnarray}
}
Keeping the term up to the leading order in $N$, we could write the embedding function $t(r)$ as follows:
\begin{eqnarray}
\label{t(r)}
t(r)=c_2-\frac{2 c_1}{3 \left(r-r_h\right){}^{3/2}}+\beta  \left(c_3-\frac{2 c_2}{3 \left(r-r_h\right){}^{3/2}}\right).
\end{eqnarray}
If $t(r=r_h)\equiv t_b$ is intended to be assessed as $t\left(r_h\biggl[1+\frac{1}{N^{n_{t_b}}}\biggr]\right), n_{t_b}\equiv{\cal O}(1)$, in the large-$N$ limit. Since (in ${\cal R}_{D5/\overline{D5}}=1$-units) $\frac{c_1}{r_h^{3/2}} \sim c_2$, i.e., $c_1 \sim r_h^{\alpha} c_2$ implying $\alpha = 3/2$, i.e. $l_p^6 \sim r_h^{3/2}$ or $l_p \sim r_h^{1/4}$ i.e.,
\begin{equation}
\label{lp-rh-relation} 
l_p \sim \left(g_s^{\frac{4}{3}}\alpha'\ ^2 \left(\frac{r_h}{{\cal R}_{D5/\overline{D5}}}\right)\right)^{\frac{1}{4}}.
\end{equation}
To identify the Hartman-Maldacena-like surface's turning point, $r_*$, we must apply the condition given below:
\begin{equation}
\label{r*-HM}
\left(\frac{1}{t'(r)}\right)_{r=r_*}=\frac{\left(r_*-r_h\right){}^{5/2}}{c_1}-\frac{\beta  c_2 \left(r_*-r_h\right){}^{5/2}}{c_1{}^2}=0,
\end{equation}
As $\frac{c_1}{r_h^{3/2}} \sim c_2$. As a result, for the purpose of determining $r_*$, (\ref{r*-HM}) could be approximated by the equation that follows:
\begin{equation}
\label{rstar-equation}
(r_*-r_h)^{3/2}-\beta  \kappa_{r_*}^\beta=0,
\end{equation}
which has the following solution,
\begin{equation}
\label{solution-rstar}
r_*=r_h+\beta ^{2/3} {\left(\kappa_{r_*}^\beta\right)}^{2/3}
\end{equation}
The entanglement entropy of the Hawking radiation corresponding to the Hartman-Maldacena-like surface is computed as given below:
 \begin{eqnarray}
 \label{SEE-HM-beta0}
 S_{\rm EE}^{\beta^0, \rm HM}= \int_{r_h}^{r_*} dr \Biggl(\lambda (r) \sqrt{\alpha (r) \left(\sigma (r)-\left(1-\frac{r_h^4}{r^4}\right) t_0'(r){}^2\right)}\Biggr),
 \end{eqnarray}
utilizing (\ref{alpha-sigma-lambda}) and (\ref{solution-t0}), (\ref{SEE-HM-beta0}) is expressed as:
{\footnotesize
\begin{eqnarray}
\label{SEE-HM-beta0-simp}
& &
S_{\rm EE}^{\beta^0, \rm HM}\sim \int_{r_h}^{r_*}\left(\frac{\sqrt{\left(r-r_h\right) r_h} \log \left(r_h\right) \left(\log (N)-9 \log \left(r_h\right)\right) \left(\log (N)-3 \log \left(r_h\right)\right){}^{5/6}}{r_h^3 \sqrt{r_h^2}}\right)\nonumber\\
& & \times \Biggl(-\frac{2^{5/6} M N^{13/10} \log (2) (\log (64)-1) N_f^{5/3} g_s^{10/3} \sqrt{\kappa _{\alpha }} \kappa _{\lambda } \sqrt{\kappa _{\sigma }}}{81\ 3^{2/3} \pi ^{11/12}}\Biggr)\nonumber\\
& & \sim \frac{2 \left(r_*-r_h\right){}^{3/2} \log \left(r_h\right) \left(\log (N)-9 \log \left(r_h\right)\right) \left(\log (N)-3 \log \left(r_h\right)\right){}^{5/6}}{3 r_h^{7/2}}\nonumber\\
& & \times \Biggl(-\frac{2^{5/6} M N^{13/10} \log (2) (\log (64)-1) N_f^{5/3} g_s^{10/3} \sqrt{\kappa _{\alpha }} \kappa _{\lambda } \sqrt{\kappa _{\sigma }}}{81\ 3^{2/3} \pi ^{11/12}}\Biggr) \nonumber\\
& &  \sim \frac{2 \left(\beta ^{2/3} {\left(\kappa^\beta_{r_*}\right)}^{2/3}\right)^{3/2} \log \left(r_h\right) \left(\log (N)-9 \log \left(r_h\right)\right) \left(\log (N)-3 \log \left(r_h\right)\right){}^{5/6}}{3 r_h^{7/2}}\nonumber\\
& & \times \Biggl(-\frac{2^{5/6} M N^{13/10} \log (2) (\log (64)-1) N_f^{5/3} g_s^{10/3} \sqrt{\kappa _{\alpha }} \kappa _{\lambda } \sqrt{\kappa _{\sigma }}}{81\ 3^{2/3} \pi ^{11/12}}\Biggr)
 \nonumber\\
& & \sim -\frac{2\ 2^{5/6} \beta  {g_s}^{10/3} \sqrt{\kappa_\alpha} \kappa^\beta_{r_*} \kappa_{\lambda} \sqrt{\kappa_{\sigma}} M N^{13/10} \log (2) (\log (64)-1) N_f^{5/3} \log \left(r_h\right)
   \left(\log (N)-9 \log \left(r_h\right)\right) \left(\log (N)-3 \log \left(r_h\right)\right){}^{5/6}}{243\ 3^{2/3} \pi ^{11/12} r_h^{7/2}} \nonumber\\
\end{eqnarray}
}
If $t_b=t(r=r_h+\epsilon)$ where $\epsilon = \frac{r_h}{N^{n_{t_b}}}, n_{t_b}\equiv{\cal O}(1)$ then
\begin{equation}
\label{tb0}
t_{b_0}=c_2-\frac{2 c_1 N^{3 n_{t_b}/2}}{3 r_h^{3/2}},
\end{equation}
we obtained the $r_h$ utilizing (\ref{tb0}) as follows:
\begin{equation}
\label{rh-tb-relation}
r_h=\frac{\left(\frac{2}{3}\right)^{2/3}}{\left(\frac{(c_2-{t_{b_0}}) \
N^{-\frac{3 n_{t_b}}{2}}}{c_1}\right)^{2/3}}
\end{equation}
We will show that $\beta$ is related to black hole horizon(see the discussion around (\ref{scalings})). When $\left|\log(r_h)\right| \gg \log(N)$ then the equation (\ref{SEE-HM-beta0-simp}), utilizing (\ref{scalings}), has been approximated by the following:
\begin{eqnarray}
\label{SEE-HM-simp-beta0}
S_{\rm EE}^{\beta^0, {\rm HM}} \sim e^{-\frac{3\kappa_{l_p}N^{\frac{1}{3}}}{2}}\frac{M N^{13/10} N_f^{5/3}\left(\left|\log(r_h)\right|\right)^{17/6}}{r_h^2}.
\end{eqnarray}
As $\frac{1}{r_h^2} \sim \left(\frac{c_2}{c_1}-\frac{t_{b_0}}{c_1}\right)^{4/3} N^{-2 n_{t_b}} = \left(\frac{c_2}{c_1}\right){}^{2/3} \left(1-\frac{t_{b_0}}{c_2}\right){}^{4/3} N^{-2 n_{t_b}}$. For $t_{b_0} \ll c_2$, \\
 $\frac{1}{r_h^2} \sim \left(\frac{c_2}{c_1}\right){}^{4/3} \left(1-\frac{4 t_{b_0}}{3 c_2}\right) N^{-2 n_{t_b}}$.
Hence,
\begin{eqnarray}
\label{SEE-HM-tb0}
& & S_{\rm EE}^{\beta^0, {\rm HM}} \sim  e^{-\frac{3\kappa_{l_p}N^{\frac{1}{3}}}{2}}M N^{13/10} N_f^{5/3}\left(1-\frac{4 t_{b_0}}{3 c_2}\right) \left(\frac{2}{3}\log \left(\frac{c_2-t_{b_0}}{c_1}\right) -n_{t_b} \log(N)\right)^{\frac{17}{6}}\nonumber\\
& & \approx e^{-\frac{3\kappa_{l_p}N^{\frac{1}{3}}}{2}}M N^{13/10} N_f^{5/3}\left(1-\frac{4 t_{b_0}}{3 c_2}\right)\left(\frac{2}{3}\log \left(\frac{c_2}{c_1}\right) -n_{t_b} \log(N)\right)^{\frac{17}{6}},\nonumber\\
\end{eqnarray}
when $c_1,c_2 <0$. The Fig. \ref{SEEHM-vs-tb0-curve-plot} for the numerical values  $N=10^{3.3}, M=N_f=3, g_s=0.1, c_1=-10^3, c_2=-10^8$, $n_{t_b}=1$ obtained from (\ref{SEEHM-tb0}), shows the linearization assumed in obtaining the last line of (\ref{SEE-HM-tb0}), can be approximately justified. In it,
\begin{eqnarray}
\label{SEEHM-tb0}
& & S_{\rm EEHM1}\sim 51.024 \left(\frac{t_{{b_0}}}{7.5\times10^7}+1\right) \left(\frac{2}{3} \log
   \left(\frac{t_{{b_0}}+10^8}{10^3}\right)-7.6\right)^{\frac{17}{6}},\nonumber\\
& & S_{\rm EEHM2}\sim51.024 \left(\frac{t_{{b_0}}}{10^8}+1\right){}^{4/3} \left(\frac{2}{3} \log
   \left(\frac{t_{{b_0}}+10^8}{10^3}\right)-7.6\right)^{\frac{17}{6}},\nonumber\\
& &  S_{\rm EEHM3}\sim 0.0354 \left(\frac{t_{{b_0}}}{7.5\times10^7}+1\right).
\end{eqnarray}
\begin{figure}
\begin{center}
\includegraphics[width=0.70\textwidth]{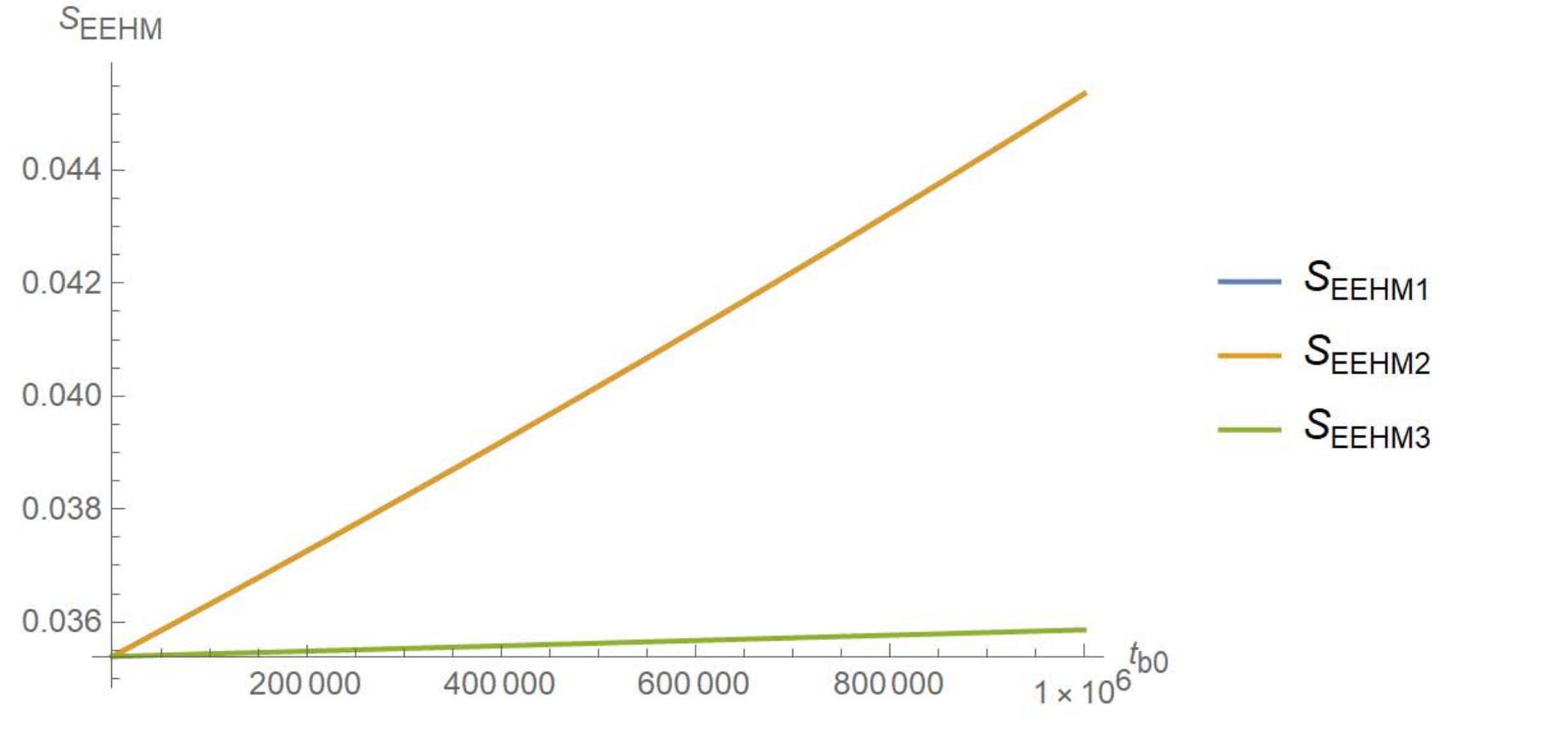}
\end{center}
\caption{Hartman-Maldacena-like surface entanglement entropy as a function of $t_{b_0}$.}
\label{SEEHM-vs-tb0-curve-plot}
\end{figure}
The entanglement entropy of the Hawking radiation originating from a Hartman-Maldacena-like surface exhibits a linear temporal dependency, as shown by equation (\ref{SEE-HM-tb0}). As a result, it grows over time and diverges at late times, i.e., when $t_{b_0} \rightarrow \infty$.

\subsection{``Swiss-Cheese'' Structure of $S_{\rm EE}^{\beta^0, {\rm HM}}$ in a Large-$N$-Scenario}
\label{Swiss-Cheese-i}
The equation for $S_{\rm EE}^{\beta^0, {\rm HM}}$ in (\ref{SEE-HM-tb0}) for the values of intergation constants $c_{1,2}$ utilized beneath the same proposes the following hierarchy: $
|c_2|\sim e^{\kappa_{c_2}|c_1|^{1/3}},\ |c_1|\sim N$. Let us assume that $c_{1,2}<0$, equation (\ref{SEE-HM-tb0}) can be re-expressed as follows:
\begin{eqnarray}
\label{SHMEEbeta0-modc12}
& & \hskip -0.2in S_{\rm EE}^{\beta^0, {\rm HM}} \sim  e^{-\frac{3\kappa_{l_p}N^{\frac{1}{3}}}{2}}M N^{13/10} N_f^{5/3}\left(1+\frac{4 t_{b_0}}{3 |c_2|}\right) \left(\frac{2}{3}\log \left(\frac{|c_2|+t_{b_0}}{|c_1|}\right) -n_{t_b} \log(N)\right)^{\frac{17}{6}},
\end{eqnarray}
implying
\begin{eqnarray}
\label{dSEEHbeta0overdmodc2-i}
& & \frac{\partial S_{\rm EE}^{\beta^0, {\rm HM}}}{\partial |c_2|} \sim  e^{-\frac{3\kappa_{l_p}N^{\frac{1}{3}}}{2}}M N^{13/10} N_f^{5/3}\Biggl[-\frac{4 t_{b_0}}{3 |c_2|^2}\left(\frac{2}{3}\log \left(\frac{|c_2|+t_{b_0}}{|c_1|}\right) -n_{t_b} \log(N)\right)^{\frac{17}{6}}\nonumber\\
& & + \frac{17}{6}\frac{2}{3|c_2|}\left(1+\frac{4 t_{b_0}}{3 |c_2|}\right)\left(\frac{2}{3}\log \left(\frac{|c_2|+t_{b_0}}{|c_1|}\right) -n_{t_b} \log(N)\right)^{\frac{11}{6}} \Biggr].
\end{eqnarray}
For $N=10^{3.3}, M=N_f=3, g_s=0.1, c_1=-10^3, c_2=-10^8$, $n_{t_b}=1$ as in \ref{EE-HM-HD}, as $t_b\leq t_{\rm Page}\sim10^6$,
\begin{eqnarray}
\label{dSEEHbeta0overdmodc2-ii}
& & \frac{\partial S_{\rm EE}^{\beta^0, {\rm HM}}}{\partial |c_2|} \sim e^{-\frac{3\kappa_{l_p}N^{\frac{1}{3}}}{2}}M N^{13/10} N_f^{5/3}\Biggl[-\leq 10^{6-16}\times{\cal O}(1)+{\cal O}(1)\times {\cal O}(1)\times 10^{-8} \Biggr]\approx  {\cal O}(1)\times 10^{-8}\nonumber\\
& & \sim e^{-\frac{3\kappa_{l_p}N^{\frac{1}{3}}}{2}}M N^{13/10} N_f^{5/3}{\cal O}(1)\times 10^{-8} >0.
\end{eqnarray}
Similarly,
\begin{eqnarray}
\label{dSEEHbeta0overdmodc1-i}
& & \frac{\partial S_{\rm EE}^{\beta^0, {\rm HM}}}{\partial |c_1|} \sim  -e^{-\frac{3\kappa_{l_p}N^{\frac{1}{3}}}{2}}M N^{13/10} N_f^{5/3}\Biggl[\left(1+\frac{4 t_{b_0}}{3 |c_2|}\right)\frac{1}{|c_1|} \Biggr]<0.
\end{eqnarray}
Further, in the $|c_2|\gg|c_1|$-limit and assume $|c_1|\sim N$, we obtained:
\begin{eqnarray}
\label{SHMEEbeta0-largemodc2}
& & S_{\rm EE}^{\beta^0, {\rm HM}} \sim  e^{-\frac{3\kappa_{l_p}N^{\frac{1}{3}}}{2}}M N^{13/10} N_f^{5/3}\left(\log |c_2|\right)^{11/6}\left(12\log |c_2|-(34 + 51 n_{t_b})\log |c_1|\right).
\end{eqnarray}
This is equivalent to a Swiss-Cheese volume defined in terms of a single ``large divisor'' volume $\log |c_2|$ and a single ``small divisor'' volume $\log |c_1|$ (with 12 and $34+51n_{t_b}$ encapsulating for some version of ``classical intersection numbers'' of these ``divisors''). Likewise, specifying $S_{\rm EE}^{\beta^0, {\rm HM}} \equiv e^{-\frac{3\kappa_{l_p}N^{\frac{1}{3}}}{2}}M N^{13/10} N_f^{5/3}\tilde{S}_{\rm EE}^{\rm HM}$, one might conceive of (\ref{SHMEEbeta0-largemodc2}) as an open Swiss-Cheese surface (in the same sense as (\ref{dSEEHbeta0overdmodc1-i}) and (\ref{SHMEEbeta0-largemodc2}), i.e., $\tilde{S}_{\rm EE}^{\rm HM}$ decreases as  $|c_1|$ grows, while $\tilde{S}_{\rm EE}^{\rm HM}$ increases as $|c_2|$ increases) in $\mathbb{R}_{\geq0}\left(\tilde{S}_{\rm EE}^{\rm HM}\right)\times\mathbb{R}^2_{+}\left(|c_1|, |c_2|\right)$ - refer to Fig. \ref{Swiss-Cheese}.  
\begin{figure}
\begin{center}
\includegraphics[width=0.70\textwidth]{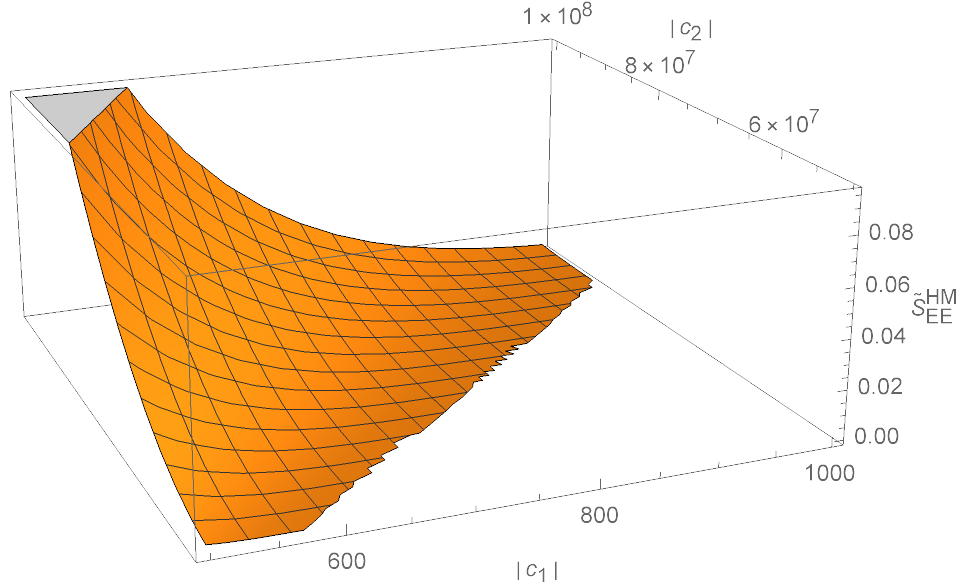}
\end{center}
\caption{Plot of $\tilde{S}_{\rm EE}^{\rm HM}$ as a bi-function of $\left(|c_1|, |c_2|\right)$ at $t=t_{\rm Page}=10^6$}
\label{Swiss-Cheese}
\end{figure}
Thus, (\ref{Swiss-Cheese-c12}) accompanied by (\ref{dSEEHbeta0overdmodc2-ii}), (\ref{dSEEHbeta0overdmodc1-i}) and (\ref{SHMEEbeta0-largemodc2}), indicates very interestingly a ``Swiss-Cheese'' structure of $S_{\rm EE}^{\beta^0, {\rm HM}}$ in the context of a Large $N$ Scenario (which is similar of the ``Large Volume Scenario'' in moduli stabilizations in string theory \cite{LVS,Swiss-Cheese-LVS}). 
\subsection{Entanglement Entropy Contribution from Island Surface}
\label{EE-IS-HD}
The island surface is a co-dimension two surface that exists at a continuous time slice, much like the Hartman-Maldacena-like surface. Therefore, we can formulate the formula for the holographic entanglement entropy as follows:
\begin{equation}
\label{HD-Entropy-IS}
S_{EE}=\int dr dx^2 dx^3  d\theta_1 d\theta_2dxdydz dx^{10} \sqrt{-g}\Biggl[\frac{\partial {\cal L}}{\partial R_{z\bar{z}z\bar{z}}}+\sum_{\alpha}\left(\frac{\partial^2 {\cal L}}{\partial R_{zizj}\partial R_{\bar{z}m\bar{z}l}}\right)_{\alpha} \frac{8 K_{zij}K_{\bar{z}ml}}{(q_{\alpha}+1)}\Biggr].
\end{equation}
Considering the action (\ref{D=11_O(l_p^6)-intro}) utilizing (\ref{HD-Entropy-IS}) and appendix (\ref{PT-IS}) the holographic entanglement entropy corresponding to island surface,  from ${\cal O}(\beta^0)$ and ${\cal O}(\beta)$ terms is obtained as\footnote{The angular integrations have been performed using (\ref{angular-integrations-2}) and (\ref{angular-integrations-1}). We have taken the most dominant terms in the leading order in $N$ from \ref{PT-IS} for the ${\cal O}(\beta^0)$ and ${\cal O}(\beta)$ contributions.}:
\begin{equation}
\label{Lag-total-IS}
S_{\rm EE}^{\rm total}=\int dr {\cal L}_{\rm {Total}}^{\rm {IS}}=\int dr \left({\cal L}_{0}^{\rm IS}+ {\cal L}_{\cal W}^{\rm IS}+ {\cal L}_{\cal A}^{\rm IS}\right).
\end{equation}
where
\begin{eqnarray}
\label{L0WA}
& & {\cal L}_{0}^{\rm IS}= \int dr\left( 4 \lambda_5(r) x(r)^2 \sqrt{\alpha (r) \left(\sigma (r)+x'(r)^2\right)}\right),\nonumber\\ & &  {\cal L}_{\cal W}^{\rm IS}=\int dr \left(4 x(r)^2 (\lambda_3(r)+\lambda_4(r)) \sqrt{\alpha (r) \left(\sigma (r)+x'(r)^2\right)}\right),\nonumber\\
& & {\cal L}_{\cal A}^{\rm IS}= \left(Z_1(r) {\cal L}_1+ x(r)^4 W_1(r){\cal L}_2 -2 x(r)^2 U_1(r){\cal L}_3 +4 x(r)^2 \left(U_1(r)+ 2 V_1(r)\right){\cal L}_4 \right),
\end{eqnarray}
where different $r$ dependent functions that exist in equation (\ref{L0WA}) are provided in (\ref{lambda3-4}) and (\ref{Z-W-U-V-1}), and
\begin{eqnarray}
\label{L1234}
& &
\lambda_5(r)=\kappa_{\lambda_5}\frac{ M N^{17/10} N_f^{4/3} g_s^{5/2} r_h^2 \log (N) \log (r) \sqrt[3]{\log (N)-3 \log (r)}}{r^4 \alpha _{\theta _1}^3 \alpha _{\theta _2}^2} \nonumber\\
& & \sim \frac{M N^{21/20} \log (2) (\log (64)-1) N_f^{4/3} g_s^{13/4} r_h^2 \kappa _{\lambda _5} \log (N) \log (r) \sqrt[3]{\log (N)-3 \log (r)}}{r^4},\nonumber\\
& &
  {\cal L}_1= {\cal L}_2=\frac{\sqrt{\alpha (r) \left(\sigma (r)+x'(r)^2\right)} \left(\alpha '(r) \left(\sigma (r)+x'(r)^2\right)+\alpha (r) \left(\sigma '(r)+2 x'(r) x''(r)\right)\right)^2}{x'(r)^2 \left(\sigma (r)+x'(r)^2\right)^4};\nonumber\\
   & &  {\cal L}_3= {\cal L}_4=\frac{\sqrt{\alpha (r) \left(\sigma (r)+x'(r)^2\right)}}{x'(r)^2},
\end{eqnarray}
Utilizing equations (\ref{dLagoverdx}), (\ref{dLagoverdxprime}) and (\ref{dLagoverdxprimexprime}), $x(r)$ EOM $\frac{\delta {\cal L}_{\rm {Total}}^{\rm {IS}}}{\delta x(r)} - \frac{d}{dr}\left(\frac{\delta{\cal L}_{\rm {Total}}^{\rm {IS}}}{dx^\prime(r)}\right) + \frac{d^2}{dr^2}\left(\frac{\delta {\cal L}_{\rm {Total}}^{\rm {IS}}}{\delta x^{\prime\prime}(r)}\right) = 0$, has been obtained as:
{\footnotesize
\begin{eqnarray}
\label{x(r)-EOM}
& & -\frac{N^{3/10} x(r) \left(F_1\left(r_h\right) \left(4 \left(r-r_h\right) x'(r)^2+x(r) \left(2 \left(r-r_h\right) x''(r)+x'(r)\right)\right)-2 N f_1\left(r_h\right)\right)}{2 \sqrt{r-r_h}} \nonumber\\
& & + \frac{\beta}{4 N^{7/10}
   (r-r_h)^{3/2} x'(r)^4}\Biggl[2 N^{7/10} F^\beta_4(r_h) x(r) \left(4 (r_h-r) x'(r)^2+x(r) \left(6 (r-r_h) x''(r)+x'(r)\right)\right) \nonumber\\
  & &  +(r-r_h) x'(r)^4 \left(3 {\cal F}^{\beta}_1(r_h)+3
   {\cal F}^{\beta}_2(r_h)+4 N^2 x(r) Y_1(r_h)\right)\Biggr]= 0.
\end{eqnarray}
}
Let us write $x(r) = x_0(r) + \beta x_1(r)$, and consider the ansatz:
\begin{equation}
\label{ansatz-x0(r)}
x_0(r)=\sqrt{\frac{{a_3}}{{a_2}}+r} \sqrt{\frac{F_1(r_h)}{N f_1(r_h)}}.
\end{equation}
Define $X(r) \equiv \frac{x_0(r) \sqrt{\frac{F_1(r)}{N f_1(r_h)}}}{\sqrt{2}}$, we found that $X(r)$ should satisfy:
{\footnotesize
\begin{eqnarray}
\label{X0(r)}
& & \hskip -0.3in X(r) \left(2 (r-r_h) X''(r)+X'(r)\right)+4 (r-r_h) X'(r)^2-1=\frac{a_3 \left(a_1^2 a_2-2\right)+{a_2} \left(2 r \left(a_1^2 a_2-1\right)-a_1^2 a_2r_h\right)}{2 ({a_2} r+{a_3})} = 0.\nonumber\\
\end{eqnarray}
}
If $a_1^2a_2=2$, the LHS of (\ref{X0(r)}) then turns into proportional to $\frac{r-r_h}{\frac{{a_3}}{{a_2}}+r}$, which in the deep IR, i.e., $r\sim r_h$ becomes unimportant if $\frac{a_3}{a_2 r}\gg1$, becomes zero. Thus,
\begin{equation}
\label{x0(r)-solution}
x_0(r)=\sqrt{\frac{{a_3}}{{a_2}}+r} \sqrt{\frac{F_1(r_h)}{N f_1(r_h)}}.
\end{equation}
Hence, we found that:
{\footnotesize
\begin{eqnarray}
\label{x1(r)EOM-i}
& & \hskip -0.3in x_1(r) = \frac{N^{3/10} }{2 \sqrt{r-r_h}}\Biggl[2 N f_1(r_h) {x_1}(r)-{F_1}(r_h) \Biggl({x_0}(r) \left(2  (r-r_h) \left(2 {x_1}(r) {x_0}''(r)+{x_0}(r) {x_1}''(r)\right) 
   +{x_0}(r)
  {x_1}'(r)\right)
  \nonumber\\
   & & +2 {x_0}(r) {x_0}'(r) \left(4 (r-r_h) {x_1}'(r)+{x_1}(r)\right)+4(r-r_h) {x_1}(r) {x_0}'(r)^2\Biggr)\Biggr] \nonumber\\
   & & =\frac{2 {a_2} N^2{f_1}(r_h)^2 {x_1}(r)-{F_1}(r_h)^2 \left({x_1}'(r) (5 a_2 r-4 {a_2}r_h+{a_3})+2 (r-r_h) ({a_2} r+{a_3}) {x_1}''(r)+{a_2} {x_1}(r)\right)}{2{a_2} N^{7/10} {f_1}(r_h) \sqrt{r-r_h}}, \nonumber\\
\end{eqnarray}
}
and,
{\footnotesize
\begin{eqnarray}
\label{x1(r)EOM-ii}
& & \frac{\beta}{4 N^{7/10}
   (r-r_h)^{3/2} x_0'(r)^4}\Biggl(2 N^{7/10} F^\beta_4(r_h) x_0(r) \left(4 (r_h-r) x_0'(r)^2+x_0(r) \left(6 (r-r_h) x_0''(r)+x_0'(r)\right)\right) \nonumber\\
  & &  +(r-r_h) x_0'(r)^4 \left(3 {\cal F}^{\beta}_1(r_h)+3
   {\cal F}^{\beta}_2(r_h)+4 N^2 x_0(r) Y_1(r_h)\right)\Biggr) \nonumber\\
 & &  =\frac{4 \left(a_2 r+a_3\right) \sqrt{\frac{a_3}{a_2}+r} \left(a_2 \left(5 r_h-4 r\right)+a_3\right) F^\beta_4\left(r_h\right)}{a_2^2 \left(r-r_h\right){}^{3/2} \sqrt{\frac{F_1\left(r_h\right)}{N
   f_1\left(r_h\right)}}}+\frac{4 N^2 \sqrt{\frac{a_3}{a_2}+r} Y_1\left(r_h\right) \sqrt{\frac{F_1\left(r_h\right)}{N f_1\left(r_h\right)}}+3 {\cal F}^{\beta}_1(r_h)+3 {\cal F}^{\beta}_2(r_h)}{4 N^{7/10}
   \sqrt{r-r_h}}.\nonumber\\
\end{eqnarray}
}
Thus, the island surface's equation of motion associated with embedding $x_1(r)$ is as follows:
{\footnotesize
\begin{eqnarray}
\label{x1(r)-EOM}
& & \frac{2 {a_2} N^2{f_1}(r_h)^2 {x_1}(r)-{F_1}(r_h)^2 \left({x_1}'(r) (5 a_2 r-4 {a_2}r_h+{a_3})+2 (r-r_h) ({a_2} r+{a_3}) {x_1}''(r)+{a_2} {x_1}(r)\right)}{2{a_2} N^{7/10} {f_1}(r_h) \sqrt{r-r_h}}\nonumber\\
& & +\frac{4 \left(a_2 r+a_3\right) \sqrt{\frac{a_3}{a_2}+r} \left(a_2 \left(5 r_h-4 r\right)+a_3\right) F^\beta_4\left(r_h\right)}{a_2^2 \left(r-r_h\right){}^{3/2} \sqrt{\frac{F_1\left(r_h\right)}{N
   f_1\left(r_h\right)}}}+\frac{4 N^2 \sqrt{\frac{a_3}{a_2}+r} Y_1\left(r_h\right) \sqrt{\frac{F_1\left(r_h\right)}{N f_1\left(r_h\right)}}+3 {\cal F}^{\beta}_1(r_h)+3 {\cal F}^{\beta}_2(r_h)}{4 N^{7/10}
   \sqrt{r-r_h}}=0,\nonumber\\
\end{eqnarray}
}
which near $r=r_h$ rewritten as:
\begin{eqnarray}
\label{x1(r)-EOM-final}
& & \frac{4 \sqrt{\frac{a_3}{a_2}+r_h} \left(a_2 r_h+a_3\right){}^2 F^\beta_4\left(r_h\right)}{a_2^2 \left(r-r_h\right){}^{3/2} \sqrt{\frac{F_1\left(r_h\right)}{N f_1\left(r_h\right)}}}+\frac{N^{13/10} x_1(r) f_1\left(r_h\right)-\frac{F_1\left(r_h\right){}^2 \left(\left(a_2 r_h+a_3\right) x_1'(r)+a_2 x_1(r)\right)}{2 a_2 N^{7/10} f_1\left(r_h\right)}}{\sqrt{r-r_h}}=0.\nonumber\\
\end{eqnarray}
(\ref{x1(r)-EOM-final})'s solution is provided as:
\begin{eqnarray}
\label{x1(r)-solution}
& & x_1(r) = \frac{8 \sqrt{\frac{a_3}{a_2}+r_h} \left(a_2 r_h+a_3\right) {F^\beta_4}\left(r_h\right) \log \left(r-r_h\right)}{a_2 N^{3/10} F_1\left(r_h\right) \left(\frac{F_1\left(r_h\right)}{N f_1\left(r_h\right)}\right){}^{3/2}},
   \end{eqnarray}
    implying,
   \begin{eqnarray}
   \label{x(r)-solution}
   & & x(r)\approx \sqrt{\frac{a_3}{a_2}+r} \sqrt{\frac{ F_1(r_h)}{N  f_1(r_h)}}+\beta N^{\frac{11}{5}}\left( \frac{8  \left(r_h+\frac{a_3}{a_2}\right)^{\frac{3}{2}} {F^\beta_4}\left(r_h\right) \log \left(r-r_h\right)}{F_1\left(r_h\right) \left(\frac{F_1\left(r_h\right)}{ f_1\left(r_h\right)}\right)^{3/2}}\right).
\end{eqnarray}
In order to find the turning point using $r_T: 1/x'(r_T) = 0$, one gets:
\begin{eqnarray}
\label{rT-2}
& & \frac{32 \beta N^{11/5} \left(a_2 r_T+a_3\right) \sqrt{\frac{a_3}{a_2}+r_h} \left(a_2 r_h+a_3\right) F^\beta_4\left(r_h\right)}{a_2^2  \left(r_T-r_h\right) F_1\left(r_h\right)
   \left(\frac{F_1\left(r_h\right)}{ f_1\left(r_h\right)}\right)^{5/2}}-\frac{2 \sqrt{\frac{a_3}{a_2}+r_T}}{\sqrt{\frac{F_1(r_h)}{N
    f_1(r_h)}}}=0. 
\end{eqnarray}
The above equation has the following solution:
\begin{eqnarray}
\label{rT-solution}
& & r_T = r_h + \frac{128 \beta^2 }{a_2^4 F_1\left(r_h\right){}^6}\Biggl({a_2  N^{17/5} \left(a_2 r_h+a_3\right)^3 f_1\left(r_h\right)^4 F^\beta_4\left(r_h\right)^2}+a_2 a_3^3  N^{17/5} f_1\left(r_h\right)^4
   F^\beta_4\left(r_h\right){}^2 \nonumber\\
  & & +a_2^4  N^{17/5} r_h^3 f_1\left(r_h\right){}^4 F^\beta_4\left(r_h\right){}^2+3 a_2^3 a_3  N^{17/5} r_h^2 f_1\left(r_h\right){}^4
   F^\beta_4\left(r_h\right){}^2+3 a_2^2 a_3^2  N^{17/5} r_h f_1\left(r_h\right){}^4 F^\beta_4\left(r_h\right){}^2\Biggr)\nonumber\\
   & & \hskip -0.2in \sim r_h +\frac{289 \left(\frac{3}{2}\right)^{2/3} \pi ^{17/3} \beta ^2 M^2 N^{17/5} (107-540 \log (2))^2 g_s^7 \kappa _{\sigma }^6 \kappa _{U_1}^2 \log ^2(N)\left(3 a_2 a_3^2 r_h+3 a_2^2 a_3 r_h^2+2 a_2^3
   r_h^3+a_3^3\right) }{50 a_2^3 \log ^2(2) (\log (64)-1)^2 N_f^{2/3} r_h^{20} \kappa _{\lambda _5}^2 \left(-\log \left(r_h\right)\right){}^{2/3}}\nonumber\\
& &  \equiv r_h + \delta \in{\rm IR},
\end{eqnarray}
in ${\cal R}_{D5/\overline{D5}}=1$-units,  ${\cal O}(\beta^0)$ contributions to the on-shell entanglement entropy is obtained as:
{\footnotesize
\begin{eqnarray}
\label{on-shell-L-beta0}
& & S_{\rm EE}^{\beta^0, {\rm IS}} = \int_{r_h}^{r_h+\delta {\cal R}_{D5/\overline{D5}}} dr\Biggl(\frac{\kappa_{\lambda_5} {\log N} M
   N^{17/10} {N_f}^{4/3} {r_h}^2 x(r)^2 \log (r) \sqrt[3]{{\log N}-3 \log(r)} \sqrt{\alpha (r) \left(\sigma (r)+x'(r)^2\right)}}{
   r^4 \alpha _{\theta _1}^3 \alpha _{\theta _2}^2}\Biggr)\nonumber\\
& & \hskip 0.5in \sim \int_{r_h}^{r_h+\delta}dr\Biggl(\frac{2\ 2^{2/3} M N^{3/10} \log (2) (\log (64)-1) N_f^{5/3} g_s^{10/3} \kappa _{\lambda _5} \log (N) }{81\ 3^{5/6} \pi ^{11/12} a_2 \sqrt{r-r_h} r_h^{5/2} f_1\left(r_h\right)}\Biggr)\nonumber\\
& & \hskip 0.6in \times\left(\left(a_2 r_h+a_3\right) F_1\left(r_h\right) \log \left(r_h\right) \sqrt{\kappa _{\alpha } \kappa
   _{\sigma }} \left(\log (N)-3 \log \left(r_h\right)\right){}^{2/3}\right)
\nonumber\\
& & \hskip 0.5in \sim \Biggl(\frac{ \sqrt{\delta } M N^{3/10} N_f^{5/3} g_s^{7/3} \sqrt{r_h} \kappa _{\alpha } \kappa _{\lambda _5} \log (N) \left(a_2 r_h+a_3\right) \log \left(r_h\right)
   \left| \log \left(r_h\right)\right| {}^{2/3}}{ a_2 \sqrt{\kappa _{\alpha } \kappa _{\sigma }}}\Biggr)\nonumber\\
& & \hskip 0.5in   \sim \frac{\beta  M^2 N^2 N_f^{4/3} g_s^{35/6}  \log ^2(N) \left(a_2 r_h+a_3\right)   \left| \log \left(r_h\right)\right| {}^{4/3}\sqrt{3 a_2 a_3^2 r_h+3 a_2^2 a_3 r_h^2+2
   a_2^3 r_h^3+a_3^3}}{a_2^{5/2} r_h^{19/2}}  .
\end{eqnarray}
}
Right now, it seems to be in conflict with itself: $ S_{\rm EE}^{\beta^0, {\rm IS}}\sim \beta$. Utilizing (\ref{lp-rh-relation}), the conclusion is the following: $\beta\propto r_h^{\frac{3}{2}}$. \par
The black hole entropy for the metric (\ref{TypeIIA-from-M-theory-Witten-prescription-T>Tc-BHIP-ch7}) is obtained as:
\begin{eqnarray}
\label{SBH}
& & S_{\rm BH} \sim 
%\frac{g_s^{5/4} M N^{9/20} N_f^3 r_h^3 \log ^4(r_h) (2-\beta
   % ({\cal C}^{\rm BH}_{zz}-2 {\cal C}^{\rm BH}_{\theta_1 z}))}{\alpha _{\theta _1}^3 \alpha _{\theta_2}^2}\nonumber\\
 \frac{g_s^{7/4} M N_f^3 r_h^3 \log (N) \log ^4(r_h) (2-\beta
    ({\cal C}^{\rm BH}_{zz}-2 {\cal C}^{\rm BH}_{\theta_1 z}))}{N^{3/4}}, 
\end{eqnarray}
From (\ref{SBH}), we found that ${\cal O}(\beta^0)$ contribution to the thermal entropy of the black hole behaves as: $S_{\rm BH}^{\beta^0} \sim r_h^3 \log ^4(r_h)$. The ${\cal O}(\beta)$-corrections to the ${\cal M}$-theory three-form potential had been made to zero in order to compute the ${\cal O}(\beta)$-corrections to the MQGP background of \cite{MQGP} as worked out in \cite{HD-MQGP} and quoted in appendix \ref{METRIC_HD_MQGP}. For this, ${\cal C}^{\rm BH}_{zz}=2 {\cal C}^{\rm BH}_{\theta_1 z}$ were needed. Therefore, there is no ${\cal O}(R^4)$ correction to the black hole thermal entropy (\ref{SBH}).\par
Using (\ref{on-shell-L-beta0}) and (\ref{SBH}), and $\beta = \kappa_\beta(g_s, N) \left(\frac{r_h}{{\cal R}_{D5/\overline{D5}}}\right)^{\frac{3}{2}}\alpha'\ ^3$ , we found that:
\begin{eqnarray}
\label{SEEISoverSBH}
 & &\frac{S^{\beta^0,\ IS}_{\rm EE}}{S_{\rm BH}} \sim \frac{\kappa_\beta(g_s, N)  {g_s}^{49/12} {\log N} M N^{11/4} ({a_2} {r_h}+{a_3}) \sqrt{3 {a_2}^2
   {a_3} {r_h}^2+2 {a_2}^3 {r_h}^3+3 {a_2} {a_3}^2 {r_h}+{a_3}^3}}{{a_2}^{5/2}
   {N_f}^{5/3} {r_h}^{11} | \log ({r_h})| ^{8/3}}.\nonumber\\
   & & 
\end{eqnarray}
When $a_2 r_h \gg a_3$, (\ref{SEEISoverSBH}) implying
\begin{eqnarray}
\label{SEEISoverSBH-arrh>>a3}
& &  \frac{S^{\beta^0,\ IS}_{\rm EE}}{S_{\rm BH}} \sim \frac{\kappa_\beta(g_s, N){g_s}^{49/12} {\log N} M N^{11/4}}{ {N_f}^{5/3} {r_h}^{17/2} | \log ({r_h})| ^{8/3}}\left[1 + \sum_{n=1}^\infty {\cal A}_n\left(\frac{a_3}{a_2 r_h}\right)^n\right], 
\end{eqnarray}
where, e.g., ${\cal A}_1 = \frac{7}{4}, {\cal A}_2 = \frac{39}{32}$, etc. Now, for a black hole in $D$ dimensions, we expect:
\begin{equation}
\label{SEEISoverSBH-D-dims}
\frac{S^{\beta^0,\ IS}_{\rm EE}}{S_{\rm BH}} = 2 + \sum_{n=1}a_n\left(\frac{G_N^D}{r_h^{D-2}}\right)^n,
\end{equation}
where the central charge associated with the conformal backgrounds has been absorbed in $a_n$'s \cite{SEEISoverSBHapprox2}; $a_3$ in (\ref{SEEISoverSBH-arrh>>a3}) being the non-conformal analogue of the central charge ``$c$'' appearing in \cite{SEEISoverSBHapprox2}. To guarantee that (\ref{SEEISoverSBH-arrh>>a3}) $\sim$ (\ref{SEEISoverSBH-D-dims}), we have to cure the large-$N$ and IR(via small $r_h$) enhancements of (\ref{SEEISoverSBH-arrh>>a3}). Utilizing the estimate in \cite{Bulk-Viscosity-McGill-IIT-Roorkee} of the $r=r_0\sim r_h: N_{\rm eff}(r_0=0)(N_{\rm eff}$ being the effective number of color $D3$-branes in \cite{metrics} - the type IIB dual of thermal QCD-like theories), and in particular the exponential $N$-suppression therein \footnote{To find an appropriate $r_0/r_h$ it would be easier to work with the type IIB side instead of its type IIA mirror as the mirror {\it a la} SYZ
keeps the radial coordinate unchanged. To proceed then, let us define an {\it effective} number of three-brane charge
as:
\begin{equation*}
\label{Neff}
N_{\rm eff}(r) = \int_{\mathbb{M}_5} F_5^{\rm IIB} + \int_{\mathbb{M}_5} B_2^{\rm IIB} \wedge F_3^{\rm IIB}, 
\end{equation*}
where $B_2^{\rm IIB}, F_3^{\rm IIB}$ and $F_5^{\rm IIB}$ are given in \cite{metrics}. The five-dimensional internal space $\mathbb{M}_5$,
with coordinates ($\theta_i, \phi_i, \psi$), is basically the base of the resolved warped-defomed conifold. As shown in \cite{Bulk-Viscosity-McGill-IIT-Roorkee}, 
\begin {eqnarray*}
& & N_{\rm eff}(r_0) = N + {3g_sM^2 \log~r\over 10 r^4} \bigg\{18\pi r (g_sN_f)^2 \log~N \sum_{k = 0}^1\left(18a^2 (-1)^k\log~r + r^2\right)\left({108a^2 \log~r\over 2k+1} + r\right) \\
 &+& 5 \left(3 a^2 ({g_s}-1)+r^2\right) (3 {g_s} {N_f} \log ~r+2\pi ) (9 {g_s} {N_f} \log ~r+4 \pi )
 \left[9 a^2 {g_s} {N_f} \log\left({e^2\over r^3}\right) +4 \pi  r^2\right]\biggr\}, \nonumber\\
 & = & N\left[1 + 6\pi\log~r\left(3 g_s N_f \log~r + 2\pi\right)\left(9 g_s N_f \log~r + 4\pi\right){g_sM^2\over N}  \right] +
 {\cal O}\left[{g_sM^2\over N}\left(g_sN_f\right)^2 \log~N\right]. \end {eqnarray*}
Note that the assumption of small $r_0$ is crucial here as the same implies the dominance of $g_sN_f|\log~r_0|$ over other constant pieces. Solving for $r_0: N_{\rm eff}(r_0)=0$ yields $r_0\sim r_h\sim e^{-\kappa_{r_h}(M, N_f, g_s)N^{\frac{1}{3}}}$.} ,  we therefore propose $\kappa_\beta(g_s, N) \sim e^{-\kappa_{l_p}N^{1/3}}$, i.e.,
\begin{eqnarray}
\label{scalings}
& & \beta \sim \left(g_s^{\frac{4}{3}}\alpha'^2e^{-\kappa_{l_p}N^{1/3}}\frac{r_h}{{\cal R}_{D5/\overline{D5}}}\right)^{3/2}.
\end{eqnarray}
Therefore, the entanglement entropy for the island surface after substitution of $\beta$ simplified as:
\begin{eqnarray}
\label{SEE-IS-simp}
S_{\rm EE}^{\beta^0, {\rm IS}} \sim \frac{M^2 e^{-\frac{3\kappa_{l_p} N^{\frac{1}{3}}}{2}}
 N_f^{4/3} g_s^{35/6} \log ^2(N)  \left| \log \left(r_h\right)\right|^{4/3} }{r_h^{11/2}}.
\end{eqnarray}
Using the numerical values $N=10^{3.3}, M=N_f=3, g_s=0.1$, from (\ref{SBH}) we found that, $11.4 r_h^3|\log r_h|^4 = S_{\rm BH}$. which results in,
\begin{equation}
\label{rh-SBH}
r_h = e^{\frac{4}{3}W(0.4 S_{\rm BH}^{\frac{1}{4}})}.
\end{equation}
 From Fig. \ref{SEEISoverSBH-vs-SBH} and utilizing (\ref{rh-SBH}), it is obvious that  (\ref{SEEISoverSBH-arrh>>a3}) $\sim$ (\ref{SEEISoverSBH-D-dims}) means that a lower constraint will be placed on $r_h$, the non-extremality parameter in the ${\cal M}$-theory dual of large-$N$ thermal QCD.
\begin{figure}
\begin{center}
\includegraphics[width=0.45\textwidth]{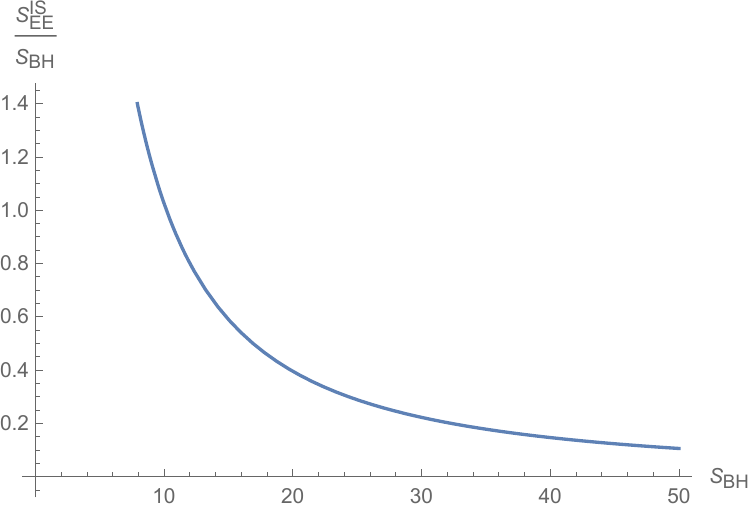}
\end{center}
\caption{$\frac{S^{\beta^0,\ IS}_{\rm EE}}{S_{\rm BH}}$-versus-$S_{\rm BH}$ for $N=10^{3.3}, M=N_f=3, g_s=0.1, \kappa_{l_p}=0.47$}
\label{SEEISoverSBH-vs-SBH}
\end{figure}

\subsection{Page Curve and An Exponential(-in-$N$) Hierarchy of Entanglement Entropies up to ${\cal O}(R^4)$ Before/After the Page Time}
\label{Page-curve-plot-HD}
Because we have all of the results in hand, we will be obtaining the Page curve now. Entanglement entropy of the Hawking radiation that corresponds to the Hartman-Maldacena-like surface is provided in equation (\ref{SEE-HM-tb0}), while for the island surface the result is provided in equation (\ref{SEE-IS-simp}). We obtained the Page curve by plotting $S_{EE}^{\beta^0, {\rm HM}}$ and $S_{EE}^{\rm IS}$ in Fig. \ref{Page-curve-plot} for $N=10^{3.3},M=N_f=3,g_s=0.1, \kappa_{l_p}=0.47, c_2=-10^{8}, c_1=-10^3$.  
%\begin{eqnarray}
%\label{S_EE^HM-approx-linear-time}
%S_{EE}^{\beta^0,\ {\rm HM}} \sim 
%\left(1-\frac{4 t_{b_0}}{3 c_2}\right)\left(\frac{2}{3}\log \left(\frac{c_2}{c_1}\right) -n_{t_b} \log(N)\right)^{17/6}.
%\end{eqnarray}
\begin{figure}
\begin{center}
\includegraphics[width=0.60\textwidth]{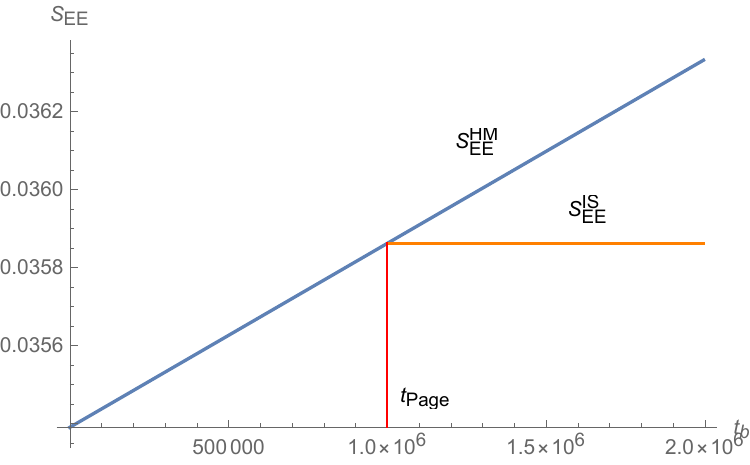}
\end{center}
\caption{The Page curve obtained from the use of Dong's proposal to compute entanglement entropies of the Hartman-Maldacena-like (blue line) and island(orange line) surfaces.}
\label{Page-curve-plot}
\end{figure}
The initially calculated entanglement entropy of the Hawking radiation for the Hartman-Maldacena-like surface increases linearly with time, as seen in picture \ref{Page-curve-plot}. After the Page time, the entanglement entropy of the Hawking radiation from the island surface governs, therefore the entanglement entropy stops rising and the Page curve is obtained. The values of $S_{\rm EE}$ in \ref{Page-Curve-Areas} and Fig. \ref{Page-curve-plot} disagree due to the fact that, for example, the latter one skipped over the factor of $\left(2\pi\right)^4$ originating from integrations w.r.t. $\phi_{1,2}, \psi, x^{10}$, and so on.\par
{\bf Page time}: At the Page time, the entanglement entropies for the Hartman-Maldacena-like surface and the Island Surface are equal. This results in:
\begin{eqnarray}
\label{Page-time}
& & t_{\rm Page} = \frac{3}{4} c_2 \left(1-\frac{9\ 3^{5/6} {g_s}^{35/6} M N^{7/10} \log ^2(N) | \log ({r_h})|
   ^{4/3}}{\sqrt[3]{{N_f}} {r_h}^{11/2} \left(-3 {n_{t_b}} \log (N)+2 \log
   \left(\frac{c_2}{c_1}\right)\right){}^{17/6}}\right).
\end{eqnarray}
Figure Fig. \ref{Page-time-vs-SBH-plot} depicts the variation of the Page time as a function of the Black-hole entropy (using (\ref{rh-SBH})). This also demonstrates that Page time positivity necessitates an upper limit on black-hole entropy and hence the IR cut-off $r_h$. Knowing that $r_h$, the non-extremality parameter, corresponds to a constant of integration \cite{Klebanov+Buchel-et-al_rh-const-of-integration} and demanding this value (in units of ${\cal R}_{D5/\overline{D5}}$) for being smaller than unity, could be carried out, e.g., by pointing out via (\ref{rh-SBH}) that $r_h$ has become a rising function of $S_{\rm BH}$, and thus: $r_h(S_{\rm BH})\rightarrow \frac{r_h(S_{\rm BH})}{r_h(S^0_{\rm BH}:t_{\rm Page}(S^0_{\rm BH})>0)}$. As a result, with the choices made of $N, M, g_s, c_1, c_2, n_{t_b}, \kappa_{l_p}$, the result would suggest $r_h\rightarrow\frac{r_h(S_{\rm BH})}{r_h(S_{\rm BH}\approx8)}$.
\begin{figure}
\begin{center}
\includegraphics[width=0.60\textwidth]{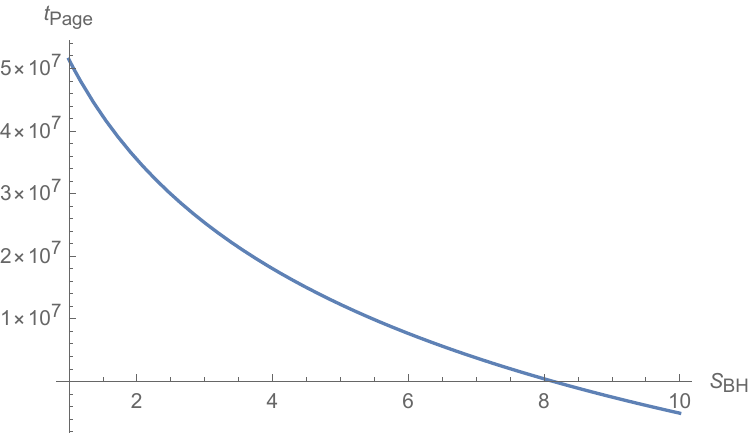}
\end{center}
\caption{Page time for $N=10^{3.3}, M=N_f=3, g_s=0.1,  c_2=-10^8, c_1=-10^{3}$}
\label{Page-time-vs-SBH-plot}
\end{figure}
\\
{\bf ${\cal O}(\beta)$ contributions to the entanglement entropies}: We come about to have completely ignored $S_{\rm EE, HM}^\beta$. Let us next look at why this seems reasonable and the way it results in an exponential-large-$N$-suppression hierarchy. We obtained the entanglement entropy for the HM-like surface at ${\cal O}(\beta)$ given as below:
\begin{eqnarray}
\label{SEEHMbeta-i}
& & S_{\rm EE, HM}^\beta \sim \frac{\beta^{4/3}x_R^2g_s^{\frac{23}{6}}M^3N_f^{\frac{2}{3}}\log ^2({r_h}) (\log (N)-12 \log ({r_h})) (\log (N)-9 \log ({r_h}))}{r_h^{\frac{5}{2}}\left(\log N - 3 \log r_h\right)^{\frac{7}{6}}} \nonumber\\
& & \times  \left(-216
   \left(16+\sqrt{2}\right) {g_s}^4 \kappa_{\lambda_1} M^4 {N_f}^2 \log (N) \log ^3({r_h})+9
   \left(16+\sqrt{2}\right) {g_s}^4 \kappa_{\lambda_1} M^4 {N_f}^2 \log ^2(N) \log ^2({r_h})\right.\nonumber\\
   & & \left.+1296
   \left(16+\sqrt{2}\right) {g_s}^4 \kappa_{\lambda_1} M^4 {N_f}^2 \log ^4({r_h})+4096
   \left(4+\sqrt{2}\right) \pi ^4 {\kappa_{\lambda_2}}\right),
\end{eqnarray}
the above in $|\log r_h|\gg \log N$-limit, corresponds to:
\begin{eqnarray}
\label{SEEHMbeta-ii}
& & \frac{\beta ^{4/3} {g_s}^{47/6} M^7 N^{3/10} {N_f}^{8/3} x_R^2 \log (2)
   (-\log ({r_h}))^{41/6}}{{r_h}^{5/2}}.
\end{eqnarray}
Utilizing (\ref{scalings}), (\ref{SEEHMbeta-ii}) implies,
\begin{eqnarray}
\label{SEEHMbeta-iii}
& & S_{\rm EE, HM}^\beta \sim \frac{{g_s}^{47/6} M^7 N^{3/10} {N_f}^{8/3} x_R^2 e^{-2 \kappa_{l_p}
   N^{\frac{1}{3}}} (-\log ({r_h}))^{41/6}}{\sqrt{{r_h}}}.
\end{eqnarray}
Similarly, for the island surface ${\cal O}(\beta)$ contributions to the entanglement entropy is obtained as: 
\begin{eqnarray}
\label{SEEISbeta-i}
& & S_{\rm EE, IS}^\beta \sim \frac{ \beta ^2 {g_s}^{15/2} \sqrt{{\kappa_\alpha}} {\kappa_\sigma}^{5/2} {\kappa_{U_1}} M^2 N \log ^2(N) ({a_2} {r_h}+{a_3}) \sqrt{3 {a_2}^2 {a_3} {r_h}^2+2 {a_2}^3
   {r_h}^3+3 {a_2} {a_3}^2 {r_h}+{a_3}^3}}{{a_2}^{5/2} {\kappa_{\lambda_5}} {r_h}^{35/2} (-\log ({r_h}))^{2/3}}\nonumber\\
& & \times  \left(6561 \sqrt[6]{2} \left(1+8 \sqrt{2}\right)
   {g_s}^{17/6} \kappa_{\lambda_3} M^6 {N_f}^{7/3} {r_h}^8 \log ^6({r_h})+1048576 \sqrt[3]{6} \pi
   ^{47/6} {\kappa_\alpha}^2 {\kappa_{Z_1}} N^2 (-\log ({r_h}))^{2/3}\right)\nonumber\\
& & \approx \frac{17 \beta ^2 {g_s}^{15/2} {\log N}^2 M^2 N^3 ({a_2} {r_h}+{a_3}) \sqrt{3 {a_2}^2 {a_3}
   {r_h}^2+2 {a_2}^3 {r_h}^3+3 {a_2} {a_3}^2 {r_h}+{a_3}^3}}{{a_2}^{5/2}
   {r_h}^{35/2}},
\end{eqnarray}
in the $a_2r_h\gg a_3$-limit, the above equation simplifies to:
\begin{eqnarray}
\label{SEEISbeta-ii}
& & 
   S_{\rm EE, IS}^\beta \sim \frac{\beta ^2 {g_s}^{15/2} {\log N}^2 M^2 N^3}{{r_h}^{15}}.
\end{eqnarray}
Utilizing (\ref{scalings}), (\ref{SEEISbeta-ii}) results in:
\begin{eqnarray}
\label{SEEISbeta-iii}
& & S_{\rm EE, IS}^\beta \sim \frac{{g_s}^{15/2} {\log N}^2 M^2 N^3 e^{-3 {\kappa_{l_p}} N^{\frac{1}{3}}}}{{r_h}^{12}}.
\end{eqnarray}
From the equations (\ref{SEE-HM-tb0}), (\ref{SEE-IS-simp}), (\ref{SEEHMbeta-iii}) and (\ref{SEEISbeta-iii}), we obtained the following hierarchy:
\begin{eqnarray}
\label{S_EE_hierarchy}
& & S_{\rm EE}^{\rm HM, \beta^0}:S_{\rm EE}^{\rm IS,\ \beta^0}:S_{\rm EE}^{\rm HM,\ \beta}:S_{\rm EE}^{\rm IS,\ \beta} \sim \gamma^{\frac{3}{2}}:\gamma^{\frac{3}{2}}:\gamma^2:\gamma^3,
\end{eqnarray}
where $\gamma\equiv e^{-\kappa_{l_p}N^{\frac{1}{3}}}$. We can infer from the preceding equation that ${\cal O}(\beta)$ corrections to entanglement entropies are more exponentially large-$N$ suppressed for both Hartman-Maldacena-like and Island surfaces. As a result, we ignored such contributions when computing the Page curve.
\section{Massless Graviton - the Physical reason for Exponentially Suppressed Entanglement Entropies}
\label{massless_graviton}
The goal of this section is to give a scientific explanation for the exponential suppression that occurs in the entanglement entropies of both HM-like and Island surfaces in \ref{Page-curve-plot-HD}.  We will prove that, in spite of the coupling of a non-gravitational bath to the ETW-``brane'', (imposition of the Dirichlet boundary condition at the horizon upon the radial profile of the graviton wave function) leads to the quantization of the graviton mass, and this can consequently be massless (for a suitable choice of the quantum number). Typically, in the scenario of AdS$_{d+1}$ gravity duals of CFT$_d$ on $\partial$AdS$_{d+1}$ at zero temperature, massless graviton indicates a vanishing angle between the ETW/KR-brane and $\partial$AdS$_{d+1}$, implying the islands stop to make a contribution \cite{GB-3,Island-IIB-2}. Nevertheless, as discussed in \ref{ETW-sub}, the ETW-``brane'' that we use ($x^1=$constant), was orthogonal (in the $x^1-r$-plane) to the thermal bath/QCD-like theory (after integrating out the angular directions of ${\cal M}_6(\theta_{1,2},\phi_{1,2},\psi,x^{10})$). Regardless of this, we showed that a massless graviton could be produced in our framework.  What's more, there's a corresponding exponential-in-$N$ suppression of the HM-like entanglement entropy, thereby explaining how both of them are possibly compared at the Page time and a Page curve is obtained.\\
Similar to \cite{C. Bachas and J. Estes [2011]}, we can write the eleven dimensional metric (\ref{TypeIIA-from-M-theory-Witten-prescription-T>Tc-BHIP-ch7}) as given below:
\begin{equation}
\label{Kounas+Estes-metric}
ds^2 = e^{2A(y)}\overline{g}_{\mu\nu}(x)dx^\mu dx^\nu + 
\hat{g}_{mn}dy^m dy^n.
\end{equation}
The perturbed metric that we considered is of the form: 
 $\widetilde{ds^2} = e^{2A(y)}\left(\overline{g}_{\mu\nu} + h_{\mu\nu}\right) dx^\mu dx^\nu + \hat{g}_{mn}dy^m dy^n$, where $h_{\mu\nu}(x,y) = h^{[tt]}_{\mu\nu}(x)\psi(y)$, as ansatz under linear perturbations: $\overline{D}^\mu h_{\mu\nu}^{[tt]} = \overline{g}^{\mu\nu}h_{\mu\nu}^{[tt]}=0$, and the equation of motion for the graviton wave function is written as\cite{C. Bachas and J. Estes [2011]}:
 \begin{equation}
 \label{EOM-psi}
 -\frac{e^{-2A(y)}}{\sqrt{|\hat{g}(y)|}}\partial_m\left(
 \sqrt{|\hat{g}(y)|}\hat{g}^{mn}e^{4A(y)}\partial_n\psi(y)\right) = m^2\psi(y).
 \end{equation}
Near, $(\theta_1,\theta_2)\sim\left(\frac{\alpha_{\theta_1}}{N^{\frac{1}{5}}},\frac{\alpha_{\theta_2}}{N^{\frac{3}{10}}}\right)$, we found that $\psi(r,\theta_{1,2},\phi_{1,2},\psi,x^{10})\rightarrow\psi(r,\theta_1)$, in the IR-UV interpolating region, the eigenvalue equation (\ref{EOM-psi}) up to leading order in $N$ simplifies to the following form:
\begin {eqnarray}
\label{EOM-LO-N}
& & \hskip -0.4in -\frac{\partial^2\psi(r,\theta_1)}{\partial r^2} +\frac{16(r^4+r_h^4)}{(r^4-r_h^4)}
\frac{\partial\psi(r,\theta_1)}{\partial r}\nonumber\\
& & \hskip -0.4in + \kappa_{r\theta_1}\frac{N^{4/5}r^6}{g_s^3M^2N_f^2\left(r^2-3a^2\right)^2\left(r^4-r_h^4\right)\left(\log N - 9 \log r\right)^2\left(\log r\right)^2}\left(\frac{\partial^2\psi(r,\theta_1)}{\partial\theta_1^2} - 2\frac{\partial\psi(r,\theta_1)}{\partial\theta_1}\right)\nonumber\\
& & \hskip -0.4in - m^2\psi(r,\theta_1) = 0,
\end {eqnarray}
we can separate the radial and angular equation via the seperation of variables, $\psi(r,\theta_1) = \mathbb{R}(r)\Theta(\theta_1)$ as follows:
\begin {eqnarray}
\label{EOM-separation_of_variables}
& & \hskip -0.4in \left(-\frac{1}{{\cal R}(r)}\mathbb{R}''(r) + \frac{1}{\mathbb{R}(r)}\frac{16(r^4+r_h^4)}{(r^4-r_h^4)}\mathbb{R}'(r) - m^2\right)\nonumber\\
& & \hskip -0.4in \left(\kappa_{r\theta_1}\frac{N^{4/5}r^6}{g_s^3M^2N_f^2\left(r^2-3a^2\right)^2\left(r^4-r_h^4\right)\left(\log N - 9 \log r\right)^2\left(\log r\right)^2}\right)^{-1}\nonumber\\
& & \hskip -0.4in = \frac{1}{\Theta(\theta_1)}\left(\Theta''(\theta_1) - 2 \Theta'(\theta_1)\right) \equiv \lambda
\end {eqnarray}
The $\Theta(\theta_1)$ equation has the following solution:
\begin {eqnarray}
\label{solution-psi}
e^{\theta_1\left(1\pm\sqrt{1+\lambda}\right)},
\end {eqnarray}
which is relevant if $\lambda=0$; we therefore get: $\Theta(\theta_1)=$Constant. The equation of motion associated with the radial profile $\mathbb{R}$ via (\ref{EOM-separation_of_variables}) is obtained as follows:
\begin{equation}
\label{EOM-R}
-\mathbb{R}''(r) + \left(\frac{8}{r-r_h} - \frac{4}{r_h} + {\cal O}(r-r_h)\right)\mathbb{R}'(r) - m^2 \mathbb{R}(r) = 0.
\end{equation}
Since the equation of motion (\ref{EOM-psi}) is homogeneous, and therefore $\psi\rightarrow m^2\psi(r)$ is also a viable solution, yielding:
\begin {eqnarray}
\label{solution-psi-TricomiU}
& & m^2c_1 U\left(-\frac{8-5 \sqrt{4-m^2 {r_h}^2}}{\sqrt{4-m^2 {r_h}^2}},10,\frac{2
   r \sqrt{4-m^2 {r_h}^2}}{{r_h}}-2 \sqrt{4-m^2 {r_h}^2}\right)\nonumber\\
& & \times \exp
   \left(\frac{r \left(-\sqrt{4-m^2 {r_h}^2}-2\right)+9 {r_h} \log
   ({r_h}-r)}{{r_h}}\right)\nonumber\\
& & +m^2c_2 L_{\frac{8-5 \sqrt{4-m^2
   {r_h}^2}}{\sqrt{4-m^2 {r_h}^2}}}^9\left(\frac{2 r \sqrt{4-m^2
   {r_h}^2}}{{r_h}}-2 \sqrt{4-m^2 {r_h}^2}\right)\nonumber\\
& & \times \exp \left(\frac{r
   \left(-\sqrt{4-m^2 {r_h}^2}-2\right)+9 {r_h} \log
   ({r_h}-r)}{{r_h}}\right).
\end {eqnarray}
When $c_2=0$, we obtained:
\begin {eqnarray}
\label{R-exp-rh}
& & m^2 U\left(-\frac{8-5 \sqrt{4-m^2 {r_h}^2}}{\sqrt{4-m^2 {r_h}^2}},10,\frac{2
   r \sqrt{4-m^2 {r_h}^2}}{{r_h}}-2 \sqrt{4-m^2 {r_h}^2}\right)\nonumber\\
& & \times \exp
   \left(\frac{r \left(-\sqrt{4-m^2 {r_h}^2}-2\right)+9 {r_h} \log
   ({r_h}-r)}{{r_h}}\right) \nonumber\\
& &  = \frac{e^{-2-\sqrt{4-m^2r_h^2}}}{\sum_{l=0}^3c_l (m r_h)^{2l}\Gamma\left(-4 - \frac{8}{\sqrt{4-m^2r_h^2}}\right)}\left[1 + {\cal O}\left((r-r_h)^2\right) \right].
\end {eqnarray}
When we enforces Dirichlet boundary condition\footnote{Neumann b.c., $\psi'(r=r_h)=0$, is identically satisfied $\forall m$.} at $r=r_h: psi(r=r_h)=0$, we found that $ -4 - \frac{8}{\sqrt{4-m^2r_h^2}} = - n, n\in\mathbb{Z}^+$ is required, i.e., 
\begin{equation}
\label{solution-m}
m = \frac{2\sqrt{n(n-8)}}{4r_h(1-n)}\stackrel{n=8}{\rightarrow}0.
\end{equation}
We found that:
\begin{eqnarray}
\label{m_0}
& & U\left(-\frac{8-5 \sqrt{4-m^2 {r_h}^2}}{\sqrt{4-m^2 {r_h}^2}},10,\frac{2
   r \sqrt{4-m^2 {r_h}^2}}{{r_h}}-2 \sqrt{4-m^2 {r_h}^2}\right)\nonumber\\
& & \times \exp
   \left(\frac{r \left(-\sqrt{4-m^2 {r_h}^2}-2\right)+9 {r_h} \log
   ({r_h}-r)}{{r_h}}\right) \stackrel{m=0}{\rightarrow}-\frac{c_1 {r_h}^9 \Gamma \left(9,\frac{4 r}{{r_h}}-4\right)}{262144 e^4}\nonumber\\
   & & = -\frac{315 \left(c_1 {r_h}^9\right)}{2048 e^4}+\frac{c_1 (r-r_h )^9}{9 e^4} + {\cal O}\left((r-{r_h})^{10}\right).
\end{eqnarray}
When $c_1=0$, then \\ $c_2 L_{\frac{8-5 \sqrt{4-m^2
   {r_h}^2}}{\sqrt{4-m^2 {r_h}^2}}}^9\left(\frac{2 r \sqrt{4-m^2
   {r_h}^2}}{{r_h}}-2 \sqrt{4-m^2 {r_h}^2}\right) \exp \left(\frac{r
   \left(-\sqrt{4-m^2 {r_h}^2}-2\right)+9 {r_h} \log
   ({r_h}-r)}{{r_h}}\right)$\\
    satisifies Dirichlet/Neumann boundary condition $\forall m$. Particularly for $m=0$, the preceding yields $c_2 e^{-\frac{4 r}{{r_h}}} ({r_h}-r)^9 L_{-1}^9\left(\frac{4 r}{{r_h}}-4\right)$. Interestingly, $\frac{d^n}{dr^n}L_{-1}^9\left(\frac{4 r}{{r_h}}-4\right)=(-4)^n\frac{1}{r_h^n}L_{-n-1}^{n+9}\left(\frac{4 r}{r_h }-4\right)$, implying $\lim_{r\rightarrow r_h}L_{-n-1}^{n+9}\left(\frac{4 r}{r_h }-4\right) = \binom{8}{n+9}=0. $ \\
The equation of motion up to leading order in $N, N_{f,\ \rm UV}, M_{\rm UV}$ in the UV is obtained as: 
\begin {eqnarray}
\label{EOM-UV}
& & -\frac{\partial^2\psi_{\rm UV}(r,\theta_1)}{\partial r^2} + \frac{16}{r}\frac{\partial\psi_{\rm UV}(r,\theta_1)}{\partial\theta_1}\nonumber\\
& & -2\kappa_{r\theta_1}\frac{N^{4/5}}{g_s^3M_{\rm UV}^2N_{f\ \rm UV}^2\left(\log N - 9 \log r\right)^2\left(\log r\right)^2r^2}\frac{\partial\psi_{\rm UV}}{\partial\theta_1}-m^2\psi_{\rm UV}=0 ,
\end {eqnarray}
the solution of above equation is:
\begin {eqnarray}
\label{EOM-UV-sol-i}
& & \psi_{\rm UV}\sim\frac{1}{m^{17/2}}\Biggl[c_1\sum_{l_1=0}^3a_{l_1} (m r)^{2l_1+1}
\cos(m r) \nonumber\\
& & + \Biggl(c_1\sum_{l_2=0}^4a_{l_2}(m r)^{2l_2} + c_2 \sum_{l_1=0}^3
(m r)^{2l_1+1}\Biggr)\sin (m r)\Biggr].
\end {eqnarray}
 Since (\ref{EOM-UV-sol-i}) is not well defined in the UV when $m\neq0$. Therefore, we need to consider $m=0$ for which the graviton wave function is:
\begin {eqnarray}
\label{EOM-UV-sol-ii}
\mathbb{R}_{\rm UV}(r)=c_1^{\rm UV}r^{17} + c_2^{\rm UV}.
\end {eqnarray}
Normalization of $\psi(r)$ requires $c_1^{\rm UV}=0$, and hence there is a constant graviton wave function in the UV.\par 
{\it The physical/intuitive evidence behind the exponentially suppressed entanglement entropy for Island Surface (\ref{SEE-IS-simp}) would be because the Laplace-Beltrami equation for internal coordinates (\ref{EOM-psi}) enables vanishing graviton mass - which is as well associated with being the case that the calculations that we perform are in the ``near-horizon'' limit ($r<\left(4\pi g_s N\right)^{1/4}$) wherein even the UV cut-off $r_{\rm UV}\stackrel{\sim}{<}(4\pi g_s N)^{1/4}$, and therefore the internal manifold is compact \cite{C. Bachas and J. Estes [2011]}. The aforementioned is the reason for the Island surface's extremely low (including exponential-in-$N$ suppression) entanglement entropy. A similar entanglement entropy associated with the Hartman-Maldacena-like surface in (\ref{SEE-HM-tb0}) at the Page time is non-trivial and hence somewhat fascinating. The diminishing graviton mass is also seen in the entanglement entropies of the IS-vs-HM-like surfaces in (\ref{S_EE_hierarchy}) at ${\cal O}(\beta)$ - the former is suppressed compared to the latter.}
%%%%%%%%%
We are able to translate equation (\ref{EOM-separation_of_variables}) into an equivalent Schr\"{o}dinger-like equation using $r=r_h e^Z$\footnote{The benefit of utilizing redefined radial coordinates is that $r \in (0,\infty)$ translates to $Z \in (-\infty,\infty)$, allowing us to observe a lovely volcano-like potential on both sides of $Z=0$.}:
\begin{eqnarray}
\label{S-like-EOM-Z}
-R''(Z)+ V(Z)R(Z) = 0,
\end{eqnarray}
where 
\begin{equation}
\label{R-calR}
R(Z) = \mathbb{R}(Z) e^{r_h\left(-8 e^Z + \frac{Z}{2r_h} + 8 \tan^{-1}(e^Z) + 8 \tanh^{-1}(e^Z)\right)}\approx \mathbb{R}(Z) e^{\frac{Z}{2}},
\end{equation}
and,
\begin{eqnarray}
\label{V[Z]}
& & \hskip -0.2in
V(Z)=-\frac{e^{2 Z} r_h^2 \left(\left(m^2-64\right) e^{8 Z}-2 \left(m^2+64\right) e^{4 Z}+m^2-64\right)}{\left(e^{4 Z}-1\right)^2}+\frac{64 e^{5 Z} r_h}{\left(e^{4 Z}-1\right)^2}+\frac{1}{4}
\end{eqnarray}
{\it Figure \ref{V-Potential} plots the aforementioned potential for massless graviton $(m=0)$; this potential is ``volcano''-like, with the massless graviton localized at the horizon on the ETW ``brane''.  This is analogous to \cite{KR1}, in which gravity can be localized on the end-of-the-world (ETW) brane with non-zero brane tension owing to the emergence of a ``crater'' in the ``volcano'' potential in the Schr\"{o}dinger-like equation of motion of the graviton wave function.}\\
\begin{figure}
\begin{center}
\includegraphics[width=0.60\textwidth]{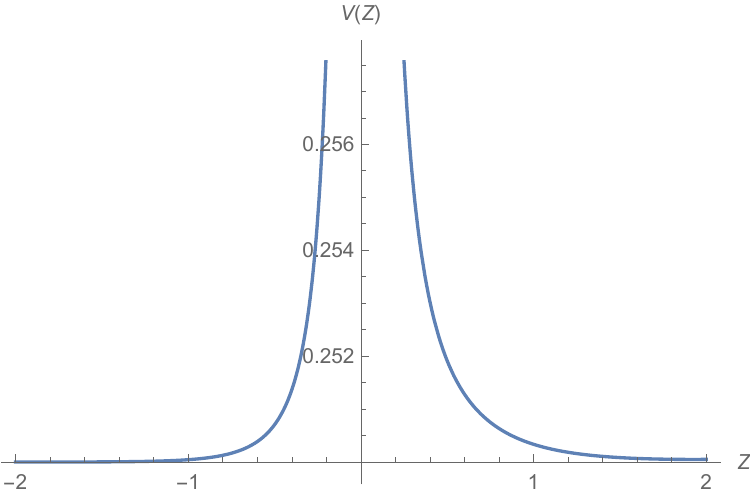}
\end{center}
\caption{Volcano potential for massless graviton from (\ref{V[Z]})}
\label{V-Potential}
\end{figure}
Because the ETW-``brane'' possesses non-zero ``tension'' (\ref{T_ETW}) within our set-up, gravity could be localized on the ETW-``brane''. Utilizing (\ref{m_0}), one can observe that the graviton wave-function is definitely localized towards the horizon in the massless-limit of the graviton, as shown in Fig. \ref{grav-wf-localized}.
\begin{figure}
\begin{center}
\includegraphics[width=0.60\textwidth]{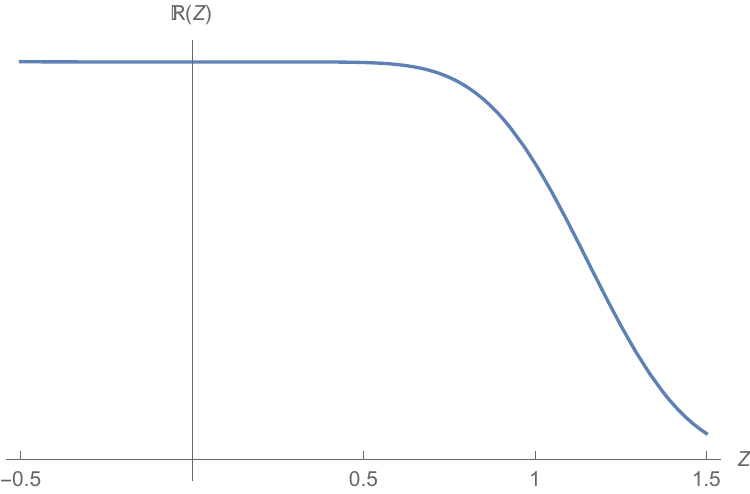}
\end{center}
\caption{Graviton Wave Function localization at the horizon via the $m=0$-limit of the solution (\ref{m_0}) of (\ref{EOM-R}).}
\label{grav-wf-localized}
\end{figure}
On the other hand, by expanding $V(Z)$ of (\ref{V[Z]}) across the horizon $Z=0$, we get: $V(Z\sim0)\sim \frac{4r_h(1+4r_h)}{Z^2} + \frac{4r_h(1+8r_h)}{Z}
-\frac{1}{12}(-3+40r_h-896r_h^2) + {\cal O}(Z)$, we obtained:
\begin{eqnarray}
\label{wave-function-Schroedinger}
& &  R(Z) = c_1 M_{-\frac{4 \sqrt{3} r_h  (8 r_h +1)}{\sqrt{896 r_h ^2-40 r_h +3}},4
   r_h +\frac{1}{2}}\left(\frac{\sqrt{896 r_h ^2-40 r_h +3} Z}{\sqrt{3}}\right) \nonumber\\
 & &  +c_2 W_{-\frac{4 \sqrt{3}
   r_h  (8 r_h +1)}{\sqrt{896 r_h ^2-40 r_h +3}},4 r_h +\frac{1}{2}}\left(\frac{\sqrt{896
   r_h ^2-40 r_h +3} Z}{\sqrt{3}}\right).
\end{eqnarray}
Now, both Whittaker functions are complex for $Z<0$, i.e., $r<r_h$. Choosing $c_1=0$, we obtained $\mathbb{R}(Z) = e^{-\frac{Z}{2}}W_{-\frac{4 \sqrt{3}
   r_h  (8 r_h +1)}{\sqrt{896 r_h ^2-40 r_h +3}},4 r_h +\frac{1}{2}}\left(\frac{\sqrt{896
   r_h ^2-40 r_h +3} Z}{\sqrt{3}}\right)$. As a result, the graph \ref{grav-wf-localized-Schroedinger} depicts a decrease of the graviton wave-function farther from the horizon.
  \begin{figure}
\begin{center}
\includegraphics[width=0.60\textwidth]{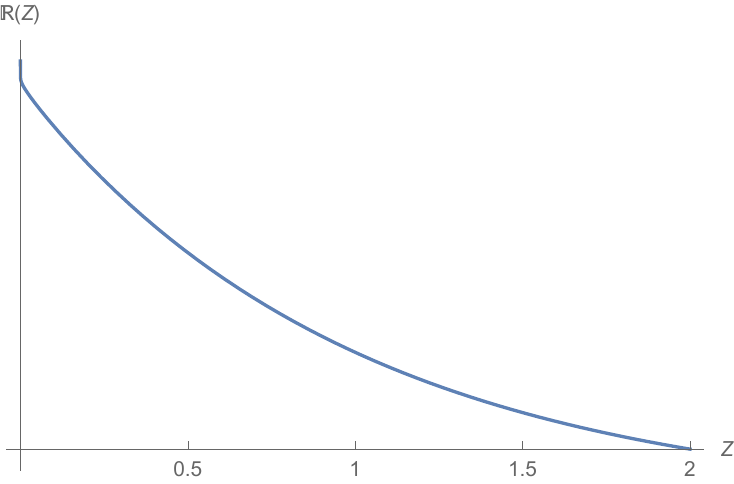}
\end{center}
\caption{Localization of the Graviton Wave Function near the horizon from the solution of the Schr\"odinger-like radial wave equation (\ref{S-like-EOM-Z}).}
\label{grav-wf-localized-Schroedinger}
\end{figure}  

\section{Summary}\label{summary}
\vspace{-0.2cm}
In this chapter, we constructed the doubly holographic setup from a top-down approach as discussed in \ref{DHS-M-Theory}. We have taken the external bath as thermal QCD in three dimension whose holographic dual is ${\cal M}$-theory inclusive of ${\cal O}(R^4)$ corrections \cite{HD-MQGP}, see chapter {\bf 1} for more detail of the same. The intermediate description of the doubly holographic setup constructed by us couples the black hole on the ETW-``brane'' at $x=0$ to thermal QCD bath via transparent boundary condition at the defect. We have addressed the effect of higher derivative terms in the context of top-down construction of double holography for the first time.
First we obtained the Page curve without inclusion of ${\cal O}(R^4)$ terms in the eleven dimensional supergavity action in \ref{Page-curve-WHD}. To achieve the same, we computed the entanglement entropies associated with the Hartman-Maldacena-like (HM-like) and island surfaces and obtained the Page curve with the help of these entropies because the HM-like surface has linear time dependence whereas the island surface's entanglement entropy is found to be constant. In ${\cal R}_{D5/\overline{D5}}=1$-units, utilizing (\ref{Page-curve-beta0}), (\ref{EE-IS-simp}), (\ref{r_T-iii}) and $ r_h\sim e^{-\kappa_{r_h}N^{\frac{1}{3}}}$ \cite{Bulk-Viscosity-McGill-IIT-Roorkee}, the above is concisely expressed via:
\begin{eqnarray}
\label{SEE-beta0-Area-Summary-i}
& & S^{\rm HM} = \frac{A_{\rm HM}}{4G_N^{(11)}}\sim \frac{{\cal O}(1)\times10^{-4}M^2  N_f^6 g_s^{15/4} N^{34/15} {\bf e^{-6\kappa_{r_h}N^{1/3}}}}{ G_N^{(11)}}t_b,\ t_b\leq t_{\rm Page};\nonumber\\
& & {\cal S}^{IS} = {\cal S}^{IS}\left(\tilde{r}_T \equiv \frac{r_T}{r_h} = (1+\delta_2); r_h\right),\  t_b\geq t_{\rm Page},
\end{eqnarray}
where $\delta_2$ is provided in (\ref{r_T-iii}). Next, we included the ${\cal O}(R^4)$ terms in the supergravity action, and obtained the Page curve in \ref{Page-curve-HD}. We used the Dong's formula to compute the entanglement entropies in the presence of higher derivative terms \cite{Dong}. In this case entanglement entropies of HM-like and island surfaces are summarized as:
\begin{eqnarray}
\label{SEE-beta0-Wald-Summary-i}
& & \hskip -0.3in S_{\rm EE}^{\beta^0, {\rm HM}} \sim   {\bf e^{-\frac{3\kappa_{l_p}N^{\frac{1}{3}}}{2}}}M N^{13/10} N_f^{5/3}\left(1-\frac{4 t_{b_0}}{3 c_2}\right)\left(\frac{2}{3}\log \left(\frac{c_2}{c_1}\right) -n_{t_b} \log(N)\right)^{\frac{17}{6}},\ \kappa_{l_p}\equiv\frac{1}{{\cal O}(1)},\nonumber\\
& & \hskip -0.3in c_{1,2}<0, |c_2|\gg |c_1|; |c_2|\sim e^{\kappa_{c_2}|c_1|^{1/3}},\ \kappa_{c_1}\equiv{\cal O}(1)\nonumber\\
& & \hskip -0.3in t\leq t_{\rm Page},\nonumber
\end{eqnarray}
\begin{eqnarray}
& & \hskip -0.3in S_{\rm EE}^{\beta^0, {\rm IS}} \sim \frac{M^2 {\bf e^{-\frac{3\kappa_{l_p} N^{\frac{1}{3}}}{2}}}
 N_f^{4/3} g_s^{35/6} \log ^2(N)  \left| \log \left(r_h\right)\right|^{4/3} }{r_h^{11/2}}, \nonumber\\
 & & \hskip -0.3in t\geq t_{\rm Page}.
\end{eqnarray}
The aforementioned has been showed to be compatible with the prior summary of RT computation. (\ref{SEE-beta0-Area-Summary-i}) and (\ref{SEE-beta0-Wald-Summary-i}) must be compatible if $\kappa_{r_h},\ \kappa_{l_p},\ \kappa_{c_2}$:\\ $M^2N_f^6g_s^{15/4}N^{34/15}e^{-6\kappa_{r_h}N^{1/3}}\sim M N^{13/10}N_f^{5/3}\left(\frac{2}{3}\biggl[\kappa_{c_2}N^{1/3} - \log |c_1|\biggr] -n_{t_b} \log(N)\right)^{\frac{17}{6}}e^{-\frac{3\kappa_{l_p}N^{1/3}}{2}}$ and $11\kappa_{r_h} < 3\kappa_{l_p}$, so that $S^{\beta^0, {\rm IS}}_{\rm EE}$ is be well-defined in the large-$N$ limit. \par
It is interesting to note that in \ref{Swiss-Cheese-i}, where $c_1$ and $c_2$ are two parameters from the group of HM-like surface embeddings, $\frac{\delta S_{\rm EE}^{\rm HM, \beta^0}}{\delta |c_2|}>0,\ \frac{\delta S_{\rm EE}^{\rm HM, \beta^0}}{\delta |c_1|}<0$. The ideal ``Swiss-Cheese'' structure (of the ``single-big-divisor-single-small-divisor variety'') is provided by delta $|c_1|0$, which together with $|c_2|\sim e^{\kappa_{c_2}|c_1|^{\frac{1}{3}}},\ |c_1|\sim N$, provides a perfect ``Swiss-Cheese'' (of the ``single-big-divisor-single-small-divisor'' variety) structure to $S_{\rm EE}^{\rm HM, \beta^0}(|c_1|, |c_2|)$ (which essentially is a co-dimension-two volume) wherein $\log|c_2|$ plays the role of the ``big divisor'' volume and $\log|c_1|$ plays the role of a solitary ``small divisor'' volume, realizing what could be dubbed as a ``Large N Scenario''(LNS). Likewise, the entanglement entropy for the Hartman-Maldacena-like surface has to be considered just like a Swiss-Cheese-like open surface in the two-dimensional (in the IR) space of family of HM-like embeddings $\mathbb{R}_+^2\left(|c_1|, |c_2|\right)$ supplemented by the entanglement entropy that coordinatizes $\mathbb{R}_{\geq0}(S_{\rm EE}^{\rm HM, \beta^0})$. \par
Based on the viability of the Islands scenario, the ${\cal M}$-theory dual of large-$N$ thermal QCD at high temperature obtained in \cite{MQGP}, \cite{HD-MQGP} (based on \cite{metrics}) produces the previously mentioned Page curve. The above-mentioned top-down ${\cal M}$-theory dual, on the other hand, provides a number of new conceptual information:
\begin{itemize}
\item Due to the lack of details of precise structures for boundary terms on the ETW-``brane'' along with the presence of higher derivative terms, there exists relatively few works (e.g., \cite{NGB}) in which the authors look into the doubly holographic setup in higher derivative theory of gravity. Remarkably we showed in \ref{ETW-sub-sub-ii} that the presence of the ${\cal O}(R^4)$ terms generates no boundary terms. 

\item As for as we know, our setup/(above-mentioned) ${\cal M}$-theory dual appears either the sole one in existence (from ${\cal M}$-theory) or one among the comparatively few top-down approaches that generate(s) the Page curve for massless graviton\footnote{See \cite{critical-islands}, where the author discusses how to obtain the Page curve for a massless graviton in the critical Randall-Sundrum II model via a bottom-up strategy, also have a look at \cite{Massless-Gravity} where the Page curve in the presence of massless graviton has been obtain via the inclusion of DGP term on the Karch-Randall brane in wedge holography.}.

\item In our top-down ${\cal M}$-theory dual, we find that ETW-``brane'' to be a fluxed hypersurface ${\cal W}$ that is a warped product of an asymptotic $AdS_4$ and a six-fold $M_6$ where $M_6$ is a warped product of the ${\cal M}$-theory circle and a non-Einstenian generalization of $T^{1,1}$; the hypersurface ${\cal W}$, can also be thought of as an effective ETW-``brane'' corresponding to fluxed intersecting $M5$-brane wrapping a homologous sum of $S^3\times[0,1]$ and
$S^2\times S^2$ in a warped product of $\mathbb{R}^2$ and an $SU(4)/Spin(7)$-structure eight-fold. The ETW-``brane'', ${\cal W}$, then has non-zero ``tension'' and a massless graviton localized near the horizon by a ``volcano''-like potential.

\item Unlike largely other works in the literature, which compute the Page curve with a CFT bath, the external bath in our model is a non-CFT bath (thermal QCD). 

\item Entanglement entropy contribution generated by a Hartman-Maldacena (HM)-like surface, which has been causing the growth of the Einstein-Rosen bridge in time, shows a Swiss-Cheese structure in the Large-$N$ scenario (\ref{Swiss-Cheese-i}) as discussed earlier.

\item By the presence of ${\cal O}(R^4)$ terms in the action, our previously discussed ${\cal M}$-theory dual produces a hierarchy in the entanglement entropies of the HM-like and Island surface (IS) with respect to a large-$N$ exponential suppression factor, resulting physically via the presence of massless graviton mode on an ETW-brane. This suppression also suggests the fact that the addition of higher derivative terms - ${\cal O}(R^4)$ in particular - has no effect on the Page curve.

\item To regulate the IR- and large-$N$ enhancement in the IS entanglement entropy per unit BH entropy, a relationship involving the Planckian length and the non-extremality parameter (the horizon radius) has been shown to occur.

\item The positivity of the Page time, calculated in (refPage-time), has been showed to set an upper constraint on the non-extremality parameter, the black-hole horizon radius $r_h$.

\end{itemize}

%m

\chapter{Black Hole Islands in Multi-Event Horizon Space-Times}
\graphicspath{{Chapter8/}{Chapter8/}}

\section{Introduction and Motivation}\label{basics}
The information paradox of black holes has been investigated for black holes that are asymptotically flat, however the most recent observation indicates that our universe is currently in an accelerated phase. Because of this, it is only natural to wonder how the presence of a positive value for the cosmological constant $\Lambda$ influences the information paradox problem. The information paradox of a stationary black hole with a positive cosmological constant $\Lambda$ or the Schwarzschild de-Sitter black hole spacetime is something that attracted our attention. The information paradox of Schwarzschild de-Sitter black holes is important due to the fact that these black holes originated during the early inflationary phase of our universe (for example, \cite{Bousso:1997wi, Chao:1997em, Bousso:1999ms, Anninos:2010gh}). It additionally offers a fantastic toy model for the global framework of isolated black holes of our universe, which is quite useful while keeping in mind the current period of accelerated expansion. Additionally, similar to the situation with black holes, there are regions of de Sitter space that are causally separated from one another. Therefore, an observer is only able to access the portions of the cosmos that are constrained by their own horizon. There are two event horizons associated with the Schwarzschild de-Sitter black hole: the cosmological event horizon (CEH) and the black hole event horizon (BEH). The global causal border of the de Sitter spacetimes has been provided by the cosmological event horizon. In comparison to $\Lambda \leq 0$ single horizon spacetime, the thermodynamics of these event horizons are distinct \cite{Lochan:2018pzs, Goheer:2002vf, Marolf:2010tg}. Gibbons-Hawking radiation is emitted and absorbed by cosmological event horizon similar to the Hawking radiation of black holes. Comparatively speaking, the entropy creation of the cosmic horizon is an observer-dependent characteristic, in contrast to the entropy generation of the black hole. It is caused by people's lack of knowledge regarding the things that exist beyond the cosmological horizon. \\
~\\
   In order to obtain the Page curve of the Schwarzschild de-Sitter black hole, we make use of the island concept. In the case of models that are not holographic, we are able to use the s-wave approximation for studying black holes in higher dimensions. Because we neglect the angular component of the metric when performing the s-wave approximation, we are left with a CFT metric that only has two dimensions. Because of this, we are able to compute the entanglement entropy of Hawking radiation using the formula for 2D CFT that is provided in \cite{CC,CC-1}. {\it The purpose of this work was to investigate the information paradox and its resolution of black holes with multiple horizons. We considered the Schwarzschild de-Sitter black hole with two horizons and obtained the Page curves for the black hole and the de-Sitter patches by placing thermally opaque membranes on the two sides of the region of study. The ``effect of temperature'' on the Page curves and the scrambling time is something else that would be of interest to us. In this chapter, we will discuss the Page curve of black hole patch only.} The entanglement entropy and complexity have been studied in \cite{S1,S2,S3} in the context of de-Sitter space.
  
\section{Preliminary}\label{S2}
In this section, we will discuss the basics of Schwarzschild de-Sitter black hole, the concept of thermal opaque membrane and the effect of gravity near the aforementioned membrane in \ref{SdS-Intro}, \ref{TOM}, and \ref{BGE} respectively.
\subsection{Schwarzschild de-Sitter Black Hole}
\label{SdS-Intro}
The Schwarzschild de-Sitter (SdS) metric has the following form in spherical polar coordinates: 
\begin{eqnarray}
ds^2=-\left(1-\frac{2M}{r}-\frac{\Lambda r^2}{3}\right)dt^2+\left(1-\frac{2M}{r}-\frac{\Lambda r^2}{3}\right)^{-1}dr^2+r^2 \left(d\theta^2 +\sin^2\theta d\phi^2 \right),
\label{l1}
\end{eqnarray}
where $M$ denotes mass parameter. Horizons of SdS black hole can be obtained by solving, $\left(1-\frac{2M}{r}-\frac{\Lambda r^2}{3}\right)=0$. There are three roots of the aforementioned equation in the range, $0<3M \sqrt{\Lambda} < 1$ \cite{Gibbons, JHT,SA} which are written below:
\begin{eqnarray}
& &
r_{H}=\frac{2}{{\sqrt \Lambda} }\cos\frac{\pi+\cos^{-1}(3M\sqrt{\Lambda})}{3},\nonumber\\
& & r_{C}=\frac{2}{{\sqrt \Lambda} }\cos\frac{\pi-\cos^{-1}(3M \sqrt{\Lambda})}{3},\nonumber\\
& & r_{U}=-(r_H+r_C),
\label{l2}
\end{eqnarray}
where $r_H$ and $r_C$ are both positive values and are referred to as the cosmology event horizon (CEH) and the black hole event horizon (BEH), respectively, and $r_U<0$ acts as unphysical horizon. (\ref{l1}) denotes a Schwarzschild black hole that is located inside the cosmic horizon and it appears when $3M\sqrt{\Lambda}<1$. Both of the horizons will merge into a single horizon in Nariai limit, where $3M\sqrt{\Lambda} \to 1$, and for  $3M\sqrt{\Lambda}>1$, space-time will have the naked curvature singularity. The surface gravities of black hole and the cosmological event horizons are calculated to be as follows:
{\footnotesize
\begin{eqnarray}\label{l3}
&& \hskip -0.25 in \kappa_H= \frac{\Lambda (2r_H+r_C)(r_C-r_H)}{6 r_H}=-\sqrt{\Lambda}\left(\cos\left[\frac{1}{3}\cos^{-1}(3M\sqrt{\Lambda})+\frac{\pi}{3}\right]-\frac{1}{4\cos\left[\frac{1}{3}\cos^{-1}(3M\sqrt{\Lambda})+\frac{\pi}{3}\right]}\right),\nonumber
\end{eqnarray}
\begin{eqnarray}
& & \hskip -0.25in -\kappa_C=\frac{\Lambda (2r_C+r_H)(r_H-r_C)}{6 r_C}=\sqrt{\Lambda}\left(\frac{1}{4\cos\left[\frac{1}{3}\cos^{-1}(3M\sqrt{\Lambda})-\frac{\pi}{3}\right]}-\cos\left[\frac{1}{3}\cos^{-1}(3M\sqrt{\Lambda})-\frac{\pi}{3}\right]\right).\nonumber\\
\end{eqnarray}
}
Because of the repulsive effects caused by the positive cosmological constant $\Lambda$, the value of $\kappa_C$ should really be written with a negative sign in front of it. It is important to keep in mind that when $r_C\geq r_H$, we get $\kappa_H\geq \kappa_C$, and within the Nariai limit, both the surface gravities $\kappa_H$ and $\kappa_C$ disappear. Equation (\ref{l1}) have two coordinate singularities at the points $r=r_H,,r_C$; hence, we require two Kruskal-like coordinates in order to get rid of them and expand spacetime beyond them. To begin, we transform the (\ref{l1}) into the form \cite{Bhattacharya:2018ltm}:
\begin{eqnarray}
ds^2=\left(1-\frac{2M}{r}-\frac{\Lambda r^2}{3}\right)\left(-dt^2+dr_{\star}^2\right) +r^2(r_{\star})\left(d\theta^2 +\sin^2\theta d\phi^2 \right),
\label{ds4}
\end{eqnarray}
where $r_{\star}$ denotes tortoise coordinate and is defined as:
\begin{eqnarray}
& &
r_{\star}=\int \left(1-\frac{2M}{r}-\frac{\Lambda r^2}{3}\right)^{-1}\,dr \nonumber\\
& & =\frac{1}{2\kappa_H}\ln \left(\frac{r}{r_H}-1\right) -\frac{1}{2\kappa_C} \ln \left(1-\frac{r}{r_C}\right) +\frac{1}{2\kappa_u}\ln \left(\frac{r}{r_U}-1\right), 
\label{tor}
\end{eqnarray}
wherein $\kappa_u= \frac{\kappa_H\kappa_C}{\kappa_H-\kappa_C}= (M/r_u^2-\Lambda r_u/3)$ being the surface gravity of unphysical horizon $r_u$.Define $u=t-r_{\star}$, $v=t+r_{\star}$, (\ref{ds4}) simplified to the following form:
\begin{eqnarray}
ds^2=-\left(1-\frac{2M}{r}-\frac{\Lambda r^2}{3}\right)\,dudv +r^2(u,v)\left(d\theta^2 +\sin^2\theta d\phi^2 \right).
\label{ds7}
\end{eqnarray}
With the help of (\ref{tor}), we were able to transform the previously mentioned metric (\ref{ds7}), into two other forms:
\begin{eqnarray}
ds^2=-\frac{2M}{r}\left\vert1-\frac{r}{r_C}\right\vert^{1+\frac{\kappa_H}{\kappa_C}} \left(1+\frac{r}{r_H+r_C}\right)^{1-\frac{\kappa_H}{\kappa_U}}\, d{U}_H d {V}_H+r^2(d\theta^2+\sin^2\theta d\phi^2),
\label{ds16}
\end{eqnarray}
and,
\begin{eqnarray}
ds^2=-\frac{2M}{r}\left\vert\frac{r}{r_H}-1\right\vert^{1+\frac{\kappa_C}{\kappa_H}} \left(1+\frac{r}{r_H+r_C}\right)^{1+\frac{\kappa_C}{\kappa_U}}\, d {U}_C d {V}_C+r^2(d\theta^2+\sin^2\theta d\phi^2),
\label{ds17}
\end{eqnarray}
where the Kruskal coordinates are:
\begin{eqnarray}
{U}_H=-\frac{1}{\kappa_H}e^{-\kappa_H u},\quad {V}_H=\frac{1}{\kappa_H}e^{\kappa_H v} \quad {\rm and} \quad
{U}_C=\frac{1}{\kappa_C}e^{\kappa_C u},\quad {V}_C=-\frac{1}{\kappa_C}e^{-\kappa_C v}.
\label{ds15}
\end{eqnarray}
Therefore (\ref{ds16}) and (\ref{ds17}) do not have any coordinate singularities at $r=r_H$ and $r=r_C$. It is not possible to remove the coordinate singularities of both event horizons simultaneously. Hence, we have to study the black hole patch by freezing the de-Sitter patch using thermal opaque membrane and vice-versa.
\subsection{Thermal Opaque Membranes in SdS Space-Time} 
\label{TOM}
The idea of a thermally opaque membrane has been utilized extensively in published works to look at one horizon while taking another as the fixed, e.g., see \cite{Saida:2009ss,Ma:2016arz,Sekiwa:2006qj,Gomberoff:2003ea}. We provide an illustration of the proposal discussed in \cite{Nitin} regarding the construction of thermally opaque membranes. The following is the form that the Klein-Gordon equation takes in $(3+1)$-dimensions. The radial equation is:
\begin{eqnarray}\label{thermalwall}
\left(-\frac{\partial^2}{\partial t^2}+\frac{\partial^2}{\partial r_{\star}^2} \right)R(r)+\left(1-\frac{2M}{r}-\frac{\Lambda r^2}{3} \right)\left(\frac{l(l+1)}{r^2} +\frac{2M}{r^3} -\frac{\Lambda}{3} \right) R(r)=0 .
\end{eqnarray}
The fact that the effective potential term disappears at both horizons and remains positive in between them is something that stands out as an interesting feature of the Schr\"{o}dinger-like equation that was given before. Therefore, this bell-shaped potential acts as a barrier between the event horizons of the black hole and de-Sitter space. To be able to depict it in Penrose diagram of the extended Schwarzschild de-Sitter space-time, we will use (\ref{ds15}) to describe the Kruskal timelike and spacelike coordinates as follows:
\begin{eqnarray}
& & \hskip -0.2in U_H = T_H -R_H, \quad V_H = T_H+R_H,\qquad {\rm and} \qquad U_C = T_C -R_C, \quad V_C = T_C+R_C,
\end{eqnarray}
implying
\begin{eqnarray}
&&-U_H V_H=R_H^2 -T_H^2= \frac{1}{\kappa_H^2} \left\vert 1-\frac{r}{r_C}\right\vert^{-\kappa_H/\kappa_C} \left\vert \frac{r}{r_U}-1\right\vert^{\kappa_H/\kappa_C}\left(\frac{r}{r_H}-1 \right),\nonumber\\
&&-U_C V_C=R_C^2 -T_C^2= -\frac{1}{\kappa_C^2} \left\vert \frac{r}{r_U}-1\right\vert^{-\kappa_C/\kappa_U} \left\vert \frac{r}{r_H}-1\right\vert^{-\kappa_C/\kappa_H}\left(1-\frac{r}{r_C} \right).
\label{ds5'}
\end{eqnarray}
Therefore, the line $r=${\it constant} is a hyperbola linking $i^{\pm}$, and it is possible to draw it with regard to any of the Kruskal coordinates listed above. This appears in the Penrose diagram of the maximally extended SdS space-time. When we have put this hyperbola or the thermal opaque membrane, modes on each side of wall will be unable to pass through it and will continue to exist in their respective regions. A natural manifestation of this effective potential could be possible with the thermally opaque membrane.
\subsection{Bulk Gravitational Effect Near Thermal Opaque Membranes}
\label{BGE}
In the context of this discussion, we are interested in the bulk gravitational influence on thermally opaque membranes. Using the island formula, the authors of \cite{Sybesma,anchor-curve} studied the information paradox of pure de-Sitter space, evaporating and eternal black holes in the weak gravity domain. These setups do not include an exterior bath and the authors included an anchor curve that produces both an interior and an exterior. The island formula can be applied despite the fact that gravity is extremely weak on the exterior region. We suggest that we may think of an ``anchor curve'' going through the locations $b_{1,2}^{\pm}$ within our setup also where we have specified the boundaries of radiation regions in Figs. \ref{fig1} and \ref{fig2}. Our justification is based on \cite{Sybesma,anchor-curve}. The interior and exterior in our setup are marked by these ``anchor curves''. The ``thermal opaque membrane'' is being able of blocking the radiation coming from the region, that is not relevant to our research in any way. Therefore, the applicability of the island formula to our situation can only be considered notional at this point. The information paradox that arises from black holes in multi-event horizon space-time can be solved using this approach. Considering the membrane that was discussed before is located at a significant distance from the region affected by the black hole, it is reasonable to suppose that the gravitational effects around these membranes is insufficiently strong. As a result, $2D$ CFT formulas can be utilized to determine the entanglement entropy of Hawking radiation. We have discussed another approach to get the Page curve of Schwarzschild de-Sitter black hole in chapter ${\bf 9}$ using wedge holography.

\section{Black Hole Islands in Schwarzschild de-Sitter Black Hole}
\label{IP-BH}
The black hole and the de-Sitter patches are the two components that make up a Schwarzschild de-Sitter black hole. Only the black hole patch will be taken into consideration in this chapter. In order to investigate the information problem associated with the black hole patch, we first freeze the de-Sitter patch by placing the thermal opaque membrane that was discussed earlier on both sides of the black hole patch.  Near thermal opaque membranes, we have chosen the boundaries of the radiation regions. In order to calculate the entanglement entropy of Hawking radiation, we will utilize the two-dimensional CFT formula in the s-wave assumption of \cite{Island-SB} (in which the observer is very distant away from the black hole, represented by  $b_{1,2}^{\pm} \gg r_{H,C}$). The following is the metric for the black hole:
\begin{eqnarray}
\label{metric-BH}
ds^2= - g_H^2(r) dU_H \ dV_H +r^2\left(d\theta^2+\sin^2\theta d\phi^2\right).
\end{eqnarray}
The conformal factor and Kruskal coordinates, both of which exist in (\ref{metric-BH}), are defined as follows:
\begin{eqnarray}
\label{conf-BH}
& &
g^2_H(r)=\frac{2 M}{r} \biggl|1-\frac{r}{r_C}\biggr|^{\left(1+\frac{\kappa_H}{\kappa_C}\right)}\left(1+\frac{r}{r_C+r_H}\right)^{\left(1-\frac{\kappa_H}{\kappa_U}\right)},\nonumber\\
& & U_H=-\frac{1}{\kappa_H}e^{-\kappa_H \left(t-r_*(r)\right)}; \ V_H=\frac{1}{\kappa_H}e^{\kappa_H \left(t+r_*(r)\right)}.
\end{eqnarray}
It has been calculated that thermal entropy of the black hole is:
\begin{eqnarray}
S_{\rm th}^{\rm BEH}=\frac{A_{\rm BEH}}{4 G_N}=\frac{\pi r_H^2}{G_N}.
\end{eqnarray} 
\subsection{No Island Phase}
\label{No-Island-BH}
In this case, the region of interest for the black hole patch is the area that is completely encompassed by the thermally opaque membrane on both sides in the $C$ and $L$ domains.
\begin{figure}[h!]
\begin{center}
  \includegraphics[width=12.0cm]{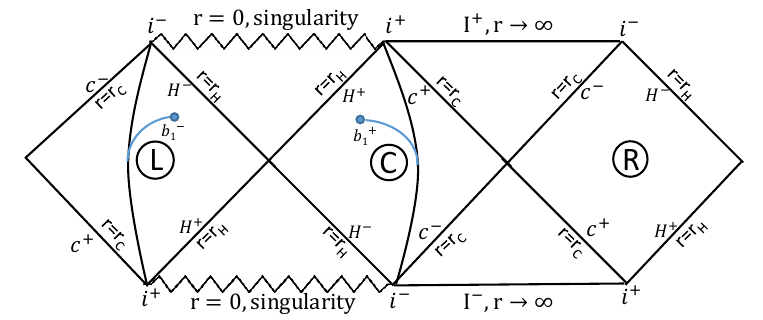}
  \caption{Carter-Penrose diagram of Schwarzschild de-Sitter spacetime. The causal connections between any two of the seven wedges are broken, and the passage of time is reversed for  $R, L$  regions in reference to the central region (C). If it serves the purpose, spacetime can have an additional extension made toward both sides that goes on indefinitely. The radiation regions $\mathcal{R}$ have been defined on both sides of the thermally opaque membranes, which are represented by the blue color curves. $H^\pm(C^\pm)$ shows the  past and future black hole horizons (cosmological horizons), $i^\pm$ depicts the past and future timelike infinities, and $I^\pm$ denotes the past and future spacelike infinities.}
  \label{fig1}
\end{center}
\end{figure}
Without island surface, we were able to calculate the entanglement entropy of Hawking radiation by utilizing the two-dimensional CFT formula that is given below \cite{CC,CC-1}: 
\begin{eqnarray}
\label{EE-formula-no-island-BH}
S_{\rm BEH}^{\mathcal{R}} = \frac{Q}{6} \log\left(d^2(b_1^+,b_1^-)\right),
\end{eqnarray}
where $b_1^+(t_{b_1},b_1)$, $b_1^-(-t_{b_1}+\iota \frac{\beta}{2},b_1)$, ($\beta=2\pi/\kappa_H$, and $t_{b_1}$ being the boundary time), are the radiation regions boundaries in the right and left wedges of the black hole patch, and $Q$ denotes the central charge of two-dimensional CFT (boson: $Q=1$, fermion: $Q=1/2$). The following expression \cite{NBH-HD}, will give us the geodesic distance between two points $l_1$ and $l_2$ for the 2D component of the metric (\ref{metric-BH}):
\begin{equation}
\label{d}
d(l_1,l_2)=\sqrt{g_H(l_1)g_H(l_2)(U_H(l_2)-U_H(l_1))(V_H(l_1)-V_H(l_2))}.
\end{equation}
By utilizing (\ref{conf-BH}), (\ref{tor}) and (\ref{d}), we are able to simplify the equation (\ref{EE-formula-no-island-BH}). The form in its refined version is written as follows:
\begin{eqnarray}
\label{HR-BH-LT}
    S_{\rm BEH}^{\mathcal{R}} = \frac{Q}{6} \log \left(\frac{4 g_H(b_1)^2}{\kappa_H^2} \cosh^2\left(\kappa_H t_{b_1}\right)\right).
\end{eqnarray}
At late times, i.e., $t_{b_1} \rightarrow \infty$, $\cosh\left(\kappa_H t_{b_1}\right) \sim e^{\kappa_H t_{b_1}}$,
(\ref{HR-BH-LT}) implying:
\begin{eqnarray}
\label{EE-R-BH}
    S_{\rm BEH}^{\mathcal{R}} \sim \frac{Q}{3}\kappa_H t_{b_1}. 
\end{eqnarray}
The above equation has a significant physical meaning that it predicts that there is going to be a limitless quantity of Hawking radiation due to its linear time dependence upon entanglement entropy. This is in contrast to the fact that the Page curve of an eternal black hole predicts that there will only be a finite amount of Hawking radiation \cite{Page}.
\subsection{Island Phase}
\label{Island-BH}
Taking into account the island surface as part of the setup, we will now calculate the entanglement entropy of Hawking radiation.
\begin{figure}[h!]
\begin{center}
  \includegraphics[width=12.0cm]{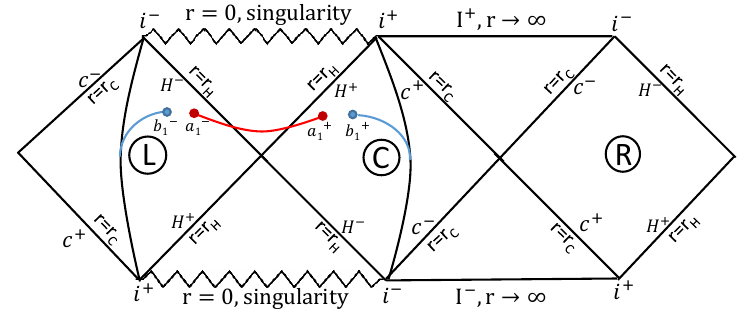}
  \caption{Carter-Penrose diagram of Schwarzschild de-Sitter spacetime. Within the black hole patch, we have included an island surface denoted by the notation $\mathcal{I}$ (the red curve) apart from the radiation regions $\mathcal{R}$.}
  \label{fig2}
\end{center}
\end{figure}
With inclusion of the island surface, the entanglement entropy of the Hawking radiation can be determined using the formula \cite{CC,CC-1}, which is the matter component of the generalized entropy (\ref{Island-proposal}).
\begin{eqnarray}
\label{EE-formula-island-BH}
S_{\rm matter}^{\rm BEH}({\cal R}\cup {\cal I}) = \frac{Q}{3} \log\left(\frac{d(a_1^+,a_1^-)d(b_1^+,b_1^-)d(a_1^+,b_1^+)d(a_1^-,b_1^-)}{d(a_1^+,b_1^-)d(a_1^-,b_1^+)}\right),
\end{eqnarray}
where $a_1^+(t_{a_1},a_1)$ and $a_1^-(-t_{a_1}+\iota \frac{\beta}{2},a_1)$ represent island surface boundaries in the right and left wedges of the black hole patch. Under the assumption of large distance (distance between observer and black hole horizon is enormous), $d(a_1^+,a_1^-)\equiv d(b_1^+,b_1^-) \equiv d(a_1^{\pm},b_1^{\mp}) \gg d(a_1^{\pm},b_1^{\pm})$ \cite{Islands-KdS},  and hence (\ref{EE-formula-island-BH}) transformed into:
\begin{eqnarray}
\label{EE-formula-island-BH-large-distance}
S_{\rm matter}^{\rm BEH}({\cal R}\cup {\cal I}) = \frac{Q}{6} \log\left(d^2(a_1^+,b_1^+)\right).
\end{eqnarray}
Simplifying the expression for the matter contribution to the generalized entropy (\ref{Island-proposal}) from the radiation and island regions by utilizing (\ref{metric-BH}), and (\ref{d}). We obtained:
{
\begin{eqnarray}
\label{EE-formula-island-BH-large-distance-simp}
& &
 S_{\rm matter}^{\rm BEH}({\cal R}\cup {\cal I}) = \frac{Q}{6} \log\left(\frac{g_H(a_1)g_H(b_1)}{\kappa_H^2}\right) \nonumber\\
 & & +\frac{Q}{6} \log\left(-\cosh\left(t_{a_1}-t_{b_1}\right) e^{\kappa_H\left(r_*(a_1)+r_*(b_1)\right)}+e^{2 \kappa_H r_*(a_1)}+e^{2 \kappa_H r_*(b_1)}\right).
\end{eqnarray}
}
As a result, the area of the boundary of the island surface and the matter component (\ref{EE-formula-island-BH-large-distance-simp}) are added together to generate the generalized entropy (\ref{Island-proposal}):
{
\begin{eqnarray}
\label{gen-BH}
& & 
   S_{\rm gen}^{\rm BEH} =\frac{2 \pi a_1^2}{G_N}+\frac{Q}{6} \log\left(\frac{g_H(a_1)g_H(b_1)}{\kappa_H^2}\right)\nonumber\\
 & &  +\frac{Q}{6} \log\left(-2\cosh\left(t_{a_1}-t_{b_1}\right) e^{\kappa_H\left(r_*(a_1)+r_*(b_1)\right)}+e^{2 \kappa_H r_*(a_1)}+e^{2 \kappa_H r_*(b_1)}\right).
\end{eqnarray}
}
The black hole patch's generalized entropy is being extremized with regard to the island coordinates $(t_{a_1},a_1)$ at the moment. To begin, let's use the following method of extremization with respect to $t_{a_1}$:
\begin{eqnarray}
\frac{\partial S_{\rm gen}^{\rm BEH}}{\partial t_{a_1}}=  -\frac{2 e^{\kappa_H\left(r_*(a_1)+r_*(b_1)\right)} \sinh\left(t_{a_1}-t_{b_1}\right)}{e^{2 \kappa_H r_*(a_1)}+e^{2 \kappa_H r_*(b_1) }-2\cosh\left(t_{a_1}-t_{b_1}\right) e^{\kappa_H\left(r_*(a_1)+r_*(b_1)\right)}} =0.
\end{eqnarray}
The previous equation has the following solution:
\begin{equation}
\label{ta-BH}
t_{a_1} = t_{b_1}+ 2 \iota \pi c_2,
\end{equation}
where $c_2 \in \mathbb{Z}$. Using the preceding equation as a reference, we were able to get the expression for the generalized entropy (\ref{gen-BH}) by substituting $t_{a_1}$.
\begin{eqnarray}
& &
    S_{\rm gen}^{\rm BEH} =\frac{2 \pi a_1^2}{G_N}+\frac{Q}{6} \log\left(\frac{g_H(a_1)g_H(b_1)}{\kappa_H^2}\right) \nonumber\\
   & & +\frac{Q}{6} \log\left(-2 e^{\kappa_H\left(r_*(a_1)+r_*(b_1)\right)}+e^{2 \kappa_H r_*(a_1)}+e^{2 \kappa_H r_*(b_1)}\right).
\end{eqnarray}
By solving for $a_1$ at an extreme, we may determine where on the black hole patch the island surface lies as follows:
\begin{eqnarray}
    \frac{\partial S_{\rm gen}^{\rm BEH}}{\partial a_1}=\frac{4 \pi  r_H}{G_N}+\frac{Q}{4 (a_1- r_H)} =0.
\end{eqnarray}
Using the above equation, we can determine that the island exists at the following location:
\begin{eqnarray}
\label{a-BH}
a_1=r_H-\frac{Q G_N}{16 \pi r_H}.
\end{eqnarray}
Since the second component in the above equation has a negative sign in front of it, the island must be found inside the black hole horizon. We have showed in the Penrose diagram where islands are beyond the black hole event horizon, but we have found that it lies within the horizon itself. Substituting $a_1$ from (\ref{a-BH}) into the definition of generalized entropy yields:
\begin{eqnarray}
\label{gen-BH-simp}
    S_{\rm total}^{\rm BEH} =\frac{2 \pi r_H^2}{G_N}+{\cal O}(G_N^0)=2 S_{\rm th}^{\rm BEH}+{\cal O}(G_N^0).
\end{eqnarray}
Because of this, we may deduce that the entanglement entropy of Hawking radiation remains constant when the island is incorporated into the black hole's interior, and is thus equal to twice the black hole's thermal entropy (to leading order in $G_N$).\\
%%%%%%%%%%%%%%%%
{\it \textbf{ Page time:}} Hawking radiation's entanglement entropy without the island surface is equivalent to its entanglement entropy with the island surface present at a point in time called ``Page time'' as discussed in chapter {\bf 6}. By equating (\ref{EE-R-BH}) and (\ref{gen-BH-simp}), we obtained the Page time as given below:
\begin{eqnarray}
\label{Page-time-BH}
t_{\rm Page}^{\rm BEH}=\frac{6 S_{\rm th}^{\rm BEH}}{Q \kappa_H}.
\end{eqnarray}
%%%%%%%%%%%%%%%%%
{\it \textbf{Scrambling time:}} The scrambling time is discussed in detail in chapter {\bf 6}. For the black hole patch in Schwarzschild de-Sitter black hole, scrambling time can be obtained using the following equation:
\begin{eqnarray}
\label{t-scr}
    t_{\rm scr} = r_*(b_1)-r_*(a_1).
\end{eqnarray}
From (\ref{tor}) and (\ref{a-BH}), the black hole patch's scrambling time is therefore obtained as:
\begin{eqnarray}
   t_{\rm scr}^{\rm BEH} \approx \frac{1}{2 \kappa_H} \log\left(\frac{\pi r_H^2}{G_N}\right)+ small\approx \frac{1}{2 \kappa_H} \log\left(S_{\rm th}^{\rm BEH}\right) + small.
\end{eqnarray}
\subsection{Page Curves}
The black hole patch Page curves are now obtained using the results of \ref{No-Island-BH} and \ref{Island-BH}. (\ref{EE-R-BH}) and (\ref{gen-BH-simp}) are the key equations. When we incorporate the island surface, the entanglement entropy of Hawking radiation changes from a time-dependent function of entanglement entropy to a constant value of $(2 S_{\rm th}^{\rm BEH})$ (\ref{gen-BH-simp}). As a result, if we plot all of these contributions together, we get the Page curves of eternal black holes that are in accord with their unitary evolution. From the use of (\ref{l2}), (\ref{l3}), (\ref{EE-R-BH}) and (\ref{gen-BH-simp}), we obtained the Page curves for black hole patch when $Q=G_N=1$, $3M\sqrt{\Lambda}=0.4$ (green) and $3M\sqrt{\Lambda}=0.5$ (magenta), shown in Fig. \ref{PC-BH-Lambda}.
\begin{figure}[h!]
\begin{center}
\includegraphics[width=10.0cm]{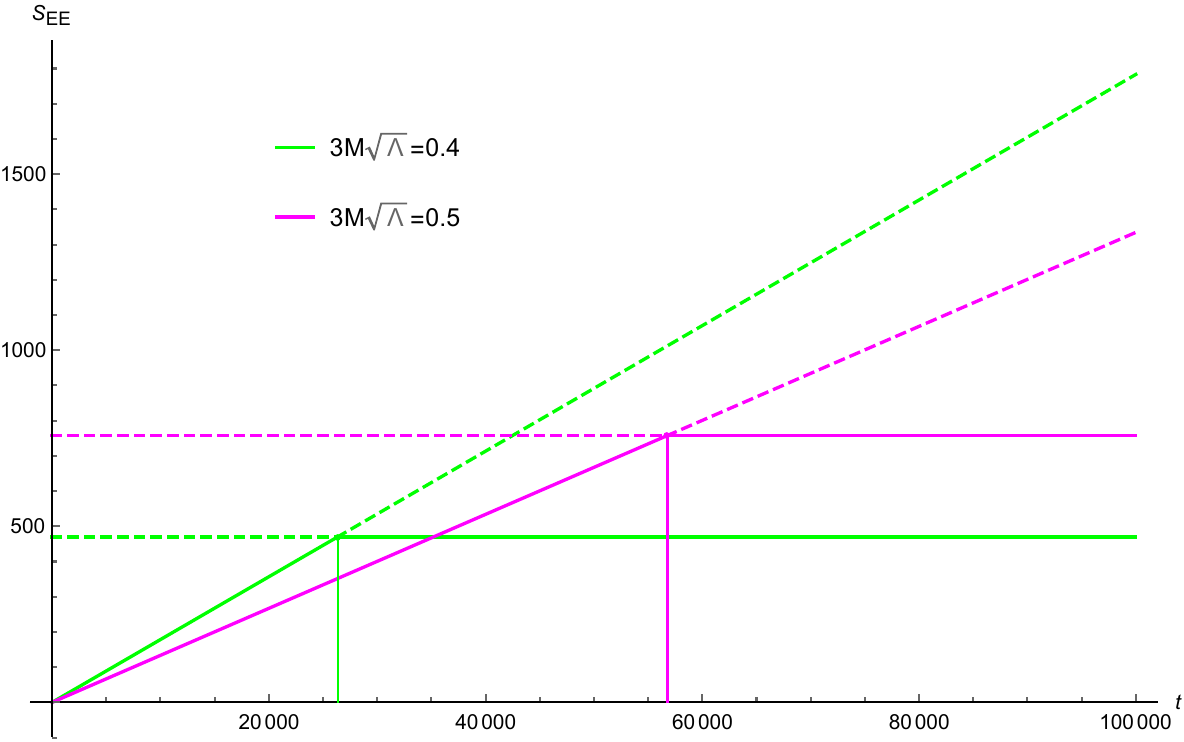}
  \caption{Black hole patch's Page curves for $3M\sqrt{\Lambda} = 0.4$(green) and $3M\sqrt{\Lambda}=0.5$(magenta). Page times for $3M\sqrt{\Lambda} = 0.4$, $0.5$ are $t_{P_1} \approx 26356.8$  and $t_{P_2} \approx 56787.5$ respectively. }
  \label{PC-BH-Lambda}
\end{center}
\end{figure}
As the mass of the black hole grows (as shown in \ref{PC-BH-Lambda}), the Page curves move later in time. As a result, information recovered from large black holes takes much more time than that from smaller black holes. This finding can also be interpreted in terms of the temperature of black holes. Information can be recovered faster from black holes with higher temperatures than from those with lower temperatures because the black hole temperature is inversely related to black hole mass. To rephrase, island emerges earlier for the black holes with higher temperatures compared to the black holes with lower temperature.
\section{Information Problem in SdS Black Hole as a Whole}\label{bhch}
Since Schwarzschild de-Sitter (SdS) spacetime has an effective equilibrium temperature \cite{Saida:2009ss,Nitin,pc,saida,urano,pappas}, the information  problem can be extended to the entire spacetime. Let us examine what exactly the difficulty of the island proposal is right now. By placing the observers in regions $L$ and $R$ for collecting the Hawking radiation, one is able to analyze the information paradox over all of the SdS spacetime. Additionally, one is able to specify the boundaries, $(b_-,b+)$, of radiation regions $({\cal R})$ there. Then, it is possible to incorporate two islands, ${\cal I}_1$ and ${\cal I}_2$, with borders, $(a_1^-,a_1^+)$ and $(a_2^-,a_2^+)$, simultaneously in black hole and de-Sitter patches. If someone adopts this technique, then the matter contributing to the entanglement entropy can be determined by applying the $2D$ CFT formula for three intervals as provided below\footnote{Without the island, the matter part is going to be similar to \ref{IP-BH}.}.
{\footnotesize
\begin{eqnarray}
\label{EE-3I}
 S_{\rm matter}({\cal R}\cup {{\cal I}_1}\cup {\cal I}_2)=\frac{Q}{3} \log \Biggl(\frac{d(b_-,a_1^-)d(a_1^+,a_1^-)d(a_2^+,a_1^-)d(b_-,a_2^-)d(a_1^+,a_2^-)d(a_2^+,a_2^-)d(b_-,b_+)d(a_1^+,b_+)d(a_2^+,b_+)}{d(a_1^+,b_-)d(a_2^+,b_-)d(a_2^+,a_1^+)d(a_2^-,a_1^-)d(b_+,a_1^-)d(b_+,a_2^-)}\Biggr). \nonumber\\
\end{eqnarray}
}
In this particular scenario, the first term in the generalized entropy (\ref{Island-proposal}) will have the form $\frac{2 \pi \left(a_1^2+a_2^2\right)}{G_N}$. It is possible to obtain the Page curves by following a process that is analogous to the one described in \ref{IP-BH}. The problem is that when we take into account the entirety of the SdS spacetime, then it's possible that we won't be capable to identify identical radiation regions on the two sides comparable to \ref{IP-BH}. This is because there's a black hole patch on one side of the system. On the other hand, the other side contains a patch known as the de-Sitter patch.
\section{Conclusion}\label{discussion}
Using the Island concept, we have investigated the information paradox that is associated with the black hole patch in the Schwarzschild-deSitter black hole spacetime. On both sides of the black hole patch, we used thermally opaque membranes to create an isolation barrier between the black hole and the de-Sitter patches, and vice versa. We calculated the entanglement entropy of Hawking radiation without and with the inclusion of an island surface and found that, as is typical, the entanglement entropy has a linear time dependency as long as there isn't an island surface, but it turns into constant value when there is an island surface present. As a result, we are able to produce the Page curve in a manner that accords with the unitary evolution of the black holes. In this particular instance, the island can be found inside the horizon of the black hole, in contrast to the universal finding, which states that the island must be found outside the horizon of the black hole in the case of eternal black holes \cite{island-o-h}. Our results are likewise comparable to those presented in \cite{I-3}, in which the researchers calculated the Page curve of a one-sided asymptotically flat black hole and found that the island lies within the horizon of the black hole. \par
In addition, we have investigated the ``effect of temperature'' on the Page curves depicted by the black hole patch. We noticed that as the temperature of the black hole patch rose, a shift toward later times occurred in the Page curves.\footnote{Due to the existence of higher derivative terms in the gravitational action, similar effects, such as the ``emergence of Page curves at later times or earlier times'', were also found in \cite{RNBH-HD,Ankit}.}. When compared to the black hole patch with the greater temperature, the one with the lower temperature will take a significantly longer amount of time to send the information to the observer. This suggests that the island of the black hole patch doesn't make its appearance until much later for the black hole patch with the lower temperature, but the island makes its debut much sooner for the black hole patch with the higher temperature. We are able to begin the process of recovering the information that was flung into the black holes horizon as soon as an island surface enters into the picture \cite{scrambling-time-1,scrambling-time-2}. As a result, we get to the conclusion that the temperature of both event horizons has a role in determining the ``dominance of islands'' and the ``scrambling time''.

%m

\chapter{Multiverse in Karch-Randall Braneworld}
\graphicspath{{Chapter9/}{Chapter9/}}
\section{Introduction}
Wedge holography is constructed by embedding two Karch-Randall (KR) branes in bulk. These branes are joined at the defect via the transparent boundary conditions so that degrees of freedom can be exchanged between them. The wedge holography is very useful for getting the Page curve of black holes. For the paper on wedge holography, see \cite{WH-i,WH-ii,Wedge-JT,Geng}. Most of the papers discuss the application of wedge holography to resolve the information paradox. We have already discussed the wedge holography in chapter {\bf 5}; for more details about the same, please see \ref{WH-intro}. As discussed earlier, the wedge holography contains two KR branes.
A natural question that came to our mind is whether it is possible to construct the wedge holography with many KR branes instead of only two KR branes. If yes, then what will describe this setup? This is what has been addressed in this chapter based on the paper \cite{Multiverse}. In the process of answering these questions, we found that the wedge holography with many KR branes describes a ``Multiverse''. We would like to make a remark that we ask the question in reverse order, i.e., can we describe the Multiverse from wedge holography? In any way, we reach the same theoretical model. We explained the Multiverse by constructing the wedge holography in such a way that there are $2n$ KR branes that are embedded in the $(d+1)$-dimensional bulk. The aforementioned branes have Einstein gravity localized on them. Therefore, we have $2n$ copies of the gravitating system, and all these gravitating systems are connected to each other via the transparent boundary condition at the defect. This model has been used to obtain the Page curve of black holes with multiple horizons and a qualitative idea to resolve the ``grandfather paradox''. Let us discuss these concepts in more detail.

\section{Emerging Multiverse from Wedge Holography}
\label{Multiverse-section}
We're going to discuss how we can construct a multiverse from wedge holography in this section. When discussing the multiverse, $\alpha$ and $\beta$ will have $2n$ values, however when discussing wedge holography, $\alpha,\beta=1,2$ as in \ref{WH-intro}.
\subsection{Anti de-Sitter Background}
\label{AdS-multiverse}
In this case, we build a multiverse using $AdS$ spacetimes. Let's start with the most basic scenario covered in \ref{WH-intro}. We require many Karch-Randall branes at $r=\pm n \rho$ such that bulk metric must satisfy the Neumann boundary condition at the locations mentioned above in order to define the multiverse. Extrinsic curvature associated with the Karch-Randall brane and related trace is calculated as:
\begin{eqnarray}
\label{Extrinsic-Curvature}
& & 
{\cal K}_{ij}^\alpha=  \frac{1}{2} \left(\partial_r g_{ij}\right)|_{r=\pm n \rho} = \tanh( r) g_{ij}|_{r=\pm n \rho} =\tanh(\pm n \rho) h_{ij}^\alpha ,\nonumber\\
& & {\cal K}^\alpha=h^{ij}_\alpha K_{ij}^\alpha= d \tanh(\pm n \rho).
\end{eqnarray}
It is evident that the Neumann boundary condition (ref. NBC) has been satisfied at $r=\pm n \rho$ provided tensions of the branes should be, $T_{\rm AdS}^\alpha=(d-1) \tanh(\pm n \rho)$ with $\alpha=-n,...,n$. Some of the branes appear to have negative tension. Let's talk about the scenario in which there are branes at $- n \rho_1$ and $ n \rho_2$ with $\rho_1 \neq \rho_2$. The tensions of the branes in the present case are $(d-1) \tanh(- n \rho_1)$ and $(d-1) \tanh( n \rho_2)$. When $\rho_1 <0$ and $\rho_2>0$ are taken into account, the negative tension problem can be handled, according to \cite{Wedge-JT}. As a result, this resolves our setup's problem with brain stability. This result can also be discussed in the scenario when $\rho_1=\rho_2$. Furthermore, bulk metric (\ref{metric-bulk}) also satisfies the Einstein equation  (\ref{Einstein-equation}), which ensures the presence of $2n$ Karch-Randall branes in this setup. The $2n$-branes are models of the embedded universes in $AdS_{d+1}$. The defect is defined as: $P=Q_\alpha \cap Q_\beta$, where $\alpha,\beta=-n, -n+1,..,1,...,n-1,n$. The Dvali-Gabadadze-Porrati (DGP) term \cite{DGP-2}, which explains massless gravity \cite{Massless-Gravity}, is now included in the gravitational action as follows:
\begin{eqnarray}
\label{bulk-action-DGP}
& &
\hskip -0.2in S=\frac{1}{16 \pi G_N^{(d+1)}}\Biggl[ \int_M d^{d+1}x \sqrt{-g} \left(R[g] + {d(d-1)}\right)+2\int_{\partial M} d^d x\sqrt{-h}K \nonumber\\
& &
+2 \int_{Q_\alpha} d^dx \sqrt{-h_\alpha}\left({\cal K}_\alpha-T_\alpha+\lambda_\alpha R_{h_\alpha}\right)\Biggr] , 
\end{eqnarray}
we have an extra term in comparison to (\ref{bulk-action}) which is $R_{h_\alpha}$ term and is known as intrinsic curvature scalars of $2n$ Karch-Randall branes. In the present scenario, the bulk metric satisfies the Neumann boundary condition at $r=\pm n \rho$ as given below:
\begin{eqnarray}
\label{NBC-DGP}
{\cal K}_{\alpha,ij}-({\cal K}_\alpha-T_\alpha+\lambda_\alpha R_{h_\alpha} )h_{\alpha,ij}+2 \lambda_\alpha R_{\alpha, {ij}} =0.
\end{eqnarray}
The Einstein equation of bulk action (\ref{bulk-action-DGP}) is going to be the identical to (\ref{Einstein-equation}), therefore the solution is as follows:
\begin{equation}
\label{metric-bulk-DGP}
ds_{(d+1)}^2=g_{\mu \nu} dx^\mu dx^\nu=dr^2+\cosh^2(r) h_{ij}^{\alpha, \rm AdS} dy^i dy^j,
\end{equation}
where $-n \rho_1 \leq r \leq n \rho_2$. The induced metric $h_{ij}^\alpha$ satisfies Einstein's equation on the brane given below:
\begin{eqnarray}
\label{Brane-Einstein-equation-DGP}
R_{ij}^\alpha-\frac{1}{2}h_{ij}^\alpha R[h_{ij}]^\alpha =\frac{(d-1)(d-2)}{2} h_{ij}^\alpha.
\end{eqnarray}
The equation mentioned above is obtained from the Einstein-Hilbert term, which includes a negative cosmological constant on the brane:
\begin{eqnarray}
\label{boundary-EH-action-AdS}
S_{\rm AdS}^{\rm EH} =\lambda_\alpha^{\rm AdS} \int d^{d}x \sqrt{-h_{\alpha}} \left(R[h_{\alpha}] - 2 \Lambda_{\rm brane}^{\rm AdS}\right),
\end{eqnarray}
where in $d$ dimensions, $\Lambda_{\rm brane}^{\rm AdS}=-\frac{(d-1)(d-2)}{2}$,  and \\ $\lambda_\alpha^{\rm AdS}  \left(\equiv \frac{1}{16 \pi G_N^{d,\ \alpha}}=\frac{1}{16 \pi G_N^{(d+1)}}\int_0^{\alpha \rho}\cosh^{d-2}(r) dr \ ; (\alpha=1,2,...,n)\right)$; (\ref{boundary-EH-action-AdS}) is produced by inserting (\ref{metric-bulk-DGP}) into (\ref{bulk-action}) and utilizing a result of ${\cal K}^{\alpha}$ from (\ref{Extrinsic-Curvature}) and branes tensions  $T_{\rm AdS}^\alpha=(d-1) \tanh(\pm n \rho)$.
\begin{figure}
\begin{center}
\includegraphics[width=0.8\textwidth]{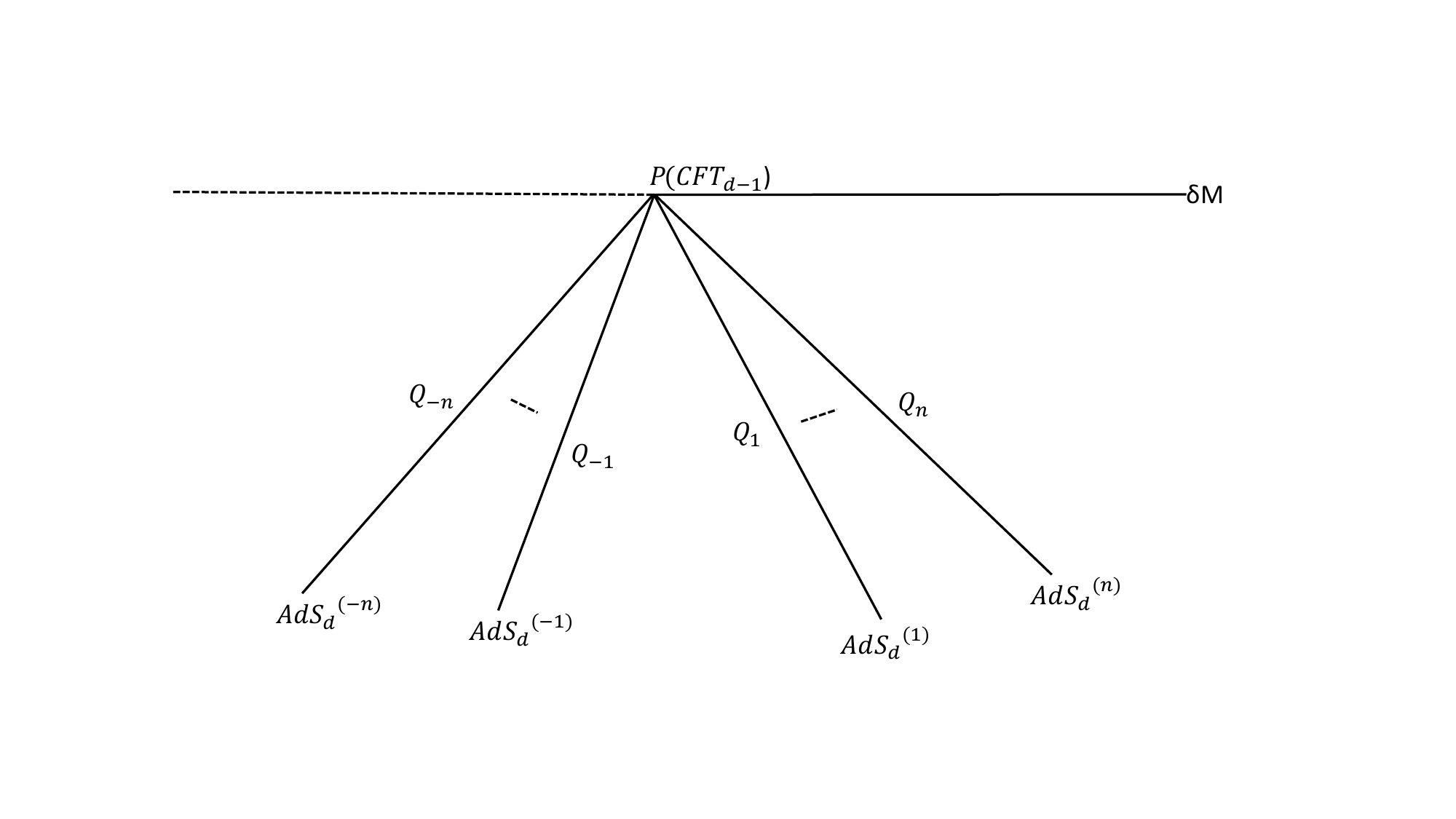}
\end{center}
\caption{Multiverse constructed from $2n$ Karch-Randall branes ($Q_{-n,-n+1,...,1,2,...,n-1,n}$) which are $d$-dimensional gravitating objects and these branes are embedded in the $(d+1)$-dimensional bulk. The dimension of defect $P$ is $(d-1)$.}
\label{Multiverse-AdS-i}
\end{figure}
\begin{figure}
\begin{center}
\includegraphics[width=0.8\textwidth]{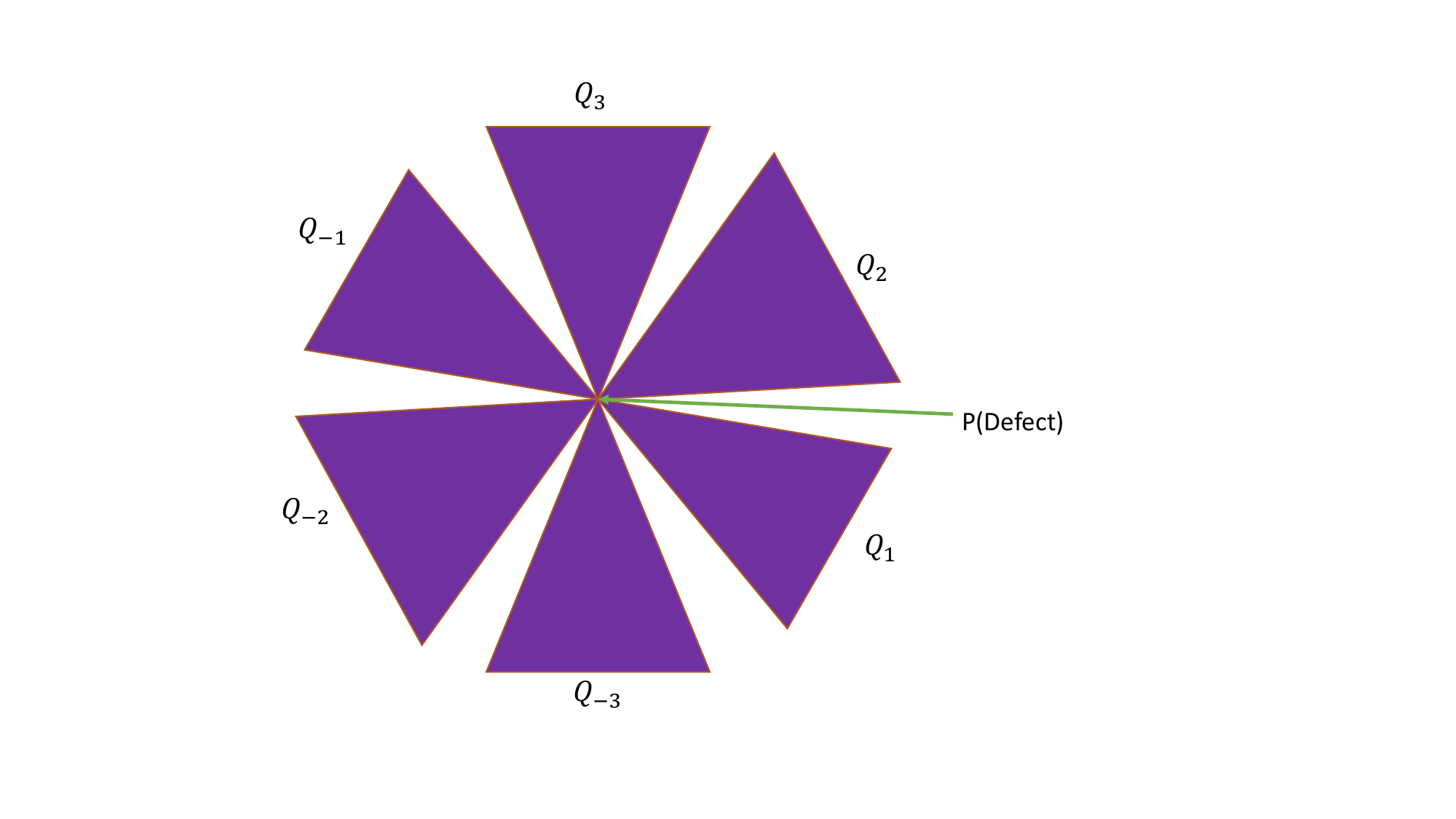}
\end{center}
\caption{Multiverse's cartoon picture when $n=3$ in AdS spacetimes.}
\label{CP-AdS-Multiverse}
\end{figure}
The multiverse in the context of wedge holography is shown in figure \ref{Multiverse-AdS-i} and a cartoon picture of the same has been shown in figure \ref{CP-AdS-Multiverse}. This setup has the following three descriptions:
\begin{itemize}
\item {\bf Boundary description:} $d$-dimensional boundary conformal field theory with $(d-1)$-dimensional boundary.
\item {\bf Intermediate description:} All $2n$ gravitating systems are connected at the interface point by transparent boundary condition.
\item {\bf Bulk description:} Einstein gravity in the $(d+1)$-dimensional bulk.
\end{itemize}
Because there exists a transparent boundary condition at the defect in the intermediate description, the multiverse created in this configuration consists of communicating universes located on Karch-Randall branes. The following is a Wedge holography dictionary for the ``multiverse'' having $2n$ AdS branes.\\ \\
\fbox{\begin{minipage}{38em}{\it 
Classical gravity in $(d+1)$-dimensional anti de-Sitter spacetime\\ $\equiv$ (Quantum) gravity on $2n$ $d$-dimensional Karch-Randall branes with metric $AdS_d$\\ $\equiv$ CFT living on $(d-1)$-dimensional defect.}
\end{minipage}}\\ \\
The braneworld holography connects the first and second lines \cite{KR1,KR2} and the AdS/CFT correspondence connects the second and third lines \cite{AdS/CFT} since there is gravity on the KR branes. Hence, {\it there exists co-dimensional two duality between the $(d+1)$-dimensional classical gravity on $AdS_{d+1}$ background and the $(d-1)$-dimensional defect conformal field theory, $CFT_{d-1}$.}

\subsection{de-Sitter Background}
\label{de-Sitter-multiverse}
Now we address the multiverse's realization in a manner in which the geometry of Karch-Randall branes corresponds to de-Sitter spacetime. Wedge holography having de-Sitter metric on the Karch-Randall branes has been studied in \cite{WH-2}, where the bulk geometry is AdS spacetime, as well as in \cite{WH-i}, where the bulk geometry is flat spacetime. Before delving further the specifics of constructing a ``multiverse'' along with de-Sitter geometry of Karch-Randall branes, let us first summarize the key concepts of \cite{WH-i}. The authors of \cite{WH-i} discussed a wedge holography with a Lorentzian signature in $(d+1)$-dimensional flat spacetime. Karch-Randall branes are built with either $d$-dimensional hyperbolic space or de-Sitter space geometry. Because we are only interested in the de-Sitter space, we will only describe the outcomes of that sector. The defect has $S^{d-1}$ geometry. In this context, according to wedge holography,\\ 
\fbox{\begin{minipage}{38em}{\it 
Classical gravity in $(d+1)$-dimensional flat spacetime\\ $\equiv$ (Quantum) gravity on two $d$-dimensional Karch-Randall branes with metric $dS_d$\\ $\equiv$ CFT living on $(d-1)$-dimensional defect $S^{d-1}$.}
\end{minipage}}\\ \\
The third line in the aforementioned duality originated from dS/CFT correspondence \cite{dS-CFT, dS-CFT-1}. The authors of \cite{WH-i} precisely computed the central charge of dual CFT  and found that central charge is imaginary implying that CFT located at the defect is non-unitary. The aforementioned explanation also works for AdS bulk too. In the present scenario, the wedge holographic dictionary is:\\ 
\fbox{\begin{minipage}{38em}{\it 
Classical gravity in $(d+1)$-dimensional anti de-Sitter spacetime\\ $\equiv$ (Quantum) gravity on two $d$-dimensional Karch-Randall branes with metric $dS_d$\\ $\equiv$ non-unitary CFT living at the $(d-1)$-dimensional defect.}
\end{minipage}}\\ \\
To begin discussing the existence of the multiverse, we will use the bulk metric \cite{WH-2}:
\begin{eqnarray}
\label{de-Sitter-metric}
ds_{(d+1)}^2=g_{\mu \nu} dx^\mu dx^\nu=dr^2+\sinh^2(r) h_{ij}^{\beta, \rm dS} dy^i dy^j,
\end{eqnarray}
equation (\ref{de-Sitter-metric}) represents the solution of (\ref{Einstein-equation}) with a negative cosmological constant provided the induced metric on the Karch-Randall branes, ($h_{ij}^\beta$) denotes the solution of the following Einstein's equation with a positive cosmological constant:
\begin{eqnarray}
\label{Brane-Einstein-equation-dS}
R_{ij}^\beta-\frac{1}{2}h_{ij}^\beta R[h_{ij}]^\beta =-\frac{(d-1)(d-2)}{2} h_{ij}^\beta.
\end{eqnarray}
By utilizing Neumann boundary condition (\ref{NBC}) for the de-Sitter branes and substitution of (\ref{de-Sitter-metric}) into (\ref{bulk-action}), one may construct Einstein-Hilbert terms with a positive cosmological constant on the Karch-Randall branes; the resultant action is provided as follows:
\begin{eqnarray}
\label{boundary-EH-action-dS}
S_{\rm dS}^{\rm EH} =\lambda_\beta^{\rm dS} \int d^{d}x \sqrt{-h_{\beta}} \left(R[h_{\beta}] - 2 \Lambda_{\rm brane}^{\rm dS}\right),
\end{eqnarray}
where  $\lambda_\beta^{\rm dS}\left(\equiv \frac{1}{16 \pi G_N^{d,\ \beta}}=\frac{1}{16 \pi G_N^{(d+1)}}\int_0^{\beta \rho}\sinh^{d-2}(r) dr \ ; (\beta=1,2,...,n)\right)$\footnote{See \cite{WH-2} where the explicit derivation is given for the two branes. Here, we have $\beta=1,2,...,n$ implying $2n$ branes.} reflects a connection with effective Newton's constant on the branes and $\Lambda_{\rm brane}^{\rm dS}=\frac{(d-1)(d-2)}{2}$. Extrinsic curvature and trace on the Karch-Randall branes for the de-Sitter embeddings in the bulk AdS spacetime (\ref{de-Sitter-metric}) are computed as:
\begin{eqnarray}
\label{Extrinsic-Curvature-dS}
& & 
{\cal K}_{ij}^\beta=  \frac{1}{2} \left(\partial_r g_{ij}\right)|_{r=\pm n \rho} = \coth( r) g_{ij}|_{r=\pm n \rho} =\coth(\pm n \rho) h_{ij}^\beta ,\nonumber\\
& & {\cal K}^\beta=h^{ij}_\beta K_{ij}^\beta= d \coth(\pm n \rho).
\end{eqnarray}
Utilizing (\ref{Extrinsic-Curvature-dS}), we found that the bulk metric (\ref{de-Sitter-metric}) satisfies the Neumann boundary condition (\ref{NBC}) at $r=\pm n \rho$ provided the branes have the tensions, $T_{\rm dS}^{\beta}=(d-1)\coth\left(\pm n\rho\right)$, where $\beta=-n,...,n$. As a result, we obtained the  $2n$ copies of Karch-Randall branes having metric de-Sitter spacetime on each brane. Therefore, {\it the multiverse is made up of $2n$ universes that are localized on the Karch-Randall branes with $dS_d$ geometry, and all of these $2n$ copies are embedded in $AdS_{d+1}$}. Figure \ref{CP-dS-Multiverse} shows a visual depiction of the same for $n=3$.  
\begin{figure}
\begin{center}
\includegraphics[width=0.8\textwidth]{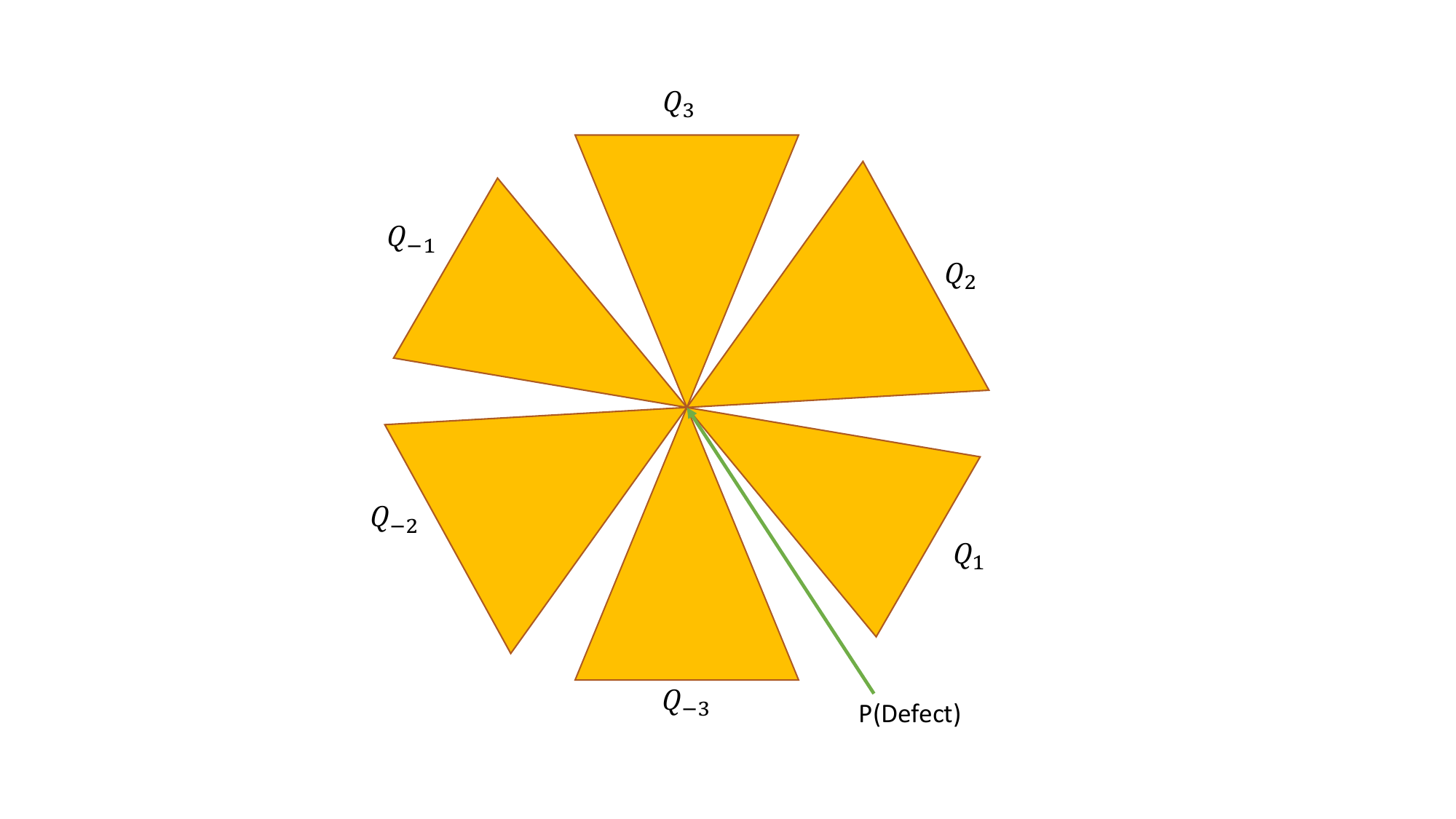}
\end{center}
\caption{The figure describing the multiverse when $n=3$ with de-Sitter branes ($Q_{-1/1,-2/2,-3/3}$) and $(d-1)$-dimensional defect $P$.}
\label{CP-dS-Multiverse}
\end{figure}
Let us now look at the three descriptions of the wedge holography in the context of multiverse with de-Sitter branes.
\begin{itemize}
\item {\bf Boundary description:} $d$-dimensional BCFT with $(d-1)$-dimensional defect.

\item {\bf Intermediate description:} $2n$ gravitating systems with de-Sitter geometry connected to each other at the $(d-1)$-dimensional defect.

\item {\bf Bulk description:} $(d+1)$-dimensional Einstein gravity with negative cosmological constant in the bulk.
\end{itemize}
The first and third descriptions are connected with one another via AdS/BCFT correspondence, and a $(d-1)$-dimensional non-unitary defect CFT arises due to dS/CFT correspondence \cite{dS-CFT,dS-CFT-1}. de-Sitter space persists for a certain period of time and then ceases to exist. Another de-Sitter space formed following the end of the preceding one \cite{mismatched-branes}. As a result, it is feasible to have a ``multiverse'' (say $M_1$) containing de-Sitter branes if they are all formed at the same ``creation time''(the ``time'' when any universe is born \cite{mismatched-branes} is termed as creation time), but this will only persist for a finite period until $M_1$ ceases to exist. Following the extinction of $M_1$, another multiverse (say, $M_2$) consisting of numerous de-Sitter branes created at exactly the same creation time as all the de-Sitter branes.

\subsection{Braneworld Consists of Anti de-Sitter and de-Sitter Spacetimes}
\label{AdS+dS-Multiverse}
In accordance with the discussions in refAdS-multiverse and refde-Sitter-multiverse, we are able to construct the two copies of the multiverse, $M_1$ and $M_2$, with the metric of Karch-Randall branes in $M_1$ having the structure of $AdS_d$ spacetime and Karch-Randall branes in $M_2$ having a structure of de-Sitter spaces in $d$ dimensions. The bulk metrics (\ref{metric-bulk}) of $M_1$ and (\ref{de-Sitter-metric}) of $M_2$ are the solutions of the Einstein's equation in bulk along with a negative cosmological constant (\ref{Einstein-equation}). For this situation, $M_1$ is made up of $2n_1$ Karch-Randall branes placed at $r=\pm n_1 \rho$ having induced metric $h_{ij}^{\alpha,\rm AdS}$ and the tensions $T_{\rm AdS}^\alpha=(d-1)\tanh(\pm n_1 \rho)$, and $M_2$ is made up of $2n_2$ Karch-Randall branes having induced metric $h_{ij}^{\beta,\rm dS}$ and the tensions $T_{\rm dS}^\beta=(d-1)\coth(\pm n_2 \rho)$ at $r=\pm n_2 \rho$, where $\alpha=-n_1,...,n_1$ and $\beta=-n_2,...,n_2$.\par
One could wonder why we are so interested in a configuration that has both anti-de-Sitter and de-Sitter branes. The reason given is that this model helps in studying of the information paradox of the Schwarzschild de-Sitter black hole with two horizons using wedge holography. To accomplish this, AdS branes in $M_1$ must be replaced by flat-space branes with $n_1=1$. Overall, we're left with two flat-space branes and two de-Sitter branes with $n_1=n_2=1$ and has been discussed in \ref{IP-SdS}.
\begin{figure}
\begin{center}
\includegraphics[width=0.8\textwidth]{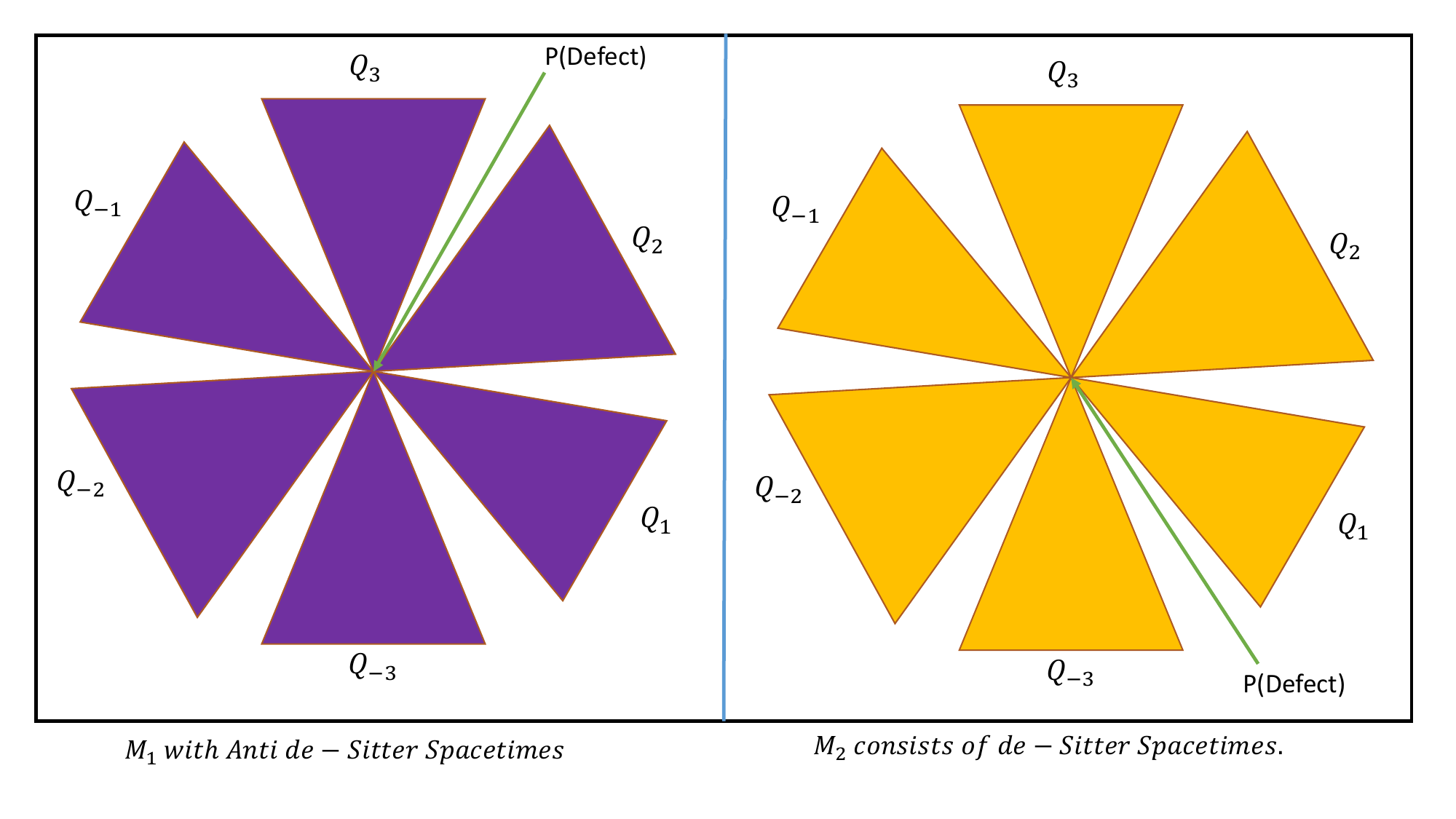}
\end{center}
\caption{Braneworld consists of $d$-dimensional anti de-Sitter and de-Sitter spacetimes. AdS spacetimes are embedded in the bulk (\ref{metric-bulk}) where as de-Sitter spacetimes are embedded in the bulk spacetime with metric (\ref{de-Sitter-metric}). We have used $n_1=n_2=3$ to draw this figure.}
\label{CP-dS+AdS-Multiverse}
\end{figure}
The question now is whether the above description makes sense. When $d$-dimensional AdS spacetimes are embedded in $AdS_{d+1}$, their branes come together at the time-like surface located at the $AdS_{d+1}$ boundary, whereas $dS_d$ Karch-Randall branes intersect at the space-like surface of the $AdS_{d+1}$ boundary. Fortunately, there is no issues in Fig. \ref{CP-dS+AdS-Multiverse} when $M_1$ and $M_2$ remain separated from themselves. This is the approach used in refIP-SdS to get the Page curve of a Schwarzschild de-Sitter black hole by considering the Schwarzschild and de-Sitter patches independently. We have explored the embedding of several types of Karch-Randall branes in distinct bulks that are not related to each other.\par
The authors of \cite{mismatched-branes} looked at the numerous options for embedding several sorts of branes, such as Minkowski, de-Sitter, and anti-de-Sitter branes, in the identical bulk. The existence of multiple branes is defined by the creation time $\tau_*$. Minkowski and de-Sitter branes have been created for a finite period of time, however anti de-Sitter branes have no creation time. Authors noted out that among the several possibilities stated in \cite{mismatched-branes}, one can find Minkowski, de-Sitter, and anti de-Sitter branes at the same moment at creation time $\tau_*=-\pi/2$ in a particular bulk. Branes have a time-dependent location in this scenario. We begin by summarizing this conclusion and then remark on its implementation using wedge holography. For more details, see cite{mismatched-branes}. The bulk $AdS_5$ metric is written as follows:
\begin{eqnarray}
\label{metric-AdS5-KR}
ds^2=\frac{1}{z^2}\left(-dt_h^2+t_h^2 dH_3^2+dz^2\right),
\end{eqnarray}
where $dH_3^2=d\theta^2+\sinh^2(\theta)d\omega_2^2$. In \ref{metric-AdS5-KR}, the Minkowski Randall-Sundrum brane is placed at $z_M(t_h)=z_0$, wherein $z_0$ being constant, $AdS_4$ branes are placed at $z_{\rm AdS,1}(t_h)=\sqrt{l^2+t_h^2}-\sqrt{l^2-1}$ (for $X_4>0$) and $z_{\rm AdS,2}(t_h)=\sqrt{l^2+t_h^2}+\sqrt{l^2-1}$ (for $X_4<0$) on the opposite sides of turninf point $X_4=0$(where $X_4$ is the one of the parametrizations of $AdS_5$ defined in \cite{mismatched-branes}). At $t_h=0$, $z_{\rm AdS,min}=l \mp \sqrt{l^2-1}$. The Minkowski and the AdS brane may coexist for constant values of $z$ greater than $z_{\rm AdS,min}$. The $AdS_4$ brane has the following metric:
\begin{eqnarray}
\label{metric-AdS4-KR}
ds^2=-d\tau_h^2+a(\tau_h) dH_3^2,
\end{eqnarray}
where $a(\tau_h)=\ \sin\left(\tau_h/l\right)$. The de-Sitter branes are located at  $z_{\rm dS,1}(t_h)=\sqrt{l^2+t_h^2}+\sqrt{l^2+1}$ and $z_{\rm dS,2}(t_h)=\sqrt{l^2+1}-\sqrt{l^2+t_h^2}$ with the following metric:
\begin{eqnarray}
\label{metric-dS4-KR}
ds^2=-d\tau_h^2+a(\tau_h) dH_3^2,
\end{eqnarray}
where $a(\tau_h)=\ \sinh\left(\tau_h/l\right)$.

{\bf Comment on the Wedge Holographic Realization of Mismatched Branes}: Using the AdS/BCFT concept, one may build a double holographic system from (\ref{metric-AdS5-KR}). Let us provide three acceptable descriptions of a double holographic setup made of (\ref{metric-AdS5-KR}).
\begin{itemize}
\item {\bf Boundary description}: $4D$ quantum field theory (QFT) at conformal boundary of (\ref{metric-AdS5-KR}).

\item {\bf Intermediate description}: Dynamical gravity localized on $4D$ end-of-the-world brane coupled to $4D$ boundary QFT.

\item {\bf Bulk description}: $4D$ QFT defined in the first description has $5D$ gravity dual whose metric is (\ref{metric-AdS5-KR}).
\end{itemize}
Because of the covariant character of the AdS/CFT duality, the duality stays the same when one is working with altered coordinates of the bulk, i.e. different AdS parametrizations don't suggest distinct dualities, and thus in the aforementioned doubly holographic setup, it is expected the defect to be $3$-dimensional conformal field theory since $4$-dimensional gravity serves as simply FRW parametrization of $AdS_4$ spacetime (\ref{metric-AdS4-KR}). This type of duality has been studied by the authors in \cite{Kostas}, wherein the bulk represents the de-Sitter parametrization of $AdS_4$ and conformal field theory was QFT on $dS_3$. As addressed thoroughly in appendix {\bf A} of \cite{mismatched-branes} and summarized within this chapter, de-Sitter and Minkowski branes can also exist in this coordinate system (\ref{metric-AdS5-KR}).\par
Let us now look at why defining wedge holography using ``mismatched branes'' is problematic. The ``defect CFT'' in wedge holography is produced via dynamical gravity on Karch-Randall branes. Assume that there are two Karch-Randall branes with differing geometries, one AdS brane and one de-Sitter brane. The defect CFT must be unitary owing to the AdS brane and non-unitary because of the de-Sitter brane. We appear to have two distinct CFTs at exactly the same defect. This condition will not alter even if four branes or $2n$ branes are considered. As a result, we may be unable to properly describe the ``multiverse'' using mismatched branes via wedge holography. Due to the ``time-dependent'' location of branes, the shared boundary of multiverses $M_1$ and $M_2$ (explained in Fig. \ref{CP-dS+AdS-Multiverse}) cannot be identical. All AdS branes in $M_1$ are capable of communicating with one another via transparent boundary conditions across the defect, and all de-Sitter branes in $M_2$ can communicate as well. However, even with the metric (\ref{metric-AdS5-KR}), there does not exist a connection between $M_1$ and $M_2$.\par
{\it As a result, we come to the conclusion that it is possible to construct a multiverse of identical branes (AdS or de-Sitter) however not as a mixture of both of them. As a result, the issue of mismatched branes does not change from a wedge holography standpoint. The multiverse of AdS branes survives indefinitely, but the multiverse of de-Sitter branes has a finite lifespan.}
\section{Application to Information Paradox}
\label{AIP}
The multiverse is made up of $2n$ Karch-Randall branes that are embedded in the $AdS_{d+1}$ bulk. As a result, just one Hartman-Maldacena surface can stretch between the defect CFTs which are thermofield double partners and the total $n$ island surfaces (${\cal I}_1$,${\cal I}_2$,.....,${\cal I}_n$) stretched $(r=\pm n \rho)$ across identical branes of the same positions with opposing sign as shown in Fig. \ref{BH+MBs}. Let us give a specific assertion about the wedge holographic dictionary with $2n$ branes.\\ \\
\fbox{\begin{minipage}{38em}{\it 
Classical gravity in $(d+1)$-dimensional AdS bulk\\ $\equiv$ (Quantum) gravity on $2n$ $d$-dimensional Karch-Randall branes with metric $AdS_d/dS_d$\\ $\equiv$ CFT living on $(d-1)$-dimensional defect.}
\end{minipage}}\\ \\
CFT is going to be non-unitary if the metric of Karch-Randall branes corresponds to the de-Sitter metric. As a result, this description is identical to the normal wedge holography involving two Karch-Randall branes, with the exception that we are now equipped with $2n$ Karch-Randall branes.
\par
Let us now construct a mathematical formula to obtain entanglement entropies. We take $Q_{1,2,....,n}$ to be black holes that produce Hawking radiation, which is contained by gravitational baths $Q_{-1,-2,....,-n}$, see Fig. \ref{BH+MBs}. In this configuration, the entanglement entropy of the island surfaces will have the following form:
\begin{eqnarray}
\label{Island-Formula-n-BHs}
& & S_{\rm Island}=S_{Q_{-1}-Q_1}^{{\cal I}_1} + S_{Q_{-2}-Q_2}^{{\cal I}_2}+.......+S_{Q_{-n}-Q_n}^{{\cal I}_{n}}.
\end{eqnarray}
When we utilize the entanglement entropy associated with the Hartman-Maldacena surface, that is, $S_{\rm HM} \propto t$ and $S_{\rm Island}=2 S^{{i=1,2,..,n}, \ \rm thermal}_{\rm BH}$, we are able to compute the Page curve, wherein $S_{\rm Island}$ and $S_{\rm HM}$ could possibly be computed utilizing the Ryu-Takayanagi formula \cite{RT}.
\begin{figure}
\begin{center}
\includegraphics[width=0.8\textwidth]{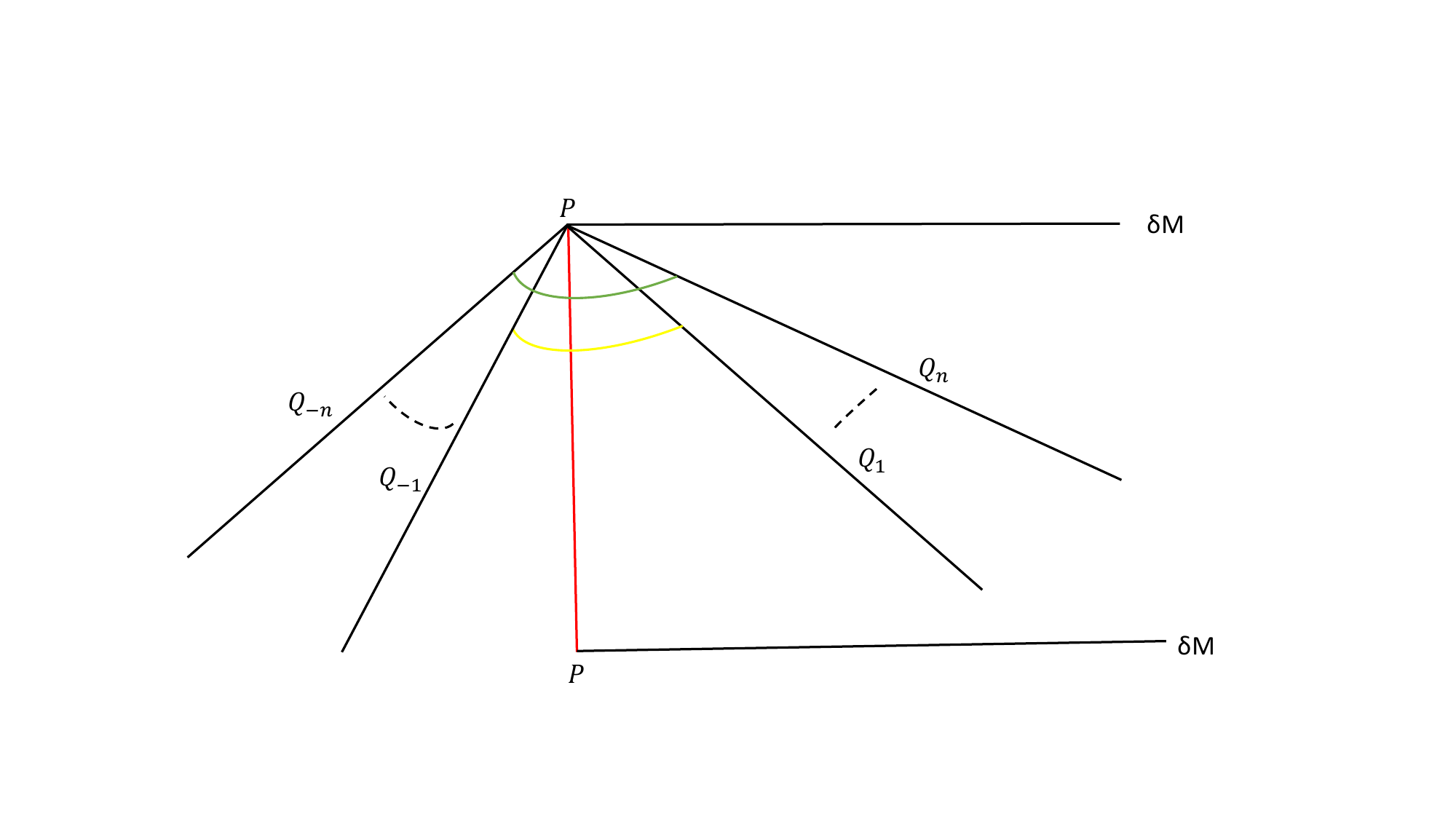}
\end{center}
\caption{The green and yellow curves indicate island surfaces across $Q_{-n}$ and $Q_n$, and $Q_{-1}$ and $Q_1$. The red curve depicts the Hartman-Maldacena surface from the defect to its thermofield double partner. The bulk AdS boundary is $\delta M$.}
\label{BH+MBs}
\end{figure}
The three descriptions pertaining to the multiverse are as follows: 
\begin{itemize}
\item {\bf Boundary Description:} BCFT is living at the $AdS_{d+1}$ boundary with $(d-1)$-dimensional boundary.
\item {\bf Intermediate Description:} $2n$ gravitating systems interact with each other via transparent boundary conditions at the  $(d-1)$-dimensional defect.
\item {\bf Bulk Description:} Gravity dual of BCFT is Einstein gravity in the bulk.
\end{itemize} 
{\bf Consistency Check:}
Let's look at the formula provided in (\ref{Island-Formula-n-BHs}) when $n=2$ whether it is giving the consistent results or not. 
\subsection{Page Curve of Eternal AdS Black Holes in $n=2$ Multiverse}
\label{PC-TEBH}
We began by calculating the thermal entropy of black hole. The black hole metric in the $AdS$ background has the following structure:
\begin{eqnarray}
\label{metric-bulk-BH}
ds_{(d+1)}^2=g_{\mu \nu} dx^\mu dx^\nu=dr^2+\cosh^2(r)\Biggl(\frac{\frac{dz^2}{f(z)}-f(z) dt^2+\sum_{i=1}^{d-2}dy_i^2 }{z^2}\Biggr),
\end{eqnarray}
 where $f(z)=1-\frac{z^{d-1}}{z_h^{d-1}}$. Thermal entropy has the following form for $z=z_h$ (we used $z_h=1$ everywhere in the computation for simplification and concentrated on $d=4$):
\begin{eqnarray}
\label{thermal}
S_{\rm AdS}^{\rm thermal} = \frac{A^{\rm BH}_{z=z_h}}{4 G_N^{(d+1)}}= \frac{1}{4 G_N^{(5)}}\int dr \cosh^{2}(r) \int dy_1 \int dy_2 = \frac{V_{2}}{4 G_N^{(5)}} \int dr \cosh^{2}(r),
\end{eqnarray} 
where $V_2=\int \int dy_1 dy_2$.  Take the $n=2$ situation, where two Karch-Randall branes that exists between $-2 \rho \leq r \leq 2 \rho$ and $ -\rho \leq r \leq \rho$ serve as a black hole and bath system. As a result, the total thermal entropies of two eternal AdS black holes are:
\begin{eqnarray}
\label{s-thermal-total-AdS}
& & \hskip -0.9in
S_{\rm AdS}^{\rm thermal, \ total} = \frac{V_{2}}{4 G_N^{(5)}} \int_{-2 \rho}^{2 \rho} dr \cosh^{2}(r)+ \frac{V_{2}}{4 G_N^{(5)}} \int_{-\rho}^{ \rho} dr \cosh^2(r)  \nonumber\\
& &=\frac{V_{2}}{4 G_N^{(5)}}\left(\frac{1}{2} (6 \rho +\sinh (2 \rho )+\sinh (4 \rho ))\right).
\end{eqnarray}
Let us now produce the Page curve of two eternal black holes utilizing the formula presented in (\ref{Island-Formula-n-BHs}).

{\bf Entanglement entropy contribution from Hartman-Maldacena surface}: The bulk metric (\ref{metric-bulk-BH}) expressed in terms of infalling Eddington-Finkelstein coordinate, $d v= dt -\frac{dz}{f(z)}$ has been rewritten as follows:
\begin{eqnarray}
\label{metric-bulk-BH-EFC}
ds_{(4+1)}^2=g_{\mu \nu} dx^\mu dx^\nu=dr^2+\cosh^2(r)\Biggl(\frac{-f(z) dv^2-2 dv dz+\sum_{i=1}^{2}dy_i^2 }{z^2}\Biggr).
\end{eqnarray}
The induced metric associated with the Hartman-Maldacena surface which has the parametrization, $r \equiv r(z)$ and $v \equiv v(z)$, is derived as:
\begin{eqnarray}
\label{induced-metric-HM}
& & ds^2= \Biggl({r'(z)^2-\frac{\cosh^2(r(z))v'(z)}{z^2} \left(2+f(z)v'(z)\right)}\Biggr) dz^2+ \frac{\cosh^2(r(z))}{z^2}\sum_{i=1}^{2}dy_i^2,
\end{eqnarray}
wherein $r'(z)=\frac{dr}{dz}$ and $v'(z)=\frac{dv}{dz}$. The area associated with the Hartman-Maldacena surface has been calculated from (\ref{induced-metric-HM}) as follows:
\begin{eqnarray}
& & A_{\rm HM}^{\rm AdS}=V_{2} \int_{z_1}^{z_{\rm max}} dz \Biggl( \frac{\cosh^2(r(z))}{z^2} \sqrt{ {r'(z)^2-\frac{\cosh^2(r(z))v'(z)}{z^2} \left(2+f(z)v'(z)\right)}}\Biggr),
\end{eqnarray}
where $z_1$ represents the gravitating bath's point and $z_{\rm max}$ being the Hartman-Maldacena surface turning point and $V_{2}=\int \int dy_1 dy_2$. In late time approximation, i.e., $t\rightarrow \infty$, $r(z) \rightarrow 0$ \cite{Massless-Gravity}. Hence,
\begin{eqnarray}
\label{AHM-AdS}
& & A_{\rm HM}^{\rm AdS}=V_{2} \int_{z_1}^{z_{\rm max}} dz\Biggl(   \frac{\sqrt{- v'(z)\left(2+f(z)v'(z)\right)}}{z^3}\Biggr).
\end{eqnarray}
The embedding $v(z)$ equation of motion is:
\begin{eqnarray}
& & 
\frac{d}{dz}\left(\frac{\partial L}{\partial v'(z)}\right)=0, \nonumber\\
& & \implies \frac{\partial L}{\partial v'(z)}=E, \nonumber\\
& &\implies v'(z)=\frac{-E^2 z^6-\sqrt{E^4 z^{12}+E^2 f(z) z^6}-f(z)}{E^2 f(z) z^6+f(z)^2}.
\end{eqnarray}
Because $v'(z)|_{z=z_{\rm max}} =0$ , $E=\frac{i\sqrt{f(z_{\rm max})}}{z_{\rm max}^3}$ and $\frac{dE}{dz_{\rm max}}=0$ results in $z_{\rm max}=\frac{7 z_h}{6}$ (i.e. the Hartman-Maldacena surface's turning point lies beyond the horizon). We can get time on the gravitating bath as follows:
\begin{eqnarray}
\label{tz1}
t_1=t(z_1)=-\int_{z_1}^{z_{\rm max}} \left(v'(z)+\frac{1}{f(z)}\right)dz.
\end{eqnarray}
Let us now look into the late-time behavior exhibited by the Hartman-Maldacena surface area:
\begin{eqnarray}
& & {\rm lim}_{t \rightarrow \infty} \frac{dA_{\rm HM}^{\ \rm AdS}}{dt}={\rm lim}_{t \rightarrow \infty}\Biggl( \frac{\frac{dA_{\rm HM}^{\rm AdS}}{dz_{\rm max} }}{\frac{dt}{dz_{\rm max} }}\Biggr) = \frac{L(z_{\rm max},v'(z_{\rm max}))+\int_{z_1}^{z_{\rm max}} \frac{\partial L}{\partial z_{\rm max} }dz}{-v'(z_{\rm max})-\frac{1}{f(z_{\rm max})}-\int_{z_1}^{z_{\rm max}}\frac{\partial v'(z)}{\partial z_{\rm max} }}.
\end{eqnarray}
Since,
\begin{eqnarray}
& & 
{\rm lim}_{t \rightarrow \infty}\frac{\partial v'(z)}{\partial z_{\rm max} }={\rm lim}_{t \rightarrow \infty}\frac{\partial v'(z)}{\partial E}\frac{\partial E}{\partial z_{\rm max} }=0, \nonumber\\
& & {\rm lim}_{t \rightarrow \infty}\frac{\partial L(z,v'(z))}{\partial z_{\rm max} } = \frac{\partial L(z,v'(z))}{\partial v'(z) } \frac{\partial v'(z)}{\partial z_{\rm max} }=0.
\end{eqnarray}
Hence,
\begin{eqnarray}
{\rm lim}_{t \rightarrow \infty} \frac{dA_{\rm HM}^{\rm AdS}}{dt} = \frac{L(z_{\rm max},v'(z_{\rm max}))}{-v'(z_{\rm max})-\frac{1}{f(z_{\rm max})}}  = \frac{\frac{\sqrt{-v'(z_{\rm max})(2+f(z_{\rm max})v'(z_{\rm max}))}}{z_{\rm max}^3}}{-v'(z_{\rm max})-\frac{1}{f(z_{\rm max})}} = constant.
\end{eqnarray}
According to the preceding equation, $A_{\rm HM}^{\rm AdS} \propto t_1$, resulting in entanglement entropy corresponding to the Hartman-Maldacena surface takes the following structure.
\begin{eqnarray}
\label{SHM-AdS}
& & S_{\rm HM}^{\rm AdS} \propto t_1.
\end{eqnarray}
This amounts to an unlimited quantity of Hawking radiation if $t_1 \rightarrow \infty$, i.e. at late times, and so results in the information paradox.

{\bf Entanglement entropy contribution from Island surfaces}: Consider the parametrized island surfaces $t=constant$ and $z \equiv z(r)$. The entanglement entropy associated with two eternal AdS black holes regarding the island surfaces was calculated using (\ref{Island-Formula-n-BHs}). Since there exist two island surfaces (${\cal I}_1$ and ${\cal I}_2$) that extend between the Karch-Randall branes at $r=\pm \rho$ (${\cal I}_1$) and $r=\pm 2 \rho$ (${\cal I}_2$), we could use (\ref{Island-Formula-n-BHs}) for the same.
\begin{eqnarray}
\label{EE-Island}
& & 
S_{\rm AdS}^{\rm Island} =S_{Q_{-1}-Q_1}^{{\cal I}_1} + S_{Q_{-2}-Q_2}^{{\cal I}_2} =  \frac{\left({\cal A}_{{\cal I}_1}+{\cal A}_{{\cal I}_2}\right)}{4 G_N^{(5)}}=
\frac{ \int d^3x \sqrt{h_1}+\int d^3x \sqrt{h_2}}{4 G_N^{(5)}}.
\end{eqnarray}
We begin by calculating ${\cal A}_{{\cal I}_1}$. The induced metric for Karch-Randall branes could be derived utilizing (\ref{metric-bulk-BH}) by parametrizing the island surface as $t=constant$ and $z=z(r)$ and constraining to $d=4$ with $f(z)=1-z^3$ (since $z_h=1$):
\begin{eqnarray}
\label{induced-metric-AdS-IS}
& & ds^2= \Biggl({1+\frac{\cosh^2(r)z'(r)^2}{z(r)^2(1-z(r)^{3})}}\Biggr) dr^2+ \frac{\cosh^2(r)}{z(r)^2}\sum_{i=1}^{2}dy_i^2,
\end{eqnarray}
The area associated with the island surface ${\cal I}_1$ resulted from (\ref{induced-metric-AdS-IS}) is given as:
\begin{eqnarray}
\label{AIS-AdS}
{\cal A}_{{\cal I}_1}=V_{2}\int_{- \rho}^{\rho} dr {\cal L}_{{\cal I}_1}\left(z(r),z'(r)\right) =V_{2}\int_{- \rho}^{\rho} dr\Biggl(\frac{\cosh^{2}(r)}{z(r)^{2}}\sqrt{1+\frac{\cosh^2(r)z'(r)^2}{z(r)^2(1-z(r)^{3})}}\Biggr),
\end{eqnarray} 
where we choose $z_h=1$ and so $0<z<1$ for $f(z)\geq 0$. Let us now analyze the variation of (\ref{AIS-AdS}).
\begin{eqnarray}
\label{var-AIS}
& & 
\delta {\cal A}_{{\cal I}_1} = V_{2}\int_{- \rho}^{\rho} dr \Biggl[\left(\frac{\delta{\cal L}_{{\cal I}_1}\left(z(r),z'(r)\right)}{\delta z(r)}\right) \delta z(r)+\left(\frac{\delta{\cal L}_{{\cal I}_1}\left(z(r),z'(r)\right)}{\delta z'(r)}\right) \delta z'(r) \Biggr]\nonumber\\
& &\hskip-0.2in = V_{2}\int_{- \rho}^{\rho} dr \left(\frac{\delta{\cal L}_{{\cal I}_1}\left(z(r),z'(r)\right)}{\delta z'(r)}\right) \delta z(r)- \int_{- \rho}^{\rho} dr \Biggl[\frac{d}{dr}\left(\frac{\delta{\cal L}_{{\cal I}_1}\left(z(r),z'(r)\right)}{\delta z'(r)}\right)-\left(\frac{\delta{\cal L}_{{\cal I}_1}\left(z(r),z'(r)\right)}{\delta z(r)}\right) \Biggr]\delta z(r). \nonumber\\
\end{eqnarray}
Only if the first term of the preceding equation disappears will certainly the variational principle become meaningful. The second term represents the EOM associated with the embedding $z(r)$. Let's have a look at what this means.
\begin{eqnarray}
\label{VP-meaningful}
& & \int_{- \rho}^{\rho} dr \left(\frac{\delta{\cal L}_{{\cal I}_1}\left(z(r),z'(r)\right)}{\delta z'(r)}\right) \delta z(r) = \int_{- \rho}^{\rho} dr \Biggl(\frac{\cosh ^4(r) z'(r)}{z(r)^4 f(z(r)) \sqrt{\frac{\cosh ^2(r) z'(r)^2}{z(r)^2 f(z(r))}+1}} \Biggr)\delta z(r),
\end{eqnarray}
(\ref{VP-meaningful}) disappears if we enforce the Dirichlet boundary condition on the branes, i.e., $\delta z(r=\pm \rho)=0$ or Neumann boundary condition on the branes, i.e., $z'(r=\pm \rho)=0$. Neumann boundary conditions enable RT surfaces to travel along branes in gravitating baths. Under this scenario, the black hole horizon \cite{GB-3} is the minimum surface. The Euler-Lagrange equation of motion for the action with embedding $z(r)$ becomes:
{\footnotesize
\begin{eqnarray}
\label{EOM}
& &
\frac{\cosh ^2(r) }{2 z(r)^4 \left(z(r)^3-1\right) \left(-\cosh ^2(r) z'(r)^2+z(r)^5-z(r)^2\right) \sqrt{\frac{\cosh ^2(r)
   z'(r)^2}{z(r)^2-z(r)^5}+1}}\Biggl(z(r)^4 \cosh ^2(r) z'(r)^2 \nonumber\\
   & &  +2 z(r) \cosh ^2(r) z'(r)^2
   -2 z(r)^5 \cosh (r)
   \left(\cosh (r) z''(r)  +4 \sinh (r) z'(r)\right)+6 \sinh (r) \cosh ^3(r) z'(r)^3\nonumber\\
   & &
   +2 z(r)^2 \cosh (r) \left(\cosh (r) z''(r)+4 \sinh (r) z'(r)\right)+4 z(r)^9-8
   z(r)^6+4 z(r)^3\Biggr)
   =0.
\end{eqnarray}
}
It's interesting to note that the black hole horizon $z(r)=1$, which is the solution to (\ref{EOM}), satisfying the Neumann boundary condition on the branes. The structure of (\ref{EOM}) confirms exactly that. The majority of the terms in the open bracket of (\ref{EOM}) are $z'(r)$ and $z''(r)$, however there is one combination that is not reliant on either of these variables: $(4 z(r)^9-8 z(r)^6+4 z(r)^3)$, that cancels for $z(r)=1$ and is therefore the solution of (\ref{EOM}). Because $z_h=1$, it is implied that the Ryu-Takayanagi surface corresponds to black hole horizon. It was mentioned in \cite{GB-3} that the Ryu-Takayanagi surface in the wedge holography corresponds to the black hole horizon if the Neumann boundary condition on gravitating branes is true. By applying an inequality condition to the island's surface area, the identical result was also found in the \cite{Massless-Gravity}. Wherever we have studied the entanglement entropy of island surfaces in this chapter, we have reached the same results.
By replacing $z(r)=1$ in (\ref{AIS-AdS}), we were able to determine the island's minimum surface area, ${\cal I}_1$:
\begin{eqnarray}
\label{AIS-AdS-i}
{\cal A}_{{\cal I}_1}=V_{2}\int_{- \rho}^{\rho} dr {\cosh^{2}(r)}.
\end{eqnarray}
The second island's minimum surface area (${\cal I}_{2}$) will end up being identical as that of the first (\ref{AIS-AdS-i}), with various integration limits resulting from different Karch-Randall brane positions ($r=\pm 2 \rho$).
\begin{eqnarray}
\label{AIS-AdS-ii}
{\cal A}_{{\cal I}_2}=V_{2}\int_{-2 \rho}^{2 \rho} dr {\cosh^{2}(r)}.
\end{eqnarray}
We obtained the total entanglement entropy of island surfaces by inserting (\ref{AIS-AdS-i}) and (\ref{AIS-AdS-ii}) into (\ref{EE-Island}), which is written below:
\begin{eqnarray}
\label{EE-Island-simp}
& & 
S_{\rm AdS}^{\rm Island}  =  \frac{2 V_{2}}{4 G_N^{(5)}}\Biggl(\int_{- \rho}^{\rho} dr {\cosh^{2}(r)}+\int_{-2 \rho}^{2 \rho} dr {\cosh^{2}(r)}\Biggr)=2 S_{\rm AdS}^{\rm thermal, \ total}.
\end{eqnarray}
The additional island surface obtained via the thermofield double partner is the source of the factor ``$2$'' in (\ref{EE-Island-simp}). The Page curve associated with the $n=2$ multiverse is obtained using (\ref{SHM-AdS}) and (\ref{EE-Island-simp}), which is illustrated in Fig. \ref{PC-AdS}. 
\begin{figure}
\begin{center}
\includegraphics[width=0.7\textwidth]{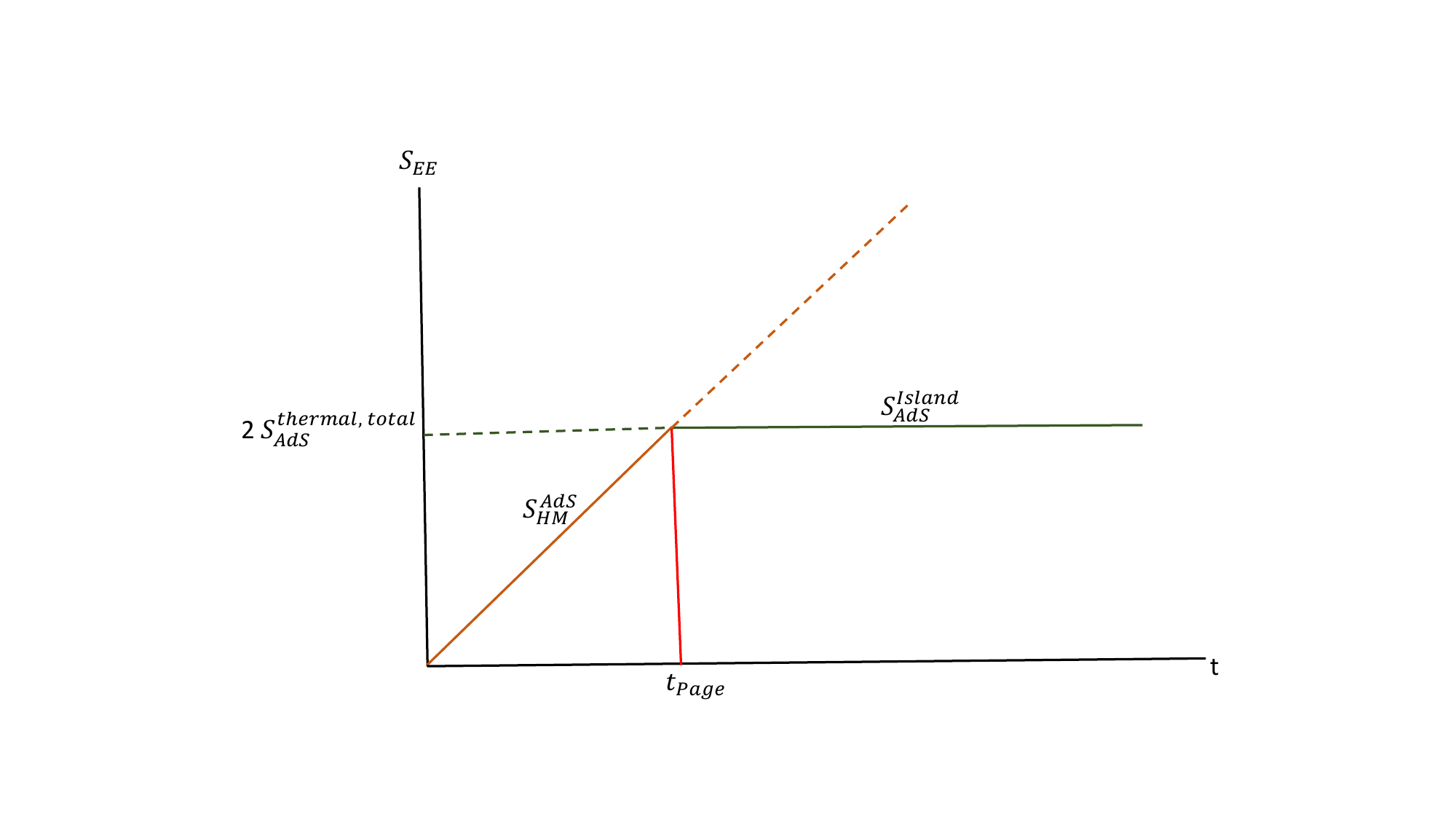}
\end{center}
\caption{Page curve of eternal AdS black holes for $n=2$ multiverse.}
\label{PC-AdS}
\end{figure}

\subsection{Page Curve of Schwarzschild de-Sitter Black Hole}
\label{IP-SdS} 
Here, we examine the Schwarzschild de-Sitter black hole's information problem. We cannot join mismatched branes at the identical defect, as stated in \ref{AdS+dS-Multiverse}. As a result, we divide our analysis of this issue into two parts and compute the Page curve corresponding to the Schwarzschild patch first, followed by the Page curve associated with the de-Sitter patch similar to \cite{Gopal+Nitin}. The next is how to do this. Two flat space branes embedded in the bulk are taken into consideration when studying the Schwarzschild patch in \ref{Sch-patch}, and two de-Sitter branes are taken into consideration when studying the de-Sitter patch in \ref{dS-patch}. The set-up has been shown in Fig. \ref{SdS-CP}. With flat space and de-Sitter branes in Schwarzschild and de-Sitter patches, respectively, the setup consists of two copies of wedge holography.
\begin{figure}
\begin{center}
\includegraphics[width=0.6\textwidth]{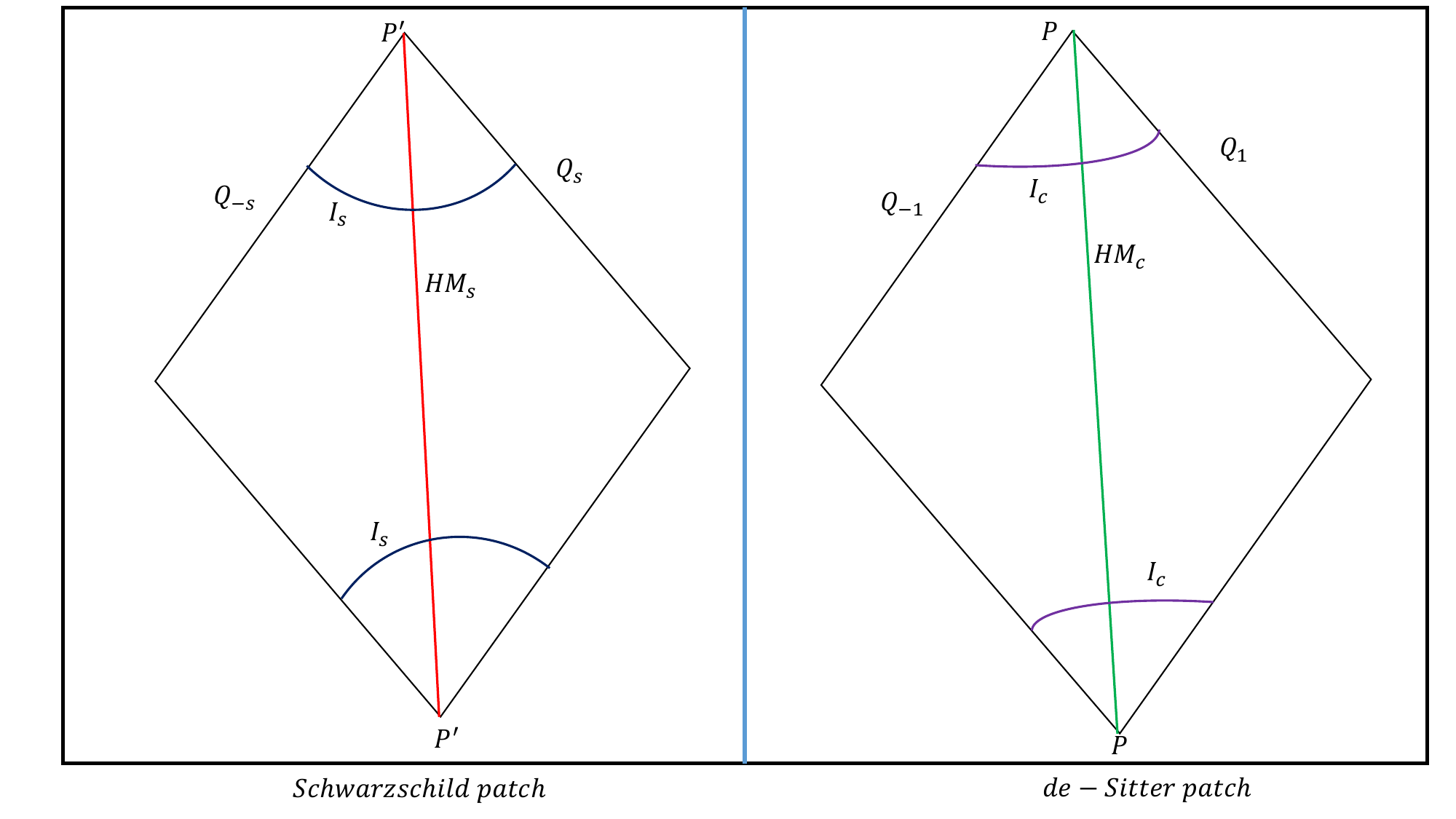}
\end{center}
\caption{Schwarzschild de-Sitter black hole realization in wedge holography. Black hole and cosmology island surfaces, or in our instance, black hole and de-Sitter horizons, are denoted by $I_s$ and $I_c$. Schwarzschild and de-Sitter patches' Hartman-Maldacena surfaces are represented by the red (${HM}_s$) and green (${HM}_c$) lines. Schwarzschild and de-Sitter patches make up the branes $Q_s$ and $Q_1$, respectively. Black hole and cosmological horizons emit Hawking and Gibbons-Hawking radiation, which is collected by the baths $Q_{-s}$ and $Q_{-1}$.}
\label{SdS-CP}
\end{figure}

\subsubsection{Schwarzschild patch} 
\label{Sch-patch}
Given that $\Lambda=0$ for the Schwarzschild black hole, we must take into account flat space branes in order to understand the Schwarzschild black hole on the Karch-Randall brane. It has been shown in \cite{WH-2} that one could acquire Karch-Randall branes with flat space black holes under the condition that bulk metric possess a particular structure:
\begin{equation}
\label{metric-bulk-flat-space}
ds_{(d+1)}^2=g_{\mu \nu} dx^\mu dx^\nu=dr^2+e^{2 r} h_{ij} dy^i dy^j =dr^2+{e^{2 r}} \Biggl(\frac{\frac{dz^2}{f(z)}-f(z) dt^2+\sum_{i=1}^{d-2}dy_i^2 }{z^2}\Biggr).
\end{equation}
Metric that was induced $h_{ij}$ in (\ref{metric-bulk-flat-space}) satisfy the Einstein equation on the brane given below:
\begin{eqnarray}
\label{Brane-Einstein-equation-vacuum}
R_{ij}-\frac{1}{2}h_{ij} R[h_{ij}] =0.
\end{eqnarray} 
(\ref{Brane-Einstein-equation-vacuum}) represents the equation of motion containing the brane Einstein-Hilbert term:
\begin{eqnarray}
\label{boundary-EH-action-FS}
S_{\rm FS}^{\rm EH} =\lambda^{\rm FS} \int d^{d}x \sqrt{-h} R[h],
\end{eqnarray}
where $\lambda^{\rm FS}\left(\equiv \frac{1}{16 \pi G_N^{d}}=\frac{1}{16 \pi G_N^{(d+1)}}\frac{e^{(d-2)a_1}}{(d-2)}\right)$ encapsulates details regarding the effective Newton's constant in $d$ dimensions, and (\ref{boundary-EH-action-FS}) was derived by substituting (\ref{metric-bulk-flat-space}) into (\ref{bulk-action}). In $d$-dimensions, for the Schwarzschild black hole $f(z)=1-\frac{z_h^{d-3}}{z^{d-3}}$ \cite{Island-SB}. Furthermore, the metric (\ref{metric-bulk-flat-space}) fulfill Neumann boundary condition having brane tension $T_{\rm flat \ space}= |d-1|$. Two Karch-Randall branes placed at $r=\pm a_1$ will provide the Schwarzchild black hole and associated bath. The thermal entropy associated with the Schwarzschild patch could be calculated using (\ref{metric-bulk-flat-space}) for $z=z_h$, and the result has been provided as:
\begin{eqnarray}
\label{TE-Sch}
S_{\rm thermal}^{\rm Schwarzschild}= \frac{ V_2 \int_{-a_1}^{a_1} dr e^{2r}}{4 G_N^{(5)}}=\frac{ V_2 \sinh (2 a_1)}{4 G_N^{(5)}}.
\end{eqnarray}
\\
{\bf Hartman-Maldacena Surface}: The infalling Eddington-Finkelstein coordinate is defined as $d v= dt -\frac{dz}{f(z)}$, and the flat space metric (\ref{metric-bulk-flat-space}) is simplified to:
\begin{eqnarray}
\label{metric-FS-branes}
& & ds^2 = dr^2+ \frac{e^{2r}}{z^2} \left(-f(z)dv^2-2 dv dz+\sum_{i=1}^2 dy_i^2\right).
\end{eqnarray}
The induced metric associated with the Hartman-Maldacena surface is given as follows for the parametrization $r=r(z)$ and $v=v(z)$.
\begin{eqnarray}
\label{induced-metric-HM-FS}
& &ds^2= \Biggl({r'(z)^2-\frac{e^{2r(z)}}{z^2} \left(2+f(z)v'(z)\right)}\Biggr) dz^2+ \frac{e^{2r(z)}}{z^2}\sum_{i=1}^{2}dy_i^2.
\end{eqnarray}
The area associated with the Hartman-Maldacena surface was calculated using (\ref{induced-metric-HM-FS}) as:
\begin{eqnarray}
\label{AHM-Sch}
& & A_{\rm HM}^{\rm Schwarzschild}=V_2 \int_{z_1}^{z_{\rm max}} dz  \Biggl(\frac{e^{2r(z)}}{z^2} \sqrt{r'(z)^2-\frac{e^{2r(z)} v'(z)}{z^2}\left(2+f(z)v'(z)\right)}\Biggr).
\end{eqnarray}
In late time approximation, i.e., $t\rightarrow \infty$, $r(z) \rightarrow 0$\footnote{We could prove this by following the methods outlined in (\ref{AHM-de-Sitter})-(\ref{soln-r(z)}). However, we must substitute the warp factor $\sinh(r(z))$ with $e^{r(z)}$.} \cite{Massless-Gravity}. Hence,
\begin{eqnarray}
\label{AHM-Sch-i}
& & A_{\rm HM}^{\rm Schwarzschild}=V_2 \int_{z_1}^{z_{\rm max}} dz   \Biggl(\frac{\sqrt{- v'(z)\left(2+f(z)v'(z)\right)}}{z^3}\Biggr).
\end{eqnarray}
Because the area associated with the Hartman-Maldacena surface has a similarity with (\ref{AHM-AdS}) with the exception of the volume factor, we're also constrained to $d=4$ for the Schwarzschild patch as well, $A_{\rm HM}^{\rm Schwarzschild} \propto t_1$. As a result, the entanglement entropy contribution that results from the Schwarzschild patch's Hartman-Maldacena surface exhibits a linear time dependency.
\begin{eqnarray}
\label{SHM-t}
S_{\rm HM}^{\rm Schwarzschild} \propto t_1.
\end{eqnarray}
\\
{\bf Island Surface}: $t=constant$ and $z=z(r)$ define the island surface. The area associated with the island surface could possibly be calculated using the induced metric in terms of embedding($z(r)$) and its derivative obtained from the bulk metric (\ref{metric-bulk-flat-space}):
\begin{eqnarray}
\label{induced-metric-Sch-IS}
& & ds^2= \Biggl({1+\frac{e^{2r}z'(r)^2}{z(r)^2\left(1-\frac{1}{z(r)}\right)}}\Biggr) dr^2+ \frac{e^{2r}}{z(r)^2}\sum_{i=1}^{2}dy_i^2.
\end{eqnarray}
where $f(z)=\left(1-\frac{1}{z}\right)$. The area associated with the island surface corresponding to the Schwarzschild patch is calculated using (\ref{induced-metric-Sch-IS}) as:
\begin{eqnarray}
\label{AIS}
A_{\rm IS}^{\rm Schwarzschild}= V_2 \int_{-a_1}^{a_1} dr  \Biggl(\frac{e^{2r}}{z(r)^2}\sqrt{1+\frac{e^{2r}z'(r)^2}{z(r)^2\left(1-\frac{1}{z(r)}\right)}}\Biggr).
\end{eqnarray}
For simplicity, we assign $z_h=1$ in the preceding equation, therefore $f(z)\geq0$ needs $z>1$. 
Substituting (\ref{AIS})'s Lagrangian in (\ref{var-AIS}), the first term that appears on the last line of (\ref{var-AIS}) for (\ref{AIS}), implies:
\begin{eqnarray}
\label{NBC-Sch}
\frac{e^{4 r} z'(r)}{\left(1-\frac{1}{z(r)}\right) z(r)^4 \sqrt{\frac{e^{2 r} z'(r)^2}{\left(1-\frac{1}{z(r)}\right) z(r)^2}+1}}=0. 
\end{eqnarray}
As a result, we are guided by the well-defined variational principle of (\ref{AIS}) assuming the embedding function meets the Neumann boundary condition on the branes, i.e., $z(r=\pm a_1)=0$, and therefore the minimal surface is going to be the black hole horizon, i.e., $z(r)=1$, as in \cite{GB-3,Massless-Gravity}. The following equation of motion of $z(r)$ can be used to reach the same result.
{
\begin{eqnarray}
\label{EOM-Sch}
& &
\frac{e^{2 r} \sqrt{\frac{e^{2 r} z'(r)^2+z(r)^2-z(r)}{(z(r)-1) z(r)}} }{2 z(r)^2 \left(e^{2 r} z'(r)^2+z(r)^2-z(r)\right)^2} \Biggl(3 e^{2 r} z'(r)^2 \left(2 e^{2 r} z'(r)-1\right)
+2 z(r)^2 \left(e^{2 r} z''(r)+4 e^{2 r} z'(r)-4\right)\nonumber\\
& & 
+2 z(r) \left(-e^{2 r}
   z''(r)+e^{2 r} z'(r)^2-4 e^{2 r} z'(r)+2\right)+4 z(r)^3\Biggr)
=0.
\end{eqnarray}
}
The black hole horizon is the solution of (\ref{EOM-Sch}), i.e. $z(r)=1$, which is compatible with the Neumann boundary condition on the branes \cite{GB-3}. As a consequence of inserting $z(r)=1$ in (\ref{AIS}), the smallest area of the island surface is produced, and the final outcome is:
\begin{eqnarray}
\label{AIS-simp-Schwarzschild}
A_{\rm IS}^{\rm Schwarzschild}= V_2 \int_{-a_1}^{a_1} dr e^{2r}=V_2 \sinh(2 a_1).
\end{eqnarray}
As a result, the entanglement entropy associated with the Schwarzschild patch's island surface becomes:
\begin{eqnarray}
\label{SIS-Sch}
S_{\rm IS}^{\rm Schwarzschild}=\frac{A_{\rm IS}^{\rm Schwarzschild}}{4 G_N^{(5)}}= \frac{2 V_2 \int_{-a_1}^{a_1} dr e^{2r}}{4 G_N^{(5)}} =\frac{2 V_2 \sinh(2 a_1)}{4 G_N^{(5)}}
= 2 S_{\rm thermal}^{\rm Schwarzschild}.
\end{eqnarray}
As a result, we obtained the Page curve by plotting (\ref{SHM-t}) and (\ref{SIS-Sch}) for the Schwarzschild patch depicted in Fig. \ref{PC-Sch}.
\begin{figure}
\begin{center}
\includegraphics[width=0.7\textwidth]{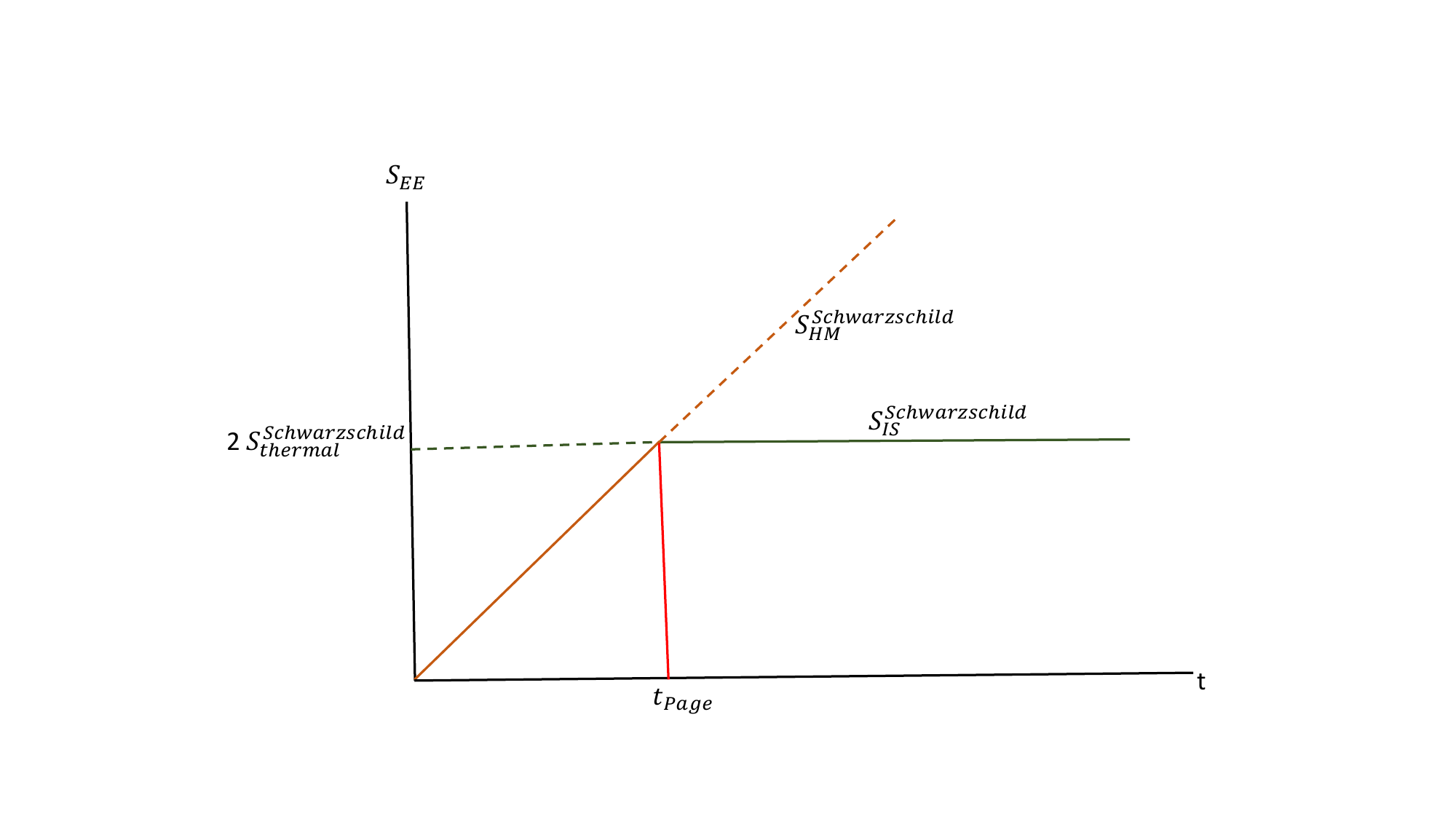}
\end{center}
\caption{Page curve of Schwarzschild patch.}
\label{PC-Sch}
\end{figure}

%%%%%%%%%%%%%%%
%%%%%%%%%%%%%
\subsubsection{de-Sitter patch}
\label{dS-patch}
The de-Sitter black hole and associated bath could possibly be placed at $r=\pm \rho$. The metric of bulk containing de-Sitter branes has emerged as: 
\begin{eqnarray}
\label{de-Sitter-metric-SdS}
& & 
ds_{(d+1)}^2=g_{\mu \nu} dx^\mu dx^\nu=dr^2+\sinh^2(r) h_{ij}^{\rm dS} dy^i dy^j \nonumber\\
& &=dr^2+\sinh^2(r) \Biggl(\frac{\frac{dz^2}{f(z)}-f(z) dt^2+\sum_{i=1}^{d-2}dy_i^2 }{z^2}\Biggr),
\end{eqnarray}
where $f(z)=1-\frac{\Lambda}{3}z^2=1-\left(\frac{z}{z_s}\right)^2$ with $z_s=\sqrt{\frac{3}{\Lambda}}$ in four dimensions. The thermal entropy associated with the de-Sitter patch could be calculated using (\ref{de-Sitter-metric-SdS}) by putting $z_s=1$\footnote{We merely used $z_s=1$ to simplify the computation. Because the cosmological constant is so tiny, in fact $z_s>>1$ but any figure that will not alter our qualitative results.} in the identical equation, and the result becomes:
\begin{eqnarray}
\label{thermal-dS}
S_{\rm dS}^{\rm thermal}=\frac{ A_{z=z_s}}{4 G_N^{(5)}} = \frac{ V_2 \int_{-\rho}^{\rho} dr \sinh^{2}(r)}{4 G_N^{(5)}}=\frac{ V_2 \left(\sinh (\rho ) \cosh (\rho )-\rho\right)}{4 G_N^{(5)}},
\end{eqnarray}
where $V_2=\int \int dy_1 dy_2$.\\
{\bf Hartman-Maldacena Surface}: We assume $d v= dt -\frac{dz}{f(z)}$ similarly to the Schwarzschild patch, and therefore (\ref{de-Sitter-metric-SdS}) turns into:
\begin{eqnarray}
\label{metric-dS-EFK}
& & ds^2 = dr^2+ \sinh^{2}(r) \left(\frac{-f(z)dv^2-2 dv dz+\sum_{i=1}^2 dy_i^2}{z^2}\right).
\end{eqnarray}
The Hartman-Maldacena surface is parametrized as $r=r(z)$ and $v=v(z)$, and therefore the area has been calculated using (\ref{metric-dS-EFK}) with the above-mentioned parametrization and expressed as follows:
\begin{eqnarray}
\label{AHM-de-Sitter}
& &  A_{\rm HM}^{\rm de-Sitter}=V_2 \int_{z_1^{\rm dS}}^{z_{\rm max}^{\rm dS}} dz {\cal L}_{\rm HM}^{\rm dS} \nonumber\\
& &
=V_2 \int_{z_1^{\rm dS}}^{z_{\rm max}^{\rm dS}} dz  \Biggl(\frac{\sinh^{2}(r(z))}{z^2}  \sqrt{r'(z)^2-\frac{\sinh^{2}(r(z))  v'(z)}{z^2}\left(2+f(z)v'(z)\right)}\Biggr),\nonumber\\
\end{eqnarray}
where  $z_1^{\rm dS}$ and $z_{\rm max}^{\rm dS}$ represent the point on the gravitating bath and the point of turning corresponding to the Hartman-Maldancena surface that defines the de-Sitter geometry, respectively. Because $v(z)$ is cyclic for (refAHM-de-Sitter), the conjugate momentum of $v(z)$ remains constant, i.e.,$\frac{\partial {\cal L}_{\rm HM}^{\rm dS}}{\partial v'(z) }=C$ ($C$ representing the constant) indicates:
{\footnotesize
\begin{eqnarray}
\label{v'(z)}
& &
v'(z)=\frac{-C z^3 \text{csch}(r(z)) \sqrt{32 C^2 z^6+15 f(z) \cosh (2 r(z))-6 f(z) \cosh (4 r(z))+f(z) \cosh (6 r(z))-10 f(z)} }{8\left(C^2 z^6 f(z)+f(z)^2 \sinh ^6(r(z))\right)} \nonumber\\
& & \hskip 0.5in  \times \left(\sqrt{2 z^2 f(z) r'(z)^2+\cosh (2 r(z))-1}-8 C^2 z^6-8 f(z) \sinh ^6(r(z))\right).
\end{eqnarray}
}
The Euler-Lagrange equation of motion of $r(z)$ given from (\ref{AHM-de-Sitter}) is:
{\footnotesize
\begin{eqnarray}
\label{EL-EOM-r(z)}
& &
\frac{\sinh ^2(r(z)) }{2 z^4 \left(z^2 r'(z)^2-\sinh ^2(r(z)) v'(z)
   \left(f(z) v'(z)+2\right)\right) \sqrt{r'(z)^2-\frac{\sinh ^2(r(z)) v'(z) \left(f(z) v'(z)+2\right)}{z^2}}}  \Biggl(z r'(z) \sinh ^2(r(z)) \nonumber\\
   & &
   \left(\left(z f'(z)+2 f(z)\right) v'(z)^2+2 v'(z) \left(z f(z) v''(z)+2\right)+2 z v''(z)\right)-\sinh ^2(r(z)) v'(z) \left(f(z) v'(z)+2\right) \nonumber\\
   & &\left(3
   f(z) \sinh (2 r(z)) v'(z)^2+2 z^2 r''(z)+6 \sinh (2 r(z)) v'(z)\right)+4 z^2 r'(z)^2 \sinh (2 r(z)) v'(z) \left(f(z) v'(z)+2\right)-4 z^3 r'(z)^3\Biggr)
   =0. \nonumber\\
\end{eqnarray}
}
By replacing $v'(z)$ for (\ref{v'(z)}) utilizing $f(z)=1-z^2$, and assign $z_s=1$ for simplicity, the EOM (\ref{EL-EOM-r(z)}) reduces to the following form:
{\footnotesize
\begin{eqnarray}
\label{EL-EOM-r(z)-simp}
& &
\frac{\sinh ^2(r(z)) }{2 z^4 \left(C^2
   z^6-\left(z^2-1\right) \sinh ^6(r(z))\right) \left(z^2 \left(z^2-1\right) r'(z)^2-\sinh ^2(r(z))\right) \sqrt{\frac{\left(z^2-z^4\right) r'(z)^2 \sinh ^6(r(z))+\sinh ^8(r(z))}{C^2
   z^8+\left(z^2-z^4\right) \sinh ^6(r(z))}}}\nonumber\\
   & & \hskip -0.3in \times \Biggl(-2 z^2 r''(z) \sinh ^2(r(z)) \left(C^2 z^6-\left(z^2-1\right) \sinh ^6(r(z))\right)+r'(z) \left(2 z \sinh ^8(r(z))-4 C^2 z^7 \sinh ^2(r(z))\right)\nonumber\\
   & &\hskip -0.3in +r'(z)^2 \left(C^2 z^8 \sinh
   (2 r(z))-8 z^2 \left(z^2-1\right) \sinh ^7(r(z)) \cosh (r(z))\right)+r'(z)^3 \left(4 z^3 \left(z^2-1\right)^2 \sinh ^6(r(z))-2 C^2 z^9\right)\nonumber\\
   & & +6 \sinh ^9(r(z)) \cosh (r(z))\Biggr)=0.
\end{eqnarray}
}
The equation above is tough to solve. One simple solution to (\ref{EL-EOM-r(z)-simp}) is obtained as:
\begin{eqnarray}
\label{soln-r(z)}
r(z)=0.
\end{eqnarray}
We are able to observe via equation (\ref{AHM-de-Sitter}) that for $r(z)=0$, $A_{\rm HM}^{\rm de-Sitter}=0$, and so the entanglement entropy associated with the Hartman-Maldacena surface is given as\footnote{The identical scenario answer $r(z)=0$ appears in \cite{Massless-Gravity} in the calculation of Hartman-Maldacena surface's area. See \cite{GB-3} for a similar approach; in our instance, the embedding is $r(z)$, but in \cite{GB-3}, the embedding was $r(\mu)$, where $\mu$ represents the angle.}:
\begin{eqnarray}
\label{SHM-dS}
& & S_{\rm HM}^{\rm de-Sitter} = \frac{A_{\rm HM}^{\rm de-Sitter}}{4 G_N^{(5)}} =0.
\end{eqnarray}
\\
{\bf Cosmological Island Surface Entanglement Entropy}:
The induced metric obtained in terms of embedding ($z=z(r)$) and its derivative (\ref{de-Sitter-metric-SdS}) has been used to calculate the area of the island surface parametrized by $t=constant$, $z=z(r)$, and the final result is:
\begin{eqnarray}
\label{action-dS-IS}
A_{\rm IS}^{\rm de-Sitter}= V_2 \int_{- \rho}^{\rho} dr  \Biggl(\frac{\sinh^{2}(r)}{z(r)^2} \sqrt{1+\frac{\sinh^{2}(r) z'(r)^2}{z(r)^2 (1-z(r)^2)}}\Biggr).
\end{eqnarray}
For a de-Sitter patch, $f(z)=1-\left(\frac{z}{z_s}\right)^2$, we used $z_s=1$ in (\ref{action-dS-IS}) to simplify the computation. As a result, $f(z)\geq 0$ when $0<z<1$. 

For the de-Sitter patch, $f(z)=1-\left(\frac{z}{z_s}\right)^2$, we have taken $z_s=1$ in (\ref{action-dS-IS}) for calculation simplification. Therefore $f(z)\geq 0$ if $0<z<1$. The Euler-Lagrange equation of motion that involves embedding $z(r)$ from (\ref{action-dS-IS}) is the following:
{\footnotesize
\begin{eqnarray}
\label{EOM-dS-IS}
& &
\frac{\sinh ^2(r) \sqrt{\frac{-\sinh ^2(r) z'(r)^2+z(r)^4-z(r)^2}{z(r)^2 \left(z(r)^2-1\right)}} }{\left(z(r) \sinh ^2(r)
   z'(r)^2-z(r)^5+z(r)^3\right)^2}\Biggl(z(r) \sinh ^2(r) z'(r)^2+3
   \sinh ^3(r) \cosh (r) z'(r)^3 \nonumber\\
  & &  -z(r)^4 \sinh (r) \left(\sinh (r) z''(r)+4 \cosh (r) z'(r)\right)+z(r)^2 \sinh (r) \left(\sinh
   (r) z''(r)+4 \cosh (r) z'(r)\right)\nonumber\\
   & &
   +2 z(r)^7-4 z(r)^5+2 z(r)^3\Biggr)
   =0.
\end{eqnarray}
}
In general, solving the given problem is difficult. remarkably, there exists a $z(r)=1$ solution for the aforementioned differential equation, that is just that the originally stated de-Sitter horizon ($z_s=1$) similar to earlier discussions on the EOM for the island surfaces. Further, this solution consistent with the Neumann boundary condition on the branes, and so the cosmological island surface embedding's EOM solution is given as:
\begin{eqnarray}
\label{soln-z(r)-dS-i}
z(r)=1.
\end{eqnarray}
One could achieve an identical result by demanding the well-defined variational principle of (\ref{action-dS-IS}) and enforcing Neumann boundary conditions on the branes, as discussed in \ref{PC-TEBH}.
\begin{eqnarray}
\label{NBC-dS}
\frac{\sinh ^4(r) z'(r)}{z(r)^4 \left(1-z(r)^2\right) \sqrt{\frac{\sinh ^2(r) z'(r)^2}{z(r)^2 \left(1-z(r)^2\right)}+1}}=0.
\end{eqnarray}
If we enforce $z'(r=\pm \rho)=0$, then we obtain that the horizon as the minimal surface, hence $z(r)=1$ \cite{GB-3}. We derive the smallest area corresponding to the cosmological island surface associated with the de-Sitter patch by inserting $z(r)=1$ in (\ref{action-dS-IS}) as given below:
\begin{eqnarray}
\label{AIS-simp-dS}
A_{\rm IS}^{\rm de-Sitter}= V_2 \int_{-\rho}^{\rho} dr \sinh^{2}(r)=V_2 \left(\sinh (\rho ) \cosh (\rho )-\rho\right).
\end{eqnarray}
Therefore, cosmological island surface has the following entanglement entropy:
\begin{eqnarray}
\label{SIS-dS}
S_{\rm IS}^{\rm dS}=\frac{2 A_{\rm IS}^{\rm de-Sitter}}{4 G_N^{(5)}}= 2 S_{\rm dS}^{\rm thermal}.
\end{eqnarray}
Due to a second cosmological island surface on the thermofield double partner side (seen in Fig. \ref{SdS-CP}), an additional numerical factor ``2'' is present. We  could get the Page curve of de-Sitter patch by plotting (\ref{SHM-dS}) and (\ref{SIS-dS}). We obtained a flat Page curve for the de-Sitter patch as in \cite{GB-3}.

%Let us summarize the results of this section. It was argued in \cite{GB-3,Massless-Gravity} that in wedge holography without DGP term, the black hole horizon is the only extremal surface and the Hartman-Maldacena surface does not exist and hence one expects the flat page curve. We also see that when we compute the entanglement entropies of island surfaces of AdS, Schwarzschild, and de-Sitter black holes then minimal surfaces turn out to be horizons of the AdS or Schwarzschild or de-Sitterblack holes. As a curiosity, we computed entanglement entropies of Hartman-Maldacena surfaces for the parametrization $r(z)$ and $v(z)$ used in the literature and we found non-trivial linear time dependence for the AdS and Schwarzschild black holes whereas Hartman-Maldacena surface entanglement entropy turns out to be zero for the de-Sitter black hole. Therefore we obtain the flat Page curve for the de-Sitter black hole not for the AdS and Schwarzschild black holes due to the non-zero entanglement entropy of Hartman-Maldacena surfaces. 
%The theme of the paper is not to discuss whether we get a flat Page curve or not. The paper aimed to construct a ``multiverse'' in Karch-Randall braneworld which we did in section \ref{Multiverse-section} and check the formula given in (\ref{Island-Formula-n-BHs}). We saw in subsection \ref{PC-TEBH} that (\ref{Island-Formula-n-BHs}) is giving consistent results.

%%%%%%%%%%%%%%%%%%%%%%%%%%%%%%%%%%%%%%%%%%%%%%%%%
%%%%%%%%%%%%%%%%%%%%%%%%%%%%%%%%%%%%%%%%%%%%
{\bf Comment on the Wedge Holographic Realization of Schwarzschild de-Sitter Black Hole with Two Karch-Randall Branes}:
We produced the Page curves of Schwarzschild and de-Sitter patches independently in \ref{IP-SdS}. There is yet another technique for us to obtain the Schwarzschild de-Sitter black hole's Page curve. Below is a summary of the concept:
\begin{itemize}
\item Take two Karch-Randall branes $Q_1$ and $Q_2$, where $Q_1$ is a Schwarzschild de-Sitter black hole and $Q_2$ is a radiation-collecting bath\footnote{In this instance, the term ``Hawking radiation'' might not be appropriate because observers might not be capable to tell the difference between ``Gibbons-Hawking radiation'' and ``Hawking radiation'' when the Schwarzschild de-Sitter black hole emits radiation as a whole.}.

\item Let's say the structure of the bulk metric is as follows:
\begin{equation}
\label{metric-SdS-induced}
ds^2 = g_{\mu \nu} dx^\mu dx^\nu=dr^2+g(r) h_{ij}^{\rm SdS} dy^i dy^j=dr^2+g(r) \Biggl(\frac{\frac{dz^2}{f(z)}-f(z) dt^2+\sum_{i=1}^{2}dy_i^2}{z^2} \Biggr),
\end{equation}
where $f(z)=1-\frac{2 M}{z}-\frac{\Lambda}{3}z^2$ in $d=4$.
\item The next thing to do is to solve the Einstein equation (\ref{Einstein-equation}) to determine $g(r)$.

\item The bulk metric (\ref{metric-SdS-induced}) has to fulfill the Neumann boundary condition (\ref{NBC}) at $r=\pm \rho$ after finding the solution.

\item To use the Ryu-Takayanagi formula, one must additionally determine if a CFT or non-CFT theory exists at the defect.

\item If the aforementioned points are correctly verified, we may compute the areas of the Hartman-Maldacena and island surfaces to derive the Page curve of the Schwarzschild de-Sitter black hole\footnote{Because we are essentially referring to the island within the Schwarzschild de-Sitter black hole in this context, the term ``island'' may become difficult to talk about. It may be difficult to determine whether the ``island'' is within the black hole horizon or the de-Sitter horizon as the SdS black hole has two horizons. It will be good to stick with the current arrangement with two black holes and two baths. To learn about a non-holographic technique, see chapter {\bf 8} based on \cite{Gopal+Nitin}.}. 
\end{itemize}
``Mathematical concept'' is essentially what the debate above is. We could be referring to \cite{mismatched-branes} since there are three potential branes: Minkowski, de-Sitter, and anti de-Sitter. Within the open bracket of (\ref{metric-SdS-induced}), there is no brane specified with the induced metric. Additionally, there is flat space holography, dS/CFT duality, or AdS/CFT correspondence. The duality among CFT and bulk, which takes the structure of a Schwarzschild de-Sitter metric, does not exist. Owing to the abovementioned reason, there won't be any defect descriptions and no ``intermediate description'' of the wedge holography. As a result, we get to the conclusion that one could represent a Schwarzschild de-Sitter black hole using wedge holography utilizing two copies of the wedge holography that define the Schwarzschild patch and the de-Sitter patch, respectively.
\section{Application to Grandfather Paradox}
\label{AGP}
The ``grandfather paradox'' is described in this section along with how our model resolves it. \par
According to the ``Grandfather paradox'', Bob is unable to travel back in time. Because he could murder his grandfather in a different universe if he could go back in time. Bob's will cease to be in the present if his grandfather is dead in another universe \cite{GP}.\par
Let's now examine how our setup might avoid this issue. In \ref{AdS-multiverse} and \ref{de-Sitter-multiverse}, we explained how a multiverse is made up of $2n$ universes that are Karch-Randall branes. These branes have AdS and de-Sitter spacetimes in  \ref{AdS-multiverse} and \ref{de-Sitter-multiverse} as their geometries. All ``universes'' have a connection at the ``defect'' in each configuration by a transparent boundary condition. Transparent boundary conditions ensure that communication exists between all of these universes.
\begin{figure}
\begin{center}
\includegraphics[width=0.8\textwidth]{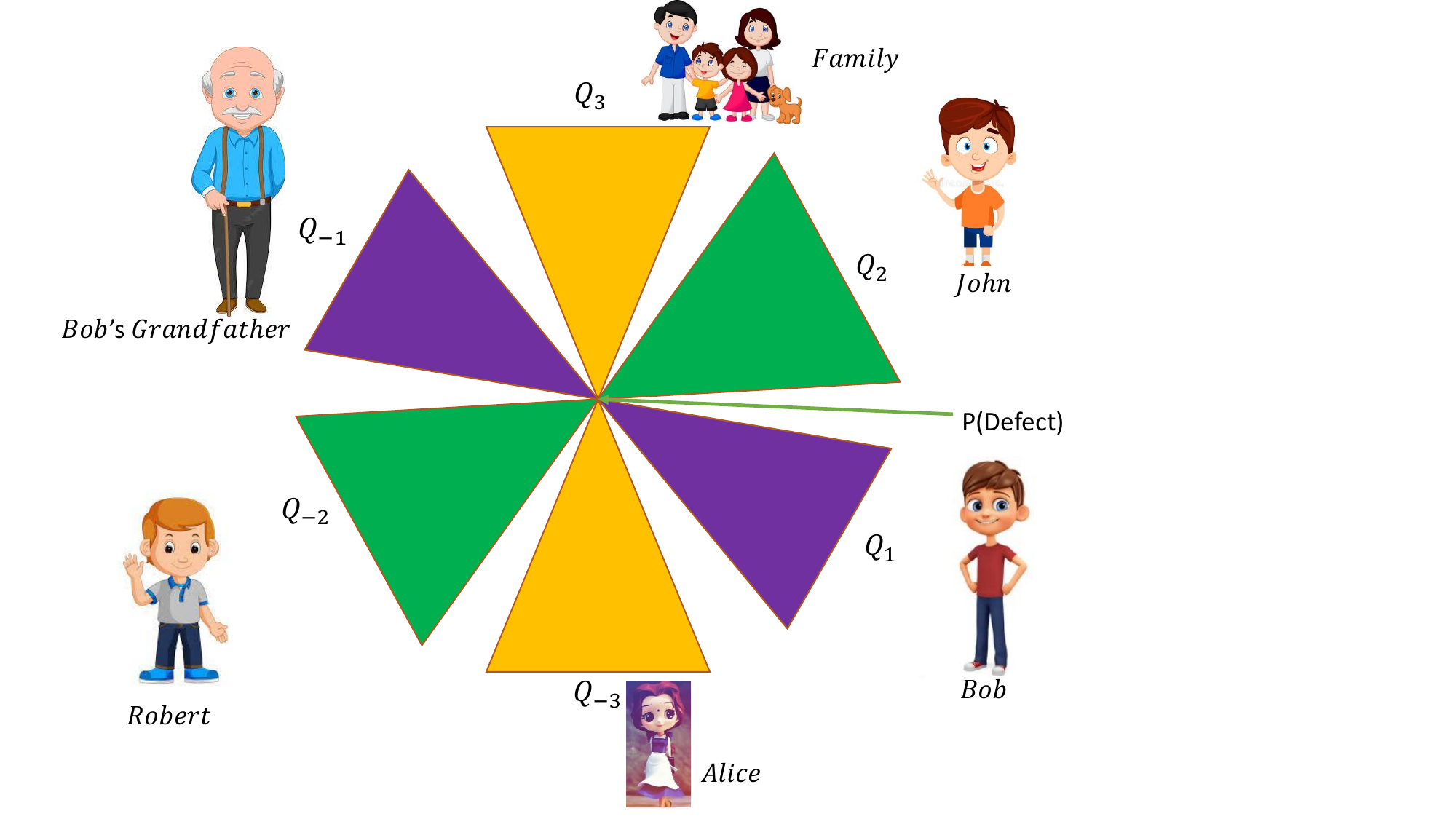}
\end{center}
\caption{Multiple universes where various individuals live is $Q_{-1,-2,-3,1,2,3}$.}
\label{RGP}
\end{figure}
Assume that Bob resides on $Q_1$ and his grandfather resides on $Q_2$. Then, in order to get around the dilemma, Bob is able to go to places like $Q_{-2}$, $Q_{-3}$, etc. where he is able to come across Robert and Alice (see Fig. \ref{RGP}). Therefore, the ``grandfather paradox'' could be resolved in this situation. The ``grandfather paradox'' has been addressed using a concept similar to that discussed in this debate, which is compatible with the ``many world theory''.
\section{Conclusion}
\label{Conclusion}
Using the concept of wedge holography, we proposed in this chapter that the Karch-Randall braneworld contains a multiverse. If we consider the $2n$ universes, then those are going to be represented as Karch-Randall branes embedded within the bulk, this is how the multiverse has been described. These branes could either contain black holes or they won't, that will be decided by the gravitational action. We looked at three distinct scenarios.
\begin{itemize}
\item In  \ref{AdS-multiverse}, we generated the multiverse using $d$-dimensional Karch-Randall branes that have been incorporated in $AdS_{d+1}$ bulk. The aforementioned branes are defined by the $AdS_d$ geometry. Transparent boundary conditions at the defect interconnect the $2n$ anti de-Sitter branes that make up the multiverse in this particular scenario to one another. As soon as constructed, the multiverse made of AdS branes is eternal.

\item In \ref{de-Sitter-multiverse}, we created a multiverse using $d$-dimensional de-Sitter spaces on the Karch-Randall branes embedded within the $(d+1)$-dimensional bulk $AdS_{d+1}$. The $2n$ de-Sitter branes that makes the multiverse possesses a short lifespan. In such a case, all of de-Sitter branes must be formed and destroyed at the identical time. As a result of the dS/CFT duality, defect CFT corresponds to non-unitary conformal field theory.

\item In \ref{AdS+dS-Multiverse}, we additionally explored why it wouldn't be conceivable to define the multiverse as a combination of $d$-dimensional de-Sitter and anti-de-Sitter spacetimes within the identical bulk. We are able to possess the multiverse with either anti de-Sitter branes ($M_1$) or de-Sitter branes ($M_2$), but it does not have both. Since AdS branes intersect on the ``time-like'' boundary of $AdS_{d+1}$ bulk and the de-Sitter branes intersect at the ``space-like'' boundary. Universes within $M_1$ are able to interact with each other; similarly, the universes in $M_2$ are able to communicate with each other, but $M_1$ is unable to communicate with $M_2$. 
\end{itemize}
We looked into if we could resolve the information paradox involving multiple black holes at the same time. This could be accomplished by creating a multiverse in which $n$ Karch-Randall branes contain black holes and the Hawking radiation from these black holes is gathered by a $n$ gravitating baths. In that case, we have a time dependency from the Hartman-Maldacena surfaces, and the constant value becomes $2 S^{{i=1,2,..,n}, \ \rm thermal}_{\rm BH}$, which comes from the $n$ island surfaces.\par
We obtained the Page curves associated with two black holes for the $n=2$ multiverse as a consistency check. We figured out that the black hole and bath systems are located at $- 2 \rho \leq r \leq 2 \rho$ and $- \rho \leq r \leq \rho$. In such a scenario, we showed that the entanglement entropy contribution generated by the Hartman-Maldacena surfaces exhibits a linear time dependence for both AdS and Schwarzschild black holes and becomes zero for the de-Sitter black hole, but the contributions that come via the island surfaces remain constant. As a result, this mimics the Page curve. Utilizing this concept, we obtained the Page curve of the Schwarzschild de-Sitter black hole. This idea could be used to compute the Page curve of black holes that have numerous horizons using wedge holography. We also addressed how we might generate a Page curve for these black holes by employing two Karch-Randall branes, one as a black hole and another as a bath. Within this situation, identifying the island surface and figuring out what type of radiation we are receiving will be difficult. As an example, if a Karch-Randall brane contains a black hole and cosmic event horizons, such as a Schwarzschild de-Sitter black hole, the observer receiving the radiation will be unable to tell whether this is Hawking radiation or Gibbons-Hawking radiation.\par
We tested our approach for very basic cases without the DGP term placed on the Karch-Randall branes, however it is also possible to discuss massless gravity by including the DGP term on the Karch-Randall branes \cite{Massless-Gravity}. Tensions associated with branes will be corrected by the additional term in (\ref{NBC-DGP}) in this situation. Furthermore, we proposed that with this system, where all universes interact via transparent boundary conditions at the junction point, one could circumvent the ``grandfather paradox''. In order to resolve the paradox, one may travel to another universe where his grandpa is not living, so preventing him from killing his grandfather. We provided a qualitative solution to the ``grandfather paradox'', although further study in this area employing wedge holography is required.

%m

\chapter{Conclusion and Future outlook }
\graphicspath{{Chapter10/}{Chapter10/}}

In this part of the thesis, we have studied the resolution of information paradox using various proposals, e.g., island proposal, doubly holographic setup, and wedge holography. In this process, we addressed the following issues:
\begin{itemize}
\item How do the higher derivative terms in the gravitational actions affect the Page curve?

\item How to obtain the Page curve of black holes with multiple horizons, e.g., Schwarzschild de-Sitter black hole?

\item Can we describe the ``Multiverse'' using wedge holography?
\end{itemize}

We started with a very simple example and considered the Reissner Nordstr\"om black hole in the presence of ${\cal O}(R^2)$ terms as higher derivative terms, which is a non-holographic model. We considered the two kinds of HD terms: Gauss-Bonnet term and general ${\cal O}(R^2)$ as considered in \cite{NBH-HD}. Following is the summary of key results obtained in chapter {\bf 6} which is based on \cite{RNBH-HD}.

\begin{itemize}
\item The Page curves of Reissner Nordstr\"om black hole are shifting towards later times or earlier times when Gauss-Bonnet coupling ($\alpha$) increases or decreases. This implies that Page time is being affected due to the presence of HD terms. As soon as islands contribute to the entanglement entropy of Hawking radiation, we get the information from the black hole. Hence, ``dominance of islands'' in the entanglement entropy of Hawking radiation to compute the Page curve is affected by the higher derivative terms.

\item We found that scrambling time is affected when we have some other general ${\cal O}(R^2)$ terms, including the Gauss-Bonnet term. In contrast, it is unaffected when we consider only the Gauss-Bonnet term as the higher derivative term.

\item We showed that our results are consistent with the literature by taking the $\alpha \rightarrow 0$ limit. We recover the results of \cite{Island-RNBH} in this limit.
\end{itemize}

We studied the black hole information problem in chapter {\bf 8} based on the paper \cite{Gopal+Nitin} and proposed a method to resolve the information paradox of black holes with multiple horizons. We focused on the Schwarzschild de-Sitter (SdS) black hole, which has two horizons: black hole and de-Sitter horizons. To obtain the Page curve of the black hole, we inserted thermal opaque membranes on both sides so that an observer living on the black hole side can access only the radiation of the black hole patch. We used the island proposal to define the radiation regions in the black hole patch. In this case, gravity is not negligible enough, but one can use the island proposal in the approximation that the observer is very far away from the black hole. Hence, we can use the island proposal. We computed the entanglement entropy of Hawking radiation in the absence and presence of the island surface. After plotting these contributions together, we obtained the Page curve of the black hole patch. We also studied the effect of temperature on the Page curves of black holes. We found that low-temperature black holes take too much time to deliver the information out of the black holes compared to high-temperature black holes. In the language of entanglement islands, this result is interpreted as follows. ``Dominance of islands'' and ``information recovery'' and hence Page time is higher for low-temperature black holes because when islands contribute to the entanglement entropy, we get information from the black hole. In this kind of black hole, it is not possible to obtain the Page curve of the Schwarzschild de-Sitter black hole as a whole due to asymmetrical regions on both sides of the SdS black hole. 

We constructed the doubly holographic setup from a top-down approach in chapter {\bf 7} based on our work \cite{HD-Page Curve-2}. In our setup, the bulk is the eleven-dimensional ${\cal M}$-theory uplift inclusive of ${\cal O}(R^4)$ corrections of type IIB string dual constructed in \cite{HD-MQGP}. The external bath to collect the Hawking radiation is a non-conformal thermal QCD bath. We obtained the Page curve of the eternal neutral black hole by computing the entanglement entropies of Hartman-Maldacena and island surfaces in the absence and presence of  ${\cal O}(R^4)$ terms. When ${\cal O}(R^4)$ terms are absent, then we obtained the entanglement entropies by computing the areas of extremal surfaces, whereas in the presence of higher derivative terms, we used Dong's formula to calculate the entanglement entropies. Let us compare the doubly holographic setup constructed from the bottom-up approach and our setup.

\begin{itemize}
\item {\bf Bottom-up double holography with CFT bath}: Three descriptions of the doubly holographic setup is given as below.
\begin{itemize}
\item {\bf Boundary Description:} $d$-dimensional BCFT living at $AdS_{d+1}$-boundary with $(d-1)$-dimensional defect.

\item {\bf Intermediate Description:} Gravity on $d$-dimensional end-of-the-world brane coupled to $d$-dimensional BCFT via transparent boundary condition at the defect.

\item {\bf Bulk Description:} $d$-dimensional BCFT has its own holographic dual which is $AdS_{d+1}$.
\end{itemize}

\item {\bf M theory brane description of top-down double holography with QCD bath}: Top-down model has three following descriptions similar to the bottom-up model.
\begin{itemize}
\item {{\bf Boundary-like Description:} $QCD_{2+1}$ is living at the tip of the conifold i.e. at $r=0$.}

\item {{\bf Intermediate Description:} Black $M_5$-brane which contains black hole coupled to  $QCD_{2+1}$ bath living at $M_2$ brane.}

\item {{\bf Bulk Description:}  $QCD_{2+1}$ has holographic dual which is eleven dimensional M theory.}
\end{itemize}
\end{itemize}

Following are the key results that we obtained in chapter {\bf 7}.
\begin{itemize}
\item In doubly holographic setups, it was found that one could get the Page curve with massive gravity on the end-of-the-world brane. In our setup, we explicitly showed that this is not the case in the top-down model. We computed the spectrum of graviton on the end-of-the-world brane and found that one could get the Page curve with massless graviton localized on the end-of-the-world brane.

\item We found that ${\cal O}(R^4)$ terms do not affect the Page curve in this setup because contributions to the entanglement entropies are large-$N$ exponentially suppressed. This exponential large-$N$ suppression exists because of massless graviton on the brane. 

\item We showed that no boundary terms arise on the end-of-the-world brane even in the presence of ${\cal O}(R^4)$ terms in the bulk, and end-of-the-world brane turns out to be a ``fluxed hypersurface'' with non-zero tension.

\item Hartman-Maldacena surface entanglement entropy also exhibits ``Swiss-Cheese'' structure in large-$N$ scenario.
\end{itemize}

In chapter {\bf 9} (which is based on the work done in \cite{Multiverse}), we used wedge holography to describe multiverse. The multiverse is constructed as follows. In wedge holography, we have two Karch-Randall branes, and these branes are joined at the defect. The setup is mathematically consistent only if the bulk metric satisfies the Neumann boundary condition (NBC) on the branes. The geometry of the branes can be anti de-Sitter, de-Sitter, or flat space, depending upon the bulk metric. We showed that one can construct a setup of $2n$ Karch-Randall branes in wedge holography, and the bulk metric still satisfies NBC on the $2n$ branes. These branes are located at $r=\pm n \rho$. We can localize the gravity on these branes using braneworld holography \cite{KR1,KR2}. Therefore, we have $2n$ branes embedded in the bulk. The geometry of these branes can be anti de-Sitter or de-Sitter or flat space but not the mixture of any two. Hence, we have a multiverse that is made up of $2n$ gravitating systems. Due to transparent boundary conditions at the defect, various universes existing in the multiverse can communicate with each other. If we consider two multiverses, then there will be the communication of the universes in a specific multiverse but not between the two multiverses. 

This model applies to the Page curve of black holes with multiple horizons. We explicitly did this for Schwarzschild de-Sitter black hole and argued that we could get the Page curve of the SdS black hole by taking two copies of wedge holography so that one copy describes the Schwarzschild patch with flat space branes and the other copy describes the de-Sitter patch with two de-Sitter branes. By doing so obtained the Page curve of Schwarzschild and de-Sitter patches separately, similar to \cite{Gopal+Nitin} and concluded that we couldn't get the Page curve of SdS black hole with two Karch-Randall branes in wedge holography. Since the multiverse consists of communicating universes and hence one could avoid the ``grandfather paradox'' by not traveling to the universe in which one's grandfather is living, similar to ``many world theory''. \\

{\bf Future Outlook}: In future, we shall work on the following issues:
\begin{itemize}
\item Using the doubly holographic setup constructed in chapter {\bf 7} from a top-down approach. We will compute the reflected entropy from the bulk point of view \cite{RE}. This will shed light on the holographic QCD via gauge-gravity duality. We are interested to see the effect of the ${\cal O}(R^4)$ terms on the reflected entropy and how the higher derivative terms affect the physics of thermal QCD.

\item We will study the complexity growth of black holes with multiple horizons using the complexity equal volume \cite{CV} and complexity equals action proposals \cite{CA}.

\item In chapter {\bf 9}, we saw that wedge holography is capable of describing the multiverse. The most interesting thing about this setup is that all the universes existing in the multiverse are capable of transferring information with each other. Using this feature, we provided a qualitative resolution of the ``grandfather paradox''. We shall work on the more concrete resolution of the ``grandfather paradox by providing a quantitative description of the ``grandfather paradox'' and its resolution. Further, using this setup, we will obtain the Page curve of the Reissner-Nordstr\"om de-Sitter black hole.
\end{itemize}

\appendix
\chapter{}
\label{appendix-MChPT}
\section{Coupling Constants $y_{1,3,5,7}$, $z_{1,....,8}$, $F_\pi^2$ and $g_{YM}^2$}
We calculate the infrared (IR) and ultraviolet (UV) contributions of the coupling constants $y_{1,3,5,7}$ and $z_{1,2,...,8}$ in this appendix.
Here, we use: $\psi_0(Z) = \int_0^Z \phi_0(Z)dZ$, and we shall divide $\int dZ$ into $\int_{\rm IR} + \int_{\rm UV}$. We found that $\int_{\rm UV}\sim \left({\cal C}_{\phi_0}^{\rm UV}\right)^m\left({\cal C}_{\psi_1}^{\rm UV}\right)^n, m, n \in \mathbb{Z}^+$, and hence we can self-consistently set ${\cal C}_{\phi_0}^{\rm UV} = {\cal C}_{\psi_1}^{\rm UV} = 0$, and therefore we have disregarded the UV contributions $\int_{\rm UV}$. In the IR (i.e., around $Z=0$), the radial profile function associated with the $rho$ vector meson is obtained as \cite{MChPT,Vikas-Thesis}:
\begin{equation}
\label{psi1(Z)}
\psi_1(Z) = \sqrt{2} {\cal C}_{\psi_1}^{(1)\ {\rm IR}} \sqrt{i \sqrt{\omega_2}} \left[1-Z \left(\beta  {\cal C}^{zz}_{\ \ \theta_1z\ \theta_1x}+\omega_1\right)\right]; {\cal C}_{\psi_1}^{(1)\ {\rm IR}} \equiv {\cal C}_{\psi_1}^{{\rm IR}}=N^{-\Omega_{\psi_1}},\ \Omega_{\psi_1}>0,
\end{equation}
wherein:
{\footnotesize
\begin{eqnarray}
\label{omega_1-and-2_defs}
& & \omega_1\equiv \frac{1}{4} \left({m_0}^2-3 b^2 \left({m_0}^2-2\right)\right)+18 b^2 {r_h} \log
   ({r_h})-\frac{3 b \gamma  {g_s} M^2 \left({m_0}^2-2\right) \log ({r_h})}{2 N}+\frac{36 b
   \gamma  {g_s} M^2 {r_h} \log ^2({r_h})}{N},\nonumber\\
& & \omega_2\equiv -\frac{4}{3}+\frac{3}{2} b^2 \left({m_0}^2+72 {r_h}-4\right)-36 b^2 {r_h} \log ({r_h})+\frac{3 b \gamma
   {g_s} M^2 \left({m_0}^2-4\right) \log ({r_h})}{N}-\frac{72 b \gamma  {g_s} M^2 {r_h}
   \log ^2({r_h})}{N},\nonumber\\
& & 
 {\cal C}^{zz}_{\ \ \theta_1z\ \theta_1x} = -\frac{\left(3 b^2-2\right) \log N  {m_0}^2 ({\cal C}_{zz}^{(1)}-2 {\cal C}_{\theta_1z}^{(1)}+2 {\cal C}_{\theta_1x}^{(1)})}{4 \left(3
   b^2+2\right) (\log N -3 \log ({r_0}))}.
\end{eqnarray}}
Similarly, the profile function of the $\pi$-meson is given as \cite{MChPT,Vikas-Thesis}:
\begin{eqnarray}
\label{phi0(Z)}
& & \phi_0(Z) =\frac{\pi ^2 {\cal C}_{\phi_0}^{\rm IR}\  N^{2/5} \alpha _{\theta _1}^3 (\log N -3 \log r_0 ) \left(\frac{27}{8 b^2 {g_s}
   \log N  (\log N +3 \log r_0 )}-\frac{81 b^2 \beta  ({\cal C}_{zz}^{(1)}-2   {\cal C}_{\theta_1z}^{(1)}+2   {\cal C}_{\theta_1x}^{(1)})}{8 \log
   ({r_0})}\right)}{{g_s} M {N_f}^2 {r_0}^3 (\log N +3 \log r_0 )}\nonumber\\
& & -\frac{\pi ^2 {\cal C}_{\phi_0}^{IR} N^{2/5}
   \alpha _{\theta _1}^3 (\log N -3 \log r_0 ) \left(\frac{9 \left(3 b^2+1\right) \beta  ({\cal C}_{zz}^{(1)}-2   {\cal C}_{\theta_1z}^{(1)}+2   {\cal C}_{\theta_1x}^{(1)})}{4 \log r_0 }+\frac{1944 b^4}{\left(3 b^2+2\right)^4}\right)}{{g_s} \log r_0  M {N_f}^2 {r_0}^2
   (\log N +3 \log r_0 )} Z^2 + {\cal O}(Z^3).\nonumber\\
\end{eqnarray}
 We showed that UV valued profile functions are \cite{MChPT,Vikas-Thesis}:
\begin{eqnarray}
\label{profile-functions-UV}
& & \psi_1^{\rm UV}(Z) = {\cal C}_{\psi_1}^{\rm UV}\frac{e^{-2Z}}{Z^{\frac{3}{2}}},\nonumber\\
& & \phi_0^{\rm UV}(Z) = {\cal C}_{\phi_0}^{\rm UV}\frac{e^{-2Z}}{Z^2}.
\end{eqnarray}
Now, let us discuss the normalization conditions on $\psi_1(Z)$ and $\phi_0(Z)$ and the resultant constraints on the integration constants, ${\cal C}_{\psi_1}^{\rm UV}$ and ${\cal C}_{\phi_0}^{\rm UV}(Z)$. The $\rho$ vector meson profile function ($\psi_1(Z)$) normalization condition: $\left.{\cal V}_{\Sigma_2}\int_{0}^\infty dZ {\cal V}_2\left(\psi_1(Z)\right)^2\right|_{b = \frac{1}{\sqrt{3}} + \epsilon} = 1$ implying:
{\footnotesize
\begin{eqnarray}
\label{Cpsi1UV}
& &  {\cal C}_{\psi_1}^{\rm UV}=\frac{\sqrt{\frac{\sqrt{7} ({f_{r_0}}-1) {f_{r_0}} {g_s}^2 M {N_f}^2 {\cal V}_{\Sigma_2}
  {\cal C}_{\psi_1}^{\rm IR}\ ^2 \log N  \left(7 \beta  {{({\cal C}_{zz}^{(1)}-2 {\cal C}_{\theta_1z}^{(1)}+2 {\cal C}_{\theta_1x}^{(1)})}} {f_{r_0}}^2 \gamma ^2 {g_s}^2 M^4 \log ^2(N)+3456
   \epsilon ^2 ({f_{r_0}}+1) N^2\right)}{2\ 3^{3/4} \epsilon ^{3/2} ({f_{r_0}}+1) N^{7/5} \alpha _{\theta _1}^3}-93312 \pi
   }}{24 \sqrt{-\frac{({f_{r_0}}-1) {g_s}^2 M N^{3/5} {N_f}^2 {\cal V}_{\Sigma_2}}{\epsilon ^2 ({f_{r_0}}+1) \alpha
   _{\theta _1}^3 \log N }}},\nonumber\\
\end{eqnarray}
}
where, as in \cite{Bulk-Viscosity-McGill-IIT-Roorkee}, the IR cut-off $r_0$ is considered that it will be provided as $r_0 = N^{-\frac{f_{r_0}}{3}}$. Demand that ${\cal C}_{\psi_1}^{\rm UV} = 0$, and this implies:
{\footnotesize
\begin{eqnarray}
\label{Vol2}
& &  {\cal V}_{\Sigma_2}=\frac{186624\ 3^{3/4} \pi  \epsilon ^{3/2} ({f_{r_0}}+1) N^{7/5} \alpha _{\theta _1}^3}{\sqrt{7} ({f_{r_0}}-1)
   {f_{r_0}} {g_s}^2 {\log N}   M {N_f}^2 {\cal C}_{\psi_1}^{\rm IR}\ ^2 \left(7 \beta  {({\cal C}_{zz}^{(1)}-2 {\cal C}_{\theta_1z}^{(1)}+2 {\cal C}_{\theta_1x}^{(1)})} {f_{r_0}}^2 \gamma ^2
   {g_s}^2 {\log N}  ^2 M^4+3456 \epsilon ^2 ({f_{r_0}}+1) N^2\right)}.\nonumber\\
\end{eqnarray}
}
The normalization condition on the $\phi_0(Z)$: $\frac{{\cal V}_{\Sigma_2}}{2}\int_{0}^\infty dZ {\cal V}_1\left(\phi_0(Z)\right)^2=1$ is providing:
{\footnotesize
\begin{eqnarray}
\label{Cphi0UV}
& & {\cal C}_{\phi_0}^{\rm UV}=\frac{243 \sqrt[4]{3} \pi ^2 {\cal C}_{\phi_0}^{\rm IR} \sqrt{\epsilon }
   N^{{f_{r_0}}+\frac{2}{5}} \sqrt{({f_{r_0}}+1) (-\beta {({\cal C}_{zz}^{(1)}-2 {\cal C}_{\theta_1z}^{(1)}+2 {\cal C}_{\theta_1x}^{(1)})}+2 {f_{r_0}}+2)}}{32 ({f_{r_0}}-1)^2
   {g_s}^2 \left(\log N\right) ^2 M {N_f}^2} + {\cal O}(\epsilon^{\frac{3}{2}}).
\end{eqnarray}
}
Since $ {\cal C}_{\phi_0}^{\rm UV} \propto \sqrt{\epsilon}$ and considering $\epsilon\ll1$ (for black-hole background, $\epsilon< r_h^2\left(\log r_h\right)^{\frac{9}{2}}N^{-\frac{9}{10}}$ \cite{HD-MQGP}), hence we considered: ${\cal C}_{\phi_0}^{\rm UV}\approx0$, i.e.,  $\psi_1^{\rm UV}(Z)\approx0, \phi_0^{\rm UV}(Z)\approx0$. The calculations of coupling constants requires the value of ${\cal V}_2$ in the IR and is provided as \cite{MChPT,Vikas-Thesis}:
{\footnotesize
\begin{eqnarray}
\label{V2-IR}
& &  {\cal V}_2(Z\in{\rm IR}) = -\frac{3 \left(3 b^2-2\right)
   {g_s}^2 M N^{4/5} {N_f}^2 \log ({r_0}) (\log N -3 \log ({r_0}))}{2 \pi  \log N  \alpha _{\theta _1} \alpha _{\theta
   _2}^2}\nonumber\\
& &  \frac{3 \left(2-3 b^2\right) \beta  {g_s}^2 \log r_0  M N^{4/5} {N_f}^2 Z ({\cal C}_{zz}^{(1)}-2   {\cal C}_{\theta_1z}^{(1)}+2   {\cal C}_{\theta_1x}^{(1)})
  }{4 \pi  \alpha _{\theta _1} \alpha _{\theta _2}^2}  \left(\frac{9 b^2 {g_s}^2 M N^{4/5} {N_f}^2 \log ({r_0}) (\log N -3 \log ({r_0}))}{\pi
   \log N  \alpha _{\theta _1} \alpha _{\theta _2}^2}\right)\nonumber\\
\end{eqnarray}
}
Hence, utilizing (\ref{LECs-RI}), (\ref{psi1(Z)}), (\ref{phi0(Z)}) and (\ref{V2-IR}) and writing $b=\frac{1}{\sqrt{3}} + \epsilon$ \cite{HD-MQGP} and considering $r_0 = N^{-\frac{f_{r_0}}{3}}$ \cite{Bulk-Viscosity-McGill-IIT-Roorkee}, we obtained the following simplified expressions for the couplings $y_{1,3,5,7}$:
{\footnotesize
\begin{eqnarray}
\label{y_i}
& &  \bullet\  y_1 = \int {\cal V}_2(Z)  \left(1 + \psi_1(Z) - \psi_0^2(Z)\right)^2\nonumber\\
& &  = \frac{177147 \pi ^6 \beta  {\cal C}_{\phi_0}^{\rm IR}\ ^4 \epsilon ^6 {f_{r_0}} ({f_{r_0}}+1)^3 \log r_0  \alpha _{\theta _1}^6
   N^{4 {f_{r_0}}+\frac{14}{5}} \left({\cal C}_{zz}^{(1)} - 2 {\cal C}_{\theta_1z}^{(1)} + 2 {\cal C}_{\theta_1x}^{(1)}\right)}{8192 ({f_{r_0}}-1)^6 {g_s}^4
   \left(\log N\right) ^7 M^2 N_f ^4}\nonumber\\
& &  -\frac{4782969 \sqrt{3} \pi ^7 {\cal C}_{\phi_0}^{\rm IR}\ ^4 \epsilon ^5 {f_{r_0}} ({f_{r_0}}+1)^4
   \alpha _{\theta _1}^9 N^{4 {f_{r_0}}+\frac{11}{5}}}{20480 ({f_{r_0}}-1)^7 {g_s}^6 \left(\log N\right) ^7 M^3 N_f ^6}\nonumber\\
& &  \bullet\  y_3 = \int dZ {\cal V}_2\psi_1^2(Z)\left(1 + \psi_1(Z)\right)^2\nonumber\\
& &   = -\frac{21 {g_s}^2 \log r_0  M N^{4/5} {N_f}^2 {\cal C}_{\psi_1}^{\rm IR}\ ^4 (\log N -3 \log r_0 )}{8 \pi
   \log N  \alpha _{\theta _1} \alpha _{\theta _2}^2}\nonumber\\
   & &  + \frac{3^{9/8} 7^{3/4} \sqrt[4]{\epsilon } {g_s}^2 \log r_0  M N^{4/5} {N_f}^2 {\cal C}_{\psi_1}^{\rm IR}\ ^3
   (\log N -3 \log r_0 )}{2 \sqrt{2} \pi  \log N  \alpha _{\theta _1} \alpha _{\theta _2}^2}\nonumber\\
& &  + 945 \sqrt[3]{3} \sqrt[6]{\frac{\pi }{2}} \beta  \epsilon  {g_s}^{3/2} \left(\frac{1}{\log N }\right)^{2/3}
   \log r_0  M N^{3/10} {N_f}^{4/3} {r_0}^2 {\cal C}_{\psi_1}^{\rm IR}\ ^4 ({\cal C}_{zz}^{(1)}-2 {\cal C}_{\theta_1z}^{(1)}+2{\cal C}_{\theta_1x}^{(1)})\nonumber\\
   & &  \bullet\  y_5 = \int dZ {\cal V}_2\psi_0^2(Z)\left(\psi_1(Z)\right)^2) =\frac{6561 \sqrt[4]{3} \sqrt{7} \pi ^3 {\cal C}_{\phi_0}^{\rm IR}\ ^2 \epsilon ^{5/2} {f_{r_0}} ({f_{r_0}}+1)^3 \alpha _{\theta _1}^5 {\cal C}_{\psi_1}^{\rm IR}\ ^2 N^{2
   {f_{r_0}}+\frac{8}{5}}}{256 ({f_{r_0}}-1)^4 {g_s}^2 \left(\log N\right) ^3 M N_f ^2 \alpha _{\theta _2}^2}
\nonumber\\
& &  -\frac{63 \sqrt[4]{3} \sqrt{7} \pi ^3 \beta  {\cal C}_{\phi_0}^{\rm IR}\ ^2 \epsilon ^{3/2} {f_{r_0}} ({f_{r_0}}+1) \alpha _{\theta _1}^3 {\cal C}_{\psi_1}^{\rm IR}\ ^2 N^{2
   {f_{r_0}}-\frac{3}{5}} \left({\cal C}_{zz}^{(1)} - 2 {\cal C}_{\theta_1z}^{(1)} + 2 {\cal C}_{\theta_1x}^{(1)}\right) \left(6 \epsilon  N+{f_{r_0}} \gamma  {g_s} \log N  M^2\right)^2}{32768
   ({f_{r_0}}-1)^3 {g_s}^2 \left(\log N\right) ^3 M N_f ^2}\nonumber\\
& &  \bullet\  y_7= \int dZ {\cal V}_2\psi_1(Z) (1 + \psi_1(Z)) (1 + \psi_1(Z) - \psi_0(Z)^2)^2\nonumber\\
& &   = \frac{4782969 \sqrt[4]{3} \sqrt{7} \pi ^7 {\cal C}_{\phi_0}^{\rm IR}\ ^4 \epsilon ^{9/2} {f_{r_0}} ({f_{r_0}}+1)^4 \alpha _{\theta _1}^9
   {\cal C}_{\psi_1}^{\rm IR}\ ^2 N^{4 {f_{r_0}}+\frac{11}{5}}}{40960 ({f_{r_0}}-1)^7 {g_s}^6 \left(\log N\right) ^7 M^3 N_f ^6}\nonumber\\
& & -\frac{3720087
   \sqrt[4]{3} \sqrt{7} \pi ^7 \beta  {\cal C}_{\phi_0}^{\rm IR}\ ^4 \epsilon ^{9/2} {f_{r_0}} ({f_{r_0}}+1)^3 \alpha _{\theta _1}^9
   {\cal C}_{\psi_1}^{\rm IR}\ ^2 N^{4 {f_{r_0}}+\frac{11}{5}} \left({\cal C}_{zz}^{(1)} - 2 {\cal C}_{\theta_1z}^{(1)} + 2 {\cal C}_{\theta_1x}^{(1)}\right)}{262144 ({f_{r_0}}-1)^7
   {g_s}^6 \left(\log N\right) ^7 M^3 N_f ^6}.
\end{eqnarray}
}
Similarly, we obtained the simplified expressions for the couplings $z_{1,...,8}$ as given below:
{\footnotesize
\begin{eqnarray}
\label{z_i}
& & \bullet\  z_1 = \int dZ {\cal V}_2 \left(1 + \psi_1(Z)\right)^2= \frac{3 \sqrt[4]{3} \sqrt{7} \sqrt{\epsilon } {g_s}^2 M N^{4/5} {N_f}^2 {\cal C}_{\psi_1}^{\rm IR}\ ^2 \log ({r_0})
   (\log N -3 \log ({r_0}))}{16 \pi  \alpha _{\theta _1} \alpha _{\theta _2}^2 \log N } \nonumber\\
& & \hskip 0.4in + \frac{7 \sqrt{7} \beta  \sqrt{\epsilon } {g_s}^2 \log r_0  M N^{4/5} {N_f}^2 {\cal C}_{\psi_1}^{\rm IR}\ ^2
   ({\cal C}_{zz}^{(1)}-2 {\cal C}_{\theta_1z}^{(1)}+2 {\cal C}_{\theta_1x}^{(1)}) }{2048\ 3^{3/4} \pi  \alpha _{\theta _1} \alpha _{\theta _2}^2 (\log N -3 \log
   ({r_0}))} \nonumber\\
   & & \hskip 0.4in \times \left(12 \log N +\left(-36+6 \log ^2(3)+\log (9) \log (27)-\log (27)
   \log (81)\right) \log ({r_0})\right)
   \nonumber\\
& & \bullet\  z_2 = \int dZ {\cal V}_2\psi_0^2(Z)  =\frac{81 \sqrt{3} \pi ^3 \beta  {\cal C}_{\phi_0}^{\rm IR}\ ^2 \epsilon ^3 ({f_{r_0}}+1)^2 \alpha _{\theta _1}^3 N^{\frac{5 {f_{r_0}}}{3}+\frac{7}{5}}
   \left({\cal C}_{zz}^{(1)} - 2 {\cal C}_{\theta_1z}^{(1)} + 2 {\cal C}_{\theta_1x}^{(1)}\right)}{128 ({f_{r_0}}-1)^2 {g_s} \left(\log N\right) ^2 M N_f ^2}\nonumber\\
& & \hskip 0.4in -\frac{81 \sqrt{3} \pi ^3
   {\cal C}_{\phi_0}^{\rm IR}\ ^2 \epsilon ^3 {f_{r_0}} ({f_{r_0}}+1)^2 \alpha _{\theta _1}^3 N^{2 {f_{r_0}}+\frac{7}{5}}}{256 ({f_{r_0}}-1)^3
   {g_s}^2 \left(\log N\right) ^3 M N_f ^2}\nonumber\\
& & \bullet\  z_3 = \int dZ {\cal V}_2\psi_1(1 + \psi_1)(Z) = \frac{3 \sqrt[4]{3} \sqrt{7} \sqrt{\epsilon } {g_s}^2 M N^{4/5} {N_f}^2 {\cal C}_{\psi_1}^{\rm IR}\ ^2 \log ({r_0})
   (\log N -3 \log ({r_0}))}{8 \pi  \alpha _{\theta _1} \alpha _{\theta _2}^2 \log N }\nonumber\\
& &\hskip 0.4in + \frac{7 \sqrt[4]{3} \sqrt{7} \beta  \sqrt{\epsilon } {g_s}^2 M N^{4/5} {N_f}^2 {\cal C}_{\psi_1}^{\rm IR}\ ^2 \log
   ({r_0}) ({\cal C}_{zz}^{(1)}-2 {\cal C}_{\theta_1z}^{(1)}+2 {\cal C}_{\theta_1x}^{(1)})}{512 \pi  \alpha _{\theta _1} \alpha _{\theta _2}^2}\nonumber\\
& & \bullet\  z_4 = \int dZ {\cal V}_2\psi_1(1 + \psi_1 - \phi_0^2)(Z)  = \frac{189\ 3^{3/8} \sqrt[4]{7} \pi ^3 \beta  {\cal C}_{\phi_0}^{\rm IR}\ ^2 \epsilon ^{3/4} {f_{r_0}} ({f_{r_0}}+1) \alpha _{\theta _1}^3
   {\cal C}_{\psi_1}^{\rm IR}\  N^{2 {f_{r_0}}+\frac{7}{5}} \left({\cal C}_{zz}^{(1)} - 2 {\cal C}_{\theta_1z}^{(1)} + 2 {\cal C}_{\theta_1x}^{(1)}\right)}{16384 \sqrt{2} ({f_{r_0}}-1)^3
   {g_s}^2 \left(\log N\right) ^3 M N_f ^2}\nonumber\\
& &\hskip 0.4in+\frac{81\ 3^{3/8} \sqrt[4]{7} \pi ^3 {\cal C}_{\phi_0}^{\rm IR}\ ^2 \epsilon ^{11/4} {f_{r_0}} ({f_{r_0}}+1)^2
   \alpha _{\theta _1}^3 {\cal C}_{\psi_1}^{\rm IR}\  N^{2 {f_{r_0}}+\frac{7}{5}}}{256 \sqrt{2} ({f_{r_0}}-1)^3 {g_s}^2 \left(\log N\right) ^3 M
   N_f ^2}\nonumber\\
& & \bullet\  z_5 = \int dZ {\cal V}_2\psi_1^2(1 + \psi_1)(Z)
  = -\frac{3 \sqrt[8]{3} 7^{3/4} \sqrt[4]{\epsilon } {g_s}^2 M N^{4/5} {N_f}^2 {\cal C}_{\psi_1}^{\rm IR}\ ^3 \log ({r_0}) (\log N -3 \log
   ({r_0}))}{4 \sqrt{2} \pi  \alpha _{\theta _1} \alpha _{\theta _2}^2 \log N }
\nonumber\\
& & \hskip 0.4in -\frac{21 \sqrt[8]{3} 7^{3/4} \beta  \sqrt[4]{\epsilon } {g_s}^2 M N^{4/5} {N_f}^2 {\cal C}_{\psi_1}^{\rm IR}\ ^3 \log ({r_0})
   ({\cal C}_{zz}^{(1)}-2 {\cal C}_{\theta_1z}^{(1)}+2 {\cal C}_{\theta_1x}^{(1)})}{256 \sqrt{2} \pi  \alpha _{\theta _1} \alpha _{\theta _2}^2}\nonumber\\
& & \bullet\  z_6 = \int dZ {\cal V}_2(1 + \psi_1)(1 + \psi_1 - \psi_0^2)(Z)  = \frac{567\ 3^{3/8} \sqrt[4]{7} \pi ^3 {\cal C}_{\phi_0}^{\rm IR}\ ^2 \epsilon ^{11/4} {f_{r_0}} ({f_{r_0}}+1) \alpha _{\theta _1}^3 {\cal C}_{\psi_1}^{\rm IR}\
   N^{2 {f_{r_0}}+\frac{7}{5}} \left({\cal C}_{zz}^{(1)} - 2 {\cal C}_{\theta_1z}^{(1)} + 2 {\cal C}_{\theta_1x}^{(1)}\right)}{8192 \sqrt{2} ({f_{r_0}}-1)^3 {g_s}^2 \left(\log N\right) ^3 M
   N_f ^2}\nonumber\\
& &\hskip 0.4in +\frac{81\ 3^{3/8} \sqrt[4]{7} \pi ^3 {\cal C}_{\phi_0}^{\rm IR}\ ^2 \epsilon ^{11/4} {f_{r_0}} ({f_{r_0}}+1)^2 \alpha _{\theta _1}^3
   {\cal C}_{\psi_1}^{\rm IR}\  N^{2 {f_{r_0}}+\frac{7}{5}}}{64 \sqrt{2} ({f_{r_0}}-1)^3 {g_s}^2 \left(\log N\right) ^3 M N_f ^2}\nonumber\\
& &\hskip 0.4in = \frac{567\ 3^{3/8} \sqrt[4]{7} \pi ^3 {\cal C}_{\phi_0}^{\rm IR}\ ^2 \epsilon ^{11/4} {f_{r_0}} ({f_{r_0}}+1) \alpha _{\theta _1}^3 {\cal C}_{\psi_1}^{\rm IR}\
   N^{2 {f_{r_0}}+\frac{7}{5}} \left({\cal C}_{zz}^{(1)} - 2 {\cal C}_{\theta_1z}^{(1)} + 2 {\cal C}_{\theta_1x}^{(1)}\right)}{8192 \sqrt{2} ({f_{r_0}}-1)^3 {g_s}^2 \left(\log N\right) ^3 M
   N_f ^2}\nonumber\\
& & \hskip 0.4in+\frac{81\ 3^{3/8} \sqrt[4]{7} \pi ^3 {\cal C}_{\phi_0}^{\rm IR}\ ^2 \epsilon ^{11/4} {f_{r_0}} ({f_{r_0}}+1)^2 \alpha _{\theta _1}^3
   {\cal C}_{\psi_1}^{\rm IR}\  N^{2 {f_{r_0}}+\frac{7}{5}}}{64 \sqrt{2} ({f_{r_0}}-1)^3 {g_s}^2 \left(\log N\right) ^3 M N_f ^2}\nonumber\\
& &  \bullet\  z_7 = \int dZ {\cal V}_2\psi_1(1 + \psi_1)^2(Z)  =
\frac{3 \sqrt[8]{3} 7^{3/4} \sqrt[4]{\epsilon } g_s ^2 M N^{4/5} {N_f}^2 {\cal C}_{\psi_1}^{\rm IR}\ ^3 \log (r_0 )
   (\log N -3 \log (r_0 ))}{4 \sqrt{2} \pi  \alpha _{\theta _1} \alpha _{\theta _2}^2 \log N }\nonumber\\
& & -\frac{21 \sqrt[8]{3} 7^{3/4} \beta  \sqrt[4]{\epsilon } g_s ^2 M N^{4/5} {N_f}^2 {\cal C}_{\psi_1}^{\rm IR}\ ^3 \log
   (r_0 ) ( {\cal C}_{zz}^{(1)} - 2 {\cal C}_{\theta_1z}^{(1)} + 2 {\cal C}_{\theta_1x})}{256 \sqrt{2} \pi  \alpha _{\theta _1} \alpha _{\theta
   _2}^2} +\frac{48\ 3^{3/4} \sqrt{\frac{1}{\epsilon }} g_s ^2 M N^{4/5} {N_f}^2 r_0  \log N  \log ^2(r_0 )
   {\cal C}_{\psi_1}^{\rm UV}\ }{\pi  \log N  \alpha _{\theta _1} \alpha _{\theta _2}^2}\nonumber\\
& &  \bullet\  z_8 = \int dZ {\cal V}_2\psi_0^2\psi_1(Z) =  \frac{567\ 3^{3/8} \sqrt[4]{7} \pi ^3 \beta  {\cal C}_{\phi_0}^{\rm IR}\ ^2 \epsilon ^{11/4} {f_{r_0}} ({f_{r_0}}+1) \alpha _{\theta
   _1}^3 {\cal C}_{\psi_1}^{\rm IR}\  N^{2 {f_{r_0}}+\frac{7}{5}} \left({\cal C}_{zz}^{(1)} - 2 {\cal C}_{\theta_1z}^{(1)} + 2 {\cal C}_{\theta_1x}^{(1)}\right)}{8192
   \sqrt{2} ({f_{r_0}}-1)^3 {g_s}^2 \left(\log N\right) ^3 M N_f ^2}\nonumber\\
& & \hskip 0.4in+\frac{81\ 3^{3/8} \sqrt[4]{7} \pi ^3
   {\cal C}_{\phi_0}^{\rm IR}\ ^2 \epsilon ^{11/4} {f_{r_0}} ({f_{r_0}}+1)^2 \alpha _{\theta _1}^3 {\cal C}_{\psi_1}^{\rm IR}\  N^{2
   {f_{r_0}}+\frac{7}{5}}}{64 \sqrt{2} ({f_{r_0}}-1)^3 {g_s}^2 \left(\log N\right) ^3 M N_f ^2}.
\end{eqnarray}
}
Further, the coupling constants appearing in $SU(3)$ chiral pertubration theory Lagrangian at ${\cal O}(p^2)$ (\ref{Lagrangian-Op2}) have the following simplified results \cite{MChPT,Vikas-Thesis}:
%%%%%%%%%%
{\footnotesize
\begin{eqnarray}
\label{Fpisq}
& &  F_\pi^2 = {\cal V}_{\Sigma_2}\Biggl(\frac{243 \pi ^2 \beta  \left({\cal C}_{zz}^{(1)} - 2 {\cal C}_{\theta_1z}^{(1)} + 2 {\cal C}_{\theta_1x}^{(1)}\right) {\cal C}_{\phi_0}^{\rm IR}\ ^2 {f_{r_0}} ({f_{r_0}}+1) \log ^2(3) \alpha _{\theta _1}^3 N^{\frac{4
   {f_{r_0}}}{3}+\frac{2}{5}}}{8192 ({f_{r_0}}-1)^3 {g_s}^3 \left(\log N\right) ^3 M N_f ^2}\nonumber\\
& &  +\frac{81 \sqrt{3} \pi ^2 {\cal C}_{\phi_0}^{\rm IR}\ ^2
   \epsilon  {f_{r_0}} ({f_{r_0}}+1)^2 \log (3) (\log (243)-6) \alpha _{\theta _1}^3 N^{\frac{4 {f_{r_0}}}{3}+\frac{2}{5}}}{2048
   ({f_{r_0}}-1)^3 {g_s}^3 \left(\log N\right) ^3 M N_f ^2} -\frac{243 \pi ^2 {\cal C}_{\phi_0}^{\rm IR}\ ^2 {f_{r_0}} ({f_{r_0}}+1)^2 \log ^2(3) \alpha
   _{\theta _1}^3 N^{\frac{4 {f_{r_0}}}{3}+\frac{2}{5}}}{4096 ({f_{r_0}}-1)^3 {g_s}^3 \left(\log N\right) ^3 M N_f ^2}\Biggr)\nonumber\\
& &
\end{eqnarray}
}
and
\begin{eqnarray}
\label{gsq}
& & g_{\rm YM}^2 =\frac{{\log N}   N \left(7 {({\cal C}_{zz}^{(1)}-2 {\cal C}_{\theta_1z}^{(1)}+2 {\cal C}_{\theta_1x}^{(1)})} {f_{r_0}}^2 \gamma ^2 {g_s}^2 M^4 \log ^2(N)+3456 ({f_{r_0}}+1)
   \lambda_{\epsilon}^2\right)}{288 \lambda_{\epsilon}^2 \alpha _{\theta _1}^2 \log N  \left(\sqrt{3} \beta ^{3/2} {({\cal C}_{zz}^{(1)}-2 {\cal C}_{\theta_1z}^{(1)}+2 {\cal C}_{\theta_1x}^{(1)})}
    \lambda_{\epsilon} m_0^2-12 ({f_{r_0}}+1)  N\right)}.
\end{eqnarray}

%\appendix
\chapter{}
\label{appendix-McTEQ}
\section{${\cal O}(R^4)$ Corrections to the ${\cal {M}}$-theory metric of \cite{MQGP} in the MQGP limit near the $\psi=2n\pi, n=0, 1, 2$-branches}
\label{METRIC_HD_MQGP}
The ${\cal O}(\beta)$-corrected ${\cal M}$-theory metric of \cite{MQGP} in the MQGP limit near the $\psi=2n\pi, n=0, 1, 2$-branches  up to ${\cal O}((r-r_h)^2)$ [and up to ${\cal O}((r-r_h)^3)$ for some of the off-diagonal components along the delocalized $T^3(x,y,z)$ was worked out in \cite{HD-MQGP} and is listed below\footnote{The components which do not receive an ${\cal O}(\beta)$ corrections, are not listed in (\ref{M-theory-metric-psi=2npi-patch}).}:
{\footnotesize
\begin{eqnarray}
\label{M-theory-metric-psi=2npi-patch}
 & &   G_{tt} =  G^{\rm MQGP}_{tt}\Biggl[1 + \frac{1}{4}  \frac{4 b^8 \left(9 b^2+1\right)^3 \left(4374 b^6+1035 b^4+9 b^2-4\right) \beta  M \left(\frac{1}{N}\right)^{9/4} \Sigma_1
   \left(6 a^2+  {r_h}^2\right) \log (  {r_h})}{27 \pi  \left(18 b^4-3 b^2-1\right)^5  \log N ^2   {N_f}   {r_h}^2
   \alpha _{\theta _2}^3 \left(9 a^2+  {r_h}^2\right)} (r-  {r_h})^2\Biggr]
\nonumber\\
& & G_{x^{1,2,3}x^{1,2,3}}  =   G^{\rm MQGP}_{x^{1,2,3}x^{1,2,3}}
\Biggl[1 - \frac{1}{4} \frac{4 b^8 \left(9 b^2+1\right)^4 \left(39 b^2-4\right) M \left(\frac{1}{N}\right)^{9/4} \beta  \left(6 a^2+{r_h}^2\right) \log
   ({r_h})\Sigma_1}{9 \pi  \left(3 b^2-1\right)^5 \left(6 b^2+1\right)^4 \log N ^2 {N_f} {r_h}^2 \left(9 a^2+{r_h}^2\right) \alpha
   _{\theta _2}^3} (r - {r_h})^2\Biggr]\nonumber\\
   & & G_{rr}  =  G^{\rm MQGP}_{rr}\Biggl[1 + \Biggl(- \frac{2 \left(9 b^2+1\right)^4 b^{10} M   \left(6 a^2+{r_h}^2\right) \left((r-{r_h})^2+{r_h}^2\right)\Sigma_1}{3 \pi
   \left(-18 b^4+3 b^2+1\right)^4 \log N  N^{8/15} {N_f} \left(-27 a^4+6 a^2 {r_h}^2+{r_h}^4\right) \alpha _{\theta
   _2}^3}\nonumber\\
& & +{\cal C}_{zz}^{\rm bh}-2 {\cal C}_{\theta_1z}^{\rm bh}+2 {\cal C}_{\theta_1x}^{\rm bh}\Biggr)\beta\Biggr]\nonumber\\
 & & G_{\theta_1x}  =  G^{\rm MQGP}_{\theta_1x}\Biggl[1 + \Biggl(
- \frac{\left(9 b^2+1\right)^4 b^{10} M  \left(6 a^2+{r_h}^2\right) \left((r-{r_h})^2+{r_h}^2\right)
   \Sigma_1}{3 \pi  \left(-18 b^4+3 b^2+1\right)^4 \log N  N^{8/15} {N_f} \left(-27 a^4+6 a^2
   {r_h}^2+{r_h}^4\right) \alpha _{\theta _2}^3}+{\cal C}_{\theta_1x}^{\rm bh}
\Biggr)\beta\Biggr]\nonumber\\
& & G_{\theta_1z}  =  G^{\rm MQGP}_{\theta_1z}\Biggl[1 + \Biggl(\frac{16 \left(9 b^2+1\right)^4 b^{12}  \beta  \left(\frac{(r-{r_h})^3}{{r_h}^3}+1\right) \left(19683
   \sqrt{3} \alpha _{\theta _1}^6+3321 \sqrt{2} \alpha _{\theta _2}^2 \alpha _{\theta _1}^3-40 \sqrt{3} \alpha _{\theta _2}^4\right)}{243
   \pi ^3 \left(1-3 b^2\right)^{10} \left(6 b^2+1\right)^8 {g_s}^{9/4} \log N ^4 N^{7/6} {N_f}^3 \left(-27 a^4 {r_h}+6 a^2
   {r_h}^3+{r_h}^5\right) \alpha _{\theta _1}^7 \alpha _{\theta _2}^6}+{\cal C}_{\theta_1z}^{\rm bh}\Biggr)\Biggr]\nonumber
\end{eqnarray}   
   \begin{eqnarray}
 & &   G_{\theta_2x}  =  G^{\rm MQGP}_{\theta_2x}\Biggl[1 + \Biggl(
   \frac{16 \left(9 b^2+1\right)^4 b^{12} \left(\frac{(r-{r_h})^3}{{r_h}^3}+1\right) \left(19683 \sqrt{3}
   \alpha _{\theta _1}^6+3321 \sqrt{2} \alpha _{\theta _2}^2 \alpha _{\theta _1}^3-40 \sqrt{3} \alpha _{\theta _2}^4\right)}{243 \pi ^3 \left(1-3
   b^2\right)^{10} \left(6 b^2+1\right)^8 {g_s}^{9/4} \log N ^4 N^{7/6} {N_f}^3 \left(-27 a^4 {r_h}+6 a^2
   {r_h}^3+{r_h}^5\right) \alpha _{\theta _1}^7 \alpha _{\theta _2}^6}+{\cal C}_{\theta_2x}^{\rm bh}\Biggr)\beta\Biggr]\nonumber\\
& & G_{\theta_2y}  =  G^{\rm MQGP}_{\theta_2y}\Biggl[1 +  \frac{3 b^{10} \left(9 b^2+1\right)^4 M \beta \left(6 a^2+{r_h}^2\right) \left(1-\frac{(r-{r_h})^2}{{r_h}^2}\right) \log
   ({r_h}) \Sigma_1}{\pi  \left(3 b^2-1\right)^5 \left(6 b^2+1\right)^4 \log N ^2 N^{7/5} {N_f} \left(9 a^2+{r_h}^2\right) \alpha
   _{\theta _2}^3}\Biggr]\nonumber\\
& & G_{\theta_2z}  =  G^{\rm MQGP}_{\theta_2z}\Biggl[1 + \Biggl(\frac{3 \left(9 b^2+1\right)^4 b^{10} M  \left(6 a^2+{r_h}^2\right) \left(1-\frac{(r-{r_h})^2}{{r_h}^2}\right) \log
   ({r_h}) \left(19683 \sqrt{6} \alpha _{\theta _1}^6+6642 \alpha _{\theta _2}^2 \alpha _{\theta _1}^3-40 \sqrt{6} \alpha _{\theta
   _2}^4\right)}{\pi  \left(3 b^2-1\right)^5 \left(6 b^2+1\right)^4 {\log N}^2 N^{7/6} {N_f} \left(9 a^2+{r_h}^2\right) \alpha
   _{\theta _2}^3}\nonumber\\
& & +{\cal C}_{\theta_2 z}^{\rm bh}\Biggr)\beta\Biggr]\nonumber\\
& & G_{xy}  =  G^{\rm MQGP}_{xy}\Biggl[1 + \Biggl(\frac{3 \left(9 b^2+1\right)^4 b^{10} M  \left(6 a^2+{r_h}^2\right) \left(\frac{(r-{r_h})^2}{{r_h}^2}+1\right) \log
   ({r_h}) \alpha _{\theta _2}^3\Sigma_1}{\pi  \left(3 b^2-1\right)^5 \left(6 b^2+1\right)^4 \log N ^2 N^{21/20} {N_f} \left(9
   a^2+{r_h}^2\right) \alpha _{\theta _{2 l}}^6}+{\cal C}_{xy}^{\rm bh}\Biggr)\beta\Biggr]\nonumber\\
& & G_{xz}   =  G^{\rm MQGP}_{xz}\Biggl[1 + \frac{18 b^{10} \left(9 b^2+1\right)^4 M \beta  \left(6 a^2+{r_h}^2\right)
   \left(\frac{(r-{r_h})^2}{{r_h}^2}+1\right) \log ^3({r_h}) \Sigma_1}{\pi  \left(3b^2-1\right)^5 \left(6 b^2+1\right)^4 \log N ^4 N^{5/4} {N_f} \left(9 a^2+{r_h}^2\right) \alpha
   _{\theta _2}^3}\Biggr]\nonumber\\
& & G_{yy}  =  G^{\rm MQGP}_{yy}\Biggl[1  - \frac{3 b^{10} \left(9 b^2+1\right)^4 M \left(\frac{1}{N}\right)^{7/4} \beta  \left(6 a^2+{r_h}^2\right) \log \left(\frac{r_h}{{\cal R}_{D5/\overline{D5}}^{\rm bh}}\right)\Sigma_1
   \left(\frac{(r-{r_h})^2}{r_h^2}+1\right)}{\pi  \left(3 b^2-1\right)^5 \left(6 b^2+1\right)^4 \log N ^2 {N_f} {r_h}^2 \left(9
   a^2+{r_h}^2\right) \alpha _{\theta _2}^3}\Biggr]\nonumber\\
& &  G_{yz}  =  G^{\rm MQGP}_{yz}\Biggl[1 + \Biggl(\frac{64 \left(9 b^2+1\right)^8 b^{22} M \left(\frac{1}{N}\right)^{29/12}  \left(6 a^2+{r_h}^2\right)
   \left(\frac{(r-{r_h})^3}{{r_h}^3}+1\right) \log \left(\frac{r_h}{{\cal R}_{D5/\overline{D5}}^{\rm bh}}\right) }{27 \pi ^4 \left(3 b^2-1\right)^{15} \left(6 b^2+1\right)^{12}
   {g_s}^{9/4} \log N ^6  {N_f}^4 {r_h}^3 \left({r_h}^2-3 a^2\right) \left(9 a^2+{r_h}^2\right)^2 \alpha
   _{\theta _1}^7 \alpha _{\theta _2}^9}\nonumber\\
& & \times \left(387420489 \sqrt{2} \alpha _{\theta _1}^{12}+87156324 \sqrt{3}
   \alpha _{\theta _2}^2 \alpha _{\theta _1}^9+5778054 \sqrt{2} \alpha _{\theta _2}^4 \alpha _{\theta _1}^6-177120 \sqrt{3} \alpha _{\theta
   _2}^6 \alpha _{\theta _1}^3+1600 \sqrt{2} \alpha _{\theta _2}^8\right)+{\cal C}_{yz}^{\rm bh}\Biggr)\beta\Biggr]\nonumber\\
& & G_{zz}  =  G^{\rm MQGP}_{zz}\Biggl[1 + \Biggl({\cal C}_{zz}^{\rm bh}-\frac{b^{10} \left(9 b^2+1\right)^4 M \left({r_h}^2-\frac{(r-{r_h})^3}{{r_h}}\right) \log \left(\frac{r_h}{{\cal R}_{D5/\overline{D5}}^{\rm bh}}\right)
   \Sigma_1}{27 \pi ^{3/2} \left(3 b^2-1\right)^5 \left(6 b^2+1\right)^4 \sqrt{{g_s}} \log N ^2 N^{23/20} {N_f} \alpha
   _{\theta _2}^5}\Biggr)\beta\Biggr]\nonumber\\
& & G_{x^{10}x^{10}}  =  G^{\rm MQGP}_{x^{10}x^{10}}\Biggl[1 -\frac{27 b^{10} \left(9 b^2+1\right)^4 M \left(\frac{1}{N}\right)^{5/4} \beta  \left(6 a^2+{r_h}^2\right)
   \left(1-\frac{(r-{r_h})^2}{{r_h}^2}\right) \log ^3({r_h}) \Sigma_1}{\pi  \left(3 b^2-1\right)^5 \left(6 b^2+1\right)^4 \log N ^4
   {N_f} {r_h}^2 \left(9 a^2+{r_h}^2\right) \alpha _{\theta _2}^3}\Biggr],
\end{eqnarray}
}
where $\Sigma_1$ has the following form:
\begin{eqnarray}
\label{Sigma_1-def}
& & \hskip -0.8in\Sigma_1 \equiv 19683
   \sqrt{6} \alpha _{\theta _1}^6+6642 \alpha _{\theta _2}^2 \alpha _{\theta _1}^3-40 \sqrt{6} \alpha _{\theta _2}^4,
 %  & & \hskip -0.8in\stackrel{\rm Global}{\longrightarrow} N^{\frac{6}{5}}\left(19683
 %  \sqrt{6} \sin^6\theta_1+6642 \sin^2{\theta _2} \sin^3{\theta _1}-40 \sqrt{6} \sin^4{\theta _2}\right),
\end{eqnarray}
and the ${\cal M}$-theory metric components in the MQGP limit at ${\cal O}(\beta^0)$ \cite{VA-Glueball-decay} are represented by $G^{\rm MQGP}_{MN}$. The following replacemements affect the more explicit dependency on  $\theta_{10,20}$ of the ${\cal M}$-theory metric components up to ${\cal O}(\beta)$, using (\ref{small-theta_12}): 
$\alpha_{\theta_1}\rightarrow N^{\frac{1}{5}}\sin\theta_{10},\ \alpha_{\theta_2}\rightarrow N^{\frac{3}{10}}\sin\theta_{20}$ in (\ref{M-theory-metric-psi=2npi-patch}). 

\section{Thermal $f_{MN}$ EOMs, their Solutions in the IR, $4D$-Limit and  ${\cal {M}}\chi$PT Compatibility}
In the following appendix, we're going to discuss the independent EOMs associated with the metric perturbations $f_{MN}$ of (\ref{TypeIIA-from-M-theory-Witten-prescription-T<Tc}) near the IR cut-off $r_0$ up to leading order in $N$, the solutions they provide and constraints as well as values for the same in the decompactification-limit of a spatial direction (that has an important role for showing proof of an all-loop non-renormalization of $T_c$ at ${\cal O}(R^4)$). Through such a manner, we are capable to calculate the values of the metric perturbations (in the deep IR) across the three-cycle $S^3(\theta_1,x,z)$ - the delocalized version of $S^3(\theta_1,\phi_1,\psi)$ -strictly speaking across the fiber $S^1(z)$ (the $S^3(\theta_1,x,z)$ is a $S^1(z)$ fibration across the vanishing two-cycle $S^2(\theta_1,x)$) as well as another two-cycle $S^2(\theta_1,z)$ (which has also a $S^1(z)$-fibration). A unique linear combination of contributions from $\left.f_{MN}\right|_{S^3(\theta_1,x,z)}$, close to the Ouyang embedding within the parent type IIB dual, shows up frequently in $T_c$ computations and the LECs of $SU(3)\chi$PT Lagrangian at ${\cal O}(p^4)$ from ${\cal M}\chi$PT in chapter {\bf 2} (from\cite{MChPT}). In chapter {\bf 2}, we found that the aforementioned combination of integration constant has a negative sign (\ref{CCsO4}). Here we will derive this constraint.\par
The EOMs of the metric perturbations $f_{MN}(r)$ are worked out as follows:
{\footnotesize
\begin{eqnarray}
\label{EOM-fMN-thermal}
& & {\rm EOM}_{tt}:\nonumber\\
& &  -\frac{\beta  \left(\frac{1}{N}\right)^{9/4} \left(19683 \sqrt{6} \alpha _{\theta _1}^6+6642 \alpha _{\theta _2}^2
   \alpha _{\theta _1}^3-40 \sqrt{6} \alpha _{\theta _2}^4\right) \log ({r_0}) }{156728328192 \pi ^3 {g_s}^5 \log N ^4 M {N_f}^3
   \epsilon ^{10} \alpha _{\theta _2}^3}\nonumber\\
& & \times \left(-\frac{98 \pi ^3 (2
   {f_{zz}}({r_0})-3 {f_{x^{10}x^{10}}}({r_0})-4 {f_{\theta_1z}}({r_0})-5 {f_{\theta_2y}}({r_0}))}{\log
   ^2({r_0})}-\frac{1728 {g_s}^3 M^2 \left(\frac{1}{N}\right)^{2/5} {N_f}^2 ({f_{yy}}({r_0})-2
   {f_{yz}}({r_0}))}{\alpha _{\theta _2}^2}\right) = 0\nonumber\\
& & \nonumber\\
& &  {\rm EOM}_{\theta_1\theta_1}:\nonumber\\
& &  \frac{81 \alpha _{\theta _1}^2 \left(\frac{1152 {g_s}^3 M^2 \left(\frac{1}{N}\right)^{2/5} {N_f}^2 \log
   ^2({r_0}) \left(\alpha _{\theta _2} {f_{zz}}({r_0})-2 \alpha _{\theta _2} {f_{yz}}({r_0})+\alpha
   _{\theta _2} {f_{yy}}({r_0})\right)}{\pi ^3}-98 \alpha _{\theta _2}^3 (2
   {f_{x^{10}x^{10}}}({r_0})+{f_{\theta_2y}}({r_0}))\right)}{1024 \alpha _{\theta _2}^5}\nonumber\\
& & +\frac{\beta  M
   \left(\frac{1}{N}\right)^{19/10} \left(-19683 \sqrt{6} \alpha _{\theta _1}^6-6642 \alpha _{\theta _2}^2 \alpha
   _{\theta _1}^3+40 \sqrt{6} \alpha _{\theta _2}^4\right)}{4782969 \pi ^{5/4} \sqrt[4]{{g_s}} \log N ^2
   {N_f} {r_0}^2 \epsilon ^7 \alpha _{\theta _2}^5}=0\nonumber\\
& & \nonumber\\
& & {\rm EOM}_{\theta_1\theta_2}:\nonumber\\
& &  -\frac{441 \sqrt{3} {r_0} \alpha _{\theta _1}^6 (2 {f_{x^{10}x^{10}}}({r_0})+{f_{\theta_2y}}({r_0}))}{\alpha _{\theta
   _2}^3}-\frac{64 \sqrt{2} {g_s}^{3/2} M {N_f} {fr}({r_0})}{\pi ^{3/2} \sqrt[5]{N}}=0\nonumber
\end{eqnarray}   
 \begin{eqnarray}  
& & \nonumber\\
& &  {\rm EOM}_{\theta_2\theta_2}:\nonumber\\
& & \frac{{f_{zz}}({r_0})-{f_{x^{10}x^{10}}}({r_0})-2 {f_{\theta_1z}}({r_0})-{fr}({r_0})}{9 \log N ^2
   \alpha _{\theta _1}^4}+\frac{9 \sqrt{6} {g_s}^{3/2} M {N_f} {r_0} \log ^2({r_0})
   ({f_{\theta_1y}}({r_0})-{f_{yz}}({r_0}))}{\pi ^{3/2} \log N ^2 N^{2/5} \alpha _{\theta _2}^3}=0\nonumber\\
& & \nonumber\\
& &  {\rm EOM}_{\theta_2y}:\nonumber\\
& & \frac{-\frac{49 \sqrt{2} \pi ^3 \alpha _{\theta _2}^3 (7 {f_{zz}}({r_0})-15 {f_{x^{10}x^{10}}}({r_0})-14
   {f_{\theta_1z}}({r_0})+5 {f_{\theta_2y}}({r_0}))}{\log ^2({r_0})}-864 \sqrt{2} {g_s}^3 M^2
   \left(\frac{1}{N}\right)^{2/5} {N_f}^2 \left(2 \alpha _{\theta _2} {f_{yz}}({r_0})-\alpha _{\theta _2}
   {f_{yy}}({r_0})\right)}{10368 \sqrt[4]{\pi } {g_s}^{13/4} \log N ^2 M^2 {N_f}^2 \epsilon ^2 \alpha
   _{\theta _1}^2 \alpha _{\theta _2}^2}\nonumber\\
& & +\frac{32 \beta  \left(\frac{1}{N}\right)^{3/2} \left(19683 \sqrt{6} \alpha
   _{\theta _1}^6+6642 \alpha _{\theta _2}^2 \alpha _{\theta _1}^3-40 \sqrt{6} \alpha _{\theta _2}^4\right)}{3486784401
   {g_s}^2 \log N ^2 {N_f}^2 \epsilon ^8 \alpha _{\theta _1}^6 \alpha _{\theta _2}}=0\nonumber\\
& &  {\rm EOM}_{xx}: \nonumber\\
& &\frac{-\frac{1024 \pi ^{3/2} {g_s}^3 \alpha _{\theta _2}^3 ({f_{zz}}({r_0})-2 {f_{\theta_1z}}({r_0})+2
   {f68}({r_0})-{fr}({r_0}))}{\alpha _{\theta _1}^4}-\frac{11907 \pi ^{9/2}
   \left(\frac{1}{N}\right)^{2/5} \alpha _{\theta _2}^5 (3 {f_{x^{10}x^{10}}}({r_0})+5
   {f_{\theta_2y}}({r_0}))}{\log N ^2 M^2 {N_f}^2 \log ^2({r_0})}}{279936 \pi ^2 {g_s}^{7/2} \epsilon ^2
   \alpha _{\theta _2}^5}\nonumber\\
& & +\frac{64 \beta  \left(\frac{1}{N}\right)^{23/20}}{43046721 \sqrt[4]{\pi } {g_s}^{9/4}
   \log N ^3 {N_f}^2 \epsilon ^7 \alpha _{\theta _1}^4 \alpha _{\theta _2}^4}=0\nonumber\\
& &  {\rm EOM}_{yy}:\nonumber\\
& & \frac{49 \pi ^{5/2} \left(16 {f_{zz}}({r_0})+3 {f_{x^{10}x^{10}}}({r_0})+4 {f_{\theta_1z}}({r_0})+\frac{18 \alpha
   _{\theta _2}^2 {f_{\theta_1y}}({r_0})}{\sqrt[5]{N} \alpha _{\theta _1}^2}+5 {f_{\theta_2y}}({r_0})-36
   {f_{yz}}({r_0})+18 {f_{yy}}({r_0})\right)}{1152 {g_s}^{7/2} \log N ^2 M^2 {N_f}^2 \epsilon
   ^2 \log ^2({r_0})}\nonumber\\
& & +\frac{2 \beta  \left(19683 \alpha _{\theta _1}^6+1107 \sqrt{6} \alpha _{\theta _2}^2 \alpha
   _{\theta _1}^3-40 \alpha _{\theta _2}^4\right)}{387420489 \sqrt[4]{\pi } {g_s}^{9/4} \log N ^3 N^{3/2}
   {N_f}^2 \epsilon ^9 \alpha _{\theta _1}^4 \alpha _{\theta _2}^2}=0.\nonumber\\   
\end{eqnarray}
}
EOMs (\ref{EOM-fMN-thermal}) have the following solutions:
\begin{eqnarray}
\label{solutions-fMN}
& & f_t(r) = f_t(r_0),\nonumber\\
& & f(r) = f(r_0),\nonumber\\
& & f_r(r) = -\frac{99 \sqrt{\frac{3}{2}} \beta  {g_s}^{3/2} M \sqrt[5]{\frac{1}{N}} {N_f} {r_0} \alpha _{\theta _1}^6
   {f_{x^{10}x^{10}}}({r_0}) \log ^2({r_0})}{2 \pi ^{3/2} \alpha _{\theta _2}^5},\nonumber\\
& & f_{\theta_1\theta_1}(r) = f_{\theta_1\theta_1}(r_0),\nonumber\\
& & f_{\theta_1\theta_2}(r) = f_{\theta_1\theta_2}(r_0),\nonumber\\
& & f_{\theta_1x}(r) = -\frac{99 \sqrt{\frac{3}{2}} {g_s}^{3/2} M \sqrt[5]{\frac{1}{N}} {N_f} {r_0} \alpha _{\theta _1}^6
   {f_{x^{10}x^{10}}}({r_0}) \log ^2({r_0})}{4 \pi ^{3/2} \alpha _{\theta _2}^5}-{f_{x^{10}x^{10}}}({r_0}),\nonumber\\
& & f_{\theta_1y}(r) =  f_{\theta_1y}(r_0),\nonumber\\
& &  f_{\theta_1z} = \frac{539 \pi ^3 N^{2/5} \alpha _{\theta _2}^2 {f_{x^{10}x^{10}}}({r_0})}{1728 {g_s}^3 M^2 {N_f}^2 \log
   ^2({r_0})}-\frac{185 {f_{x^{10}x^{10}}}({r_0})}{108},\nonumber
\end{eqnarray}
\begin{eqnarray}
& & f_{\theta_2\theta_2}(r) = f_{\theta_2\theta_2}(r_0),\nonumber\\
& & f_{\theta_2x}(r) = f_{\theta_2x}(r_0),\nonumber\\
& & f_{\theta_2y}(r) = \frac{352 {g_s}^3 M^2 \left(\frac{1}{N}\right)^{2/5} {N_f}^2 {f_{x^{10}x^{10}}}({r_0}) \log ^2({r_0})}{49 \pi
   ^3 \alpha _{\theta _2}^2}-2 {f_{x^{10}x^{10}}}({r_0}),\nonumber\\
& & f_{\theta_2z}(r) = f_{\theta_2z}(r_0),\nonumber\\
& & f_{xx}(r) = f_{xx}(r_0),\nonumber\\
& & f_{xy}(r) = f_{xy}(r_0),\nonumber\\
& & f_{xz}(r) = f_{xz}(r_0),\nonumber\\
& & f_{yy}(r) = \frac{N^{2/5} {f_{x^{10}x^{10}}}({r_0}) \left(32 \sqrt{6} \pi ^{3/2} {g_s}^{3/2} M {N_f} \alpha _{\theta
   _2}^3-4851 \pi ^3 {r_0} \alpha _{\theta _1}^4 \alpha _{\theta _2}^2\right)}{7776 {g_s}^3 M^2 {N_f}^2
   {r_0} \alpha _{\theta _1}^4 \log ^2({r_0})}+\frac{55 {f_{x^{10}x^{10}}}({r_0})}{27} \nonumber\\
   & & \hskip 0.6in +{f_{\theta_1y}}({r_0}),\nonumber\\
& &  f_{yz}(r) = \frac{\pi ^{3/2} N^{2/5} \alpha _{\theta _2}^3 {f_{x^{10}x^{10}}}({r_0})}{81 \sqrt{6} {g_s}^{3/2} M {N_f}
   {r_0} \alpha _{\theta _1}^4 \log ^2({r_0})}+\frac{{f_{\theta_1y}}({r_0})}{2},\nonumber\\
& & f_{zz}(r) = \frac{539 \pi ^3 N^{2/5} \alpha _{\theta _2}^2 {f_{x^{10}x^{10}}}({r_0})}{864 {g_s}^3 M^2 {N_f}^2 \log
   ^2({r_0})}-\frac{77 {f_{x^{10}x^{10}}}({r_0})}{54}(r_0),\nonumber\\
& & f_{x^{10}x^{10}}(r) = f_{x^{10}x^{10}}(r_0).
\end{eqnarray}
We are going to show the compatibility between the solutions that correspond to the EOMs for ${\cal O}(R^4)$ metric perturbations provided in (\ref{solutions-fMN}) and those derived by taking the $\tilde{g}(r)\rightarrow1$-limit of the ${\cal O}(R^4)$ corrections to (\ref{TypeIIA-from-M-theory-Witten-prescription-T<Tc}) given in  \cite{MChPT} and listed in \ref{M-theory-metric-psi=2npi-patch} (for thermal background $r_h \rightarrow r_0$). In the decompactification limit, i.e., $M_{\rm KK}\rightarrow0$, The compatibility of the abovementioned set of solutions necessitate:
\begin{eqnarray}
& & f_t(r) = f_t(r_0) = \left.- \frac{1}{4} \frac{4 b^8 \left(9 b^2+1\right)^4 \left(39 b^2-4\right) M \left(\frac{1}{N}\right)^{9/4} \beta  \left(6 a^2+{r_0}^2\right) \log
   ({r_0})\Sigma_1}{9 \pi  \left(3 b^2-1\right)^5 \left(6 b^2+1\right)^4 \log N ^2 {N_f} {r_0}^2 \left(9 a^2+{r_0}^2\right) \alpha
   _{\theta _2}^3} (r - {r_0})^2\right|_{r=r_0} \nonumber\\
& & = f(r) = f(r_0) = f_{x^3x^3}\nonumber\\
& &  = \left. \frac{1}{4}  \frac{4 b^8 \left(9 b^2+1\right)^3 \left(4374 b^6+1035 b^4+9 b^2-4\right) \beta  M \left(\frac{1}{N}\right)^{9/4} \Sigma_1
   \left(6 a^2+  {r_0}^2\right) \log (  {r_0})}{27 \pi  \left(18 b^4-3 b^2-1\right)^5  \log N ^2   {N_f}   {r_0}^2
   \alpha _{\theta _2}^3 \left(9 a^2+  {r_0}^2\right)} (r-  {r_0})^2\right|_{r=r_0},\nonumber
\end{eqnarray}
\begin{eqnarray*}
& & f_r(r) = -\frac{99 \sqrt{\frac{3}{2}} \beta  {g_s}^{3/2} M \sqrt[5]{\frac{1}{N}} {N_f} {r_0} \alpha _{\theta _1}^6
   {f_{x^{10}x^{10}}}({r_0}) \log ^2({r_0})}{2 \pi ^{3/2} \alpha _{\theta _2}^5} \nonumber\\
& &  = - \frac{2 \left(9 b^2+1\right)^4 b^{10} M   \left(6 a^2+{r_0}^2\right){r_0}^2\Sigma_1}{3 \pi
   \left(-18 b^4+3 b^2+1\right)^4 \log N  N^{8/15} {N_f} \left(-27 a^4+6 a^2 {r_0}^2+{r_0}^4\right) \alpha _{\theta
   _2}^3}\nonumber\\
& & +{\cal C}_{zz}^{\rm th}-2 {\cal C}_{\theta_1z}^{\rm th}+2 {\cal C}_{\theta_1x}^{\rm th},\nonumber\\
& & f_{\theta_1\theta_1}(r) = f_{\theta_1\theta_1}(r_0) = 0,\nonumber\\
& & f_{\theta_1\theta_2}(r) = f_{\theta_1\theta_2}(r_0) = 0,\nonumber\\
& & f_{\theta_1x}(r) = -\frac{99 \sqrt{\frac{3}{2}} {g_s}^{3/2} M \sqrt[5]{\frac{1}{N}} {N_f} {r_0} \alpha _{\theta _1}^6
   {f_{x^{10}x^{10}}}({r_0}) \log ^2({r_0})}{4 \pi ^{3/2} \alpha _{\theta _2}^5}-{f_{x^{10}x^{10}}}({r_0}) \nonumber\\
& &  = - \frac{\left(9 b^2+1\right)^4 b^{10} M  \left(6 a^2+{r_0}^2\right){r_0}^2
   \Sigma_1}{3 \pi  \left(-18 b^4+3 b^2+1\right)^4 \log N  N^{8/15} {N_f} \left(-27 a^4+6 a^2
   {r_0}^2+{r_0}^4\right) \alpha _{\theta _2}^3}+{\cal C}_{\theta_1x}^{\rm th},\nonumber\\
& & f_{\theta_1y}(r) =  f_{\theta_1y}(r_0) = 0,\nonumber\\
& &  f_{\theta_1z} = \frac{539 \pi ^3 N^{2/5} \alpha _{\theta _2}^2 {f_{x^{10}x^{10}}}({r_0})}{1728 {g_s}^3 M^2 {N_f}^2 \log
   ^2({r_0})}-\frac{185 {f_{x^{10}x^{10}}}({r_0})}{108} \nonumber\\
& &  = \frac{16 \left(9 b^2+1\right)^4 b^{12}  \beta  \left(19683
   \sqrt{3} \alpha _{\theta _1}^6+3321 \sqrt{2} \alpha _{\theta _2}^2 \alpha _{\theta _1}^3-40 \sqrt{3} \alpha _{\theta _2}^4\right)}{243
   \pi ^3 \left(1-3 b^2\right)^{10} \left(6 b^2+1\right)^8 {g_s}^{9/4} \log N ^4 N^{7/6} {N_f}^3 \left(-27 a^4 {r_0}+6 a^2
   {r_0}^3+{r_0}^5\right) \alpha _{\theta _1}^7 \alpha _{\theta _2}^6}+{\cal C}_{\theta_1z}^{\rm th},\nonumber\\
& & f_{\theta_2\theta_2}(r) = f_{\theta_2\theta_2}(r_0) = 0,\nonumber\\
& & f_{\theta_2x}(r) = f_{\theta_2x}(r_0) \nonumber\\
& & =  \frac{16 \left(9 b^2+1\right)^4 b^{12}\left(19683 \sqrt{3}
   \alpha _{\theta _1}^6+3321 \sqrt{2} \alpha _{\theta _2}^2 \alpha _{\theta _1}^3-40 \sqrt{3} \alpha _{\theta _2}^4\right)}{243 \pi ^3 \left(1-3
   b^2\right)^{10} \left(6 b^2+1\right)^8 {g_s}^{9/4} \log N ^4 N^{7/6} {N_f}^3 \left(-27 a^4 {r_0}+6 a^2
   {r_0}^3+{r_0}^5\right) \alpha _{\theta _1}^7 \alpha _{\theta _2}^6}+{\cal C}_{\theta_2x}^{\rm th},\nonumber\\
   & & f_{\theta_2y}(r) = \frac{352 {g_s}^3 M^2 \left(\frac{1}{N}\right)^{2/5} {N_f}^2 {f_{x^{10}x^{10}}}({r_0}) \log ^2({r_0})}{49 \pi
   ^3 \alpha _{\theta _2}^2}-2 {f_{x^{10}x^{10}}}({r_0}) \nonumber\\
& & = \frac{3 b^{10} \left(9 b^2+1\right)^4 M \beta \left(6 a^2+{r_0}^2\right)  \log
   ({r_0}) \Sigma_1}{\pi  \left(3 b^2-1\right)^5 \left(6 b^2+1\right)^4 \log N ^2 N^{7/5} {N_f} \left(9 a^2+{r_0}^2\right) \alpha
   _{\theta _2}^3} ,\nonumber\\
& & f_{\theta_2z}(r) = f_{\theta_2z}(r_0) = \frac{3 \left(9 b^2+1\right)^4 b^{10} M  \left(6 a^2+{r_0}^2\right)  \log
   ({r_0})}{\pi  \left(3 b^2-1\right)^5 \left(6 b^2+1\right)^4 {\log N}^2 N^{7/6} {N_f} \left(9 a^2+{r_0}^2\right) \alpha
   _{\theta _2}^3}\nonumber\\
& & \hskip 1.5in \times \left(19683 \sqrt{6} \alpha _{\theta _1}^6+6642 \alpha _{\theta _2}^2 \alpha _{\theta _1}^3-40 \sqrt{6} \alpha _{\theta
   _2}^4\right) +{\cal C}_{\theta_2 z}^{\rm th},\nonumber\\
& & f_{xx}(r) = f_{xx}(r_0) = 0,\nonumber\\
& & f_{xy}(r) = f_{xy}(r_0) = \frac{3 \left(9 b^2+1\right)^4 b^{10} M  \left(6 a^2+{r_0}^2\right)  \log
   ({r_0}) \alpha _{\theta _2}^3\Sigma_1}{\pi  \left(3 b^2-1\right)^5 \left(6 b^2+1\right)^4 \log N ^2 N^{21/20} {N_f} \left(9
   a^2+{r_0}^2\right) \alpha _{\theta _{2 l}}^6}+{\cal C}_{xy}^{\rm th},\nonumber\\
   & & f_{xz}(r) = f_{xz}(r_0) = \frac{18 b^{10} \left(9 b^2+1\right)^4 M \beta  \left(6 a^2+{r_0}^2\right)
   \log ^3({r_0}) \Sigma_1}{\pi  \left(3b^2-1\right)^5 \left(6 b^2+1\right)^4 \log N ^4 N^{5/4} {N_f} \left(9 a^2+{r_0}^2\right) \alpha
   _{\theta _2}^3},\nonumber
\end{eqnarray*}
\begin{eqnarray}
\label{consistency-f_MN's}
& & f_{yy}(r) = \frac{N^{2/5} {f_{x^{10}x^{10}}}({r_0}) \left(32 \sqrt{6} \pi ^{3/2} {g_s}^{3/2} M {N_f} \alpha _{\theta
   _2}^3-4851 \pi ^3 {r_0} \alpha _{\theta _1}^4 \alpha _{\theta _2}^2\right)}{7776 {g_s}^3 M^2 {N_f}^2
   {r_0} \alpha _{\theta _1}^4 \log ^2({r_0})}+\frac{55 {f_{x^{10}x^{10}}}({r_0})}{27}+{f_{\theta_1y}}({r_0})\nonumber\\
& &  =  - \frac{3 b^{10} \left(9 b^2+1\right)^4 M \left(\frac{1}{N}\right)^{7/4} \beta  \left(6 a^2+{r_0}^2\right) \log ({r_0})\Sigma_1
   }{\pi  \left(3 b^2-1\right)^5 \left(6 b^2+1\right)^4 \log N ^2 {N_f} {r_0}^2 \left(9
   a^2+{r_0}^2\right) \alpha _{\theta _2}^3},\nonumber\\
& &  f_{yz}(r) = \frac{\pi ^{3/2} N^{2/5} \alpha _{\theta _2}^3 {f_{x^{10}x^{10}}}({r_0})}{81 \sqrt{6} {g_s}^{3/2} M {N_f}
   {r_0} \alpha _{\theta _1}^4 \log ^2({r_0})}+\frac{{f_{\theta_1y}}({r_0})}{2}\nonumber\\
& &  = \frac{64 \left(9 b^2+1\right)^8 b^{22} M \left(\frac{1}{N}\right)^{29/12}  \left(6 a^2+{r_0}^2\right)
   \log ({r_0}) }{27 \pi ^4 \left(3 b^2-1\right)^{15} \left(6 b^2+1\right)^{12}
   {g_s}^{9/4} \log N ^6  {N_f}^4 {r_0}^3 \left({r_0}^2-3 a^2\right) \left(9 a^2+{r_0}^2\right)^2 \alpha
   _{\theta _1}^7 \alpha _{\theta _2}^9}\nonumber\\
& & \hskip -0.3in \times \left(387420489 \sqrt{2} \alpha _{\theta _1}^{12}+87156324 \sqrt{3}
   \alpha _{\theta _2}^2 \alpha _{\theta _1}^9+5778054 \sqrt{2} \alpha _{\theta _2}^4 \alpha _{\theta _1}^6-177120 \sqrt{3} \alpha _{\theta
   _2}^6 \alpha _{\theta _1}^3+1600 \sqrt{2} \alpha _{\theta _2}^8\right) \nonumber\\
   & & +{\cal C}_{yz}^{\rm th},\nonumber\\
& & f_{zz}(r) = \frac{539 \pi ^3 N^{2/5} \alpha _{\theta _2}^2 {f_{x^{10}x^{10}}}({r_0})}{864 {g_s}^3 M^2 {N_f}^2 \log
   ^2({r_0})} = -\frac{b^{10} \left(9 b^2+1\right)^4 M{r_0}^2 \log ({r_0})
   \Sigma_1}{27 \pi ^{3/2} \left(3 b^2-1\right)^5 \left(6 b^2+1\right)^4 \sqrt{{g_s}} \log N ^2 N^{23/20} {N_f} \alpha
   _{\theta _2}^5} \nonumber\\
   & & \hskip 2.5in + {\cal C}_{zz}^{\rm th},\nonumber\\
& & f_{x^{10}x^{10}}(r) = f_{x^{10}x^{10}}(r_0) \nonumber\\
& & = -\frac{27 b^{10} \left(9 b^2+1\right)^4 M \left(\frac{1}{N}\right)^{5/4} \beta  \left(6 a^2+{r_0}^2\right)
    \log ^3({r_0}) \Sigma_1}{\pi  \left(3 b^2-1\right)^5 \left(6 b^2+1\right)^4 \log N ^4
   {N_f} {r_0}^2 \left(9 a^2+{r_0}^2\right) \alpha _{\theta _2}^3}.
\end{eqnarray}
Utilizing (\ref{consistency-f_MN's}), around $r=r_0$, we obtained:
\begin{equation}
\label{f_MN's-zero}
f_t = f = f_{\theta_i\theta_j} = f_{\theta_1y} = f_{xx} = 0,
\end{equation}
and for $r_0$-dependent values of ${\cal C}_{\theta_2x}^{\rm th}, {\cal C}_{\theta_2z}^{\rm th}, {\cal C}_{xy}^{\rm th}$ determined by (\ref{consistency-f_MN's}),
\begin{equation}
\label{f_MN's-zero-ii}
f_{\theta_2x} = f_{\theta_2z} = f_{xy} = 0.
\end{equation}
This is what we utilized in our computations. When comparing around $r=r_0$ of $f_{r}(r), f_{\theta_1x}, f_{\theta_1z}, f_{zz}$, we can find a solution for ${\cal C}_{\theta_1x}^{\rm th}, {\cal C}_{\theta_1z}^{\rm th}, {\cal C}_{zz}^{\rm th}, f_{x^{10}x^{10}}(r_0)$. The expression $f_{x^{10}x^{10}}$ could be calculated separately from:
\begin{itemize}
\item
{\it matching $f_{\theta_2y}$}: Putting $b = \frac{1}{\sqrt{3}} + \epsilon$ \cite{HD-MQGP}, \cite{MChPT} implies
\begin{equation}
\label{f1111-theta2y}
 f_{x^{10}x^{10}}(r_0) \sim - \frac{M \Sigma_1(\alpha_{\theta_1}, \alpha_{\theta_2}) \log r_0}{N^{\frac{7}{5}}\log^2N N_f\epsilon^5\alpha_{\theta_2}^3}.
\end{equation}

\item
{\it matching $f_{yy}$}: We obtained:
\begin{equation}
\label{f1111-yy}
 f_{x^{10}x^{10}}(r_0) \sim -\frac{g_s^{\frac{3}{2}}M^2\log^3r_0\alpha_{\theta_1}^4\Sigma_1(\alpha_{\theta_1}, \alpha_{\theta_2})}{\epsilon^5\log^Nr_0\alpha_{\theta_2}^6N^{\frac{43}{20}}}.
\end{equation}
We have shown that (\ref{f1111-theta2y}) and (\ref{f1111-yy}) are compatible if the IR cut-off $r_0$ is used:
\begin{equation}
\label{r_0}
r_0 = W\left(\frac{1}{2 \sqrt{a}}\right)\approx a e^{-\frac{4 \log ^2(-\log (a))}{\log ^2(a)}} \log
   ^2(a),
\end{equation}
$W$ represents the Lambert's product log function and $a\equiv \frac{81 \sqrt{6} {g_s}^{3/2} M
   \left(\frac{1}{N}\right)^{3/4} {N_f} \alpha
   _{\theta _1}^4}{\pi ^{3/2} \alpha _{\theta
   _2}^3}$. 

\item
{\it matching $f_{x^{10}x^{10}}$}: We obtained:
\begin{equation}
\label{f1111-1111}
f_{x^{10}x^{10}}(r_0) \sim -\frac{M \log^3r_0\Sigma_1(\alpha_{\theta_1}, \alpha_{\theta_2})}{\epsilon^5 r_0^2\log^N N_f \alpha_{\theta_2}^3 N^{\frac{5}{4}}}.
\end{equation}
Numerically, e.g., for $M=N_f=3, N=10^2,g_s=0.1$ and ${\cal O}(1)$ $\alpha_{\theta_{1,2}}$, we showed that (\ref{f1111-1111}) is compatible with (\ref{f1111-theta2y}) and (\ref{f1111-yy}). 

\item
{\it matching $f_t(r)$}
\begin{eqnarray}
\label{f_r}
& & -\frac{2 \left(9 b^2+1\right)^4 b^{10} M {r_0}^2 \Sigma _1 \left(6 a^2+{r_0}^2\right)}{3 \pi  \left(-18 b^4+3
   b^2+1\right)^4 \log N  N^{8/15} {N_f} \alpha _{\theta _2}^3 \left(6 a^2 {r_0}^2-27
   a^4+{r_0}^4\right)}+{\cal C}_{zz}^{\rm th}-2 {\cal C}_{\theta_1z}^{\rm th}+2 {\cal C}_{\theta_1x}^{\rm th}  \nonumber \\
 & &
 =\frac{99 \sqrt{\frac{3}{2}} {g_s}^{3/2}
   M \sqrt[5]{\frac{1}{N}} {N_f} {r_0} \alpha _{\theta _1}^6 {f_{x^{10}x^{10}}}({r_0}) \log ^2({r_0})}{2 \pi
   ^{3/2} \alpha _{\theta _2}^5}.
\end{eqnarray}

\item
{\it matching $f_{\theta_1x}$}
\begin{eqnarray}
\label{f_theta1x}
& & {{\cal C}^{\rm th}_{\theta_{1} x}}-\frac{b^{10} \left(9 b^2+1\right)^4 M {r_0}^2 \Sigma _1 \left(6 a^2+{r_0}^2\right)}{3 \pi 
   \left(-18 b^4+3 b^2+1\right)^4 \log N  N^{8/15} {N_f} \alpha _{\theta _2}^3 \left(6 a^2 {r_0}^2-27
   a^4+{r_0}^4\right)} \nonumber \\
 & &
   =-\frac{99 \sqrt{\frac{3}{2}} {g_s}^{3/2} M \sqrt[5]{\frac{1}{N}} {N_f} {r_0}
   \alpha _{\theta _1}^6 {f_{x^{10}x^{10}}}({r_0}) \log ^2({r_0})}{4 \pi ^{3/2} \alpha _{\theta
   _2}^5}-{f_{x^{10}x^{10}}}({r_0}).
\end{eqnarray}

\item
{\it matching $f_{\theta_1z}$}
\begin{eqnarray}
\label{f_theta1z}
& & \frac{8 \sqrt{2} \left(9 b^2+1\right)^4 b^{12} \Sigma _1}{243 \pi ^3 \left(1-3 b^2\right)^{10} \left(6 b^2+1\right)^8
   {g_s}^{9/4} \log N ^4 N^{7/6} {N_f}^3 \alpha _{\theta _1}^7 \alpha _{\theta _2}^6 \left(6 a^2
   {r_0}^3-27 a^4 {r_0}+{r_0}^5\right)}+{{\cal C}^{\rm th}_{\theta_1 z}} \nonumber \\
   & &
   =\frac{539 \pi ^3 N^{2/5} \alpha _{\theta _2}^2
   {f_{x^{10}x^{10}}}({r_0})}{1728 {g_s}^3 M^2 {N_f}^2 \log ^2(r)}.
\end{eqnarray}

\item
{\it matching $f_{zz}$}
\begin{eqnarray}
\label{f_zz}
& & \hskip -0.7in  {\cal C}_{zz}^{\rm th}-\frac{b^{10} \left(9 b^2+1\right)^4 M {r_0}^2 \Sigma _1 \log ({r_0})}{27 \pi ^{3/2} \left(3
   b^2-1\right)^5 \left(6 b^2+1\right)^4 \sqrt{{g_s}} \log N ^2 N^{23/20} {N_f} \alpha _{\theta
   _2}^5}=\frac{539 \pi ^3 N^{2/5} \alpha _{\theta _2}^2 {f_{x^{10}x^{10}}}({r_0})}{864 {g_s}^3 M^2 {N_f}^2 \log
   ^2({r_0})}.
\end{eqnarray}

\item
We solved (\ref{f_theta1x}) - (\ref{f_zz}) and obtained:
\begin{eqnarray}
\label{f_zz-f_theta1x-f_theta1z}
& & {\cal C}_{zz}^{\rm th} \sim \frac{ \left(\frac{1}{N}\right)^{3/4} {r_0}^2 \Sigma _1}{\epsilon ^5
   {g_s}^{7/2} \log N ^2 M {N_f}^3 \alpha _{\theta _2}^3 \log ({r_0})}\nonumber\\
& & {\cal C}_{\theta_1z}^{\rm th} \sim \frac{ \left(\frac{1}{N}\right)^{3/4} {r_0}^2 \Sigma _1}{2 \epsilon ^5
   {g_s}^{7/2} \log N ^2 M {N_f}^3 \alpha _{\theta _2}^3 \log ({r_0})}\nonumber\\
& &{\cal C}_{\theta_1x}^{\rm th} \sim \frac{\left(\frac{1}{N}\right)^{7/6} \Sigma _1}{ \epsilon ^{11} {g_s}^{9/4}
   \log N ^4 {N_f}^3 {r_0}^5 \alpha _{\theta _1}^7 \alpha _{\theta _2}^6}. 
\end{eqnarray}
Hence, ${\cal C}_{zz}^{\rm th} - 2 {\cal C}_{\theta_1z}^{\rm th} + 2{\cal C}_{\theta_1x}^{\rm th}  = 2{\cal C}_{\theta_1x}^{\rm th}< 0$ (as $\Sigma_1<0$). This was argued in (\ref{CCsO4}) by demanding compliance with the phenomenological value of the 1-loop renormalized LEC occurring in the ${\cal O}(p^4)$ $SU(3)\ \chi$PT $\left(\nabla_\mu U^\dagger\nabla_\mu U\right)^2$.
\end{itemize}
\section{{${\cal M}$-Theory Metric Associated with Rotating Cylindrical Black Hole and Thermal Backgrounds}}
\label{metric}
\begin{itemize}
\item The following are ${\cal O}(\beta^0)$ components associated with the ${\cal M}$-theory dual involving a rotating cylindrical black hole background in the MQGP limit differ from those derived in \cite{MQGP,NPB}. Apart from these components, rest of the components are same as given in (\ref{M-theory-metric-psi=2npi-patch}) at ${\cal O}(\beta^0)$.
{\scriptsize
\begin{eqnarray}
& & 
 G^{\cal M}_{tt}(r)= \frac{\left(\frac{1}{N}\right)^{3/2} r^2 \left(a_1(r) (N-B(r))-a_2(r)\right)}{2 \sqrt{\pi } \sqrt[3]{a_1(r)} \sqrt{g_s}}\left(\frac{g(r)-l^2\omega^2}{1-l^2\omega^2}\right), \nonumber\\
 & & G^{\cal M}_{t \phi}(r)= G^{\cal M}_{\phi t}(r)=\frac{\left(\frac{1}{N}\right)^{3/2} r^2  \left(a_1(r) (N-B(r))-a_2(r)\right)}{2 \sqrt{\pi } \sqrt[3]{a_1(r)} \sqrt{g_s}}\left(\frac{l^2\omega(1-g(r))}{1-l^2\omega^2}\right),\nonumber\\
& & G^{\cal M}_{\phi \phi}(r)=\frac{\left(\frac{1}{N}\right)^{3/2} r^2 \left(a_1(r) (N-B(r))-a_2(r)\right)}{2 \sqrt{\pi } \sqrt[3]{a_1(r)} \sqrt{g_s}}\left(\frac{-g(r)l^4\omega^2+l^2}{1-l^2\omega^2}\right),
\end{eqnarray}
}
where
{\scriptsize
\begin{eqnarray}
& & 
a_1(r)=\frac{3 \left(-N_f \log \left(9 a^2 r^4+r^6\right)-4 N_f \log \left(\alpha _{\theta _1}\right)-4 N_f \log \left(\alpha _{\theta _2}\right)+2 N_f \log (N)+4 \log (4) N_f+\frac{8 \pi }{g_s}\right)}{8 \pi },\nonumber\\
& & a_2(r)=\frac{12 a^2 M^2 N_f g_s \left(c_2 \log \left(r_h\right)+c_1\right)}{9 a^2+r^2}
,\nonumber\\
& &  a_3(r)=\frac{r^2 a_2(r)}{2 {N_f} \left(6 a^2+r^2\right)}
,\nonumber\\
& & a_4(r)=\frac{6 a^2+r^2}{9 a^2+r^2}
,\nonumber\\
& & b_1(r)=\frac{M g_s^{3/4}}{6 \sqrt{2} \pi ^{5/4} r^2 \alpha _{\theta _1} \alpha _{\theta _2}^2}\Biggl[\left(\log (r) \left(4 \left(r^2-3 a^2\right) N_f g_s \log \left(\frac{1}{4} \alpha _{\theta _1} \alpha _{\theta _2}\right)+8 \pi  \left(r^2-3 a^2\right)-3 {g_s} r^2 N_f\right) \right)\nonumber\\
& & \  +\left( \left(r^2-3 a^2\right) N_f g_s \log \left(\frac{1}{4} \alpha _{\theta _1} \alpha _{\theta _2}\right)+18 \left(r^2-3 a^2 (6 r+1)\right) N_f g_s \log ^2(r)-\left(r^2-3 a^2\right) N_f g_s \log (N) (2\log (r)+1)\right)\Biggr],\nonumber\\
& & b_2(r)=\frac{M N_f g_s^{7/4} \log (r) \left(36 a^2 \log (r)+r\right)}{3 \sqrt{2} \pi ^{5/4} r \alpha _{\theta _2}^3},\nonumber\\
& & B(r)=\frac{3 g_s M^2 \log (r)}{32 \pi ^2}\times \Biggl[\left(12 N_f g_s \log (r)+2 N_f g_s \log \left(\alpha _{\theta _1}\right)+2 N_f g_s \log \left(\alpha _{\theta _2}\right)+6 N_f g_s\right)   +\left(N_f g_s (-\log (N))-2 \log (4) N_f g_s+8\pi \right)\Biggr],\nonumber\\
& & B_1(r)=\frac{2 \sqrt{\pi } \sqrt{N} \sqrt{g_s}}{r^2},\nonumber\\
& & g(r)=1-\frac{r_h^4}{r^4}.
\end{eqnarray}
}
\item The ${\cal O}(\beta)$ terms in the small $\omega$-limit, remain the same as worked out in \cite{HD-MQGP} and has been provide in (\ref{M-theory-metric-psi=2npi-patch}).

\item The ${\cal M}$-theory metric containing a cylindrical thermal gravitational background for the MQGP limit could be produced from the ${\cal M}$-theory metric involving a thermal background provided in \cite{McTEQ} simply by replacing $x^3$ with $ l \phi$.
\end{itemize}
\section{${\cal O}(\beta)$ Contribution to the On-Shell Action Densities for the Rotating QGP $T_c$ Calculation}
\label{beta}
To derive the contribution that comes from the ${\cal O}(\beta)$ contributions in the on-shell action density for the black hole backdrop, write the metric in diagonal basis (\ref{TypeIIA-from-M-theory-Witten-prescription-T>Tc-Rotation-Canonical}). The $t-\phi$ subspace's unwarped metric could be expressed as:
\begin{eqnarray}
\label{metric-tphi-diagonal}
ds^2=-\left(1-\frac{r_h^4}{r^4 \gamma^2}\right)dT^2+l^2 d\Phi^2,
\end{eqnarray}
where
\begin{eqnarray}
\label{coordinate-transformations}
& & 
dT=dt - \frac{l^2 r_h^4 \omega d\phi}{r^4-r_h^4}; \ d\Phi=\frac{l^2 r_h^4 \omega dt}{r^4-r_h^4}+ d\phi .
\end{eqnarray}
In the diagonal basis, the ${\cal O}(\beta)$ implemented metric is represented as: $G_{MN}^{\cal M}=G_{MN}^{\rm MQGP}\left(1+\beta f_{MN}(r)\right)$, where $f_{MN}(r)$ are given in \cite{HD-MQGP} which can also be read from (\ref{M-theory-metric-psi=2npi-patch}). Since we are concentrating at ${\cal O}(\beta)$, we dropped the terms of ${\cal O}(\omega^2)$ in (\ref{coordinate-transformations})-(\ref{metric-tphi-diagonal-small-l-omega}). At ${\cal O}(\beta)$, finite contributions corresponding to the on-shell action densities for the rotating cylindrical black hole and thermal backgrounds were calculated via an identical method just like done in the computation of ${\cal O}(\beta^0)$ contributions, see \ref{O-beta0-Tc}, and the final results are given as:
\begin{eqnarray}
\label{bh-action-beta}
& & \left(1+\frac{r_h^4}{2{\cal R}_{\rm UV}^4}\right)\frac{S_{D=11,\ {\rm on-shell \ UV-finite}}^{\beta, \rm BH}}{{\cal V}_4} = \Biggl[{-2 {\cal C}_{\theta_1x} \kappa_{\sqrt{G^{(1)}}R^{(0)}}^{\rm IR} } +\frac{20  \left(-{\cal C}_{zz}^{\rm bh} + 2 {\cal C}_{\theta_1z}^{\rm bh} - 3 {\cal C}_{\theta_1x}^{\rm bh}\right) \kappa_{\rm EH}^{\beta,\ \rm IR} }{11 }\Biggr]\nonumber\\
& & \times\frac{ b^2   {g_s}^{3/2}  M {N_f}^3 {r_h}^4 \log ^3(N) \log
   \left(\frac{{r_h}}{{{\cal R}_{D5/\overline{D5}}}}\right) \log
   \left(1 - \frac{{r_h}}{{{\cal R}_{D5/\overline{D5}}}}\right)}{ N^{1/2} {{\cal R}_{D5/\overline{D5}}}^4}\beta,
\end{eqnarray}
and
\begin{eqnarray}
\label{th-action-beta}
& & \frac{S_{D=11,\ {\rm on-shell \ UV-finite}}^{\beta,\rm thermal}}{{\cal V}_4} =
 -\frac{20  \kappa_{\rm EH, th}^{{\rm IR},\ \beta} {r_0}^3 N^{1/2}{f_{x^{10}x^{10}}}({r_0})}{11{g_s}^{3/2} M  {N_f}^{5/3} {{\cal R}_{D5/\overline{D5}}^{\rm th}}^3  \log ^{\frac{2}{3}}(N) \log
   \left(\frac{{r_0}}{{\cal R}_{D5/\overline{D5}}^{\rm th}}\right)}\beta.  
\end{eqnarray} 
We can now write the metric in $(t,\phi)$ subspace and ${\cal O}(R^4)$ correction to the same, i.e., $f_{MN}(r)$, where $(M,N=t,\phi)$ (we are concentrating exclusively on the $t,\phi)$ subspace). The metric's $tt$ component could be written as:
\begin{eqnarray}
& & 
g_{tt}=\left(\frac{dT}{dt}\right)^2 G_{TT}^{\cal M}+\left(\frac{d\Phi}{d\phi}\right)^2 G_{\Phi \Phi}^{\cal M},
\end{eqnarray}
implying
\begin{eqnarray}
& & f_{tt}=\left(\frac{dT}{dt}\right)^2 f_{TT}+\left(\frac{d\Phi}{d\phi}\right)^2 f_{\Phi \Phi}= f_{TT}+ f_{\Phi \Phi}.
\end{eqnarray}
Similarly,the $\phi t$ component of the metric is written as:
\begin{eqnarray}
& & g_{\phi t}=\left(\frac{dT}{d\phi}\right)\left(\frac{dT}{dt}\right) G_{TT}^{\cal M}+\left(\frac{dT}{d\phi}\right)\left(\frac{dT}{dt}\right) G_{\Phi \Phi}^{\cal M},
\end{eqnarray}
and therefore, 
\begin{eqnarray}
& & f_{\phi t}=\left(\frac{dT}{d\phi}\right)\left(\frac{dT}{dt}\right) f_{TT}+\left(\frac{dT}{d\phi}\right)\left(\frac{dT}{dt}\right) f_{\Phi \Phi} = -\frac{l^2 \omega  r_h^4}{r^4-r_h^4} \left(f_{TT}-f_{\Phi \Phi}\right).
\end{eqnarray}
The same is also applicable for the $\phi \phi$ component:
\begin{eqnarray}
& & g_{\phi \phi}=\left(\frac{dT}{d\phi}\right)^2 G_{TT}^{\cal M}+\left(\frac{d\Phi}{d\phi}\right)^2 G_{\Phi \Phi}^{\cal M},
\end{eqnarray}
hence $f_{\phi \phi}$ is obtained as:
\begin{eqnarray}
& & f_{\phi \phi}=\left(\frac{dT}{d\phi}\right)^2 f_{TT}+\left(\frac{d\Phi}{d\phi}\right)^2 f_{\Phi \Phi} =-\left(\frac{l^2 \omega  r_h^4}{r^4-r_h^4}\right) f_{TT}+  f_{\Phi \Phi}.
\end{eqnarray}
In the small-$\omega$ limit, $\gamma=1$, hence the equation (\ref{metric-tphi-diagonal}) simplified as:
\begin{eqnarray}
\label{metric-tphi-diagonal-small-l-omega}
ds^2=-\left(1-\frac{r_h^4}{r^4}\right)dT^2+ l^2 d\Phi^2.
\end{eqnarray}
As a result, the structure of the ${\cal M}$-theory metric for the black hole background in the small $\omega$-limit in rotating quark-gluon plasma is the same as the structure of the ${\cal M}$-theory metric without rotation. As a result, in the small $\omega$-limit, the higher derivative corrections to the ${\cal M}$-theory metric for the rotating cylindrical black hole and thermal backgrounds will be the same as obtained in \cite{HD-MQGP}. As a result, the on-shell action densities associated with the rotating cylindrical thermal and black hole background will be identical to \cite{McTEQ} (i.e. similar to \ref{DPT-WR-i}) augmented by a factor of $l$, which is going to cancel out at the UV cut-off from both sides of the equation (\ref{SBH=Sth-i}).

\section{Holographic Renormalization of Rotating Gravitational Backgrounds}
\label{HRN-R-Tc}
Here, we have discussed the holographic renormalization of rotating cylindrical black hole and thermal backgrounds used in \ref{DPT-WR-ii}.
\begin{itemize}
\item {\bf Rotating Cylindrical Black Hole Background:} Divergent parts of the ${\cal M}$-theory on-shell action of the rotating cylindrical black hole background at ${\cal O}(\beta^0)$ have the following structure:
\begin{eqnarray}
\label{UV-div-beta0-BH}
& & S_{\rm UV-div}^{\beta^0, \rm BH} =
-\frac{1}{2}\Biggl(-2 S_{\rm EH}^{(0)} + 2 S_{\rm GHY}^{(0)}\Biggr) \nonumber\\
& & = \left(\tilde{\lambda}_{\rm GHY,BH}^{ {\rm UV-div}}+\tilde{\lambda}_{\rm UV-div}^{EH}\right)
\ \left(\frac{{{\cal R}_{\rm UV}}}{{{\cal R}_{D5/\overline{D5}}^{\rm BH}}}\right)^4 \log \left(\frac{{{\cal R}_{\rm UV}}}{{{\cal R}_{D5/\overline{D5}}^{\rm BH}}}\right).
\end{eqnarray}
 UV-divergences appearing in the ${\cal M}$-theory on-shell action of the rotating cylindrical black hole background at ${\cal O}(\beta^0)$ will be cancelled by the following term:
 {\footnotesize
\begin{eqnarray}
\label{CT-BH-beta0}
& & \hskip -0.2in
S_{\rm CT}^{\beta^0, \ \rm BH}=\frac{\int \sqrt{-h^{\cal M}}R|^{r={\cal R}_{\rm UV}}}{{\cal V}_4}=-\frac{\left(\tilde{\lambda}_{\rm GHY,BH}^{ {\rm UV-div}}+\tilde{\lambda}_{\rm UV-div}^{EH}\right)}{\tilde{\lambda}_{\rm \sqrt{-h^{\cal M}}R@\partial M_{11}}}\left(\frac{{{\cal R}_{\rm UV}}}{{{\cal R}_{D5/\overline{D5}}^{\rm BH}}}\right)^4 \log \left(\frac{{{\cal R}_{\rm UV}}}{{{\cal R}_{D5/\overline{D5}}^{\rm BH}}}\right),
\end{eqnarray}
}
where, $\tilde{\lambda}_{\rm GHY,BH}^{ {\rm UV-div}}, \ \tilde{\lambda}_{\rm UV-div}^{EH}$ and $\tilde{\lambda}_{\rm \sqrt{-h^{\cal M}}R@\partial M_{11}}$ are the prefactors defined in terms of the parameters of the model, e.g., $N,M^{\rm UV},N_f^{\rm UV},g_s^{\rm UV},\alpha_{\theta_{1,2}}$ etc. Divergent part in the on-shell action for the rotating cylindrical black hole background at ${\cal O}(\beta)$ in the UV is: 
 \begin{eqnarray}
\label{int-sqrtGiGdeltaJ0-beta0-UV-div}
& &  \hskip -0.4inS_{\rm UV-div}^{\beta, \ \rm BH}=  \frac{\left.\int \sqrt{-G^{\cal M}}G^{MN}_{\cal M}\frac{\delta J_0}{\delta G^{MN}_{\cal M}}\right|_{\rm UV-div}}{{\cal V}_4} = \tilde{\lambda}^{(0)}_{\sqrt{-G^{\cal M}}G^{MN}_{\cal M}\frac{\delta J_0}{\delta G^{MN}_{\cal M}}}\frac{{{\cal R}_{\rm UV}}^4 }{ \log\left(\frac{{{\cal R}_{\rm UV}}}{{{\cal R}_{D5/\overline{D5}}^{\rm BH}}}\right){N_f^{\rm UV}}^{8/3}}
\end{eqnarray}
We found that UV-divergence appearing in equation (\ref{int-sqrtGiGdeltaJ0-beta0-UV-div}) will be cancelled by the following counter term:
\begin{eqnarray}
S_{\rm CT}^{\beta, \ \rm BH}= -\left(\frac{\tilde{\lambda}^{(0)}_{\sqrt{-G^{\cal M}}G^{MN}_{\cal M}}\frac{\delta J_0}{\delta G^{MN}_{\cal M}}}{\tilde{\lambda}^{(0)}_{\sqrt{-h^{\cal M}}h^{MN}_{\cal M}\frac{\delta J_0}{\delta G^{MN}_{\cal M}}}}\right)\left.\frac{\left.\int \sqrt{-h^{\cal M}}h^{mn}\frac{\delta J_0}{\delta h^{mn}_{\cal M}}\right|_{\rm UV-div}}{{\cal V}_4}\right|_{r=\rm constant},
\end{eqnarray} 
provided 
\begin{equation}
\label{Nf-LogRUV}
N_f^{\rm UV}\sim\frac{\left(\log\left(\frac{{{\cal R}_{\rm UV}}}{{{\cal R}_{D5/\overline{D5}}^{\rm bh}}}\right)\right)^{\frac{15}{2}}}{\log^{1/2}N}, 
\end{equation}
where $\tilde{\lambda}^{(0)}_{{\sqrt{-G^{\cal M}}G^{MN}_{\cal M}}\frac{\delta J_0}{\delta G^{MN}_{\cal M}}}$ and $\tilde{\lambda}^{(0)}_{{\sqrt{-h^{\cal M}}h^{MN}_{\cal M}\frac{\delta J_0}{\delta G^{MN}_{\cal M}}}}$ are depending on $N,M^{\rm UV},g_s^{\rm UV},\alpha_{\theta_{1,2}}$ etc.\\
%%%%%%%%%%
%%%%%%%
\item {\bf Rotating Cylindrical Thermal Background:}
Divergent part of the ${\cal M}$-theory on-shell action at ${\cal O}(\beta^0)$ in the UV have the following form:
\begin{eqnarray}
\label{UV-div-beta0-thermal}
& &  S_{\rm UV-div}^{\beta^0, \rm th} =-
\frac{1}{2}\Biggl(-2 S_{\rm EH}^{(0)} + 2 S_{\rm GHY}^{(0)}\Biggr)\nonumber\\
& &  =- \tilde{\lambda}_{\rm GHY}^{ {\rm UV-div}}\ \left(\frac{{{\cal R}_{\rm UV}}}{{{\cal R}_{D5/\overline{D5}}^{\rm th}}}\right)^4 \log \left(\frac{{{\cal R}_{\rm UV}}}{{{\cal R}_{D5/\overline{D5}}^{\rm th}}}\right) - \tilde{\lambda}_{\rm UV-div}^{\beta^0,2}\ \left(\frac{{{\cal R}_{\rm UV}}}{{{\cal R}_{D5/\overline{D5}}^{\rm th}}}\right)^4,
\end{eqnarray}
where $\tilde{\lambda}_{\rm GHY}^{ {\rm UV-div}}$ and $\tilde{\lambda}_{\rm UV-div}^{\beta^0,2}$ are the prefactors defined in terms of $N,M^{\rm UV},N_f^{\rm UV},g_s^{\rm UV},\alpha_{\theta_{1,2}}$ etc. Counter terms to cancel the UV-divergences for the cylindrical thermal background at ${\cal O}(\beta^0)$ are:
\begin{eqnarray}
\label{CT-thermal}
& & \hskip -0.4in S_{\rm CT}^{\beta^0, \rm th} =\left( \frac{\tilde{\lambda}_{\rm GHY}^{ {\rm UV-div}}}{{\lambda}_{\sqrt{-h^{\cal M}}@\partial M_{11}}^{(0)}}\right)\frac{\int_{r={\cal R}_{\rm UV}}\sqrt{-h^{\cal M}}}{{\cal V}_4} +\left( \frac{\tilde{\lambda}_{\rm UV-div}^{\beta^{0,2}}}{{\lambda}_{\sqrt{-h^{\cal M}}R@\partial M_{11}}^{(0)}}\right)\frac{\int_{r={\cal R}_{\rm UV}} \sqrt{-h^{\cal M}}R}{{\cal V}_4}.
\end{eqnarray}
where first and second terms in equation (\ref{CT-thermal}) are counter terms for the first and second terms in equation (\ref{UV-div-beta0-thermal}) and are given as below:
\begin{eqnarray}
& &\int_{r={\cal R}_{\rm UV}} \sqrt{-h^{\cal M}}={\lambda}_{\sqrt{-h^{\cal M}}@\partial M_{11}}^{(0)} \left(\frac{{{\cal R}_{\rm UV}}}{{{\cal R}_{D5/\overline{D5}}^{\rm th}}}\right)^4 \log \left(\frac{{{\cal R}_{\rm UV}}}{{{\cal R}_{D5/\overline{D5}}^{\rm th}}}\right),  \nonumber \\
& &\int_{r={\cal R}_{\rm UV}} \sqrt{-h^{\cal M}} R={\lambda}_{\sqrt{-h^{\cal M}}R@\partial M_{11}}^{(0)}\left(\frac{{{\cal R}_{\rm UV}}}{{{\cal R}_{D5/\overline{D5}}^{\rm th}}}\right)^4,
\end{eqnarray}
where ${\lambda}_{\sqrt{-h^{\cal M}}@\partial M_{11}}^{(0)}$ and ${\lambda}_{\sqrt{-h^{\cal M}}R@\partial M_{11}}^{(0)}$ are defined in terms of the parameters \\($N,M^{\rm UV},N_f^{\rm UV},g_s^{\rm UV},\alpha_{\theta_{1,2}}$ etc.) of the model. UV divergence at ${\cal O}(\beta)$ coming from $\frac{\sqrt{-G^{\cal M}}G^{MN}\frac{\delta J_0}{\delta G^{MN}}}{{\cal V}_4}$ will be cancelled by $\frac{\int\sqrt{-h^{\cal M}}G^{mn}\frac{\delta J_0}{\delta G^{mn}}}{{\cal V}_4}$ provided $N_f^{\rm UV}=\frac{1}{\log N} + \epsilon_1$, where $0<\epsilon_1\ll1$ .
\end{itemize}

\chapter{}
\section{Complex Extrinsic Curvatures and Riemann Tensors}
\label{complex-to-real}
Let $z=x e^{it}$ and $\bar{z}=x e^{-it}$. Then
\begin{eqnarray}
& & 
\partial_z=-\left(\frac{i}{z}\right)\partial_t + \left(\frac{x}{z}\right) \partial_x \nonumber\\
& & \partial_{\bar{z}}=\left(\frac{i}{\bar{z}}\right)\partial_t + \left(\frac{x}{\bar{z}}\right) \partial_x \nonumber
\end{eqnarray}
Since $K_{zij}=\frac{1}{2}\partial_z g_{ij}(r)$ and $K_{\bar{z}ml}=\frac{1}{2}\partial_{\bar{z}} g_{ml}(r)$. Therefore,
\begin{equation}
\label{KK-C-to-R}
K_{zij}K_{\bar{z}ml}=\left(\frac{1}{x^2}K_{tij}K_{tml}+K_{xij}K_{xml} -\frac{i}{x} K_{tij} K_{xml} +\frac{i}{x} K_{xij} K_{tml}\right)
\end{equation}
For the Hartman-Maldacena surface induced metric is $t(r)$ dependent and for the island surface induced metric is $x(r)$ dependent. Therefore $K_{xij}=0$ for Hartman-Maldacena surface and 
$K_{tij}=0$ for the island surface and hence in either case imaginary part of (\ref{KK-C-to-R}) will be zero. \par
We can go from complex Riemann tensors to real Riemann tensors by using the following coordinate transformations:
 \begin{eqnarray}
 \label{Reimann-C-to-R}
 R_{z\overline{z}z\overline{z}}=\frac{\partial x^{\alpha}}{\partial z}\frac{\partial x^{\beta}}{\partial \overline{z}}\frac{\partial x^{\gamma}}{\partial z}\frac{\partial x^{\eta}}{\partial \overline{z}} R_{x^{\alpha}x^{\beta}x^{\gamma}x^{\eta}},
 \end{eqnarray}
 where $x^{\alpha /\beta /\gamma /\eta }=t,r$. Therefore,
 \begin{eqnarray}
 \label{GCT-SD-J0}
 \frac{\delta {\cal L}}{R_{z\overline{z}z\overline{z}}}=\frac{\partial z}{\partial x^{\alpha}}\frac{\partial \overline{z}}{\partial x^{\beta}}\frac{\partial z}{\partial x^{\gamma}}\frac{\partial \overline{z}}{\partial x^{\eta}} \left( \frac{\delta {\cal L}}{\delta R_{x^{\alpha}x^{\beta}x^{\gamma}x^{\eta}}}\right).
 \end{eqnarray}
Similarly to calculate the second term is holographic entanglement entropy formula we use the following coordinate transformations:
\begin{eqnarray}
\label{GCT-DD-J0}
& & 
\frac{\delta }{\delta R_{zizj}} = \frac{\partial z}{\partial x^{\alpha}}\frac{\partial z}{\partial x^{\beta}} \left(\frac{\delta }{\delta R_{x^{\alpha}ix^{\beta}j}}\right), \nonumber\\
& & \frac{\delta }{\delta R_{\overline{z}k\overline{z}l}} = \frac{\partial \overline{z}}{\partial x^{\gamma}}\frac{\partial \overline{z}}{\partial x^{\eta}} \left(\frac{\delta }{\delta R_{x^{\gamma}k x^{\eta}l}}\right).
\end{eqnarray} 
 Hence by using equation (\ref{GCT-DD-J0}) we can obtain the second term appearing in holographic entanglement entropy formula in terms of the real coordinates $(t,x)$ using the coordinate transformations given below:
 \begin{eqnarray}
 \frac{\delta^2 {\cal L}}{\delta R_{zizj} \delta R_{\overline{z}k\overline{z}l}}=\frac{\partial z}{\partial x^{\alpha}}\frac{\partial z}{\partial x^{\beta}}\frac{\partial \overline{z}}{\partial x^{\gamma}}\frac{\partial \overline{z}}{\partial x^{\eta}} \left(\frac{\delta^2 {\cal L}}{\delta R_{x^{\alpha}ix^{\beta}j}\delta R_{x^{\gamma}k x^{\eta}l}}\right).
 \end{eqnarray}
\section{HM-Like/IS Analytics/Numerics}
\label{HM+IS-analyt+num}
In the present appendix, we describe (i) the angular integrations utilized throughout the computation and the equation describing the area of a Hartman-Maldacena-like surface, and (ii) the process by which we estimate the turning point in the context of Island Surface entanglement entropy.
\subsection{HM-Like Surface Area}
\label{a-1}
The angular integrations in changing the delocalized results near $(\theta_1,\theta_2)=\left(\frac{\alpha_{\theta_1}}{N^{\frac{1}{5}}},\frac{\alpha_{\theta_2}}{N^{\frac{3}{10}}}\right)$, to global results, ignoring the entirety of ${\cal O}\left(\frac{{\cal O}(1)}{N^\alpha}\right), \alpha>1$, were carried out as follows. Transforming local to global coordinates, i.e., $(x, y, z)\rightarrow (\phi_1, \phi_2, \psi)$ using \cite{MQGP}:
\begin{eqnarray}
\label{local-to-global}
& &
dx = \sqrt{h_2}\left(g_sN\right)^{\frac{1}{4}}\Biggl[1 + {\cal O}\left(\frac{g_sM^2}{N}\right) + {\cal O}\left(\frac{(g_sM^2)(g_sN_f)}{N}\right)\Biggr]\sin\theta_1d\phi_1, \nonumber\\
& & dy = \sqrt{h_4}\left(g_sN\right)^{\frac{1}{4}}\Biggl[1 + {\cal O}\left(\frac{g_sM^2}{N}\right) + {\cal O}\left( \frac{(g_sM^2)(g_sN_f)}{N}\right)\Biggr]\sin\theta_2d\phi_2, \nonumber\\
& & dz = \sqrt{h_1}\left(g_sN\right)^{\frac{1}{4}}\Biggl[1 + {\cal O}\left(\frac{g_sM^2}{N}\right) + {\cal O}\left( \frac{(g_sM^2)(g_sN_f)}{N}\right)\Biggr]d\psi,
\end{eqnarray}
where $h_1 = \frac{1}{9} + {\cal O}\left(\frac{g_sM^2}{N}\right),\ h_2 = \frac{1}{6} + {\cal O}\left(\frac{g_sM^2}{N}\right),\ h_4 = h_2 + \frac{4a^2}{r^2}$ \cite{metrics}, \cite{MQGP}, 
\begin{eqnarray}
\label{angular-integrations-2}
& & \frac{1}{\alpha_{\theta_1}^3\alpha_{\theta_2}^2}\rightarrow\lim_{\epsilon_{1,2}\rightarrow0}
\int_{\epsilon_2}^{\pi-\epsilon_2}\sqrt{g_sN}d\theta_2\sin\theta_2\int_{\epsilon_1}^{\pi-\epsilon_1}d\theta_1\sin\theta_1\frac{1}{\left(N^{\frac{1}{5}}\sin{\theta_1}\right)^3\left(N^{\frac{3}{10}}\sin{\theta_2}\right)^2}  \sim \lim_{\epsilon_{1,2}\rightarrow0}\frac{|\log \epsilon_2|}{N^{\frac{7}{10}}\epsilon_1}.
\nonumber\\
\end{eqnarray}
The principal (${\cal P}$) corresponding to (\ref{angular-integrations-2}) is found by reducing $|\log \epsilon_2|$ to $\log \mathbb{P} + |\log \epsilon_2|, \mathbb{P}\in\mathbb{Z}^+$ and requiring $\log \mathbb{P} + \log \epsilon_2 = - \epsilon_1$, or
$\epsilon_2=\frac{e^{-\epsilon_1}}{\sqrt{\mathbb{P}}}, \mathbb{P}>1$.
Analogously,
\begin{eqnarray}
\label{angular-integrations-1}
& &  \frac{1}{\alpha_{\theta_1}\alpha_{\theta_2}^6}\rightarrow\lim_{\epsilon_{1,2}\rightarrow0}
\int_{\epsilon_2}^{\pi-\epsilon_2}\sqrt{g_sN}d\theta_2\sin\theta_2\int_{\epsilon_1}^{\pi-\epsilon_1}d\theta_1\sin\theta_1\frac{1}{\left(N^{\frac{1}{5}}\sin{\theta_1}\right)\left(N^{\frac{3}{10}}\sin{\theta_2}\right)^6} \nonumber\\
& & \stackrel{{\cal P}}{\rightarrow} \frac{\pi  (540 \log (2)-107) \sqrt{g_s}}{720 N^{3/2}}.
\nonumber\\
\end{eqnarray}
Assume that $r_*\in$IR, approximating $\log r\approx\log r_h$ when $r = \upsilon r_h\forall r\in$IR, $\upsilon={\cal O}(1)$, 
we obtained the result given below after radial integration of (\ref{AHM_i}):
{\footnotesize
\begin{eqnarray}
\label{AHM_ii}
& &  A_{\rm HM} \sim \Biggl[E M^2 \sqrt[10]{N} N_f^6 g_s^{17/4} \log ^2\left(r_h\right) \left(\log (N)-3 \log \left(r_h\right)\right){}^4   \Biggl(3 r_*^5 \left(\log (N)-9 \log \left(r_h\right)\right){}^2 +15 r_* r_h^4 \log ^2(N)\nonumber\\
& &-10 r_*^3 r_h^2 \log (N) \left(\log (N)-9 \log \left(r_h\right)\right)-r_h^5 \left(36 \log (N) \log \left(r_h\right)+243 \log ^2\left(r_h\right)+8 \log ^2(N)\right)\Biggr) \Biggr].
\end{eqnarray}
}
\subsection{IS Turning Point}
\label{a-2}
To determine the point of turning $tilder_T$ in the scenario of island surface's entanglement entropy (refEE-IS-simp), observe that:  $\frac{H(r_T)}{\sigma(r_T)}={\cal C}^2$, up to leading order in $N$, that acquires:
\begin{eqnarray}
\label{r_T-i}
& & 1458{g_s}^{5/4} M^2 N^{2/5} {N_f}^6 r_T^4 \left(r_T^4-{r_h}^4\right) \log ^7(r_T)
   \left(7\log N  r_T^2-3 a^2 \log N \right)\nonumber\\
   & & -6561{g_s}^{5/4} M^2 N^{2/5} {N_f}^6 r_T^6 \left(r_T^4-{r_h}^4\right)
   \log ^8(r_T)   + {\cal O}\left(\frac{1}{(\log r_h)^6}\right) = \tilde{\cal C}^2.
\end{eqnarray}
Let us write $\tilde{r}_T = 1 + \delta_2$, insertion in (\ref{r_T-i}), considering $0<\delta_2\ll1$ and so approximating $\log r\approx \log r_h$, produces the result that follows:
\begin{eqnarray} 
\label{r_T-ii}
& & 1458 \delta_2  {r_h}^{10} \log ^7({r_h}) (4 (47 \delta_2 +6) \log N -9 (15
   \delta_2 +2) \log ({r_h})) \nonumber\\
 & &  -5832 \sqrt{3} \delta_2  (11 \delta_2 +2)
   \epsilon  \log N  {r_h}^{10} \log ^7({r_h})=\frac{M^2 N^{2/5} {N_f}^6
   \tilde{\cal C}^2}{{g_s}^{5/4}}.
\end{eqnarray}
Assuming $|\log r_h|\gg \log N$ again, the appropriate solution to (\ref{r_T-ii}) is obtained as:
\begin{eqnarray}
\label{r_T-iii}
& & \delta_2\approx \frac{M^2 N^{2/5} {N_f}^6 \tilde{\cal C}^2}{26244 {g_s}^{5/4} {r_h}^{10} \log
   ^8({r_h})}-\frac{2}{15}.
\end{eqnarray}
Based on the estimate of $r=r_0\sim r_h: N_{\rm eff}(r_0)=0$ obtained in \cite{Bulk-Viscosity-McGill-IIT-Roorkee}, writing $r_h\sim e^{-\kappa_{r_h}N^{\frac{1}{3}}}$, and assuming $\tilde{\cal C}^2=e^{-\kappa_{\cal C}N^{\frac{1}{3}}}$. Numerically, for $M=N_f=3, g_s=0.1, N=10^{3.3}$ and assuming $\kappa_{\cal C}=1.37, \kappa_{r_h}=0.1, \tilde{\cal C}=3\times10^{-8}$, we obtained $\delta_2=5.6\times10^{-4}$.

\section{Hartman-Maldacena-like Surface Miscellania}
\label{HM}
We present numerous $r$ dependent functions that appear in the entanglement entropy formula to describe the Hartman-Maldacena-like surface at ${\cal O}(\beta^0)$ and ${\cal O}(\beta)$ in this appendix. We also calculated the equation of motion pertaining to the embedding function associated with the Hartman-Maldacena-like surface.
\begin{itemize}
\item
The $r$ dependent functions appearing in equation (\ref{Lag0waldA}) are given as:
\begin{eqnarray}
\label{alpha-sigma-lambda}
& & \alpha(r)= \kappa_{\alpha}\frac{ r^2 \left(N_f g_s (2 \log (N)-6 \log (r))\right){}^{2/3}}{ \sqrt{N} g_s^{3/2}},\nonumber\\
& & \sigma(r)=\kappa_{\sigma}\frac{ N g_s-3 M^2 N_f g_s^3 \log (r) (\log (N)-6 \log (r))}{ \left(r^4-{r_h}^4\right)}, \nonumber\\
& & \lambda (r)=\kappa_{\lambda}\Biggl(-\frac{ M N^{17/10} N_f^{4/3} g_s^{5/2} \left(r^2-{r_h}^2\right) \log (r) (\log (N)-9 \log (r)) \sqrt[3]{\log (N)-3 \log (r)}}{ r^4 \alpha _{\theta _1}^3 \alpha_{\theta _2}^2}\Biggr) \nonumber\\
& & \sim -\frac{M N^{21/20} N_f^{4/3} g_s^{13/4} \left(r^2-r_h^2\right) \kappa _{\lambda } \log (r) (\log (N)-9 \log (r)) \sqrt[3]{\log (N)-3 \log (r)}}{3 r^4},\nonumber\\
& & 
 \lambda_1(r)= \kappa_{\lambda_1}\left(\frac{ M^7 N^{7/10} N_f^{7/3} g_s^7 \log ^4(r) (\log (N)-12 \log (r))^3 (\log (N)-9 \log (r))}{ r^2 \alpha _{\theta _1}^3 \alpha
   _{\theta _2}^2 (\log (N)-3 \log (r))^{5/3}}\right) \nonumber\\
   & & \sim \frac{M^7 \sqrt[20]{N} N_f^{7/3} g_s^{31/4} \kappa _{\lambda _1} \log ^4(r) (\log (N)-12 \log (r))^3 (\log (N)-9 \log (r))}{3 r^2 (\log (N)-3 \log (r))^{5/3}}
   ,\nonumber\\
   & & 
  \lambda_2(r)=\kappa_{\lambda_2}\left(\frac{ M^3 N^{7/10} \sqrt[3]{N_f} g_s^3 \log ^2(r) (\log (N)-12 \log (r)) (\log (N)-9 \log (r))}{ r^2 \alpha _{\theta _1}^3 \alpha _{\theta _2}^2 (\log(N)-3 \log (r))^{5/3}}\right) \nonumber\\
  & & \sim \frac{M^3 \sqrt[20]{N} \sqrt[3]{N_f} g_s^{15/4} \kappa _{\lambda _2} \log ^2(r) (\log (N)-12 \log (r)) (\log (N)-9 \log (r))}{3 r^2 (\log (N)-3 \log (r))^{5/3}},\nonumber\\
  & &    {\cal L}_1= {\cal L}_2=\frac{\sqrt{\alpha (r) \left(\sigma (r)-\left(1-\frac{r_h^4}{r^4}\right) t'(r)^2\right)} }{t'(r)^2 \left(\sigma (r)-\left(1-\frac{r_h^4}{r^4}\right) t'(r)^2\right)^4} \nonumber\\
& & \times \left(\alpha '(r) \left(\sigma (r)-\left(1-\frac{r_h^4}{r^4}\right) t'(r)^2\right)+\alpha (r) \left(\sigma '(r)-t'(r) \left(\frac{4 r_h^4}{r^5} t'(r)+2
   \left(1-\frac{r_h^4}{r^4}\right) t''(r)\right)\right)\right)^2, \nonumber\\
   & &   {\cal L}_3= {\cal L}_4=\frac{\sqrt{\alpha (r) \left(\sigma (r)-\left(1-\frac{r_h^4}{r^4}\right) t'(r)^2\right)}}{t'(r)^2},\nonumber\\
    & & Z(r)=-\kappa_Z \frac{  M N^{37/10} g_s^{35/6} \log (r) (\log (N)-9 \log (r))}{ r^{14} \alpha _{\theta _1}^3 \alpha _{\theta _2}^2 N_f^{4/3} (\log (N)-3 \log (r))^{7/3}} \nonumber\\
& & \sim -\frac{\pi ^{23/6} M N^{61/20} g_s^{79/12} \kappa _Z \log (r) (\log (N)-9 \log (r))}{r^{14} N_f^{4/3} (\log (N)-3 \log (r))^{7/3}}
, \nonumber\\
& &  W(r)=-\kappa_W \frac{ M N^{37/10} g_s^{35/6} \log (r) (\log (N)-9 \log (r))}{ r^{14} \alpha _{\theta _1}^3 \alpha _{\theta _2}^2 N_f^{4/3} (\log (N)-3 \log (r))^{7/3}} \nonumber\\
& & \sim -\frac{M N^{61/20} g_s^{79/12} \kappa _W \log (r) (\log (N)-9 \log (r))}{r^{14} N_f^{4/3} (\log (N)-3 \log (r))^{7/3}}
, \nonumber
\end{eqnarray}

\begin{eqnarray}
& & U(r)=-\kappa_U \frac{ M^2 N N_f g_s^3 \log (r) (\log (N)-45 \log (r))}{r^4 \alpha _{\theta _1} \alpha _{\theta _2}^6 (\log (N)-3 \log (r))^3} \nonumber\\
& & \sim -\frac{M^2 N_f g_s^{15/4} \kappa _U \log (r) (\log (N)-45 \log (r))}{\sqrt[4]{N} r^4 (\log (N)-3 \log (r))^3}
, \nonumber\\
& & V(r)=\kappa_V\frac{ M^2 N N_f g_s^3}{ r^4 \alpha _{\theta _1} \alpha _{\theta _2}^6 \log ^2(r)}  \sim \frac{M^2 N_f g_s^{15/4} \kappa _V}{\sqrt[4]{N} r^4 \log ^2(r)},
\end{eqnarray}
where $\kappa_{\alpha},\kappa_{\sigma}$, $\kappa_{\lambda}$,  $\kappa_{\lambda_1}$ and $\kappa_{\lambda_2}$  etc. are numerical pre-factors. Further, $\kappa_Z,\kappa_W,\kappa_U$ and $\kappa_V$ are the numerical factors including $\left(\frac{8}{q_\alpha +1}\right)$.
\item
In the case of the existence of higher derivative terms, the following have been used to compute the first and second terms of holographic entanglement entropy computation:
{\footnotesize
\begin{eqnarray}
\label{dJ0overdR+d2J0overdR2}
& &
\frac{\delta J_0}{\delta R_{M_1 N_1 P_1 Q_1}}=G^{M_1 M_1}G^{N_1 N_1}G^{P_1 P_1}G^{Q_1 Q_1}\left( R_{P N_1 P_1 Q}  R^Q_{\ \ RSQ_1}  +\frac{1}{2}R_{PQP_1 Q_1}  R^Q_{\ \ RSN_1}\right)R_{M_1}^{\ \ RSP} \nonumber\\
& &
+\left(\delta^{Q_1}_Q R^{HN_1 P_1 K}+\frac{1}{2}\delta^{N_1}_Q R^{HKP_1 Q_1}\right)R_{H}^{\ \ RSM_1} R^Q_{\ \ RSK} \nonumber\\
& & + G^{N_1 N_1}G^{P_1 P_1}G^{Q_1 Q_1} \left( R_{Q_1 MNQ}R^{M_1 MNK}+\frac{1}{2}R_{Q_1 QMN} R^{M_1 KMN}\right)R^Q_{\ \ N_1 P_1 K}
 \nonumber\\
& & +G^{M_1 M_1}\left( R_{PMNM_1} R^{HMN Q_1}+\frac{1}{2}  R_{PM_1 MN} R^{H Q_1 MN}\right)R_H^{\ \ N_1 P_1 P}
\end{eqnarray}
}
and 
\begin{eqnarray}
\label{dJ0overdR+d2J0overdR2-i}
& &
\frac{\delta^2 J_0}{\delta R_{M_2 N_2 P_2 Q_2} \delta R_{M_1 N_1 P_1 Q_1}}=A_1+A_2+A_3+A_4,
\end{eqnarray}
where
{\footnotesize
\begin{eqnarray}
\label{dJ0overdR+d2J0overdR2-ii}
& &
A_1=G^{M_1 M_1}G^{N_1 N_1}G^{P_1 P_1}G^{Q_1 Q_1}\Biggl[\delta^{M_2}_{M_1} G^{N_2 N_2}G^{P_2 P_2}G^{Q_2 Q_2}\left(R_{Q_2 N_1 P_1 Q}R^Q_{\ \ N_2 P_2 Q_1}+\frac{1}{2}R_{Q_2 Q P_1 Q_1}R^Q_{\ \ N_2 P_2 N_1}\right)\nonumber\\
& & + \left(\delta^{M_2}_P\delta^{N_2}_{N_1}\delta^{P_2}_{P_1}\delta^{Q_2}_{Q} R^Q_{\ \ RSQ_1}R_{M_1}^{\ \ RSP}+\frac{1}{2}\delta^{M_2}_P\delta^{N_2}_{Q}\delta^{P_2}_{P_1}\delta^{Q_2}_{Q_1} R^Q_{\ \ RSN_1}R_{M_1}^{\ \ RSP} \right)\nonumber\\
& & + \left(G^{M_2M_2}\delta^{N_2}_{R}\delta^{P_2}_{S}\delta^{Q_2}_{Q_1} R_{M_1}^{\ \ RSP}R_{PN_1P_1M_2}+\frac{1}{2}G^{M_2M_2}\delta^{N_2}_{R}\delta^{P_2}_{S}\delta^{Q_2}_{N_1} R_{M_1}^{\ \ RSP}R_{PM_2P_1Q_1}\right)\Biggr] \nonumber
\end{eqnarray}
\begin{eqnarray}
& & A_1 \equiv G^{M_1 M_1}G^{N_1 N_1}G^{P_1 P_1}G^{Q_1 Q_1}\Biggl[\delta^{M_2}_{M_1} G^{N_2 N_2}G^{P_2 P_2}G^{Q_2 Q_2}\left(R_{Q_2 N_1 P_1 Q}R^Q_{\ \ N_2 P_2 Q_1}+\frac{1}{2}R_{Q_2 Q P_1 Q_1}R^Q_{\ \ N_2 P_2 N_1}\right)\nonumber\\
& & + \left(\delta^{N_2}_{N_1}\delta^{P_2}_{P_1} R^{Q_2}_{\ \ RSQ_1}R_{M_1}^{\ \ RSM_2}+\frac{1}{2}\delta^{P_2}_{P_1}\delta^{Q_2}_{Q_1} R^{N_2}_{\ \ RSN_1}R_{M_1}^{\ \ RSM_2} \right)\nonumber\\
& & + \left(G^{M_2M_2}\delta^{Q_2}_{Q_1} R_{M_1}^{\ \ N_2 P_2 P}R_{PN_1P_1M_2}+\frac{1}{2}G^{M_2M_2}\delta^{Q_2}_{N_1} R_{M_1}^{\ \ N_2 P_2 P}R_{PM_2P_1Q_1}\right)\Biggr],
\nonumber\\
& & A_2= \Biggl[\left(G^{M_2M_2}G^{Q_2Q_2}G^{N_2N_1}G^{P_2P_1}R^{Q_1}_{\ \ RS Q_2}+\frac{1}{2}G^{M_2M_2}G^{N_2N_2}G^{P_2P_1}G^{Q_2Q_1}R^{N_1}_{\ \ RS N_2} \right)R_{M_2}^{\ \ RS M_1} 
\nonumber\\
& & +\left(G^{Q_2M_1}G^{N_2N_2}G^{P_2P_2}R^{M_2 N_1 P_1 K}+\frac{1}{2}G^{Q_2M_1}G^{N_2N_2}G^{P_2P_2}R^{M_2 K P_1 Q_1} \right)R^{N_1}_{\ \ N_2 P_2 K}\nonumber\\
& & +\left(G^{M_2Q_1}R^{H N_1 P_1 Q_2}+\frac{1}{2}G^{M_2N_1}R^{H Q_2 P_1 Q_1} \right)R_{H}^{\ \ N_2 P_2 M_1}\Biggr],\nonumber\\
& & 
A_3= \Biggl[G^{N_1 N_1}G^{P_1 P_1}G^{Q_1 Q_1}\Biggl[\Biggl(\delta^{M_2}_{Q_1}\delta^{Q_2}_Q R^{M_1 N_2 P_2 K}+G^{M_2 M_1}G^{N_2 N_2}G^{P_2 P_2}G^{Q_2 K} R_{Q_1 N_2 P_2 Q} \nonumber\\
& & + \frac{1}{2}\delta^{M_2}_{Q_1}\delta^{N_2}_Q R^{M_1 K P_2 Q_2}+\frac{1}{2}G^{M_2 M_1}G^{N_2 K}G^{P_2 P_2}G^{Q_2 Q_2} R_{Q_1 Q P_2 Q_2}\Biggr)R^Q_{\ \ N_1 P_1 K} \nonumber\\
& & + G^{M_2 M_2} \delta^{N_2}_{N_1}\delta^{P_2}_{P_1}\left(R_{Q_1 MN M_2}R^{M_1 MN Q_2}+\frac{1}{2}R_{Q_1 M_2 MN}R^{M_1 Q_2 MN}\right)
\Biggr] \Biggr], \nonumber\\
& & A_4=G^{M_1 M_1}\Biggl[\Biggl(\delta^{M_2}_P \delta^{Q_2}_{M_1}R^{H N_2 P_2 Q_1}+G^{M_2 H}G^{N_2 N_2}G^{P_2 P_2}G^{Q_2 Q_1}R_{P N_2 P_2 M_1}\nonumber\\
& & + \frac{1}{2} \delta^{M_2}_P \delta^{N_2}_{M_1}R^{HQ_1 P_2 Q_2}+\frac{1}{2}G^{M_2 H}G^{N_2 Q_1}G^{P_2 P_2}G^{Q_2 Q_2}R_{P M_1 P_2 Q_2}\Biggr)R_H^{\ \ N_1 P_1 P} \nonumber\\
& & + G^{N_2 N_1}G^{P_2 P_1}G^{Q_2 Q_2}\left(R_{Q_2 MN M_1}R^{M_2 MN Q_1}+\frac{1}{2}R_{Q_2  M_1 MN}R^{M_2 Q_1 MN} \right)\Biggr].
\end{eqnarray}
}
\item To derive the equation of motion associated with embedding $t(r),$ we must compute the derivatives that result from equation (\ref{Lag-total-HM}):
{\footnotesize
\begin{eqnarray}
\label{DLag-tprime}
& &\frac{\delta {\cal L}^{\rm HM}_{\rm Total}}{\delta t'(r)} =\beta  \Biggl(\frac{\left(r-r_h\right){}^{3/2} p_4^{\beta }\left(r_h\right) \left({{\cal A}_1} t'(r)+{{\cal A}_2} t''(r)\right)}{N^{7/10} t'(r)^2}+\frac{p_5^{\beta }\left(r_h\right) \left({{\cal A}_3}
   t'(r)+{{\cal A}_4} t''(r)\right)}{N^{7/10} t'(r)^2}+\frac{\left(r-r_h\right){}^{13/2} p_7^{\beta }\left(r_h\right)}{N^{57/10} t'(r)^4}\nonumber\\& &
 \hskip 0.6in  +\frac{ p_6^{\beta
   }\left(r_h\right)}{\left(r-r_h\right)N^{7/10} t'(r)} +\frac{\left(r-r_h\right){}^{3/2} p_1^{\beta }\left(r_h\right)}{N^{5/4} t'(r)}+\frac{N^{3/10} p_8^{\beta }\left(r_h\right)}{\sqrt{r-r_h} t'(r)^3}+\frac{N^{3/10}
   p_9^{\beta }\left(r_h\right)}{\sqrt{r-r_h} t'(r)^3}+\frac{p_2^{\beta }\left(r_h\right)}{\sqrt{r-r_h} t'(r)^3}\nonumber\\& & \hskip 0.6in
   +\frac{p_3^{\beta }\left(r_h\right)}{\sqrt{r-r_h} t'(r)^3}
   +N^{3/10}
   \left(r-r_h\right){}^{3/2} {Y_5}\left(r_h\right) t'(r)\Biggr)+N^{3/10} \left(r-r_h\right){}^{5/2} p_1\left(r_h\right) t'(r),
\end{eqnarray}
}
and 
\begin{eqnarray}
\label{DLag-tprimeprime}
& &\frac{\delta {\cal L}^{\rm HM}_{\rm Total}}{\delta t''(r)} =\frac{\beta  {\cal F}^{\beta}({r_h}) (r-{r_h})^{5/2}}{N^{7/10} t'(r)},
\end{eqnarray}
where 
{\footnotesize
\begin{eqnarray}
\label{fi[rh]-HM}
& &
p_1(r_h)=\kappa_{p_1}\left(\frac{M N_f^{5/3} g_s^{7/3} \sqrt{\kappa _{\alpha }} \kappa _{\lambda } \log \left(r_h\right) \left(\log (N)-9 \log \left(r_h\right)\right) \left(\log (N)-3 \log\left(r_h\right)\right){}^{5/6}}{r_h^{3/2} \sqrt{\kappa _{\sigma }}}\right),\nonumber\\
   & & p_1^\beta(r_h)=-\kappa_{p_1^\beta}\left(\frac{M^2 \sqrt[4]{N} x_R^2 N_f^{4/3} g_s^{17/6} \sqrt{\kappa _{\alpha }} \kappa _U \log \left(r_h\right) \left(\log (N)-45 \log \left(r_h\right)\right)}{r_h^{5/2} \sqrt{\kappa _{\sigma }} \left(\log
   (N)-3 \log \left(r_h\right)\right){}^{5/2}}\right),\nonumber\\
   & & p_2^\beta(r_h)= \kappa_{p_2^\beta} \left(\frac{M^2 x_R^2 N_f^{4/3} g_s^{23/6} \kappa _U \log \left(r_h\right) \sqrt{\kappa _{\alpha } \kappa _{\sigma }} \left(\log (N)-45 \log \left(r_h\right)\right)}{r_h^{9/2} \left(\log (N)-3 \log
   \left(r_h\right)\right){}^{5/2}}\right),\nonumber\\
   & & p_3^\beta(r_h)=-\kappa_{p_3^\beta}\Biggl(\frac{M^2 x_R^2 (540 \log (2)-107) N_f^{4/3} g_s^{23/6} \sqrt{\kappa _{\alpha } \kappa _{\sigma }}}{r_h^{9/2} \log ^2\left(r_h\right) \left(\log (N)-3 \log\left(r_h\right)\right){}^{5/2}} \nonumber\\
   & & \times \left(28 \kappa _V \left(\log (N)-3 \log \left(r_h\right)\right){}^3-27 \kappa _U \log ^3\left(r_h\right) \left(\log (N)-45 \log \left(r_h\right)\right)\right)\Biggr),\nonumber\\
   & & p_4^\beta(r_h)=\kappa_{p_4^\beta}F(g_s,r_h,N_f,M); \ p_5^\beta(r_h)=\kappa_{p_5^\beta}F(g_s,r_h,N_f,M),\nonumber\\
   & & p_6^\beta(r_h)=-\kappa_{p_6^\beta}\left(\frac{M x_R^4 \sqrt[3]{N_f} g_s^2 \kappa _W \log \left(r_h\right) \left(\kappa _{\alpha } \kappa _{\sigma }\right){}^{5/2} \left(\log (N)-9 \log \left(r_h\right)\right) \sqrt[6]{\log (N)-3 \log
   \left(r_h\right)}}{ r_h^{5/2} \kappa _{\sigma }^5}\right),\nonumber\\
   & & p_7^\beta(r_h)=-\kappa_{p_7^\beta}\left(\frac{M \sqrt[3]{N_f} r_h^{25/2} \kappa _Z \log \left(r_h\right) \left(\kappa _{\alpha } \kappa _{\sigma }\right){}^{5/2} \left(\log (N)-9 \log \left(r_h\right)\right) \sqrt[6]{\log (N)-3 \log
   \left(r_h\right)}}{g_s^3 \kappa _{\sigma }^{10}}\right),\nonumber\\
   & & p_8^\beta(r_h)=\kappa_{p_8^\beta}\left(\frac{M x_R^4 \sqrt[3]{N_f} g_s^3 \kappa _W \log \left(r_h\right) \left(\kappa _{\alpha } \kappa _{\sigma }\right){}^{5/2} \left(\log (N)-9 \log \left(r_h\right)\right) \sqrt[6]{\log (N)-3 \log
   \left(r_h\right)}}{r_h^{9/2} \kappa _{\sigma }^4}\right),\nonumber\\
   & & p_9^\beta(r_h)=\kappa_{p_9^\beta}\left(\frac{M \sqrt[3]{N_f} g_s^3 \kappa _Z \log \left(r_h\right) \left(\kappa _{\alpha } \kappa _{\sigma }\right){}^{5/2} \left(\log (N)-9 \log \left(r_h\right)\right) \sqrt[6]{\log (N)-3 \log
   \left(r_h\right)}}{r_h^{9/2} \kappa _{\sigma }^4}\right),\nonumber\\
   & & {\cal F}^\beta=-\kappa_{{\cal F}^\beta}\left(\frac{M \sqrt[3]{N_f} g_s^2 \kappa _{\alpha }^{5/2} \log \left(r_h\right) \left(\log (N)-9 \log \left(r_h\right)\right) \sqrt[6]{\log (N)-3 \log \left(r_h\right)} \left(x_R^4 \kappa _W+2 \kappa
   _Z\right)}{r_h^{5/2} \kappa _{\sigma }^{5/2}}\right),\nonumber\\
\end{eqnarray}
}
where $\kappa_{p_1},\kappa_{p_{i=1,..,9}^\beta}$ and $\kappa_{{\cal F}^\beta}$ correspond to the numerical factors. For the action (\ref{Lag-total-HM}) and utilizing the equations (\ref{DLag-tprime}) and (\ref{DLag-tprimeprime}), the equation of motion associated with the embedding $t(r)$ is obtained as: 
{\scriptsize
\begin{eqnarray}
\label{EOM-HM}
& &\hskip -0.3in \beta  \left(-\frac{N^{3/10} p_8^{\beta }\left(r_h\right)}{2 (r-r_h)^{3/2} t'(r)^3}-\frac{N^{3/10} p_9^{\beta }\left(r_h\right)}{2 (r-r_h)^{3/2} t'(r)^3}-\frac{3 N^{3/10} t''(r) p_8^{\beta
   }\left(r_h\right)}{\sqrt{r-r_h} t'(r)^4}-\frac{3 N^{3/10} t''(r) p_9^{\beta }\left(r_h\right)}{\sqrt{r-r_h} t'(r)^4}\right)+N^{3/10} (r-r_h)^{5/2} p_1\left(r_h\right)
   t''(r) \nonumber\\
 & & \hskip -0.3in  +\frac{5}{2} N^{3/10} (r-r_h)^{3/2} p_1\left(r_h\right) t'(r)=-\frac{\beta  M \sqrt[3]{N_f} g_s^2 \sqrt{r-r_h} \kappa _{\alpha }^{5/2} \log \left(r_h\right) \left(\log (N)-9 \log
   \left(r_h\right)\right) \sqrt[6]{\log (N)-3 \log \left(r_h\right)} \left(x_R^4 \kappa _W+2 \kappa _Z\right)}{4 N^{7/10} r_h^{5/2} \kappa _{\sigma }^{5/2} t'(r)^3}\nonumber\\
   & & \times \left(8 \left(r-r_h\right){}^2 t''(r)^2-4 \left(r-r_h\right) t'(r) \left(\left(r-r_h\right)
   t^{(3)}(r)+5 t''(r)\right)+15 t'(r)^2\right).
\end{eqnarray}
} 
\end{itemize}
where EOM=$\frac{d \left(\frac{\delta {\cal L}}{\delta t'(r)}\right)}{dr} = \frac{d^2\left(\frac{\delta{\cal L}}{\delta t''(r)}\right)}{dr^2}$

\section{Island Surface Miscellania}
\label{ISM-PCM}
We enumerate the many functions that emerge from the entanglement entropy of the island surface at ${\cal O}(\beta)$ in this appendix. We additionally calculated the derivatives of the Lagrangian associated with the island surface here, which were utilized to calculate the equation of motion concerning the island surface's embedding.
\begin{itemize}
\item
The following terms ($\lambda_{3,4}(r)$) exist in the Wald entanglement entropy term (first term of holographic entanglement entropy (\ref{HD-Entropy-IS})) of the island surface: 
{\footnotesize
\begin{eqnarray}
\label{lambda3-4}
& & 
 \lambda_3(r)= \kappa_{\lambda_3}\frac{ M^7 N^{7/10} N_f^{7/3} g_s^7 r_h^{14} \log (N) \log ^7(r) (5 \log (N)-12 \log (r))^3}{ r^{16} \alpha _{\theta _1}^3 \alpha
   _{\theta _2}^2 (\log (N)-3 \log (r))^{14/3}} \nonumber\\
   & & \sim \frac{M^7 \sqrt[20]{N} \log (2) (\log (64)-1) N_f^{7/3} g_s^{31/4} r_h^{14} \kappa _{\lambda _3} \log (N) \log ^7(r) (5 \log (N)-12 \log (r))^3}{r^{16} (\log (N)-3 \log (r))^{14/3}},\nonumber\\
   & & 
   \lambda_4(r)=\kappa_{\lambda_4}\frac{ M^3 N^{7/10} \sqrt[3]{N_f} g_s^3 r_h^{14} \log (N) \log ^3(r) (5 \log (N)-12 \log (r))}{ r^{16} \alpha _{\theta _1}^3 \alpha _{\theta _2}^2
   (\log (N)-3 \log (r))^{8/3}} \nonumber\\
 & & \sim  \frac{M^3 \sqrt[20]{N} \log (2) (\log (64)-1) \sqrt[3]{N_f} g_s^{15/4} r_h^{14} \kappa _{\lambda _4} \log (N) \log ^3(r) (5 \log (N)-12 \log (r))}{r^{16} (\log (N)-3 \log (r))^{8/3}},
\end{eqnarray}
}
where $\kappa_{\lambda_3}$ and $\kappa_{\lambda_4}$ being the numerical factors. 

\item 
The $r$ dependent functions appeared in the anomaly term (the second term of holographic entanglement entropy (\ref{HD-Entropy-IS})) are as follows:
{\footnotesize
\begin{eqnarray}
\label{Z-W-U-V-1}
& & 
Z_1(r)=\kappa_{Z_1}\frac{ M N^{37/10} g_s^{35/6} r_h^2 \log (N) \log (r)}{ r^{16} \alpha _{\theta _1}^3 \alpha _{\theta _2}^2 N_f^{4/3} (\log (N)-3 \log (r))^{7/3}} \nonumber\\
& & \sim \frac{M N^{61/20} \log (2) (\log (64)-1) g_s^{79/12} r_h^2 \log (N) \log (r)  \kappa _{Z_1}}{r^{16} N_f^{4/3} (\log (N)-3 \log
   (r))^{7/3}},\nonumber\\
   & & W_1(r)=\kappa_{W_1}\frac{ M N^{37/10} g_s^{35/6} \log (N) \log (r)}{ r^8 \alpha _{\theta _1}^3 \alpha _{\theta _2}^2 N_f^{4/3} r_h^6 (\log (N)-3 \log (r))^{7/3}} \nonumber\\
  & & \sim \frac{ M N^{61/20} \log (2) (\log (64)-1) g_s^{79/12} \kappa _{W_1} \log (N) \log (r)}{ r^8 N_f^{4/3} r_h^6 (\log (N)-3 \log (r))^{7/3}},
   \nonumber
   \end{eqnarray}
   \begin{eqnarray}
& & U_1(r)=\kappa_{U_1}\frac{M^2 N N_f g_s^3 r_h^{12} \log ^2(N) \log ^3(r) (\log (N)-17 \log (r)) (\log (N)-9 \log (r))}{ r^{15} \alpha _{\theta _1} \alpha _{\theta _2}^6 (\log (N)-3 \log (r))^4}\nonumber\\
& & \sim \frac{M^2 N_f g_s^{15/4} r_h^{12} \kappa _{U_1} \log ^2(N) \log ^3(r) (\log (N)-17 \log (r)) (\log (N)-9 \log (r))}{  \sqrt[4]{N} r^{15} (\log (N)-3 \log (r))^4},\nonumber\\
& & V_1(r)=-\kappa_{V_1}\frac{M^2 N N_f g_s^3 r_h^8 \log ^5(N) \log ^3(r) (\log (N)-6 \log (r))}{ r^{11} \alpha _{\theta _1} \alpha _{\theta _2}^6 (\log (N)-9 \log (r))^3 (\log (N)-3 \log (r))^3} \nonumber\\
& & \sim -\frac{M^2  N_f g_s^{15/4} r_h^8 \kappa _{V_1} \log ^5(N) \log ^3(r) (\log (N)-6 \log (r))}{\sqrt[4]{N} r^{11} (\log (N)-9 \log (r))^3 (\log (N)-3 \log (r))^3},
\end{eqnarray}
}
where $\kappa_{Z_1},\kappa_{W_1},\kappa_{U_1}$ and $\kappa_{V_1}$ are the numerical pre-factors including $\left(\frac{8}{q_\alpha +1}\right)$.
\item
For the action associated with the island surface's embedding (\ref{Lag-total-IS}), we obtained:
{\footnotesize
\begin{eqnarray}
\label{dLagoverdx}
& & \frac{\delta {\cal L}_{\rm {Total}}^{\rm {IS}}}{\delta x(r)} = \beta  \Biggl(\frac{N^{3/10} f^\beta_1(r_h) x(r)}{\sqrt{r-r_h}}+\left(\frac{ f^\beta_2(r_h)}{\sqrt{r-r_h}}+\frac{
   f^\beta_3(r_h)}{N^{1/4}\sqrt{r-r_h}}\right)\frac{x(r)}{x'(r)^2}+\frac{N^{3/10} f^\beta_4(r_h)
   x(r)^3}{\sqrt{r-r_h} x'(r)^2}\nonumber\\
  & & +\frac{N^{13/10} x(r) Y_1(r_h)}{\sqrt{r-r_h}}\Biggr)  +\frac{N^{13/10}
   f_1(r_h) x(r)}{\sqrt{r-r_h}}, 
\end{eqnarray}
} 
{\footnotesize
\begin{eqnarray}
\label{dLagoverdxprime}
& & \frac{\delta{\cal L}_{\rm {Total}}^{\rm {IS}}}{\delta x'(r)} = \beta  \Biggl[\frac{{F^\beta_1}(r_h) \sqrt{r-r_h} x(r)^2 x'(r)}{N^{7/10}} +\frac{F^\beta_{10}(r_h) \sqrt{r-r_h} x(r)^4}{N^{7/10} x'(r)}+\frac{F^\beta_{11}(r_h)
   \sqrt{r-r_h}}{N^{7/10} x'(r)}+\frac{N^{3/10} F^\beta_{12}(r_h) x(r)^4}{\sqrt{r-r_h}
   x'(r)^3}\nonumber\\
& &+\frac{N^{3/10} F^\beta_{13}(r_h)}{\sqrt{r-r_h} x'(r)^3}   +\frac{F^\beta_2(r_h)
   \sqrt{r-r_h} x(r)^2}{N x'(r)} +\frac{ {F^\beta_3}(r_h) \sqrt{r-r_h}
   x(r)^2}{N^{5/4} x'(r)}+\frac{{F^\beta_4}(r_h) x(r)^2}{\sqrt{r-r_h} x'(r)^3}+\frac{
   {F^\beta_5}(r_h) x(r)^2}{N^{1/4}\sqrt{r-r_h} x'(r)^3}\nonumber\\
  & & +\frac{{F^\beta_6}(r_h) (r-r_h)^{3/2}
   x(r)^4 \left(2 x'(r)+r x''(r)\right)}{N^{7/10} x'(r)^2}
   +\frac{{F^\beta_7}(r_h) (r-r_h)^{3/2} \left(2
   x'(r)+r x''(r)\right)}{N^{7/10} x'(r)^2}\nonumber\\
   & &+\frac{{F^\beta_8}(r_h) \sqrt{r-r_h} x(r)^4}{N^{7/10}
   x'(r)}+\frac{{F^\beta_9}(r_h) \sqrt{r-r_h}}{N^{7/10} x'(r)} +N^{3/10} x(r)^2{Y_3}(r_h) x'(r)\Biggr]+N^{3/10} {F_1}(r_h) \sqrt{r-r_h} x(r)^2 x'(r),\nonumber\\
\end{eqnarray}
}
and,
\begin{eqnarray}
\label{dLagoverdxprimexprime}
& & \frac{\delta{\cal L}_{\rm {Total}}^{\rm {IS}}}{\delta x^{\prime\prime}(r)} = \beta   \left(\frac{{\cal F}^\beta_1(r_h) (r-r_h)^{3/2}}{N^{7/10}}+\frac{{\cal F}^\beta_2(r_h)
   (r-r_h)^{3/2}}{N^{7/10}}\right),
   \end{eqnarray}
 where $F_1(r_h),F^\beta_{i=1,..,13}(r_h),f_1(r_h),f^\beta_{i=1,..,4}(r_h)$, ${\cal F}^\beta_{i=1,2}(r_h)$ and $Y_{i=1,3}(r_h)$ appearing in (\ref{dLagoverdx}), (\ref{dLagoverdxprime}) and (\ref{dLagoverdxprimexprime}) are:
 {\footnotesize
 \begin{eqnarray}
 \label{Fbetai[rh]s}
 & & F_1(r_h)=\frac{\kappa_{F_1} M \log (2) (\log (64)-1) N_f^{5/3} g_s^{7/3} \sqrt{r_h} \kappa _{\alpha } \kappa _{\lambda _5} \log (N) \log \left(r_h\right) \left(\log (N)-3 \log \left(r_h\right)\right){}^{2/3}}{\sqrt{\kappa _{\alpha } \kappa _{\sigma }}},\nonumber\\
 & & F_1^\beta(r_h) =\frac{\kappa_{F^\beta_1}M^3 \log (2) (\log (64)-1) N_f^{2/3} g_s^{17/6} \sqrt{r_h} \kappa _{\alpha } \log (N) \log ^3\left(r_h\right) \left(5 \log (N)-12 \log \left(r_h\right)\right)}{\sqrt{\kappa _{\alpha } \kappa _{\sigma }} \left(\log (N)-3 \log \left(r_h\right)\right){}^{13/3}} \nonumber\\
   & & \times \left( 81 \left(16+\sqrt{2}\right) M^4 N_f^2 g_s^4 \kappa _{\lambda _3} \log ^4\left(r_h\right) \left(5 \log (N)-12 \log \left(r_h\right)\right){}^2+4096 \left(4+\sqrt{2}\right) \pi ^4 \kappa _{\lambda _4}
   \left(\log (N)-3 \log \left(r_h\right)\right){}^2\right),\nonumber\\
 & & F_2^\beta(r_h) = -\frac{\kappa_{F^\beta_2}M^2 g_s^{3/2} \kappa _{\alpha } \kappa _{U_1} \log ^2(N) \log ^3(r) \left(\log (N)-17 \log \left(r_h\right)\right) \left(\log (N)-9 \log \left(r_h\right)\right) }{\sqrt{r_h} \sqrt{\kappa _{\alpha } \kappa _{\sigma }} \left(\log (N)-3 \log
   \left(r_h\right)\right){}^5}, \nonumber\\
   & & \times \left(N_f g_s
   \left(\log (N)-3 \log \left(r_h\right)\right)\right){}^{4/3}
   \nonumber\\
 & & F_3^\beta(r_h) =\frac{\kappa_{F^\beta_3}M^4 N_f^{7/3} g_s^{79/12} \kappa _{\alpha } \log ^4(N) \log ^6\left(r_h\right)}{ r_h^{7/2} \sqrt{\kappa _{\alpha } \kappa _{\sigma }}
   \left(\log (N)-9 \log \left(r_h\right)\right){}^6 \left(\log (N)-3 \log \left(r_h\right)\right){}^{23/3}} \nonumber\\
   & & \times \Biggl(\kappa _{V_1} \log ^3(N) \left(-9 \log (N) \log \left(r_h\right)+18 \log ^2\left(r_h\right)+\log ^2(N)\right)\nonumber\\
   & &-6 \kappa _{U_1} \left(\log (N)-17 \log \left(r_h\right)\right) \left(\log (N)-9 \log
   \left(r_h\right)\right){}^4\Biggr){}^2,\nonumber\\
 & & F_4^\beta(r_h) =\frac{\kappa_{F^\beta_4}M^2 N_f^{4/3} g_s^{23/6} \kappa _{U_1} \log ^2(N) \log ^3\left(r_h\right) \sqrt{\kappa _{\alpha } \kappa _{\sigma }} \left(\log (N)-17 \log \left(r_h\right)\right) \left(\log
   (N)-9 \log \left(r_h\right)\right)}{r_h^{7/2} \left(\log (N)-3 \log \left(r_h\right)\right){}^{11/3}},\nonumber\\
   & &  F_5^\beta(r_h) = -\frac{\kappa_{F^\beta_5}M^4  g_s^{21/4} \log ^4(N) \sqrt{\kappa _{\alpha } \kappa _{\sigma }} \left(N_f g_s \left(\log (N)-3 \log \left(r_h\right)\right)\right){}^{7/3}}{r_h^{13/2} \left(\log (N)-9 \log \left(r_h\right)\right){}^6 \left(\log (N)-3 \log \left(r_h\right)\right){}^{10}}  \nonumber\\
   & & \Biggl(\kappa _{V_1} \log ^3(N) \log ^3\left(r_h\right) \left(-9 \log (N) \log \left(r_h\right)+18 \log ^2\left(r_h\right)+\log ^2(N)\right)\nonumber\\
& &   -6 \kappa _{U_1} \log ^3(r) \left(\log (N)-17 \log
   \left(r_h\right)\right) \left(\log (N)-9 \log \left(r_h\right)\right){}^4\Biggr){}^2;\nonumber\\
   & & F_6^\beta(r_h) =-\frac{\kappa_{F^\beta_6} M  \sqrt[3]{N_f} g_s^2 \kappa _{W_1} \log (N) \log \left(r_h\right) \left(\kappa _{\alpha } \kappa _{\sigma }\right){}^{5/2}}{r_h^{5/2} \kappa _{\sigma }^5 \left(\log (N)-3 \log \left(r_h\right)\right){}^{2/3}},\nonumber\\
    & &   F_7^\beta(r_h) =-\frac{\kappa_{F^\beta_7} M \log (2) (\log (64)-1) \sqrt[3]{N_f} g_s^2 \kappa _{Z_1} \log (N) \log \left(r_h\right) \left(\kappa _{\alpha } \kappa _{\sigma }\right){}^{5/2}}{r_h^{5/2} \kappa _{\sigma }^5 \left(\log (N)-3 \log \left(r_h\right)\right){}^{2/3}}, \nonumber\\   
 & & F_8^\beta(r_h) =\frac{\kappa_{F^\beta_8} M \log (2) (\log (64)-1) \sqrt[3]{N_f} g_s^2 \kappa _{W_1} \log (N) \log \left(r_h\right) \left(\kappa _{\alpha } \kappa _{\sigma }\right){}^{5/2}}{r_h^{3/2} \kappa _{\sigma }^5 \left(\log (N)-3 \log \left(r_h\right)\right){}^{2/3}},\nonumber\\
 & & F_9^\beta(r_h) = \frac{\kappa_{F^\beta_9} M \log (2) (\log (64)-1) \sqrt[3]{N_f} g_s^2 \kappa _{Z_1} \log (N) \log \left(r_h\right) \left(\kappa _{\alpha } \kappa _{\sigma }\right){}^{5/2}}{ r_h^{3/2} \kappa _{\sigma }^5 \left(\log (N)-3 \log \left(r_h\right)\right){}^{2/3}},\nonumber\\
 & & F_{10}^\beta(r_h) = -\frac{\kappa_{F^\beta_{10}} M \log (2) (\log (64)-1) \sqrt[3]{N_f} g_s^2 \kappa _{W_1} \log (N) \log \left(r_h\right) \left(\kappa _{\alpha } \kappa _{\sigma }\right){}^{5/2}}{r_h^{3/2} \kappa _{\sigma }^5 \left(\log (N)-3 \log \left(r_h\right)\right){}^{2/3}},\nonumber\\
 & & F_{11}^\beta(r_h) = -\frac{\kappa_{F^\beta_{11}}M \log (2) (\log (64)-1) \sqrt[3]{N_f} g_s^2 \kappa _{Z_1} \log (N) \log \left(r_h\right) \left(\kappa _{\alpha } \kappa _{\sigma }\right){}^{5/2}}{r_h^{3/2} \kappa _{\sigma }^5 \left(\log (N)-3 \log \left(r_h\right)\right){}^{2/3}}, \nonumber\\
   & &F_{12}^\beta(r_h) =-\frac{\kappa_{F^\beta_{12}} M \log (2) (\log (64)-1) \sqrt[3]{N_f} g_s^3 \kappa _{W_1} \log (N) \log \left(r_h\right) \left(\kappa _{\alpha } \kappa _{\sigma }\right){}^{5/2}}{r_h^{9/2} \kappa _{\sigma }^4 \left(\log (N)-3 \log \left(r_h\right)\right){}^{2/3}},
   \nonumber
 \end{eqnarray}
 }
 {\footnotesize
 \begin{eqnarray} 
   & & F_{13}^\beta(r_h) =-\frac{\kappa_{F^\beta_{13}} M \log (2) (\log (64)-1) \sqrt[3]{N_f} g_s^3 \kappa _{Z_1} \log (N) \log \left(r_h\right) \left(\kappa _{\alpha } \kappa _{\sigma }\right){}^{5/2}}{r_h^{9/2} \kappa _{\sigma }^4 \left(\log (N)-3 \log \left(r_h\right)\right){}^{2/3}};
 \end{eqnarray}
 }
{\footnotesize
\begin{eqnarray}
\label{fi[rh]s}
& & f_1(r_h)=\frac{\kappa_{f_1} M \log (2) (\log (64)-1) N_f^{5/3} g_s^{10/3} \kappa _{\lambda _5} \log (N) \log \left(r_h\right) \sqrt{\kappa _{\alpha } \kappa _{\sigma }} \left(\log (N)-3 \log
   \left(r_h\right)\right){}^{2/3}}{r_h^{5/2}},\nonumber\\
   & & f^\beta_1(r_h)=\frac{\kappa_{f^\beta_1}M^3  N_f^{2/3} g_s^{23/6} \log (N) \log ^3\left(r_h\right) \sqrt{\kappa _{\alpha } \kappa _{\sigma }} \left(5 \log (N)-12 \log \left(r_h\right)\right)}{ r_h^{5/2} \left(\log (N)-3 \log \left(r_h\right)\right){}^{13/3}} \nonumber\\
   & & \times \left(\frac{M^3 \log (2) (\log (64)-1) N_f^{2/3} g_s^{23/6} \log (N) \log ^3\left(r_h\right) \sqrt{\kappa _{\alpha } \kappa _{\sigma }} \left(5 \log (N)-12 \log \left(r_h\right)\right)}{1572864 \sqrt[3]{2}
   3^{5/6} \pi ^{77/12} r_h^{5/2} \left(\log (N)-3 \log \left(r_h\right)\right){}^{13/3}}\right),\nonumber\\
   & & f^\beta_2(r_h)=-\frac{\kappa_{f^\beta_2}M^2  N_f^{4/3} g_s^{23/6} \kappa _{U_1} \log ^2(N) \log ^3\left(r_h\right) \sqrt{\kappa _{\alpha } \kappa _{\sigma }} \left(\log (N)-17 \log \left(r_h\right)\right) \left(\log
   (N)-9 \log \left(r_h\right)\right)}{r_h^{7/2} \left(\log (N)-3 \log \left(r_h\right)\right){}^{11/3}},\nonumber\\
   & & f^\beta_3(r_h)=\frac{\kappa_{f^\beta_3}M^4 g_s^{21/4} \log ^4(N) \sqrt{\kappa _{\alpha } \kappa _{\sigma }} \left(N_f g_s \left(\log (N)-3 \log \left(r_h\right)\right)\right){}^{7/3}}{r_h^{13/2} \left(\log (N)-9 \log \left(r_h\right)\right){}^6 \left(\log (N)-3 \log \left(r_h\right)\right){}^{10}} \nonumber\\
   & & \times \Biggl(\kappa _{V_1} \log ^3(N) \log ^3\left(r_h\right) \left(-9 \log (N) \log \left(r_h\right)+18 \log ^2\left(r_h\right)+\log ^2(N)\right) \nonumber\\
  & & -6 \kappa _{U_1} \log ^3(r) \left(\log (N)-17 \log
   \left(r_h\right)\right) \left(\log (N)-9 \log \left(r_h\right)\right){}^4\Biggr){}^2,\nonumber\\
   & & f^\beta_4 (r_h)=\frac{\kappa_{f^\beta_4} M \log (2) (\log (64)-1) \sqrt[3]{N_f} g_s^3 \kappa _{W_1} \log (N) \log \left(r_h\right) \left(\kappa _{\alpha } \kappa _{\sigma }\right){}^{5/2}}{r_h^{9/2} \kappa _{\sigma }^4 \left(\log (N)-3 \log \left(r_h\right)\right){}^{2/3}}, \nonumber\\
   & & {\cal F}^\beta_1(r_h)=-\frac{\kappa_{{\cal F}^\beta_1}M \log (2) (\log (64)-1) \sqrt[3]{N_f} g_s^2 \kappa _{\alpha }^{5/2} \kappa _{W_1} \log (N) \log \left(r_h\right)}{ r_h^{3/2} \kappa _{\sigma }^{5/2}
   \left(\log (N)-3 \log \left(r_h\right)\right){}^{2/3}},\nonumber\\
   & &{\cal F}^\beta_2(r_h) =-\frac{\kappa_{{\cal F}^\beta_1} M \log (2) (\log (64)-1) \sqrt[3]{N_f} g_s^2 \kappa _{\alpha }^{5/2} \kappa _{Z_1} \log (N) \log \left(r_h\right)}{r_h^{3/2} \kappa _{\sigma }^{5/2}\left(\log (N)-3 \log \left(r_h\right)\right){}^{2/3}},
\end{eqnarray}
}
and,  
{\footnotesize
 \begin{eqnarray}
 \label{Yi[rh]s}
 & & Y_1(r_h) = \frac{\kappa_{Y_1}{\cal C}_{\theta_1 x}M N_f^{5/3} g_s^{10/3} \log ^2\left(r_h\right) \sqrt{\kappa _{\alpha } \kappa _{\sigma }} \left(\log (N)-3 \log \left(r_h\right)\right){}^{2/3}}{ r_h^{3/2}},\nonumber\\
   & & Y_3(r_h) =\frac{\kappa_{Y_3} {\cal C}_{\theta_1 x} M  N_f^{5/3} g_s^{7/3} r_h^{3/2} \kappa _{\alpha } \log ^2\left(r_h\right) \left(\log (N)-3 \log \left(r_h\right)\right){}^{2/3}}{ \sqrt{\kappa _{\alpha } \kappa _{\sigma }}} ,
 \end{eqnarray}
 }
 where $\kappa_{F_1},\kappa_{F^\beta_{i=1,..,13}},\kappa_{f_1},\kappa_{f^\beta_{i=1,..,4}},\kappa_{{\cal F}^\beta_{1,2}},\kappa_{Y_{1,3}}$ represent the numerical factors and ${\cal C}_{\theta_1 x}$ being the integration constant.
 \end{itemize}

\section{Possible Terms Appearing in Holographic Entanglement Entropies}
\label{PT}
Within this appendix, we provided all of the terms that could possibly be obtained by differentiating the Lagrangian of the ${\cal M}$-theory dual, including ${\cal O}(R^4)$ corrections. We have included all of the words associated with the Hartman-Maldacena-like surface in \ref{PT-HM}, and all of the available terms of the island surface in \ref{PT-IS}. 
\subsection{Hartman-Maldacena-like Surface}
\label{PT-HM}
\begin{itemize}
\item {\bf $\sqrt{-g}$ Associated with the Induced Metric (\ref{metric-HM-t(r)})}:
{\scriptsize
\begin{eqnarray}
\label{sqrtminusg-HM}
& &
\sqrt{-g}\sim \frac{M N^{7/10} \sqrt{g_s} \left(N_f (\log (N)-3 \log (r))\right){}^{5/3} \sqrt{\alpha (r) \left(\sigma (r)-\left(1-\frac{r_h^4}{r^4}\right) t'(r)^2\right)} }{\alpha _{\theta _1}^3 \alpha _{\theta _2}^2}\nonumber\\
& & \times \Biggl(N_f g_s \left(r^2 (\log (N) (2 \log
   (r)+1)+3 (1-6 \log (r)) \log (r))-r_h^2 \left(\log (N) (2 \log (r)+1)-18 (6 r+1) \log ^2(r)\right)\right)\nonumber\\
   & &+8 \pi  \left(r_h^2-r^2\right) \log (r)\Biggr).
\end{eqnarray}
}
\item {\bf $\int d{\cal V}_9 \sqrt{-g}\frac{\partial J_0}{\partial R_{txtx}}$}:
{\footnotesize
\begin{eqnarray}
\label{Wald-J0-i}
& & \hskip -0.3in (i)\int d{\cal V}_9 \sqrt{-g}\left( \left(G^{xx}\right)^2 \left(G^{tt}\right)^2\left(R_{PxtQ} + \frac{1}{2}R_{PQtx}\right)R_{t}^{\ \ RSP}R^Q_{\ \ RSx}\right)\nonumber\\
& & \sim \int d{\cal V}_9 \sqrt{-g} \Biggl(-\frac{r_h^8 \left(-9 \log ^2(N) (\log (N)-6 \log (r))^2-13 (\log (N)-3 \log (r))^4-21 (\log (N)-3 \log (r))^2\right)}{\sqrt{N} r^{12} N_f^{10/3} \sqrt{g_s} (\log (N)-3 \log (r))^{22/3}}\Biggr);\nonumber\\
& & \hskip -0.3in (ii)\int d{\cal V}_9 \sqrt{-g}\left( R^{HxtK}R_H^{\ \ RSt}R^x_{\ \ RSK} + \frac{1}{2}R^{HKtx}R_H^{\ \ RSt}R^Q_{\ \ RSK} \right)\nonumber\\
& & \sim \int d{\cal V}_9 \sqrt{-g} \Biggl(\frac{1}{\sqrt{N} r^{12} N_f^{22/3} \sqrt{g_s} r_h^8 (\log (N)-3 \log (r))^{22/3}}\Biggr) ;
\nonumber\\
& & \hskip -0.3in (iii)\int d{\cal V}_9 \sqrt{-g}  \left(\left(G^{xx}\right)^2 G^{tt}\left(R_{PxtQ} + \frac{1}{2}R_{PQtx}\right)R_{t}^{\ \ RSP}R^Q_{\ \ RSx}\right)\nonumber\\
& & \hskip -.5in \sim \int d{\cal V}_9 \Biggl(\lambda_1(r) \sqrt{\alpha (r) \left(\sigma (r)-\left(1-\frac{r_h^4}{r^4}\right) t'(r)^2\right)}\Biggr) ;\nonumber\\
& & \hskip -0.3in (iv)\int d{\cal V}_9 \sqrt{-g} \left( G^{tt} \left(R^{HMNx}R_{PMNt} + \frac{1}{2}R^{HxMN}R_{PtMN}\right)R_H^{\ \ xtP}\right)\nonumber\\
& & \sim \int d{\cal V}_9 \Biggl(\lambda_2(r) \sqrt{\alpha (r) \left(\sigma (r)-\left(1-\frac{r_h^4}{r^4}\right) t'(r)^2\right)}\Biggr).
\end{eqnarray}
}
The equation (\ref{Wald-J0-i}) suggests that in the large-$N$ limit, (iii) and (iv) terms correspond to the most dominant terms.
\item {\bf $\int d{\cal V}_9 \sqrt{-g}\left(\frac{\partial^2 J_0}{\partial R_{xixj}\partial R_{xmxl}}\right) K_{tij}K_{tml}$}:
{\scriptsize
\begin{eqnarray}
& & \hskip -0.3in (i) \int d{\cal V}_9 \sqrt{-g} \left(G^{xx}\right)^2 \left(G^{mm}\right)\left(G^{ll}\right) \delta^i_m R_r^{~TSx} R^j_{~TSl}K_{tij}K_{tml}\nonumber\\ 
& & \sim \int d{\cal V}_9 \sqrt{-g}\Biggl(-\frac{M^2 N N_f g_s^3 \log ^3(r) (23 \log (N)-72 \log (r))^2 (29 \log (N)-72 \log (r)) (2 \log (N)-3 \log (r)) \sqrt{\alpha (r) \left(\sigma (r)-t'(r)^2\right)}}{r^4 \alpha _{\theta _1} \alpha _{\theta
   _2}^6 (\log (N)-9 \log (r))^3 (\log (N)-3 \log (r))^6 t'(r)^2} \Biggr),\nonumber\\
& & \hskip -0.3in (ii)\int d{\cal V}_9 \sqrt{-g} \left(G^{xx}\right)^3 \left(G^{ii}\right)\left(G^{jj}\right)\left(G^{mm}\right)\left(G^{ll}\right) R_{jmxQ} R^Q_{~~ixl}K_{tij}K_{tml} \sim 0, \nonumber\\
& & \hskip -0.3in (iii)\int d{\cal V}_9 \sqrt{-g}\frac{1}{2}\left(G^{xx}\right)^2 \left(G^{mm}\right)\left(G^{jl}\right)R_r^{~TSx} R^i_{~TSx}K_{tij}K_{tml} \sim 0,\nonumber\\
& & \hskip -0.3in (iv)\int d{\cal V}_9 \sqrt{-g}\frac{1}{2}\left(G^{xx}\right)^2\left(G^{ii}\right)\left(G^{jl}\right) R_r^{~TSx} R^m_{~~TSi} K_{tij}K_{tml}\nonumber\\
& &\sim \int d{\cal V}_9 \sqrt{-g}\Biggl( \frac{M^3 N^{13/10} \alpha _{\theta _1} N_f^2 g_s^{9/2} \log ^3(N) (\log (N)-23 \log (r))^3 \left(\log (N)+18 \log ^2(r)\right) \sqrt{\alpha (r) \left(\sigma (r)-t'(r)^2\right)}}{r^4 \alpha _{\theta
   _2}^{10} \log ^2(r) (\log (N)-9 \log (r))^3 (\log (N)-3 \log (r))^5 t'(r)^2}\Biggr),\nonumber\\
& & \hskip -0.3in (v)\int d{\cal V}_9 \sqrt{-g}\frac{1}{2} \left(G^{xx}\right)^3\left(G^{ii}\right)\left(G^{ll}\right)\left(G^{mm}\right)\left(G^{jj}\right) R_{jQxl} R^Q_{~~ixm}K_{tij}K_{tml} \sim \int d{\cal V}_9 \left(\frac{Z(r)}{2}{\cal L}_1 \right),\nonumber\\
& & \hskip -0.3in (vi)\int d{\cal V}_9 \sqrt{-g}\frac{1}{2}  \left(G^{xx}\right)^3\left(G^{jm}\right)\left(G^{ll}\right) R_{Pxxl} R_x^{~ixP}K_{tij}K_{tml} 
\nonumber\\
& & \sim \int d{\cal V}_9 \sqrt{-g}\Biggl(-\frac{M^2 N N_f g_s^3 \log ^6(N) (\log (N)-51 \log (r)) (\log (N)-23 \log (r)) \sqrt{\alpha (r) \left(1-\sigma (r) r'(t)^2\right)}}{r^4 \alpha _{\theta _1} \alpha _{\theta _2}^6 \log ^3(r) (\log (N)-9
   \log (r))^3 (\log (N)-3 \log (r))^5 t'(r)^2} \Biggr),\nonumber\\
   & & \hskip -0.3in (vii)\int d{\cal V}_9 \sqrt{-g} \left(G^{xx}\right)^2\left(G^{im}\right)\left(G^{ll}\right) R^{xMNj} R_{lMNx}K_{tij}K_{tml} \sim 0, \nonumber\\
   & & \hskip -0.3in (viii)\int d{\cal V}_9 \sqrt{-g}\left(G^{xx}\right)^2 \left(G^{jj}\right)\left(G^{im}\right) R^{xMNl} R_{jMNx}K_{tij}K_{tml} \sim 0,\nonumber\\
& & \hskip -0.3in (ix)\int d{\cal V}_9 \sqrt{-g} \left(G^{xx}\right)^3 \left(G^{ii}\right)\left(G^{jj}\right)\left(G^{ll}\right)\left(G^{mm}\right) R^Q_{~~mxj} R_{lixQ}K_{tij}K_{tml}\sim 0, \nonumber\\
& & \hskip -0.3in (x)\int d{\cal V}_9 \sqrt{-g}\frac{1}{2} \left(G^{xx}\right)^2 \left(G^{im}\right)\left(G^{ll}\right) R^{xjMN} R_{lxMN}K_{tij}K_{tml}\nonumber\\
& & \sim \int d{\cal V}_9 \sqrt{-g}\Biggl[-\frac{M N^{7/10} r^2 N_f^{8/3} g_s^{3/2} \log (r) (\log (N)-9 \log (r)) (\log (N)-3 \log (r))^{5/3} \sqrt{\alpha (r) \left(\sigma (r)-t'(r)^2\right)}}{\alpha _{\theta _1}^3 \alpha _{\theta _2}^2
   t'(r)^2} \nonumber\\
   & & \times \biggl( 76800 {G_1}(r)-\frac{29 \sqrt[3]{6} M^4 g_s^4 \log ^2(N) \log ^2(r) (\log (N)-12 \log (r))^2 (\log (N)-6 \log (r))^2 }{\pi ^{4/3} r^4 \alpha (r)^2 N_f^{2/3} (\log (N)-3 \log (r))^{20/3} \left(\sigma (r)-t'(r)^2\right)^2} \nonumber\\
   & & \hskip 0.5in \times \left(\sigma (r) \alpha '(r)-\alpha '(r) t'(r)^2+\alpha (r) \left(\sigma '(r)-2
   t'(r) t''(r)\right)\right)^2 \Biggr) \Biggr],\nonumber\\
& & \hskip -0.3in (xi)\int d{\cal V}_9 \sqrt{-g}\frac{1}{2} \left(G^{xx}\right)^2 \left(G^{im}\right)\left(G^{jj}\right) R^{xlMN} R_{jxMN}K_{tij}K_{tml}\nonumber\\
& & \sim \int d{\cal V}_9 \sqrt{-g}\Biggl[-\frac{M N^{7/10} r^2 N_f^{8/3} g_s^{3/2} \log (r) (\log (N)-9 \log (r)) (2 \log (N)-\log (r))^{5/3} \sqrt{\alpha (r) \left(\sigma (r)-t'(r)^2\right)}}{129600 \sqrt{2} 3^{5/6} \pi ^{19/6} \alpha _{\theta
   _1}^3 \alpha _{\theta _2}^2 t'(r)^2}\nonumber\\
   & &  \times \Biggl( 450 {G_2}(r)-\frac{87 \sqrt[3]{3} M^4 g_s^4 \log ^2(N) \log ^2(r) (\log (N)-12 \log (r))^2 (\log (N)-6 \log (r))^2 (\log (N)-3 \log (r))^2 }\nonumber\\
   & & \times \left(\sigma (r) \alpha '(r)-\alpha '(r) t'(r)^2+\alpha (r)
   \left(\sigma '(r)-2 t'(r) t''(r)\right)\right)^2{\pi^{4/3}} r^4 \alpha (r)^2 N_f^{2/3} (2 \log (N)-\log (r))^{26/3} \left(\sigma (r)-t'(r)^2\right)^2\Biggr) \Biggr],\nonumber
\end{eqnarray}
\begin{eqnarray}
& & \hskip -0.3in (xii)\int d{\cal V}_9 \sqrt{-g}\frac{1}{2}\left(G^{xx}\right)^3 \left(G^{ii}\right)\left(G^{jj}\right)\left(G^{ll}\right)\left(G^{mm}\right)R^Q_{~~mxi} R_{lQxj}K_{tij}K_{tml} \sim \int d{\cal V}_9 \left(\frac{Z(r)}{2}{\cal L}_1 \right),\nonumber\\
& & \hskip -0.3in (xiii)\int d{\cal V}_9 \sqrt{-g}\frac{1}{2}\left(G^{xx}\right)^3 \left(G^{jj}\right)\left(G^{il}\right)R_x^{~mxP} R_{Pxxj}K_{tij}K_{tml} \nonumber\\
& & \sim \int d{\cal V}_9 \sqrt{-g}\Biggl(-\frac{M^2 N N_f g_s^3 \log ^5(N) (5 \log (N)-211 \log (r)) (\log (N)-23 \log (r)) \sqrt{\alpha (r) \left(\sigma (r)-t'(r)^2\right)}}{r^4 \alpha _{\theta _1} \alpha _{\theta _2}^6 \log (r) (\log (N)-9
   \log (r))^3 (\log (N)-3 \log (r))^5 t'(r)^2} \Biggr),
\end{eqnarray}
}
where there is no $N$ dependence in $G_1(r)$ and $G_2(r)$. The terms (v) and (xii) are the most prominent in the large-$N$ limit among all the terms stated above.

\item {\bf $\int d{\cal V}_9 \sqrt{-g}\left(\frac{\partial^2 J_0}{\partial R_{titj}\partial R_{tmtl}}\right) K_{tij}K_{tml}$}:
{\scriptsize
\begin{eqnarray}
& & \hskip -0.3in (i)\int d{\cal V}_9 \sqrt{-g}\left(G^{tt}\right)^2 \left(G^{mm}\right)\left(G^{ll}\right) \delta^i_m R_t^{~TSt} R^j_{~TSl}K_{tij}K_{tml}\nonumber\\
& & \sim \int d{\cal V}_9 \sqrt{-g}\Biggl(-\frac{M^2 N N_f g_s^3 \log ^5(N) (23 \log (N)-72 \log (r))^2 (\log (N)-27 \log (r)) \sqrt{\alpha (r) \left(\sigma (r)-t'(r)^2\right)}}{r^4 \alpha _{\theta _1} \alpha _{\theta _2}^6 (\log (N)-9 \log
   (r))^4 (\log (N)-3 \log (r))^6 t'(r)^2} \Biggr),\nonumber\\
& & \hskip -0.3in (ii)\int d{\cal V}_9 \sqrt{-g} \left(G^{tt}\right)^3 \left(G^{ii}\right)\left(G^{jj}\right)\left(G^{mm}\right)\left(G^{ll}\right) R_{jmtQ} R^Q_{~~itl}K_{tij}K_{tml}\sim 0,\nonumber\\
& & \hskip -0.3in (iii)\int d{\cal V}_9 \sqrt{-g}\frac{1}{2}\left(G^{tt}\right)^2 \left(G^{mm}\right)\left(G^{jl}\right)R_t^{~TSt} R^i_{~TSt}K_{tij}K_{tml} \sim 0,\nonumber\\
& & \hskip -0.3in (iv)\int d{\cal V}_9 \sqrt{-g}\frac{1}{2}\left(G^{tt}\right)^2\left(G^{ii}\right)\left(G^{jl}\right) R_t^{~TSt} R^m_{~~TSi} K_{tij}K_{tml}\nonumber\\
& & \sim \int d{\cal V}_9 \sqrt{-g}\Biggl(\frac{M^3 N^{13/10} \alpha _{\theta _1} N_f^2 g_s^{9/2} \log ^3(N) (\log (N)-23 \log (r))^3 \left(\log (N)+18 \log ^2(r)\right) \sqrt{\alpha (r) \left(\sigma (r)-t'(r)^2\right)}}{1728 \sqrt{6} \pi ^{7/2}
   r^4 \alpha _{\theta _2}^{10} \log ^2(r) (\log (N)-9 \log (r))^3 (\log (N)-3 \log (r))^5 t'(r)^2} \Biggr),\nonumber\\
& & \hskip -0.3in (v)\int d{\cal V}_9 \sqrt{-g}\frac{1}{2} \left(G^{tt}\right)^3\left(G^{ii}\right)\left(G^{ll}\right)\left(G^{mm}\right)\left(G^{jj}\right) R_{jQtl} R^Q_{~~itm}K_{tij}K_{tml} \sim \int d{\cal V}_9\left(\frac{W(r)}{2}{\cal L}_2 \right),\nonumber\\
& & \hskip -0.3in (vi)\int d{\cal V}_9 \sqrt{-g}\frac{1}{2}  \left(G^{tt}\right)^3\left(G^{jm}\right)\left(G^{ll}\right) R_{Pttl} R_t^{~itP} K_{tij}K_{tml}
\nonumber\\
& & \sim \int d{\cal V}_9 \sqrt{-g}\Biggl(\frac{M^2 N N_f g_s^3 \log ^5(N) (5 \log (N)-211 \log (r)) (\log (N)-23 \log (r)) \sqrt{\alpha (r) \left(\sigma (r)-t'(r)^2\right)}}{ r^4 \alpha _{\theta _1} \alpha _{\theta _2}^6
   \log (r) (\log (N)-9 \log (r))^3 (\log (N)-3 \log (r))^5 t'(r)^2} \Biggr),\nonumber\\
& & \hskip -0.3in (vii)\int d{\cal V}_9 \sqrt{-g} \left(G^{tt}\right)^2\left(G^{im}\right)\left(G^{ll}\right) R^{tMNj} R_{lMNt} K_{tij}K_{tml} \sim 0,\nonumber\\
& & \hskip -0.3in (viii)\int d{\cal V}_9 \sqrt{-g}\left(G^{tt}\right)^2 \left(G^{jj}\right)\left(G^{im}\right) R^{tMNl} R_{jMNt}K_{tij}K_{tml} \sim 0,\nonumber\\
& & \hskip -0.3in (ix)\int d{\cal V}_9 \sqrt{-g} \left(G^{tt}\right)^3 \left(G^{ii}\right)\left(G^{jj}\right)\left(G^{ll}\right)\left(G^{mm}\right) R^Q_{~~mtj} R_{litQ}K_{tij}K_{tml}\sim 0,\nonumber\\
& & \hskip -0.3in (x)\int d{\cal V}_9 \sqrt{-g}\frac{1}{2} \left(G^{tt}\right)^2 \left(G^{im}\right)\left(G^{ll}\right) R^{tjMN} R_{ltMN}K_{tij}K_{tml}\nonumber\\
& & \sim \int d{\cal V}_9 \sqrt{-g}\Biggl[\frac{M^5 N^{7/10} N_f^2 g_s^{11/2} {G_3}(r) \log ^3(N) \log ^3(r) (\log (N)-24 \log (r)) (\log (N)-9 \log (r)) (\log (N)-6 \log (r))^2}{r^2 \alpha _{\theta _1}^3 \alpha _{\theta _2}^2 \alpha (r)^2
   (\log (N)-3 \log (r))^5}\nonumber\\
   & & \times \frac{\sqrt{\alpha (r) \left(\sigma (r)-t'(r)^2\right)} \left(\sigma (r) \alpha '(r)-\alpha '(r) t'(r)^2+\alpha (r) \left(\sigma '(r)-2 t'(r) t''(r)\right)\right)^2}{t'(r)^4 \left(\sigma
   (r)-t'(r)^2\right)^2} \Biggr],\nonumber
\end{eqnarray}
}
{\scriptsize
\begin{eqnarray}
& & \hskip -0.3in (xi)\int d{\cal V}_9 \sqrt{-g}\frac{1}{2} \left(G^{tt}\right)^2 \left(G^{im}\right)\left(G^{jj}\right) R^{tlMN} R_{jtMN}K_{tij}K_{tml}\nonumber\\
& & \sim \int d{\cal V}_9 \sqrt{-g}\Biggl[-\frac{M N^{7/10} r^2 N_f^{8/3} g_s^{3/2} \log (r) (\log (N)-9 \log (r)) (2 \log (N)-\log (r))^{5/3} \sqrt{\alpha (r) \left(\sigma (r)-t'(r)^2\right)}}{\alpha _{\theta _1}^3 \alpha _{\theta _2}^2
   t'(r)^2} \nonumber\\
& & \times \left( {G_4}(r)-\frac{52 \sqrt[3]{6} M^4 g_s^4 \log ^2(r) (\log (N)-12 \log (r))^2 \left(\sigma (r) \alpha '(r)-\alpha '(r) t'(r)^2+\alpha (r) \left(\sigma '(r)-2 t'(r) t''(r)\right)\right)^2}{\pi
   ^{4/3} r^4 \alpha (r)^2 N_f^{2/3} (\log (N)-3 \log (r))^{8/3} \left(\sigma (r)-t'(r)^2\right)^2} \right)\Biggr],\nonumber\\
& & \hskip -0.3in (xii)\int d{\cal V}_9 \sqrt{-g}\frac{1}{2}\left(G^{tt}\right)^3 \left(G^{ii}\right)\left(G^{jj}\right)\left(G^{ll}\right)\left(G^{mm}\right)R^Q_{~~mti} R_{lQtj}K_{tij}K_{tml}\sim \int d{\cal V}_9 \left(\frac{W(r)}{2}{\cal L}_2 \right),\nonumber\\
& & \hskip -0.3in (xiii)\int d{\cal V}_9 \sqrt{-g}\frac{1}{2}\left(G^{tt}\right)^3 \left(G^{jj}\right)\left(G^{il}\right)R_t^{~mtP} R_{Pttj}K_{tij}K_{tml} \nonumber\\
& & \sim \int d{\cal V}_9 \sqrt{-g}\Biggl(-\frac{M^2 N N_f g_s^3 \log ^6(N) (\log (N)-51 \log (r)) (\log (N)-23 \log (r)) \sqrt{\alpha (r) \left(\sigma (r)-t'(r)^2\right)}}{ r^4 \alpha _{\theta _1} \alpha _{\theta _2}^6 \log
   ^3(r) (\log (N)-9 \log (r))^3 (\log (N)-3 \log (r))^5 t'(r)^2}\Biggr),
\end{eqnarray}
}
where $G_3(r)$ and $G_4(r)$ have no $N$ dependence. The terms (v) and (xii) are the most dominant in the large-$N$ limit.
\item {\bf $\int d{\cal V}_9 \sqrt{-g}\left(\frac{\partial^2 J_0}{\partial R_{titj}\partial R_{xmxl}}\right)K_{tij}K_{tml}$}:
{\footnotesize
\begin{eqnarray}
& & \hskip -0.3in (i)\int d{\cal V}_9 \sqrt{-g}\left(G^{xx}\right)^2 \left(G^{mm}\right)\left(G^{ll}\right) \delta^i_m \delta^t_P \delta^t_R \delta^j_Q R_x^{~TSP} R^Q_{~TSl}K_{tij}K_{tml} \sim \int d{\cal V}_9 \left(U(r){\cal L}_3 \right),\nonumber\\
& & \hskip -0.3in (ii)\int d{\cal V}_9 \sqrt{-g} \left(G^{xx}\right)^2 \left(G^{ii}\right)\left(G^{jj}\right)\left(G^{mm}\right)\left(G^{ll}\right)\left(G^{tt}\right)\delta^t_x R_{jmxQ} R^Q_{~~itl}K_{tij}K_{tml} \sim 0,\nonumber\\
& & \hskip -0.3in (iii)\int d{\cal V}_9 \sqrt{-g}\left(G^{xx}\right)^2 \left(G^{mm}\right)\left(G^{ll}\right)\left(G^{tt}\right) \delta^i_T \delta^t_S \delta^j_l  R_{Pmxt} R_x^{~TSP}K_{tij}K_{tml} \sim 0,\nonumber\\
& & \hskip -0.3in (iv)\int d{\cal V}_9 \sqrt{-g}\frac{1}{2}\left(G^{xx}\right)^2\left(G^{mm}\right)\left(G^{ll}\right)\left(G^{tt}\right)\delta^i_T \delta^t_S \delta^j_m  R_{Ptxl} R_x^{~~TSP} K_{tij}K_{tml} \sim 0,\nonumber\\
& & \hskip -0.3in (v)\int d{\cal V}_9 \sqrt{-g} \left(G^{xx}\right)\left(G^{tt}\right)^2\left(G^{ii}\right)\left(G^{jl}\right) R_t^{mxP} R_{Pitx}K_{tij}K_{tml} \sim 0,\nonumber\\
& & \hskip -0.3in (vi)\int d{\cal V}_9 \sqrt{-g}\frac{1}{2}  \left(G^{tt}\right)^2\left(G^{xx}\right)\left(G^{jj}\right)\left(G^{iK}\right)\delta^l_K R_t^{mxP} R_{Pxtj} K_{tij}K_{tml} \sim 0.
\end{eqnarray}
}
Since, only first term is non-zero, and hence this contributes to the Lagrangian.
\item {\bf $\int d{\cal V}_9 \sqrt{-g}\left(\frac{\partial^2 J_0}{\partial R_{tixj}\partial R_{xmtl}}\right)K_{tij}K_{tml}$}:
{\footnotesize
\begin{eqnarray}
\label{list-second-der-J0-mix}
& & \hskip -0.3in (i)\int d{\cal V}_9 \sqrt{-g}\left(G^{xx}\right)^2 \left(G^{mm}\right)\left(G^{ll}\right) \delta^i_m  R_x^{~TSx} R^j_{~TSl}K_{tij}K_{tml}\sim \int d{\cal V}_9 \left(U(r){\cal L}_4 \right),\nonumber\\
& & \hskip -0.3in (ii)\int d{\cal V}_9 \sqrt{-g} \left(G^{xx}\right)^3 \left(G^{ii}\right)\left(G^{jj}\right)\left(G^{ll}\right)\left(G^{mm}\right) R_{jmxQ} R^Q_{~~ixl}K_{tij}K_{tml}
 \sim 0,\nonumber\\
& & \hskip -0.3in (iii)\int d{\cal V}_9 \sqrt{-g}\frac{1}{2}\left(G^{xx}\right) \left(G^{tt}\right)^2\left(G^{jm}\right)\left(G^{ll}\right)  R_x^{~ixP} R_{Pttl}K_{tij}K_{tml} \sim \int d{\cal V}_9 \left( V(r){\cal L}_4 \right),\nonumber\\
& & \hskip -0.3in (iv)\int d{\cal V}_9 \sqrt{-g}\frac{1}{2}\left(G^{xx}\right)^2\left(G^{tt}\right)\left(G^{jj}\right)\left(G^{il}\right)  R_t^{~mtP} R_{Pxxj} K_{tij}K_{tml} \sim \int d{\cal V}_9 \left( V(r){\cal L}_4 \right)
\end{eqnarray}
}
For the above equation, we found that all the three non-zero terms scale as $N$, and hence we kept all the non-zero terms in the Lagrangian.
\end{itemize}

\subsection{Island Surface}
\label{PT-IS}
\begin{itemize}
\item {\bf $\sqrt{-g}$ Associated with the Induced Metric (\ref{induced-metric-IS})}:
{\footnotesize
\begin{eqnarray}
& &
\sqrt{-g}=\frac{M N^{7/10} \sqrt{g_s} \left(N_f (\log (N)-3 \log (r))\right){}^{5/3} \sqrt{\alpha (r) \left(\sigma (r)+x'(r)^2\right)}}{144\ 6^{5/6} \pi ^{19/6} \alpha _{\theta _1}^3 \alpha _{\theta
   _2}^2} \nonumber\\
   & & \times  \Biggl(18 N_f g_s \log ^2(r) \left(r^2-3 b^2 (6 r+1) r_h^2\right)+\log (r)
   \left(8 \pi  \left(r^2-3 b^2 r_h^2\right)-3 r^2 N_f g_s\right)\nonumber\\
   & & -N_f g_s \log (N) (2 \log (r)+1) \left(r^2-3 b^2 r_h^2\right)\Biggr).
\end{eqnarray}
}
\item {\bf $\int d{\cal V}_9 \sqrt{-g}\frac{\partial J_0}{\partial R_{txtx}}$}:{\footnotesize
\begin{eqnarray}
\label{Wald-J0-i-IS}
& & \hskip -0.3in (i)\int d{\cal V}_9 \sqrt{-g}\left( \left(G^{xx}\right)^2 \left(G^{tt}\right)^2\left(R_{PxtQ} + \frac{1}{2}R_{PQtx}\right)R_{t}^{\ \ RSP}R^Q_{\ \ RSx}\right)\nonumber\\
& & \sim\int d{\cal V}_9 \sqrt{-g} \Biggl(\frac{r_h^8 \left(13 (\log (N)-3 \log (r))^4+21 (\log (N)-3 \log (r))^2+9 \log (r) (6 \log (N)-9 \log (r))^2\right)}{\sqrt{N} r^{12} N_f^{10/3} \sqrt{g_s} (\log (N)-3 \log (r))^{22/3}}\Biggr) ;\nonumber\\
& & \hskip -0.3in (ii)\int d{\cal V}_9 \sqrt{-g}\left( R^{HxtK}R_H^{\ \ RSt}R^x_{\ \ RSK} + \frac{1}{2}R^{HKtx}R_H^{\ \ RSt}R^Q_{\ \ RSK} \right)\nonumber\\
& & \sim \int d{\cal V}_9 \sqrt{-g}\Biggl(\frac{r_h^8 \left(-2 g_s \log ^2(N) (\log (N)-12 \log (r))^2-5 (\log (N)-9 \log (r))^2 (\log (N)-3 \log (r))^2\right)}{\sqrt{N} r^{12} N_f^{10/3} \sqrt{g_s} (\log (N)-9 \log (r))^2 (\log (N)-3 \log
   (r))^{16/3}}\Biggr) ;
\nonumber\\
& & \hskip -0.3in (iii)\int d{\cal V}_9 \sqrt{-g}  \left(\left(G^{xx}\right)^2 G^{tt}\left(R_{PxtQ} + \frac{1}{2}R_{PQtx}\right)R_{t}^{\ \ RSP}R^Q_{\ \ RSx}\right)\nonumber\\
& &  \sim \int d{\cal V}_9 \left( \lambda_3(r)  \sqrt{\alpha (r) \left(\sigma (r)+x'(r)^2\right)}\right);\nonumber\\
& & \hskip -0.3in (iv)\int d{\cal V}_9 \sqrt{-g} \left( G^{tt} \left(R^{HMNx}R_{PMNt} + \frac{1}{2}R^{HxMN}R_{PtMN}\right)R_H^{\ \ xtP}\right)\nonumber\\
& & \sim \int d{\cal V}_9\left( \lambda_4(r) \sqrt{\alpha (r) \left(\sigma (r)+x'(r)^2\right)}\right).
\end{eqnarray}
}
In the large $N$ limit, terms (iii) and (iv) turn out to be the most dominant.

\item \textbf{$\int d{\cal V}_9 \sqrt{-g}\left(\frac{\partial^2 J_0}{\partial R_{xixj}\partial R_{xmxl}}\right) K_{xij}K_{xml}$}:
{\scriptsize
\begin{eqnarray}
& & \hskip -0.3in (i) \int d{\cal V}_9 \sqrt{-g} \left(G^{xx}\right)^2 \left(G^{mm}\right)\left(G^{ll}\right) \delta^i_m R_r^{~TSx} R^j_{~TSl}K_{xij}K_{xml}\nonumber\\
& & \sim  \int d{\cal V}_9 \sqrt{-g} \Biggl(-\frac{M^2 N N_f g_s^3 \log ^2(r) (23 \log (N)-72 \log (r))^3}{ r^4 \alpha _{\theta _1} \alpha _{\theta _2}^6 (\log (N)-9 \log (r))^2 (\log (N)-3 \log (r))^6 x'(r)^2} \Biggr), \nonumber\\
& & \hskip -0.3in (ii)\int d{\cal V}_9 \sqrt{-g} \left(G^{xx}\right)^3 \left(G^{ii}\right)\left(G^{jj}\right)\left(G^{mm}\right)\left(G^{ll}\right) R_{jmxQ} R^Q_{~~ixl}K_{xij}K_{xml} \sim 0, \nonumber\\
& & \hskip -0.3in (iii)\int d{\cal V}_9 \sqrt{-g}\frac{1}{2}\left(G^{xx}\right)^2 \left(G^{mm}\right)\left(G^{jl}\right)R_r^{~TSx} R^i_{~TSx}K_{xij}K_{xml}\sim 0, \nonumber\\
& & \hskip -0.3in (iv)\int d{\cal V}_9 \sqrt{-g})\frac{1}{2}\left(G^{xx}\right)^2\left(G^{ii}\right)\left(G^{jl}\right) R_r^{~TSx} R^m_{~~TSi} K_{xij}K_{xml}\nonumber\\
& & \sim  \int d{\cal V}_9 \sqrt{-g} \Biggl(\frac{M^3 N^{13/10} \alpha _{\theta _1} N_f^5 g_s^{15/2} \log (r) (23 \log (N)-72 \log (r))^3 \left(\log (N)+18 \log ^2(r)\right) \sqrt{\alpha (r) \left(\sigma (r)+x'(r)^2\right)}}{ r^4 \alpha _{\theta _2}^{10} (\log (N)-3 \log (r))^5 x'(r)^2 \left(4 \pi -N_f g_s (\log (N)-9 \log (r))\right){}^3} \Biggr),\nonumber\\
& & \hskip -0.3in (v)\int d{\cal V}_9 \sqrt{-g}\frac{1}{2} \left(G^{xx}\right)^3\left(G^{ii}\right)\left(G^{ll}\right)\left(G^{mm}\right)\left(G^{jj}\right) R_{jQxl} R^Q_{~~ixm}K_{xij}K_{xml} \sim  \int d{\cal V}_9 \left(\frac{Z_1(r)}{2}{\cal L}_1\right),\nonumber\\
& & \hskip -0.3in (vi)\int d{\cal V}_9 \sqrt{-g}\frac{1}{2}  \left(G^{xx}\right)^3\left(G^{jm}\right)\left(G^{ll}\right) R_{Pxxl} R_x^{~ixP} K_{xij}K_{xml}\nonumber\\
& & \sim  \int d{\cal V}_9 \sqrt{-g} \Biggl(\frac{M^2 N N_f g_s^3 \log ^3(N) (5 \log (N)-196 \log (r)) (23 \log (N)-72 \log (r)) \sqrt{\alpha (r) \left(\sigma (r)+x'(r)^2\right)}}{r^4 \alpha _{\theta _1} \alpha _{\theta _2}^6
   (\log (N)-9 \log (r))^3 (\log (N)-3 \log (r))^4 x'(r)^2} \Biggr),
\nonumber\\
& & \hskip -0.3in (vii)\int d{\cal V}_9 \sqrt{-g} \left(G^{xx}\right)^2\left(G^{im}\right)\left(G^{ll}\right) R^{xMNj} R_{lMNx} K_{xij}K_{xml} \sim  0,\nonumber\\
& & \hskip -0.3in (viii)\int d{\cal V}_9 \sqrt{-g}\left(G^{xx}\right)^2 \left(G^{jj}\right)\left(G^{im}\right) R^{xMNl} R_{jMNx}K_{xij}K_{xml}\sim  0, \nonumber\\
& & \hskip -0.3in (ix)\int d{\cal V}_9 \sqrt{-g} \left(G^{xx}\right)^3 \left(G^{ii}\right)\left(G^{jj}\right)\left(G^{ll}\right)\left(G^{mm}\right) R^Q_{~~mxj} R_{lixQ}K_{xij}K_{xml} \sim  0, \nonumber\\
& & \hskip -0.3in (x)\int d{\cal V}_9 \sqrt{-g}\frac{1}{2} \left(G^{xx}\right)^2 \left(G^{im}\right)\left(G^{ll}\right) R^{xjMN} R_{lxMN}K_{xij}K_{xml}\nonumber\\
& & \sim  \int d{\cal V}_9 \sqrt{-g} \Biggl[\frac{M N^{7/10} r^2 N_f^{8/3} g_s^{3/2} \log (r) (\log (N)-9 \log (r)) (\log (N)-3 \log (r))^{5/3} \sqrt{\alpha (r) \left(\sigma (r)+x'(r)^2\right)}}{ \alpha _{\theta _1}^3
   \alpha _{\theta _2}^2 x'(r)^2} \nonumber\\
   & & \times \left(\frac{ M^4 g_s^4 \log ^2(r) (\log (N)-12 \log (r))^2 \left(\sigma (r) \alpha '(r)+\alpha '(r) x'(r)^2+\alpha (r) \left(\sigma '(r)+2 x'(r) x''(r)\right)\right)^2}{64\ 6^{2/3} \pi ^{4/3} r^4 \alpha
   (r)^2 N_f^{2/3} (\log (N)-3 \log (r))^{8/3} \left(\sigma (r)+x'(r)^2\right)^2}-200 {G_5}(r)\right) \Biggr],\nonumber\\
& & \hskip -0.3in (xi)\int d{\cal V}_9 \sqrt{-g}\frac{1}{2} \left(G^{xx}\right)^2 \left(G^{im}\right)\left(G^{jj}\right) R^{xlMN} R_{jxMN}K_{xij}K_{xml}\nonumber\\
& & \sim  \int d{\cal V}_9 \sqrt{-g} \Biggl[ \frac{M N^{7/10} r^2 N_f^{8/3} g_s^{3/2} \log (r) (\log (N)-9 \log (r)) (\log (N)-3 \log (r))^{5/3} \sqrt{\alpha (r) \left(\sigma (r)+x'(r)^2\right)}}{14400\ 6^{5/6} \pi ^{19/6} \alpha _{\theta _1}^3\alpha _{\theta _2}^2 x'(r)^2} \nonumber\\
& & \times \left(\frac{29 M^4 g_s^4 \log ^2(r) (\log (N)-12 \log (r))^2 \left(\sigma (r) \alpha '(r)+\alpha '(r) x'(r)^2+\alpha (r) \left(\sigma '(r)+2 x'(r) x''(r)\right)\right)^2}{64\ 6^{2/3} \pi ^{4/3} r^4 \alpha
   (r)^2 N_f^{2/3} (\log (N)-3 \log (r))^{8/3} \left(\sigma (r)+x'(r)^2\right)^2}-200 {G_6}(r)\right)\Biggr],\nonumber\\
& & \hskip -0.3in (xii)\int d{\cal V}_9 \sqrt{-g}\frac{1}{2}\left(G^{xx}\right)^3 \left(G^{ii}\right)\left(G^{jj}\right)\left(G^{ll}\right)\left(G^{mm}\right)R^Q_{~~mxi} R_{lQxj}K_{xij}K_{xml} \sim  \int d{\cal V}_9 \left(\frac{Z_1(r)}{2}{\cal L}_1\right),\nonumber\\
& & \hskip -0.3in (xiii)\int d{\cal V}_9 \sqrt{-g}\frac{1}{2}\left(G^{xx}\right)^3 \left(G^{jj}\right)\left(G^{il}\right)R_x^{~mxP} R_{Pxxj}K_{xij}K_{xml}\nonumber\\
& & \sim  \int d{\cal V}_9 \sqrt{-g} \Biggl(-\frac{M^2 N N_f g_s^3 \log ^3(N) (5 \log (N)-196 \log (r)) (23 \log (N)-72 \log (r)) \sqrt{\alpha (r) \left(\sigma (r)+x'(r)^2\right)}}{ r^4 \alpha _{\theta _1} \alpha _{\theta _2}^6
   (\log (N)-9 \log (r))^3 (\log (N)-3 \log (r))^4 x'(r)^2} \Biggr),
\end{eqnarray}
}
where $G_5(r)$ and $G_6(r)$ do not depend on $N$. We found that, terms (v) and (xii) dominate in comparison to other terms in the large $N$ limit. 

\item {\bf $\int d{\cal V}_9 \sqrt{-g}\left(\frac{\partial^2 J_0}{\partial R_{titj}\partial R_{tmtl}}\right) K_{xij}K_{xml}$}:
{\scriptsize
\begin{eqnarray}
& & \hskip -0.3in (i)\int d{\cal V}_9 \sqrt{-g}\left(G^{tt}\right)^2 \left(G^{mm}\right)\left(G^{ll}\right) \delta^i_m R_t^{~TSt} R^j_{~TSl}K_{xij}K_{xml}\nonumber\\
& & \sim  \int d{\cal V}_9 \sqrt{-g} \Biggl(-\frac{M^2 N N_f g_s^3 \log ^4(N) \log ^2(r) (23 \log (N)-72 \log (r))^2 (\log (N)-27 \log (r)) \sqrt{\alpha (r) \left(\sigma (r)+x'(r)^2\right)}}{ r^4 \alpha _{\theta _1} \alpha
   _{\theta _2}^6 (\log (N)-9 \log (r))^4 (\log (N)-3 \log (r))^6 x'(r)^2}\Biggr),\nonumber\\
& & \hskip -0.3in (ii)\int d{\cal V}_9 \sqrt{-g} \left(G^{tt}\right)^3 \left(G^{ii}\right)\left(G^{jj}\right)\left(G^{mm}\right)\left(G^{ll}\right) R_{jmtQ} R^Q_{~~itl}K_{xij}K_{xml} \sim  0, \nonumber\\
& & \hskip -0.3in (iii)\int d{\cal V}_9 \sqrt{-g}\frac{1}{2}\left(G^{tt}\right)^2 \left(G^{mm}\right)\left(G^{jl}\right)R_t^{~TSt} R^i_{~TSt}K_{xij}K_{xml}\sim  0, \nonumber\\
& & \hskip -0.3in (iv)\int d{\cal V}_9 \sqrt{-g}\frac{1}{2}\left(G^{tt}\right)^2\left(G^{ii}\right)\left(G^{jl}\right) R_t^{~TSt} R^m_{~~TSi} K_{xij}K_{xml}\nonumber\\
& & \sim  \int d{\cal V}_9 \sqrt{-g} \Biggl(-\frac{M^3 N^{13/10} \alpha _{\theta _1} N_f^2 g_s^{9/2} \log (r) (23 \log (N)-72 \log (r))^3 \left(\log (N)+18 \log ^2(r)\right) \sqrt{\alpha (r) \left(\sigma (r)+x'(r)^2\right)}}{ r^4 \alpha _{\theta _2}^{10} (\log (N)-9 \log (r))^3 (\log (N)-3 \log (r))^5 x'(r)^2}\Biggr),\nonumber\\
& & \hskip -0.3in (v)\int d{\cal V}_9 \sqrt{-g}\frac{1}{2} \left(G^{tt}\right)^3\left(G^{ii}\right)\left(G^{ll}\right)\left(G^{mm}\right)\left(G^{jj}\right) R_{jQtl} R^Q_{~~itm}K_{xij}K_{xml}\sim  \int d{\cal V}_9 \left(\frac{W_1(r)}{2}{\cal L}_2\right),\nonumber\\
& & \hskip -0.3in (vi)\int d{\cal V}_9 \sqrt{-g}\frac{1}{2}  \left(G^{tt}\right)^3\left(G^{jm}\right)\left(G^{ll}\right) R_{Pttl} R_t^{~itP} K_{xij}K_{xml}\nonumber\\
& & \hskip -0.2in \sim  \int d{\cal V}_9 \sqrt{-g} \Biggl(\frac{M^2 N N_f g_s^3 \log ^3(N) (5 \log (N)-196 \log (r)) (23 \log (N)-72 \log (r)) (\log (N)-3 \log (r)-1)^2 \sqrt{\alpha (r) \left(\sigma (r)+x'(r)^2\right)}}{ r^4 \alpha _{\theta
   _1} \alpha _{\theta _2}^6 (\log (N)-9 \log (r))^3 (\log (N)-3 \log (r))^6 x'(r)^2} \Biggr),
\nonumber\\
& & \hskip -0.3in (vii) \int d{\cal V}_9 \sqrt{-g}\left(G^{tt}\right)^2\left(G^{im}\right)\left(G^{ll}\right) R^{tMNj} R_{lMNt}K_{xij}K_{xml} \sim 0, \nonumber\\
& & \hskip -0.3in (viii)\int d{\cal V}_9 \sqrt{-g}\left(G^{tt}\right)^2 \left(G^{jj}\right)\left(G^{im}\right) R^{tMNl} R_{jMNt}K_{xij}K_{xml}\sim  0, \nonumber\\
& & \hskip -0.3in (ix)\int d{\cal V}_9 \sqrt{-g} \left(G^{tt}\right)^3 \left(G^{ii}\right)\left(G^{jj}\right)\left(G^{ll}\right)\left(G^{mm}\right) R^Q_{~~mtj} R_{litQ}K_{xij}K_{xml} \sim  0, \nonumber\\
& & \hskip -0.3in (x)\int d{\cal V}_9 \sqrt{-g}\frac{1}{2} \left(G^{tt}\right)^2 \left(G^{im}\right)\left(G^{ll}\right) R^{tjMN} R_{ltMN}K_{xij}K_{xml}\nonumber\\
& & \sim  \int d{\cal V}_9 \sqrt{-g} \Biggl[\frac{M N^{7/10} r^2 N_f^{8/3} g_s^{3/2} \log (r) (\log (N)-9 \log (r)) (\log (N)-3 \log (r))^{5/3} \sqrt{\alpha (r) \left(\sigma (r)+x'(r)^2\right)}}{ \alpha _{\theta _1}^3
   \alpha _{\theta _2}^2 x'(r)^2} \nonumber\\
   & & \times \left(\frac{13 M^4 g_s^4 \log ^2(r) (\log (N)-12 \log (r))^2 \left(\sigma (r) \alpha '(r)+\alpha '(r) x'(r)^2+\alpha (r) \left(\sigma '(r)+2 x'(r) x''(r)\right)\right)^2}{64\ 6^{2/3} \pi ^{4/3} r^4 \alpha
   (r)^2 (\log (N)-3 \log (r))^2 \left(N_f (\log (N)-3 \log (r))\right){}^{2/3} \left(\sigma (r)+x'(r)^2\right)^2}-25 {G_7}(r)\right)\Biggr],\nonumber\\
& & \hskip -0.3in (xi)\int d{\cal V}_9 \sqrt{-g}\frac{1}{2} \left(G^{tt}\right)^2 \left(G^{im}\right)\left(G^{jj}\right) R^{tlMN} R_{jtMN}K_{xij}K_{xml}\nonumber\\
& & \sim  \int d{\cal V}_9 \sqrt{-g} \Biggl[\frac{M N^{7/10} r^2 N_f^{8/3} g_s^{3/2} \log (r) (\log (N)-9 \log (r)) (\log (N)-3 \log (r))^{5/3} \sqrt{\alpha (r) \left(\sigma (r)+x'(r)^2\right)}}{\alpha _{\theta _1}^3
   \alpha _{\theta _2}^2 x'(r)^2} \nonumber\\
   & & \times \left(\frac{13 M^4 g_s^4 \log ^2(r) (\log (N)-12 \log (r))^2 \left(\sigma (r) \alpha '(r)+\alpha '(r) x'(r)^2+\alpha (r) \left(\sigma '(r)+2 x'(r) x''(r)\right)\right)^2}{64\ 6^{2/3} \pi ^{4/3} r^4 \alpha
   (r)^2 (\log (N)-3 \log (r))^2 \left(N_f (\log (N)-3 \log (r))\right){}^{2/3} \left(\sigma (r)+x'(r)^2\right)^2}-25 {G_8}(r)\right)\Biggr],\nonumber\\
& & \hskip -0.3in (xii)\int d{\cal V}_9 \sqrt{-g}\frac{1}{2}\left(G^{tt}\right)^3 \left(G^{ii}\right)\left(G^{jj}\right)\left(G^{ll}\right)\left(G^{mm}\right)R^Q_{~~mti} R_{lQtj}K_{xij}K_{xml} \sim  \int d{\cal V}_9  \left(\frac{W_1(r)}{2}{\cal L}_2\right),\nonumber
\end{eqnarray}
\begin{eqnarray}
& & \hskip -0.3in (xiii)\int d{\cal V}_9 \sqrt{-g}\frac{1}{2}\left(G^{tt}\right)^3 \left(G^{jj}\right)\left(G^{il}\right)R_t^{~mtP} R_{Pttj}K_{xij}K_{xml}\nonumber\\
& & \sim  \int d{\cal V}_9 \sqrt{-g} \Biggl(\frac{M^2 N N_f g_s^3 \log ^3(N) (5 \log (N)-196 \log (r)) (23 \log (N)-72 \log (r)) \sqrt{\alpha (r) \left(\sigma (r)+x'(r)^2\right)}}{ r^4 \alpha _{\theta _1} \alpha _{\theta
   _2}^6 (\log (N)-9 \log (r))^3 (\log (N)-3 \log (r))^4 x'(r)^2} \Biggr).
\end{eqnarray}
}
where $G_7(r)$ and $G_8(r)$ independent of $N$ and the most dominant terms are  (v) and (xii) in the larg-$N$ limit.
 
\item {\bf $\int d{\cal V}_9 \sqrt{-g}\left(\frac{\partial^2 J_0}{\partial R_{titj}\partial R_{xmxl}}\right)K_{xij}K_{xml}$}:  The only term (i) turns out be non-zero in this case as follows.
{\footnotesize
\begin{eqnarray}
& & \hskip -0.3in (i)\int d{\cal V}_9 \sqrt{-g}\left(G^{xx}\right)^2 \left(G^{mm}\right)\left(G^{ll}\right) \delta^i_m \delta^t_P \delta^t_R \delta^j_Q R_x^{~TSP} R^Q_{~TSl}K_{xij}K_{xml} \sim  \int d{\cal V}_9 \left(U_1(r) {\cal L}_3\right),\nonumber\\
& & \hskip -0.3in (ii)\int d{\cal V}_9 \sqrt{-g} \left(G^{xx}\right)^2 \left(G^{ii}\right)\left(G^{jj}\right)\left(G^{mm}\right)\left(G^{ll}\right)\left(G^{tt}\right)\delta^t_x R_{jmxQ} R^Q_{~~itl}K_{xij}K_{xml}\sim 0,\nonumber\\
& & \hskip -0.3in (iii)\int d{\cal V}_9 \sqrt{-g}\left(G^{xx}\right)^2 \left(G^{mm}\right)\left(G^{ll}\right)\left(G^{tt}\right) \delta^i_T \delta^t_S \delta^j_l  R_{Pmxt} R_x^{~TSP}K_{xij}K_{xml}\sim 0,\nonumber\\
& & \hskip -0.3in (iv)\int d{\cal V}_9 \sqrt{-g}\frac{1}{2}\left(G^{xx}\right)^2\left(G^{mm}\right)\left(G^{ll}\right)\left(G^{tt}\right)\delta^i_T \delta^t_S \delta^j_m  R_{Ptxl} R_x^{~~TSP}K_{xij}K_{xml}\sim 0, \nonumber\\
& & \hskip -0.3in (v)\int d{\cal V}_9 \sqrt{-g} \left(G^{xx}\right)\left(G^{tt}\right)^2\left(G^{ii}\right)\left(G^{jl}\right) R_t^{mxP} R_{Pitx}K_{xij}K_{xml}\sim 0,\nonumber\\
& & \hskip -0.3in (vi)\int d{\cal V}_9 \sqrt{-g}\frac{1}{2}  \left(G^{tt}\right)^2\left(G^{xx}\right)\left(G^{jj}\right)\left(G^{iK}\right)\delta^l_K R_t^{mxP} R_{Pxtj}K_{xij}K_{xml}\sim 0. 
\end{eqnarray}
}
\item {\bf $\int d{\cal V}_9 \sqrt{-g} \left(\frac{\partial^2 J_0}{\partial R_{tixj}\partial R_{xmtl}}\right)K_{xij}K_{xml}$}: In this case all the non-zero three terms contribute equally and scale as $N$.
{\footnotesize
\begin{eqnarray}
& & \hskip -0.3in (i)\int d{\cal V}_9 \sqrt{-g}\left(G^{xx}\right)^2 \left(G^{mm}\right)\left(G^{ll}\right) \delta^i_m  R_x^{~TSx} R^j_{~TSl}K_{xij}K_{xml} \sim  \int d{\cal V}_9 \left(U_1(r) {\cal L}_4\right),\nonumber\\
& & \hskip -0.3in (ii)\int d{\cal V}_9 \sqrt{-g} \left(G^{xx}\right)^3 \left(G^{ii}\right)\left(G^{jj}\right)\left(G^{ll}\right)\left(G^{mm}\right) R_{jmxQ} R^Q_{~~ixl}K_{xij}K_{xml} \sim 0, \nonumber\\
& & \hskip -0.3in (iii)\int d{\cal V}_9 \sqrt{-g}\frac{1}{2}\left(G^{xx}\right) \left(G^{tt}\right)^2\left(G^{jm}\right)\left(G^{ll}\right)  R_x^{~ixP} R_{Pttl}K_{xij}K_{xml} \sim  \int d{\cal V}_9 \left(V_1(r) {\cal L}_4\right),\nonumber\\
& & \hskip -0.3in (iv)\int d{\cal V}_9 \sqrt{-g}\frac{1}{2}\left(G^{xx}\right)^2\left(G^{tt}\right)\left(G^{jj}\right)\left(G^{il}\right)  R_t^{~mtP} R_{Pxxj} K_{xij}K_{xml}\sim  \int d{\cal V}_9 \left(V_1(r) {\cal L}_4\right).
\end{eqnarray}
}
Each term contain a factor $\frac{8}{(q_{\alpha}+1)}$ absorbed in the numerical factors.
 
 \end{itemize}
%\input{AppendixD}

%\chapter{Appendix}
%\input{Appendix}
%\appendix
%\chapter{Appendix}
%\begin{appendices}
%\begin{subappendices}

%\end{subappendices}
%\end{appendices}
%\input{empty1}
%\begin{appendices}
%\begin{subappendices}
\
%\end{subappendices}
%\end{appendices}
%\input{empty1}
%\begin{appendices}
%\begin{subappendices}
%\input{AppendixC}
%\input{AppendixE}
%\end{subappendices}
%\end{appendices}
%\input{green}
%\input{redgreen}
%\input{seven}
%\input{six}

\begin{spacing}{1}
\end{spacing}

%\begin{spacing}{1}
%\bibliographystyle{unsrt}
%\cleardoublepage
%\addcontentsline{toc}{chapter}{Bibliography}
%\bibliography{sarita_23911}

\clearpage
\addcontentsline{toc}{chapter}{Bibliography}

\begin{thebibliography}{400}
%%%%%%%%%%%%%%%%%%
\bibitem{HD-MQGP} V.~Yadav and A.~Misra, {\it On ${\cal M}$-Theory Dual of Large-$N$ Thermal QCD-like Theories up to ${\cal O}(R^4)$ and $G$-structure classification of Underlying Non-Supersymmetric Geometries}, to appear in Advances in Theoretical and Mathematical Physics (2023)) [arXiv:2004.07259[hep-th]].
%%%%%%%%%%%%
\bibitem{MChPT} V.~Yadav, G.~Yadav and A.~Misra, {\it (Phenomenology/Lattice-Compatible) $SU(3)$ M$\chi$PT HD up to ${\cal O}(p^4)$ and the ${\cal O}\left(R^4\right)$-Large-$N$ Connection}, JHEP {\bf 08} (2021) 151 [arXiv:2011.04660 [hep-th]].
%%%%%%%%%%1
\bibitem{McTEQ} G.~Yadav, V.~Yadav and A.~Misra, {\it $\cal {M}$cTEQ ($\cal {M}$ chiral perturbation theory-compatible deconfinement Temperature and Entanglement entropy up to terms Quartic in curvature) and FM (Flavor Memory)}, JHEP {\bf 10} (2021) 220 [arXiv:2108.05372 [hep-th]].
%%%%%%%%%%%%%%%%%%%%%%%% 
\bibitem{Witten-Hawking-Page-Tc} E.~Witten, {\it Anti-de Sitter space, thermal phase transition, and confinement in gauge theories}, Adv. Theor. Math. Phys. {\bf 2}, 505 (1998) [arXiv:hep-th/9803131].
%%%%%%%%%%
\bibitem{Rotation-Tc-M-Theory} G.~Yadav, {\it Deconfinement Temperature of Rotating QGP at Intermediate coupling from ${\cal M}$-Theory}, Phys. Lett. B {\bf 841} (2023) 137925 [arXiv:2203.11959[hep-th]]. 
%%%%%%%%%%%%%%%%%
\bibitem{AMMZ} A.~Almheiri, R.~Mahajan, J.~Maldacen and Y.~Zhao, {\it The Page curve of Hawking radiation from semiclassical geometry}, JHEP {\bf 03} (2020) 143 [arXiV:1908.10996[hep-th]].
%%%%%%%%%%%%%%%%
\bibitem{PBD} E.~Caceres, A.~Kundu, Ayan K.~Patra and S. Shashi, {\it Page Curves and Bath Deformations}, SciPost Phys. Core 5, 033 (2022) [arXiv:2107.00022 [hep-th]].
%%%%
\bibitem{WH-1} I.~Akal, Y.~Kusuki, T.~Takayanagi and Z.~Wei, {\it Codimension two holography for wedges}, Phys. Rev. D 102, 126007 (2020) [arXiv:2007.06800[hep-th]].
%%%%%%%%%
\bibitem{WH-2} R.~X.~Miao, {\it An exact construction of codimension two holography}, JHEP {\bf 01} (2021) 150 [arXiv:2009.06263[hep-th]].
%%%%%%%%%%
 \bibitem{RNBH-HD} G.~Yadav, {\it Page curves of Reissner-Nordstr{\"o}m black hole in HD gravity}, Eur. Phys. J. C {\bf 82} (2022) 904 [arXiv:2204.11882[hep-th]].
%%%%%%%%%%%
\bibitem{HD-Page Curve-2} G.~Yadav and A.~Misra, {\it Entanglement entropy and Page curve from the ${\cal M}$-theory dual of thermal QCD above $T_c$ at intermediate coupling}, to appear in Physical Review D [arXiv:2207.04048[hep-th]].
%%%%%%%
\bibitem{Gopal+Nitin} G.~Yadav and N.~Joshi, {\it Cosmological and black hole islands in multi-event horizon spacetimes}, Phys. Rev. D {\bf 107}, 026009 (2023) [arXiv:2210.00331[hep-th]].
%%%%%%%%%%%%%%%%%%%
\bibitem{Multiverse} G.~Yadav, {\it Multiverse in Karch-Randall Braneworld}, JHEP {\bf 03} (2023) 103 [arXiv:2301.06151[hep-th]].
%%%%%%%%%%%%15
\bibitem{metrics}  M.~Mia, K.~Dasgupta, C.~Gale and S.~Jeon, {\it Five Easy Pieces: The Dynamics of Quarks in Strongly Coupled Plasmas}, Nucl.\ Phys.\ B {\bf 839}, 187 (2010) [arXiv:hep-th/0902.1540].
%%%%%%%%%%%%%%%%%%%%%%%%%%%%%%
\bibitem{MQGP} M.~Dhuria and A.~Misra, {\it Towards MQGP}, JHEP 1311 (2013) 001 [arXiv:hep-th/1306.4339].
%%%%%%%
\bibitem{Pich} A.~Pich, {\it Chiral Perturbation Theory}, Rept. Prog. Phys. 58:563-610(1995), [arxiv:hep-th/9502366]. 
%%%%%%%%
\bibitem{AdS/CFT} J. M.~Maldacena, {\it The Large N Limit of Superconformal Field Theories and Supergravity}, Adv. Theor. Math. Phys. 2, 231 (1998) [arXiv:hep-th/9711200].
%%%%%%%
\bibitem{Ganor-D3} O.~J.~Ganor, U.~Varadarajan, {\it Nonlocal Effects on D-branes in Plane-Wave Backgrounds}, JHEP 0211:051,2002 [arXiv:hep-th/0210035].
%%%%%%%%%%%%
\bibitem{S.Roy-1}K.~Nayek and S.~Roy, {\it Decoupling limit and throat geometry of non-susy D3 brane}, Phys.\ Lett.\ B {\bf 766}, 192 (2017)   [arXiv:1608.05036 [hep-th]].
    %%%%%%%%%%%%3
\bibitem{S.Roy-2} S.~Chakraborty, K.~Nayek and S.~Roy, {\it Wilson loop calculation in QGP using non-supersymmetric AdS/CFT}, arXiv:1710.08631 [hep-th].
%%%%%%%%%%%%%%4
\bibitem{S.Roy-3} S.~Chakraborty, N.~Haque and S.~Roy, {\it Wilson loops in noncommutative Yang-Mills theory using gauge/gravity duality},   Nucl.\ Phys.\ B {\bf 862}, 650 (2012)[arXiv:1201.0129 [hep-th]].
%%%%%%%%%%%%%%%%5
\bibitem{S.Roy-4} S.~Chakraborty and S.~Roy, {\it Calculating the jet quenching parameter in the plasma of NCYM theory from gauge/gravity duality},  Phys.\ Rev.\ D {\bf 85}, 046006 (2012)  [arXiv:1105.3384 [hep-th]].
%%%%%%%%%%%%%%%%%%6
\bibitem{S.Roy-5}  S.~Chakraborty and S.~Roy, {\it Wilson loops in (p+1)-dimensional Yang-Mills theories using gravity/gauge theory correspondence},   Nucl.\ Phys.\ B {\bf 850}, 463 (2011)  [arXiv:1103.1248 [hep-th]].
%%%%%%%%%%%%%%%%%%%7
\bibitem{S.Roy-6} K.~L.~Panigrahi and S.~Roy,{\it Drag force in a hot non-relativistic, non-commutative Yang-Mills plasma}, JHEP {\bf 1004}, 003 (2010)  [arXiv:1001.2904 [hep-th]].
%%%%%%%%%%%%%%%%%%%8
\bibitem{S.Roy-7}  S.~Roy, {\it Holography and drag force in thermal plasma of non-commutative Yang-Mills theories in diverse dimensions},  Phys.\ Lett.\ B {\bf 682}, 93 (2009)  [arXiv:0907.0333 [hep-th]].
%%%%%%%%%%%%16
\bibitem{KW} Igor R. Klebanov and Edward Witten, {\it Superconformal Field Theory on Threebranes at a Calabi-Yau Singularity},  Nucl. Phys. B {\bf 536}, 199 (1998)[arXiv:hep-th/9807080].
%%%%%%%%%%%18
\bibitem{KS} I.~R.~Klebanov and  M.~J.~Strassler, {\it Supergravity and a Confining Gauge Theory: Duality Cascades and $X$SB-Resolution of Naked Singularities}, JHEP {\bf 08} (2000) 052 [arXiv:hep-th/0007191].
 %%%%%%%11
\bibitem{Nunez-6} J.~M.~Maldacena and C.~Nunez, {\it Towards the large N limit of pure N=1 superYang-Mills}, Phys.\ Rev.\ Lett.\  {\bf 86}, 588 (2001) [arXiv:hep-th/0008001].
    %%%%%%%%%%%12
\bibitem{Kruczenski:2003uq} M.~Kruczenski, D.~Mateos, R.~C.~Myers and D.~J.~Winters, {\it Towards a holographic dual of large N(c) QCD},
JHEP \textbf{05}, 041 (2004) [arXiv:hep-th/0311270].
%%%%%%%%%%%%%%%%%%%%%
\bibitem{Bergman:2001rw} A.~Bergman, K.~Dasgupta, O.~J.~Ganor, J.~L.~Karczmarek and G.~Rajesh, {\it Nonlocal field theories and their gravity duals}, Phys. Rev. D \textbf{65}, 066005 (2002) [arXiv:hep-th/0103090].
%%%%%%%%%%%%%%%%%%
\bibitem{Ballon-Bayona:2018ddm} A.~Ballon-Bayona, H.~Boschi-Filho, L.~A.~H.~Mamani, A.~S.~Miranda and V.~T.~Zanchin,
{\it An effective holographic approach to QCD}, arXiv:1804.01579 [hep-th].
%%%%%%%
\bibitem{MK-ii} U.~H.~Danielsson, M.~Kruczenski, A.~G\"{u}ijosa and B.~Sundborg, {\it D3-brane holography}, JHEP {\bf 05} (2000) 028 [arXiv:hep-th/0004187].
%%%%%%%%%%%%
\bibitem{H-QCD} S.~Hands, T.~J.~Hollowood and J.~C.~Myers, {\it QCD with Chemical Potential in a Small Hyperspherical Box}, JHEP {\bf 07} (2010) 086 [arXiv:1003.5813[hep-th]].
%%%%%%%%%%%%
\bibitem{H-QCD-1} T.~J.~Hollowood and J.~C.~Myers, {\it Deconfinement transitions of large N QCD with chemical potential at weak and strong coupling}, JHEP {\bf 10} (2012) 067 [arXiv:1207.4605[hep-th]].
%%%%%%%%%%%%
\bibitem{H-QCD-2} T.~J.~Hollowood and J.~C.~Myers, {\it Overview of large N QCD with chemical potential at weak and strong coupling}, Proceedings for XQCD 2012 [arXiv:1301.5750[hep-th]].
%%%%%%%%%%%%
\bibitem{H-QCD-3} T.~J.~Hollowood and J.~C.~Myers, {\it Phase diagram of adjoint QCD at weak coupling and finite volume}, PoS (LAT2009) 046.
%%%%%%%%%%%%
\bibitem{H-QCD-4} S.~Hands, T.~J.~Hollowood and J.~C.~Myers, {\it QCD with chemical potential on $S^1\times S^3$}, PoS (Lattice 2010) 204.
%%%%%%%
\bibitem{MK-i} M.~Kruczenski, {\it Towards a dual description of QCD}, AIP Conf.Proc. 917 (2007) 1, 154-160.
%%%%%%%%
\bibitem{A-QCD} R.~McNees, R.~C.~Myers and A.~Sinha, {\it On quark masses in holographic QCD}, JHEP {\bf 11} (2008) 056 [arXiv:0807.5127[hep-th]].
%%%%%%%%%
\bibitem{Filho-1} R.~C. L. Bruni, E.~F.~Capossoli, and H.~B.~Filho, {\it Quark-antiquark potential from a deformed AdS/QCD}, Adv.High Energy Phys. 2019 (2019) 1901659 [arXiv:1806.05720[hep-th]].
%%%%%%%%
\bibitem{Filho-2} E.~F.~Capossoli, J.~P.~M.~Graca, and H.~B.~Filho, {\it AdS/QCD oddball masses and Odderon Regge trajectory from a twist-five operator approach}, Phys. Rev. D {\bf 105}, 026026 (2022) [arXiv:2110.12498[hep-th]].
%%%%%%%%%
\bibitem{Filho-3} A.~B.~Bayona, H.~B.~Filho, L.~A.~H.~Mamani, A.~S.~Miranda, and V.~T.~Zanchin, {\it Effective holographic models for QCD: glueball spectrum and trace anomaly}, Phys. Rev. D {\bf 97}, 046001 (2018) [arXiv:1708.08968[hep-th]].
%%%%%%%%
\bibitem{Filho-4} E.~F.~Capossoli, M.~A.~M.~Contreras, D.~Li, A.~Vega, and H.~B.~Filho, {\it Proton Structure Functions from an AdS/QCD model with a deformed background}, Phys. Rev. D {\bf 102}, 086004 (2020) [arXiv:2007.09283[hep-th]].
%%%%%%%
\bibitem{SP-1} S.~Panda, M.~Sami and I.~Thongkool, {\it Reheating the D-brane universe via instant preheating}, Phys. Rev. D {\bf 81}, 103506 (2010) [arXiv:0905.2284[hep-th]].
%%%%%%%
\bibitem{SP-2} A.~Ali, R.~Chingangbam, S.~Panda and M.~Sami, {\it Prospects of inflation with perturbed throat geometry}, Phys. Lett. B {\bf 674}, 131-136 (2009) [arXiv:0809.4941 [hep-th]].
%%%%%%%%%17
\bibitem{KT} I.R. Klebanov and A. Tseytlin, {\it Gravity Duals of Supersymmetric $SU(N)\times SU(M+N)$ Gauge Theories}, Nucl. Phys. B {\bf 578} (2000) 123-138, [arXiv:hep-th/0002159].
%%%%%%%%%%%%%%%%%%19
\bibitem{Zayas+Tseytlin}L.~A.~Pando Zayas and A.~A.~Tseytlin, {\it 3-branes on resolved conifold},  JHEP {\bf 0011}, 028 (2000) [arXiv:hep-th/0010088].
%%%%%%%%%%%%%%%%%%%%%
\bibitem{ouyang} P.~Ouyang, {\it Holomorphic D7-Branes and Flavored N=1 Gauge Theories}, Nucl.\ Phys.\ B {\bf 699}, 207 (2004) [arXiv:hep-th/0311084].
%%%%%%%%%15
\bibitem{Nunez-2} C.~Nunez, A.~Paredes and A.~V.~Ramallo,{\it Unquenched Flavor in the Gauge/Gravity Correspondence}, Adv.\ High Energy Phys.\  {\bf 2010}, 196714 (2010)  [arXiv:1002.1088 [hep-th]].
%%%%%%16
\bibitem{Nunez-3} F.~Benini, F.~Canoura, S.~Cremonesi, C.~Nunez and A.~V.~Ramallo, {\it Backreacting flavors in the Klebanov-Strassler background},  JHEP {\bf 0709}, 109 (2007)  [arXiv:0706.1238 [hep-th]].
%%%%%%%%17
\bibitem{Nunez-4} F.~Benini, F.~Canoura, S.~Cremonesi, C.~Nunez and A.~V.~Ramallo, {\it Unquenched flavors in the Klebanov-Witten model}, JHEP {\bf 0702}, 090 (2007)  [hep-th/0612118].
%%%%%%%%%%%%25
 \bibitem{K. Dasgupta  et al [2012]} M.~Mia, F.~Chen, K.~Dasgupta, P.~Franche and S.~Vaidya, {\it Non-Extremality, Chemical Potential and the Infrared limit of Large N Thermal QCD}, Phys.\ Rev.\ D {\bf 86}, 086002 (2012)[arXiv:1202.5321 [hep-th]].
%%%%%%%%%%%%%%%%%%%
%%%%%%%%%%%%%24
 \bibitem{arnabkundu0709.1547} T. Albash, V. G. Filev, C. V. Johnson and A. Kundu, {\it Finite temperature large N gauge theory with quarks in an external magnetic field}, JHEP {\bf 0807},080 (2008) [arXiv:0709.1547 [hep-th]]
%%%%%%%%%%%%%%225
\bibitem{arnabkundu0709.1554} T. Albash, V. G. Filev, C. V. Johnson, and A. Kundu, {\it Quarks in an External Electric Field in Finite Temperature Large N Gauge Theory}, JHEP {\bf 08}, 092 (2008), arXiv:0709.1554[hep-th].
%%%%%%%%%%%%%%%%%%26
\bibitem{Minwalla-3} S.~Minwalla, {\it Black holes in large N gauge theories}, Class.\ Quant.\ Grav.\  {\bf 23}, S927 (2006), Lectures from the European RTN Winter School on Strings, Supergravity and Gauge Theories, CERN, January 2006.
%%%%%%%%%%27
\bibitem{Obers:2008pj} N.~A.~Obers, {\it Black Holes in Higher-Dimensional Gravity}, Lect. Notes Phys. \textbf{769}, 211-258 (2009) doi:10.1007/978-3-540-88460-6\_6 [arXiv:0802.0519 [hep-th]].
%%%%%%%%%%
\bibitem{SS-KJ} C.~V.~Johnson and A.~Kundu, {\it External fields and chiral symmetry breaking in the Sakai-Sugimoto model}, JHEP {\bf 12} (2008) 053 [arXiv:0803.0038[hep-th]].
%%%%%%%%%%%%%%%%%%%%%40%%%%%%%%%%%%%%%%%
\bibitem{syz} A.~Strominger, S.~T.~Yau and E.~Zaslow, {\it Mirror symmetry is T duality},  Nucl.\ Phys.\ B {\bf 479}, 243 (1996)  [hep-th/9606040].
%%%%%%%%%%%%23
 \bibitem{M.Ionel and M.Min-Oo(2008)} M.~Ionel and M.~Min-Oo, {\it Cohomogeneity One Special Lagrangian 3-Folds in the Deformed and the Resolved Conifolds},  Illinois Journal of Mathematics, Vol. 52, Number 3 (2008).
%%%%%%%%%%24
 \bibitem{EPJC-2}K.~Sil and A.~Misra,
  {\it New Insights into Properties of Large-N Holographic Thermal QCD at Finite Gauge Coupling at (the Non-Conformal/Next-to) Leading Order in N}, Eur.\ Phys.\ J.\ C {\bf 76}, 618 (2016)
  [arXiv:1606.04949 [hep-th]].
%%%%%%%%%%%%%%
\bibitem{DM-transport-2014}M.~Dhuria and A.~Misra, {\it Transport Coefficients of Black MQGP M3-Branes},
Eur. Phys. J. C {\bf 75}, 16 (2015) [arXiv:1406.6076 [hep-th]].  
 %%%%%%%%%%%%%%%%%%%%%42%%%%%%%%%%%%%%%%%
\bibitem{M. Becker et al [2004]}S.~Alexander, K.~Becker, M.~Becker, K.~Dasgupta, A.~Knauf and R.~Tatar, {\it In the realm of the geometric transitions,}  Nucl.\ Phys.\ B {\bf 704}, 231 (2005) [hep-th/0408192].
%%%%%%%%%%%%%%%%%%%%%43%%%%%%%%%%%%%%%%%
\bibitem{F. Chen et al [2010]}F.~Chen, K.~Dasgupta, P.~Franche, S.~Katz and R.~Tatar, {\it Supersymmetric Configurations, Geometric Transitions and New Non-Kahler Manifolds}, Nucl.\ Phys.\ B {\bf 852}, 553 (2011) [arXiv:hep-th/1007.5316].
%%%%%%%%%%%
\bibitem{NAB-1} T.~Harmark and N.~A.~Obers, {\it Thermodynamics of the Near-Extremal NS5-brane}, Nucl. Phys. B {\bf 742} (2006) 41-58 [arXiv:hep-th/0510098].
%%%%%%%%
\bibitem{NAB-2} T.~Harmark and N.~A.~Obers, {\it Hagedorn Behaviour of Little String Theory from String Corrections to NS5-Branes}, Phys. Lett. B {\bf 485} (2000) 285-292 [arXiv:hep-th/0005021].
%%%%%%%%%%20
\bibitem{NPB}K.~Sil and A.~Misra, {\it On Aspects of Holographic Thermal QCD at Finite Coupling},
  Nucl.\ Phys.\ B {\bf 910}, 754 (2016) [arXiv:1507.02692 [hep-th]].  
%%%%%%%%
\bibitem{NAB-4} J.~Armas, N.~Nguyen, V.~Niarchosc and N.~A.~Obers, {\it Thermal transitions of metastable M-branes}, JHEP {\bf 08} (2019) 128 [arXiv:1904.13283[hep-th]].
%%%%%%%%%%%
\bibitem{Ganor-1} J.~Brown, O.~J.~Ganor and C.~Helfgott, {\it M-theory and E10: Billiards, Branes, and Imaginary Roots}, JHEP 0408:063,2004 [arXiv:hep-th/0401053].
%%%%%%%%%%%%
\bibitem{Ganor-2} O.~J.~Ganor and J.~Sonnenschein, {\it On the strong coupling dynamics of heterotic string theory on $C^3/Z_3$},  JHEP {\bf 05} (2002) 018 [arXiv:hep-th/0202206].
%%%%%%%
\bibitem{Ganor-3} O.~J.~Ganor, S.~Ramgoolam and W.~Taylor IV, {\it Branes, Fluxes and Duality in M(atrix)-Theory},  Nucl. Phys. B 492:191-204,1997 [arXiv:hep-th/9611202].
%%%%%%%%%%%%19
\bibitem{Misra+Gale} A.~Misra and C.~Gale, {\it The QCD trace anomaly at strong coupling from  M-theory }, Eur. Phys. J. C {\bf 80}, no.7, 620 (2020)
[arXiv:1909.04062 [hep-th]].
%%%%%%%%%21
 \bibitem{M(r)N_f(r)-Dasgupta_et_al} M.~Mia, K.~Dasgupta, C.~Gale and S.~Jeon,
  {\it Toward Large N Thermal QCD from Dual Gravity: The Heavy Quarkonium Potential},
  Phys.\ Rev.\ D {\bf 82}, 026004 (2010) [arXiv:1004.0387 [hep-th]].
  %%%%%%%%%16
\bibitem{Green and Gutperle} M.~B.~Green and M.~Gutperle, {\it Effects of D instantons},
Nucl. Phys. B {\bf 498}, 195 (1997) [arXiv:hep-th/9701093].
  %%%%%%%%%%
\bibitem{NAB-3} C.~Bachas, C.~Fabre, E.~Kiritsis, N.~A.~Obers, P.~Vanhove, {\it Heterotic/type I duality and D-brane instantons}, Nucl. Phys. B {\bf 509} (1998) 33-52 [arXiv:hep-th/9707126].
%%%%%%%%%
\bibitem{A-n/s-Weyl} R.~C.~Myers, M.~F.~Paulos and A.~Sinha, {\it Quantum corrections to $\eta/s$}, Phys. Rev. D {\bf 79}:041901,2009 [arXiv:0806.2156[hep-th]].
%%%%%%%%%%%%%%%%%%%23-28
\bibitem{Green and Vanhove} M.~B.~Green and P.~Vanhove, {\it D instantons, strings and ${\cal M}$ theory},
Phys. Lett. B {\bf 408}, 122 (1997) [arXiv:hep-th/9704145].
%%%%%%%%%%%%%%
\bibitem{Tseytlin} A.~A.~Tseytlin, {\it Heterotic type I superstring duality and low-energy effective actions},
Nucl.\ Phys.\ B {\bf 467}, 383 (1996) [arXiv:hep-th/9512081].
%%%%%%%%%%%%%%%%%%%%%%%%%%%%%%%%%%%
\bibitem{Duff} M.~J.~Duff, J.~T.~Liu and R.~Minasian, {\it Eleven-dimensional origin of string-string duality: A One loop test}, Nucl. Phys. B \textbf{452}, 261-282 (1995)
[arXiv:hep-th/9506126 [hep-th]].
%%%%%%%%%%%%%%%%%%%%25-29
\bibitem{Horava and Witten} P.~Horava and E.~Witten,
{\it Eleven-dimensional supergravity on a manifold with boundary},
Nucl. Phys. B {\bf 475}, 94-114 (1996)
[arXiv:hep-th/9603142].
%%%%%%%%%%%%%%%%%%%%24-30
\bibitem{Vafa and Witten} C.~Vafa and E.~Witten, {\it A One loop test of string duality}, Nucl. Phys. B {\bf 447}, 261-270 (1995) [arXiv:hep-th/9505053].
%%%%%%%%%%%%%%%%%%%%%%%%
\bibitem{Becker-sisters-O(R^4)}K.~Becker and M.~Becker, {\it Supersymmetry breaking, ${\cal M}$ theory and fluxes}, JHEP \textbf{07}, 038 (2001) [arXiv:hep-th/0107044].
%%%%%%%%%%%%%%%%%%%%%%%%%%%%%%%%%%%%%%%%%%%%%%%%%%%%%
\bibitem{Bulk-Viscosity-McGill-IIT-Roorkee}A.~Czajka, K.~Dasgupta, C.~Gale, S.~Jeon, A.~Misra, M.~Richard and K.~Sil,
{\it Bulk Viscosity at Extreme Limits: From Kinetic Theory to Strings}, JHEP \textbf{07}, 145 (2019)[arXiv:1807.04713 [hep-th]].
%%%%%%%%%%%%
\bibitem{Vikas-Thesis} V.~Yadav, {\it String/${\cal M}$-theory Dual of Large-$N$ Thermal QCD-Like Theories at Intermediate Gauge/'t Hooft Coupling and Holographic Phenomenology}, arXiv:2111.12655[hep-th].
%%%%%%%%%%
\bibitem{HARADA}  M.~Harada, S.~Matsuzaki and K.~Yamawaki, {\it Holographic QCD Integrated back to Hidden Local Symmetry}, Phys.Rev.D82:076010,2010 [arXiv:hep-th/1007.4715].
%%%%%%%%%%%%3
\bibitem{MG} A.~Manohar and H.Georgi, {\it Chiral Quarks and the Nonrelativistic Quark Model}, Nucl. Phys. B 234, 189 (1984).
%%%%%%%%%%%%4
%%%%%%%%%%%%4
\bibitem{GLF} J.~Gasser and H.~Leutwyler, {\it Chiral perturbation theory to one loop}, Annals of Physics  158, 142 (1984).
%%%%%%%%%%5
\bibitem{GL} J.~Gasser and H.~Leutwyler, {\it Chiral Perturbation Theory: Expansions in the Mass of the Strange Quark}, Nucl. Phys. B250(1985) 465.
%%%%%%%%%6
\bibitem{MILC} C. ~Bernard et al., {\it Low energy constants from the MILC Collaboration} [arxiv:hep-lat/0611024].
%%%%%%%%%%1
\bibitem{Ecker-2015} G.~Ecker, {\it Status of Chiral Perturbation Theory for Light Mesons}, PoS CD15 (2015) 011 [arXiv:hep-ph/1510.01634].
%%%%%%%%%%12
\bibitem{SS1} T.~Sakai and S.~Sugimoto, {\it Low energy hadron physics in holographic QCD},
Prog. Theor. Phys. \textbf{113}, 843-882 (2005) [arXiv:hep-th/0412141 [hep-th]]; {\it More on a holographic dual of QCD}, Prog. Theor. Phys. \textbf{114}, 1083-1118 (2005) [arXiv:hep-th/0507073 [hep-th]].
%%%%%%%%%%%%14
\bibitem{HMM} M.~Harada, Y.-Liang~Ma and S.~Matsuzaki, {\it Chiral effective theories from holographic QCD with scalars}, Phys. Rev. D {\bf 89}, 115012 (2014) [arxiv:hep-th/1404.4532].
%%%%%%%%%%%34
\bibitem{BE14}J.~Bijnens and G.~Ecker, {\it Mesonic low-energy constants}, Ann. Rev. Nucl. Part. Sci. 64 (2014) 149 [arXiv:1405.6488 [hep-ph]].
%%%%%%%%%%%%%%%%%%%%%%%%%%%%%%%%%%
\bibitem{Yadav+Misra+Sil-Mesons} V.~Yadav, A.~Misra and K.~Sil, {\it Delocalized SYZ Mirrors and Confronting Top-Down $SU(3)$-Structure Holographic Meson Masses at Finite g and $N_c$ with P(article) D(ata) G(roup) Values}, Eur. Phys. J. C {\bf 77}, no. 10, 656 (2017) [arXiv:1707.02818[hep-th]]. 
%%%%%%%%%
\bibitem{meson-N=2} M.~Kruczenski, D.~Mateos, R.~C.~Myers, D.~J.~Winters, {\it Meson Spectroscopy in AdS/CFT with Flavour}, JHEP {\bf 07} (2003) 049 [arXiv:hep-th/0304032].
%%%%%%
\bibitem{A-N=2} R.~C.~Myers and A.~Sinha, {\it The fast life of holographic mesons}, JHEP {\bf 06} (2008) 052 [arXiv:0804.2168[hep-th]].
%%%%%%%%%%%%
\bibitem{Nunez-Mesons} G.~Itsios, C.~Nunez and D.~Zoakos, {\it Mesons from (non) Abelian T-dual backgrounds}, JHEP {\bf 01} (2017) 011 [arXiv:1611.03490[hep-th]].
 %%%%%%%%%%%%%%%%%%%%%%%%%%
\bibitem{HLS-Physics-Reports} M.~Harada and K.~Yamawaki, {\it Hidden local symmetry at loop: A New perspective of composite gauge bosons and chiral phase transition}, Phys. Rept. {\bf 381} (2003) 1 [arXiv:hep-ph/0302103].
%%%%%%%%%%%%%%%%%%%%%%%%%%%%%%%%%%%%%
\bibitem{Relationship-Li-yi-zi} M.~Tanabashi, {\it Chiral perturbation to one loop including the rho meson},
Phys. Lett. B \textbf{316}, 534-541 (1993) [arXiv:hep-ph/9306237 [hep-ph]].
 %%%%%%%%%%%36
\bibitem{Wilson-matching} M.~Harada and K.~Yamawaki, {\it Wilsonian matching of effective field theory with underlying QCD}, Phys. Rev. D \textbf{64}, 014023 (2001) [arXiv:hep-ph/0009163 [hep-ph]].
%%%%%%%%%%%%%
\bibitem{SGS} J.~Solana, S.~Grozdanov and A.~Starinets,  {\it Transport peak in thermal spectral function of $\cal N$ = 4 supersymmetric Yang-Mills plasma at intermediate coupling},
Phys. Rev. Lett. 121, 191603 (2018) [arXiv:hep-th/1806.10997].
%%%%%%%%%%%%%%%%%%%%
\bibitem{IC} J.~O.~Andersen, L.~E.~Leganger, M.~Strickland and N.~Su,
{\it NNLO hard-thermal-loop thermodynamics for QCD}, Phys. Lett. B \textbf{696}, 468-472 (2011) [arXiv:1009.4644 [hep-ph]].
%%%%%%%%%%%%%%
\bibitem{Wald-Entropy}R. M.~Wald, {\it Black Hole Entropy is Noether Charge}, Phys. Rev. D {\bf 48}, 3427 (1993) [arXiv:gr-qc/9307038].
%%%%%%%%%%%%%%%%%%%
\bibitem{Wald-Entropy-2}V.~Iyer and R. M.~Wald, {\it Some Propeties of Noether Charge and a Proposal for Dynamical Black Hole Entropy}, Phys. Rev. D {\bf 50}, 846 (1994) [arXiv:gr-qc/9403028].
%%%%%%%%%%%%%%%%%%%%%%%%
\bibitem{UV_IR} N.~Craig and S.~Koren, {\it IR Dynamics from UV Divergences:UV/IR Mixing, NCFT, and the Hierarchy Problem} JHEP 03(2020) 037 [arXiv:hep-th/1909.01365].
%%%%%%%%%%%%%%%%%%%
\bibitem{Susskind+Witten} L.~Susskind and E.~Witten, {\it The Holography Bound in Anti-de Sitter Space} [arXiv:hep-th/9805114].
%%%%%%%%%%%
\bibitem{Tc-EE}I.~R.~Klebanov, D.~Kutasov and A.~Murugan, {\it Entanglement as a probe of confinement},
Nucl.\ Phys.\ B {\bf 796}, 274 (2008) [arXiv:0709.2140 [hep-th]].
%%%%%%%%%%%%%%%%%%%%%%%%%%%%%%%%%%%%%%%%%%%%%%%
\bibitem{RT} S.~Ryu, T.~Takayanagi, {\it Holographic Derivation of Entanglement Entropy from AdS/CFT}, Phys. Rev. Lett. {\bf 96}, 181602 (2006) [arXiv:hep-th/0603001].
%%%%%%%%%%12
\bibitem{omega-simulation} Y.~Jiang, Z.W.~Lin, and J.~Liao, {\it Rotating quark-gluon plasma in relativistic heavy ion collisions}, Phys. Rev. C {\bf 94} (2016) 044910 [arXiv:1602.06580[hep-ph]].
%%%%%%%%%%%13
\bibitem{R-Tc} Nelson R.F.~Bragga, Luiz F.~Faulhaber, Octavio C.~Junqueira, {\it Confinement/Deconfinement temperature for a rotating quark-gluon plasma}, [arXiV:2201.05581[hep-th]].
%%%%%%%%%14
\bibitem{G-D-Rotation} X.~Chen, L.~Zhang, D.~Li, D.~Hou, and M.~Huang, {\it Gluodynamics and deconfinement phase transition under rotation from holography}, JHEP {\bf 07} (2021) 132 [arXiv:2010.14478[hep-ph]].
%%%%%%%%%%%
\bibitem{Kerr-AdS-1}I.Y.~Arefeva, A.A.~Golubtsova, and E.~Gourgoulhon, {\it  Holographic drag force in 5d Kerr-AdS black hole}, JHEP {\bf 04} (2021)169 [arXiv:2004.12984[hep-th]].
\bibitem{Kerr-AdS-2} A.A.~Golubtsova, E.~Gourgoulhon, and M.K.~Usova, {\it Heavy quarks in rotating plasma via holography} [arXiv:2107.11672[hep-th]]. 
%%%%%%%%%%%%%%%%%%%%%%%%%%%
\bibitem{VA-Glueball-decay}V.~Yadav and A.~Misra, {\it M-Theory Exotic Scalar Glueball Decays to Mesons at Finite Coupling}, JHEP {\bf 09}, 133 (2018) [arXiv:1808.01182 [hep-th]].
%%%%%%%%%%%%%%%%%%%%%%%%%%%%%%
 %
%%%%%%%%%%%%%%%%%%%%%
\bibitem{Glueball-Roorkee} K.~Sil, V.~Yadav and A.~Misra, {\it Top-down holographic G-structure glueball spectroscopy at (N)LO in $N$ and finite coupling}, Eur. Phys. J. C {\bf 77}, no.6, 381 (2017) [arXiv:hep-th/1703.01306].
%%%%%%%%%%%%%%%%%%%%%
\bibitem{wittengross} D.~J.~Gross and E.~Witten, {\it Superstring modifications of Einstein's equations}, Nucl. Phys. B {\bf 277} (1986) 1.
%%%%%%%%%%21
\bibitem{Lorentz-boost-1} M.~Bravo~Gaete, L.~Guajardo, and M.~Hassaine, {\it A Cardy-like for rotating black holes with planar horizon}, JHEP {\bf 04} (2017) 092 [arXiv:1702.02416[hep-th]].
%%%%%%%%%%22
\bibitem{Lorentz-boost-2} C.~Erices~Gaete, and C.~Martinez, {\it Rotating hairy black holes in arbitrary dimensions}, Phys. Rev. D {\bf 97} (2018) 024034 [arXiv:1707.03483[hep-th]].
%%%%%%%%%%%%%%%%%%%%%%%%10
\bibitem{Tc-HW_SW} S.S.~Afonin, {\it A holographic relation between the deconfinement temperature and gluon condensate}, Phys. Lett. B {\bf 809} (2020) 135780 [arXiv:2005.01550[hep-th]].
%%%%%%%%%%%
\bibitem{Tc-L} S.~Borsanyi et al., JHEP {\bf 09}, 073 (2010); S.~Borsanyi et al., Phys. Lett. B 730, 99-104 (2014);  A.~Bazavov et al., Phys. Rev. D 90, 094503 (2014).
%%%%%%%%%%%%%%%%%%%
\bibitem{R-Tc3}Y.~Jiang and J.~Liao, {\it Pairing Phase Transitions of Matter under Rotation}, Phys. Rev. Lett. 117, 192302 (2016) [arXiv:1606.03808[hep-ph]].
%%%%%%%%%%%%%%%
\bibitem{Lattice-1} V.~V.~Braguta, A.~Y.~Kotov, D.~D.~Kuznedelev, and A.~A.~Roenko, Phys. Rev.
D 103, 094515 (2021).
%%%%%%%%%%%%%
\bibitem{Lattice-2} V.~V.~Braguta, A.~Y.~Kotov, D.~D.~Kuznedelev, and A.~A.~Roenko, in Proceedings of the 38th International Symposium on Lattice Field Theory
(2021).
%%%%%%%%%11
\bibitem{RHIC-omega} L.~Adamczyk et al. {\it (STAR Collaboration) Global $\Lambda$ hyperon polarization in nuclear collisions: Evidence for the most vortical fluid}, Nature (London) {\bf 548}, 62 (2017). 
%%%%%%%%%%%%%%%%%%%%%%%%%%%%%%%%%%%%%%%%%%%%%%%%%
\bibitem{Op6} H.~W.~Fearing and S.~Scherer, {\it Extension of the chiral perturbation theory meson Lagrangian to order p(6)}, Phys. Rev. D \textbf{53}, 315-348 (1996) [arXiv:hep-ph/9408346 [hep-ph]].
%%%%%%%%%%%
\bibitem{HRT} V.~E.~Hubeny, M.~Rangamani and T.~Takayanagi, {\it A Covariant Holographic Entanglement Entropy Proposal}, JHEP {\bf 07} (2007) 062 [arXiv:0705.0016[hep-th]].
%%%%%%%%%%%%%%%%%%
\bibitem{EW} N~Engelhardt and A.C.~Wall, {\it Quantum Extremal Surfaces: Holographic Entanglement Entropy beyond the Classical Regime}, JHEP {\bf 01} (2015) 073 [arXiv:1408.3203[hep-th]].
%%%%%%%%%
\bibitem{Dong} X.~Dong, {\it Holographic Entanglement Entropy for General Higher Derivative Gravity}, JHEP {\bf 01} (2014) 044 [arXiv:1310.5713[hep-th]].
%%%%%%%%%%%
\bibitem{A-HEE} A.~Bhattacharyya and A.~Sinha, {\it Entanglement entropy from the holographic stress tensor}, Class. Quantum Grav. {\bf 30} (2013) 235032 [arXiv:1303.1884[hep-th]].
%%%%%%%
\bibitem{A-HEE-1} A.~Bhattacharyya and A.~Sinha, {\it Entanglement entropy from surface terms in general relativity}, IJMPD 22, 12 (2013) 1342020 [arXiv:1305.3448[gr-qc]].
%%%%%%%%%%%%%%%
\bibitem{Hawking} S.W.~Hawking, {\it Particle Creation by Black Holes}, Commun. Math. Phys. 43, 199 (1975) Erratum: [Commun. Math. Phys. 46, 206 (1976)].
%%%%%
\bibitem{Hawking1} S.W.~Hawking,{\it Breakdown of Predictability in Gravitational Collapse}, Phys. Rev. D {\bf 14} (1976) 2460-2473.
%%%%%%%%%%
\bibitem{Page} D.N.~Page, {\it Information in Black Hole Radiation}, Phys. Rev. Lett. {\bf 71} (1993) 3743-3746 [arXiv:hep-th/9306083 [hep-th]].
%%%%%%%%%%%%%%%%%%%%%
\bibitem{rw-1} G.~Penington, S.~H.~Shenker, D.~Stanford and Z.~Yang, {\it Replica wormholes and the black hole interior}, [arXiv:1911.11977[hep-th]].
%%%%%%%%%%%
\bibitem{rw-2}A.~Almheiri, T.~Hartman, J.~Maldacena, E.~Shaghoulian and A.~Tajdini, {\it Replica Wormholes and the Entropy of Hawking Radiation}, JHEP {\bf 05} (2020) 013 [arXiv:1911.12333[hep-th]].
%%%%%%%%%%%
\bibitem{Swansea} T.~J. Hollowood and S.P. Kumar, {\it Islands and Page curves for evaporating black holes in JT gravity}, JHEP {\bf 08} (2020) 094 [arXiv:2004.14944[hep-th]].
%%%%
\bibitem{Swansea1} T.~J. Hollowood, S.~P.~Kumar and A.~Legramandi, {\it Hawking radiation correlations of evaporating black holes in JT gravity}, J.Phys.A 53 (2020) 47, 475401 [arXiv:2007.04877[hep-th]].
%%%%%%
\bibitem{Swansea5} Z. Gyongyosi, T.~J. Hollowood, S.~P. Kumar, A. Legramandi and N.~Talwar, {\it Black Hole Information Recovery in JT Gravity}, JHEP {\bf 03} (2023) 139 [arXiv:2209.11774 [hep-th]].
%%%%%%%
\bibitem{Swansea2} T.~J. Hollowood, S.~P.~Kumar, A.~Legramandi and N.~Talwar, {\it Islands in the stream of Hawking radiation}, JHEP {\bf 11} (2021) 067 [arXiv:2104.00052 [hep-th]].
%%%%
\bibitem{Swansea3} T.~J.~Hollowood, S.~P.~Kumar, A.~Legramandi and N.~Talwar, {\it Ephemeral islands, plunging quantum extremal surfaces and BCFT channels}, JHEP {\bf 01} (2022) 078 [arXiv:2109.01895[hep-th]].
%%%%%%
\bibitem{Swansea4} T.~J. Hollowood, S.~P.~Kumar, A.~Legramandi and N.~Talwar, {\it Grey-body factors, irreversibility and multiple island saddles}, JHEP {\bf 03} (2022) 110 [arXiv:2111.02248 [hep-th]].
%%%%%
\bibitem{NBH-HD} M.~Alishahiha, A.F.~Astaneh and A.~Naseh, {\it Island in the presence of higher derivative terms}, JHEP {\bf 02} (2021) 035 [arXiv:2005.08715[hep-th]]. 
%%%%%%%%%%%%%%%%%
\bibitem{KR1} A.~Karch and L.~Randall, {\it Locally localized gravity}, JHEP {\bf 05} (2001) 008 [arXiv:hep-th/0011156].
%%%%%%%%%%%%%
\bibitem{KR2} A.~Karch and L.~Randall, {\it Open and closed string interpretation of SUSY CFT's on branes with boundaries}, JHEP {\bf 06} (2001) 063 [arXiv:hep-th/0105132].
%%%%%%%%%%%%%
\bibitem{Hartman-Maldacena} T.~Hartman and J.~Maldacena, {\it Time Evolution of Entanglement Entropy from Black Hole Interiors}, JHEP {\bf 05} (2013) 014 [arXiv:1303.1080[hep-th]].
%%%%%%%%%%%%
\bibitem{B-N-P-1} A.~Bhattacharya, A.~Bhattacharyya, P.~Nandy and A.K.~Patra, {\it Islands and complexity of eternal black hole and radiation subsystems for a doubly holographic model}, JHEP {\bf 05} (2021) 135 [arXiv:2103.15852[hep-th]].
%%%%%%%%%%%%%%
\bibitem{B-N-P-2} A.~Bhattacharya, A.~Bhattacharyya, P.~Nandy and A.K.~Patra, {\it Bath deformations, islands and holographic complexity}, Phys. Rev. D {\bf 105} (2022) 066019 [arXiv:2112.06967[hep-th]].
%%%%%%%%%%%%%
\bibitem{NGB} Q.L.~Hu, D.~Li, R.X.~Miao and Y.Q.~Zeng, {\it AdS/BCFT and Island for curvature-squared gravity}, J. High Energ. Phys. 2022, 37 (2022) [arXiv:2202.03304[hep-th]].
%%%%%%%%%%
\bibitem{Island-IIB-1}  C.F.~Uhlemann, {\it Islands and Page curves in 4d from Type IIB}, JHEP {\bf 08} (2021) 104 [arXiv:2105.00008[hep-th]].
%%%%%%%%%%%%
\bibitem{Island-IIB-2} S.~Demulder, A.~Gnecchi, I.~Lavdas and D.~Lust, {\it Island and Light Gravitons in type IIB String Theory}, J. High Energ. Phys. 2023, 16 (2023) [arXiv:2204.03669[hep-th]].
%%%%%%%%
\bibitem{Island-IIB-3} A.~Karch, H.~Sun and C.F.~Uhlemann, {\it Double holography in string theory}, J. High Energ. Phys. 2022, 12 (2022) [arXiv:2206.11292[hep-th]].
%%%%
%%%%%%%%%
\bibitem{Ling+Liu+Xian} Y.~Ling, Y.~Liu and Z.~Y.~Xian, {\it Island in Charged Black Holes}, JHEP {\bf 03} (2021) 251 [arXiv:2010.00037[hep-th]].
%%%%%%
\bibitem{Omiya+Wei} H.~Omiya and Z.~Wei, {\it Causal Structures and Nonlocality in Double Holography},  J. High Energ. Phys. 2022, 128 (2022) [arXiv:2107.01219[hep-th]].
%%%%%%%%%%%
\bibitem{Phase-BCFT} H.~Geng, A.~Karch, C.P.~Pardavila, S.~Raju and L.~Randall, {\it Entanglement Phase Structure of a Holographic BCFT in a Black Hole Background}, J. High Energ. Phys. 2022, 153 (2022) [arXiv:2112.09132[hep-th]].
%%%%%%%%%%%
\bibitem{critical-islands} C.~Krishnan, {\it Critical Islands}, JHEP {\bf 01} (2021) 179 [arXiv:2007.06551[hep-th]].
%%%%%%%%%%%%
\bibitem{IITK1} M.~Afrasiar, J.K.~Basak, A.~Chandra and G.~Sengupta, {\it Islands for Entanglement Negativity in Communicating Black Holes} [arXiv:2205.07903[hep-th]].
%%%%%%%%
\bibitem{IITK2} D.~Basu, H.~Parihar, V.~Raj and G.~Sengupta, {\it Defect extremal surfaces for entanglement negativity} [arXiv:2205.07905[hep-th]].
%%%%%%%%%%%%%%%%%
\bibitem{Liu et al} Y.~Liu, Z.Y.~Xian, C.~Peng and Y.~Ling, {\it Black holes Entangled by Radiation}, J. High Energ. Phys. 2022, 179 (2022) [arXiv:2205.14596[hep-th]].
%%%%%%%%%%%%%%%
\bibitem{Li+Yang} Z.~Li and R.Q.~Yang, {\it Upper bounds of holographic entanglement entropy growth rate for thermofield double states}, JHEP {\bf 10} (2022) 072 [arXiv:2205.15154[hep-th]].
%%%%%%%%%%%%%%%%%%%%%%%%%%
\bibitem{Deng+An+Zhou} F.~Deng, Y.S.~An and Y.~Zhou, {\it JT Gravity from Partial Reduction and Defect Extremal Surface}, JHEP {\bf 02} (2023) 219 [arXiv:2206.09609[hep-th]].
%%%%%%%%%
\bibitem{Geng+Nomura+Sun} H.~Geng, Y.~Nomura and H.~Y.~Sun, {\it Information paradox and its resolution in de Sitter holography}, Phys.Rev.D 103 (2021) 12, 126004 [arXiv:2103.07477[hep-th]].
%%%%%%%%%%%%
%%%%%%%%%
\bibitem{JIITK} M.~Afrasiar, J.~K.~Basak, A.~Chandra and G.~Sengupta, {\it Reflected Entropy for Communicating Black Holes I: Karch-Randall Braneworlds}, J. High Energ. Phys. 2023, 203 (2023) [arXiv:2211.13246[hep-th]].
%%%%%%%%%%%%%%%%%%%%%%%%%%%%
\bibitem{GB-3} H.~Geng, A.~Karch, C.~P.~Pardavila, S.~Raju and L.~Randall, {\it Information transfer with a gravitating bath}, SciPost Phys. 10 (2021) 5, 103 [arXiv:2012.04671[hep-th]].
%%%%%%%%
\bibitem{massive-gravity} O.~Aharony, O.~DeWolfe, D.Z.~Freedman and A.~Karch, {\it Defect conformal field theory and
locally localized gravity}, JHEP {\bf 07} (2003) 030 [hep-th/0303249].
%%%%%%%
\bibitem{GB-2} H.~Geng and A.~Karch, {\it Massive islands}, JHEP {\bf 09} (2020) 121 [arXiv:2006.02438[hep-th]].
%%%%%%%
\bibitem{GB-4} H.~Geng, A.~Karch, C.~P.~Pardavila, S.~Raju and L.~Randall, {\it Inconsistency of islands in theories with long-range gravity}, JHEP 01 (2022) 182 [arXiv:2107.03390[hep-th]].
%%%%%%%%%%%%%%
\bibitem{Massless-Gravity} R.~X.~Miao, {\it Massless Entanglement Island in Wedge Holography}, [arXiv:2212.07645[hep-th]].
%%%%%%%%%%%%%
\bibitem{C1} K.~Ghosh and C.~Krishnan, {\it Dirichlet baths and the not-so-fine-grained Page curve}, JHEP 08 (2021) 119 [arXiv:2103.17253[hep-th]].
%%%%%%%%%%%%%%%
 \bibitem{DGP-2} G.~R.~Dvali, G.~Gabadadze and M.~Porrati, {\it 4D gravity on a brane in 5D Minkowski space}, Phys. Lett. B 485 (2000) 208 [arXiv:hep-th/0005016].
%%%%%%%%%%1
\bibitem{Wedge-JT} H.~Geng, A.~Karch, C.~P.-Pardavila, S.~Raju, L.~Randall, M.~Riojas and S.~Shashi {\it Jackiw-Teitelboim Gravity from the Karch-Randall Braneworld}, Phys. Rev. Lett. {\bf 129}, 231601 [arXiv:2206.04695 [hep-th]].
%%%%%%%%%%%%
\bibitem{Massless-Gravity-1} R.X.~Miao, {\it Entanglement Island and Page Curve in Wedge Holography}, JHEP {\bf 03} (2023) 214 [arXiv:2301.06285[hep-th]].
%%%%%%%%%
\bibitem{Massless-Gravity-2} D.Q.~Li and R.X.~Miao, {\it Massless Entanglement Islands in Cone Holography}, arXiv:2303.10958[hep-th].
%%%%%%%%
\bibitem{Island-RNBH} X.~Wang, R.~Li and J.~Wang, {\it Islands and Page curves of Reissner-Nordstr{\"o}m black holes}, JHEP {\bf 04} (2021) 103, [arXiv:2101.06867[hep-th]].
%%%%%%  
\bibitem{Yu-Ge} M.H.~Yu and X.H.~Ge, {\it Page Curves and Islands in Charged Dilaton Black Holes} Eur.Phys.J.C 82 (2022) 2, 167 [arXiv:2107.03031[hep-th]].
%%%%%%
\bibitem{Island-SB} K.~Hoshimoto, N.~Iizuka and N.Matsuo, {\it Islands in Schwarzschild black holes}, JHEP {\bf 06} (2020)085 [arXiv:2004.05863[hep-th]].
%%%%
\bibitem{Omidi} F.~Omidi, {\it Entropy of Hawking Radiation for Two-Sided Hyperscaling Violating Black Branes}, JHEP {\bf 04} (2022) 022 [arXiv:2112.05890[hep-th]].
%%%%%%
\bibitem{CLDBH} B.~Ahn, S.~E.~Bak, H.~S.~Jeong, K.~Y.~Kim and Y.~W.~Sun, {\it Islands in charged linear dilaton black holes}, Phys. Rev. D {\bf 105}, no.4, 046012 (2022) [arXiv:2107.07444 [hep-th]].
%%%%%%%
\bibitem{Islands-KdS} S.~Azarnia and R.~Fareghbal, {\it Islands in Kerr-de Sitter spacetime and their flat limit}, Phys. Rev. D 106, 026012 (2022) [arXiv:2204.08488[hep-th]].
%%%%%%%%
\bibitem{Tian} J.~Tian, {\it Islands in Generalized Dilaton Theories}, [arXiv:2204.08751[hep-th]].
%%%%%%%%
\bibitem{Zhang-Li-Guo} C.Y.~Zhang, S.J.~Zhang, P.C.~Li, and M.~Guo, {\it Superradiance and stability of the regularized 4D charged Einstein-Gauss-Bonnet black hole}, JHEP {\bf 08} (2020) 105 [arXiv:2004.03141[gr-qc]].
%%%%%%%%%%%
\bibitem{BBB} M.~Bousder, K.~El~Bourakadi and M.~Bennai, {\it Charged 4D Einstein-Gauss-Bonnet Black Hole: Vacuum Solutions, Cauchy Horizon, Thermodynamics} [arXiv:2107.00463[gr-qc]].
%%%%%%%
\bibitem{ZZZY} M.~Zhang, C.M.~Zhang, D.C.~Zou and R.H.~Yue, {\it Phase transition and Quasinormal modes for Charged black holes in 4D Einstein-Gauss-Bonnet gravity}, Chinese Physics C Vol. 45, No. 4 (2021) 045105 [arXiv:2009.03096[hep-th]].
%%%%%%%%%%%
\bibitem{CGLY} D.~Chen, C.~Gao, X.~Liu and C.~Yu, {\it The correspondence between shadow and test field in a four-dimensional charged Einstein-Gauss-Bonnet black hole}, Eur. Phys. J. C {\bf 81} (2021)700 700 [arXiv:2103.03624[gr-qc]].
%%%%%%%%%%
\bibitem{BB} M.~Bousder and M.~Bennai, {\it Particle-antiparticle in 4D charged Einstein-Gauss-Bonnet black hole}, Physics Letters B 817C (2021) 136343 [arXiv:2105.05038[gr-qc]].
%%%%%%%%%
\bibitem{LNZ} P.~Liu, C.~Niu and C.Y.~Zhang, {\it Instability of regularized 4D charged Einstein-Gauss-Bonnet de-Sitter black holes}, Chin.Phys.C 45 (2021) 2, 025104 [arXiv:2004.10620[gr-qc]].
%%%%%%%%%%%%
\bibitem{Charged-GB-BH-App} F.~Atamurotov, S.~Shaymatov, P.~Sheoran and S.~Siwach, {\it Charged black hole in $4D$ Einstein-Gauss Bonnet Gravity: Particle motion, plasma effect on weak gravitational lensing and centre-of-mass energy}, JCAP 
{\bf 08} (2021) 045 [arXiv:2105.02214[hep-th]].
%%%%%%%%%
\bibitem{EGB-Review} Pedro~G.S.~Fernandes, P.~Carrilho, T.~Clifton and David~J.~Mulryne, {\it The 4D Einstein-Gauss-Bonnet Theory of Gravity: A Review}, Class. Quantum Grav. 39 (2022) 063001 [arXiv:2202.13908[gr-qc]].
%%%%%%%%%%%
\bibitem{Kim+Nam} W.~Kim and M.~Nam, {\it Entanglement entropy of asymptotically flat non-extremal and extremal black holes with an island} Eur. Phys. J. C {\bf 81}, 869 (2021) [arXiv:2103.16163[hep-th]].
%%%%%%%%%%%%%
\bibitem{Yu et al} M.H.~Yu, C.Y.~Lu, X.H.~Ge and S.J.~Sin ,{\it Island, Page Curve and Superradiance of Rotating BTZ Black Holes} Phys. Rev. D {\bf 105} (2022) 6, 066009 [arXiv:2112.14361[hep-th]].
%%%%%%%%%%%%
\bibitem{Charged-GB-BH} Pedro~G.S.~Fernandes, {\it Charged Black Holes in AdS Spaces in 4D Einstein Gauss-Bonnet Gravity}, Phys. Lett. B {\bf 805} (2020) 135468 [arXiv:2003.05491[hep-th]].
%%%%%%%
\bibitem{Glavan+Lin} D.~Glavan and C.~Lin, {\it Einstein-Gauss-Bonnet Gravity in Four-Dimensional Spacetime}, Phys.
Rev. Lett. 124 (2020) 081301 [arXiv:1905.03601[gr-qc]].
%%%%%
\bibitem{AGM} K.~Aoki, M.~A.~Gorji and S.~Mukohyama, {\it A consistent theory of $D  \rightarrow 4$ Einstein-Gauss-Bonnet
gravity}, Phys. Lett. B 810 (2020) 135843 [arXiv:2005.03859[gr-qc]].
%%%%%%
\bibitem{CC} P.~Calabrese and J.~Cardy, { \it Entanglement entropy and quantum field theory: a non-technical introduction}, Int.J.Quant.Inf. 4 (2006) 429 [arXiv:quant-ph/0505193].
%%%%%%%%%%
\bibitem{CC-1} P.~Calabrese, J.~Cardy and E.~Tonny, { \it Entanglement entropy of two disjoint intervals in conformal field theory}, J. Stat. Mech P11001, 2009 [arXiv:0905.2069[hep-th]].
%%%%%%%%%%%%%%
\bibitem{scrambling-time-1} P.~Hayden and J.~Preskill, {\it Black holes as mirrors: quantum information in random subsystems} JHEP, 0709:120,2007 [arXiv:0708.4025[hep-th]].
%%%%%%%%%%%%%%
\bibitem{scrambling-time-2} Y.~Sekino and L.~Susskind, {\it Fast Scramblers}, JHEP 0810:065,2008 [arXiv:0808.2096[hep-th]].
%%%%%%%%%%%%
\bibitem{scrambling-time-EW} G.~Penington, {\it Entanglement Wedge Reconstruction and the Information Paradox}, [arXiv:1905.08255[hep-th]].
%%%%%%%%%%%%%%%%%%%%%
\bibitem{island-o-h} A.~Almheiri, R.~Mahajan and J.~Maldacena, {\it Islands outside the horizon}, [arXiv:1910.11077[hep-th]].
%%%%%%%%%%%%%%%%%%%
\bibitem{QFC} R.~Bousso, Z.~Fisher, S.~Leichenauer and Aron C.~Wall, {\it A Quantum Focussing Conjecture}, Phys. Rev. D 93, 064044 (2016) [arXiv:1506.02669[hep-th]].
%%%%%%%%%%%%%%%%
\bibitem{island-coupled} A.~Almheiri, A.~Milekhin and B.~Swingle, {\it Universal Constraints on Energy Flow and SYK Thermalization}, [arXiv:1912.04912[hep-th]].
%%%%%%%%%%
\bibitem{BSS} A.~Bhattacharyya, M.~Sharma and A.~Sinha, {\it On generalised gravitational entropy, squashed cones and holography}, JHEP 1401 (2014) 021 [arXiv:1308.5748[hep-th]].
%%%%%%%%%
\bibitem{BS} A.~Bhattacharyya and M.~Sharma, {\it On entanglement entropy functionals in higher-derivative gravity theories}, JHEP {\bf 10} (2014) 130 [arXiv:1405.3511[hep-th]].
%%%%%%%%%%%%15
\bibitem{Island-HD} A. Almheiri, R. Mahajan and J.E. Santos, {\it Entanglement islands in higher dimensions}, SciPost Phys. {\bf 9} (2020) 001[arXiv:1911.09666[hep-th]].
%%%%%%%%%%%%%%
\bibitem{Bath-WCFT} E.~Caceres, A.~Kundu, Ayan K.~Patra and S. Shashi, {\it Warped information and entanglement islands in AdS/WCFT}, JHEP {\bf 07} (2021) 004 [arXiv:2012.05425 [hep-th]].
%%%%%%%%%%%%%%%%%%%%%%%%%%%%%%%%%%%%%%%%%%%%%%%
\bibitem{Takayanagi-ETW}T.~Takayanagi, Holographic Dual of BCFT, Phys. Rev. Lett. \textbf{107}, 101602 (2011) [arXiv:1105.5165 [hep-th]].
%%%%%%%%%%%%%%%%%%%%%%%%%%%%%%%%%%
\bibitem{LVS} V.~Balasubramanian, P.~Berglund, J.~P.~Conlon and F.~Quevedo, {\it Systematics of moduli stabilisation in Calabi-Yau flux compactifications}, JHEP \textbf{03}, 007 (2005)
[arXiv:hep-th/0502058].
%%%%%%%%%%%%%%%%%%%%%%%%%%%%%%%%%%%%%
\bibitem{Swiss-Cheese-LVS} J.~P.~Conlon, F.~Quevedo and K.~Suruliz,
{\it Large-volume flux compactifications: Moduli spectrum and D3/D7 soft supersymmetry breaking}, JHEP \textbf{08}, 007 (2005)
[arXiv:hep-th/0505076].
%%%%%%%%%%%%%%%%%%%%%%%%%%%%%%
\bibitem{SEEISoverSBHapprox2}A.~Almheiri, T.~Hartman, J.~Maldacena, E.~Shaghoulian and A.~Tajdini, {\it The entropy of Hawking radiation}, Rev. Mod. Phys. \textbf{93}, no.3, 035002 (2021) [arXiv:2006.06872 [hep-th]]; K.~Hashimoto, N.~Iizuka and Y.~Matsuo, {\it Islands in Schwarzschild black holes}, JHEP \textbf{06}, 085 (2020) [arXiv:2004.05863 [hep-th]]; X.~Wang, R.~Li and J.~Wang, {\it Islands and Page curves of Reissner-Nordstr\"om black holes}, JHEP \textbf{04}, 103 (2021)
[arXiv:2101.06867 [hep-th]]. 
%%%%%%%%%%%%%%%%%%%%%%%%%%%%%%%%%
\bibitem{Klebanov+Buchel-et-al_rh-const-of-integration} 
A.~Buchel, C.~P.~Herzog, I.~R.~Klebanov, L.~A.~Pando Zayas and A.~A.~Tseytlin,
{\it Nonextremal gravity duals for fractional D-3 branes on the conifold}, JHEP \textbf{04}, 033 (2001)
[arXiv:hep-th/0102105].
%%%%%%%%%%%%%%%%%%%%%%%%%%%%
\bibitem{C. Bachas and J. Estes [2011]} C.~Bachas and J.~Estes,
{\it Spin-2 spectrum of defect theories}, JHEP \textbf{06}, 005 (2011)[arXiv:1103.2800 [hep-th]].
\bibitem{Bousso:1997wi} 
  R.~Bousso and S.~W.~Hawking,
  {\it (Anti)evaporation of Schwarzschild de-Sitter black holes},
  Phys.\ Rev.\ D {\bf 57}, 2436 (1998)
  [arXiv:hep-th/9709224].
  %%%%%%%%%%%%%%%%%%%%%%%%%%%%%%%%%%%%%
   \bibitem{Chao:1997em} 
  W.~Z.~Chao,
  {\it Quantum fields in Schwarzschild de-Sitter space},
  Int.\ J.\ Mod.\ Phys.\ D {\bf 07}, 887 (1998)
  [arXiv:gr-qc/9712066].
  %%%%%%%%%%%%%%%%%%%%%%%%%%%%%%%%%%%%%
  \bibitem{Bousso:1999ms} 
  R.~Bousso,
  {\it Quantum global structure of de Sitter space},
  Phys.\ Rev.\ D {\bf 60}, 063503 (1999)
  [arXiv:hep-th/9902183].
  %%%%%%%%%%%%%%%%%%%%%%%%%%%%%%%%%%%%% 
  \bibitem{Anninos:2010gh} 
  D.~Anninos and T.~Anous,
  {\it A de Sitter hoedown},
  JHEP {\bf 1008}, 131 (2010)
[arXiv:1002.1717 [hep-th]].
%%%%%%%%%%%%%%%%%%%%%%%%%%%%%%%%%%%%%
\bibitem{Lochan:2018pzs} 
  K.~Lochan, K.~Rajeev, A.~Vikram and T.~Padmanabhan,
  {\it Quantum correlators in Friedmann spacetimes -The omnipresent de Sitter and the invariant vacuum noise}, Phys. Rev. D {\bf 98}, 105015 (2018) [arXiv:1805.08800 [gr-qc]].
%%%%%%%%%%%%%%%%%%%%%%%%%%%%%%%%%%%%%
 \bibitem{Goheer:2002vf} 
  N.~Goheer, M.~Kleban and L.~Susskind,
  {\it The trouble with de Sitter space},
  JHEP {\bf 0307}, 056 (2003)
  [arXiv:hep-th/0212209].
  %%%%%%%%
  \bibitem{Marolf:2010tg} 
  D.~Marolf, M.~Rangamani and M.~Van Raamsdonk,
  {\it Holographic models of de Sitter QFTs},
  Class.\ Quant.\ Grav.\  {\bf 28}, 105015 (2011)
  [arXiv:1007.3996 [hep-th]].
  %%%%%%%%%%%%%%
\bibitem{S1} S.~Choudhury, S.~Panda, {\it Entangled de Sitter from Stringy Axionic Bell pair I: An analysis using Bunch Davies vacuum}, Eur.Phys.J. C78 (2018) no.1, 52 [arXiv:1708.02265[hep-th]].
%%%%%%%%%%%
\bibitem{S2} S.~Choudhury, S.~Panda, {\it Quantum entanglement in de Sitter space from Stringy Axion: An analysis using $\alpha$ vacua}, Nucl. Physic. B 943 (2019) 114606 [arXiv:1712.08299[hep-th]].
%%%%%%%%%%%
\bibitem{S3} S.~Choudhury et al, {\it Circuit Complexity From Cosmological Islands}, Symmetry 13 (2021) no. 7, 1301 [arXiv:2012.10234[hep-th]].
%%%%%%%%%%%
%%%%%%%%%%%  
 \bibitem{Gibbons} 
G.~W.~Gibbons and S.~W.~Hawking, {\it Cosmological event horizons, thermodynamics, and particle creation},
Phys.~Rev.~D{\bf15}, 2738 (1977)
%%%%%%%%%%%%%%%%%%%%%%%%%%%%%%%%%%%%%
\bibitem{JHT}
J.~H.~Traschen, {\it An Introduction to black hole evaporation}, [arXiv:gr-qc/0010055]
%%%%%%%%%%
\bibitem{SA} S.~Bhattacharya and A.~Lahiri, {\it Mass function and particle creation in Schwarzschild-de Sitter spacetime}, Eur. Phys. J. C 73, 2673 (2013) [arXiv:1301.4532[gr-qc]].
%%%%%%%%%%%
\bibitem{Bhattacharya:2018ltm}
S.~Bhattacharya,
{\it Particle creation by de Sitter black holes revisited},
Phys. Rev. D \textbf{98}, no.12, 125013 (2018)
[arXiv:1810.13260 [gr-qc]].
%%%%%%%%%%%%%%%%%%%%%%%%%%%%%%%%%%%%%%%%
\bibitem{Saida:2009ss}
H.~Saida,
{\it To what extent is the entropy-area law universal?: Multi-horizon and multi-temperature spacetime may break the entropy-area law},
Prog. Theor. Phys. \textbf{122}, 1515-1552 (2010) [arXiv:0910.2510 [gr-qc]].
%%%%%%%%%%%%%%%%%%%%%%%%%%%%%%%%%%%%%
\bibitem{Ma:2016arz}
M.~S.~Ma, R.~Zhao and Y.~Q.~Ma,
{\it Thermodynamic stability of black holes surrounded by quintessence},
Gen. Relativ. Grav. \textbf{49}, no.6, 79 (2017)
[arXiv:1606.06070 [gr-qc]].
%%%%%%%%%%%%%%%%%%%%%%%%%%%%%%%%%%%%%
\bibitem{Sekiwa:2006qj}
Y.~Sekiwa,
{\it Thermodynamics of de Sitter black holes: Thermal cosmological constant},
Phys. Rev. D \textbf{73}, 084009 (2006)
[arXiv:hep-th/0602269].
%%%%%%%%%%%%%%%%%%%%%%%%%%%%%%%%%%%%%
\bibitem{Gomberoff:2003ea}
A.~Gomberoff and C.~Teitelboim,
{\it de Sitter black holes with either of the two horizons as a boundary},
Phys. Rev. D \textbf{67}, 104024 (2003) [arXiv:hep-th/0302204].
%%%%%%%%%%%%%%%%%%%%%%%%%%%%%%%%%%%%%
\bibitem{Nitin} S.~Bhattacharya and N.~Joshi, {\it Entanglement degradation in multi-event horizon spacetimes}, Phys. Rev. D 105, 065007 [arXiv:2105.02026[hep-th]].
%%%%%%%%%%%
\bibitem{Sybesma} W.~Sybesma, {\it Pure de Sitter space and the island moving back in time}, Class. Quantum Grav. 38 (2021) 145012 [arXiv:2008.07994[hep-th]].
%%%%%%%%
\bibitem{anchor-curve} F.~F.~Gautason, L.~Schneiderbauer, W.~Sybesmab and L.~Thorlaciusb, {\it Page curve for an evaporating black hole}, JHEP {\bf 05} (2020) 091 [arXiv:2004.00598[hep-th]].
%%%%%%%%%%%%%%%%%%%%%
\bibitem{pc}
P. C. W. Davies and T. M. Davis, {\it How far can the generalized second law be generalized}, Found. Phys.32, 1877 (2002)
[astro-ph/0310522]
%%%%%%%%%%%%%%%%%%%%%%%%%
\bibitem{urano}
M. Urano, A. Tomimatsu and H. Saida, {\it The mechanical first law of black hole spacetimes with cosmological constant and
its application to Schwarzschild de-Sitter spacetime}, Class. Quant. Grav.26, 105010 (2009) [arXiv:0903.4230[gr-qc]]
%%%%%%%%%%%%%%%%%%%%%%%
\bibitem{saida}
H. Saida, {\it de Sitter thermodynamics in the canonical ensemble}, Prog. Theor. Phys.122, 1239 (2010) [arXiv:0908.3041[gr-qc]]
%%%%%%%%%%%%%%%%%%%%
\bibitem{pappas}
T. Pappas, P. Kanti, {\it Schwarzschild de-Sitter spacetime: The role of temperature in the emission of Hawking radiation},
Phys. Lett. B {\bf 775}, 140 (2017) arXiv:1707.04900[hep-th].
%%%%%%%%%%
\bibitem{I-3} W.~C.~Gan,  D.~H.~Du and F.~W.~Shu, {\it Island and Page curve for one-sided asymptotically flat black hole}, JHEP {\bf 07} (2022) 020 [arXiv:2203.06310[hep-th]].
%%%%%%%%%%%%%{}
 \bibitem{Ankit} A.~Anand, {\it Page curve and Island in EGB gravity}, [arXiv:2205.13785[hep-th]].
%%%%%%%
\bibitem{WH-ii} P.~J.~Hu and R.~X.~Miao, {\it Effective action, spectrum and first law of wedge holography}, JHEP {\bf 03} (2022)145 [arXiv:2201.02014[hep-th]].
%%%%%
\bibitem{WH-i} N.~Ogawa, T.~Takayanagi, T.~Tsuda and T.~Waki, {\it Wedge Holography in Flat Space and Celestial Holography}, Phys. Rev. D {\bf 107} (2023), 026001 [arXiv:2207.06735[hep-th]].
%%%%%%%%%%%%
\bibitem{Geng} H.~Geng, {\it Aspects of $AdS_2$ quantum gravity and the Karch-Randall braneworld}, J. High Energy Phys. {\bf 09} (2022) 024 [arXiv:2206.11277[hep-th]].
%%%%%%
\bibitem{dS-CFT} A.~Strominger, {\it The dS/CFT correspondence}, JHEP {\bf 10} (2001) 034 [arXiv:hep-th/0106113].
%%%%%%%%
\bibitem{dS-CFT-1} J.M.~Maldacena, {\it Non-Gaussian features of primordial fluctuations in single field inflationary
models}, JHEP {\bf 05} (2003) 013 [astro-ph/0210603].
%%%%%%%%%%%%%%%
\bibitem{mismatched-branes} A.~Karch and L.~Randall, {\it Geometries with mismatched branes}, J. High Energ. Phys. 2020, 166 (2020) [arXiv:2006.10061[hep-th]].
%%%%%%%%%
\bibitem{Kostas} J.~M.~Penín, K.~Skenderis and B.~Withers, {\it Massive holographic QFTs in de Sitter}, SciPost Phys. 12, 182 (2022) [arXiv:2112.14639[hep-th]].
%%%%%
\bibitem{GP} O.~Y.~Kupervasser, {\it Grandfather Paradox in Non-Quantum and Quantum Gravitation Theories}, http://dx.doi.org/10.4236/ns.2014.611079.
%%%%%%%%%%%%%%
\bibitem{RE} S.~Dutta and T.~Faulkner, {\it A canonical purification for the entanglement wedge cross-section}, JHEP {\bf 03}, 178 (2021) [arXiv:1905.00577 [hep-th]].
%%%%%%%%%%
\bibitem{CV} D. Stanford and L. Susskind, {\it Complexity and Shock Wave Geometries}, Phys. Rev. D {\bf 90}, 126007 (2014) [arXiv:1406.2678[hep-th]].
%%%%%%%%%%%%%
\bibitem{CA} A.R. Brown et al, {\it Holographic Complexity Equals Bulk Action?}, Phys. Rev. Lett. {\bf 116}, 191301  (2016) [arXiv:1509.07876 [hep-th]];  A.R. Brown et al, {\it Complexity, action, and black holes}, Phys. Rev. D {\bf 93}, 086006 (2016) [arXiv:1512.04993[hep-th]].







\end{thebibliography}
%\end{spacing}

\end{document}